\documentclass[11pt,twoside,a4paper,notitlepage]{report}
\pdfoutput=1

\usepackage{definitions}
\usepackage{amsfonts, amscd, amsmath, amssymb}
\usepackage{color}
\usepackage{xcolor}
\usepackage{graphics}
\usepackage{graphicx}
\usepackage[affil-it]{authblk}
\usepackage[normalem]{ulem}
\usepackage{ulinedefs}
\usepackage{mathrsfs}
\usepackage{bbm}
\usepackage{fancyhdr}
\usepackage{slashed}
\usepackage{subfigure}
\usepackage{arydshln}
\usepackage{cancel}
\usepackage[T1]{fontenc}
\usepackage{titlesec, blindtext}
\definecolor{gray75}{gray}{0.75}
\newcommand{\hsp}{\hspace{20pt}}
\titleformat{\chapter}[hang]{\huge\bfseries}{\thechapter\hsp\textcolor{gray75}{$|$}\hsp}{0pt}{\huge\bfseries}
\usepackage[OT2,OT1]{fontenc}
\newcommand\cyr{%
\renewcommand\rmdefault{wncyr}%
\renewcommand\sfdefault{wncyss}%
\renewcommand\encodingdefault{OT2}%
\normalfont
\selectfont}
\DeclareTextFontCommand{\textcyr}{\cyr}
\usepackage[small,bf]{caption}

\fancypagestyle{plain}{%
\fancyhead{} % get rid of headers on plain pages
}

\usepackage{hyperref}
\hypersetup{
   colorlinks,
   citecolor=black,
   filecolor=black,
   linkcolor=black, %for printing
   urlcolor=black
}
\hypersetup{linktocpage}

\definecolor{NavyBlue}{cmyk}{0.94,0.54,0,0}
\definecolor{MyPurple2}{cmyk}{.75,1,0,0.1}

\def\NF{{N_f}}
\def\NA{{N_a}}

\title{\vspace{-60pt}\hfill\small{HU-EP-13/53}\\\hfill\small{HU-MATH-18-2013}\\ \hfill\small{ITP-UU-13/16}\\ \hfill\small{SPIN-13/19}\vspace{60pt}\\
\vspace{22pt}
\LARGE Integrability of the $\ads$ superstring and its deformations\\
\vspace{33pt}}
\author{Stijn J. van Tongeren\thanks{\tt svantongeren $|$ s.j.vantongeren @ physik.hu-berlin.de $|$ gmail.com}\hspace{5pt}\thanks{This article is based primarily on research performed by the author while at the Institute for Theoretical Physics and Spinoza Institute of Utrecht University.}}
\affil{
{\small Institut f\"ur Mathematik und Institut f\"ur Physik, Humboldt-Universit\"at zu Berlin}
\\
{\small IRIS Geb\"aude, Zum Grossen Windkanal 6, 12489 Berlin}\vspace{10pt}}
\date{21 October 2013\\
\vspace{11pt}}

\begin{document}

\begin{titlepage}
\maketitle
\begin{abstract}
\noindent This article reviews the application of integrability to the spectral problem of strings on $\ads$ and its deformations. We begin with a pedagogical introduction to integrable field theories culminating in the description of their finite-volume spectra through the thermodynamic Bethe ansatz. Next, we apply these ideas to the $\ads$ string and in later chapters discuss how to account for particular integrable deformations. Through the AdS/CFT correspondence this gives an exact description of anomalous scaling dimensions of single trace operators in planar $\mathcal{N}=4$ supersymmetry Yang-Mills theory, its `orbifolds', and $\beta$ and $\gamma$-deformed supersymmetric Yang-Mills theory. We also touch upon some subtleties arising in these deformed theories. Furthermore, we consider complex excited states (bound states) in the $\mathfrak{su}(2)$ sector and give their thermodynamic Bethe ansatz description. Finally we discuss the thermodynamic Bethe ansatz for a quantum deformation of the $\ads$ superstring S-matrix, with close relations to among others Pohlmeyer reduced string theory, and briefly indicate more recent developments in this area.
\end{abstract}
\end{titlepage}

%\begin{center}
%\vskip 9cm
%{\Large{Integrability of the $\ads$ superstring and its deformations}}
%
%\vskip 5cm
%
%{\large{Stijn Jurri\"en van Tongeren}}
%
%\end

\thispagestyle{empty} \tableofcontents

\thispagestyle{empty}

%\pagevalues

\chapter{Introduction}

Exact results are any scientist's dream, providing an unequivocal testing ground of concepts, ideas, and at times even hopes. Here we describe one particular instance where it is possible to get admittedly complicated but nonetheless exact results, of a kind that can normally only be approximated.

The big picture here is quantum field theory, one of the cornerstones of modern theoretical physics essential in describing elementary particle physics as we know it. Despite a quite elegant structure, complications rear their ugly head in many computations we would like to do in for example quantum chromodynamics (QCD) to directly compare with data coming from the LHC. In fact, we know that the way we typically approach our calculations is simply doomed to failure in some situations. For one we may be attempting to approximate things in a series around a point where this does not converge, or we may not even be able to determine whether what we are trying to compute is a sensible quantity. Essentially, we are unable to compute generic observable quantities in interacting quantum field theories without approximating them in some fashion, and many times even then without having good control over the approximation. It would be a great thing to have clean insight into at least some interacting quantum field theories, providing us with concrete and correct answers to confront with our general ideas.

There are many well known and important cases where certain observables of a quantum field theory can be understood exactly, typically owing to special symmetries of the theory under consideration. Supersymmetry for one can `protect' various quantities against quantum corrections and yield non-renormalization theorems by which (semi-)classical computations can give all-loop answers. Moreover, the famous results by Seiberg and Witten \cite{Seiberg:1994rs,Seiberg:1994aj} give exact low energy effective actions of theories with extended supersymmetry. Also, supersymmetry can be powerfully employed in localization, for example to derive exact partition functions \cite{Pestun:2007rz}. More relevant than supersymmetry in the present context, there are also cases such as the famous two dimensions conformal field theories \cite{DiFrancesco:1997nk} where the symmetry algebra of a theory becomes infinite dimensional. As a result the theory can become `exactly solvable', as happens in the minimal models as well as in a related way in integrable field theories. We will come back to the latter soon.

In this paper we will be considering a quantum field theory that despite being four dimensional appears to be at least partly `exactly solvable'. This special field theory is so-called $\mathcal{N}=4$ supersymmetric Yang-Mills theory (SYM) in four dimensions with gauge group $\mbox{SU}(N)$ in the limit where $N$ is large. While considerably different from QCD, it is definitely a nontrivial gauge theory, and the fact that we may be able to understand \emph{any} nontrivial gauge theory non-perturbatively should be very exciting. So what is required to understand $\mathcal{N}=4$ SYM in the large $N$ limit? Firstly, because of conformal symmetry the two-point correlation functions in this model are necessarily of the form
\begin{equation}
\langle \mathcal{O}_i (x) \mathcal{O}_j (y)\rangle = \frac{\delta_{ij}}{|x-y|^{2 \Delta_i}}\, ,
\end{equation}
where $\Delta_i$ is the so-called scaling dimension of the operator $\mathcal{O}_i$ which is generically a nontrivial function of the coupling constant $\lambda$.\footnote{In the large $N$ limit the effective coupling constant is not $g_{\scriptscriptstyle YM}$ but rather the 't Hooft coupling constant $\lambda = g_{\scriptscriptstyle YM}^2 N$.}  Secondly, three-point correlation functions are fixed by this symmetry up to a constant function of $C_{ijk}(\lambda)$
\begin{equation}
\langle \mathcal{O}_i (x) \mathcal{O}_j (y)\mathcal{O}_k (z)\rangle = \frac{C_{ijk}}{|x-y|^{\Delta_i+\Delta_j -\Delta_k}|y-z|^{\Delta_j+\Delta_k -\Delta_i}|x-z|^{\Delta_i+\Delta_k -\Delta_j}}\, .
\end{equation}
Now if we can compute these constants and scaling dimensions (the `conformal data') we can in principle compute any correlation function we like through the operator product expansion. At that point we can say we have understood the theory as we can compute any observable.

Now if we give someone the action for $\mathcal{N}=4$ SYM and expect them to compute these scaling dimensions or normalization constants exactly as a function of $\lambda$ without some further insight, we will likely be waiting for an answer indefinitely. Remarkably enough, it turns out that we can write down equations that give the $\Delta_i$ exactly, i.e. \emph{non-perturbatively}, as a function of $\lambda$.\footnote{The hope is that integrability can be used to `solve' planar $\mathcal{N}=4$ SYM completely. The story for the $C_{ijk}$ is more complicated and less developed however, and is not part of this review.} It is clearly very impressive that we can obtain such highly nontrivial dynamical information in an interacting quantum field theory,\footnote{This goes far beyond the protected operators that we can describe exactly thanks to supersymmetry.} which becomes possible because of two concepts, integrability and AdS/CFT. Let us start with integrability, noting that many of the concepts we will only mention below are explained in some detail in the following two chapters.

\subsection*{Integrability in gauge theory}

The type of integrability that appears directly in $\mathcal{N}=4$ SYM is that of particular one dimensional spin chains, the most famous being the Heisenberg XXX spin chain. It might seem odd at first that one-dimensional discrete quantum systems have something concrete to do with gauge theories, but it has for example been known for many years that integrable spin chains play a role in the description of Reggeized gluon scattering in high-energy QCD \cite{Lipatov:1993yb}. In $\mathcal{N}=4$ SYM spin chains enter the game when computing scaling dimensions of gauge invariant operators \cite{Minahan:2002ve}.

\begin{figure}[h]
\begin{center}
\includegraphics{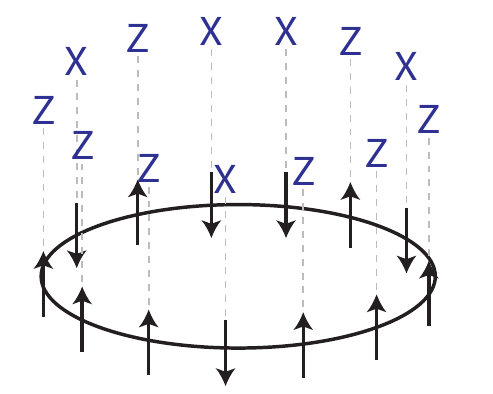}
\caption{Single trace operators as states in a spin chain. By cyclicity of the trace we can picture the fields inside of it as living on a circle (here illustrated for $\mbox{Tr}(ZXZXXZXZZZXZZ)$). Identifying $Z$ with spin up, and $X$ with spin down, we get a periodic chain of spins.}
\label{fig:spinchaintrace}
\end{center}
\end{figure}

To be slightly more concrete, we can consider two of the three complex scalar fields of $\mathcal{N}=4$ SYM, say $X$ and $Z$ out of $X$, $Y$ and $Z$, and ask ourselves what the scaling dimension of generic single trace\footnote{Mixing between different multi-trace operators is suppressed at large $N$.} operators like
\begin{equation}
\mathcal{O} = \mbox{Tr}(XZXX \ldots ZXZXZ)
\end{equation}
is. Classically, the scaling dimension of such an operator is just $L$, the total number of scalar fields in the trace (with scaling dimension one each). If we consider quantum corrections however, the dilatation operator acts as a matrix in the space of these operators, and its eigenstates are in general mixed combinations of these operators. The structure of this matrix is very special. Indeed, doing an explicit one loop computation it turns out that this matrix is precisely the Heisenberg XXX spin chain Hamiltonian acting on a periodic chain of spins of length $L$, upon appropriately identifying states in the spin chain with operators with given numbers and orderings of $X$ and $Z$ fields. Perhaps this is best explained graphically as in figure \ref{fig:spinchaintrace}. This is a very nice result because the XXX spin chain is an integrable spin chain, meaning that we can use the Bethe ansatz to explicitly diagonalize this matrix efficiently, also when this matrix is very large. By diagonalizing this matrix we compute the quantum corrected scaling dimensions of these mixed eigenstates; $\Delta_{c} \rightarrow \Delta_{c} + \gamma(\lambda)$, where $\gamma$ is known as the anomalous dimension.

The really magical thing is that doing similar computations with other (more complicated) field content of SYM in the trace, the idea of identifying the problem with a particular spin chain still yields an integrable model, albeit a significantly more complicated one. If we then believe this structure to be universal also at higher loops, it is in fact possible to fix the associated Bethe equations uniquely from symmetry up to a single phase factor \cite{Beisert:2005tm} (see also \cite{Beisert:2004hm,Beisert:2005fw}). So if we could fix this scalar factor in some fashion, we would have an all-loop prediction for the (anomalous) scaling dimensions of operators in $\mathcal{N}=4$ SYM. This would mean we understand the two point functions of an interacting quantum field theory exactly at any coupling! It turns out however that the picture is not quite this simple. The Heisenberg XXX spin chain that describes the one loop anomalous dimensions of operators containing $X$ and $Z$ has only nearest neighbour interactions between the spins of the spin chain, but at higher loops this interaction becomes first next-to-nearest neighbour then next-to-next-to-nearest neighbour and so on \cite{Beisert:2003tq}. At some loop order, this interaction becomes of such a long range that the interaction reaches, or \emph{wraps}, all the way around the spin chain \cite{Serban:2004jf,Beisert:2004hm,Sieg:2005kd}. It turns out that from precisely this order the `all-loop' Bethe ansatz for $\mathcal{N}=4$ SYM stops correctly accounting for all contributions to the dilatation operator (regardless of the phase factor we are skipping over for the moment). The extra contributions are known as wrapping interactions, leading to wrapping corrections to the energy. It is not obvious at all how to generically incorporate such wrapping interactions in the spin chain picture, and so it seems we are unfortunately still not doing non-perturbative quantum field theory.\footnote{We should note however that for long operators, i.e. operators made out of many fields, the all-loop Bethe ansatz should hold up to high loop order and still gives us results we could never hope to obtain by direct perturbative calculations.}

The way to get around these problems is to approach the problem along a completely different route, using string theory.

\subsection*{AdS/CFT and integrable strings}

Before any of the integrable structures in $\mathcal{N}=4$ SYM were found it was in fact already possible to get non-perturbative results in planar $\mathcal{N}=4$ SYM at strong coupling thanks to the AdS/CFT correspondence \cite{Maldacena:1997re}. The AdS/CFT correspondence is a conjecture of a duality (equivalence) between string theories on anti-de Sitter spaces and certain gauge theories on the conformal boundary of these anti-de Sitter spaces, the canonical example being the conjecture that $\mathcal{N}=4$ SYM with gauge group $\mbox{SU}(N)$ at 't Hooft coupling $\lambda$ is dual to type IIB (closed) string theory on $\ads$ with string coupling $g_s = \lambda/4\pi N$ and (effective) string tension $g=\sqrt{\lambda}/2\pi$, at least in the large $N$ limit. The conformal $\mbox{SO}(4,2)$ and $\mbox{SO}(6)$ R-symmetry of $\mathcal{N}=4$ SYM correspond to the isometries of $\ads$; including the supersymmetries of $\mathcal{N}=4$ SYM the total symmetry group is $\mbox{PSU}(2,2|4)$, which is of course the isometry group of the full supergeometry of the string.

We can approximate string theory by supergravity when the radius of curvature of $\ads$ is large compared to the string length and $g_s$ is small. Since the first of these translates to the requirement that $g_s N$ is large, this means in particular that both $\lambda$ and $N$ are large; we are in the strong coupling regime of planar $\mathcal{N}=4$ SYM. Conversely, when $N$ is large and $\lambda$ is small we can do perturbative planar gauge theory, but this is a highly non-perturbative regime for string theory since the string length is too big compared to the radius of curvature. So we see that at this stage we can consider planar $\mathcal{N}=4$ SYM perturbatively at weak coupling directly, and at strong coupling via supergravity. Still we do not know how to compute something when $\lambda$ is neither large nor small.

To describe planar $\mathcal{N}=4$ SYM at finite $\lambda$, the above discussion shows us that we need to consider string theory on $\ads$, with $g_s\rightarrow 0$ but very much finite string tension $g$. This means we are considering a free superstring at finite tension, and according to AdS/CFT the energy spectrum of this string (in global AdS coordinates) is equal to the spectrum of scaling dimensions of $\mathcal{N}=4$ SYM.\footnote{The string Hamiltonian and the SYM dilatation operator are part of the symmetry algebra on either side of the correspondence. They are not quite one and the same generator but are related by a similarity transformation. Then if AdS/CFT is correct the spectra of the corresponding operators simply have to agree.} Were we considering our string theory on flat space we could readily solve this problem since it essentially amounts to finding the energy spectrum of a free Hamiltonian on a circle (the closed string). When dealing with a generic curved background however, it is not even clear how to write down the full string action. In the present case, because of the self-dual Ramond-Ramond five form flux that supports the $\ads$ background in supergravity, it is not clear how to use the standard Neveu-Schwarz-Ramond approach to construct a superstring action. Fortunately, there is a natural way to construct a full Green-Schwarz type action for the superstring on $\ads$ \cite{Metsaev:1998it}. Unfortunately, this action looks slightly more complicated than one for a bunch of free particles, at least in its manifest form.

Given the perturbative integrability in $\mathcal{N}=4$ SYM which extends to arbitrary loop order for appropriately long operators, we may expect that there is something special about the $\ads$ superstring as well. This is indeed the case; the $\ads$ superstring as described by the action of \cite{Metsaev:1998it} is manifestly classically integrable \cite{Bena:2003wd}, meaning that we can represent its equations of motion via a Lax pair and can construct infinitely many classically conserved quantities. If we could show that these infinitely many conserved charges survive quantization we would clearly be dealing with a very special quantum field theory. It would be a so-called integrable quantum field theory, in the sense that the many conservation laws prevent macroscopic particle production and force any scattering event to factorize into a product of two body scattering events \cite{Zamolodchikov:1978xm}. Given the complexity of the action however, there is no particularly clear way to proceed to try and quantize the theory directly (even in a light-cone gauge), let alone prove that many conserved charges survive quantization. One promising direction in this regard is to recast our Green-Schwarz string in the pure spinor formalism \cite{Berkovits:2000fe} where it can be argued that the $\ads$ string is quantum integrable to all-loop order \cite{Berkovits:2004xu} (classical integrability is discussed in \cite{Vallilo:2003nx,Berkovits:2004jw}).\footnote{Since the pure spinor formalism allows conformal-covariant quantization of the string, here quantum integrability refers directly to the existence of infinitely many conserved quantities; integrability in the sense of S-matrix factorization requires a particle theory such as the world-sheet theory in a light-cone gauge.} However, explicitly demonstrating quantum integrability in the sense of concretely constructing (finite coupling) quantum transfer matrices appears to be a subtle question also there \cite{Mikhailov:2007mr,Mikhailov:2007eg,Puletti:2008ym}. Still, knowing of the promising structures on the gauge theory side, we can proceed by \emph{assuming} our quantum field theory to be quantum integrable, and comparing the consequences with explicit computations in both string and gauge theory. In particular, we will assume that the massive quantum field theory describing the light-cone gauge fixed Green-Schwarz string is integrable at the quantum level.

Under this assumption we can fix the S-matrix of our (light-cone) superstring on a line, since there are enough symmetries to constrain the two body S-matrix up to a phase factor \cite{Arutyunov:2006ak,Arutyunov:2006yd,Beisert:2005tm}. Of course, since our string is not a line but rather a circle we can wonder what use this S-matrix is. The idea is that if the circle is large enough we can approximately talk about an S-matrix. Due to the lack of macroscopic particle production we can then construct approximate wave-functions for multi-particle states in our quantum field theory on a large circle, and find the associated energy spectrum via a set of equations that look exactly like Bethe ansatz equations \cite{Zamolodchikov:1978xm}. Of course these equations are only correct when the length of the circle is very large, and when properly analyzing the units this length is essentially the length of the spin chain we encountered in $\mathcal{N}=4$ SYM. In fact, the full sets of Bethe equations are perfectly compatible up to one subtlety. The subtlety sits in the overall phase factor of the string S-matrix known as the dressing phase \cite{Arutyunov:2004vx}. In well known integrable quantum field theories such as the sine-Gordon model, such phase factors are fixed non-perturbatively by insisting on crossing symmetry and unitarity. We did not mention it before, but in the light-cone gauge the string is a non-relativistic model, so that it is not immediately clear what principle should fix our phase. Still, as the (non-)relativistic nature of our string is a gauge dependent statement, it may be appropriate to ask for a non-relativistic generalization of crossing, and indeed this appears to bring everything together nicely \cite{Janik:2006dc}. Comparing the gauge and string theory equations, this phase is precisely the phase factor we skipped over in gauge theory earlier, and indeed it is required to match perturbative computations there. In this way the arbitrary gauge theory phase is fixed through the dual string theory where the fundamentally quantum field theoretical concept of crossing exists.\footnote{We should mention however that there is an alternative way of deriving the crossing equation which can be applied directly in the spin chain \cite{Beisert:2005tm}.}

Unfortunately, it appears that at this stage we have not gained anything beyond a way to fix a phase factor, since we find the all-loop Bethe equations for $\mathcal{N}=4$ SYM we know to be missing something. The important difference is that just as how we could fix the last piece of the S-matrix by crossing, from the point of quantum field theory we understand precisely what the Bethe equations are missing; putting a quantum field theory on a circle, even in an integrable field theory there are quantum corrections due to virtual particles that loop around the circle. It seems natural that this should explain wrapping effects in $\mathcal{N}=4$ SYM \cite{Ambjorn:2005wa}. In fact, as advocated in \cite{Ambjorn:2005wa}, rather than treating this as inherently perturbative we should be able to use well-known ideas from relativistic integrable field theories where such effects can be `summed up' by the thermodynamic Bethe ansatz \cite{Zamolodchikov:1989cf}.

\subsection*{Integrable quantum field theory in finite volume}

The idea of using the thermodynamic Bethe ansatz (TBA) in quantum field theory is based on the rather ingenious observation that we can compute the exact ground state energy of a quantum field theory on a circle as the free energy density of the quantum field theory obtained by interchanging our notion of space and (Euclidean) time, living on a line (a decompactified circle) but at a finite temperature (the radius of the circle) \cite{Zamolodchikov:1989cf}. Since the resulting model is defined on a non-compact dimension we can talk of a definite S-matrix, which we may be able to find exactly if the model is also integrable. The key point is that the circumference of the spatial circle is now infinite, so that virtual corrections to the Bethe equations can be safely ignored. The problem is then simply how to compute the free energy, and this can be done by the thermodynamic Bethe ansatz as originally intended \cite{Yang:1968rm}.

Applying these ideas to the $\ads$ superstring \cite{Ambjorn:2005wa,Arutyunov:2007tc} we can get a set of TBA equations that describe the ground state energy of our superstring exactly \cite{Arutyunov:2009ur,Bombardelli:2009ns,Gromov:2009bc}; we can find a very fancy way to write zero. Fortunately it is possible to extend the resulting TBA equations beyond the ground state and use them to find the exact energies of specific excited string states as well \cite{Gromov:2009bc,Arutyunov:2009ax}. By AdS/CFT this means we can obtain equations that describe the scaling dimensions as functions of $\lambda$ \emph{exactly}. Of course, actually solving the equations requires numerics (outside of special limits), but these problems are of a purely technical and not a conceptual nature. In this way then, we are doing non-perturbative quantum field theory. Viewed in another light this is a big nontrivial precision test of the AdS/CFT correspondence, upon which so much current research builds.

Of course there is a host of hypotheses going into this story, and everything builds on one another in a rather intricate fashion. Fortunately at the end of the day we have certain nice checks on the whole approach. We already saw that the all-loop Bethe equations of the gauge and string theory agree, and although the fact that they agree between themselves is purely a consequence of symmetries and assumptions, the fact that they agree with perturbative gauge and string theory is not (see e.g. \cite{Sieg:2010jt} and \cite{McLoughlin:2010jw}). Moreover, there are explicit perturbative results for scaling dimensions which go beyond wrapping order \cite{Fiamberti:2007rj,Velizhanin:2008jd,Eden:2012fe} and these wrapping effects indeed precisely match with explicit computations based on the idea of virtual particles wrapping around the circle \cite{Bajnok:2008bm,Bajnok:2009vm} and the TBA equations of our integrable superstring in finite volume \cite{Arutyunov:2010gb,Balog:2010xa}.

Since $\mathcal{N}=4$ SYM is a highly symmetric field theory and $\ads$ a highly symmetric space, we might wonder whether any of this applies in less symmetric cases as well. It seems that we can at least deform $\mathcal{N}=4$ SYM and the associated superstring (in relatively mild ways) to obtain less symmetric cases where these ideas still apply with equal power. The most famous example of this is $\beta$-deformed SYM \cite{Leigh:1995ep}, which has $\mathcal{N}=1$ superconformal symmetry. Other examples of where this applies are non-supersymmetric generalizations of $\beta$-deformed SYM we will call $\gamma$-deformed SYM, and various `orbifolds' of $\mathcal{N}=4$ SYM, the field theories dual to strings on orbifolds of $\ads$. We should also note that the whole story we described above for the canonical example of AdS/CFT basically goes through for the duality between strings on ${\rm AdS}_4 \times \mathbb{CP}^3$ and ABJM theory as well (see e.g. \cite{Klose:2010ki}).

\subsection*{This paper}

After a general review of the story described above, the first point of this paper is to carefully describe how to account for the deformations described just above at the level of the TBA equations for the $\ads$ superstring. By doing so, the integrability of any of these deformed theories is essentially put on the same footing as that of the undeformed parent theory. As a very concrete check, we will see that these general results match perfectly with explicit perturbative computations of wrapping effects in $\beta$-deformed SYM. Secondly, while simply stated above, we will see how to actually extend the ground state TBA equations to excited states, and that this can in fact be quite tricky. On the one hand by means of example we will see that considering deformed theories simply enlarges the parameter space of the problem of finding excited state equations. On the other hand we will see that considering excited states with particles that form approximate bound states (complex momenta) can complicate the structure of the excited state TBA equations greatly, and lead to interesting structures that are presumably due to the non-relativistic lattice-like dispersion relation of our model and have hence not been observed before. Thirdly, we will take a look at a particular deformation of the integrable structure of the superstring (its S-matrix) rather than the theory itself. This deformation is related to a relativistic cousin of the superstring known as the Pohlmeyer reduced superstring, but is definitely also interesting purely from the point of view of integrable models. We will see that the TBA equations for this model take a very interesting form, and the so-called Y-system behaves in a slightly unexpected new way.

Rather than jumping straight into the details of the superstring, this review starts with a short but largely self-contained pedagogical introduction to integrable quantum field theory, covering the ingredients used in the story above. We will see how factorization of the S-matrix comes about, what it exactly means, and how this allows us to find approximate energy levels in an integrable quantum field theory. For concreteness this is illustrated in detail on the simple example of the chiral Gross-Neveu model. Then we will see how to precisely implement the ideas of Zamolodchikov to obtain ground state energies through the TBA, explicitly illustrated for the chiral Gross-Neveu model, and how to extend these ideas to excited states. Chapter \ref{chapter:AdS5string} contains the specific structure and notation needed to describe the $\ads$ superstring as an integrable model, and gives the ground state TBA equations and necessary information required to extend them to any excited state. The rest of this review covers work the author contributed to, beginning in chapter \ref{chapter:twistedspectrum} with a discussion of how to exactly extend the TBA equations for strings on $\ads$ to the deformed string theories mentioned above. We also briefly touch upon more recent developments regarding the non-conformality of $\gamma$-deformed SYM and the pre-wrapping effect in $\beta$-deformed SYM. Chapter \ref{chapter:excstatesandwrapping} contains explicit computations of wrapping effects in $\beta$-deformed SYM from the TBA side which match nicely with known perturbative gauge theory results, and discusses how to find the excited state TBA equations for (a descendant of) the so-called Konishi state in a particularly orbifolded theory. This is followed by chapter \ref{chapter:boundstates} which discusses the complicated excited state TBA equations for states with complex momenta. Finally chapter~\ref{chapter:quantumTBA} discusses the TBA equations that arise when deforming the S-matrix of the superstring. The first part of this review is intended to introduce integrable quantum field theory and the thermodynamic Bethe ansatz to readers unfamiliar with the topic, and then present the necessary data for our superstring. The later parts of this review each cover a well-defined and relatively self-contained topic. As such, provided the relevant material in chapters~\ref{chapter:finitevolumeIQFT} and \ref{chapter:AdS5string} is familiar to the reader, chapters~\ref{chapter:twistedspectrum} and \ref{chapter:excstatesandwrapping}, chapter~\ref{chapter:boundstates}, and chapter~\ref{chapter:quantumTBA} can be read independently.

It was not the goal of this introduction to provide an exhaustive list of references for many of the topics only briefly touched upon. The references that are of direct relevance are given in the individual chapters. For a general review of integrability in AdS/CFT we refer the reader to the review \cite{Beisert:2010jr}. In particular, integrable deformations in AdS/CFT are reviewed in the chapter \cite{Zoubos:2010kh}. For a pedagogical review of integrability of the $\ads$ superstring stopping just short of the thermodynamic Bethe ansatz we refer the reader to \cite{Arutyunov:2009ga}, and note that (the remainder of) the superstring story directly relevant to us is reviewed in chapter \ref{chapter:AdS5string}. Reviews of $\ads$ integrability from the pure spinor point of view can be found in \cite{Puletti:2010ge,Mazzucato:2011jt}. As it will not come back in further chapters, let us note that localization is originally discussed in \cite{Duistermaat:1982vw,Atiyah84localization,Berline82localization}, and \cite{Witten:1988ze,Witten:1991zz} in a physical context. A recent pedagogical introduction to its application to partition functions can be found in \cite{Marino:2011nm}. 

\chapter[Integrable quantum field theory in finite volume]{Integrable QFT in finite volume}
\label{chapter:finitevolumeIQFT}

This chapter introduces integrable quantum field theories and their description in terms of factorized scattering, culminating in the description of their finite volume spectra by means of the thermodynamic Bethe ansatz. The subsequent chapters will present a technically complicated version of the general picture described here, with important applications in AdS/CFT. In this chapter we will come across a few integrable models whose integrability can basically be proven. We will however use them to illustrate general features and to introduce the factorized scattering and bootstrap ideas emphasized by Zamolodchikov which can then be applied to the light-cone gauge superstring whose quantum integrability is `harder to prove'. Let us begin by introducing the notion of classical integrability.

\section{Classical integrability}
\label{sec:classint}

Both classical and quantum integrability rely on the existence of many conserved charges which mutually commute with each other. This is most well known in the case of classical mechanical systems, where there is a theorem by Liouville which states that if we have a dynamical system with a phase space of dimension $2N$ and we can find $N$ functionally independent conserved charges (including the Hamiltonian) which mutually Poisson commute, then we can solve the equations of motion by quadratures - solving finitely many algebraic equations and doing finitely many integrals.

A convenient way of generating such conserved quantities can be found through a so-called Lax pair representation of the equations of motion. For a classical mechanical system, a Lax pair consists of a pair of matrices $L$ and $M$ with entries depending on the phase space variables, such that the equations of motion are equivalent to the Lax equation
\begin{equation}
\partial_t L - [L,M]=0\, .
\end{equation}
Provided we have such a construction, the trace of any power of the L matrix will be a conserved quantity
\begin{equation}
\label{eq:classmechQs}
\partial_t \mbox{tr}(L^n) = n \mbox{tr}([L,M]L^{n-1}) = 0\, .
\end{equation}
If we can then prove that these conserved quantities Poisson commute and there are $N$ of them, we have proven the classical integrability of the system.\footnote{If there exists a classical r-matrix satisfying the classical Yang-Baxter equation, this guarantees the commutativity of the conserved charges, see for example \cite{BBTclassicalint}.} Note that the representation of a set of equations of motion by a Lax pair is by no means unique. Without further structure we at least need the dimension of the Lax matrices to be bigger than $N$ since otherwise our construction can never give enough independent conserved charges.

As a simple example we can consider the pendulum with Hamiltonian
\begin{equation}
\label{eq:pendulumham}
H = \frac{1}{2}I^2 - \cos\theta + 1\, ,
\end{equation}
and equations of motion
\begin{equation}
\dot{I} = -\sin\theta\, , \hspace{20pt} \dot{\theta} = I\, .
\end{equation}
These equations of motion are equivalent to the Lax equation for the Lax pair
\begin{equation}
L = I S_z + 2 \sin \frac{\theta}{2} S_x \, , \hspace{20pt} M= - i \cos \frac{\theta}{2} S_y\, ,
\end{equation}
where $S_x$, $S_y$ and $S_z$ are the generators of $\mathfrak{su}(2)$ in some finite dimensional nontrivial representation. The only independent conserved charge we can construct this way is
\begin{equation}
2 \mbox{tr}L^2 = I^2 + 2 \sin^2\frac{\theta}{2}\, ,
\end{equation}
where for concreteness we have taken the spin $1/2$ representation. This indeed reproduces the Hamiltonian \eqref{eq:pendulumham}.

If we allow for a little more structure, the Lax pair representation of a classical mechanical system is something that translates readily into two dimensional classical field theory as well. There we call a system classically integrable if the equations of motion can be represented in terms of the zero curvature condition of a Lax pair
\begin{equation}
\partial_\alpha L_\beta - \partial_\beta L_\alpha - [L_\alpha, L_\beta] =0\, ,
\end{equation}
where $L_\alpha (\sigma,\tau,z)$ depends on an auxiliary parameter $z$ called the spectral parameter. Now the conserved quantities can be generated by considering the monodromy matrix $T(z)$, defined as the following path ordered exponential
\begin{equation}
T(z) \equiv \,\,\stackrel{\leftarrow}{\exp} \int_0^{2\pi} d\sigma L_\sigma (z)\,,
\end{equation}
where we take the model to be defined on a spatial circle of circumference $2\pi$. Using the zero curvature condition  it readily follows that\footnote{See for example section 3.7 of \cite{BBTclassicalint} or page 24 of \cite{Arutyunov:2009ga}.}
\begin{equation}
\partial_\tau T(z) = [L_\tau(0,\tau,z),T(z)]\,,
\end{equation}
provided all fields are periodic. This implies that the eigenvalues of $T$ are independent of time, and typically by expanding them in a series in the spectral parameter about some special point we can generate infinitely many local conserved charges.

As an example of a classical integrable field theory we can consider the two-dimensional sine-Gordon model with Lagrangian
\begin{equation}
\mathcal{L}_{sG} = \frac{1}{2}\partial_\mu \phi \partial^\mu \phi + \frac{m^2}{\beta^2}(\cos\beta\phi-1)\, ,
\end{equation}
The corresponding equations of motion are equivalent to the zero curvature condition of the Lax pair\footnote{Note that this gives a Lax pair with spectral parameter for the pendulum as well by setting $m=\beta=1$ and $\phi=\phi(t)$. In fact the given Lax pair is just this one at $z=1$. Of course the sine-Gordon model is nothing but the continuum limit of a string of pendula connected by springs.}
\begin{align}
i L_\tau &= \frac{\beta \phi^\prime}{2} S_z + c_1 \sin\frac{\beta\phi}{2} S_x + c_0 \cos\frac{\beta\phi}{2} S_y\, ,\\
i L_\sigma &= \frac{\beta \dot{\phi}}{2} S_z + c_0\sin\frac{\beta\phi}{2} S_x + c_1\cos\frac{\beta\phi}{2} S_y\, ,
\end{align}
where
\begin{equation}
c_0 = \frac{m}{2}\left(z+\frac{1}{z}\right)\, , \hspace{20pt} c_1 =
\frac{m}{2}\left(z-\frac{1}{z}\right)\, .
\end{equation}
For the moment we have considered constructing infinitely many classically conserved charges, leading us to wonder what can happen if these survive quantization.

\section{Quantum integrability as factorized scattering}
\label{sec:factscat}

Proving that the infinite number of conserved charges of an integrable classical field theory survive quantization can be a complicated problem. Provided they do however, they can have some very interesting consequences for the field theory. Namely, there can be no (on-shell) particle production, and many-body scattering events naturally factorize into the product of two-body scattering events. In relativistic theories, these constraints follow from standard analyticity assumptions about the S-matrix and having at least two local conserved charges which transform in higher spin (tensor) representations of the Lorentz group \cite{Parke:1980ki}. Although they are not required to get the desired constraints, typically these conserved charges are part of a set of infinitely many conserved charges of various Lorentz spins. Infinitely many independent conserved charges immediately imply lack of particle production as follows.

We will consider a state in the far past where we have $N_i$ well-separated and hence non-interacting particles, and let this evolve into the asymptotic future. After a complicated scattering process, in the asymptotic future we will end up with a final set of $N_f$ well-separated particles again. Now as we all know very well, there are at least some constraints on these particles in any quantum field theory: energy and momentum should be conserved. In the case of our two dimensional quantum field theory however, we were supposed to have many more conserved quantities. Labelling the conserved quantities as $Q_s$, where $s$ runs over the infinitely many conserved quantities, diagonalized in the one particle basis (labeled by momentum), charge conservation tells us that
\begin{equation}
\sum_{j=1}^{N_i} Q_s (p_j^{i}) = \sum_{l=1}^{N_f} Q_s (p_l^{f}) \, .
\end{equation}
Since there are infinitely many independent equations to be satisfied for a given set of incoming momenta (one for each $s$) for only finitely many outgoing momenta, the only possible solutions are the trivial solutions where the right hand side is a rearrangement of the terms on the left hand side; $\{p^{i} \} = \{p^{f} \}$.\footnote{By inserting a reflecting boundary it is possible to get $\{p^{i} \} = \{-p^{f} \}$, which is allowed since the conserved charges other than $p$ are typically even functions of $p$.} In other words, the scattering process can not produce particles and can only lead to particles possibly exchanging momenta; the scattering is non-diffractive.

Now let us consider the consequences of having charges which transform in higher spin representations of the Lorentz group. To people not familiar with integrability or a two-dimensional world this might seem like a silly thing to consider, as the Coleman-Mandula theorem \cite{Coleman:1967ad} colloquially put supposedly tells us that higher spin conserved quantities imply a trivial S-matrix. Slightly more specific, the Coleman-Mandula theorem states that the symmetries of a nontrivial S-matrix can be at most a product of the Poincar\'e group and internal symmetry groups.\footnote{As is of course well known, supersymmetry evades this theorem by considering anti-commuting generators. In that case the super-Poincar\'e group would be the maximal symmetry group containing the space-time symmetries of the theory.} Of course, precisely in two dimensions the Coleman-Mandula theorem does not apply in this strict sense and it is its two dimensional analogue that we are trying to describe here.

In many integrable models these conserved charges can actually be chosen such that they roughly correspond to entire powers of the momenta, $Q_s \sim p^s$ \cite{Zamolodchikov:1978xm}, which for the sake of intuition is what we will base the following discussion on \cite{Shankar:1977cm,Dorey:1996gd}. In order to understand what a conserved charge of the schematic form $p^s$ corresponds to we begin by recalling that ordinary momentum conservation follows from translational invariance, or in other words that $p$ is the generator of translations. In this picture a higher spin conserved charge $p^s$ would roughly correspond to a momentum-dependent translation. We can show this rather more explicitly for a wave packet localized in both coordinate and momentum space as follows \cite{Shankar:1977cm,Dorey:1996gd}. We start with a wave packet with a Gaussian momentum distribution\footnote{This discussion directly follows  \cite{Shankar:1977cm}.}
\begin{equation}
\psi(x) = \int d p e^{-a^2 (p-p_0)^2} e^{i p(x-x_0)}\,.
\end{equation}
By construction this wave packet is localized in momentum space around $p_0$, and we can determine its average position in coordinate space by the stationary phase approximation: the integral is dominated by the part where $p \sim p_0$, where the phase is then stationary when $x = x_0$. Acting with $e^{i b p^s}$ on this state we end up with the transformed wave packet
\begin{equation}
\tilde{\psi}(x) = \int d p e^{-a^2 (p-p_0)^2} e^{i p(x-x_0)} e^{i b p^s}\,,
\end{equation}
which is still localized in momentum space around $p_0$. However now the phase will be stationary when $x=x_0-s b p_0^{s-1}$. From this we see that indeed for $s=1$ we have translated the wave-packet by $b$, but also that for $s>1$ we in fact translate the wave-packet by an amount dependent on its average momentum.

%\begin{figure}%
%\centering
%\subfigure[]{\includegraphics[width=4cm]{3to3as2a.pdf}\label{fig:3bodyscatteringa}}\quad
%\subfigure[]{\includegraphics[width=4cm]{3to3.pdf}\label{fig:3bodyscatteringb}}\quad
%\subfigure[]{\includegraphics[width=4cm]{3to3as2b.pdf}\label{fig:3bodyscatteringc}}
%\caption{The three possible forms of three body scattering. The left and right scattering pictures represent the two cases of three body scattering as a sequence of two body scattering events (time runs in the upward direction), while the middle illustrates the case where the three particles interact simultaneously. In an integrable quantum field theory these pictures are all equivalent; the higher spin conserved charges can be used to turn each into another ($e^{\pm i p^2}$ was used to generate figures (a) and (c) from (b)).}
%\label{fig:3bodyscattering}
%\end{figure}

\begin{figure}%
\centering
\subfigure[]{\includegraphics[width=4cm]{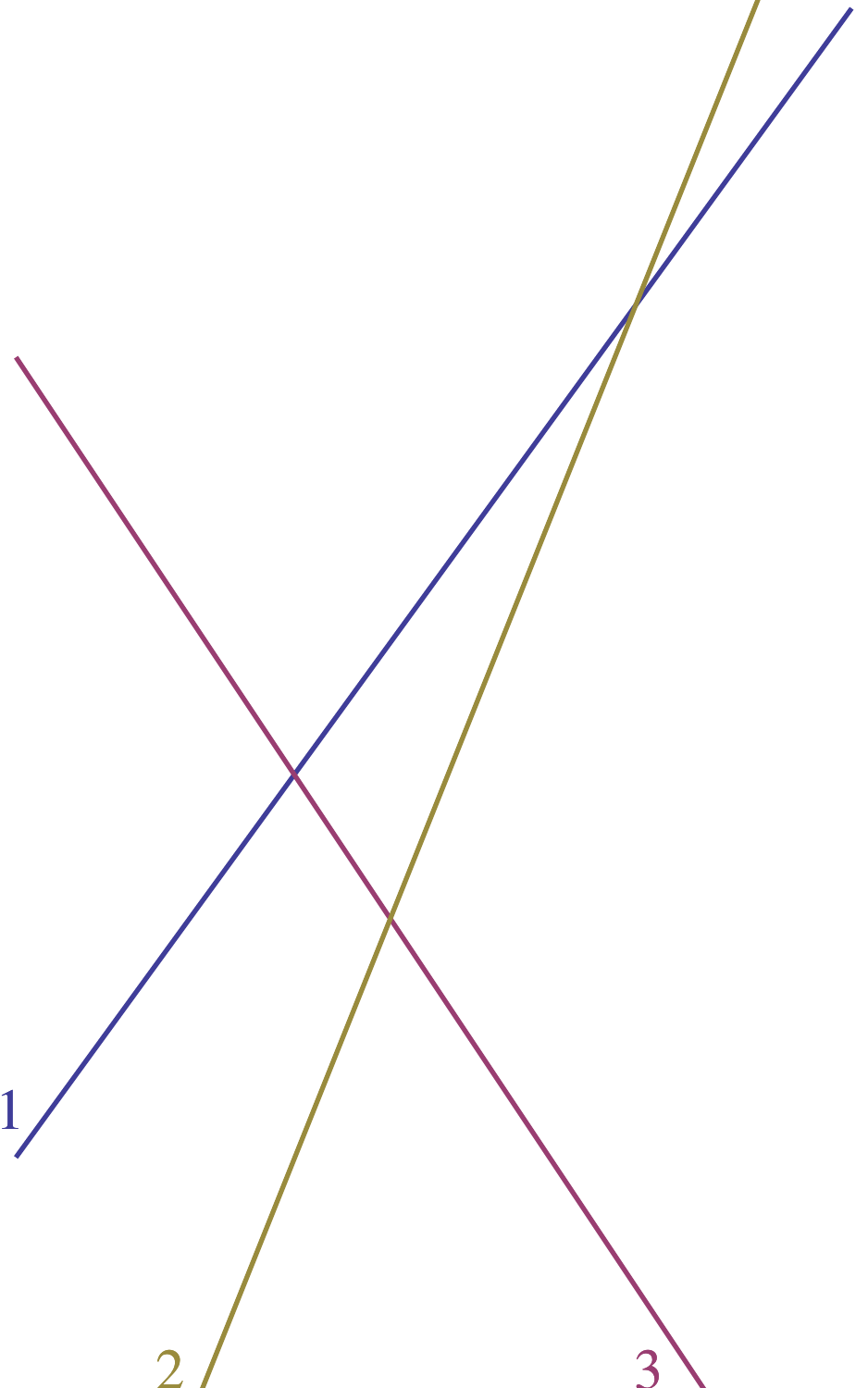}\label{fig:3bodyscatteringa}}\quad
\subfigure[]{\includegraphics[width=4cm]{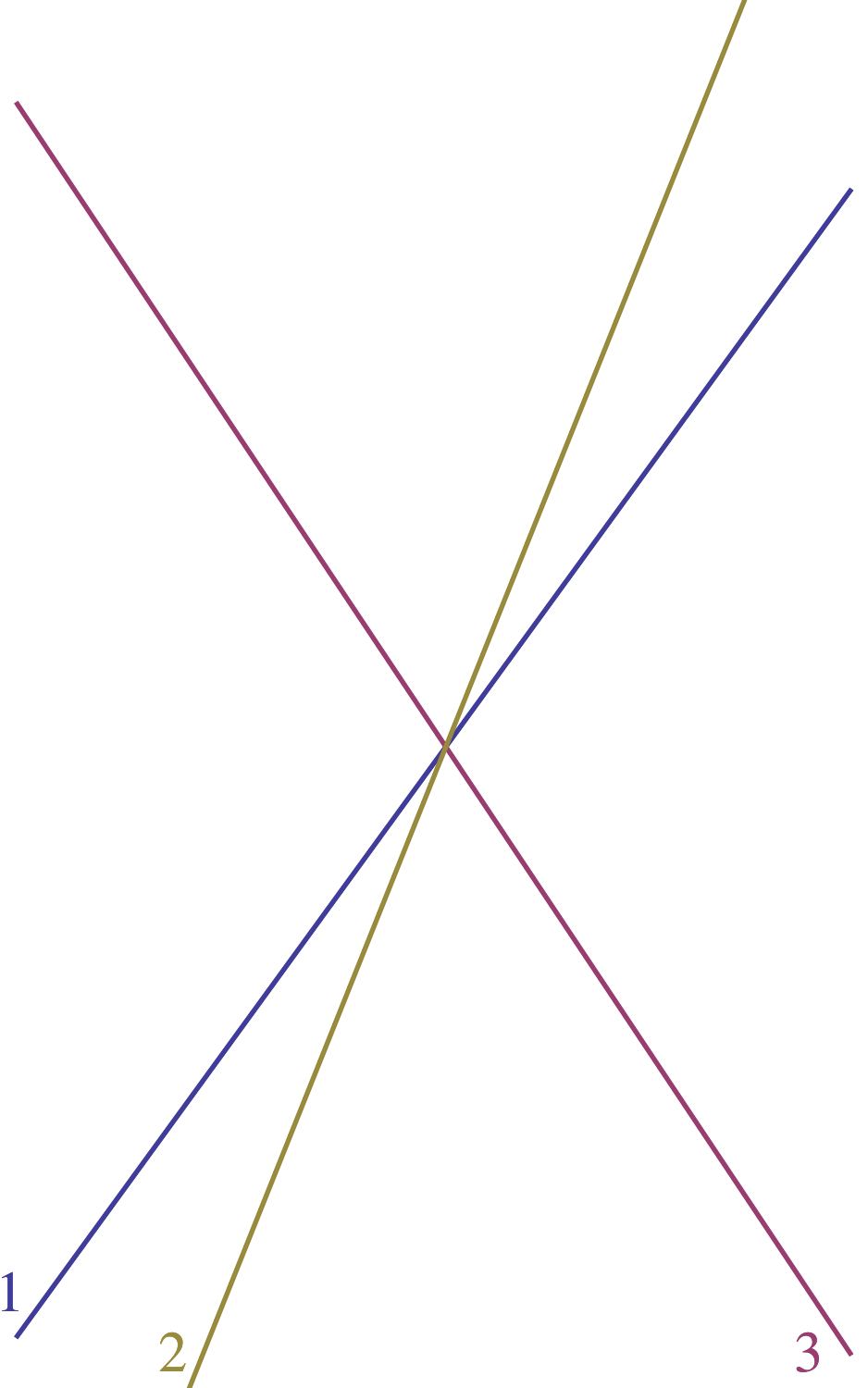}\label{fig:3bodyscatteringb}}\quad
\subfigure[]{\includegraphics[width=4cm]{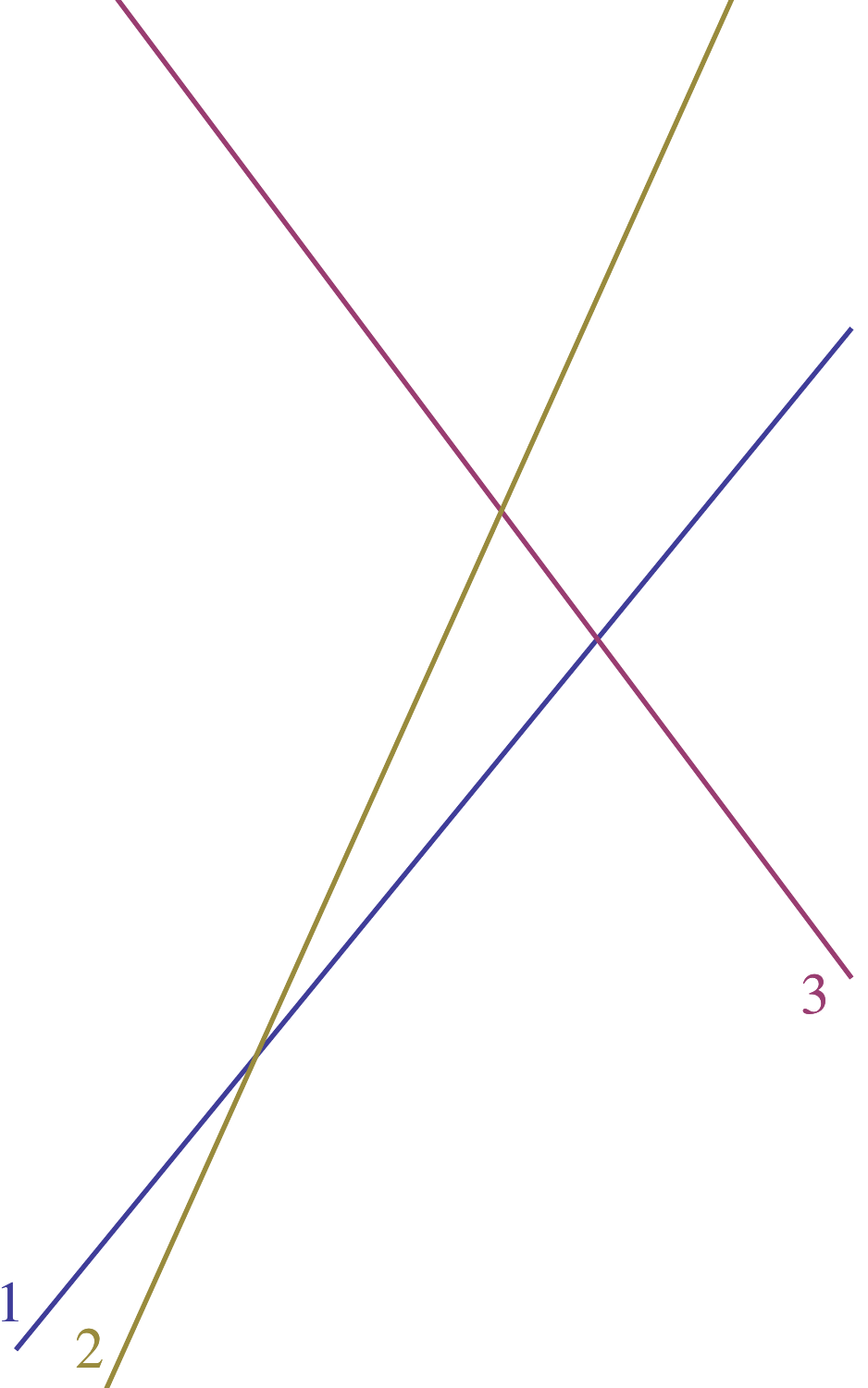}\label{fig:3bodyscatteringc}}
\caption{The three possible forms of three body scattering. The left and right scattering pictures represent the two cases of three body scattering as a sequence of two body scattering events (time runs in the upward direction), while the middle illustrates the case where the three particles interact simultaneously. In an integrable quantum field theory these pictures are all equivalent; the higher spin conserved charges can be used to turn each into another ($e^{\pm i p^2}$ was used to generate figures (a) and (c) from (b)).}
\label{fig:3bodyscattering}
\end{figure}

\begin{figure}%
\centering
\includegraphics[width=8cm]{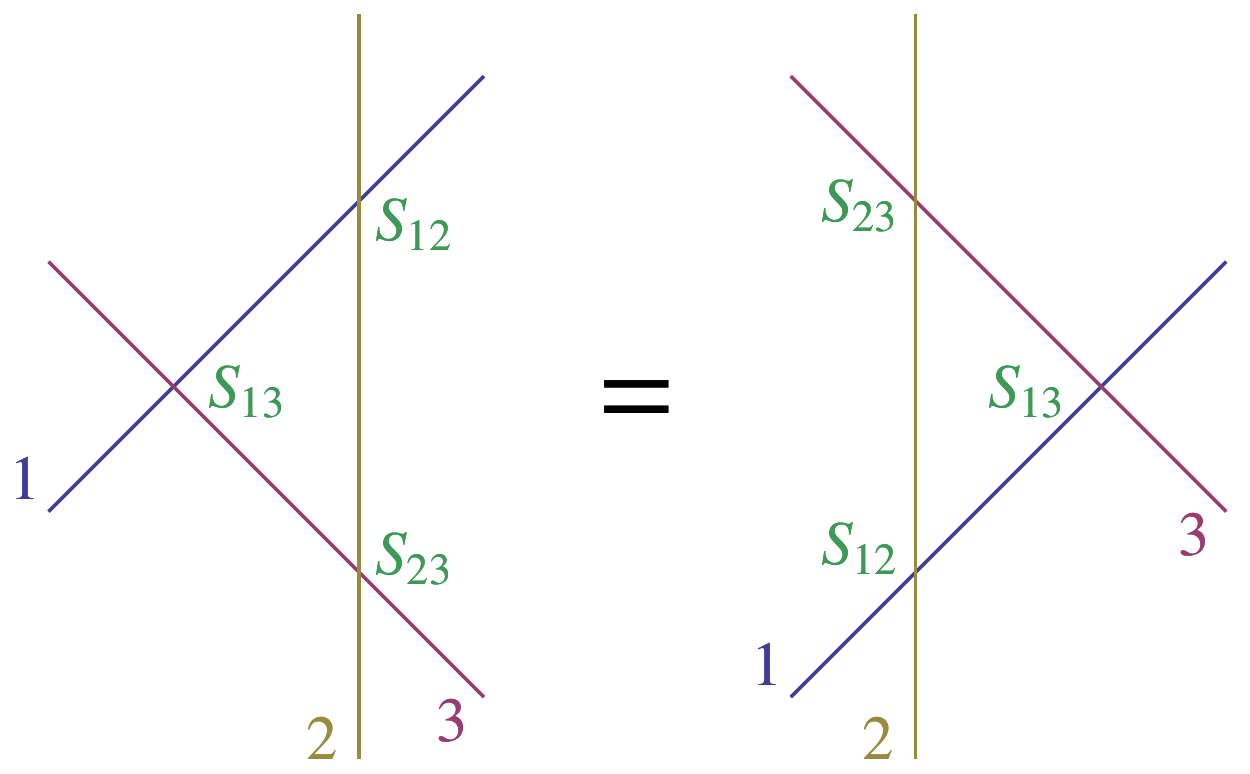}
\caption{The Yang-Baxter equation in diagrammatic form. Reading the diagram upwards, we obtain the Yang-Baxter equation \eqref{eq:YB}.}
\label{fig:YB}
\end{figure}

Now we are in a position to understand the origin of factorized scattering. Let us consider the three possible a priori different types of three body scattering illustrated in figure \ref{fig:3bodyscattering} (recall that there is no particle production). The higher spin conserved charges allow us to start with a given scattering process and consider a physically equivalent scattering process where the particles have been translated proportional to (powers of) their momenta. This means that if we were to start with the simultaneous three body scattering depicted in figure \ref{fig:3bodyscatteringb}, we can act with the higher spin conserved charges to obtain either the scattering picture in figure \ref{fig:3bodyscatteringa} or \ref{fig:3bodyscatteringc}, since the particles get translated by different amounts depending on their momenta.\footnote{If any of the particles have equal momenta they were either not well separated in the initial state or will never scatter with each other.} In other words, in an integrable model these three types of scattering are physically equivalent. Hence, by considering the problem in either pictures \ref{fig:3bodyscatteringa} or \ref{fig:3bodyscatteringc} it is clear that the three to three body scattering matrix can be written as a product of two body scattering matrices; it \emph{factorizes}. The equivalence of scattering pictures \ref{fig:3bodyscatteringa} and \ref{fig:3bodyscatteringc} then gives us a consistency constraint on the two body scattering matrix which is known as the Yang-Baxter equation
\begin{equation}
\label{eq:YB}
S_{12} S_{13} S_{23} = S_{23} S_{13} S_{12} \, ,
\end{equation}
a constraint on the matrix structure of the two body S-matrix. Writing the amplitudes for scattering processes $a(p_1) + b(p_2) \rightarrow d(p_2) + c(p_1)$ where $p_1>p_2$ as $S_{ab}^{cd}(p_1,p_2)$, it explicitly reads
\begin{equation}
S_{a b}^{lm}(p_1,p_2) S_{p c}^{a n}(p_1,p_3) S_{qr}^{b c}(p_2,p_3) = S_{e f}^{mn}(p_2,p_3) S_{dr}^{l f}(p_1,p_3) S_{pq}^{de}(p_1,p_2)\, .
\end{equation}
Note that this is an overdetermined set of equations; for $n$ particle types there are $n^6$ equations for $n^4$ amplitudes. For future reference, this relation can be represented diagrammatically as in figure \ref{fig:YB}. Finally, it should be clear that by the same considerations the scattering of more than three particles also factorizes. The Yang-Baxter equation guarantees equivalence of the multiple ways of factoring any many body scattering event.

\begin{figure}%
\centering
\includegraphics[width=8cm]{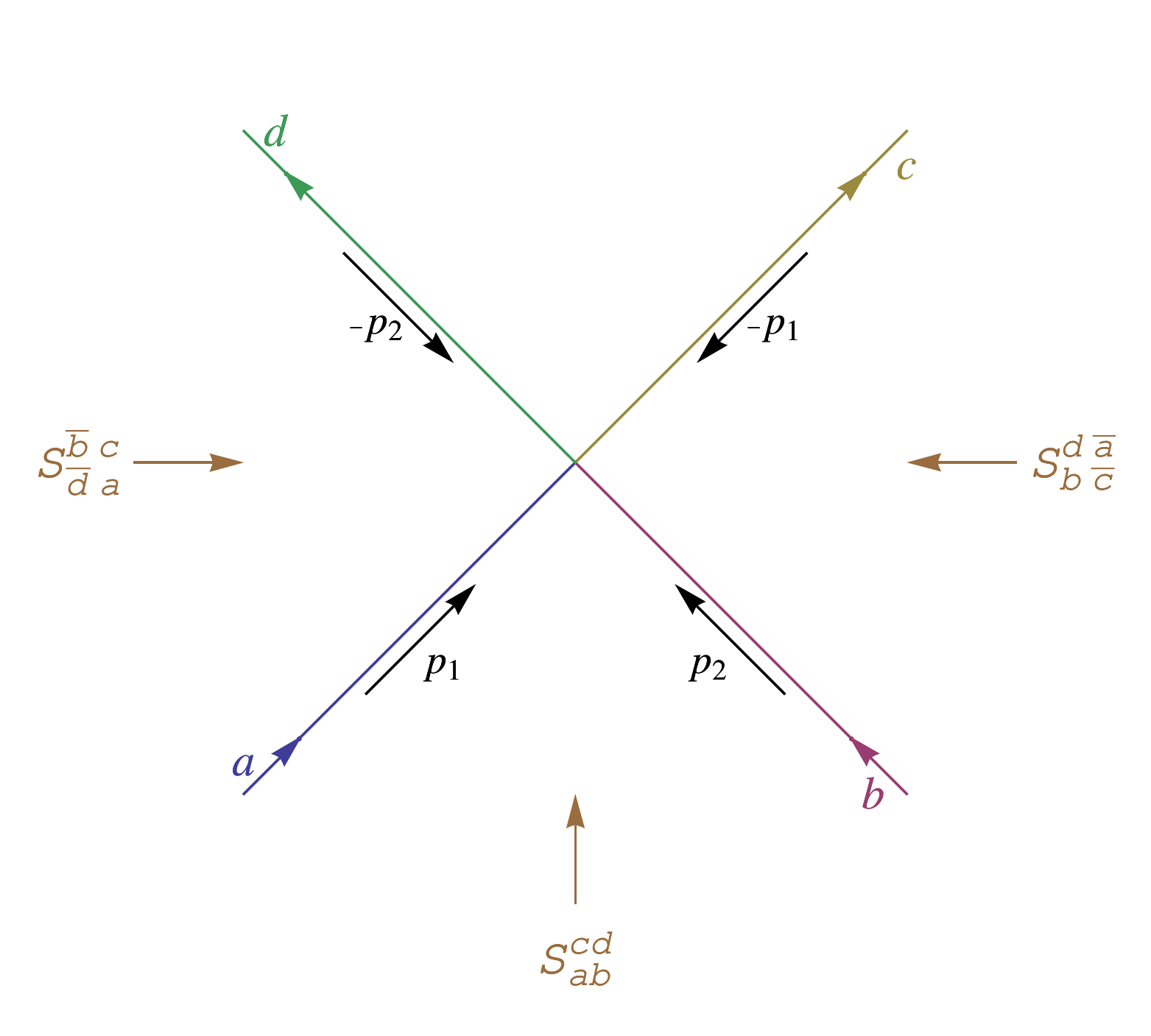}
\caption{Crossing states that analytically continuing the momenta in a scattering amplitude appropriately we can obtain a crossed scattering amplitude. If we cross one particle from ingoing to outgoing and one from outgoing to ingoing we obtain the same amplitude except viewed sideways (where time runs in the direction of the arrow), with particles travelling backwards in time interpreted as anti-particles with opposite momenta. For example, viewing the above scattering event $S_{ab}^{cd}$ from the left instead of the bottom we scatter $\bar{d}(-p_2)$ ($d(p_2)$ moving backwards in time) with $a(p_1)$ to get $\bar{b}(-p_2)$ and $c(p_1)$.}
\label{fig:crossing}
\end{figure}

The above considerations tell us that we can determine the complete S-matrix of an integrable quantum field theory from its two body S-matrix. This two-body S-matrix is then constrained by the Yang-Baxter equation \eqref{eq:YB}. It should also satisfy a number of physical constraints, namely\footnote{We have and will keep an explicit dependence on individual momenta rather than the relativistic Mandelstam variables, as we are going to apply the same ideas to a non-relativistic model in the next chapter.}
\begin{itemize}
\item{(Physical) unitarity\\
\begin{equation}
\label{eq:physicalunitarity}
S^\dagger S = 1 \, ,
\end{equation}}
\item{Hermitian analyticity\\
\begin{equation}
\label{eq:heritiananalyticity}
S_{12}^\dagger(p_1,p_2) = S_{21}(p_2^*,p_1^*)\, ,
\end{equation}}
\item{Crossing symmetry\footnote{This equation should be understood with the usual $i \epsilon$ prescription in mind.}\\
\begin{equation}
\label{eq:crossinggeneneral}
S_{\bar{d}a}^{\bar{b}c}(-p_2,p_1) = S_{ab}^{cd}(p_1,p_2) = S_{b\bar{c}}^{d\bar{a}}(p_2,-p_1) \, ,
\end{equation}}
\item{Compatibility with internal symmetries\\
\begin{equation}
\label{eq:compatibilitywithsymmetries}
S J = J S\, .
\end{equation}}
\end{itemize}
The first of these constraints follows from insisting on unitary time evolution in our theory, and takes this particularly simple form for the S-matrix of an integrable field theory because there are only two-particle intermediate states. Hermitian analyticity is a fact from relativistic S-matrix theory that can be readily seen in perturbative calculations; the Feynman amplitudes manifestly have this property. Crossing symmetry is the statement that by analytically continuing the momenta in a scattering amplitude appropriately we get the scattering amplitude where some of the incoming (outgoing) particles have been crossed to outgoing (incoming) anti-particles and vice versa. As a mnemonic for the formulae we can say that we can look at a scattering process from any side we please, as illustrated in figure \ref{fig:crossing}.
Of course the S-matrix should be compatible with the symmetries of the theory, and hence must intertwine the action of the symmetry generators $J$ on two-particle states before and after scattering.\footnote{This can of course be familiarly written as a commutator when the action of $J$ on two-particle states is symmetric.} Finally, we note that physical unitarity and hermitian analyticity together imply
\begin{itemize}
\item{Unitarity\\
\begin{equation}
\label{eq:unitarity}
S_{12}(p_1,p_2) S_{21}(p_2,p_1) = 1 \, .
\end{equation}}
\end{itemize}
This last property signifies a type of unitarity at the level of the algebra of scatterings; scattering two particles and immediately scattering them back does nothing. Of course this second step requires analytically continuing the S-matrix $S(p_i,p_j)$ to values $p_i<p_j$, which is why it is equivalent to physical unitarity provided the S-matrix is hermitian analytic.

\begin{figure}%
\centering
\includegraphics[width=8cm]{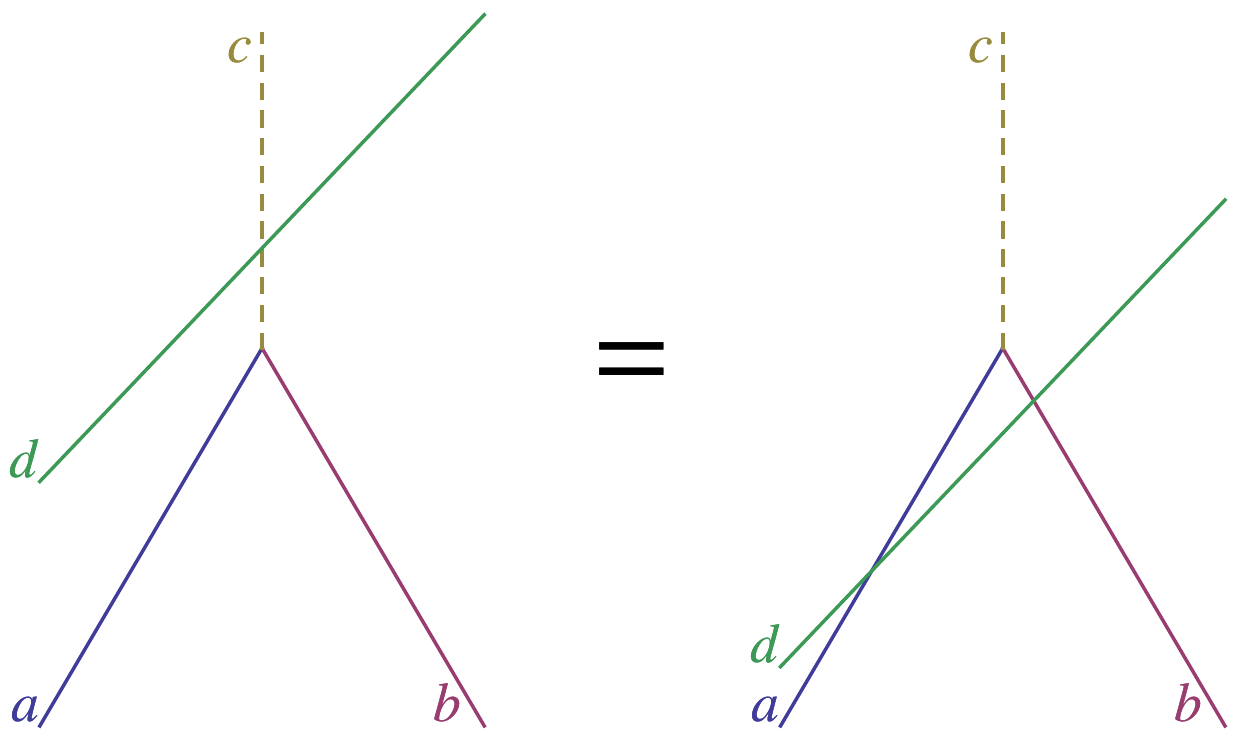}
\caption{Bound state scattering as constituent scattering. If two particles $a$ and $b$ form a bound state $c$, thanks to factorized scattering (and analytic continuation) we can view the scattering of a particle $d$ with the bound state $c$ (left) equivalently as the sequential scattering of particle $d$ with the bound state constituents $a$ and $b$.}
\label{fig:Boundstatescat}
\end{figure}

Typically we can fix the matrix structure of the S-matrix uniquely by (a subset of) these relations, leaving a scalar factor to be determined. This phase is constrained by crossing symmetry and we can fix it uniquely under some mild physical assumptions about the pole structure of the S-matrix, itself closely related to but not uniquely fixed by the particle spectrum of the theory. Furthermore, if we have physical input that the fundamental particles can form bound states, there should be corresponding poles in their S-matrix. Conversely, if the S-matrix has appropriate poles they may signal bound states whose presence we did not fully appreciate before. We can determine the S-matrices for such bound states by appealing to factorized scattering again. If we consider two particles that will join to form a bound state, by considering two equivalent scattering processes, one before and after this has happened, we can obtain the bound state S-matrix as a product of the constituent S-matrices. This procedure goes under the name of bootstrap, where the bound state S-matrix is said to be obtained as the fusion of constituent S-matrices, and can be schematically illustrated as in figure \ref{fig:Boundstatescat}. Note however that most of the particles that fuse to form a bound state will necessarily have complex momenta, meaning that the physically intuitive picture we just painted is not very physical at all. Still, if our theory behaves nicely under analytic continuation this should give the correct result.\footnote{In section \ref{subsec:TBAgeneral} we will discuss how this picture of bound states arises also at the level of the Bethe-Yang equations, offering a different and in some sense more physical point of view.}
Putting all this together leaves us in the remarkable position of knowing the S-matrix of a quantum field theory exactly.

\section{Integrable quantum field theory on a big circle}

Provided we can indeed fix the two body S-matrix exactly, we can say that we are done, at least as far as our theory on two dimensional Minkowski space is concerned; we have a continuous spectrum of scattering states whose scattering we can describe exactly. However as mentioned in the introduction already, we are interested in the spectrum of our quantum field theory when it is defined on a cylinder. When the particles move on a circle there are strictly speaking no asymptotic states, since particles cannot be infinitely separated from each other. Still, we would like to use the powerful results we obtained in the previous section.

To be able to talk about scattering and asymptotic states, we will first consider the theory in the asymptotically large volume limit. By taking the circumference $L$ to infinity, up to periodic boundary conditions the cylinder decompactifies to a plane and we can apply the ideas of factorized scattering. At the same time the periodic boundary conditions will result in momentum quantization, no matter how large $L$. This can be immediately seen for free particles (standing waves), where the periodicity of an $N$-particle wave-function in each argument implies that the momenta of the particles are quantized
\begin{equation}
e^{i p_j L} = 1 \implies p_j = \frac{2 \pi n_j}{L} \, .
\end{equation}

In an integrable quantum field theory multi particle states are still sensible objects to consider because there is no particle production. Putting the theory on a circle, the momenta of the individual particles will be quantized just as happens for free theories. To see this let us consider an $N$-particle state with totally ordered momenta $p_1>p_2>\ldots>p_N$ where the associated particles have positions $x_1\ll x_2 \ll \ldots \ll x_N$ in the infinite past.\footnote{More precisely we assume there is a finite interaction range between the particles and that their separation is large with respect to it.} As long as these particles remain well-separated we can describe this state by a wave-function for $N$ free particles. As time passes and the particles approach each other this wave-function picture breaks down completely;\footnote{While there is no real particle creation, virtual pair creation is not prohibited.} the particles scatter. However, due to the factorized scattering structure in our integrable model we know that this process can be described as a sequence of \emph{pair-wise} scattering events \emph{without particle production}. This means that between each of these pair-wise scattering events we again have $N$-particle states of approximately free particles we can describe by a wave-function, the transition between the regions given by the appropriate two-body S-matrix. There are $N!$ possible states corresponding to the orderings of the particles on the line, and the wave-function for each of these can be obtained by multiplication of the wave-function of the incoming state by a product of two-body S-matrices that accomplishes the desired ordering via pairwise permutations (scatterings). Note that the Yang-Baxter equation guarantees that the resulting wave-function is unique.\footnote{We are constructing a wave-function for $N!$ orderings even though in a given physical scattering event we will only encounter $N(N-1)/2$ of them by pairwise scattering. This is like one dimensional quantum mechanical scattering off of a potential, where in the time-independent setting we set up a wave-function on both sides of the potential, while physically it may be zero on either side.} As usual the time-independent wave-function in each regime contains both incident (transmitted) and scattering (reflected) contributions.

Concretely, the wave-function is a superposition of free waves in the different separated and ordered regimes. To take care of (anti-)symmetrization for identical particles while not obscuring the physical scattering picture above, let us work in terms of creation and annihilation operators $a$ and $a^\dagger$ and describe the wave-function as
\begin{equation}
\label{eq:wavefngen}
\psi_{i_1 \ldots i_N} (x_1,\ldots,x_N) = \langle  0 | a_{i_1} (x_1) \ldots a_{i_N}(x_N) | \psi(p_1,\ldots,p_N) \rangle\, ,
\end{equation}
where
\begin{align}
\label{eq:wavefnket}
| \psi(\{p\}) \rangle =\int d^N y \hspace{-2pt} \sum_{\mathcal{P} \in S_N}\hspace{-2pt} A^{\mathcal{P}}_{j_1\ldots j_N}(\{p\})\, e^{i p_{\mathcal{P}_m} y_m} \theta(y_1 \ll \ldots \ll y_N)  a^\dagger_{j_1}(y_1) \ldots a^\dagger_{j_N}(y_N) |0\rangle\, ,
\end{align}
and the indices on $a$ and $a^\dagger$ label the various possible particle types, summed over where repeated. The coefficients $A$ are related between each other as $A^{\mathcal{P}^\prime}=S_{\mathcal{P}_i,\mathcal{P}_{i+1}} \cdot A^\mathcal{P}$ when $\mathcal{P}^\prime$ is obtained from $\mathcal{P}$ by the permutation of elements $i$ and $i+1$, corresponding to particles $\mathcal{P}_i$ and $\mathcal{P}_{i+1}$. As a matrix $S$ of course acts on the tensor structure of $A$ appropriately. The above relations between the coefficients $A$ represent precisely the scattering picture described above.

For two identical particles that only scatter between themselves this gives for example
\begin{align}
\label{eq:twoidpartwavefn}
\begin{split}
\psi_{x_1\ll x_2} & = e^{i (p_1 x_1+p_2 x_2)} + S(p_1,p_2) e^{i (p_2 x_1+p_1 x_2)} \, ,\\
\psi_{x_1\gg x_2} & = e^{i (p_2 x_1+p_1 x_2)} + S(p_1,p_2) e^{i (p_1 x_1+p_2 x_2)} \, ,
\end{split}
\end{align}
where $S$ is simply the scattering phase, as there is no distinction between reflection and transmission for identical particles. Note that while it may not manifestly look like it because of symmetrization, the wave-functions in the two regimes are most definitely obtained from one another by multiplication by the S-matrix, as is clear from \eqref{eq:wavefnket}.

At this point we should recall that we mean to consider this picture on a circle, albeit a very large one. On a circle, the absolute ordering of particles of course becomes meaningless, since we can only distinguish between cyclically inequivalent orderings. Concretely, let us consider wave-function for the ordering $x_1 \ll x_2 \ll \ldots \ll x_N$. As the separation between $x_1$ and $x_N$ is by definition smaller than $L$ we can write $y_1 = x_1+L$ where $x_2 \ll \ldots \ll x_N \ll y_1$. In this regime we have an alternate description of the wave-function, but since these two domains are now equivalent we should identify the wave-functions. The resulting equations are known as the Bethe-Yang equations.\footnote{While these equations and wave functions are of Bethe ansatz form, here there is strictly speaking no ansatz and the result is only correct in the asymptotic large volume limit, hence the different name.} For a theory with a single type of particle this means
\begin{equation}
\label{eq:BEphase}
e^{i p_m L}\prod_{n\neq m} S_{mn} = 1\, ,
\end{equation}
as follows by writing out the wave-function in eqs. (\ref{eq:wavefngen},\ref{eq:wavefnket}) explicitly, taking into account the fixed ordering. The same conditions can be obtained by shifting any of the other coordinates. For a theory with multiple kinds of excitations, the particle content on both to-be-identified ends of the line should be matched as this is no longer trivially the case. In line with eqn. \eqref{eq:wavefnket}, eqn. \eqref{eq:BEphase} becomes
\begin{equation}
\label{eq:BEmat}
e^{i p_m L} S_{m,m+1} \ldots S_{m,M} S_{m,1} \ldots S_{m,m-1} A^{\mathcal{I}} = A^{\mathcal{I}}\, ,
\end{equation}
where $\mathcal{I}$ denotes the identity permutation. Note that since $S$ is now a matrix the ordering in the product is essential.\footnote{Up to a redefinition of the arbitrary tensor $A$, the product of S-matrices in the above formula is just the cyclic product of S-matrices starting at $m$.}
This is a natural point to introduce some further notions and notations to work efficiently with this matrix structure. The product of S-matrices in the above equation can be expressed through the so-called transfer matrix $T$, the (super-)trace of the monodromy matrix $\mathcal{T}$
\begin{equation}
\label{eq:transfermatrixdef}
T(q|\{p_j\}) \equiv \mbox{Tr}_a \mathcal{T}_a(q|\{p_j\}) \equiv \mbox{Tr}_a \prod_{i=1}^N S_{a,i}(q,p_i)\, ,
\end{equation}
where we have introduced an auxiliary particle living in an auxiliary space $a$ with momentum $q$, the trace is over this auxiliary space, and the product runs from left to right. Noting that the S-matrix at coincident arguments typically reduces to minus the permutation operator, $S_{ab}(p,p) = -P_{ab}$, it follows that
\begin{equation}
\label{eq:transferphysical}
T(p_1|\{p_j\}) = -\mbox{Tr}_a P_{a1} \prod_{i=2}^N S_{a,i}(p_1,p_i) = -\mbox{Tr}_a \prod_{i=2}^N S_{1,i}(p_1,p_i) P_{a1} =  -\prod_{i=2}^N S_{1,i}(p_1,p_i)\, ,
\end{equation}
by noting that $P_{ab}^2 = P_{ab} P_{ba} = 1_{ab}$ and $\mbox{Tr}_a P_{ab} = 1_{b}$. The right most expression is just minus the product of S-matrices we have in eqs. \eqref{eq:BEmat} for the first particle. Similarly, by cyclicity of the trace and the fact that the product in eqn. \eqref{eq:transfermatrixdef} runs over distinct particles (spaces), we find that in general the product of S-matrices in eqs. \eqref{eq:BEmat} is nothing but $-T(p_m|\{p_j\})$. Hence we see that the transfer matrix indeed represents the transfer of a particle around the circle.

\begin{figure}%
\centering
\includegraphics[width=0.5\textwidth]{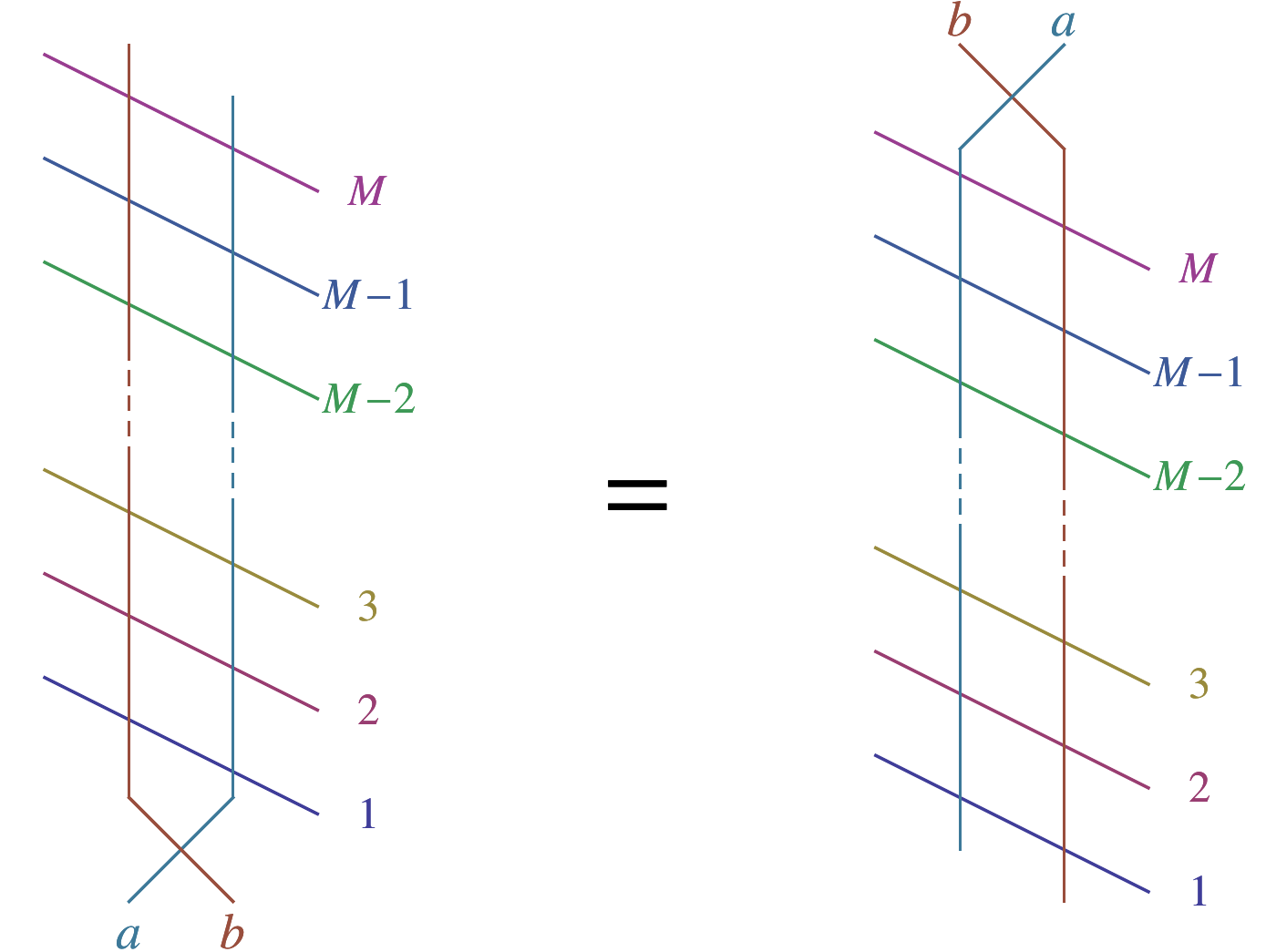}
\caption{A diagrammatic form of the fundamental commutation relations. Reading this diagram with time going up and associating an S-matrix $S_{xy}$ to the crossing of lines $x$ and $y$ (as in figure \ref{fig:YB}), the left diagram represents $S_{ab}\mathcal{T}_a \mathcal{T}_b$ (noting that $[S_{ai},S_{bj}]=0$ for $a \neq b$ and $i \neq j$). Pulling the numbered lines through the crossing of $a$ and $b$ - giving an equivalent picture by the Yang-Baxter equation (see figure \ref{fig:YB}) - we obtain the right diagram which reads $\mathcal{T}_b \mathcal{T}_a S_{ab}$.}
\label{fig:FCR}
\end{figure}

The essential property of the transfer matrix it that it commutes with itself for different values of the auxiliary momentum
\begin{equation}
\label{eq:Transfermatrixcomm}
[T(q|\{p_j\}),T(\tilde{q}|\{p_j\})] = 0\, .
\end{equation}
This is a purely algebraic property following from the Yang-Baxter equation, and does not depend on whether the $p_j$ satisfy the Bethe equations. Explicitly, it follows from the \emph{fundamental commutation relations}
\begin{equation}
\label{eq:FCR}
S_{ab}(q,\tilde{q}) \mathcal{T}_a(q|\{p_j\}) \mathcal{T}_b(\tilde{q}|\{p_j\}) =  \mathcal{T}_a(q|\{p_j\}) \mathcal{T}_b(\tilde{q}|\{p_j\}) S_{ab}(q,\tilde{q})\, ,
\end{equation}
by taking a trace over the auxiliary spaces $a$ and $b$. The fundamental commutation relations are a direct consequence of the Yang-Baxter equation as illustrated in figure \ref{fig:FCR}.
Finally then, we see that by commutativity of the transfer matrices for arbitrary auxiliary momentum, they can be simultaneously diagonalized in a fashion independent of this momentum. In other words, if we can diagonalize $T(q|\{p_j\})$, the Bethe-Yang equations \eqref{eq:BEmat} become
\begin{equation}
\label{eq:BEeigenval}
e^{i p_m L} \Lambda(p_m|\{p_j\}) =-1\, ,
\end{equation}
where $\Lambda(p_m|\{p_j\})$ denotes any of the eigenvalues of $T(p_m|\{p_j\})$. As it turns out, $T(q|\{p_j\})$ can be diagonalized by a Bethe ansatz for a discrete integrable system (no asymptotic volume limits required). If we do this by making an explicit ansatz for the wave-function of the form (\ref{eq:wavefngen},\ref{eq:wavefnket}), except on a lattice and with a one rank lower S-matrix of course, this goes under the name of the nested coordinate Bethe ansatz \cite{Yang:1967bm}. By going down in rank multiple times if necessary, eventually we end up with a simple scattering phase so that the coordinate Bethe ansatz directly applies. Alternatively we can approach the problem in a algebraic fashion based on the fundamental commutation relations \eqref{eq:FCR}; commutation relations between elements of the monodromy matrix. This goes under the name of the algebraic Bethe ansatz \cite{Faddeev:1979gh}, which we will explicitly demonstrate on a simple example just below.

Before moving on, let us note that we do not have to restrict ourselves to periodic boundary conditions. In fact we will soon be interested in quasi-periodic (also called twisted) boundary conditions, however completely general quasi-periodic boundary conditions are not necessarily compatible with factorized scattering. Intuitively we would like the wave-function obtained by considering scattering before crossing the quasi-periodic boundary to be the same as the one obtained by scattering afterwards since in a factorized scattering picture the location of a scattering event can be shifted. In other words if the quasi-periodic boundary conditions are given by
\begin{equation}
\label{eq:genquasiperbcs}
g_j \cdot \psi(x_1,\ldots,x_j+L,\ldots,x_N) = \psi(x_1,\ldots,x_j,\ldots,x_N) \, , \hspace{20pt} j =1,\ldots,N\, .
\end{equation}
where $g_j$ denotes the operator $g$ acting on the $j$th particle, we would like these operators to commute with the S-matrix,
\begin{equation}
\label{eq:BCsymmetryofS}
[g_i \otimes g_j,S_{ij}] =0\, .
\end{equation}
In this case the transfer matrix can be naturally generalized to the \emph{twisted} transfer matrix
\begin{equation}
T^g(q|\{p_j\}) \equiv \mbox{Tr}_a \mathcal{T}^g_a(q|\{p_j\}) = \mbox{Tr}_a \, g_a \prod_{i=1}^N S_{a,i}(q,p_i)\, ,
\end{equation}
where $g_a$ acts in the auxiliary space $a$ so that at $q=p_k$ this indeed gives the desired boundary condition (compare to \eqref{eq:transferphysical}). The condition \eqref{eq:BCsymmetryofS} precisely guarantees that also these twisted monodromy matrices satisfy the fundamental commutation relations \eqref{eq:FCR} and the transfer matrices commute \cite{Sklyanin:1988yz}.\footnote{This follows immediately by drawing an extra (dotted) line below line $1$ in figure \ref{fig:FCR} representing the tensor $g_a g_b$.} There are more general types of boundary conditions that are compatible with integrability, but we will not need them here.

In order to derive the Bethe-Yang equations we assumed that our cicle is big so that the notion of an S-matrix makes approximate sense. These equations capture the $1/L$ corrections to the energy of a state in finite volume, just as the energy of a free particle on a circle depends on $1/L$. Unlike the free particle however, when there are interactions there are corrections to these quantization equations, and this is not the whole story. Still, since these corrections are exponentially suppressed in the volume\footnote{We will discuss this in some detail in section \ref{sec:luscherandexcitedstatesgeneral}.} the Bethe-Yang equations provide a good approximation to the exact result for large volumes.

\section{The large volume spectrum of the chiral Gross-Neveu model}

In this section we will introduce the chiral Gross-Neveu model as a simple example of an integrable field theory by means of which we can clearly illustrate important features we will encounter in the later chapters.

\subsection{The chiral Gross-Neveu model}

The $\mbox{SU}(N)$ chiral Gross-Neveu model is a model of $N$ interacting Dirac fermions with Lagrangian\footnote{Our $\gamma$ matrices are defined as $\gamma_0 = \sigma_1$, $\gamma_1 = i \sigma_2$, $\gamma_5 = \gamma_0 \gamma_1$, where $\gamma_{0,1}$ form the Clifford algebra $\{ \gamma_\mu, \gamma_\nu \} = 2 \eta^{\mu \nu}$ with $\eta = \mbox{diag}(1,-1)$. Note that $\gamma_5$ is Hermitian. As usual $\bar{\psi} = \psi^\dagger \gamma_0$ and $\slashed{\partial} = \gamma^\mu \partial_\mu$.}
\begin{equation}
\mathcal{L}_{cGN} = \bar{\psi}_a i\slashed{\partial} \psi^a + \frac{1}{2} g_s^2\left( (\bar{\psi}_a \psi^a)^2- (\bar{\psi}_a \gamma_5 \psi^a)^2\right) - \frac{1}{2} g_v^2 (\bar{\psi}_a \gamma_\mu \psi^a)^2\, ,
\end{equation}
where $a=1,\ldots,N$ labels the $N$ Dirac spinors. This Lagrangian has $\mbox{U}(N) \times \mbox{U}(1)_c$ symmetry, where viewed as an $N$-component vector the spinors transform in the fundamental representation of $\mbox{U}(N)$, and $\mbox{U}(1)_c$ denotes the chiral symmetry $\psi \rightarrow e^{i \theta \gamma_5} \psi$. Typically the chiral Gross-Neveu Lagrangian is written with $g_v=0$.\footnote{The $\mbox{SU}(N)$ chiral Gross-Neveu model is often used interchangeably with the $\mbox{SU}(N)$ invariant Thirring model. The latter model has Lagrangian
\begin{equation}
\mathcal{L}_{ITM} = \bar{\psi}_a i\slashed{\partial} \psi^a - \frac{1}{4} g^2 (\bar{\psi}_a \lambda_{ab} \gamma_\mu \psi^b)^2 - \frac{1}{2} h^2 (\bar{\psi}_a \gamma_\mu \psi^a)^2\, ,
\end{equation}
where $\lambda_{ab}$ are the generalized Gell-Man matrices for $\mbox{SU}(N)$. Like the chiral Gross-Neveu model, the Lagrangian for this model is typically written with $h=0$. These models are physically equivalent. For $h$ nonzero the Lagrangians are manifestly equivalent by Fierz transformations, under the identifications $g_s^2 = g^2$ and $g_v^2 = h^2 - \frac{1}{N} g^2$ \cite{Forgacs:1991nk}. As mentioned in the main text below, the massless modes associated to the $\mbox{U}(1)$ symmetry decouple from the massive interacting fermions. Because of this both models are physically equivalent even with $h=0$ and $g_v=0$, which is presumably the reason for the sometimes cavalier treatment of the corresponding terms and nomenclature.}

This model was initially studied as it shares some features with QCD, namely it is asymptotically free and has dynamical mass generation. The subtle point however is that chiral symmetry naively prevents the fermions from gaining mass, as there is no spontaneous symmetry breaking in two dimensions \cite{Coleman:1973ci}. Yet this was precisely what is observed in the large $N$ limit \cite{Gross:1974jv}, with even a massless particle present in the spectrum. The resolution of this apparent mismatch is nicely discussed in \cite{Witten:1978qu}, the upshot being that the physical fermions in this model carry no chirality, and that while there is a chiral massless particle in the spectrum it is not a Goldstone boson. The full spectrum of this theory contains $N-1$ $\mbox{SU}(N)$ multiplets of interacting massive fermions, and massless excitations with $\mbox{U}(1)$ charge that decouple completely \cite{Witten:1978qu,Andrei:1979sq,Moreno:1987np}. Interestingly, this model is related to the Sine-Gordon model we briefly mentioned in section \ref{sec:classint} at $\beta=8\pi$ by bosonization \cite{Banks:1975xs}. This model possesses an infinite number of conserved charges \cite{Abdalla:1991vua}, and is integrable in the sense we just discussed.

Before moving on we should introduce a bit of notation. The conventional variables in relativistic scattering theory are the Mandelstam variables $s$, $t$ and $u$, but only one of these is independent in two dimensions. Hence we can write the S-matrix as a function of $s=(p_1+p_2)^2$. Without going into details, continuing $s$ into the complex plane the S-matrix has branch cuts from $(m_1+m_2)^2$ to positive infinity, and $(m_1-m_2)^2$ to negative infinity (see e.g. \cite{Dorey:1996gd}). By definition, the physical scattering values lie just above the first of these branch cuts, thereby singling out this copy of the $s$-plane as the physical one, known as the physical sheet. It turns out to be possible to resolve these branch cuts by introducing a (relative) rapidity variable $u$ as\footnote{We choose this unconventional normalization of $u$ for uniformity between the relativistic case here and the non-relativistic spin chains and superstring we will treat later.}
\begin{equation}
s = m_1^2 +m_2^2 + 2m_1m_2 \cosh\frac{\pi u}{2}\, ,
\end{equation}
in terms of which the physical copy of the $s$-plane is mapped to the physical strip $\mbox{Im}(u) = (0,2i)$. This variable $u=u_1-u_2$ is nothing but the difference of the physical rapidity variables of the particles, related to energy and momentum as
\begin{equation}
E_i = m_i \cosh{\frac{\pi u_i}{2}}\, , \, \, \,
p_i = m_i \sinh{\frac{\pi u_i}{2}}\, ,
\end{equation}
where $m$ is the mass of the fermions. Branch cut issues aside, another reason to introduce these variables is that Lorentz boosts act additively on the rapidity, and therefore by Lorentz invariance the two-body S-matrix is a function of the difference of the particles' rapidities only. Furthermore, note that this parametrization uniformizes the relativistic dispersion relation
\begin{equation}
E^2 - p^2 = m^2\, ,
\end{equation}
meaning that it is completely parametrized by $u$. Note that in these variables two-particle bound states have $u \in (0,2i)$, corresponding to $s \in ((m_1-m_2)^2,(m_1+m_2)^2)$ on the \emph{physical} sheet. We should also mention that crossing should take us from physical values of $s$ to physical values of $t$ which by the usual $i \epsilon$ prescription corresponds to $u \rightarrow 2i-u$.\footnote{Using unitarity we can write crossing for the first particle as $S_{12}(u_1,u_2) S_{\bar{1}2}(u_1+2i,u_2) =1$ and as $S_{12}(u_1,u_2) S_{1\bar{2}}(u_1,u_2-2i) =1$. for the second.}

Coming back to our model, for simplicity we will focus on the case $N=2$ where we have only one $\mbox{SU}(2)$ multiplet of interacting fermions of mass $m$. Unitarity, $\mbox{SU}(2)$ symmetry, the crossing relations, and requiring a minimal number of poles and zeros of the S-matrix,\footnote{This means there are no superfluous poles or zeroes for the S-matrix in the physical strip. Note that in accordance with the fact that there are no bound states for $N=2$ (the bound states correspond to the other $N-2$ multiplets) there are no poles within the physical strip.} together fix the two body S-matrix of this model to be \cite{Zamolodchikov:1978xm}\footnote{We write $R^{-1}$ so that $R$ is the canonical rational $\mbox{SU}(2)$ invariant R-matrix.}
\begin{equation}
\label{eq:cGNSmatrix}
S_{ab}(u) = S^{ff}(u) R^{-1}_{ab}(u)\, ,
\end{equation}
where
\begin{equation}
\, S^{ff}(u) = -\frac{\Gamma(1-\frac{u}{4i})\Gamma(\frac{1}{2}+\frac{u}{4i})}{\Gamma(1+\frac{u}{4i})\Gamma(\frac{1}{2}-\frac{u}{4i})}\, , \, \, \, \mbox{and} \, \, \, R(u)=  \frac{1}{u+2i}\left(u 1_{ab} + 2i P_{ab} \right),
\end{equation}
where $P$ is the permutation operator.\footnote{For a nice discussion of how these typical ratios of $\Gamma$-functions arise by an interplay of unitarity and crossing we refer the reader to \cite{Shankar:1977cm}.} As we saw above, to get the Bethe-Yang equations from this S-matrix we will first need to diagonalize the associated transfer matrix
\begin{equation}
\label{eq:GNtransfer}
T^g(u|\{u_i\}) = \mbox{Tr}_a \mathcal{T}^g_a(u|\{u_i\}) = \mbox{Tr}_a \, g_a  \prod_{i=1}^{\NF} R^{-1}_{ai}(u-u_i)  = \mbox{Tr}_a  \, g_a \prod_{i=1}^{\NF} R_{ai}(u_i-u) \, ,
\end{equation}
where we take out the scalar factor $S^{ff}$ for soon-to-be-obvious reasons, we note that $R_{ab}(u) = R_{ba}(u)$ for this particular model, and we have allowed ourselves quasi-periodic boundary conditions for generality. $\NF$ denotes the number of physical fermions in a given state. As our S-matrix is $\mbox{SU}(2)$ symmetric any quasi-periodic boundary condition associated to a $g \in \mbox{SU}(2)$ is compatible with integrability. Without loss of generality we can restrict to elements of the Cartan subgroup since the eigenvalues of the transfer matrix depend only on these (see section \ref{sec:TTM} for details). As such we take $g$ to be of the form
\begin{equation}
\label{eq:cGNtwist}
g = \left(\begin{array}{cc} e^{i \alpha} & 0 \\ 0& e^{-i \alpha} \end{array} \right)\, .
\end{equation}
Note that these boundary conditions break the $\mbox{SU}(2)$ symmetry down to $\mbox{U}(1)$. We already mentioned that the transfer matrix for an integrable quantum field theory has an associated discrete integrable model. In this case \eqref{eq:GNtransfer} is the transfer matrix for the spin-$\frac{1}{2}$ Heisenberg XXX spin chain. As we will encounter more complicated integrable spin chains later, let us discuss this simple example in some detail.

\subsection{The XXX spin chain and the algebraic Bethe ansatz}

The Heisenberg XXX spin chain is described by the Hamiltonian
\begin{equation}
\label{eq:XXXham}
H = - \frac{J}{4} \sum_{i=1}^{\NF} \left(\vec{\sigma}_i \cdot \vec{\sigma}_{i+1}-1\right) \, ,
\end{equation}
where $\vec{\sigma}$ is the vector of Pauli matrices, and $\sigma_{\NF+1}=g^{-1}\sigma_{1}g$. This Hamiltonian acts on a Hilbert space given by $\NF$ copies of $\mathbb{C}^2$, one for each site $i$. Identifying $(1,0)$ as $|\hspace{-2pt} \uparrow \hspace{1pt} \rangle$  and $(0,1)$ as $|\hspace{-2pt} \downarrow \hspace{1pt} \rangle$ , states in this Hilbert space can be viewed as chains of spins, in this case quasi-periodic. This Hamiltonian \eqref{eq:XXXham} can be obtained from the transfer matrix \eqref{eq:GNtransfer} as \footnote{This follows by taking the derivative of $T$, using $P_{ij} P_{jk} = P_{ik} P_{ij}$ and $P_{ij}P_{ji}=1$, and representing the Hamiltonian as $H = -\frac{J}{2} \left( \sum_n^{\NF-1} (P_{n,n+1}  -1) + (g_1^{-1} P_{\NF,1} g_1 -1)\right)$.}
\begin{equation}
\label{eq:HfromT}
H = i J \frac{d}{du} \log T^g(u|\{0\})|_{u=0} \, .
\end{equation}
From the definition of the R-matrix and the transfer matrix it is not hard to see that the transfer matrix is a polynomial in $u$. Combining this with the commutativity of the transfer matrix, we can expand the transfer matrix in powers of $u$ to get a set of manifestly mutually commuting operators, all furthermore commuting with the Hamiltonian \eqref{eq:HfromT}. This makes contact with the notion of integrability we discussed for continuous systems in sections \ref{sec:classint} and \ref{sec:factscat}. The transfer matrix and hence the Heisenberg Hamiltonian can be diagonalized by the Bethe ansatz in either coordinate or algebraic form. Here we will briefly but explicitly illustrate the algebraic Bethe ansatz, as we will encounter it again. Note that we will be keeping nonzero $u_i$s in our transfer matrix since we are really considering the chiral Gross-Neveu model; this corresponds to an inhomogeneous spin chain.

The algebraic Bethe ansatz is based on the fundamental commutation relations \eqref{eq:FCR}. To write them out explicitly, we will first write the monodromy matrix $\mathcal{T}_g$ of equation \eqref{eq:GNtransfer} explicitly as a matrix in auxiliary space
\begin{equation}
\mathcal{T}^g(u|\{u_i\}) = \left( \begin{array}{cc} A(u) & B(u) \\ C(u) & D(u) \end{array}\right)\, ,
\end{equation}
where we suppressed the dependence on $u_i$ on the right hand side. Explicitly writing out the fundamental commutation relations directly gives (recall that we are working with $R^{-1}$)
\begin{equation}
\label{eq:FCRXXX}
\begin{aligned}
\mbox{$[$} B(v) , B(w) \mbox{$]$}  & =0 \, ,\\ %bs interference of [] with something in aligned it seems.
A(v)B(w) = f(v-w) B(w) &A(v) + g(v-w) B(v) A(w)\, ,\\
D(v)B(w) = f(w-v) B(w) &D(v) - g(v-w) B(v) D(w)\, ,
\end{aligned}
\end{equation}
where
\begin{equation}
f(u) = \frac{u+2i}{u} \, , \, \, \mbox{and}\,\, \, g(u)= -\frac{2 i}{u}\,.
\end{equation}
Note that these commutation relations do not contain $u_i$s. The crucial point is now to view these commutation relations as an analogue of the harmonic oscillator algebra $N a^\dagger = a^\dagger (N+1)$, $N a = a (N-1)$, with the number operator $N=a^\dagger a$. Concretely, we will make the ansatz that $B$ creates eigenstates of the transfer matrix $A+D$ by acting on a vacuum $|0\rangle$ which is annihilated by $C$. To show that such a vacuum exists, we note that the $R$-matrix acts upper triagonally in auxiliary space on the vector $|\hspace{-2pt} \uparrow \hspace{1pt} \rangle$ in space $i$ since
\begin{equation*}
R_{ai}(u) = R_{ia}(u) = \frac{1}{u+2i} \left(\begin{array}{cc:cc} u+2i & 0 & 0 & 0 \\ 0 & u & 2i & 0 \\ \hdashline 0 & 2i &u & 0 \\ 0 & 0 & 0 & u+2i \end{array}\right) = \frac{1}{u + 2 i}\left( \hspace{-2pt} \begin{array}{cc} u + i + i\sigma_3 & 2i \sigma_- \\ 2i \sigma_+ & u + i - i\sigma_3  \end{array} \hspace{-2pt} \right) ,
\end{equation*}
and $\sigma_+|\hspace{-2pt} \uparrow \hspace{1pt} \rangle = 0$.\footnote{Here as usual $\sigma_\pm = \frac{1}{2} (\sigma_1 \pm i \sigma_2)$} This means that if we take $|0\rangle = |\hspace{-2pt} \uparrow \ldots \uparrow \hspace{1pt} \rangle$ the monodromy matrix $\mathcal{T}$ acts as
\begin{equation}
\label{eq:XXXTonvacuum}
\mathcal{T} |0 \rangle = \left( \begin{array}{cc} A & B \\ C & D \end{array}\right) |0\rangle =  \left( \begin{array}{cc} e^{i \a}& \bullet \\ 0 &  e^{-i \a}\prod_i (f(u_i - u))^{-1} \end{array}\right) |0\rangle \, ,
\end{equation}
from which we see that $C |0 \rangle =0$, $A|0\rangle = e^{i \a}|0 \rangle$, and $ D | 0 \rangle = e^{-i \a}\prod_i (f(u_i - u))^{-1} |0\rangle $; the bullet point $\bullet$ denotes terms irrelevant for our considerations. In terms of field theory the existence of this vacuum simply means that there is a one-particle-type sector which is closed under scattering, which is of course obvious from the form of the S-matrix \eqref{eq:cGNSmatrix}. We will act with creation operators on this vacuum to construct the other eigenstates of the transfer matrix. Let us consider an arbitrary state obtained by acting with $\NA$ creation operators
\begin{equation}
| \{v_1,\ldots,v_\NA\} \rangle \equiv B(v_1)\ldots B(v_\NA) |0 \rangle\, ,
\end{equation}
and insist that this is in fact an eigenstate of the transfer matrix $A(u)+D(u)$. Using the commutation relations \eqref{eq:FCRXXX} we can commute $A(u)$ and $D(u)$ through to the vacuum where we understand their action. When we do so however, there will also be many contributions from the terms in the commutation relations which exchange rapidities. This means that unless these terms cancel between $A$ and $D$, we are not creating an eigenstate. Explicitly we have
\begin{equation}
\begin{aligned}
A(u)| \{v_1,\ldots,v_\NA\} \rangle &=  e^{i \a}\prod_{i=1}^\NA f(u-v_i)  B(v_1)\ldots B(v_\NA) |0 \rangle \\
& \hspace{30pt} +
 e^{i \a}\sum_j M_j B(v_1)\ldots \hat{B}(v_j) \ldots B(v_\NA) B(u) |0 \rangle \, ,\\
D(u)| \{v_1,\ldots,v_\NA\} \rangle &=  e^{-i \a}\prod_{i=1}^\NA f(v_i-u) \prod_{j=1}^\NF(f(u_j - u))^{-1} B(v_1)\ldots B(v_\NA) |0 \rangle  \\
& \hspace{30pt} + e^{-i \a}
\sum_j \tilde{M}_j B(v_1)\ldots \hat{B}(v_j) \ldots B(v_\NA) B(u) |0 \rangle \, ,
\end{aligned}
\end{equation}
where $M_j$ and $\tilde{M}_j$ are coefficients of the terms where $A$ respectively $D$ was commuted with the $j$th $B$ using the term in the commutation relations that exchanges the rapidities; the hat on $B$ denotes that the operator is removed from the product. Noting that $M_1$ and $\tilde{M}_1$ can be easily computed, and that by commutativity of the $B$s we can simply replace $v_1$ by $v_j$ in the result to generalize it, we get
\begin{equation}
\begin{aligned}
M_j & = g(u-v_j) \prod_{i\neq j}^\NA f(v_j-v_i)  \, ,\\
\tilde{M}_j & = -g(u-v_j) \prod_{i\neq j}^\NA f(v_i-v_j) \prod_{m=1}^\NF (f(u_m - v_j))^{-1}\, .
\end{aligned}
\end{equation}
Now we can cancel the unwanted terms against each other by insisting that the $v_j$ satisfy a set of so-called Bethe equations
\begin{equation}
\prod_{i\neq j}^\NA f(v_j-v_i)   =  e^{-2 i \a}\prod_{i\neq j}^\NA f(v_i-v_j)\prod_{m=1}^\NF (f(u_m - v_j))^{-1}\, .
\end{equation}
Shifting $\tilde{v}_l = v_l + i$ and dropping the tilde this can be rewritten as
\begin{equation}
\label{eq:XXXbethe}
\prod_{m=1}^\NF \frac{v_j-u_m-i}{v_j-u_m+i} = e^{-2 i \a} \prod_{i\neq j}^\NA \frac{v_j - v_i - 2i}{v_j-v_i + 2i}\, .
\end{equation}
The corresponding eigenvalue of the transfer matrix is given by
\begin{equation}
\label{eq:TXXXeigenvalues}
\lambda(u|\{v_j\},\{u_m\})=e^{i \a}\prod_{i=1}^\NA \frac{u-v_i+i}{u-v_i-i}+e^{-i \a}\prod_{m=1}^\NF \frac{u-u_m}{u-u_m-2i} \prod_{i=1}^\NA \frac{u-v_i-3i}{u-v_i-i}\, ,
\end{equation}
where again the $v_l$ have been shifted. Note that the Bethe equations also follow by insisting that the residues of the superficial poles in $\lambda$ at $u=v_j + i$ vanish, which must be the case since by construction $T$ has no poles there.\footnote{Obtaining the Bethe equations this way goes under the name of the analytic Bethe ansatz.} Solutions to the Bethe equations with repeating rapidities $u_i=u_j$ would require further constraints to be consistent and should in fact be discarded. In the framework of the coordinate Bethe ansatz this would be obvious as the ansatz for the wave-function manifestly vanishes in such situations.

We see that we can create eigenstates of the transfer matrix by acting with a number of creation operators on the vacuum, provided the associated rapidities satisfy eqn. \eqref{eq:XXXbethe}. To give a physical interpretation to these eigenstates let us consider their quantum numbers. Using eqn. \eqref{eq:HfromT} we can easily find the spin chain energy of a general eigenstate to be
\begin{equation}
\scE(\{v_l\}) = -i J \frac{d}{du} \log \lambda(u|\{v_j\},\{0\})|_{u=0} = 2 J \sum_i \frac{1}{v_i^2+1}\, ,
\end{equation}
where we note that the vacuum has zero energy. We see that the energy is a sum of different contributions dependent on $v_i$ only. Similarly, we can consider the spin chain momentum identified through\footnote{As defined $e^{i\scp}$ translates the spins by one site in the positive direction; $e^{i\scp} \vec{a}_n e^{-i \scp} = \vec{a}_{n+1}$ with $\vec{a}_j$ a vector in $\mathbb{C}^2$ in space $j$.}
\begin{equation}
\scp = i \log \lambda (0|\{0\})\, ,
\end{equation}
which gives
\begin{equation}
\label{eq:XXXmomentum}
\scp(\{v_j\}) = i \sum_j \log \frac{v_j-i}{v_j+i}\, ;
\end{equation}
again a sum of individual contributions.\footnote{Note also how this allows us to write the left hand side of eqn. \eqref{eq:XXXbethe} as $e^{-i \scp_j \NF}$ in the homogeneous limit $u_l =0$.} Finally we can consider the spin of these eigenstates. In the periodic limit $\a \rightarrow 0$ it immediately follows from the $\mbox{SU}(2)$ symmetry of the S-matrix that the monodromy matrix commutes with the total action of $\mathfrak{su}(2)$\footnote{Total spin is a conserved quantity and is of course contained in the family of commuting charges generated by $T$ \cite{Faddeev:1996iy}, hence this relation is also contained in the fundamental commutation relations.}
\begin{equation}
\label{eq:SU2transfercomm}
[\mathcal{T}^g_a(v|\{v_1, \ldots, v_\NA\}), \sigma_{i;a} + \sum_{j=1}^\NA \sigma_{i;j}] =0\, ,
\end{equation}
where $\sigma_{i;j}$ denotes the $i$th pauli matrices acting in space $j$. For our general quasi-periodic boundary conditions we can still sensibly talk about the $\mbox{U}(1)$ charge or total $z$ spin, and defining $S_3 = \frac{1}{2}\sum_{h=1}^\NA \sigma_{3;j}$ and writing things out we get
\begin{equation}
[S_3,B] = -B\, ,
\end{equation}
showing that $B$ lowers the spin by one unit. Since the vacuum has spin $\NF/2$, this means that an eigenstate with $\NA$ $v_j$s has spin $\NF/2-\NA$. This shows that by all accounts these eigenstates can be interpreted as states of $\NA$ particles, the energy, momentum and charge being a sum of the individual particles' contributions, where $B$ now really creates these particles out of the vacuum. These particles are called magnons, and their dispersion relation is
\begin{equation}
\scE = J(1-\cos(\scp))\, ,
\end{equation}
where we have introduced $\scp$ for the spin chain momentum of a single magnon cf. eqn. \eqref{eq:XXXmomentum}. From the point of view of our field theory these excitations do not carry energy or momentum. They do carry $\mbox{SU}(2)$ spin however, and correspond to `exciting' a spin down fermion in a sea of spin up fermions. Since they only excite the auxiliary quantum numbers we will call them auxiliary excitations with associated auxiliary rapidities.

Let us note that in the periodic limit $\alpha \rightarrow 0$ the states generated in this fashion are highest weight states with respect to $\mathfrak{su}(2)$. For the vacuum this follows immediately while for excited states this follows by the (fundamental) commutation relations \eqref{eq:SU2transfercomm} and the Bethe equations \eqref{eq:XXXbethe}, see for example \cite{Faddeev:1996iy}. Note that this means we cannot act with more than $M/2$ creation operators to generate an eigenstate since highest weight states have positive spin. We do not need to either since flipping all spins is a symmetry of the spin chain Hamiltonian. Similarly, at the level of the chiral Gross-Neveu model it is clear that the labeling of the two fermionic species is inconsequential. Including the descendants of our highest weight states given by the algebraic Bethe ansatz we obtain a complete set of eigenstates for the Heisenberg Hamiltonian and our transfer matrix. These descendants can also be seen in the Bethe ansatz by allowing ourselves to consider irregular rapidities; adding an extra particle with infinite rapidity immediately satisfies its own equation, and does not change any of the others. In the quasi-periodic case there are no descendants, and correspondingly these infinite rapidity `solutions' have turned into extra regular solutions with finite rapidities.

\subsection{The large volume spectrum of the chiral Gross-Neveu model}

Now we are ready to write the Bethe-Yang equations \eqref{eq:BEeigenval} for our specific model. By the algebraic Bethe ansatz we just found a description of the eigenvalues of our transfer matrix, and setting $u=u_j$ in \eqref{eq:TXXXeigenvalues} we obtain the Bethe-Yang equations
\begin{equation}
\label{eq:cGNMainBethe}
e^{i p_j L} \prod_{m=1}^\NF S^{ff} (u_j-u_m)\prod_{i=1}^\NA \frac{u_j-v_i+i}{u_j-v_i-i} = -1\, ,
\end{equation}
where we have re-instated the scalar factor $S^{ff}$,\footnote{Note the correct combination of $\NF$ scalar factors and a minus sign on the right hand side.} and the $v_i$ satisfy the auxiliary Bethe equations
\begin{equation}
\label{eq:cGNAuxBethe}
\prod_{m=1}^\NF \frac{v_i-u_m-i}{v_i-u_m+i} = e^{-2i \a}\prod_{j\neq i}^\NA \frac{v_i - v_j - 2i}{v_i-v_j + 2i}\, .
\end{equation}
For future reference we would like to introduce the notation
\begin{equation}
\label{eq:cGNSmatdef}
S^{11}(w) \equiv \frac{w-2i}{w+2i} \, , \, \, \, S^{f1}(w) \equiv \frac{w+i}{w-i}  \, , \, \, \, S^{1f}(w) \equiv \frac{w+i}{w-i} \, ,
\end{equation}
so that these Bethe-Yang equations read
\begin{align}
\label{eq:cGNfullBethe1}
&e^{i p_j L} \prod_{m=1}^\NF S^{ff} (u_j-u_m)\prod_{i=1}^\NA S^{f1}(u_j-v_i) = -1\, ,\\
\label{eq:cGNfullBethe2}
&e^{-2i \a}\prod_{m=1}^\NF S^{1f}(v_i-u_m)\prod_{j=1}^\NA S^{11}(v_i-v_j)= -1\, .
\end{align}
The relevance of this notation will become apparent in section \ref{subsec:TBAgeneral}.

\begin{figure}%
\centering
\subfigure[]{\includegraphics[width=7cm]{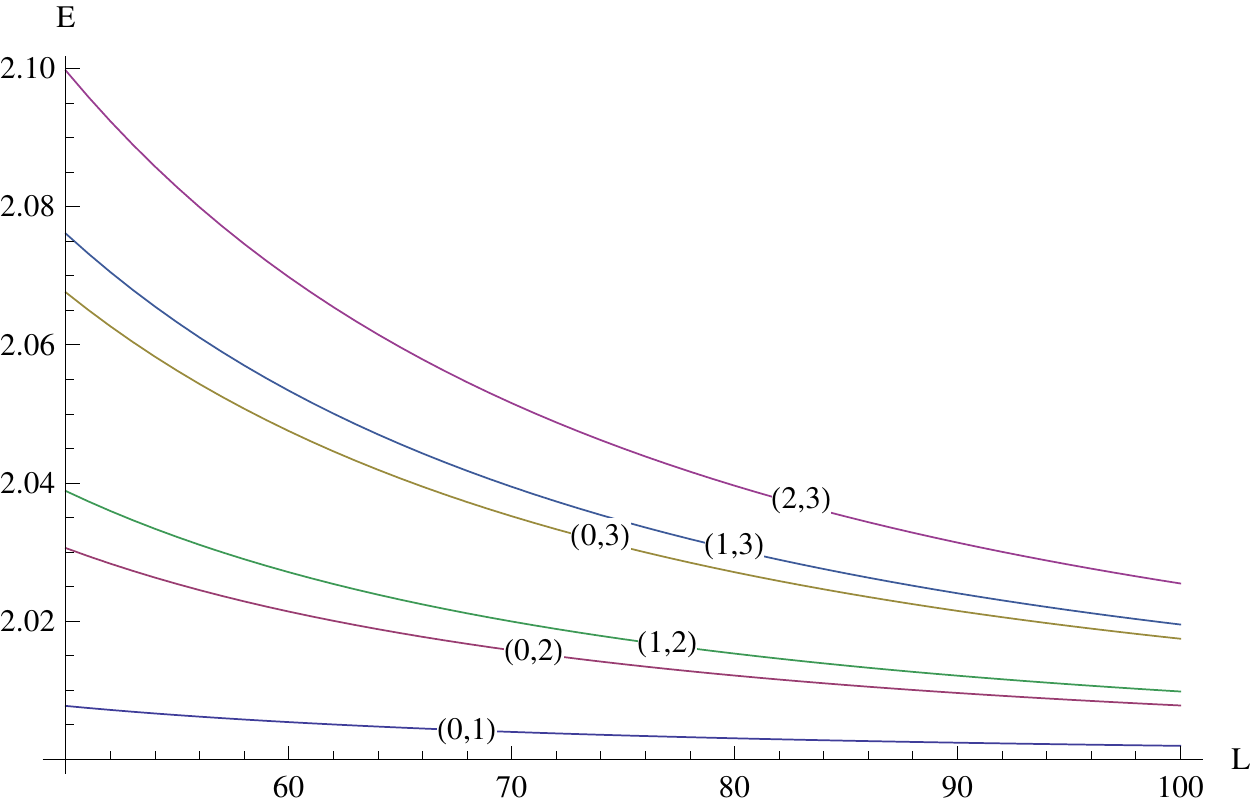} \label{fig:cGNenergies0aux}} \quad
\subfigure[]{\includegraphics[width=7cm]{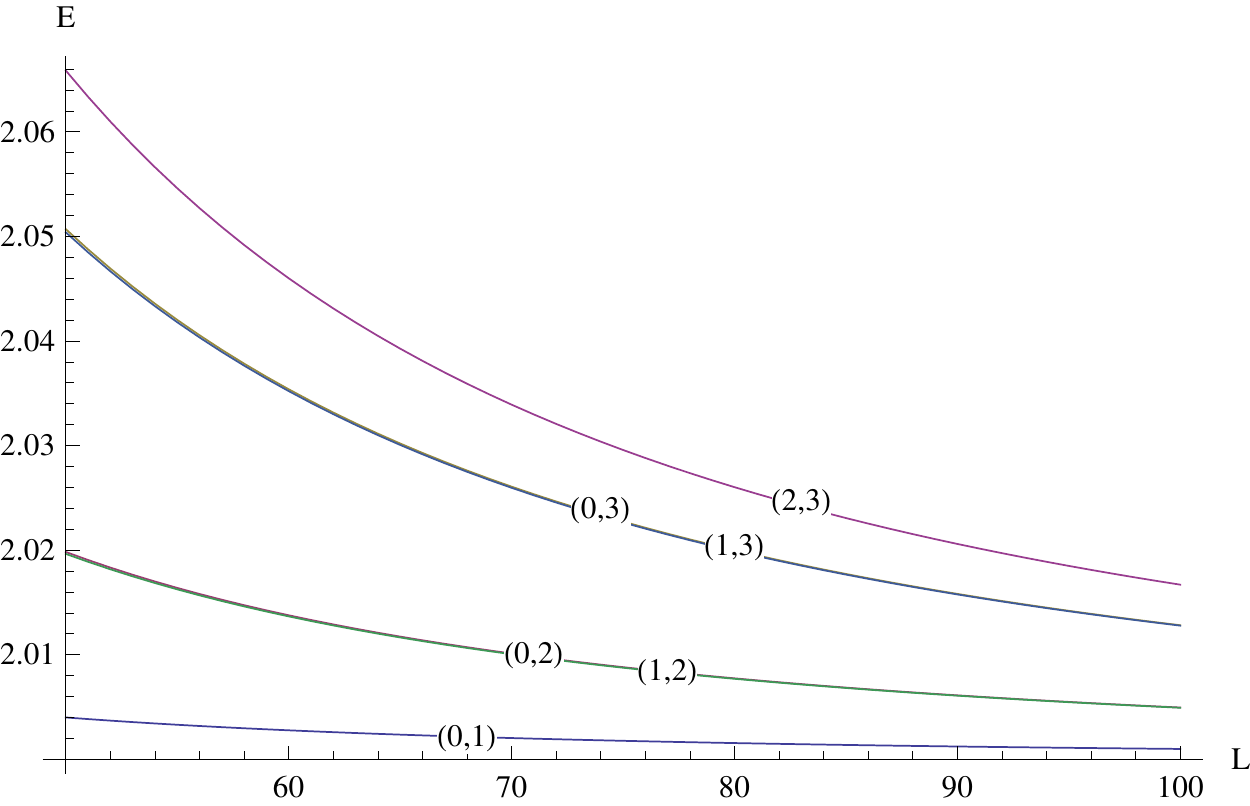} \label{cGNenergies1aux}}
\caption{Asymptotic two-particle energies in the chiral Gross-Neveu model. The left respectively right figure shows the energies of various periodic two-particle states as a function of length taking $m=1$, where both particles have spin up respectively one has spin down (corresponding to one auxiliary excitation in the equations). The states (solutions) are labeled by sets of integers which will be properly introduced in section \ref{subsec:TBAgeneral}. While some of the energy levels in the right figure come very close to each other, there is no real degeneracy.}
\label{fig:cGNenergies}
\end{figure}

The solutions of these auxiliary equations simply parametrize the possible eigenvalues of the transfer matrix; once we fix the $\mbox{SU}(2)$ spin there is a fixed number of possible eigenvalues for the transfer matrix depending on the physical rapidities only, concretely obtained by substituting the solution of eqs. \eqref{eq:cGNAuxBethe} in eqs. \eqref{eq:cGNMainBethe}. Choosing one of these amounts to choosing one of the possible states with given spin. In this way we can think of auxiliary excitations as exciting the spin of the fermion. Note that as discussed above we should only consider states with $\NA\leq \NF/2$. Solving these equations for a given state gives its energy as $E = \sum_j m \cosh \frac{\pi u_j}{2}$. Solving these equations for some two-particle states of spin one and zero gives the energy values shown in figure \ref{fig:cGNenergies}.

\section{Integrability in finite volume}

\label{sec:intfinvol}

At this stage we might be worried that what we were considering above is slightly pointless. Ultimately we are interested in describing the finite size spectrum of our model, and when the system size is truly finite the notion of an S-matrix - let alone factorized scattering - does not exist, making our approach fundamentally inapplicable. Of course there turns out to be a way around this while still exploiting factorized scattering, albeit with a twist.

\subsection{The mirror trick}
\label{subsec:mirrortrick}

\begin{figure}%
\centering
\includegraphics[width=0.7\textwidth]{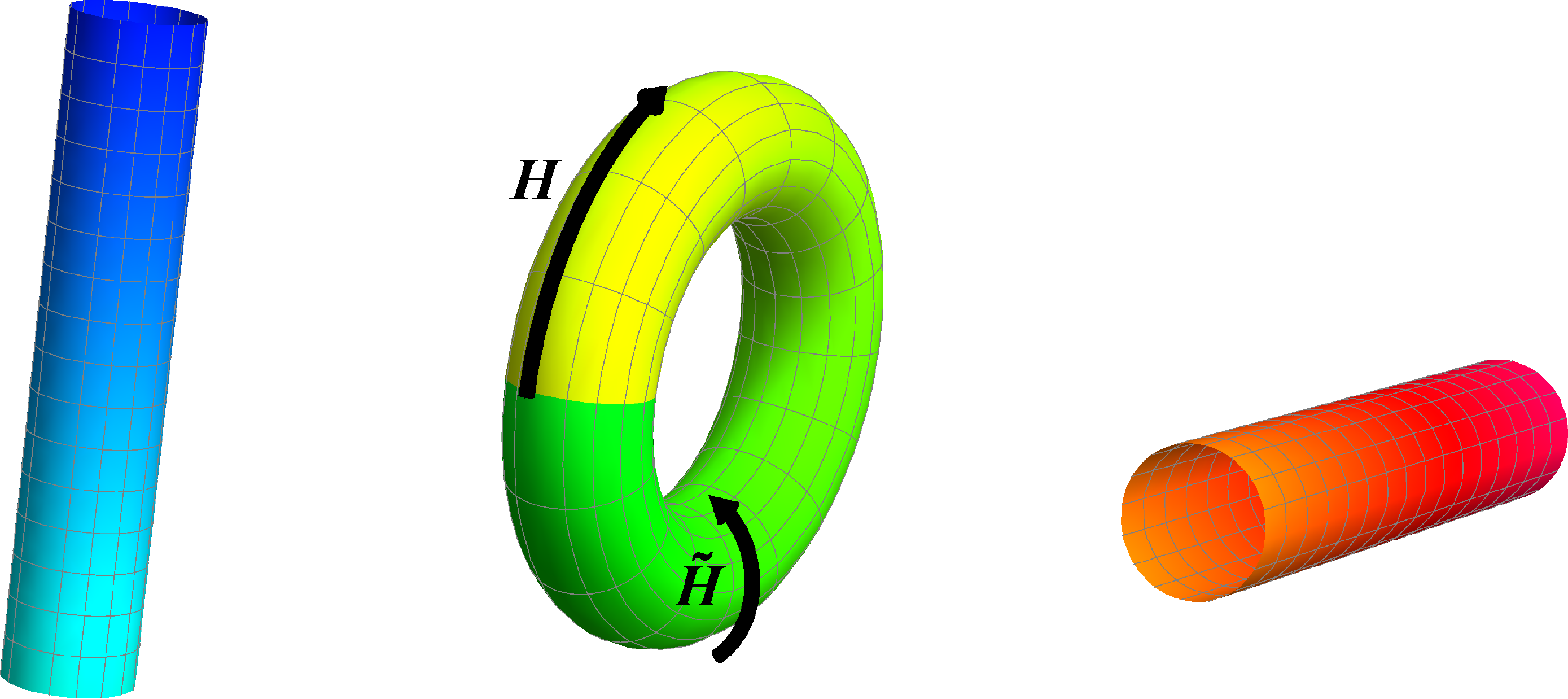}
\caption{The mirror trick. The partition function for a theory on a finite circle at finite temperature lives on a torus (middle). In the zero temperature limit this torus degenerates and gives the partition function on a circle at zero temperature (left), dominated by the ground state energy. Interchanging space and time we obtain a mirrored view of this degeneration as the partition function of the mirror theory at finite temperature but on a decompactified circle, determined by the infinite volume mirror free energy (or Witten's index).}
\label{fig:mirrortrick}
\end{figure}

The first step towards computing the finite size spectrum of our model will be to compute the ground state energy. This can be done exactly thanks to a clever idea by Zamolodchikov \cite{Zamolodchikov:1989cf}. To describe this idea we begin by recalling that the ground state energy is the leading low temperature contribution to the Euclidean partition function
\begin{equation}
\label{eq:ZlimittoGSE}
Z(\beta,L) = \sum_n e^{-\beta E_n} \sim e^{-\beta E_0} \, , \, \hspace{20pt} \mbox{as} \, \, \beta \equiv \frac{1}{T} \rightarrow \infty\, .
\end{equation}
This partition function can be computed from our original quantum field theory by Wick rotating $\tau \rightarrow \tilde{\sigma} =i \tau $ and considering a path integral over fields periodic in $\tilde{\sigma}$ with period $\beta$. In geometrical terms we are putting the theory on a torus which in the zero temperature limit degenerates to the cylinder we started with. Analytically continuing $\tilde{\sigma}$ back to $\tau$ takes us back to our original Lorentzian theory, however we could also analytically continue $\sigma \rightarrow \tilde{\tau} = -i \sigma$. This gives us a Lorentzian theory where the role of space and time have been interchanged with respect to the original theory; it gives us its \emph{mirror model}.\footnote{We should mention that the term mirror model and mirror transformation were introduced in \cite{Arutyunov:2007tc}. Since the mirror transformation does not really affect a relativistic model, there was no need for a careful distinction in the preexisting literature.} Putting it geometrically, we could consider Hamiltonian evolution along either of the two cycles of the torus. Note that at the level of the Hamiltonian and the momentum the mirror transformation corresponds to
\begin{equation}
H \rightarrow i \tilde{p}\, , \hspace{20pt} p \rightarrow i \tilde{H}\,,
\end{equation}
where mirror quantities are denoted with a tilde.\footnote{In fact, the double Wick rotation of $\tau$ and $\sigma$ as just discussed requires an additional sign on $\tilde{H}$. However, we will always use the above (conventional) mirror transformation of the energy and momentum. This matter of conventions is of course completely inconsequential for parity invariant theories.} To emphasize its role as the mirror volume, let us from now on denote the inverse string temperature by $R (=\beta)$. In principle we can compute the Euclidean partition function both through our original model at size $L$ and temperature $1/R$ and through the mirror model at size $R$ and temperature $1/L$. These ideas are illustrated in figure \ref{fig:mirrortrick}.

To compute the ground state energy of our model then, we can equivalently compute the infinite volume partition function of our mirror model at finite temperature or just its (generalized) (Helmholtz) free energy ($\tilde{F}$) since
\begin{equation}
Z = e^{- L \tilde{F}}\, .
\end{equation}
More precisely, cf. eqn. \eqref{eq:ZlimittoGSE} the ground state energy is the (generalized) free energy density of the mirror model
\begin{equation}
E_0 = \frac{\tilde{F}}{R}\, .
\end{equation}
The key point is that we are considering the mirror model in the infinite volume limit where we can use factorized scattering and the asymptotic Bethe ansatz of the previous section, since any exponential corrections to them can be safely neglected.\footnote{Note that the mirror of a relativistic model is equal to the original (up to the specific boundary conditions required to compute the same partition function), and therefore the mirror model is immediately integrable as well. In general we can Wick rotate the large number of conservation laws responsible for factorized scattering, so that the mirror theory has many conserved quantities and mirror scattering should factorize. (Also, we can obtain the S-matrix from four point correlations functions via the LSZ reduction formula, and correlation functions can be computed by Wick rotations.)} The price we have to pay is dealing with a finite temperature. Fortunately the thermodynamic Bethe ansatz will allow us to do precisely this.

\begin{figure}[h]
\begin{center}
\includegraphics[width=0.8\textwidth]{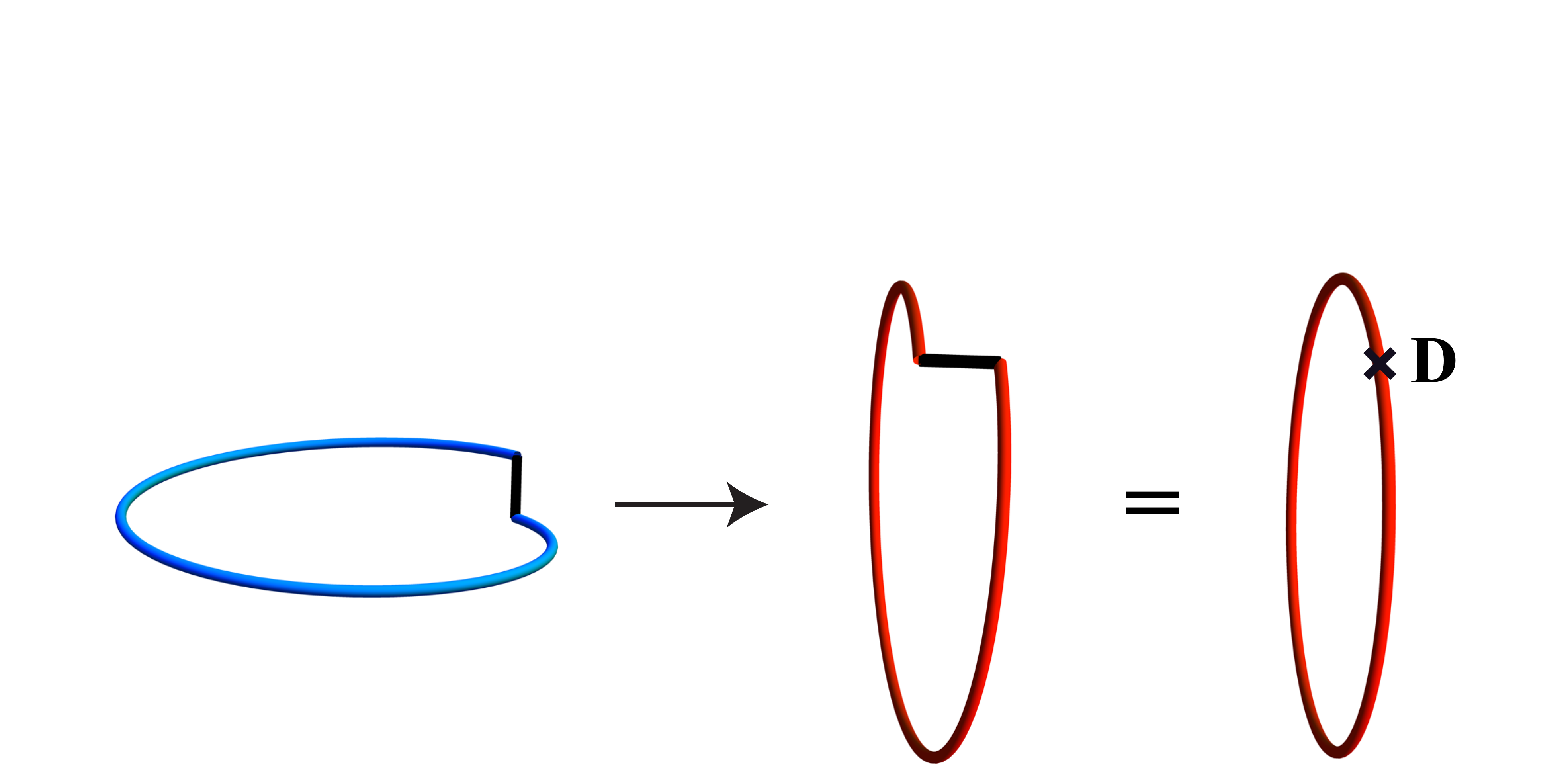}
\caption{Quasi-periodic boundary conditions give a defect operator in the mirror theory. The quasi-periodic boundary condition denoted by the black discontinuity on the circle of the original theory (blue), turns into a discontinuity in the mirror time evolution (red), which is equivalent to inserting a defect operator $D$ in the partition function.}
\label{fig:BCtoD}
\end{center}
\end{figure}

Before moving on to a computation of the free energy we should be a little careful about the boundary conditions in our model. Firstly, where fermions are concerned we should note that the Euclidean partition function is only the proper statistical mechanical partition function used above provided the fermions are anti-periodic in imaginary time. Turning things around, if the fermions are periodic on the circle then from the mirror point of view they will be periodic in imaginary time, so that our goal in the mirror theory is not to compute the standard statistical mechanical partition function but rather what is known as Witten's index
\begin{equation}
Z_W = \mbox{Tr} \left( (-1)^F e^{-L \tilde{H}}\right)\, ,
\end{equation}
where $F$ is the fermion number operator. Continuing along these lines, if we consider quasi-periodic boundary conditions instead of (anti-)periodic boundary conditions we can expect a more general operator to enter in the trace. In fact we can view quasi-periodic boundary conditions as a line defect running along the space-time cylinder of our theory, and upon a double Wick rotation this induces a so-called defect operator in the partition function of the mirror theory \cite{Zamolodchikov:1989cf,Bajnok:2004jd} as illustrated in figure \ref{fig:BCtoD}.
%\begin{figure}[h!]
%\begin{center}
%\includegraphics[width=9cm]{BCtoD3.pdf}
%\caption{Boundary conditions under the mirror trick. A boundary condition in the spatial direction of the physical theory ($\sigma$) becomes a defect operator in the time direction ($\tilde{\tau}$) of the mirror model, here both illustrated by a color discontinuity.}
%\end{center}
%\end{figure}
This defect operator is of course just the inverse\footnote{Let us quickly check the signs by analogy. Consider the following quantum mechanical transition amplitude $\langle \psi_f,x+L|\psi_i,x\rangle$ (in the Heisenberg picture). Then we have $\langle \psi_f,x+L|\psi_i,x\rangle  = \langle e^{i L P}\psi_f,x|\psi_i,x\rangle$ since the momentum operator $P$ generates translations. If $\psi_f = \psi_i$ this gives $\langle \psi_f | e^{-i L P} | \psi_f \rangle$, but for our twisted boundary condition $g \psi_f = \psi_i$ this gives $\langle  \psi_f |e^{-i L P} g | \psi_f \rangle$. From the point of view of the mirror theory $(P=i\tilde{H})$ this looks like $\langle e^{L \tilde{H}} \rangle\rightarrow\langle e^{L \tilde{H}} g \rangle$ when introducing the twisted boundary condition in our fashion.} of the operator $g$ we introduced in eqn. \eqref{eq:genquasiperbcs}, meaning we want to compute
\begin{equation}
Z = \mbox{Tr} \left( g^{-1} \, (-1)^F e^{-L\tilde{H}}\right)\, .
\end{equation}
Since the boundary conditions we will consider are compatible with integrability, this defect operator commutes with the mirror $S$-matrix and Hamiltonian. As such we can arrange the excitations of the Bethe ansatz to be eigenstates of this operator, so that the defect operator will effectively add (imaginary) chemical potential terms to the mirror Hamiltonian. Concretely, denoting the eigenvalue of $g$ on a particle of type $\chi$ as $e^{i \a_\chi}$, we need to add $\sum_\chi i \a_\chi N_\chi/L$ to the free energy, where $N_\chi$ is the number operator for particle type $\chi$.\footnote{To make the partition function well defined we may have to regularize these imaginary chemical potentials with a small real part. We will come back to this in chapter \ref{chapter:twistedspectrum}.}

\subsection{The thermodynamic Bethe ansatz}
\label{subsec:TBAgeneral}

We just saw that we would like to compute the free energy of our mirror model in the infinite size limit. In this limit, the ideas of factorized scattering and the asymptotic Bethe ansatz apply, so in principle we know the possible momentum distributions of the particles. As we also know their dispersion, we ought to be able to compute the free energy $F = E - T S$. The procedure to actually do this computation goes under the name of the thermodynamic Bethe ansatz (TBA) \cite{Yang:1968rm}.

Let us denote the system size of our (mirror) model by $R (=\beta)$ and the number of particles by $K$. Then in the limit $R \rightarrow \infty$ the partition function is dominated by states with a finite density $K/R$. Since in a nested system the type of particle is determined by the auxiliary excitations, we should also keep their densities finite to allow all particle types to contribute to the partition function. Hence in order to proceed we need to understand what happens to the Bethe-Yang equations when the system size and excitation numbers become large. In turns out that in this limit the solutions to the Bethe-Yang equations arrange themselves in certain patterns in the complex plane which have an interpretation in terms of (auxiliary) bound states.

To understand this intuitively, let us consider a hypothetical theory we know to have physical bound states. At large but finite lengths there is no reason for the momenta solving the Bethe-Yang equations to be real and in general there are in fact solutions with complex momenta. At the same time our equations should reproduce the continuum of scattering states in the infinite volume limit. This means that unless all momenta become real in this limit, the resulting configuration has to have a different particle content than we naively thought - some particles have formed one or more bound states. As we will see the auxiliary excitations can form similar configurations, which by analogy we interpret as auxiliary bound states. Another motivation for this interpretation comes from the spin chain interpretation of the auxiliary system, where these configurations really have the interpretation of physical bound states.

Determining the precise spectrum of excitations in the thermodynamic limit is the nontrivial step in the derivation of the TBA equations, differing from model to model. Rather than attempting a semi-general discussion we will discuss these ideas in detail for the simple example of the chiral Gross-Neveu model. By considering the XXX spin chain as both the corresponding auxiliary problem as well as a physical model on its own, we can kill two birds with one stone by discussing the bound states of magnons in the XXX spin chain.\footnote{While this model is not a field theory, for the present argument there is no problem in pretending that its Bethe equations arose from some field theory (the only S-matrix property missing is crossing symmetry).} After this we hope that the general story will be quite readily accepted. Note that the mirror Bethe-Yang equations for the chiral Gross-Neveu model are given by \eqref{eq:cGNMainBethe} and \eqref{eq:cGNAuxBethe} with $L \rightarrow R$ and $\a \rightarrow 0$.

\subsubsection*{The string hypothesis}

\begin{figure}%
\centering
\includegraphics[width=8cm]{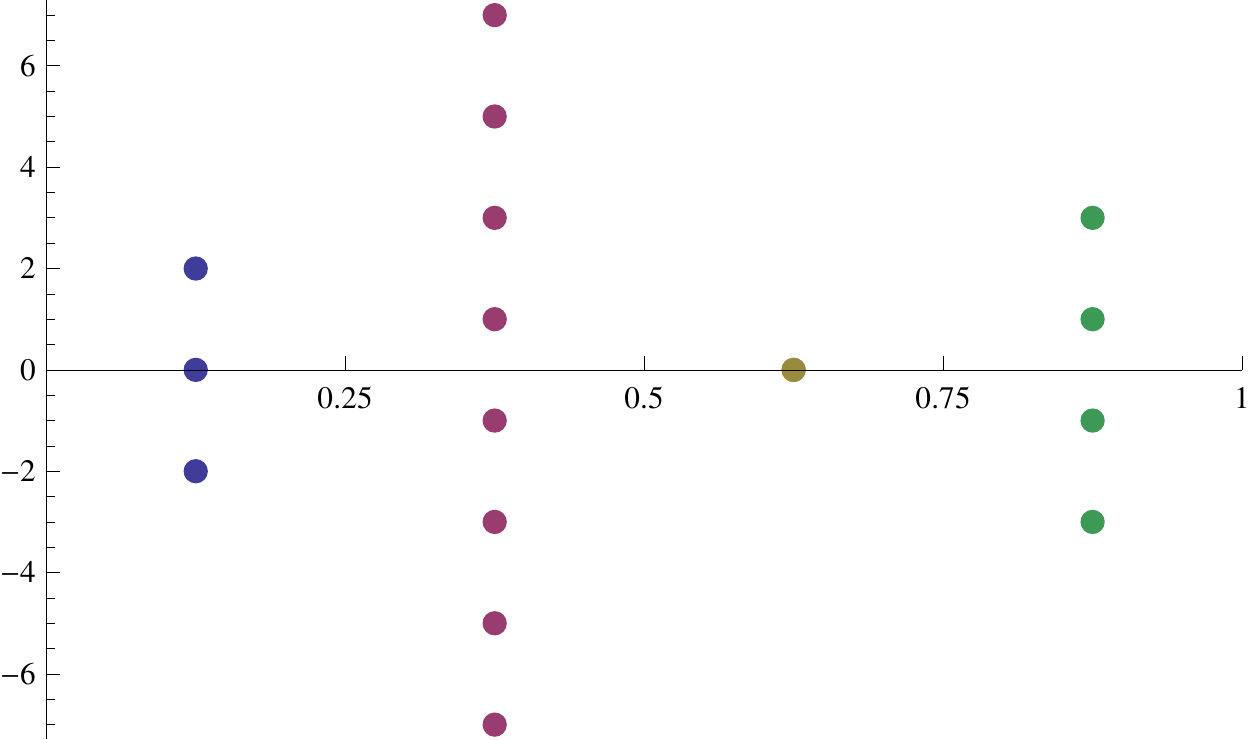}
\caption{An illustration of Bethe strings of length three, eight, one and four, with center $1/8$, $3/8$, $5/8$ and $7/8$ respectively.}
%\caption{Bethe strings. Bethe strings are complex solutions of the large volume Bethe equations, where the constituents of the solution are spaced apart evenly in the imaginary direction and symmetric about the real axis (here the spacing is $2i$. The overall rapidity of the string is the center of this configuration. Here we illustrate strings of length three, eight, one and four, with center $1/8$, $3/8$, $5/8$ and $7/8$ respectively.}
\label{fig:bethestrings}
\end{figure}

We will begin our discussion with magnon bound states in the XXX spin chain. In the $u_m=0$ limit the Bethe equations \eqref{eq:cGNfullBethe2} can be written as
\begin{equation}
\label{eq:homogeneousXXXbetheequations}
e^{i \scp_i \NF} \prod_{j=1}^\NA S^{11}(v_i-v_j) = -1\, ,
\end{equation}
and we have identified the spin chain momentum $\scp$ via eqn. \eqref{eq:XXXmomentum};  $S^{1f}(v_i)=e^{i \scp_i}$. We would like to understand the type of solutions these equations can have, specifically as we take the system size $\NF$ to infinity. For real momenta nothing particular happens in these equations, and we simply get many more possible solutions as $\NF$ grows. If we consider a solution with complex momenta however,\footnote{Solutions with complext momenta exist for small chains already, as we can readily determine from e.g. the Bethe equations for $\NF=5$, $\NA=2$. Note that in integrable field theory Hermitian analyticity together with the specific form of (physical) unitarity following from lack of particle production immediately imply that if a theory has bound states (some of) the constituent particles have complex momenta.} say a state with $\mbox{Im}(\scp_1)>0$, we have an immediate problem:
\begin{equation}
e^{i \scp_q \NF} \rightarrow 0 \, , \, \, \,\,\,\, \mbox{as} \,\,\, \NF \rightarrow \infty\, .
\end{equation}
We see that the only way a solution containing $p_1$ can exist in this limit is if this zero is compensated by a pole in one of the $S^{11}$ (eqn. \eqref{eq:cGNSmatdef}), which can be achieved by setting
\begin{equation}
v_2 = v_1 + 2 i\, .
\end{equation}
At this point we have doctored up the equation for $\scp_1$, but we have introduced potential problems in the equation for $\scp_2$. Whether there is a problem can be determined by multiplying the equations for $\scp_1$ and $\scp_2$ so that the singular contributions of their relative S-matrix cancel out
\begin{equation*}
%\label{eq:XXXtwoparticleproductBE}
e^{i (\scp_1+\scp_2) \NF} \prod_{i\neq1}^\NA S^{11}(v_1-v_i) \prod_{i\neq2}^\NA S^{11}(v_2-v_i)  = e^{i (\scp_1+\scp_2) \NF} \prod_{i\neq1,2}^\NA S^{11}(v_1-v_i) S^{11}(v_2-v_i) = 1\,.
\end{equation*}
If the sum of their momenta is real this equation is fine, and this set of momenta can be part of a solution to the Bethe equations. In terms of the rapidities this solution looks like
\begin{equation}
\label{eq:XXXtwostring}
v_1 = v - i \, , \, \, \, v_2 = v + i \, , \, \, \, v \in \mathbb{R}\,.
\end{equation}
On the other hand, if the sum of our momenta has positive imaginary part we are still in trouble.\footnote{By rearranging the order of our argument (particles considered) we do not have to consider the case where the remaining imaginary part is of different sign.} In this case, since we should avoid coincident rapidities, the only way to fix the equation is to have a third particle in the solution, with rapidity
\begin{equation}
v_3 = v_2 + 2 i\, .
\end{equation}
As before, if now the total momentum is real the equations are consistent and these three rapidities can be part of a solution. If not, we continue this process and create a bigger configuration or run off to infinity. These configurations in the complex rapidity plane are known as Bethe strings, illustrated in figure \ref{fig:bethestrings}.
Since in our spin chain $\scp$ has positive imaginary part in the lower half of the complex rapidity plane and vice versa, strings of any size can be generated in this fashion by starting appropriately far below the real line.\footnote{In other models the pattern of possible string configurations can be quite complicated; in chapter \ref{chapter:quantumTBA} we will see examples of models where the string are bounded in length.} Concretely, a Bethe string with $Q$ constituents and rapidity $v$ is given by the configuration
\begin{equation}
\label{eq:XXXstringconfiguration}
\{v_Q\} \equiv \{v-(Q+1-2j)i|j=1,\ldots,Q\}\, ,
\end{equation}
where $v \in \mathbb{R}$ is called the center of the string. Full solutions of the Bethe equation in the limit $M\rightarrow \infty$ can be built out of these string configurations.

We would like to give these (Bethe) strings the interpretation of bound states as they have less energy than sets of individual real magnons.\footnote{The corresponding Bethe wave-function also shows an exponential decay in the separation of string constituents.} For example, the energy of the two-string \eqref{eq:XXXtwostring} is given by
\begin{equation}
\scE_2(v) = \scE(v_1)+\scE(v_2) = -2 J \left( \frac{1}{(v-i)^2+1}+\frac{1}{(v+i)^2+1}\right) = -2 J \frac{2}{v^2+2^2}\, ,
\end{equation}
which is less than that of any two-particle state made out of real particles;
\begin{equation}
\scE_2(v)<\scE(\tilde{v}_1)+\scE(\tilde{v}_2) \, \, \,\mbox{ for }\, v,\tilde{v}_{1,2} \in \mathbb{R} \, \,\,\mbox{ (real momenta).}
\end{equation}
Similarly, the energy of a $Q$-string is lower than that of $Q$ separate particles and is given by
\begin{equation}
\scE_Q(v) = \sum_{v_j \in \{v_Q\}} \scE(v_j) = -2 J \frac{Q}{v^2+Q^2}\, .
\end{equation}
This is most easily shown by noting that
\begin{equation}
\scE(v) = J \frac{d\scp(v)}{dv}\, ,
\end{equation}
and the fact that the expression for the momentum of a $Q$-string is particularly simple
\begin{equation}
\scp_Q(v) = i \log \frac{v-Q i}{v+Q i}\, ,
\end{equation}
as follows by cancelling numerators and denominators in the product $\frac{v_1-i}{\cancel{v_1+i}}\frac{\cancel{v_2-i}}{v_2+i}\ldots \frac{v_Q-i}{v_Q+i}$ as indicated.

We have just determined that the possible solutions of the Bethe equations in the limit $\NF\rightarrow \infty$ are built out of elementary objects called Bethe strings (a one-string being a normal magnon). Interpreting them as bound states, we see that the spectrum thus obtained is reflected by an appropriate pole in the two-particle S-matrix. This example is not a field theory, but in the next chapter we will see that the bound state spectrum of the (mirror) light-cone superstring has the exact same pattern in appropriate variables. So far so good, but ultimately we are interested in thermodynamic limits, meaning we should take $\NF\rightarrow \infty$ with $\NA/\NF\leq1/2$ fixed; the number of magnons goes to infinity as well. In this limit the analysis above is clearly no longer even remotely rigorous since an ever growing product of magnon S-matrices with complex momenta can perfectly mimic the role of the pole in our story for example. Still, since such solutions seem rather atypical and at least low magnon density solutions should essentially conform to the string picture, we can hypothesize that the string complexes should contain `most' of the possible solutions, in the sense that they are the ones that give measurable contributions to the free energy. Indeed in the XXX spin chain there are examples of solutions that do not approach string complexes in the thermodynamic limit \cite{Woynarovich:1981ca,Woynarovich:1982,Babelon:1982mc}, but nonetheless the free energy is captured correctly by taking only string configurations into account \cite{Tsvelick:1983}. The assumption that all thermodynamically relevant solutions to the Bethe equations are built up out of such string configurations, and which form these configurations take, goes under the name of the \emph{string hypothesis}. More details and references on the string hypothesis can for example be found in chapter four of \cite{Korepin}.

Now we would like to come to the chiral Gross-Neveu model where the XXX spin chain is an auxiliary system. Because the chiral Gross-Neveu model has no bound states the only solutions to the Bethe equations correspond to fundamental excitations with real momenta with correspondingly real rapidities, also in the thermodynamic limit.\footnote{This is the physical statement based on knowledge of the spectrum; of course this can be confirmed by analyzing the Bethe-Yang equations similarly to the XXX ones.} However, in the thermodynamic limit $R\rightarrow \infty$ keeping $\NF/R$ and $\NA/\NF$ fixed, we are also effectively taking the infinite length limit of the spin chain, in fact its thermodynamic limit. In this case the auxiliary equations become
\begin{equation}
\left(S^{1f}(v_i)\right)^{\NF} \prod_{j\neq i}^\NA S^{11}(v_i-v_j) =1 \rightarrow \prod_{m=1}^\NF S^{1f}(v_i-u_m) \prod_{j\neq i}^\NA S^{11}(v_i-v_j) =1\, ,
\end{equation}
but since all $u_m$ are real $|S^{1f}(v_i-u_m)|>1$ when $|S^{1f}(v_i)|>1$ and vice versa, the above analysis for the spin chain is not affected. Therefore will make the string hypothesis that the solutions of the Bethe-Yang equations of the chiral Gross-Neveu model in the thermodynamic limit consist of
\begin{itemize}
\item{Fermions with real momenta}
\item{Strings of auxiliary magnons of any length with real center}
\end{itemize}
The constituent rapidities of a $Q$ string with center $v$ are given by eqn. \eqref{eq:XXXstringconfiguration}. While they carry no energy in the chiral Gross-Neveu model and hence cannot have a 'binding energy', we will still refer to and think of these patterns of auxiliary excitations as bound states of magnons.

In general the string hypothesis for a given model may contain bound states of physical particles, bound states of auxiliary particles like in the chiral Gross-Neveu model, and even bound states of physical and auxiliary particles as happens in the Hubbard model \cite{Takahashi:1972hub}. Whatever the type of bound state, it appears to always be possible to find some appropriate rapidity type variable in terms of which the bound states look like Bethe strings. The allowed lengths for typical string complexes may be constrained in quite complicated ways however \cite{Takahashi:1972}. We will encounter examples of all this in the later chapters.

\subsubsection*{The Bethe-Yang equations for strings}

We would like to group terms in the (auxiliary) Bethe equations \eqref{eq:cGNMainBethe} and \eqref{eq:cGNAuxBethe} in accordance with the string hypothesis; having a thermodynamic limit in mind, the $N$ magnons of a given solution of the Bethe equations should arrange themselves into combinations of string complexes. In other words, denoting the number of bound states of length $Q$ occurring in a given configuration by $N_Q$ we have
\begin{equation}
\prod_{j=1}^{\NA} \rightarrow \prod_{Q=1}^{\infty} \prod_{l=1}^{N_{Q}} \prod_{j \in \{v_{Q,l}\}}\, ,
\end{equation}
under the constraint
\begin{equation}
\sum_{Q=1}^\infty Q N_Q = \NA\, .
\end{equation}
Taking this into account we can appropriately represent the Bethe equations \eqref{eq:cGNfullBethe1} and  \eqref{eq:cGNfullBethe2} as
\begin{align}
e^{i p_j L} \prod_{m=1}^\NF S^{ff} (u_j-u_m) \prod_{Q=1}^{\infty} \prod_{l=1}^{N_{Q}} S^{fQ} (u_j-v_{Q,l}) & = -1\, ,\\
\prod_{m=1}^\NF S^{1f}(v_i-u_m)\prod_{Q=1}^{\infty} \prod_{l=1}^{N_{Q}} S^{1Q} (v_i-v_{Q,l}) & = -1\, ,
\end{align}
where
\begin{align}
&S^{\chi Q}(v-w_Q) \equiv \prod_{w_j \in \{w_Q\}} S^{\chi 1}(v - w_j) \, , \hspace{20pt} \chi = f,1\,.
\end{align}
At this point not all $\NA$ auxiliary Bethe equations are independent anymore, as some magnons are bound in strings. We already saw that we can get the Bethe equation for the center of a bound state by taking a product over the Bethe equations of its constituents, so that our complete set of Bethe equations becomes
\begin{align}
\label{eq:cGNBethestrings1}
e^{i p_j L} \prod_{m\neq j}^\NF S^{ff} (u_j-u_m) \prod_{Q=1}^{\infty} \prod_{l=1}^{N_{Q}} S^{fQ} (u_j-v_{Q,l}) & = -1\, ,\\
\label{eq:cGNBethestrings2}
\prod_{m=1}^\NF S^{Pf}(v_{P,r}-u_m)\prod_{Q=1}^{\infty} \prod_{l=1}^{N_{Q}} S^{PQ} (v_{P,r}-v_{Q,l}) & = -1\, ,
\end{align}
where
\begin{align}
&S^{P\chi}(v_P-w) \equiv \prod_{v_i \in \{v_P\}} S^{1\chi}(v_i-w) \, , \hspace{20pt} \chi = f,Q\, .
\end{align}
The physical interpretation of these expressions is of course that they are the scattering amplitudes between the particles indicated by superscripts. These products of constituent S-matrices typically simplify, but their concrete expression is not important for our considerations (yet); what is important is that they exist and only depend on the centers of the strings, or in other words, the bound state momenta. The process of obtaining the bound state S-matrices from fundamental ones in this way of course corresponds to the discussion of fusion in section \ref{sec:factscat}, just applied at the diagonalized level.

\subsubsection*{The thermodynamic Bethe ansatz equations}

To proceed, we will work with the Bethe-Yang equations in logarithm form. Through a fixed choice of branch for the logarithm this introduces an integer $I$ in each equation which labels the possible solutions of the equation
\begin{align}
2 \pi I^f_j &= R p_j  - i \sum_{m=1}^{\NF} \log{S^{ff}(u_j-u_m)}- i \sum_{Q=1}^{\infty} \sum_{l=1}^{N_{Q}} \log{S^{fQ}(u_j-v_{Q,l})} \, , \\
\label{eq:logBYcGNmagnons}
- 2 \pi I^P_r &= -i \sum_{m=1}^\NF \log{S^{Pf}(v_{P,r}-u_m)} -i \prod_{Q=1}^{\infty} \prod_{l=1}^{N_{Q}} \log{S^{PQ}(v_{P,r}-v_{Q,l})} \, .
\end{align}
We choose to define the integer in the second equation with a minus sign for reasons we will explain shortly. In the thermodynamic limit the solutions to these equations become dense
\begin{equation}
u_i-u_j \sim \mathcal{O}(1/R)\, , \,\,\,\,\, u_i-v_j \sim \mathcal{O}(1/R) \, \, \, \,\, \mbox{and} \,\,\,\,\, v_i-v_j \sim \mathcal{O}(1/R)\, .
\end{equation}
With this in mind it may be sensible to generalize the integers $I$ to functions of the relevant rapidity (momentum), known as \emph{counting functions}, interpolating between the integers corresponding to different solutions of the Bethe-equations. Concretely
\begin{align}
R c^f (u) &= \frac{R}{2\pi} p(u) + \frac{1}{2\pi i}  \sum_{m=1}^{\NF} \log{S^{ff}(u-u_m)}+ \frac{1}{2\pi i} \sum_{Q=1}^{\infty} \sum_{l=1}^{N_{Q}} \log{S^{fQ}(u-v_{Q,l})} \, , \\
R c^P (u) &= -\frac{1}{2\pi i} \sum_{m=1}^\NF \log{S^{Pf}(u-u_m)} - \frac{1}{2\pi i} \sum_{Q=1}^{\infty} \sum_{l=1}^{N_{Q}} \log{S^{PQ}(u-v_{Q,l})} \, ,
\end{align}
so that
\begin{equation}
R c^f(u_k) = I^f_k \, , \,\,\,\,\, \mbox{and} \, \,\,\,\, R c^P(v_l) = I^P_l\, .
\end{equation}
Importantly we assume that these counting functions are monotonically increasing functions of $u$ provided their leading terms are\footnote{We cannot prove that these functions are monotonically increasing for given excitation numbers without knowing the precise root distribution, which is actually what we are ultimately trying to determine. We may consider it part of the string hypothesis by saying we are not making a mistake in treating the thermodynamic limit as the \emph{ordered} limits $R\rightarrow\infty$, then $M\rightarrow \infty$, then $N\rightarrow \infty$, in which case the statement does clearly hold. Similar statements are made on the first page of section six in \cite{Faddeev:1996iy}.}, and here indeed we have
\begin{equation}
\frac{dp}{du}>0\, , \hspace{20pt} \mbox{and} \,  \hspace{20pt} \frac{1}{2\pi i} \frac{d\log{S^{Pf}}}{du}<0\, .
\end{equation}
This is the reason for our sign choice above. Clearly in general we have
\begin{equation}
c(w_i)-c(w_j) = \frac{I_i-I_j}{R}\, .
\end{equation}
For a given solution of the Bethe-Yang equations some of the integers are said to be `occupied' and some to be `vacant' if there is, respectively there is not, a corresponding rapidity in the solution. Since the solutions become dense we will introduce an associated particle density $\rho$ for the rapidities that are taken in a solution, and a hole density $\bar{\rho}$ for the ones that are not. The density of states in `integer space' is one (in units of $R^{-1}$), which in rapidity space means
\begin{align}
\rho^f(u) + \bar{\rho}^f(u) & = \frac{d c^f (u)}{du}\, , \\
\rho^P(v) + \bar{\rho}^P(v) & = \frac{d c^P (v)}{dv}\, .
\end{align}
Explicitly taking the derivative of the counting functions gives us the thermodynamic analogue of the Bethe-Yang equations
\begin{align}
\label{eq:cGNdensityeqn1}
\rho^f(u) + \bar{\rho}^f(u) &= \frac{1}{2\pi} \frac{dp(u)}{du} + K^{ff} \star \rho^f(u) - K^{fQ} \star \rho_Q(u) \, ,\\
\label{eq:cGNdensityeqn2}
\rho^P(v) + \bar{\rho}^P(v) &= K^{Pf} \star \rho^f(u) - K^{PQ} \star \rho^Q(u)\, ,
\end{align}
where we implicitly sum over repeated indices and we have replaced the sums over roots by integrals over densities, resulting in convolutions defined as
\begin{equation}
K\star f(u) \equiv \int_{-\infty}^\infty dv\,  K(u,v)f(v)\, .
\end{equation}
The kernels $K$ are defined as the logarithmic derivatives of the scattering amplitudes
\begin{equation}
K^\chi(u,v) \equiv \pm \frac{1}{2\pi i}\frac{d}{du} \log S^\chi (u,v)\, ,
\end{equation}
where $\chi$ denotes an arbitrary set of particle labels. The sign is chosen such that the kernels are positive, in this case requiring minus signs for $K^{fP}$ and $K^{Mf}$.\footnote{Unfortunately there is no way to define a notation which uniformizes both the Bethe-Yang equations in the way we did and automatically gives positive kernels.} From this point on we will distinguish the two arguments even though the S-matrices and kernels we have seen so far are all of difference form.

The remainder of the derivation simply amounts in finding an expression for the generalized free energy given these constraints on the density of states. The energy density (per unit length) is just
\begin{equation}
e = \int_{-\infty}^\infty du\,  \mathcal{E}(u) \rho^f (u)\, ,
\end{equation}
where in this case
\begin{equation}
\mathcal{E}(u) = m \cosh\frac{\pi u}{2}\, ,
\end{equation}
while the entropy in an interval $\Delta u$ for each particle type is of course
\begin{equation}
\Delta S(\rho) = \log \frac{(R(\rho(u) + \bar{\rho}(u))\Delta u)!}{(R \rho(u) \Delta u)!(R \bar{\rho}(u) \Delta u)!}
\end{equation}
which by use of Stirling's formula ($R \rho (\bar{\rho}) \Delta u\gg 1$) gives us the entropy density as
\begin{equation}
s = \sum_{i\in\{f,P\}} \int_{-\infty}^\infty du
\, \rho^i \log\left( 1 + \frac{\bar{\rho}^i}{\rho^i}\right) + \bar{\rho}^i \log\left( 1 + \frac{\rho^i}{\bar{\rho}^i}\right) \, .
\end{equation}
Next, since we are considering twisted boundary conditions we should add chemical potentials corresponding to the defect operator $g$, in this case given by eqn. \eqref{eq:cGNtwist}. Also, we should not forget about the $(-1)^F=e^{i \pi N_f}$ in our trace. The eigenvalue of $g$ on the fundamental fermions is just $e^{i\a}$, so taking into account also their fermionic nature we should add $i (\alpha+\pi) N_f T$ to the free energy. As auxiliary particles represent flipping the spin of a fermion they carry \emph{double} the opposite $\mbox{SU}(2)$ charge, meaning we should add $-2 i \alpha N_1 T$ for them. Bound states of auxiliary particles of course carry the charge of their constituents, so that for these we add $-2 i \alpha Q N_Q T$. Of course the particle densities are just given by
\begin{equation}
n^i = \int_{-\infty}^\infty du \, \rho^i\, ,
\end{equation}
Putting this together we get an expression for the generalized free energy
\begin{equation}
F = R(e - Ts - \mu_f n^f - \mu_Q n^Q)\, ,
\end{equation}
where $\mu_f = -i (\a+\pi) T$ and $\mu_Q = 2 i Q \a T$. Being in thermodynamic equilibrium however means
\begin{equation}
\delta F = 0\, .
\end{equation}
Here the Bethe-Yang equations come in by giving us the hole densities as functions of the particle densities. Varying eqs. \eqref{eq:cGNdensityeqn1} and \eqref{eq:cGNdensityeqn2} we get
\begin{align}
\label{eq:cGNvardensityeq1}
\delta \rho^f + \delta \bar{\rho}^f &=  K^{ff} \star \delta \rho_f - K^{fQ} \star \delta \rho_Q\, , \\
\label{eq:cGNvardensityeq2}
\delta \rho^P + \delta \bar{\rho}^P &=  K^{Pf} \star \delta \rho_f - K^{PQ}\star \delta \rho_Q\, ,
\end{align}
Writing this schematically as
\begin{equation}
\label{eq:vardensityeqgen}
\delta \rho^i + \delta \bar{\rho}^i = K^{ij}\star \delta \rho^j\, ,
\end{equation}
after a little algebra we get the variation of the entropy
\begin{equation}
\frac{\delta s}{\delta\rho^j(u)} = \log \frac{\bar{\rho}^j}{\rho^j}(u) + \log \left(1 + \frac{\rho^i}{\bar{\rho}^i}\right) \tilde{\star}\, K^{ij} (u)\, ,
\end{equation}
where $\tilde{\star}$ denotes `convolution' from the right
\begin{equation}
f \, \tilde{\star}\, K(u) \equiv \int_{-\infty}^\infty dv f(v)K(v-u)\, .
\end{equation}
The variation of the other terms is immediate, and $\delta F=0$ results in the \emph{thermodynamic Bethe ansatz equations}
\begin{equation}
\log \frac{\bar{\rho}^j}{\rho^j} =\frac{\mathcal{E}_j - \mu_j}{T} - \log \left(1 + \frac{\rho^i}{\bar{\rho}^i}\right)\star K^{ij}\, ,
\end{equation}
where by conventional abuse of notation we dropped the tilde on the `convolution'. We will henceforth denote the combination $\frac{\bar{\rho}^j}{\rho^j}$ by the \emph{Y-functions} $Y_j$, meaning the TBA equations read
\begin{equation}
\label{eq:TBAgen}
\log Y_j =\frac{\mathcal{E}_j - \mu_j}{T} - \log \left(1 + \frac{1}{Y_i}\right)\star K^{ij}\, .
\end{equation}
Taking into account the generalized form of eqs. \eqref{eq:cGNdensityeqn1} and \eqref{eq:cGNdensityeqn2}, on a solution of the TBA equations the free energy of the mirror model is given by
\begin{equation}
F = -R \int_{-\infty}^\infty du \frac{1}{2\pi} \frac{dp_j}{du} \log\left(1+\frac{1}{Y_j}\right)\, ,
\end{equation}
giving the ground state energy of our original model as
\begin{equation}
\label{eq:TBAgroundstateenergy}
E_0 = -\int_{-\infty}^\infty du \frac{1}{2\pi} \frac{dp_j}{du} \log\left(1+\frac{1}{Y_j}\right)\, .
\end{equation}

Specifying our schematic notation to the chiral Gross-Neveu model by comparing eqs. \eqref{eq:cGNvardensityeq1}, \eqref{eq:cGNvardensityeq2} and \eqref{eq:vardensityeqgen} gives
\begin{align}
\label{eq:cGNcanonicalTBA1}
\log Y_f &= L \mathcal{E} + i (\a+\pi) - \log \left(1 + \frac{1}{Y_f}\right)\star K^{ff} - \log \left(1 + \frac{1}{Y_Q}\right)\star K^{Qf}\, ,\\
\label{eq:cGNcanonicalTBA2}
\log Y_P &= -2 i \a P + \log \left(1 + \frac{1}{Y_Q}\right)\star K^{QP} + \log \left(1 + \frac{1}{Y_f}\right)\star K^{fP}\, ,
\end{align}
where we note that in the setting of the mirror trick the temperature $T=1/L$. The ground state energy is given by
\begin{equation}
E_0 = -\int_{-\infty}^\infty du \frac{1}{2\pi} \frac{dp}{du} \log\left(1+\frac{1}{Y_f}\right)\, .
\end{equation}

At this point the generalization to an arbitrary model is hopefully almost obvious, with the exception of the string hypothesis which depends on careful analysis of the Bethe-Yang equations for a particular model. If we have this however, we can readily determine the complete set of Bethe-Yang equations analogous to the procedure to arrive at eqs. \eqref{eq:cGNBethestrings1} and \eqref{eq:cGNBethestrings2}. From there we immediately get the analogue of eqs. \eqref{eq:cGNvardensityeq1} and \eqref{eq:cGNvardensityeq2} by a logarithmic derivative.\footnote{Since we like to think of densities as positive we may have to invert the Bethe-Yang equations for a specific particle type to make sure the counting function is defined to be monotonically increasing, just like we did for the magnons (cf. eqn. \eqref{eq:logBYcGNmagnons}).} This is all we need to specify the general TBA equations \eqref{eq:TBAgen} to a given model.

\subsubsection*{Simplified TBA equations and the Y-system}

\begin{figure}%
\centering
\includegraphics[width=6cm]{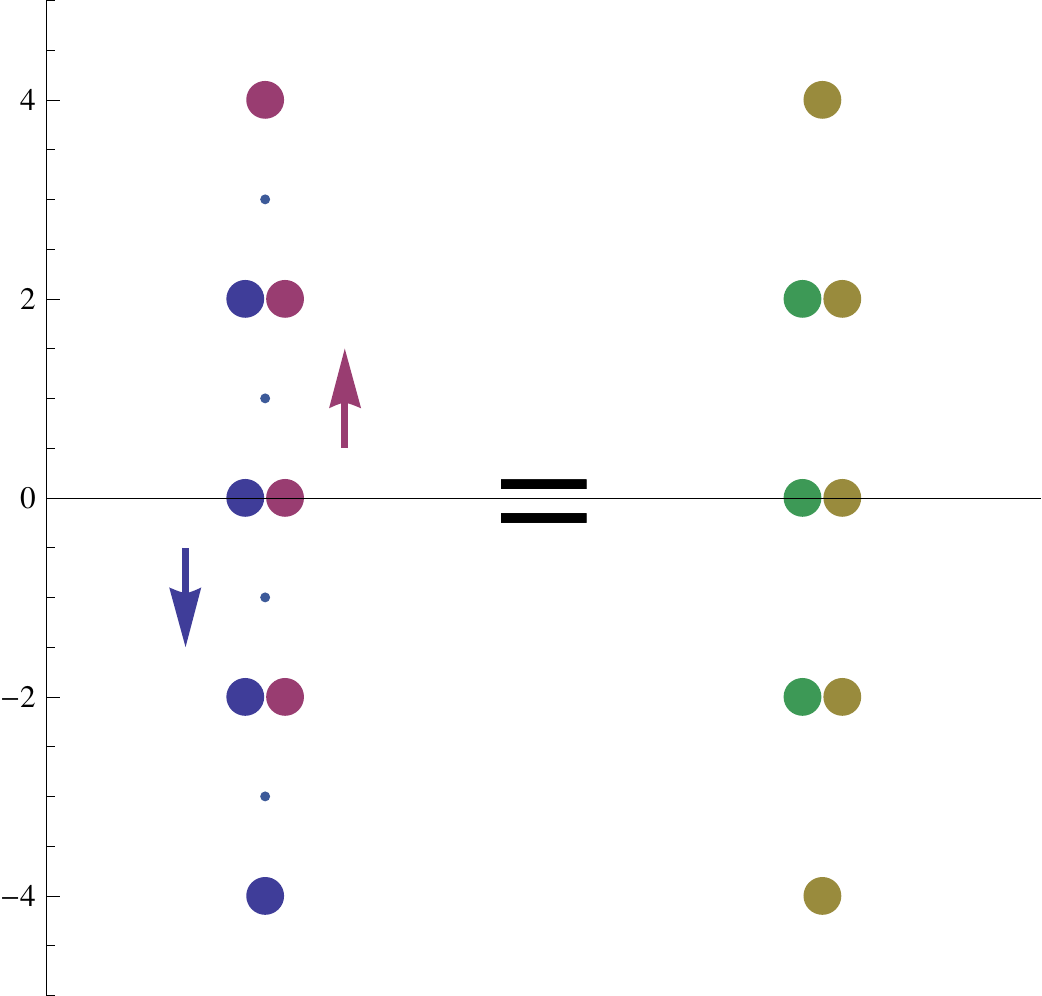}
\caption{The discrete Laplace equation for strings. Shifting a length $Q$ string configuration up by $i$ and another down by $i$ gives a configuration equivalent to two unshifted strings, one of length $Q+1$ and another of length $Q-1$, here illustrated for $Q=4$. The small dots indicate the position of the rapidities before shifting.}
\label{fig:stringdiscretelaplace}
\end{figure}

In problems where there are bound states the TBA equations can typically be rewritten in a simpler fashion. The intuitive picture why this should be possible is illustrated in figure \ref{fig:stringdiscretelaplace}.
Since we obtain all S-matrices by fusing over the constituents, provided $S$ has no branch cuts the figure tells us
\begin{equation}
\frac{S^{\chi Q+1}(v,u)S^{\chi Q-1}(v,u)}{S^{\chi Q}(v,u+i)S^{\chi Q}(v,u-i)} = 1\,,
\end{equation}
where $\chi $ is any particle type and we have reinstated a dependence on two arguments for future reference. In other words, (the logs of) our S-matrices satisfy a discrete Laplace equation.  The associated kernels should then satisfy
\begin{equation}
K^{\chi Q}(v,u+i) + K^{\chi Q}(v,u-i) - (K^{\chi Q+1}(v,u) + K^{\chi Q-1}(v,u)) = 0\, .
\end{equation}
However, when we shift $u$ by $\pm i$ we may generate a pole in $K(v,u+i)$ for some real value of $v$. This can lead to a discontinuity in integrals involving K such as those in the TBA equations. Therefore we need to understand what exactly we mean by this equation. The way this is done is to introduce the kernel $s$
\begin{equation}
s(u)=\frac{1}{4\cosh\frac{\pi u}{2}}\, ,
\end{equation}
and the operator $s^{-1}$ that in hindsight will properly implement our shifts
\begin{equation}
f \star s^{-1} (u) = \lim_{\epsilon\rightarrow 0} \left(f(u+i-i\epsilon) + f(u-i+i\epsilon)\right)\, ,
\end{equation}
which satisfy
\begin{equation}
s\star s^{-1}(u) = \delta(u)\,.
\end{equation}
Note that $s^{-1}$ has a large null space, so that $f \star s^{-1} \star s \neq f$ in general (we will see examples of this soon). This kernel can now be used to define
\begin{equation}
(K+1)^{-1}_{PQ} = \delta_{P,Q} - I_{PQ} s\, ,
\end{equation}
where $I_{PQ} = \delta_{P,Q+1} +  \delta_{P,Q-1}$. In other words, the kernel $K^{PQ}$ introduced above satisfies
\begin{equation}
\label{eq:KQMidentity}
K^{PQ} - (K^{PQ+1}+K^{PQ-1} )\star s = s \, I_{PQ}\,,
\end{equation}
which we can easily prove by Fourier transformation (see appendix \ref{app:qdefSmatricesandkernels}). Similarly we have
\begin{equation}
K^{fQ} - (K^{fQ+1}+K^{fQ-1} )\star s = s \, \delta_{Q1}\,.
\end{equation}
If the TBA equations contain other types of kernels these typically also reduce to something nice under the action of $(K+1)^{-1}$; we will see examples of this in later chapters.

With these identities we can rewrite the auxiliary TBA equations \eqref{eq:cGNcanonicalTBA2} for the chiral Gross-Neveu model as
\begin{equation}
\log Y_Q = \log (1 + Y_{Q+1})(1 + Y_{Q-1}) \star s+ \delta_{Q,1}\log \left(1 + \frac{1}{Y_f}\right) \star s \, ,
\end{equation}
so that the infinite sums have disappeared. The chemical potentials (twists) have disappeared completely from these equations as well, since $1\star s=1/2$ and hence $c \,Q$ is in the null-space of $(K+1)^{-1}_{PQ}$ for any constant $c$. This shows us that the canonical TBA equations carry more information than the simplified TBA equations. As we will discuss in more detail in chapter \ref{chapter:twistedspectrum} the twists modify the large $u$ asymptotics of the Y-functions, and once these are specified the simplified and canonical TBA equations are equivalent. The infinite sum in the main TBA equation can now be removed by noting that similarly to $K^{fQ}$, $K^{Qf}$ satisfies
\begin{equation}
\label{eq:KMfidentity}
K^{Qf} - I_{QP} s \star K^{Pf}= s \delta_{Q1}\,.
\end{equation}
Then by rewriting the above simplified equations as
\begin{equation}
\log Y_Q - I_{QP} \log Y_P \star s = I_{QP} \log \left(1 + \frac{1}{Y_{P}}\right) \star s + \delta_{Q,1}\log \left(1 + \frac{1}{Y_f}\right) \star s\, ,
\end{equation}
integrating with $K^{Qf}$ and using eqn. \eqref{eq:KMfidentity} we get
\begin{equation}
\log Y_1 \star s + i \a  = \log \left(1 + \frac{1}{Y_{Q}}\right) \star K^{Qf} - \log \left(1 + \frac{1}{Y_{1}}\right) \star s  + \log \left(1 + \frac{1}{Y_f}\right) \star s \star K^{1f}\, ,
\end{equation}
or in other words
\begin{equation}
\label{eq:infmagnonsumidentity}
\log \left(1 + \frac{1}{Y_{Q}}\right) \star K^{Qf} = i \a + \log \left(1 + Y_{1}\right) \star s  - \log \left(1 + \frac{1}{Y_f}\right) \star s \star K^{1f}\, .
\end{equation}
We will for the moment skip over the technical details leading to the $i\a$ contribution in this identity; they will be discussed in chapter \ref{chapter:twistedspectrum} and appendix \ref{app:hybridTBA}. The main TBA equation \eqref{eq:cGNcanonicalTBA1} then becomes
\begin{equation}
\log Y_f = L \mathcal{E} + i\pi - \log \left(1 + Y_{1}\right) \star s \,,
\end{equation}
upon noting that magically enough the $Y_f$ contribution drops out completely thanks to $K^{ff}=s \star K^{1f}$.\footnote{To show this we can for example compute the integral in the second term by residues. The fact that the complicated scalar factor cancels out of the simplified TBA equations appears to be ubiquitous, an observation first made in \cite{Zamolodchikov:1991et}, and might in fact be used to reverse-engineer the scalar factor \cite{Janik:2008hs}.} For uniformity we can define $Y_0 \equiv Y_f^{-1}$ and get
\begin{equation}
\label{eq:cGNsTBA}
\log Y_M = \log (1+Y_{M+1})(1+Y_{M-1}) \star s - \delta_{M,0}(L \mathcal{E} + i\pi)
\end{equation}
with $Y_{M}\equiv 0$ for $M<0$. This form of the TBA equations goes under the name of \emph{simplified TBA equations}. To finish what we started, we can now apply $s^{-1}$ to these equations to get
\begin{equation}
\label{eq:cGNYsys}
Y^+_M \, Y^-_M  = (1+Y_{M+1})(1+Y_{M-1})\,,
\end{equation}
where the $\pm$ denote shifts in the argument by $\pm i$; $f^\pm(u) \equiv f(u\pm i)$ (note that the energy and $i\pi$ are in the null space of $s^{-1}$, at least mod $2\pi i$). These equations are known as the \emph{Y-system} \cite{Zamolodchikov:1991et}.

\begin{figure}%
\centering
\includegraphics[width=6cm]{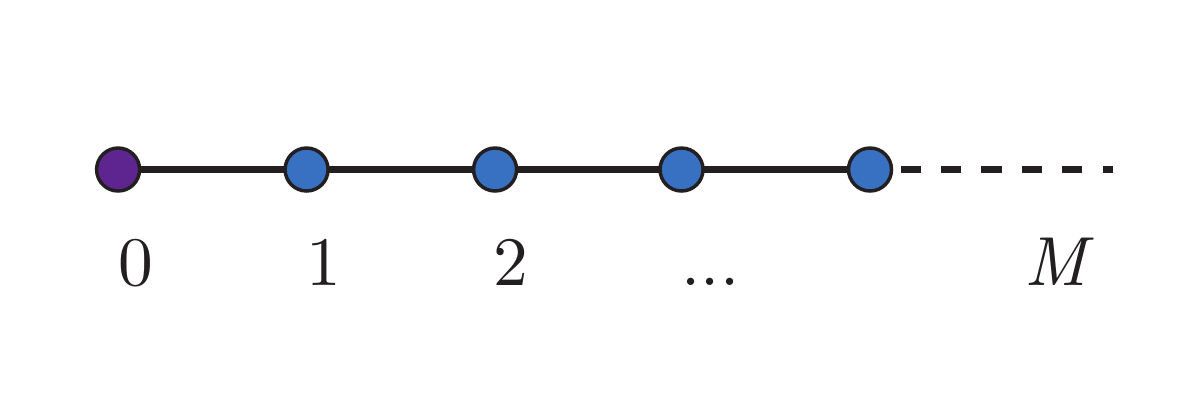}
\caption{The TBA structure for the chiral Gross-Neveu model in diagrammatic form. This graph illustrates the coupling between nearest neighbours in the simplified TBA equations \eqref{eq:cGNsTBA} or Y-system \eqref{eq:cGNYsys}, where the different colour on the first node signifies the fact that it is `massive' corresponding to the $\delta_{M,0}$ term in the simplified equations (this is also frequently denoted by putting a $\times$ in the open circle).}
\label{fig:cGNYsys}
\end{figure}

In general, the structure of simplified TBA equations and Y-systems can be represented diagrammatically by graphs. For example, in this case eqs. \eqref{eq:cGNsTBA} and \eqref{eq:cGNYsys} can be represented by figure \ref{fig:cGNYsys}.

It is important to note that in this process we lose information at each step along the way; both $(K+1)^{-1}$ and $s^{-1}$ have null-spaces. Therefore the simplified TBA equations are only equivalent to the canonical TBA equations provided we specify additional information on the Y-functions such as their large $u$ asymptotics. An alternative but equivalent specification often encountered in the literature is to give the large $Q$ asymptotics of the $Y_Q$ functions. We will discuss both of these in more detail in chapter \ref{chapter:twistedspectrum}. The Y-system requires even further specifications to really correspond to a particular model. For example the Y-system for the XXX spin chain is given by dropping $Y_0$ altogether, but this is nothing but the chiral Gross-Neveu Y-system again, just shifting the label $M$ by one unit.

\subsection{Excited states}
\label{sec:excitedstatesgeneral}

At this point we have actually done something quite impressive; we have found a system of equations we can solve (admittedly numerically) to find the exact finite volume ground state energy of a two dimensional field theory. It would be great if we could extend this approach to the entire spectrum. If we look back at our arguments however, we are immediately faced with a big conceptual problem; the mirror trick and infinite volume limit work nicely precisely for the ground state and the ground state only! Still it is hard to believe that a set of complicated TBA equations knows about the ground state only, especially since they are derived from the mirror Bethe-Yang equations which are just an analytic continuation away from describing the complete large volume spectrum. In this section we will take an approach often taken in physics; we will analytically continue from one part of a problem to another, in this case from the ground state energy to excited state energies. The idea that excited states can be obtained by analytic continuation is an old one, see for example \cite{Bender:1969si} in the case of the quantum anharmonic oscillator.

\subsubsection*{A simple example}

\begin{figure}%
\centering
\subfigure[]{\includegraphics[width=7cm]{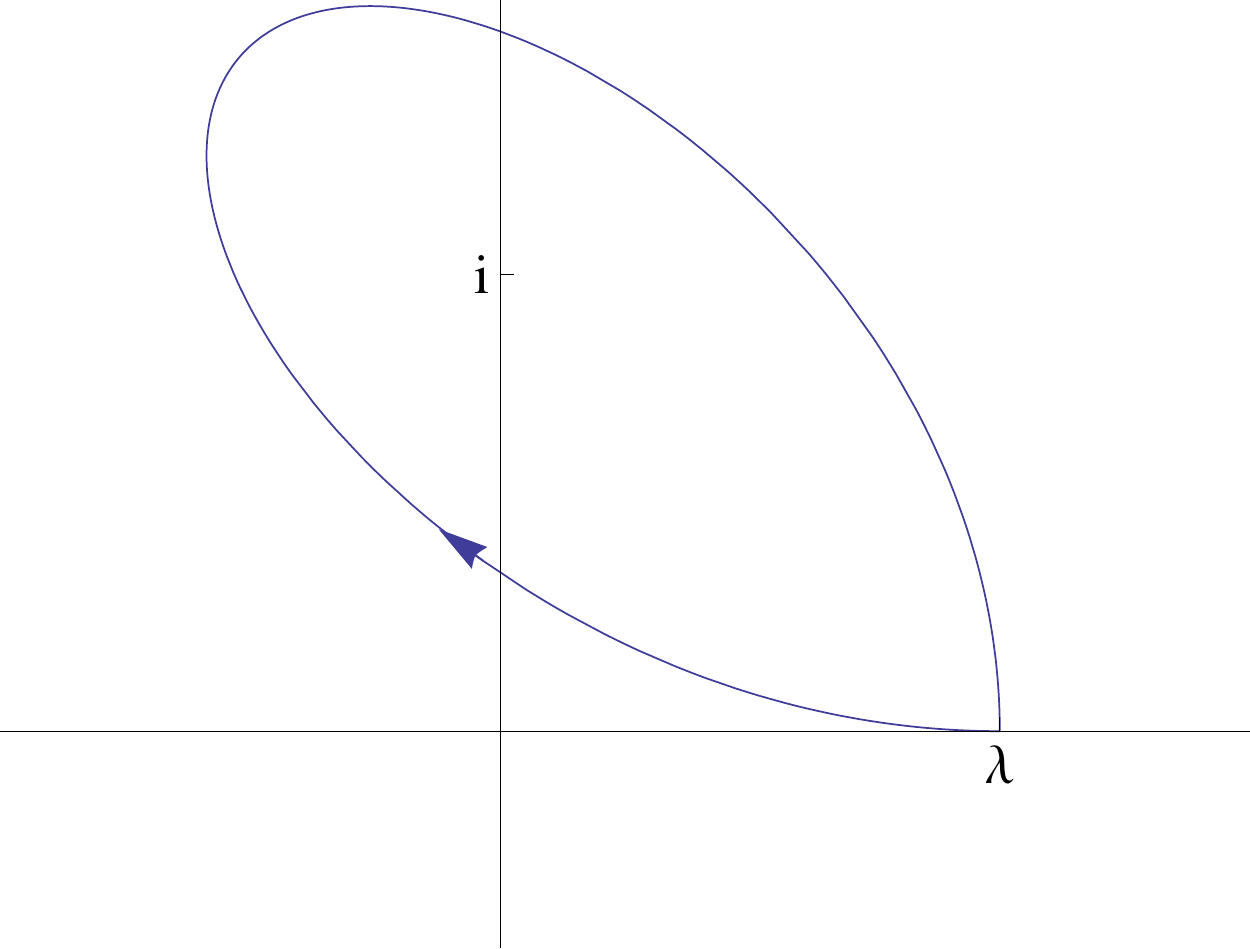} \label{fig:aroundthebranchpoint}} \quad
\subfigure[]{\includegraphics[width=7cm]{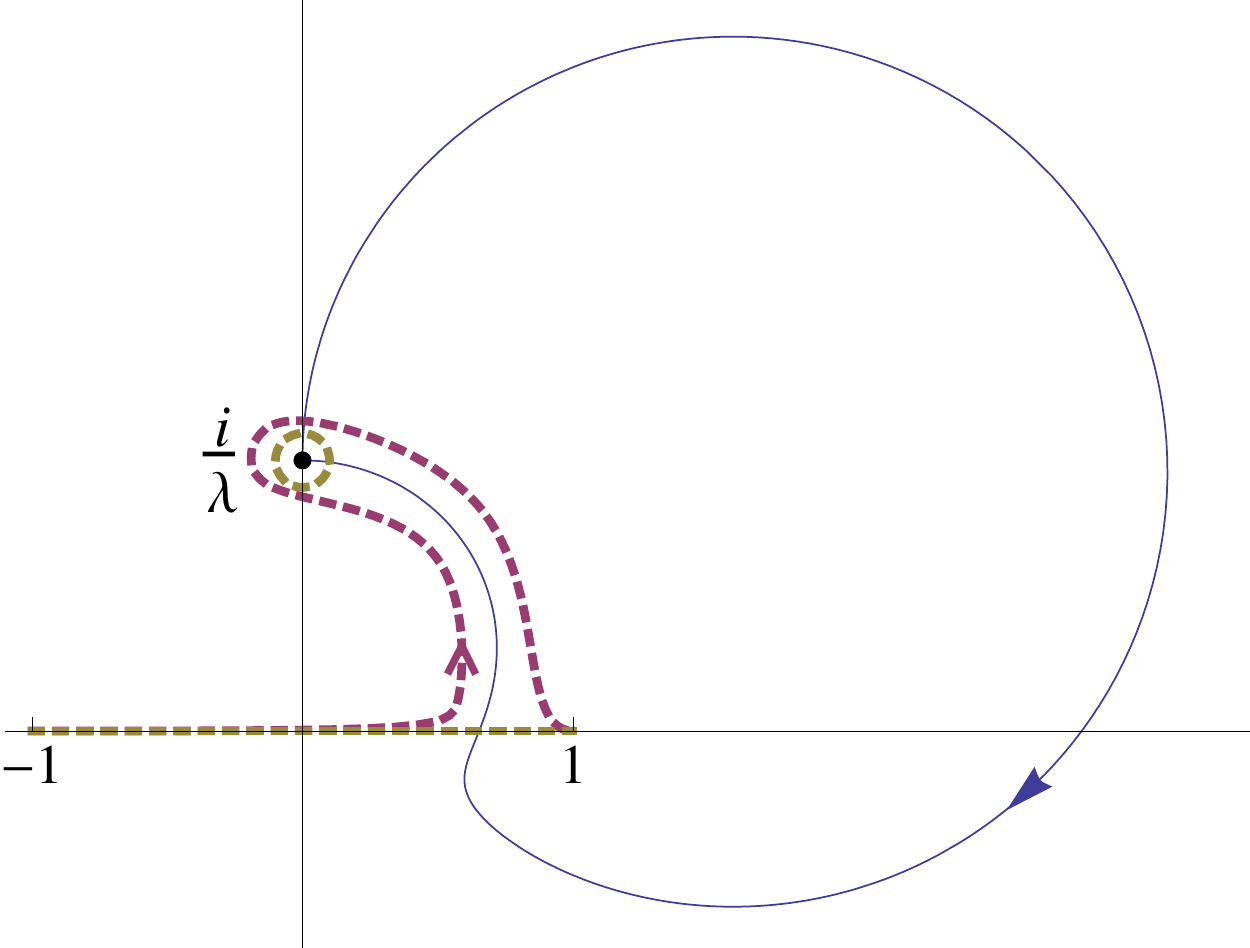} \label{fig:throughthecontour}}
\caption{Analytic continuation. The left figure shows the analytic continuation of $\lambda$ (blue) around the branch point at $i$, which corresponds to flipping the sign of $-\sqrt{1+\lambda^2}$ upon returning to the real line. The right figure shows the corresponding movement of the pole at $i/\lambda$ (blue) which drags the integration contour (red, dashed) in eqn. \eqref{eq:simpleenergyasintegral} along with itself for continuity. Upon taking the integration contour back to the real line we retain a residual contribution (yellow, dashed).}
\label{fig:analyticcont}
\end{figure}

Before moving on, we would like to motivate and illustrate these ideas on the simple quantum mechanical problem\footnote{This nice example can be found in slides of a talk by P. Dorey at IGST08 \cite{doreyIGST08}.}
\begin{equation}
H \psi = E \psi \, , \,\,\,\,\, \mbox{with} \, \, \, \,\, H = \left(\begin{array}{cc} 1 & 0 \\ 0 & -1 \end{array}\right) + \lambda \left(\begin{array}{cc}0 & 1 \\ 1 & 0 \end{array}\right)\, .
\end{equation}
After considerable effort we realize that the spectrum in this model is given by
\begin{equation}
E(\lambda) = \pm\sqrt{1+\lambda^2}\, ,
\end{equation}
and hence the ground state energy is $-\sqrt{1+\lambda^2}$. Allowing ourselves to analytically continue in the coupling constant we realize that the equation for the ground state energy has branch points at $\lambda = \pm i$. This has the interesting consequence that by analytically continuing around either of these branch points and coming back to the real line we do not quite get back the ground state energy, but rather the energy of the excited state. This is illustrated in figure \ref{fig:aroundthebranchpoint}.
The message we can take away from this, as stated in \cite{doreyIGST08}, is that by analytically continuing a parameter around a `closed contour'\footnote{We come back to the `same' value, though not necessarily on the same sheet.} we end up back at the same problem although our eigenvalue may have changed. Because it is still the same problem, if the continued eigenvalue is changed it must be one of the other eigenvalues of the problem. Note that this does not imply all eigenvalues can be found this way; the spectrum may well split into distinct sectors closed under analytic continuation.

Let us now forget about the description of this simple example in terms of linear algebra, and suppose for the sake of the argument that in solving our spectral problem we had obtained
\begin{equation}
\label{eq:simpleenergyasintegral}
E(\lambda) = -\int_{-1}^{1} dz \, \frac{1}{2\pi i} f(z) g(z)-1\, .
\end{equation}
where
\begin{equation}
f(z) = \frac{1}{z - i/\lambda}\, ,\,\,\,\, \mbox{and} \,\,\,\, g(z) = 2 \lambda \sqrt{1-z^2}\, .
\end{equation}
We can determine that this integral has branch points at $\lambda = \pm i$ without knowing anything about $f(z)$ other than that it is meromorphic with a single pole at $i/\lambda$ (with finite residue). Conceptually we consider $g(z)$ to be some nice known function, while $f$ is not explicitly known. Analytically continuing the integral in $\lambda$ we get a function that is well defined everywhere except for the half-lines $i \lambda>1$ and $i \lambda<-1$ where this pole moves into the domain of integration. Continuing around the point $\lambda = i$ as in figure \ref{fig:aroundthebranchpoint}, nothing happens when we first cross the line $\mbox{Re}(\lambda)=0$ but when we cross the second time, the pole moves through our integration contour on the real line, dragging the contour along as illustrated in figure \ref{fig:throughthecontour}. We can rewrite the resulting contour integral in terms of our original one by picking up the residue, resulting in
\begin{equation}
E^c(\lambda) = -\int_{-1}^{1} dz \, \frac{1}{2\pi i} f(z) g(z) + g(i/\lambda)-1\, .
\end{equation}
Since $E^c(\lambda) - E(\lambda) = g(i/\lambda) \neq 0$ there must be a branch point inside this contour. In this integral picture we do not need to know the precise analytic expression of $E$ or equivalently $f$ to determine the excited state expression for the energy. All we need to know is the pole structure of $f$ relative to the integration contour.

\subsubsection*{Analytic continuation of TBA equations}

Inspired by this simple example, we can try to analytically continue our expression for the ground state energy, eqn. \eqref{eq:TBAgroundstateenergy}, in some appropriate variable and see whether we encounter any changes in the description. We could try continuing in the mass variable of the chiral Gross-Neveu model for example. This method of obtaining excited state TBA equations was proposed and successfully applied to the scaling Lee-Yang model in \cite{Dorey:1996re}. They observed that in the process of analytic continuation the Y-functions solving the TBA equations undergo nontrivial monodromies. They also noted that changes in the form of the TBA equations are possible if singular points of $1+1/Y$ move in the complex plane during this analytic continuation. These changes are analogous to the changes in the energy formula of our simple example. In this case the integral is a typical term on the right hand side of the TBA equations
\begin{equation}
y(u) \equiv \log\left(1 + \frac{1}{Y}\right)\star K(u)\, .
\end{equation}
where we recall that $\star$ denotes (right) convolution on the real line. If there is a singular point
\begin{equation}
Y(u^*) = -1\, ,
\end{equation}
and its location $u^*$ crosses the real line during the analytic continuation, we can pick up the residue just as in our simple example to get
\begin{equation}
y^c(u) = \log\left(1 + \frac{1}{Y}\right)\star K(u) \pm \log S(u^*,u)\, ,
\end{equation}
where we recall that $K(v,u) = \frac{1}{2\pi i} \frac{d}{dv} \log S(v,u)$ and the sign is positive for singular points that cross hit the contour from below and negative for those that hit it from above. If Y vanishes at a particular point, this leads to the same considerations, just resulting in an opposite sign. If we wanted to do this at the level of the simplified equations, all we need is the S-matrix associated to $s$, which is given by
\begin{equation}
\label{eq:canonicalSmatrix}
S(u) = - \tanh \frac{\pi}{4}(u-i)\, .
\end{equation}
Note that the energy itself is also determined by an integral equation in the TBA approach, meaning it can change explicitly as well as implicitly through the solution of a changed set of TBA equations.

The upshot of this is that the excited state TBA equations obtained this way differ from those of the ground state by the addition of $\log S$ terms, which we will call \emph{driving terms}.\footnote{Interestingly, it appears to be possible to obtain the same terms by viewing an excited state as a momentum dependent generalized defect operator \cite{Bombardelli:2013yka}.} It should not matter whether we consider this procedure at the level of the canonical equations or at the level of the simplified equations, and indeed the results agree because of the S-matrix analogue of identities like eqn. \eqref{eq:KQMidentity}. The case of the Y-system is a bit more peculiar however, since there the distinguishing features of an excited state completely disappear. This follows directly from the fact that the S-matrix \eqref{eq:canonicalSmatrix} naturally vanishes under application of $s^{-1}$. From this we see that whatever excited state TBA equations we obtain this way, the Y-system equations are the same as those of the ground state; the Y-system is \emph{universal}.\footnote{As we will see in chapter \ref{chapter:quantumTBA}, there are very specific circumstances in which the situation is a little more subtle.}

This last point is closely related to other approaches of obtaining equations that describe excited state energies. In some cases it is possible to construct a functional analogue of the Y-system directly. If we can then get satisfactory insight into the analytic structure of the corresponding objects, we can `integrate' these functional relations to obtain integral equations describing the energy of excited states \cite{Klumper:1992vt,Klumper:1993,Bazhanov:1996aq,Fioravanti:1996rz}. Depending on how these functional equations are integrated we can obtain equations of TBA form but also various other forms that can be more efficient in actual calculations. The latter equations generically go under the name of `non-linear integral equations' \cite{Klumper:1992vt}, but depending on the context are also called `Kl\"umper-Pearce' \cite{Klumper:1991-1,Klumper:1991-2,Klumper:1992vt,Klumper:1993vq} or `Destri-de Vega' \cite{Destri:1992qk,Destri:1994bv} equations. While not obvious from their form, these different types of equations should be equivalent \cite{Juttner:1997tc,Kuniba:1998}.

We will employ an amalgamation of these ideas in the next chapters to find excited state TBA equations, going under the name of the contour deformation trick. We will come back to this again, but the basic idea goes as follows. It turns out we will have a candidate solution of the Y-system for an excited state with some limited regime of applicability. We then assume that the form of the TBA equations for an excited state is uniform and does not change outside of the regime of applicability of our candidate solution. Next, drawing lessons from the analytic continuation story above we expect that the only changes in the equation should be the addition of possible driving terms. Furthermore, although our limited solution only gives us a static picture, we expect that we can qualitatively view these terms as if coming from singular points that crossed the integration contour. Since in this picture such singular points would have dragged the contour along with them, we expect that an excited state TBA equation should be of the same form as the ground state, except with modified integration contours. Analyzing the analytic structure of the candidate solution will allow us to consistently define these contours in such a way that the TBA equations are satisfied, and by taking the integration contours back to the real line we can explicitly pick up the corresponding driving terms. Coming back to our simple example, it would be as if internal consistency of the problem (perhaps in the form of some other equation) told us that the natural integration contour for the excited state was not $(-1,1)$, but a contour that starts at one and finishes at minus one while enclosing $i/\lambda$ between itself and the real line. Such a contour is of course equivalent to the red contour in figure \ref{fig:throughthecontour} obtained by direct analytic continuation.

In the next chapter we will discuss how the program of this chapter applies to the $\ads$ superstring.

\chapter[The \texorpdfstring{$\ads$}{} superstring as an integrable model]{The \texorpdfstring{$\boldsymbol \ads$}{} superstring\\ \texorpdfstring{$\mbox{\hspace{10pt}}$}{} as an integrable model}
\label{chapter:AdS5string}

As we mentioned in the introduction, the $\ads$ superstring is a classically integrable model, but quantum integrability of the light-cone gauge fixed superstring is considerably harder to establish. In this chapter we will therefore briefly describe the results obtained in the ``top-to-bottom" approach of \emph{assuming} the light-cone superstring to be a quantum integrable field theory. Under this assumption the S-matrix of the string can be determined exactly, which allows us to go through the program outlined in the previous chapter. Doing so we obtain TBA equations that describe the exact energy of excited states. As already mentioned, the subtlety in this story is the interesting (non-relativistic) dispersion relation of the light-cone string.

\section{The S-matrix of the \texorpdfstring{$\ads$}{} superstring}
\label{sec:AdS5stringtheory}

Fixing a light-cone gauge on the time direction $t$ in ${\rm AdS}_5$ and an angle $\phi$ parametrizing an equator in ${\rm S}^5$ the $\ads$ superstring becomes a quantum field theory defined on a circle of circumference $J$, where $J$ is the angular momentum on the sphere associated to $\phi$. If we then consider the limit $J\rightarrow \infty$ the cylinder decompactifies to a plane, resulting in a quantum field theory with massive excitations, eight bosonic and eight fermionic as appropriate for a supersymmetric string in ten dimensions. By this gauge fixing, the global $\mathfrak{psu}(2,2|4)$ symmetry algebra of the string is broken to a manifest $\mathfrak{psu}(2|2) \oplus \mathfrak{psu}(2|2) \oplus \mathbb{H}$, where $\mathbb{H}$ is the light-cone Hamiltonian central with respect to both $\mathfrak{psu}(2|2)$ factors. If we go off-shell by giving up the level-matching condition that the total world-sheet momentum $\mathbb{P}$ is zero mod $2\pi$, each copy of the $\mathfrak{psu}(2|2)$ algebra picks up the same two further central extensions $\mathbb{C}(g,\mathbb{P})$ and $\mathbb{C}^\dagger(g,\mathbb{P})$ depending on the string tension $g$ and the world-sheet momentum. The sixteen excitations of our model form a so-called short (atypical) representation of this algebra, transforming in the four dimensional short representations of each copy of the triply centrally extended $\mathfrak{psu}(2|2)$, henceforth denoted $\mathfrak{psu}(2|2)_{c.e.}$. The condition for these short representations to exist is a constraint on the central elements known as the shortening condition. By expressing the central elements through the light-cone Hamiltonian and the world-sheet momentum, this condition turns into the dispersion relation
\begin{equation}
\label{eq:dispersion}
\mathcal{E}(p) = \sqrt{1 + 4 g^2 \sin^2 \frac{p}{2}}\, ,
\end{equation}
where $\mathcal{E}$ and $p$ are the energy and momentum of an elementary excitation. This dispersion relation looks like the dispersion relation for a lattice discretization of a relativistic model,\footnote{This is perhaps most easily seen from the Klein-Gordon equation on a spatial lattice with spacing $a$, $\partial_t^2 \phi_n - (\phi_{n+1} - 2\phi_{n} +\phi_{n-1})/a^2 + m^2 \phi_n = 0$, which after a (spatially discrete) Fourier transform becomes $(\mathcal{E}^2 - 4 \sin^2(p/2) /a^2 - m^2)\hat{\phi}(p) = 0$. For $m=1$ and $a=1/g$ the resulting dispersion relation looks identical to the light-cone dispersion relation.} and shows that the light-cone gauge fixed string is a \emph{non-relativistic} model. Nonetheless we can proceed and consider the scattering of our particles assuming quantum integrability in the form of factorized scattering. Insisting that the S-matrix is compatible with the $\mathfrak{psu}(2|2)_{c.e.}^{\oplus2}$ symmetry, is unitary, and satisfies the Yang-Baxter equation, fixes it uniquely up to a phase as the tensor product of two `S'-matrices each invariant under one copy of $\mathfrak{psu}(2|2)_{c.e.}$.\footnote{For a review see e.g. \cite{Arutyunov:2009ga,deLeeuw:2010nd}. Note also that since the full string Hamiltonian is part of this symmetry algebra, the resulting S-matrix is not the canonical S-matrix as typically defined in QFT; the canonical S-matrix commutes with the \emph{free} Hamiltonian instead. Still, \emph{on the space of scattering states} these two objects are unitarily equivalent. See section 2.3.1 of \cite{Arutyunov:2009ga} for more details. We should also point out that this `S'-matrix is closely related to the R-matrix of the Hubbard model, originally found by rather different means \cite{Shastry:1986zz}. See \cite{Mitev:2012vt} for a discussion in the present context.} In relativistic theories this phase would be further constrained by crossing, a property we do not immediately have for our non-relativistic theory. Still, lacking some other obvious physical condition to impose on this phase, we can try to insist that the S-matrix satisfies some non-relativistic generalization of crossing \cite{Janik:2006dc}.\footnote{As argued in \cite{Janik:2006dc}, this is not a particularly strange thing to consider. First of all, there is nothing principally wrong with quantizing our superstring in a manifestly relativistically invariant fashion without fixing a light-cone gauge. In that case, something of the crossing symmetry of the relativistic description should remain in the physically equivalent light-cone gauge fixed model. Furthermore, in certain sectors even the light-cone string theory is relativistic \cite{Alday:2005jm}.} This generalized form can be obtained by phrasing crossing in algebraic terms \cite{Janik:2006dc} (see also \cite{Arutyunov:2006yd}) and results in the direct non-relativistic generalization of crossing already written in eqn. \eqref{eq:crossinggeneneral}, of course upon appropriately identifying the particle to anti-particle transformation. The natural solution to the resulting crossing equation gives an S-matrix that matches perfectly with all known explicit computations. Hence the scattering theory of the light-cone superstring can be considered solved. An explicit form for this S-matrix is given in appendix \ref{app:qdefmatrixSmatrix}. The entire problem is most conveniently parametrized in terms of a set of variables $x^\pm$ related to the energy and momentum as
\begin{equation}
\mathcal{E} = i g (x^- -x^+)-1\, , \, \, \, \, \, \, \frac{x^+}{x^-} = e^{ip} \, .
\end{equation}
The shortening condition, or dispersion relation, becomes the simple relation
\begin{equation}
\label{eq:xpmdispersion}
x^+ + \frac{1}{x^+} - x^- - \frac{1}{x^-} = \frac{2i}{g}\, .
\end{equation}

%Of course this S-matrix is defined up to unitary equivalence only; we can always do a unitary transformation in the one particle basis of states.

Because our total S-matrix naturally factorizes into the tensor product of two smaller $\mathfrak{psu}(2|2)_{c.e.}$ invariant S-matrices, the full transfer matrix factorizes into a product of two $\mathfrak{psu}(2|2)_{c.e.}$ invariant transfer matrices. In line with this, the eight bosons and eight fermions of our light-cone gauge fixed model can be naturally identified with pairs of excitations making up the fundamental representation of each of the two copies of $\mathfrak{su}(2|2)$, consisting of two bosons denoted $w_1$ and $w_2$ and two fermions denoted $\theta_3$ and $\theta_4$ respectively. The smaller transfer matrices can be diagonalized by the (algebraic) Bethe ansatz \cite{Martins:2007hb}. In chapter \ref{chapter:finitevolumeIQFT} we identified one particular particle type (spin up fermions) to form the vacuum in the (algebraic) Bethe ansatz for the chiral Gross-Neveu model. For our $\mathfrak{psu}(2|2)_{c.e.}$ invariant transfer matrices we will have two natural choices of vacuum due to the two $\mathfrak{su}(2)$ factors; we can choose either spin up bosons ($w_1$) or spin up fermions ($\theta_3$).\footnote{We do not consider choosing $w_2$ instead of $w_1$ or $\theta_4$ instead of $\theta_3$ as this gives not only equivalent physics, but also manifestly equivalent formulae.} In either case the transfer matrix will have length
\begin{equation}
K^{\rm I} \equiv n(w_1) + n(w_2) + n(\theta_3) + n(\theta_4)\,,
\end{equation}
which is just the total number of excitations; $n(\chi)$ denotes the number of excitations of type $\chi$.  In addition the numbers
\begin{equation}
K^{\rm II}  \equiv 2 n( w_2)+ n(\theta_3) + n(\theta_4)\,,
\end{equation}
and
\begin{equation}
K^{\rm III}  \equiv n(w_2)+ n(\theta_4) \,,
\end{equation}
are conserved under scattering and denote the numbers of auxiliary excitations introduced in the Bethe ansatz over the bosonic vacuum. $K^{\rm II}$ has the interpretation of the number of (`fermionic') excitations over the bosonic vacuum, where the boson $w_2$ effectively counts double since on a vacuum of $w_1$ bosons it can `decay' (scatter) into two fermions. $K^{\rm III}_{(a)}$ counts the number of fermions of a specific type (say $\theta_4$). Over the fermionic vacuum the relevant excitation numbers are instead given by
\begin{equation}
K^{\rm II}  \equiv n(w_1) +  n(w_2) + 2n(\theta_4)\,,
\end{equation}
and
\begin{equation}
K^{\rm III}  \equiv n(w_2) + n(\theta_4) \,.
\end{equation}
These numbers have a similar interpretation to the ones before, just in terms of bosons rather than fermions.

For the full transfer matrix we have two $\mathfrak{psu}(2|2)_{c.e.}$ factors which we will label by $\llab$ and $\rlab$ respectively. This results in four possible choices for the full vacuum - $w_1^\llab w_1^\rlab$, $w_1^\llab \theta_3^\rlab$, $\theta_3^\llab w_1^\rlab$, and $\theta_3^\llab \theta_3^\rlab$. For the light-cone string it turns out to be most convenient to choose a purely bosonic vacuum over which we have a simple bound state pole structure.\footnote{This vacuum also goes under the name of the $\mathfrak{su}(2)$ vacuum, hence the label on the S-matrix.} Just as we excited $\NA$ down spins over a vacuum of $\NF$ up spins for the auxiliary XXX spin chain of the chiral Gross-Neveu model, here we can consider $K^{\rm II}$ fermionic excitations called $y$-particles over the bosonic vacuum of length $K^{\rm I}$, which we recall is the total number of physical excitations. In this case the problem is nested and to diagonalize the transfer matrix we need to go down one more level and consider the $K^{\rm II}$ $y$-particles to make up the vacuum for the second auxiliary problem, over which we excite $K^{\rm III}$ $w$-particles. Taking the corresponding auxiliary Bethe equations and the eigenvalue of the transfer matrix that gives us the main Bethe-Yang equation we get the following set of equations \cite{Beisert:2005fw,Beisert:2005tm,Martins:2007hb,deLeeuw:2007uf}
\begin{align}
1=&e^{ip_{k} J} \prod_ {l\neq k}
^{K^{\mathrm{I}}}S_{\mathfrak{su}(2)}(x_{k},x_{l})
\prod_{a=1}^{2}\prod_{l=1}^{K^{\mathrm{II}}_{(a)}}\frac{{x_{k}^{+}-y_{l}^{(a)}}}{x_{k}^{-}-y_{l}^{(a)}}
\sqrt{\frac{x_k^-}{x_k^+}} \\
(-1)^m=&\prod_{l=1}^{K^{\mathrm{I}}}\frac{y_{k}^{(a)}-x^{-}_{l}}{y_{k}^{(a)}-x^{+}_{l}}\sqrt{\frac{x_l^+}{x_l^-}}
\prod_{l=1}^{K^{\mathrm{III}}_{(a)}}\frac{v_{k}^{(a)}-w_{l}^{(a)}-\frac{i}{g}}{v_{k}^{(a)}-w_{l}^{(a)}+\frac{i}{g}} \\
1=&\prod_{l=1}^{K^{\mathrm{II}}_{(a)}}\frac{w_{k}^{(a)}-v_{l}^{(a)}+\frac{i}{g}}{w_{k}^{(a)}-v_{l}^{(a)}-\frac{i}{g}}
\prod_ {l\neq
k}^{K^{\mathrm{III}}_{(a)}}\frac{w_{k}^{(a)}-w_{l}^{(a)}-\frac{2i}{g}}{w_{k}^{(a)}-w_{l}^{(a)}+\frac{2i}{g}}\, ,
\end{align}
where
\begin{equation}
\label{eq:su2smatrix}
S_{\mathfrak{su}(2)}(x_k,x_l)=\frac{1}{\sigma_{kl}^2}\, \frac{x_k^+}{x_k^-}\frac{x_l^-}{x_l^+}\,
\frac{x_k^--x_l^+}{x_k^+-x_l^-}\frac{1-\frac{1}{x_k^-x_l^+}}{1-\frac{1}{x_k^+x_l^-}}\, ,
\end{equation}
and $\sigma$ is the so-called dressing phase \cite{Arutyunov:2004vx} which we will discuss in more detail below once some required structure has been introduced. In the above equations $m$ is the winding number of the string appearing in the level matching condition for the total world-sheet momentum $p$ as $p = \sum_k p_k = 2\pi m$. The winding number appears here since the fermions in the light-cone gauge have periodicity $\psi(\tau,2\pi) = (-1)^m \psi (\tau,0)$ due to a field redefinition \cite{Arutyunov:2007tc}.

The light cone superstring also has bound state excitations that can be analyzed as we did for the XXX spin chain in chapter \ref{chapter:finitevolumeIQFT} \cite{Dorey:2006dq,Chen:2006gq}, and just like for the XXX spin chain there can be bound states with any number of constituents, or in other words bound states of any length. The resulting $Q$-particle bound states correspond to poles in the scalar factor $S_{\mathfrak{su}(2)}$ where the momenta of pairs of particles satisfy\footnote{This of course refers to a specific ordering of momenta and the singular behaviour of the associated exponentials. The choice is such that this pole compensates the zero that $e^{i p_i J}$ gives in the limit $J\rightarrow \infty$, see e.g. \cite{Arutyunov:2007tc}.}
\begin{equation}
x^+(p_i)=x^-(p_{i+1})\, ,
\end{equation}
by which the resulting bound state central charges precisely satisfy the bound state BPS condition
\begin{equation}
\mathcal{E}(p) = \sqrt{Q^2 + 4 g^2 \sin^2 \frac{p}{2}}\, .
\end{equation}
Formed out of bosons, these BPS bound states form totally symmetric short representations of $\mathfrak{psu}(2|2)_{c.e.}$. Other poles in the scalar factor do not correspond to bound states but rather to anomalous thresholds \cite{Coleman:1978kk,Dorey:2007xn}.

%I have nowhere to move this naturall, so I decided to remove it.. : In fact, it is quite a nontrivial check that the pole structure of the `natural' solution to the crossing equation is consistent with these thresholds.

The full set of (bound state) Bethe-Yang equations can be used to compute energies of various string states as long as the string length is asymptotically large, however to describe exact energies at finite volume we will have to go to the mirror theory.

\section{The mirror theory}
\label{sec:AdS5mirrortheory}

Now that we understand the exact S-matrix and dispersion of the $\ads$ superstring, we should be able to follow the general considerations described in chapter \ref{chapter:finitevolumeIQFT} to determine the TBA equations describing the desired free energy of the mirror theory. For a relativistic model we can describe the mirror transformation in terms of a shift of the particle rapidities in the imaginary direction of the complex plane, which manifestly leaves the S-matrix invariant as it depends on the difference of rapidities. In our non-relativistic theory the mirror transformation does not leave the S-matrix invariant. In this section we will briefly review the mirror transformation for the light-cone superstring carefully worked out in \cite{Arutyunov:2007tc}.

Starting from the dispersion relation \eqref{eq:dispersion} of the string
\begin{equation}
\label{eq:dispersionaltform}
\mathcal{E}^2 - 4g^2\sin^2\frac{p}{2}= 1\, ,
\end{equation}
we can do the mirror transformation
\begin{equation}
\mathcal{E} \rightarrow i \tilde{p} \, , \, \, \, p \rightarrow i \tilde{\mathcal{E}}\, ,
\end{equation}
to obtain the mirror dispersion relation between the mirror energy $\tilde{\mathcal{E}}$ and momentum $\tilde{p}$ as
\begin{equation}
\label{eq:mirrordispersion}
\tilde{\mathcal{E}} = 2 {\mathop \mathrm{arcsinh}} \frac{\sqrt{1 + \tilde{p}^2}}{2 g}\, .
\end{equation}
Then we can directly obtain the mirror S-matrix from the string S-matrix by the substitution
\begin{equation}
p \rightarrow i \tilde{\mathcal{E}} =2 i {\mathop \mathrm{arcsinh}} \frac{\sqrt{1 + \tilde{p}^2}}{2 g}\, .
\end{equation}
It turns out that there is a unique choice of one particle basis for which this yields a unitary S-matrix for real mirror momenta; the S-matrix of our mirror theory. In principle we could try to work directly in terms of this S-matrix as a function of the mirror momenta, but it is more convenient to introduce variables analogous to the rapidity introduced for relativistic models.

Firstly, we would like to introduce a variable which uniformizes the dispersion relation of the string \eqref{eq:dispersionaltform}, which is achieved by taking \cite{Janik:2006dc}
\begin{equation}
p= 2 \, {\rm am}\,z\,,~~\quad~~ \sin\frac{p}{2} = \sn\, z\,,~~\quad~~ \mathcal{E} =
\dn\, z\, ,
\end{equation}
where here and below we always leave the elliptic modulus $\kappa=-4g^2<0$ implicit. The variable $z$ takes values on a torus with real and respectively imaginary periods $2 \omega_1$ and $2 \omega_2$
\begin{equation}
2\omega_1=4{\rm K}(\kappa)\, , ~~~~~~~~~ 2\omega_2=4i{\rm K}(1-\kappa)-4{\rm
K}(\kappa)\, ,
\end{equation}
where $K(\kappa)$ is the complete elliptic integral of the first kind. This torus is the analogue of the periodicity strip in the rapidity plane for a relativistic theory. Nonetheless we prefer to reserve the name rapidity for another parametrization which will follow in a while.

The real $z$ line corresponds to real values of energy and momentum and noting that the variables $x^\pm$ should have $|x^\pm|>1$ there, we can express them on the torus as
\begin{equation}
x^{\pm}=\frac{1}{2g}\Big(\frac{\cn\, z}{\sn\, z} \pm i \Big)(1+\dn\, z)
\end{equation}
Since both the dispersion relation and the $x^\pm$ variables are periodic with period $\omega_1$, we can restrict the $z$ variable to $-\omega_1/2 \leq \mbox{Re}(z) \leq \omega_1/2$, corresponding to $-\pi \leq p \leq \pi$.  Note that while the dispersion relation is uniformized on the torus we should not expect the same to hold for the S-matrix. For example, while the relativistic dispersion is uniformized on the strip $\mbox{Im}(u) \in (0,4)$, the chiral Gross-Neveu S-matrix is not periodic in the imaginary direction of the $u$-plane. Still, if we disregard the scalar factor the S-matrix becomes a meromorphic function on the torus. As we will see shortly, the scalar factor itself is not periodic however, so that the full S-matrix lives on the universal cover of (two copies of) our torus.

To identify the mirror transformation in terms of the $z$ variable, we can consider for which values of $z$ the mirror momentum $\tilde{p} = - i\mathcal{E}$ is real. In close analogy to the relativistic case where the mirror momentum is real on the lines $\mbox{Im}(u) = 1$ and $\mbox{Im}(u) = 3$ , here the mirror momentum is real on the lines $\mbox{Im}(z) = |\omega_2/2|$ and $\mbox{Im}(z) = |3\omega_2/2|$. As the mirror energy is positive on the line $\mbox{Im}(z) = |\omega_2/2|$, this line should correspond to the real momentum line of the mirror theory; shifting $z$ by $\omega_2/2$ we go from the physical to the mirror theory.

For the mirror theory it is convenient to pick a vacuum made of purely fermions rather than bosons, with the mirror S-matrix resulting in the following set of mirror Bethe-Yang equations
\begin{align}
1=&e^{i\tp_{k} R} \prod_ {l\neq k}
^{K^{\mathrm{I}}}S_{\sl(2)}(x_{k},x_{l})
\prod_{a=1}^{2}\prod_{l=1}^{K^{\mathrm{II}}_{(a)}}\frac{{x_{k}^{-}-y_{l}^{(a)}}}{x_{k}^{+}-y_{l}^{(a)}}
\sqrt{\frac{x_k^+}{x_k^-}} \label{eq:mirrorBYmain}\\
-1=&\prod_{l=1}^{K^{\mathrm{I}}}\frac{y_{k}^{(a)}-x^{-}_{l}}{y_{k}^{(a)}-x^{+}_{l}}\sqrt{\frac{x_l^+}{x_l^-}}
\prod_{l=1}^{K^{\mathrm{III}}_{(a)}}\frac{v_{k}^{(a)}-w_{l}^{(a)}-\frac{i}{g}}{v_{k}^{(a)}-w_{l}^{(a)}+\frac{i}{g}} \label{eq:mirrorBYy}\\
1=&\prod_{l=1}^{K^{\mathrm{II}}_{(a)}}\frac{w_{k}^{(a)}-v_{l}^{(a)}+\frac{i}{g}}{w_{k}^{(a)}-v_{l}^{(a)}-\frac{i}{g}}
\prod_ {l\neq
k}^{K^{\mathrm{III}}_{(a)}}\frac{w_{k}^{(a)}-w_{l}^{(a)}-\frac{2i}{g}}{w_{k}^{(a)}-w_{l}^{(a)}+\frac{2i}{g}}\, ,\label{eq:mirrorBYw}
\end{align}
where
\begin{equation}
\label{eq:sl2smatrix}
 S_{\sl(2)}(x_k,x_l) = \frac{1}{\sigma_{kl}^2}\,  \frac{x_k^+ - x_l^-}{x_k^-  - x_l^+}\frac{1 - \frac{1}{x_k^- x_l^+}}{1 - \frac{1}{x_k^+ x_l^-}} \, ,
\end{equation}
and $\sigma$ is now the mirror dressing phase \cite{Arutyunov:2009kf} (see also \cite{Volin:2009uv}). Also in this case we call the auxiliary excitations $y$ and $w$-particles, but we should keep in mind that they are now excitations over a fermionic vacuum contrary to situation for the string theory equations presented above. Note that when we compute the partition function the original fermions need to be anti-periodic in the imaginary time direction so that the mirror fermions are always anti-periodic in the spatial direction, resulting in the minus sign in the second equation. The auxiliary Bethe equations of our mirror model are precisely the Bethe equations describing the inhomogeneous Hubbard model, the inhomogeneities being determined by the momenta of the mirror excitations \cite{Arutyunov:2009zu}.

\subsection{Crossing}

At this point we have introduced the necessary notation to discuss crossing in a little more detail. When we consider crossing, we are considering analytically continuing some of the particles involved in our scattering process to their corresponding anti-particles. At the level of the torus, similar to the relativistic case this corresponds to the shift $z \rightarrow z + \omega_2$ if we cross the first particle and $z \rightarrow z - \omega_2$ if we cross the second \cite{Janik:2006dc}. Indeed, the energy and momentum change sign under this transformation.\footnote{Note that $x^\pm(z\pm \omega_2) =\frac{1}{x^\pm(z)}$.} The crossing equation \eqref{eq:crossinggeneneral} (using unitarity) turns into a constraint on the overall phase of our S-matrix, i.e. $S_{\sl(2)}$ or $S_{\su(2)}$ depending on our choice of vacuum. In terms of the dressing phase $\sigma$ it reads
\begin{equation}
\sigma(z_k,z_l)\sigma(z_k,z_l-\omega_2)=  \frac{x_k^+}{x_k^-}h(z_k,z_l)\, ,
\end{equation}
or by unitarity
\begin{equation}
\sigma(z_k+\omega_2,z_l)\sigma(z_k,z_l)=  \frac{x_l^-}{x_l^+}h(z_k,z_l)\, ,
\end{equation}
where
\begin{equation}
h(z_k,z_l) = \frac{x_k^- - x_l^+}{x_k^+ - x_l^+}\frac{1-\frac{1}{x_k^- x_l^-}}{1-\frac{1}{x_k^+ x_l^-}}\, .
\end{equation}
From this equation we manifestly see that the dressing phase is not periodic, since
\begin{equation}
\sigma(z_k,z_l-2\omega_2) = \frac{h(z_k,z_l-\omega_2)}{h(z_k,z_l)}\sigma(z_k,z_l)\, ,
\end{equation}
and
\begin{equation}
\sigma(z_k+2\omega_2,z_l) = \frac{h(z_k+\omega_2,z_l)}{h(z_k,z_l)}\sigma(z_k,z_l)\, .
\end{equation}
We should note that the dressing `phase' $\sigma$ is not unitary in the mirror theory, while of course the total scalar factor is. The nice object to split off from the mirror scalar factor (particularly with regard to bound states discussed below) is the so-called improved dressing phase
\begin{equation}
\Sigma(z_k,z_l)\equiv \frac{1-\frac{1}{x_k^+ x_l^-}}{1-\frac{1}{x_k^- x_l^+}} \sigma(z_k,z_l)\,.
\end{equation}

\subsection{Bound states}

Like the string theory, the mirror theory has bound state excitations of any length \cite{Arutyunov:2007tc}. Contrary to the string theory bound states however, they are bound states of fermions and transform in totally anti-symmetric short representations.\footnote{When discussing bound states in the superspace formalism of \cite{Arutyunov:2008zt} we can switch between symmetric and anti-symmetric bound state representations by swapping the role of the superspace bosons $w$ and fermions $\theta$. This way we can avoid doing double work and directly obtain mirror bound state objects from the corresponding string object (this applies in particular to the transfer matrices we will come to below). In this setting, the $\mathfrak{sl}(2)$ vacuum fermions gets described in terms of superspace bosons \emph{as a computational trick}; physically the $\mathfrak{sl}(2)$ vacuum is always built up with the canonical fermions in the fundamental of $\mathfrak{su}(2|2)$.} They are supported by poles in the mirror S-matrix which occur when\footnote{As for string bound states, this refers to the case where $e^{i \tp_i R}$ goes to zero as $R\rightarrow \infty$. Note the consequent interchange of $x^+$ and $x^-$ are between the string and mirror theory.}
\begin{equation}
x^-(\tp_i)=x^+(\tp_{i+1})\, .
\end{equation}
Similarly to the string S-matrix, other poles in the mirror S-matrix do not correspond to BPS configurations and should not correspond to bound states.

We saw above that the torus plays a role analogous to (twice) the physical strip in relativistic models as far as the dispersion is considered. However, as the physical strip is defined in terms of a relative rapidity it does not immediately have an analogue in our non-relativistic model. To define what we will call the physical region on the torus we should consider bound states \cite{Arutyunov:2007tc}. For relativistic models all bound states lie on the imaginary axis of the rapidity plane within in the physical strip, once in the $s$ and once in the $t$-channel. In other words, the constituent rapidities of a two-particle $s$-channel bound state lie between the lines $\mbox{Im}(u_j) = \pm 1$, a region we can call the physical strip for the individual rapidities of a two-particle bound state. In this spirit we will try to define the physical region of the string and mirror theory as the region on the torus which contains the constituents of all bound states of the string, respectively mirror theory.

\begin{figure}[h]
\begin{center}
$ \vcenter{\hbox{\includegraphics{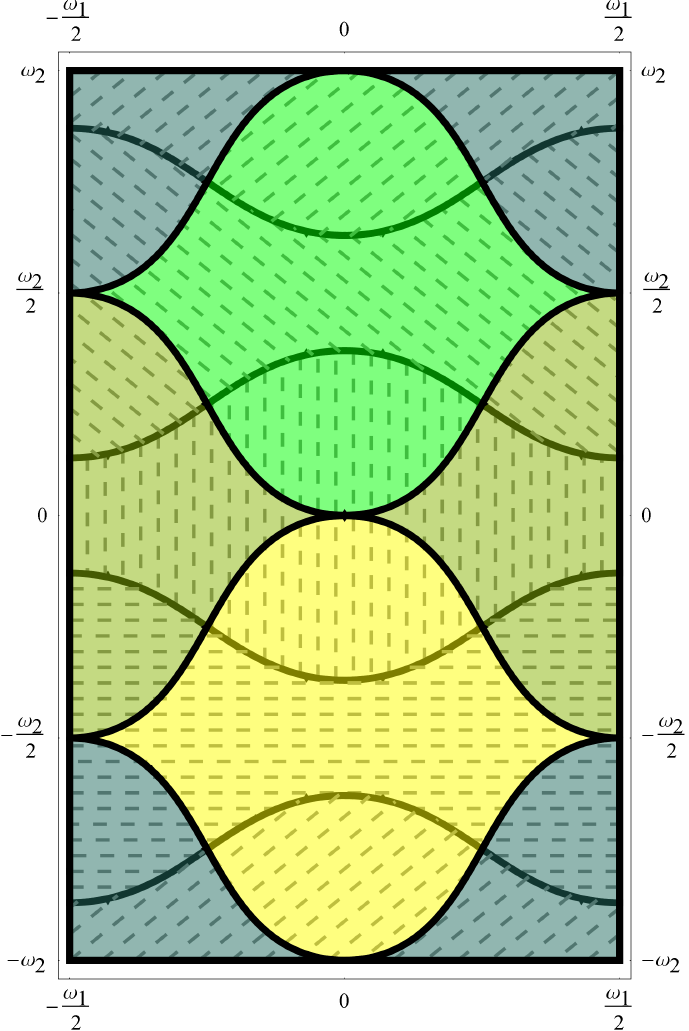}}}$ \quad $\vcenter{\hbox{\includegraphics{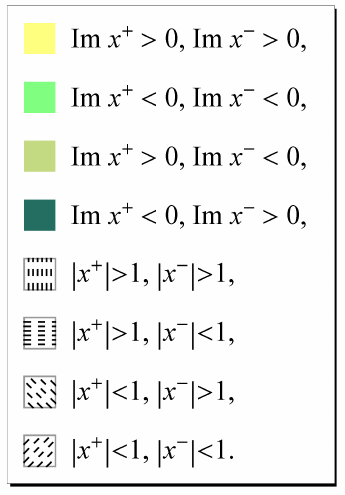}}}$
\caption{The torus and its division in terms of $|x^\pm|=1$ and $\mbox{Im}(x^\pm)=0$.}
\label{fig:torusundef}
\end{center}
\end{figure}

\begin{figure}[h]
\begin{center}
\includegraphics{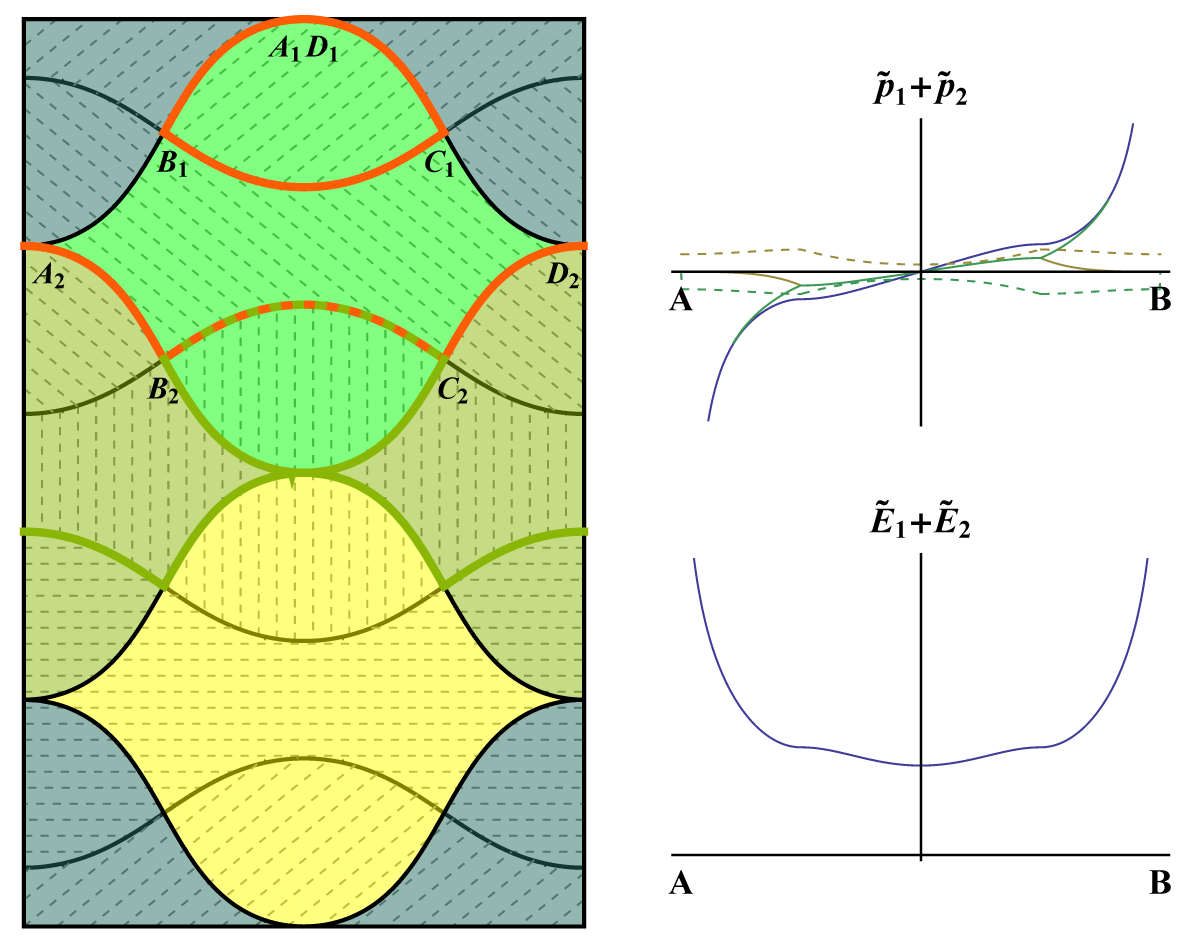}
\caption{Two-particle bound states on the torus. The constituents of a two-particle bound state in the mirror theory move along the orange paths $A_i B_i C_i D_i$ as their total momentum moves from $-\infty$ to $\infty$. The kinks corresponds to critical values of the momentum, beyond which the two constituent momenta are no longer complex conjugate. This is shown in the upper right graph where the green and yellow lines show the real (solid) and imaginary (dashed) parts of the constituent momenta. The analogous curves for a two-particle bound state in the string theory are drawn in olive-green.}
\label{fig:2partboundstateundef}
\end{center}
\end{figure}

If we were to consider the locations of string and mirror bound states on the torus we would quickly realize that it is natural to divide the torus in regions separated by the contours $|x^\pm|=1$ and $\mbox{Im}(x^\pm)=0$. Concretely then, we can illustrate the torus with these divisions as in figure \ref{fig:torusundef}.
In fact, the upshot of a careful analysis of the bound state equations for both the string and mirror theory \cite{Arutyunov:2007tc} shows that the constituents of a two-particle mirror bound state move on the torus along the curves $|x^{\pm}|=1$ inside the region $\mbox{Im}(x^\pm)<0$ as the bound state momentum $\tilde{p}$ grows. The point where the bound state constituents hit the lines $\mbox{Im}(x^\pm)=0$ corresponds to a critical value of  the momentum, after which the bound state constituents move along the lines $\mbox{Im}(x^\pm)=0$ , but in an asymmetric fashion. We have illustrated this by the orange curves in figure \ref{fig:2partboundstateundef}.
In fact, beyond the critical value of momentum there are two possible bound state solutions on the torus; the other solution corresponds to vertically flipping the light-green region. If we insist there should be only one bound state solution in the physical region of the mirror theory, it becomes very natural to consider the lines $\mbox{Im}(x^\pm)=0$ as the boundary of this region. Analyzing multi-particle bound states we would moreover find that there are $2^{Q-1}$ bound state solutions for a $Q$-particle bound state on the torus (on top of multiple choices beyond critical values), but that restricting ourselves to $\mbox{Im}(x^\pm)=0$ we find precisely one \cite{Arutyunov:2007tc}. Similar comments apply to bound states of the string theory if we interchange the role of $|x^\pm|$ and $\mbox{Im}(x^\pm)$; one of the two possible choices for a two-particle bound state is illustrated by the lime-green curves in figure \ref{fig:2partboundstateundef}. This leads us to identify the regions $|x^\pm|>1$ and $\mbox{Im}(x^\pm)<0$ as the physical regions of the string and mirror theory respectively.\footnote{Unlike relativistic models where both forward and crossed poles lie in the same physical region however, here the crossed poles of the string and mirror theory clearly do not lie in the same regions.}

We can cover each of these regions with a simple parametrization in terms of what we will call a rapidity $u$ by introducing the Zhukowski maps
\begin{equation}
\label{eq:xmirror}
x(u) = \frac{1}{2}(u - i \sqrt{4-u^2}) \, ,
\end{equation}
and
\begin{equation}
\label{eq:xstring}
x_s(u) =\frac{u}{2}(1 + \sqrt{1-\frac{4}{u^2}}) \, ,
\end{equation}
which are the explicit solutions to the dispersion relation \eqref{eq:xpmdispersion} upon identifying $x^\pm$ with $x(u\pm i/g)$ or $x_s(u\pm i/g)$ when restricting to $\mbox{Im}(x^\pm)<0$ or $|x^\pm|>1$ respectively. These functions satisfy
\begin{equation}
x_{(s)}(u) + \frac{1}{x_{(s)}(u) } = u\, .
\end{equation}
Note that $x_s$ has a branch cut on the interval $(-2,2)$ while $x$ has the complementary one on the real line, and that apart from these $|x_s|>1$ and $\mbox{Im}(x)<0$ on the complex $u$ plane. Moreover these functions are equal on the lower half plane and inverse on the upper half plane; they are each others' analytic continuation through their respective branch cuts. We would like to emphasize that the rapidity variable $u$ is more similar to the rapidity variable of a magnon than that of a relativistic particle.

In terms of this rapidity then, the mirror bound state condition $x^-(\tp_i)=x^+(\tp_{i+1})$ becomes the condition
\begin{equation}
x(u_i -i/g) = x(u_{i+1}+i/g) \, \Rightarrow \, \, u_i = u_{i+1} +2i/g\, ,
\end{equation}
exactly taking the form of the Bethe strings of section \ref{subsec:TBAgeneral}! Of course, up to an interchange of $u_i$ and $u_{i+1}$ string bound states also take this form in terms of $x_s(u)$. The energy, momentum, and S-matrices for these bound states can naturally be found by fusion as discussed in chapter \ref{chapter:finitevolumeIQFT}. Concretely for example the energy and momentum of $Q$-particle string and mirror bound states are given by
\begin{equation}
\mathcal{E}_Q(u) = i g (x_s(u-i Q/g) -x_s(u+i Q/g))-Q\, , \, \, \, \, \, \, p_Q(u) = - i \log \frac{x_s(u+i Q/g)}{x_s(u-i Q/g)} \,
\end{equation}
and
\begin{equation}
\tH_Q(u) = \log \frac{x(u+i Q/g)}{x(u-i Q/g)} \, , \, \, \, \, \, \, \tilde{p}_Q(u) = g (x(u-i Q/g) -x(u+i Q/g))+i \,
\end{equation}
respectively. The explicit mapping between the rapidity and the physical momentum $p$ is given by
\begin{equation}
\label{eq:uofp}
u = \frac{1}{g} \cot\frac{p_Q}{2}\,\sqrt{Q^2+4g^2\sin^2\frac{p_Q}{2}}\,.
\end{equation}

Within the physical regions we have defined above we can try to define an analogue of the (two-particle) physical strip. The physical string and mirror region overlap on the torus, as do the string and anti-mirror region and vice versa.\footnote{The anti-string and anti-mirror region stand for the regions obtained by crossing the string and mirror region respectively.} Clearly, this isolates a smaller region within the string and mirror regions which does not overlap with the (anti-)mirror or (anti-)string regions respectively. These regions correspond to the strip $\mbox{Im}(u) \in (-1/g,1/g)$ on the $u$-plane.\footnote{Transitioning from a string type region to a mirror type region corresponds to crossing a cut of $x^\pm$ or $x_s^\pm$. The strip $\mbox{Im}(u) \in (-1/g,1/g)$ precisely fits between these cuts.} Furthermore, as we will see below the analogue of the shift by $\pm i$ in the relativistic TBA of chapter \ref{chapter:finitevolumeIQFT} is played by a shift of $u$ by $\pm i/g$. For these reasons the strip $\mbox{Im}(u) \in (-1/g,1/g)$ deserves a special name, and we will call it our physical or analyticity strip. We will see more of its relevance especially in chapter \ref{chapter:boundstates}.

Now that we have understood the scattering theory of the mirror model we can finally analyze its thermodynamics.

\section{The string hypothesis and TBA equations}

As we just discussed, in the infinite volume limit the mirror theory has bound states of fundamental excitations of any length. In order to take a thermodynamic limit however, as we saw in section \ref{subsec:TBAgeneral} we should analyze the full set of mirror Bethe-Yang equations (\ref{eq:mirrorBYmain}-\ref{eq:mirrorBYw}) and determine whether there are other possible `bound states' contributing to the free energy. Now as it turns out, bound states of physical particles and $y$-particles do not appear to be consistent, and we are left with analyzing two copies of Bethe equations which each essentially correspond to those of the inhomogeneous Hubbard model \cite{Arutyunov:2009zu}. For this model the string hypothesis has been known for many years \cite{Takahashi:1972hub}, the spectrum consisting of $y$-particles (of two types), bound states of $y$-particles and twice as many $w$-particles, and $w$-particles and their bound states. Put together, this gives us the hypothesis that the solutions to the mirror Bethe-Yang equations which contribute to the thermodynamic limit are captured by \cite{Arutyunov:2009zu}

\paragraph{\underline{$Q$-particles}} which are bound states of fundamental particles of any length $Q$, where the associated rapidities are given by
\begin{equation}
\{ u \} = \{u +\frac{i}{g}(Q+1-2j) \, | \, j=1,\ldots,Q \} \, .
\end{equation}

\paragraph{\underline{$vw$-strings}} which are complexes of $y$ and $w$-particles of any length $M$, with rapidities
\begin{eqnarray}
\{w\}&=&\{ v + \frac{i}{g}(M+1-2j)\, | \, j=1,\ldots,M\}\, ,
\nonumber \\
\{v\}&=&\{v + \frac{i}{g}(M-2j)\, | \, j=1,\ldots,M-1\}\cup\{v + \frac{i}{g} M , v - \frac{i}{g} M \}\, ,
\end{eqnarray}
To each of the rapidities $v$ in the first set of the second line we associate two $y$-roots, one with $|y|>1$ and one with $|y|<1$, while the single $y$-roots associated to $v+ i M/g$ and $v- i M/g$ have $|y|<1$ and $|y|>1$ respectively. In terms of the mirror $x$ function we have $y_{\pm M} = x(v \pm i M/g)$.

\paragraph{\underline{$w$-strings}} which are complexes of $w$-particles of any length $M$
\begin{equation}
\{w\}=\{w + i (M+1-2j)/g\}\, ,~~~~j=1,\ldots, M\, .
\end{equation}

\paragraph{\underline{$y$-particles}} which are characterized by the property that $|y|=1$.  This means their associated rapidities $v$ run over the interval $(-2,2)$. In terms of the mirror $x$ function we have $y^- = x(v)$ while $y^+ = \frac{1}{x(v)}$, where the $\pm$ denotes the sign of the imaginary part of $y$.

The centers of these strings are all real. The interested reader can find more details on the consistency of this string hypothesis as a limiting case of its $q$-deformed version discussed in detail in chapter \ref{chapter:quantumTBA}.

\subsection{The mirror TBA equations}

Based on this string hypothesis, following the procedure described in section \ref{chapter:finitevolumeIQFT} we get the following TBA equations \cite{Arutyunov:2009ur,Bombardelli:2009ns,Gromov:2009bc} (we work in the conventions of \cite{Arutyunov:2009ur})
\begin{align}
\log Y_Q =& \, - J \, \tilde{\mathcal{E}}_{Q} + \log(1+Y_{M}) \star K_{\mathfrak{sl}(2)}^{MQ}
\\[1mm]
&\quad + \sum_{a} \log\left(1+ \tfrac{1}{Y^{(a)}_{M|vw}} \right) \star  K^{MQ}_{vwx} + \log \left(1- \tfrac{1}{Y^{(a)}_\beta}\right) \hstar  K^{yQ}_\beta   \, , \nonumber\\
\log Y^{(a)}_{M|vw} =& \log\left( 1+\tfrac{1}{Y^{(a)}_{N|vw}}\right)\star K_{NM} + \log \frac{1-\tfrac{1}{Y^{(a)}_-}}{1-\tfrac{1}{Y^{(a)}_+}}\hstar  K_M \\ & \quad - \log (1+Y_Q)\star  K^{QM}_{xv}  \, ,\nonumber \\
\log Y^{(a)}_{M|w} =& \log\left(1+\tfrac{1}{Y^{(a)}_{N|w}}\right)\star K_{NM} + \log \frac{1-\tfrac{1}{Y^{(a)}_-}}{1-\tfrac{1}{Y^{(a)}_+}} \hstar K_M \, ,\\
\log {Y^{(a)}_\pm} =& - \log\left(1+Y_Q \right)\star  K^{Qy}_\pm + \log \frac{1+\tfrac{1}{Y^{(a)}_{M|vw}}}{1+\frac{1}{Y^{(a)}_{M|w}}}\star  K_M\,,
\end{align}
where there is an implicit sum over $\beta=\pm$ in the first equation and we note that the fermionic $y$-particles have chemical potential $i \pi$ (by convention absorbed in the definition of their Y-functions) because the original physical fermions are periodic for for the ground state which has no winding (cf. the discussion of section \ref{subsec:mirrortrick}), as well as that the counting function for the $y$-particles is defined via the inverse of their Bethe equations \eqref{eq:mirrorBYy}.\footnote{When describing (excited) states with nontrivial winding however, we should keep in mind the changing boundary conditions for the fermions in string theory, and adjust the chemical potential for the mirror fermions appropriately to $i\pi(m+1)$. If we absorb this chemical potential in the fermionic Y-functions this simply amounts to multiplying these functions by $(-1)^m$ on the right-hand side of the TBA equations, a change that can be easily implemented at any moment. This distinction disappears completely in the simplified TBA equations.} Moreover, just like we defined $Y_0 = Y_f^{-1}$ in chapter \ref{chapter:finitevolumeIQFT}, for future convenience the $Y_Q$ functions are defined inversely from the standard way Y-functions were introduced in chapter \ref{chapter:finitevolumeIQFT}. The convolutions in the above equations are defined as
\begin{align}
f\star h(u,v)=\,&\int_{-\infty}^{\infty}\, dt\, f(u,t)h(t,v) \, , \label{eq:stdconv}\\
f\, \hstar\,  h(u,v)=\,&\int_{-2}^{2}\, dt\, f(u,t)h(t,v)\, , \label{eq:hatconv}
\end{align}
where the $\hstar$ convolution naturally appears since $y$-particles only exist on the interval $(-2,2)$. Let us also define the complementary convolution
\begin{equation}
f\, \check{\star}\,  h(u,v) = \int_{-\infty}^{-2}\, dt\, f(u,t)h(t,v)+\int_{2}^{\infty}\, dt\, f(u,t)h(t,v)\label{eq:checkconv}\, .
\end{equation}

The free of the mirror model gives us the ground state energy of the string as
\begin{equation}
\label{eq:groundstateEnergy}
E(L) =-\int {\rm d}u\, \sum_{Q=1}^{\infty}\frac{1}{2\pi}\frac{d\tilde{p}^Q}{du}\log\left(1+Y_Q \right)\,.
\end{equation}

We can directly read off the kernels and associated S-matrices appearing here from the Bethe-Yang equations just like we did in section \ref{subsec:TBAgeneral}; they are concretely listed in appendix \ref{app:qdefSmatricesandkernels}. By applying various identities involving these kernels summarized in this appendix, these equations can be simplified resulting in the set of equations
\begin{align}
\log Y_1=\, &\sum_a \log \left(1-\tfrac{1}{Y_-^{(a)}}\right) \hstar s - \log\left(1+\tfrac{1}{Y_2}\right)\star s -\check{\Delta}\check{\star} s\, ,\label{eq:undefsimpTBAQ1}\\
\log Y_Q = \, & \log\frac{{Y_{Q+1}Y_{Q-1}}}{(1+Y_{Q-1})(1+Y_{Q+1})}\star s +  \sum_a \log \left(1+\tfrac{1}{Y_{Q-1|vw}^{(a)}}\right) \star s\,  \hspace{20pt} \mbox{for } Q>1\, ,\label{eq:undefsimpTBAQg1}\\
\log{Y^{(a)}_{M|vw}} = \, & \log{(1+Y^{(a)}_{M+1|vw})(1+Y^{(a)}_{M-1|vw})} \star s -
\log\left(1+Y_{M+1}\right)\star s \nonumber \\
& \hspace{10pt} + \delta_{M,1} \log{\frac{1-Y^{(a)}_-}{1-Y^{(a)}_+}} \hstar  s \label{eq:undefsimpTBAvw} \\
\log{Y^{(a)}_{M|w}}  = \, &\log{(1+Y^{(a)}_{M+1|w})(1+Y^{(a)}_{M-1|w})} \star s + \delta_{M,1}
\log{\frac{1-(Y^{(a)}_-)^{-1}}{1-(Y^{(a)}_+)^{-1}}}  \hstar  s \label{eq:undefsimpTBAw}\, ,
\end{align}
where
\begin{align}
\check{\Delta}=&L\check{\cal E}+\sum_a \log \left(1-\tfrac{1}{Y_-^{(a)}}\right)\left(1-\tfrac{1}{Y_+^{(a)}}\right)\hstar \check{K}+2\log(1+Y_Q)\star
\check{K}^{\Sigma}_Q\\
\nonumber
&\quad\quad\quad +\sum_a \log \left(1-\tfrac{1}{Y_{M|vw}^{(a)}}\right)\star \check{K}_M+ \sum_a \log \left(1-Y_{k-1|vw}^{(a)}\right)\star \check{K}_{k-1}\, .
\end{align}
The equations for $y$-particles cannot be directly put in truly simplified form. However, we can remove the infinite sums of $(v)w$-strings similarly to how we did this for $Y_f$ in section \ref{subsec:TBAgeneral} (see eqs. (\ref{eq:KMfidentity}-\ref{eq:infmagnonsumidentity})). Doing so we obtain their equations in so-called hybrid form \cite{Arutyunov:2009ax}
\begin{align}
\label{eq:yhybrid1}
\log\frac{Y_{+}^{(a)}}{Y_{-}^{(a)}} = &
\log(1+Y_{Q})\star K_{Qy}\,,\\
\label{eq:yhybrid2}
\log Y_{-}^{(a)}Y_{+}^{(a)}  = & \,-\log(1+Y_{Q})\star K_{Q}+
2\log(1+Y_{Q})\star K_{xv}^{Q1}\star s+2\log\frac{1+Y_{1|vw}^{
{(a)}}}{1+Y_{1|w}^{(a)}}\star s\,,
\end{align}
Let us for completeness note that we can similarly find a set of hybrid equations for $Q$-particles given by
\begin{align}
\log Y_Q  = & \,-L\tilde{\mathcal{E}}_Q+\log(1+Y_P)\star
\left(K_{\sl(2)}^{P Q}+2s\star K_{vx}^{P -1,Q}\right)\nonumber \\
& +\sum_a \big[\log(1+Y_{1|vw}^{(a)})
\star s\,\hstar K_{yQ}+\log(1+Y_{Q-1|vw}^{(a)})\star s \nonumber \\
& \hspace{50pt} -\log
\frac{1-Y_{-}^{(a)}}{1-Y_{+}^{(a)}}\hstar s\star K_{vwx}^{1Q}
 +\frac{1}{2}\log
\frac{1-\tfrac{1}{Y_{-}^{(a)}}}{1-\tfrac{1}{Y_{+}^{(a)}}}
\hstar K_Q
\nonumber\\
& \hspace{80pt} +\frac{1}{2}\log(1-\tfrac{1}{Y_{-}^{(a)}})
(1-\tfrac{1}{Y_{+}^{(a)}})\hstar K_{yQ}\big]\,,
\end{align}
which will be useful in particular cases we will encounter later.

\subsection{The Y- and T-systems}
\label{subsec:YandTundef}

To derive the Y-system we should act on the simplified equations with $s^{-1}$. However, since $\check{\Delta}$ has internal convolutions involving Y-functions, there is no true Y-system for $Y_1$ outside the interval $(-2,2)$. Explicitly we get the following Y-system

\paragraph{$Q$-particles}

%\begin{equation}
%\frac{Y_1^+ Y_1^-}{Y_2}  = \frac{\displaystyle{\prod_a}
%\left(1-\tfrac{1}{Y^{(a)}_{-}}\right)}{1+Y_2}\,  \, \, \, \, \, \mbox{for} \, \,
%|u|<2\, , \hspace{20pt} \frac{Y_Q^+ Y_Q^-}{Y_{Q+1}Y_{Q-1}} = \frac{\displaystyle{\prod_a} \left(
%1+\tfrac{1}{Y_{Q-1|vw}^{(a)}}\right)}{(1+Y_{Q-1})(1+Y_{Q+1})} \, ,
%\end{equation}

\begin{equation}
\begin{aligned}
\frac{Y_1^+ Y_1^-}{Y_2}  = &\,\frac{\displaystyle{\prod_a}
\left(1-\tfrac{1}{Y^{(a)}_{-}}\right)}{1+Y_2}\,  \, \, \, \, \, \mbox{for} \, \,
|u|<2\, ,\\
\frac{Y_Q^+ Y_Q^-}{Y_{Q+1}Y_{Q-1}} = &\,\frac{\displaystyle{\prod_a} \left(
1+\tfrac{1}{Y_{Q-1|vw}^{(a)}}\right)}{(1+Y_{Q-1})(1+Y_{Q+1})} \, ,
\end{aligned}
\end{equation}

\paragraph{$vw$-strings}

%\begin{equation}
%Y_{1|vw}^+ Y_{1|vw}^-  =\frac{1+Y_{2|vw}}{1+Y_2}\left(\frac{1-Y_-}{1-Y_+}\right)^{\hspace{-1pt}\theta(2-|u|)} \hspace{-4pt}, \hspace{1pt}
%Y_{M|vw}^+ Y_{M|vw}^- =\frac{(1+Y_{M-1|vw})(1+Y_{M+1|vw})}{1+Y_{M+1}} ,\hspace{-7pt}
%\end{equation}

\begin{equation}
\begin{aligned}
Y_{1|vw}^+ Y_{1|vw}^-   =&\,\frac{1+Y_{2|vw}}{1+Y_2}\left(\frac{1-Y_-}{1-Y_+}\right)^{\theta(2-|u|)} \,,\\
Y_{M|vw}^+ Y_{M|vw}^- =&\,\frac{(1+Y_{M-1|vw})(1+Y_{M+1|vw})}{1+Y_{M+1}}\, ,
\end{aligned}
\end{equation}

%\begin{equation}
%Y_{1|vw}^+ Y_{1|vw}^-  =\frac{1+Y_{2|vw}}{1+Y_2}\left(\frac{1-Y_-}{1-Y_+}\right)^{\theta(2-|u|)} , \hspace{15pt}
%Y_{M|vw}^+ Y_{M|vw}^- =\frac{(1+Y_{M-1|vw})(1+Y_{M+1|vw})}{1+Y_{M+1}} ,
%\end{equation}

\paragraph{$w$-strings}

%\begin{equation}
%Y_{1|w}^+ Y_{1|w}^-  =(1+Y_{2|w})\left(\frac{1-Y_-^{-1}}{1-Y_+^{-1}}\right)^{\theta(2-|u|)} ,\hspace{20pt}
%Y_{M|w}^+ Y_{M|w}^- =(1+Y_{M-1|w})(1+Y_{M+1|w}) \, ,
%\end{equation}

\begin{equation}
\begin{aligned}
Y_{1|w}^+ Y_{1|w}^-  =& \,(1+Y_{2|w})\left(\frac{1-Y_-^{-1}}{1-Y_+^{-1}}\right)^{\theta(2-|u|)} ,\\
Y_{M|w}^+ Y_{M|w}^- =&\,(1+Y_{M-1|w})(1+Y_{M+1|w}) \, ,
\end{aligned}
\end{equation}

\paragraph{$y$-particles}

\begin{equation}
Y_{-}^{+}Y_{-}^{-} = \frac{1+Y_{1|vw}}{1+Y_{1|w}}\frac{1}{1+Y_1}\, .
\end{equation}
We can find this equation by acting with $s^{-1}$ on the hybrid equations \eqref{eq:yhybrid1} and \eqref{eq:yhybrid2} and using the identity\footnote{Note that the `integrated form' of this equation does \emph{not} hold.}
\begin{equation}
(K_{Qy}+K_Q) \circ s^{-1} = 2 K^{Q1}_{xv} + 2\delta_{Q1}\, .
\end{equation}
It is not possible to reduce the TBA equations for $Y_+$ to a local Y-system form \cite{Arutyunov:2009ur}. Still there is something interesting to be said about $Y_+$. From the TBA equations we can easily deduce that the Y-functions have cuts on the $u$-plane by analyzing the cut structure of the integration kernels. In particular the fermionic Y-functions $Y_\pm$ have cuts on the real line on the complement of the interval $(-2,2)$. At the same time from eqn. \eqref{eq:yhybrid2} we can see that their product has no cut on the real line, while by eqn. \eqref{eq:yhybrid1} their ratio has precisely the same cut structure and flips sign across the cut; put together we see that $Y_+$ and $Y_-$ are each others analytic continuation through this cut.

\begin{figure}[h]
\begin{center}
\includegraphics[width=0.85\textwidth]{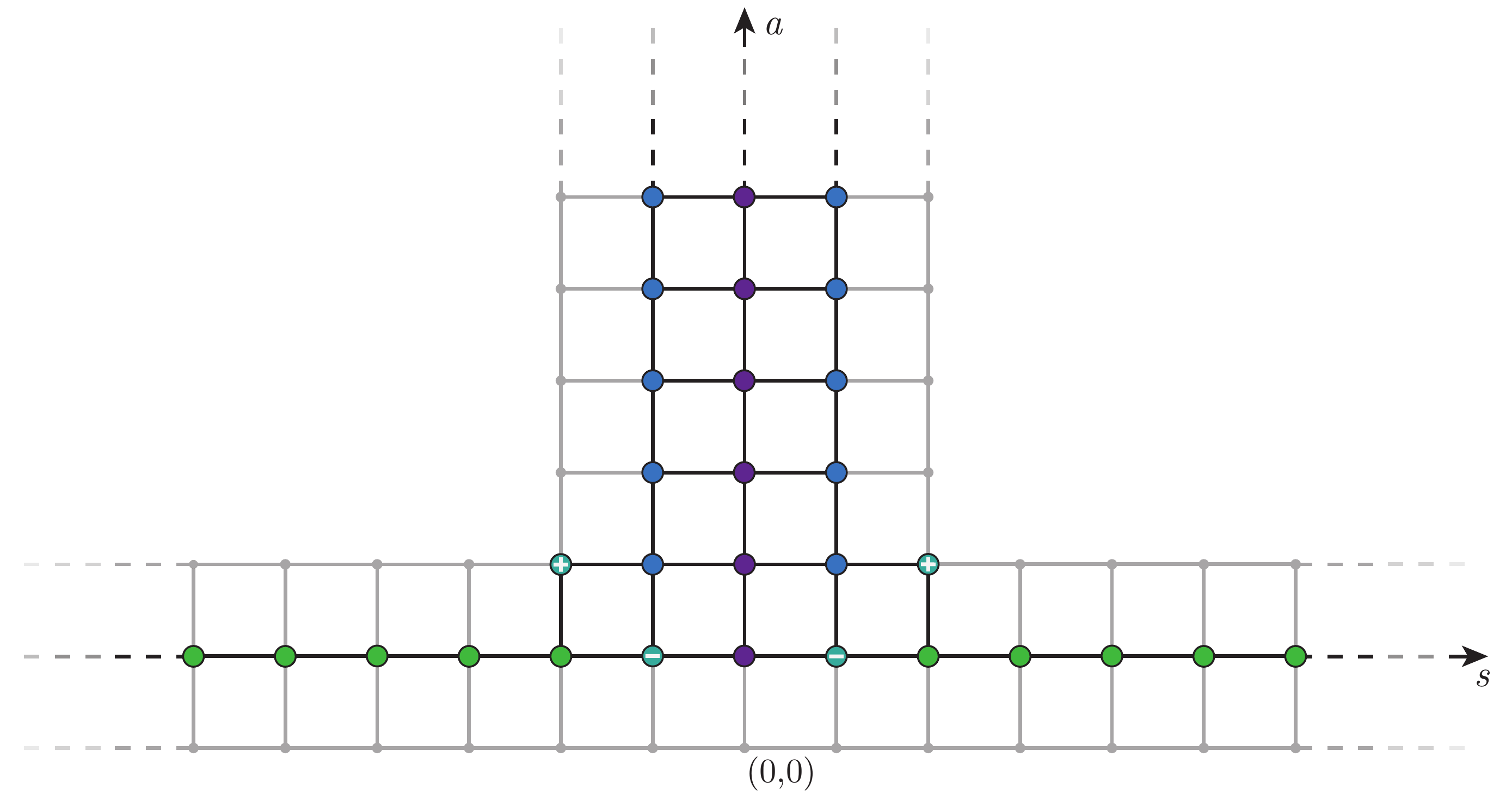}
\caption{The T-hook for the $\ads$ Y-functions. The Y-functions live on the $(a,s)$ lattice formed by the colored nodes. The extra gray nodes indicate the enlarged lattice for the T-functions introduced below. In terms of the notation of the TBA equations the $Y_Q$ are purple, $Y_{M|vw}$ blue, $Y_{M|w}$ green, and finally $Y_\pm$ teal.}
\label{fig:Thookundef}
\end{center}
\end{figure}

We should note that before the ground state TBA equations were found, there was already a conjecture for the Y-system that should describe the spectrum of the light-cone superstring \cite{Gromov:2009tv}. This Y-system is of the usual universal form
\begin{equation}
\label{eq:ysysgeneral}
Y_{a,s}^+ Y_{a,s}^- = \frac{(1+Y_{a,s+1})(1+Y_{a,s-1})}{\left(1+\tfrac{1}{Y_{a+1,s}}\right)\left(1+\tfrac{1}{Y_{a-1,s}}\right)}\,,
\end{equation}
and the Y-functions lie on the so-called T-hook lattice \cite{Gromov:2009tv} illustrated in figure \ref{fig:Thookundef}, which precisely corresponds to the structure of the TBA equations above.
Indeed, provided we make the identification
\begin{alignat}{8}
Y_{M|vw}^{(l)} =&\, 1/Y_{M+1,-1}\, , &Y_Q =&\, Y_{Q,0} & Y_{M|vw}^{(r)} =&\, 1/Y_{M+1,1}\, ,\nonumber\\
Y_{+}^{(l)} =&\, (-1)^{m+1} Y_{2,-2}\, , \hspace{5pt} & Y_{-}^{(l)} =&\, (-1)^{m+1}/Y_{1,-1}\, , & Y_{+}^{(r)} =&\, (-1)^{m+1} Y_{2,2} \, , \\
Y_{M|w}^{(l)} =&\, Y_{1,M+1}\, , & Y_{-}^{(r)} =&\, (-1)^{m+1}/ Y_{1,1} \, , \hspace{2pt} & Y_{M|w}^{(r)} =&\, Y_{1,-(M+1)}\, ,\nonumber
\end{alignat}
we find that the Y-system derived from our ground state TBA equations agrees with the conjectured Y-system of \cite{Gromov:2009tv} for $u \in (-2,2)$, and with the exception that the $Y^{(a)}_+$ ($Y_{2,\pm2}$) functions have no local Y-system equation at all.

From the ground state TBA equations we can determine that for $u\in(-\infty,-2)\cup(2,\infty)$ the Y-functions have branch cuts at
\begin{align}
&Y_\pm^{(a)}(u + i 2m/g)\, , \hspace{20pt} m \in \mathbb{Z}\, ,\\
&Y_M(u\pm i(M+2n)/g) \, ,\, Y^{(a)}_{M|vw}(u\pm i(M+2n)/g) \, ,\,Y^{(a)}_{M|w}(u\pm i(M+2n)/g) \, , \hspace{15pt} n \in \mathbb{N}\, ,\nonumber
\end{align}
and find their discontinuities. In other words, the Y-functions have branch cuts spaced by $2i/g$ outward from the real line, starting from $\pm i M/g$, where $M$ is their index which we can define as $0$ for $Y_\pm$. The precise discontinuities of the Y-functions across these cuts are explicitly given in \cite{Cavaglia:2010nm}, where it was also shown that (only) when supplemented with these discontinuity relations and conditions on the asymptotics of $Y_1$ the Y-system is equivalent to the TBA equations and can be explicitly `integrated' to find them.\footnote{By this last point we mean that we can derive the simplified TBA equations from the Y-system by relatively straightforward if perhaps tedious complex analysis given discontinuity relations and asymptotics (this general idea can already be found in e.g. \cite{Klumper:1991-1,Klumper:1991-2}). Of course, we can just view this as the statement that when supplemented with sufficient analyticity data (restrictions) on the space of functions under consideration, $s$ is the proper inverse of $s^{-1}$.}

It will also be useful for us to map the Y-system equations \eqref{eq:ysysgeneral} to the so-called Hirota equations, or T-system,
\begin{equation}
\label{eq:Hirota}
T_{a,s}^+T_{a,s}^- = T_{a+1,s}T_{a-1,s} + T_{a,s+1}T_{a,s-1}\, ,
\end{equation}
which we can do via the identification \cite{Klumper:1992vt,Kuniba:1993cn,Krichever:1996qd}
\begin{equation}
Y_{a,s} = \frac{T_{a,s+1} T_{a,s-1}}{T_{a+1,s}T_{a-1,s}}\,.
\end{equation}
However, we should note that the T-system has a `gauge' invariance
\begin{equation}
\label{eq:Tgaugetf}
T_{a,s} \rightarrow g_1^{[a+s]}g_2^{[a-s]}g_3^{[-a+s]}g_4^{[-a-s]}T_{a,s}\, ,
\end{equation}
where $f^{[m]}(u) = f(u+i m/g)$, while the Y-functions are gauge invariant quantities. In terms of our parametrization, the mapping between T-functions and Y-functions is
\begin{align}
\label{eq:YtoTgeneral}
& Y^{(l)}_{M|vw} = \frac{T_{M+2,-1}T_{M,-1}}{T_{M+1,0}T_{M+1,-2}}\, , &&
Y_{Q} = \frac{T_{Q,1}T_{Q,-1}}{T_{Q+1,0}T_{Q-1,0}}\, , &&Y^{(r)}_{M|vw} = \frac{T_{M+2,1}T_{M,1}}{T_{M+1,0}T_{M+1,2}}\, ,\nonumber
\\
& Y^{(l)}_{+} = -\frac{T_{2,-1}T_{2,-3}}{T_{1,-2}T_{1,-3}}\,, &&
 Y^{(l)}_{-} = - \frac{T_{0,-1}T_{2,-1}}{T_{1,0}T_{1,-2}}\, , &&  Y^{(r)}_{+} = -\frac{T_{2,1}T_{2,3}}{T_{1,2}T_{1,3}}\, ,\\
 & Y^{(l)}_{M|w} = \frac{T_{1,-(M+2)}T_{1,-M}}{T_{0,-(M+1)}T_{2,-(M+1)}}\, , && Y^{(r)}_{-} = - \frac{T_{0,1}T_{2,1}}{T_{1,0}T_{1,2}}\, ,  && Y^{(r)}_{M|w} = \frac{T_{1,M+2}T_{1,M}}{T_{0,M+1}T_{2,M+1}}\, .\nonumber
\end{align}
These T-functions live on the gray fat T-hook of figure \ref{fig:Thookundef}. Up to first nontrivial order at large string tension, this T-system has been derived directly from the string sigma model as well, by identifying the T-functions directly as transfer matrices and using the pure spinor formalism \cite{Benichou:2011ch}.

The TBA equations of this section allow us to compute the ground state energy of the light-cone superstring. Since the ground state is a half-BPS state its energy is protected and in fact zero. In light of the energy formula \eqref{eq:groundstateEnergy} the corresponding solution of the TBA equations should have $Y_Q=0$ \cite{Frolov:2009in}; we will discuss the corresponding solution in the generalized setting of twisted theories in the next chapter. The next natural question is how to generalize these equations to excited states, as we did not go through all these details just to confirm that our ground state indeed has zero energy. Because of the complicated structure of these equations and lack of simple analytic results to compare to it does not seem very promising to attempt direct analytic continuation in the spirit of section \ref{sec:excitedstatesgeneral}. Fortunately, by the time these equations were developed there already existed a natural candidate solution for these equations in the limit $Y_Q \sim 0$.

\section{\texorpdfstring{L\"uscher}{} formulae and excited states}
\label{sec:luscherandexcitedstatesgeneral}

\begin{figure}[h]
\begin{center}
\includegraphics{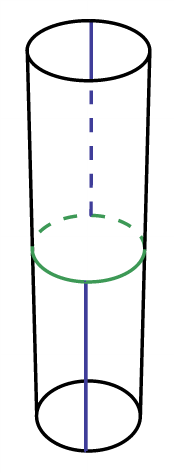}\hspace{50pt}\includegraphics{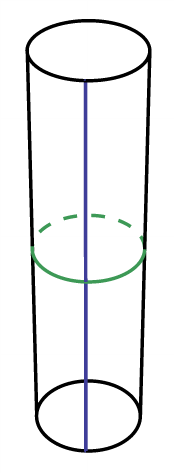}
\caption{The L\"uscher $\mu$- and F-term. The left figure shows the decay of a physical particle (blue) into a pair of virtual particles (green) which fuse to a physical particle on the other side of the cylinder, while the right figure shows the scattering of a virtual particle with the physical particle in the process of looping around the cylinder.}
\label{fig:luscherprocesses}
\end{center}
\end{figure}

To introduce the candidate solution of the TBA equations already mentioned in chapter \ref{chapter:finitevolumeIQFT} we will go through a bit of physical reasoning. If we really take our string theory at face value and simply view it as a field theory on a cylinder it is natural to expect the energy of states to get corrections due to virtual particles travelling around the circle. Of course such phenomena have been studied before. Concretely, in \cite{Luscher:1985dn} L\"uscher showed how polarization effects lead to general mass corrections for a standing particle in massive quantum field theory in a periodic box, and computed their effect to leading order in $e^{-m L}$ where $m$ is the mass of the particle and $L$ the size of the periodic box \cite{Luscher:1985dn}. These (leading order) corrections come in two types illustrated in figure \ref{fig:luscherprocesses}, the so-called $\mu$ term corresponding to the particle decaying into a pair of virtual particle which move around the circle (in two dimensions) and recombine, while the F-term corresponds to a virtual particle loop around the circle which of course involves scattering with the physical particle.

Generalizing these ideas to moving particles or further to interacting multi-particle states based on the original diagrammatic methods of \cite{Luscher:1985dn} seems a daunting task. However, in the context of simple relativistic integrable models L\"uscher's formulae readily follow by explicitly expanding the TBA equations in the large volume limit. As such we might hope that by carefully generalizing the expansions in such models to interacting multi-particle states we can obtain a type of generalized L\"uschers formulae. Motivated to find an explanation for the four loop wrapping correction to the anomalous dimension of the Konishi state, this is precisely what was done in \cite{Bajnok:2008bm}. The conjecture of \cite{Bajnok:2008bm} is that the general leading order correction to the momenta and energy of a ${\KK{I}}$-particle state in our string theory is given by
\begin{align}
\label{eq:deltaEluschergeneral}
\Delta E =&\, \sum_{j,k} \frac{d \mathcal{E}(p_k)}{d p_k} \left(\frac{\delta lBY_k}{\delta p_j}\right)^{-1} \Phi_j \\
&\,- \sum_Q \int_{-\infty}^{\infty} \frac{d\tilde{p}}{2\pi} \mbox{Str} \left( \prod_{i=1}^{\KK{I}} S^{Q1}(\tilde{p},p_i) \right) e^{-\tH_Q(\tilde{p})J}
\end{align}
where $lBY_k$ denotes the logarithm of the (nontrivial side of the) $k$th BY equation (divided by $i$), $S^{Q1}$ is the (full) S-matrix between a $Q$-particle \emph{mirror} bound state and a fundamental \emph{string} excitation (cf. the $\tilde{p}$ and $p$ respectively), and
\begin{equation}
\label{eq:LuscherPhidef}
\Phi_k = - \sum_Q \int_{-\infty}^{\infty} \frac{d\tilde{p}}{2\pi} \mbox{Str} \left( \prod_{i=1}^{k-1} S^{Q1_*}(\tilde{p},p_i) \, \frac{\partial}{\partial \tilde{p}} S^{Q1}(\tilde{p},p_k) \prod_{i=k+1}^{\KK{I}} S^{Q1}(\tilde{p},p_i)\right)\, ,
\end{equation}
is the conjectured leading order correction to the Bethe equations, i.e.
\begin{equation}
\label{eq:luscherPcorrection}
2 \pi n_k = lBY_k +\Phi_k\, .
\end{equation}
The words leading order in this case refer to leading order in $e^{-\tH J}$.\footnote{Next-to-leading order generalized L\"uscher formulae have recently been conjectured in \cite{Bombardelli:2013yka}.} In this formula we have written out the product of S-matrices, as it illustrates physical origin of the energy correction in the scattering of all possible virtual (mirror) single particles with the physical particles in our state. The fact that this formula involves a sum over mirror (or anti-symmetric) bound states is essential in matching with perturbative field theory \cite{Bajnok:2008bm}. These formulae apply in the large volume limit of our string theory, but also in the weak coupling limit\footnote{When concretely working with rapidities at weak coupling we need to rescale by $g$ to keep them finite.} where we have
\begin{equation}
e^{-\tH(v) J} = \frac{g^{2J}}{(Q^2 + v^2)^J} + \mathcal{O}(g^{4J})\, .
\end{equation}
At leading order in weak \emph{coupling}, $\Phi$ is typically of the same order as the integral term in eqn. \eqref{eq:deltaEluschergeneral} while $\frac{d\mathcal{E}}{dp}$ introduces extra factors of $g$. Therefore at leading order in weak coupling, the correction to the energy of a particular state is just given by
\begin{equation}
\label{eq:deltaEluscherleading}
\Delta E = - \sum_Q \int_{-\infty}^{\infty} \frac{d\tilde{p}}{2\pi} e^{-\tH J}T^{(l)}_{Q,1}(\tilde{p}|\{p_j\}) T^{(r)}_{Q,1}(\tilde{p}|\{p_j\}) \prod_{i=1}^{\KK{I}} S_{\sl(2)}^{Q 1}(\tilde{p},p_i)
\end{equation}
expanded to lowest order in $g$, where we have identified parts of the supertrace over the product of S-matrices in eqn. \eqref{eq:deltaEluschergeneral} with the transfer matrix\footnote{The superstrace is natural here because the S-matrix reduces to the \emph{graded} permutation at coincident arguments, resulting in a product of physical S-matrices just as in the ungraded case of eqn. \eqref{eq:transferphysical}.}
\begin{equation}
\label{eq:stringtransfermatrixdefinition}
T^{(a)}_{Q,1} (\tilde{p}|\{p_j\})\equiv \mbox{Str}_Q \left( \prod_{i=1}^{\KK{I}} \mathbb{S}^{(a)}_{Q,1}(\tilde{p},p_i) \right)\, ,
\end{equation}
where $\mathbb{S}^{(a)}_{Q,1}$ is one of the matrix factors in the full S-matrix $S^{Q,1} \equiv S_{\sl(2)}^{Q 1} \mathbb{S}^{(l)}_{Q,1} \otimes \mathbb{S}^{(r)}_{Q,1}$. Explicitly, the eigenvalues of this transfer matrix are given by \cite{Arutyunov:2009iq}\footnote{These eigenvalues can also be computed from a generating function as originally used in \cite{Beisert:2006qh}. See \cite{Arutyunov:2009iq} for a detailed comparison.}
\begin{align}
\label{eq:FullTransfer}
T_{a,1}(v\,|\,\vec{u}) =&
\left.\prod_{i=1}^{K^{\rm{II}}} {\textstyle{\frac{y_i-x^-}{y_i-x^+} \sqrt{\frac{x^+}{x^-}} }}\right[
1 + \prod_{i=1}^{K^{\rm{II}}} {\textstyle{\frac{v - \nu_i + \frac{ia }{g} }{v-\nu_i-\frac{ia }{g}}}}
\prod_{i=1}^{K^{\rm{I}}}{\textstyle{\frac{(x^--x^-_i)(1-x^- x^+_i)x^+}{(x^+-x^-_i)(1-x^+ x^+_i)x^-}+ }}\nonumber\\
&{\textstyle{+}}\sum_{k=1}^{a-1} \prod_{i=1}^{K^{\rm{II}}}
{\textstyle{\frac{v-\nu_i + \frac{ia }{g} }{v - \nu_i + \frac{i(a-2k)}{g}}}}
\left\{\prod_{i=1}^{K^{\rm{I}}}{\textstyle{\lambda_+(v,u_i,k)+}}\right.\left.\prod_{i=1}^{K^{\rm{I}}}{\textstyle{\lambda_-(v,u_i,k)}}\right\}\\
& - \sum_{k=0}^{a-1} \prod_{i=1}^{K^{\rm{II}}} {\textstyle{\frac{v - \nu_i + \frac{ia }{g} }{v-\nu_i + \frac{i(a-2k)}{g}}}}
\prod_{i=1}^{K^{\rm{I}}}{\textstyle{\frac{x^+-x^+_i}{x^+-x^-_i}\sqrt{\frac{x^-_i}{x^+_i}} \left[1-\frac{\frac{2ik}{g}}{v-u_i+\frac{i(a-1)}{g}}\right]}}\times\nonumber\\
&\times\left. \left\{\prod_{i=1}^{K^{\rm{III}}}{\textstyle{\frac{w_i-v - \frac{i(a-2k+1)}{g}}{w_i- v - \frac{i(a-2k-1)}{g}} + }}
\prod_{i=1}^{K^{\rm{II}}}{\textstyle{\frac{v-\nu_i + \frac{i(a-2k)}{g}}{v-\nu_i+\frac{i(a-2k-2)}{g}}}}
\prod_{i=1}^{K^{\rm{III}}}{\textstyle{\frac{w_i - v - \frac{i(a-2k-3)}{g}}{w_i-v - \frac{i(a-2k-1)}{g}}}}\right\}\right],\nonumber
\end{align}
where we have now labeled the auxiliary mirror bound state momentum by $v$ and the string momenta by $u_j$, meaning that in the above formula we have $x^\pm=x(u\pm Qi/g)$ and $x_j^\pm = x_s(u_j\pm i/g)$ respectively. The rapidities $\nu$ are related to $y$ via $\nu = y+1/y$, and the parameters $\lambda_\pm$ are given by
\begin{align}\nonumber \hspace{-1cm}
\lambda_\pm(v,u_i,k)=&\,\frac{1}{2}\left[1-\frac{(x^-_ix^+-1)
  (x^+-x^+_i)}{(x^-_i-x^+)
  (x^+x^+_i-1)}+\frac{2ik}{g}\frac{x^+
  (x^-_i+x^+_i)}{(x^-_i-x^+)
  (x^+x^+_i-1)}\right.\\ \label{eq:lambda-pm}
&~~~~~~~~~~~~\left.\pm\frac{i x^+
  (x^-_i-x^+_i)}{(x^-_i-x^+)
 (x^+x^+_i-1)}\sqrt{4-\left(v-\frac{i(2k-a)}{g}\right)^2}\right]\,.
\end{align}
Each specific eigenvalue is of course determined from the solution of the auxiliary Bethe equations \eqref{eq:mirrorBYy} and \eqref{eq:mirrorBYw}. Note that if we want to put the physical particles in the mirror region all we have to do is replace $x_s$ by $x$, but that replacing $x$ by $x_s$ is only sensible in this expression when the auxiliary particle is elementary ($a=1$) since the bound state structure is different.

\subsection{The asymptotic solution}

Let us now come to the TBA description of excited states. We argued in section \ref{sec:excitedstatesgeneral} that the excited state TBA equations should differ from the ground state ones by a set of driving terms, but should otherwise be of the exact same form. Considering the energy formula \eqref{eq:groundstateEnergy} in this light, we realize that at weak coupling or very large $J$ the $Y_Q$ functions should be small due to the $-J \tH$ term in their canonical TBA equations. Expanding the energy formula for small $Y_Q$ and comparing this to the leading weak coupling correction \eqref{eq:deltaEluscherleading} it is natural to identify
\begin{equation}
\label{eq:BJforYQ}
Y_Q^o = \Upsilon_Q \, T^{(l)}_{Q,1}(v|\{u_j\})\, T^{(r)}_{Q,1}(v|\{u_j\}) \, ,
\end{equation}
where
\begin{align}
\label{eq:upsilondef}
\Upsilon_Q \equiv e^{-J\tH_Q}\prod_{i=1}^{\KK{I}}S^{Q 1_*}_{\alg{sl}(2)}(v,u_i)\,,
\end{align}
and we now use an asterix $*$ to denote that the corresponding rapidity lies in the string region to avoid confusion. We have put a superscript $o$ on $Y_Q$ to indicate this holds for small $Y_Q$. This naturally fits with the fact that the Y-system in the limit of small $Y_Q$ naturally decouples in two wings which each have a solution in terms of the $\mathfrak{psu}_{c.e.}(2|2)$ transfer matrices \cite{Gromov:2009tv}, and in addition tells us we should take the `string-mirror' transfer matrix as the appropriate solution rather than for example the `string-string' transfer matrix. Concretely, from eqn. \eqref{eq:BJforYQ} we can find that the solution of the Y-system in the small $Y_Q$ limit is \cite{Gromov:2009tv,Arutyunov:2011uz}
\begin{align}
\label{eq:YtoTasympt}
& Y^{(a)o}_{M|vw} = \frac{T^{(a)}_{M+2,1}T^{(a)}_{M,1}}{T^{(a)}_{M+1,2}}\, , && Y^{(a)o}_{M|w} = \frac{T^{(a)}_{1,M+2}T^{(a)}_{1,M}}{T^{(a)}_{2,M+1}}\, ,\\
& Y^{(a)o}_{+} = -\frac{T^{(a)}_{2,1}T^{(a)}_{2,3}}{T^{(a)}_{1,2}T^{(a)}_{1,3}}\,, &&
 Y^{(a)o}_{-} = - \frac{T^{(a)}_{2,1}}{T^{(a)}_{1,2}} \, .
\end{align}
where the $T^{(a)}_{Q,M}$ are given in terms of the transfer matrix $T^{(a)}_{Q,1}$ via the Bazhanov-Reshetikhin formula \cite{Bazhanov:1989yk}\footnote{Note that the first of the indices on $T_{Q,M}$ is naturally associated with the mirror theory, corresponding to anti-symmetric bound state representations, while the second of these indices corresponds to the string theory, with symmetric bound state representations. Indeed, we could have also fused the fundamental transfer matrix over string bound states to find $T_{1,M}$ and use this to generate all $T_{Q,M}$.}
\begin{align}
\label{eq:BR}
T_{a,s}(u) = \det_{1\leq m,n\leq s} T_{a+m-n,1}(u +i(s+1-m-n)/g)\, .
\end{align}
In fact by a redefinition of the general T functions we can explicitly put a factor of $\Upsilon_Q$ in the parametrization \eqref{eq:YtoTgeneral} for $Y_Q$ without changing the parametrization of the other Y-functions.\footnote{See for example \cite{Suzuki:2011dj}.} Doing so in the asymptotic limit the $T_{Q,\pm1}$ \emph{directly} reduce to the transfer matrices $T^{(r/l)}_{Q,1}$ while the $T_{Q,0}$ reduce to one.

Strictly speaking we have now argued what the leading order weak coupling and leading order large volume expression for the Y-functions is.\footnote{This distinction is relevant since the next-to-leading order correction that we can self-consistently derive from the TBA equations given this input is not manifestly of the form of eqn. \eqref{eq:deltaEluschergeneral} (the difference sits in the expression for $\Phi$). Of course the final results should agree.} However, in weak coupling applications it is natural to practically consider the above equations without expanding them in $g$ as long as the resulting $Y_Q$ functions are still numerically small. One reason for doing so is the following. Based on experience in other integrable models it is natural to expect that on the exact solution of the TBA equations $Y_1=-1$ when evaluated at the exact rapidity of a particle in the excited state under consideration. In this case this means we should analytically continue $Y_1$ which has its argument in the mirror region, to the string region where the physical rapidity lies. This means that the proper finite volume analogue of the Bethe-Yang equation for an excited state should be the \emph{exact Bethe equation}
\begin{equation}
\label{eq:exactBethegeneral}
Y_1^*(u_j) = -1\, .
\end{equation}
If we consider this at the level of the asymptotic formula \eqref{eq:BJforYQ} we see \cite{Gromov:2009tv} that this is indeed precisely equivalent to the Bethe-Yang equations (cf. the general eqn. \eqref{eq:BEeigenval})
\begin{equation}
Y_1^*(u_j) = -1\, \Leftrightarrow \,e^{i p_j J} \Lambda(u_j|\{u_i\})= -1 \,,
\end{equation}
at any value of $g$.

In fact eqn. \eqref{eq:exactBethegeneral} is very natural indeed from the point of view of section \ref{sec:excitedstatesgeneral}; the points where $Y_1=-1$ are singular points that can lead to driving terms, and including these driving terms appropriately precisely results in an energy formula of the form
\begin{equation}
\label{eq:TBAgeneralstateenergyformula}
E = \sum_{i=1}^{\KK{I}} \mathcal{E}(p_i) -\int {\rm d}u\, \sum_{Q=1}^{\infty}\frac{1}{2\pi}\frac{d\tilde{p}^Q}{du}\log\left(1+Y_Q \right)\,,
\end{equation}
where $\mathcal{E}(p_i)$ is the asymptotic energy of the $i$th particle (recall that $\tilde{p}$ evaluated on string momenta is just $-i \mathcal{E}$). Of course there can be further modifications of this energy formula as we will see in chapter \ref{chapter:boundstates}.

\subsection{The contour deformation trick}

Let us now briefly come back to the contour deformation trick described in section \ref{sec:excitedstatesgeneral}. Essentially the generalized L\"uscher formulae of \cite{Bajnok:2008bm} tell us what form solution of the excited state TBA equations in the \emph{asymptotic limit} $Y_Q \sim 0$ is supposed to take. Concretely we can readily construct the asymptotic Y-functions of eqs. \eqref{eq:BJforYQ} and \eqref{eq:YtoTasympt} via the eigenvalues of the transfer matrix \eqref{eq:FullTransfer}. As discussed we can naturally assume that the excited state TBA equations differ from the ground state ones only in their integration contours, and by substituting the asymptotic Y-functions in the asymptotic limit of the TBA equations we can find particular contours for a given excited state such that the equations are satisfied. As a net effect, upon taking the integration contours back to the real line we find a set of driving terms in the TBA equations which are such that they are solved by the asymptotic Y-functions in the asymptotic limit. If we finally assume that the solution to these equations can be smoothly deformed away from the asymptotic one, we end up with a concrete set of excited state TBA equations as well as a concrete starting point for their solution at weak coupling.

The driving terms we introduce this way depend on the location of the singular points typically specified by $Y=-1$,\footnote{$Y=(-1)^m$ for $Y_\pm$ because of the sign in their definition.} which can in principle exist for any of the Y-functions, not just the explicit example of $Y_1$ at the physical rapidities. This means that beyond the asymptotic limit where we can explicitly determine these numerically from the asymptotic solution, we should impose the `quantization conditions' or exact Bethe equations
\begin{equation}
Y(r) =-1\, ,
\end{equation}
on any relevant Y-function with corresponding root $r$, through its TBA equation. Of course these roots generically move as we change the coupling. In particular there is the possibility that such roots (whether having already induced a driving term or not) can move through the integration contour of our equations at some finite value of $g$, leading to pictures very similar to figure \ref{fig:analyticcont} of the original analytic continuation story. If this happens, the explicit form of the driving terms (now for a fixed excited state) clearly changes as the coupling crosses this so-called critical value. We can of course explicitly investigate whether the roots of the asymptotic solution show this behaviour as we increase the coupling constant, but since the asymptotic solution becomes less and less appropriate as we do so it is not clear that this gives a conclusive answer. Still, if we take the asymptotic properties to be truly representative of the exact solution this should give the qualitative answer.\footnote{Critical values have been concretely observed for states in the $\mathfrak{sl}(2)$ sector. They were pointed out at the asymptotic level in \cite{Arutyunov:2009ax} and have since been confirmed explicitly in the exact solution \cite{Frolov:2012zv}. For the Konishi state such critical values presumably exist, but their values are at least at thus far prohibitively large values of coupling \cite{Frolov:2010wt}.}

We should mention that as an alternative to the contour deformation trick we could explicitly `integrate' the Y-system for a excited states similar to how it is done for the ground state. We can do this by assuming that the discontinuity relations of \cite{Cavaglia:2010nm} universally apply to any state in our theory, and also here assuming that the analytic properties of the asymptotic Y-functions are representative of the exact solution. Carefully taking into account the effect of singular points, this approach explicitly reproduces the excited state TBA equations found by the contour deformation trick, at least in the $\mathfrak{sl}(2)$ sector \cite{Balog:2011nm}. Given the typical types of singular points of Y-functions in the $\mathfrak{sl}(2)$ sector the two approaches are basically identical at a practical level. For more complicated states however, this is no longer clearly the case.

As mentioned in the introduction, this whole procedure has most notably been successfully applied to the Konishi state. We will discuss this procedure for various other states in chapters \ref{chapter:excstatesandwrapping} and \ref{chapter:boundstates}. While we are using the TBA to find scaling dimensions of trace operators, we should also mention that more recently the TBA approach has also been applied to Wilson loops in $\mathcal{N}=4$ SYM \cite{Drukker:2012de,Correa:2012hh}, and to baryonic (determinant-type) operators \cite{Bajnok:2012xc,BajnokIGST13talk}. In both cases the basic structure of the TBA equations is essentially identical to the one we are describing here, while the source terms are complicated.

\subsection{Non-linear integral equations}

In chapter \ref{chapter:finitevolumeIQFT} we mentioned `non-linear integral equations' \cite{Klumper:1991-1,Klumper:1991-2,Klumper:1992vt,Klumper:1993vq,Destri:1992qk,Destri:1994bv} as a possible alternative to the TBA approach. Of course, even if it is not possible to derive a set of functional relations required as input for this approach (such as a Y-system) directly, we may derive it from the TBA equations and use this as a starting point. This way we may be able to reexpress the TBA equations (Y-system) in a way that is practically very useful.\footnote{Sometimes the equations can be very elegant indeed \cite{Takahashi:2001}, but perhaps also rather hard to work with in practice.} Although quite involved this has been worked out for the superstring TBA equations \cite{Gromov:2011cx,Balog:2012zt} (see also \cite{Suzuki:2011dj,Suzuki:2012mq} and \cite{Leurent:2012xc}), at least for states in the $\mathfrak{sl}(2)$ sector.\footnote{The application of these methods to the chiral Gross-Neveu model we discussed in the previous chapter can be found in \cite{Gromov:2008gj}.} Thus far these equations have been used particularly efficiently at weak coupling to find the anomalous dimension of the Konishi operator analytically up to eight-loop order \cite{Leurent:2012ab,Leurent:2013mr}, matching exactly with generalized L\"uscher results available up to seven loops \cite{Bajnok:2012bz}.\footnote{With the next-to-leading order generalized L\"uscher formulae proposed in \cite{Bombardelli:2013yka} it should be possible to verify these results at higher loop orders as well.} Very recently this approach was refined further, resulting in a so-called quantum spectral curve or $P\mu$-system \cite{Gromov:2013pga}, reformulating the problem as a Riemann-Hilbert--type problem. With a tiny bit of external help, with this approach even the nine-loop anomalous dimension for the Konishi operator can be computed in the space of a single talk \cite{VolinIGST13talk}.

\subsection{\texorpdfstring{$\mbox{PSU}(2,2|4)$}{} symmetry}
\label{sec:psu224symmetry}

As the final point of this chapter we should address the symmetries of our model, and how they arise at the level of the spectrum. The superstring originally has $\mbox{PSU}(2,2|4)$ symmetry, of which a manifest $\mbox{PSU}(2|2)^2$ is left at the level of the light-cone S-matrix and hence the above TBA equations. Of course the spectrum should be $\mbox{PSU}(2,2|4)$ invariant and so the missing symmetry should be hiding somewhere in our equations. In fact, they are built into the TBA equations by construction  \cite{Arutyunov:2011uz}. We already discussed in chapter \ref{chapter:finitevolumeIQFT} that the Bethe ansatz describes highest weight states, or in other words describes whole multiplets at a time. In the $\ads$ context this means that the Bethe-Yang equations describe entire superconformal multiplets in one go. Now since we build our excited state TBA equations based on the asymptotic solution which has this extra symmetry, they will necessarily have this same symmetry. One curious consequence of this fact is that because supersymmetry transformations can change the $J$-charge of a state, if we happen to consider a descendant instead of a superconformal primary, the length $J$ that enters the TBA equations of the descendant is not its $J$-charge, but rather the maximal $J$-charge within its multiplet. Still, we can construct the asymptotic solution based on any state in the multiplet as the change simply corresponds to a gauge transformation on the asymptotic T-functions (transfer matrices) and leaves the Y-functions invariant.\footnote{To make this slightly more precise, the change in the transfer matrices is a gauge transformation from the point of view of the auxiliary problem and at the level of $Y_Q$ is compensated by $\Upsilon_Q$. Of course we can alternatively absorb $\Upsilon_Q$ in the T-functions, in which case they are invariant themselves.}

In the next chapter we will consider various deformations of our superstring background that result in theories with reduced (super)symmetry. In line with the above discussion, the reduction of symmetry is of course appropriately reflected in the asymptotic Bethe equations and asymptotic Y-functions, and consequently the TBA.

\chapter{The spectrum in twisted AdS/CFT}
\label{chapter:twistedspectrum}

There are various deformations of the  background the string propagates in which preserve the integrability of the two dimensional field theory on the world sheet. These include but are not limited to strings on orbifolds of $\ads$ and strings on the Lunin-Maldacena background dual to $\beta$-deformed SYM, for real $\beta$. In this chapter we discuss these deformations and how they are accounted for in terms of the integrable structure of the superstring discussed in the previous chapter. The upshot is that the deformations are in one-to-one correspondence with quasi-periodic (twisted) boundary conditions linked to Cartan elements of the symmetry algebra of the $\ads$ superstring, which as we already saw in chapter \ref{chapter:finitevolumeIQFT} should result in twists in the Bethe-Yang equations, and chemical potentials in the thermodynamic Bethe ansatz equations for the mirror model. Let us begin by discussing the deformations in some detail.

\section{Integrable deformations of \texorpdfstring{$\ads$}{}}
\label{sec:introdef}

There are two classes of known integrable deformations of the $\ads$ superstring, and they are obtained by deforming the background space-time the superstring moves in. The first of these is a class of backgrounds obtained by orbifolding the space-time by a discrete subgroup of the isometry group of $\ads$. Secondly, we can also perform a sequence of so-called $T$-duality - shift - $T$-duality (TsT) transformations, giving a string theory on a TsT transformed background. The nice feature of these deformations is that both can be described in terms of the original $\ads$ string theory, where the deformation is accounted for by quasi-periodic boundary conditions for the string fields as we will discuss below. Viewed in this way we can expect the resulting theories to be integrable since the scattering properties of the deformed theories are the same as those of the original.\footnote{Disregarding potential quantization issues associated purely to the deformation (it is not a priori clear that this boundary condition picture must survive quantization), the quantum integrability of these theories is then on an \emph{equally} subtle footing as that of the undeformed parent theory.}

When we deform the anti-de Sitter part of space-time there are certain caveats we should keep in mind, and we will address them below for the specific deformations. Nonetheless, in general deformations of $\mbox{AdS}_5$ have merit since they allow us to partly restore supersymmetry broken by deformations of the sphere, as well as providing a nice regularization of the ground state energy of the undeformed theory as we will discuss in section \ref{sec:wrappingforgroundstate}. Moreover a completely general deformation breaks all symmetries of the original model, lifting the $\mbox{PSU}(2,2|4)$ degeneracy of the spectrum completely, giving the richest possible structure. In this section we will first discuss strings on orbifolds, followed by strings on TsT transformed backgrounds, and finally we will briefly summarize the link between these boundary conditions and elements of the bosonic symmetry group of the light-cone gauge fixed superstring at the level of the coset model.

\subsection{Orbifolds}

Starting with a (super-)string theory on (super-)space-time $\mathcal{M}$ and the action of a discrete group $\Gamma$ on $\mathcal{M}$, we can orbifold the original string theory by $\Gamma$ and obtain a string theory on $\mathcal{M}/\Gamma$ \cite{Dixon:1985jw}. The orbifolded background is described in terms of the fields $X \in \mathcal{M}$ of the original string theory as
\begin{equation}
X(\tau,\sigma) \simeq g X(\tau,\sigma) \, , \, \, g \in \Gamma \, . \label{eq:orbequiv}
\end{equation}
Were we considering point particles on this space, we would take the Hilbert space of the original model, and project it onto its $\Gamma$ invariant subspace. However in the case of strings we have more structure; because of the orbifold equivalence (\ref{eq:orbequiv}) the string does not have to close on itself. Rather it can also close modulo an orbifold group element
\begin{equation}
\label{eq:orbequivBC}
X(\tau,2 \pi ) \simeq h X(\tau,0) \, , \, \, h \in \Gamma \, .
\end{equation}
In the orbifolded theory, we need to consider all independent sectors of this type, which are known as twisted sectors. These twisted sectors are in one to one correspondence with the conjugacy classes of $\Gamma$, as follows from (\ref{eq:orbequiv},\ref{eq:orbequivBC}). As such, in each twisted sector we should project onto its $\Gamma - [g]$ invariant subspace, where $[g]$ is the corresponding conjugacy class. In other words, we should project on the subspace invariant under the stabilizer of $g$, $\Gamma_g$.

For the $\ads$ superstring, this means we want to consider strings on $\mbox{AdS}_5/\Gamma^a \times \mbox{S}^5 /\Gamma^s$ by following these considerations. Let us start by discussing the boundary conditions obtained by orbifolding the sphere.

\subsubsection*{Orbifolding the sphere - $S^5/\Gamma$}

The possible orbifolds of the sphere are obtained by the action of the discrete subgroups of $\mbox{SU}(4)$. Fortunately enough, these have been classified completely in \cite{Hanany:1999sp}. Of course, all elements of the abelian subgroups can be simultaneously diagonalized, and hence immediately put into correspondence with elements of the Cartan subgroup of $\mbox{SU}(4)$. This simply corresponds to one and the same field redefinition in every twisted sector; explicitly
\begin{equation}
X(\tau,\sigma) \rightarrow \tilde{X}(\tau,\sigma) = g X(\tau,\sigma)\, , \, \, g \in \mbox{SU}(4) \, ,
\end{equation}
such that each of the representatives $h_k$ of the conjugacy classes is diagonalized
\begin{equation}
\tilde{X}(\tau,2 \pi ) \simeq d_k \tilde{X}(\tau,0) \, ,
\end{equation}
where $d_k$ is in the Cartan subgroup of $\mbox{SU}(4)$. Simultaneous diagonalization is not possible within a non-abelian subgroup; still we can diagonalize the representative conjugacy class element in each twisted sector by an independent field redefinition, as already used in this context in \cite{Solovyov:2007pw}. Picking up a different basis in each twisted sector is not a problem since each sector is closed.\footnote{This holds as long as the quantum orbifold symmetry is not broken. Non-supersymmetric orbifolds can break this symmetry; we will come back to this point shortly.} In short, in every twisted sector we can restrict ourselves to boundary conditions corresponding to elements of the Cartan subgroup of $\mbox{SU}(4)$; boundary conditions on the isometry angles of the sphere. Compared to abelian orbifolds, the only complication for non-abelian orbifolds is in the bookkeeping of the twisted sectors and corresponding orbifold invariance as clearly explained in \cite{Solovyov:2007pw}; there also the quiver structure of these orbifolds is carefully worked out.

The string theory constructed in this fashion is dual to an orbifolded version of $\mathcal{N}=4$ SYM, with field content $\mathcal{G}/\Gamma$ where $\mathcal{G}$ denotes the field content of $\mathcal{N}=4$ SYM. Correspondingly, $\Gamma$ is a discrete subgroup of the $R$-symmetry group of the field theory, meaning that these orbifolds will break a certain amount of supersymmetry. The amount of supersymmetry that is preserved after orbifolding can be found by considering the embedding of the orbifold in $\mbox{SU}(4)$, in other words finding the residual $R$-symmetry; we obtain an $\mathcal{N}=2$ theory if $\Gamma \subset \mbox{SU}(2)$, whereas $\Gamma \subset \mbox{SU}(3)$ gives us an $\mathcal{N}=1$ theory. Of course, taking $\Gamma \subset \mbox{SU}(4)$ gives a non-supersymmetric theory.

On the string theory side we can expect tachyons to be present in the non-supersymmetric theories, and as a result to suffer from instabilities \cite{Tseytlin:1999ii,Dymarsky:2005uh,Dymarsky:2005nc} related to tachyon condensation.\footnote{The presence of tachyons in string theory on these non-supersymmetric orbifolds precisely matches the \emph{unstable} running of double trace couplings in the dual field theories (at least in flat space) \cite{Dymarsky:2005uh,Dymarsky:2005nc}.} The corresponding vacuum expectation values of the dual twisted operators break the quantum orbifold symmetry \cite{Adams:2001jb,Dymarsky:2005nc}.\footnote{This should happen at intermediate values of the coupling; at large coupling there are no tachyons in the spectrum.} Apart from the instabilities themselves, this also presents us with a practical problem; without orbifold symmetry the Hilbert space of the theory formally no longer necessarily decomposes over the various twisted sectors. For non-abelian orbifolds this would greatly complicate the problem by preventing us from choosing a different field basis in each twisted sector. We might wish to disregard such theories and focus on orbifolds with $\Gamma \subset \mbox{SU}(3)$ at most. However, there are no practical obstacles in studying non-supersymmetric abelian orbifolds. In line with this, non-supersymmetric abelian orbifolds of $\mathcal{N} = 4$ SYM have been considered before \cite{Beisert:2005he}, while non-abelian ones have not \cite{Solovyov:2007pw}. Moreover, it may in fact be possible to use the exact description of the spectrum we will develop below to get some concrete (quantitative) insight into the presence and effects of tachyons in these orbifold theories. As we already mentioned, in certain cases part of the supersymmetry can be restored by simultaneously orbifolding $\mbox{AdS}_5$.

\subsubsection*{Orbifolding anti-de Sitter space - $\mbox{AdS}_5/\Gamma$}

Exactly as we did for the sphere we can consider orbifolds of $\mbox{AdS}_5$. However, since $\mbox{AdS}_5$ contains the time-like direction of our background space-time we should proceed cautiously to avoid possible time-like orbifolds. With this restriction in mind, we are left to consider orbifolds by $\Gamma \subset \mbox{SU}(2)\times \mbox{SU}(2) \subset \mbox{SU}(2,2)$. For these orbifolds we can proceed as we did for the sphere, again reducing the boundary conditions to those corresponding to diagonal elements of $\mbox{SU}(2)\times \mbox{SU}(2)$ in each twisted sector. Rather than breaking supersymmetry, now we break part of the conformal symmetry. From the point of view of integrability of the string theory this is no real objection as we will see below.\footnote{Formally as far as integrability is concerned we could even consider time-like orbifolds of $\mbox{AdS}_5$, but the corresponding spectral problem is not very exciting.} While the details of the corresponding dual field theories have not been worked out in full detail, at the level of the Bethe ansatz describing the spectrum of the dilatation operator such orbifolds can also be readily accounted for \cite{Beisert:2005he}. Naturally it is possible to simultaneously orbifold $\mbox{AdS}_5$ and $\mbox{S}^5$ in any of the ways we discussed.

\subsection{TsT transformed backgrounds}
\label{subsec:tst}

The second type of deformation we can apply to our background is a sequence of so-called TsT transformations which can be applied to backgrounds with more than one commuting $\mbox{U}(1)$ isometries. A TsT transformation takes a pair of angles corresponding to two isometry directions, say $(\phi_1, \phi_2)$, and acts on them via a $T$-duality along the $\phi_1$ direction, followed by a shift in the $\phi_2$ direction,\footnote{This variable is not affected by the $T$-duality.} $\phi_2 \rightarrow \phi_2 + \hat{\gamma} \tilde{\phi}_1$,\footnote{The shifts $\hat{\gamma}$ are related to the parameters $\gamma$ used below as $\gamma = \hat{\gamma}/\sqrt{\lambda}$.} and finally $T$-dualizing back in the $\tilde{\phi}_1$ direction, where $\tilde{\phi}_1$ is the $T$-dual variable to $\phi_1$. By sequentially applying TsT transformations on a background with $d$ commuting $\mbox{U}(1)$ isometries we can obtain a $d(d-1)/2$ parametric deformation. Note that these transformations require all other fields to be uncharged under the relevant $\mbox{U}(1)$s, specifically in this context the fermions. Fortunately they can be discharged by a field redefinition \cite{Alday:2005ww}.

Very important for our present considerations, the resulting equations of motion are in one-to-one correspondence with the untransformed equations of motion with twisted boundary conditions imposed on the $\mbox{U}(1)$ isometry fields \cite{Frolov:2005dj}. In fact, since by definition these $\mbox{U}(1)$ isometries are nothing but elements of the Cartan subgroup of the isometry group of $\ads$, it should not be surprising that the resulting boundary conditions can be put into correspondence with elements of the Cartan subgroup,\footnote{The fact that this works also for fermions is the result of the non-trivial field redefinition of \cite{Alday:2005ww}.} exactly as we could for orbifolds. Let us now briefly discuss the various possible TsT transformations.

Deformations of the sphere are very sensible from the point of view of AdS/CFT. The primary example of a deformation that can be achieved through TsT transformations is the Lunin-Maldacena background \cite{Lunin:2005jy} for real $\beta$, with $\beta$-deformed SYM as its field theory dual. This deformation preserves $\mathcal{N}=1$ supersymmetry. Restricted to the sphere, a generic sequence of TsT transformations gives a three parameter deformed background \cite{Frolov:2005dj} (see also
\cite{Frolov:2005ty,Frolov:2005iq}), which we will refer to as a $\gamma$-deformed background. The corresponding dual field theories in general have no supersymmetry.

To describe string propagation on these backgrounds through the original theory, we have to implement the following boundary conditions on the three angles on the sphere
\begin{equation}
\phi_i(2\pi) - \phi_i(0) = 2\pi(n_i -\epsilon_{ijk}\gamma_j J_k)\, , \, \, \, n_i \in \mathbb{Z}\, .
\end{equation}
Here the $J_k$ are the angular momenta of $S^5$ and $\gamma_j$ the three parameters of the deformation.

\subsubsection*{$\gamma$-deformations and conformality}

Just as we discussed for orbifolds above, without supersymmetry various subtleties can enter the game, and it appears they do; conformal symmetry of the generically $\gamma$-deformed non-supersymmetric proposed dual field theories \cite{Frolov:2005dj,Frolov:2005iq} is broken at the quantum level \cite{Fokken:2013aea}, despite previous results \cite{Ananth:2007px}. From the point of view of AdS/CFT this can have various interpretations briefly discussed in \cite{Fokken:2013aea}. One interesting possibility closely related to the situation for non-supersymmetric orbifolds is that the resulting non-supersymmetric string background is unstable due to the presence of tachyons (which are at least manifestly present in string theory on $\gamma$-deformed flat space \cite{Spradlin:2005sv}). As in the orbifold case, here we may similarly hope to get some concrete insight with an exact description of the string spectrum on these backgrounds. Also, the effects of this non-conformality appears to be limited to particular subsets of states in the theory \cite{Fokken:2013aea,Fokken:2013aea}, outside of which we can still talk sensibly of scaling dimensions. Indeed, outside of these subsets, explicit results \cite{deLeeuw:2011rw,Ahn:2011xq} obtained in the string theory picture we are describing here can again be successfully matched with perturbative gauge theory \cite{Fokken:2013aea}. Finally, as far as adapting the boundary conditions of our integrable string is concerned there are of course no immediate issues.

\subsubsection*{$\beta$-deformations and pre-wrapping}

Even when we preserve some supersymmetry there is still a subtlety to be mentioned. The asymptotic Bethe ansatz for $\beta$-deformed SYM \cite{Frolov:2005iq,Beisert:2005if} is constructed based on the dilatation operator density for the $\mbox{U}(N)$ gauge theory. In the case of $\mathcal{N}=4$ there is no difference between the dilatation operators of the $\mbox{U}(N)$ and $\mbox{SU}(N)$ theories, but for $\beta$-deformations there is a difference for certain short operators \cite{Frolov:2005iq,Fokken:2013mza,Freedman:2005cg}.\footnote{For example, $\mbox{Tr}(XZ)$ has a nonzero anomalous dimension in the $\mbox{U}(N)$ theory, but appears to be protected in the $\mbox{SU}(N)$ theory \cite{Freedman:2005cg}, at least up to two loops \cite{Penati:2005hp}.} This turns out to make the asymptotic Bethe ansatz more asymptotic than in $\mathcal{N}=4$ SYM, meaning there exists what has been dubbed a pre-wrapping effect \cite{Fokken:2013mza} accounting for the difference between the $\mbox{U}(N)$ and $\mbox{SU}(N)$ theory, entering one loop order before conventional wrapping corrections. Since only the $\mbox{SU}(N)$ theory is conformal \cite{Dorey:2004xm}, we may hope that the dual string theory knows about this effect as it is expected to describe the $\mbox{SU}(N)$ theory \cite{Frolov:2005iq}. We will briefly come back to this point in the next chapter, when we consider the operator $\mbox{Tr}(XZ)$ which is affected by pre-wrapping.

\subsubsection*{TsT transformations involving $\mbox{AdS}_5$}

Less studied but clearly possible from the point of view of string theory, we can do TsT transformations on the three commuting $\mbox{U}(1)$ isometries of anti-de Sitter space \cite{McLoughlin:2006cg,Swanson:2007dh}. However, as always we should be proceed cautiously when time-like directions are involved. Also, while they give sensible string backgrounds these deformations have a less clear interpretation from the point of view of AdS/CFT.

Let us briefly address the issues regarding TsT transformations involving $\mbox{AdS}_5$, starting with issues regarding TsT transformations involving time-like directions. First of all, strictly speaking our string theory is defined on the universal cover of $\mbox{AdS}_5$ which does not have a compact time direction, such that $T$-dualizing in this direction is not an option. Secondly, if we do a TsT transformation involving the time direction we will generate space-time directions with mixed signature. To circumvent the first problem we could initially consider TsT transformations on the space-time, and only subsequently pass to its universal cover. Also, the second problem can pragmatically be avoided by restricting study of the corresponding string theory to regions of the geometry where the signature is the same as the parent theory \cite{McLoughlin:2006cg}. In fact, we could proceed in the spirit of \cite{McLoughlin:2006cg} and formally apply a general sequence of TsT transformations, not necessarily viewing the resulting geometry as a deformation of any particular parent geometry but rather studying it at face value.

Furthermore, any TsT transformation involving an angle from $\mbox{AdS}_5$ necessarily breaks part of the conformal symmetry. The corresponding free string theory should be sensible, even integrable, but again the effect of breaking conformal symmetry on the dual field theory is not obvious. By analogy to how $\beta$-deformed SYM can be obtained by introducing a $\star$-product related to the deformation in the $R$-symmetry group, the duals are expected to be certain non-commutative dual field theories \cite{Beisert:2005if,McLoughlin:2006cg}. Despite these difficulties, TsT transformations of this type have been investigated in the setting of AdS/CFT. In fact as far as integrability is concerned we can readily implement any of these deformations, so while keeping the above concerns in mind we will discuss generic TsT transformations of $\ads$. As we can expect, the interesting features of these deformations turn to rather peculiar ones as soon as time-like directions are involved.

The boundary conditions corresponding to the most generic sequence of TsT transformations are given by
\begin{equation}\label{eqn;genTsT}
\phi_i (2\pi) -\phi_i(0) =  2\pi(n_i + \gamma_{ik} J_k) \, , \, \, \, n_i \in \mathbb{Z}\,
\end{equation}
where $i$ is a generalized index labeling the six $\mbox{U}(1)$ isometry fields of $\ads$, $J_i$ the corresponding angular momentum and $\gamma_{ij}$ an antisymmetric six by six matrix containing the $15$ deformations parameters. Clearly we can also combine orbifolds and TsT transformations \cite{Beisert:2005he}.

\subsection{Deformations in the light-cone gauge}

We have seen that strings on both orbifolds of, and TsT transformed $\ads$ can be directly described through the original string theory with modified boundary conditions. Let us parametrize a completely generic deformation of the above type by the boundary conditions
\begin{equation}
\label{eq:genBCs}
\psi_i (2\pi) = \psi_i (0) + \alpha_i \, , \, \, \, \phi_j (2\pi) = \phi_j (0) + \vartheta_i \,,
\end{equation}
where $\psi_1 \equiv t$ and $\phi_1 \equiv \phi$ with associated Noether charges $E$ and $J$ respectively, and $\psi_{2,3}$ and $\phi_{2,3}$ are the remaining $\mbox{U}(1)$ isometry fields in $\mbox{AdS}_5$ and $\mbox{S}^5$ respectively. The boundary condition on $t$ in the form of $\alpha_1$ is a formal bookkeeping device and is not meant to be interpreted in a direct sense; we introduce it for uniformity since it can appear for general TsT transformed backgrounds \textit{cf.} \eqref{eqn;genTsT} with all their caveats, for orbifolds it is absent. We would like to explicitly mention here that the parameters $\alpha_i$ and $\vartheta_i$ can be taken between zero and $2 \pi$ since only their values modulo $2 \pi$ have physical relevance.

Up to now the boundary conditions on the isometry fields were all treated on an equal footing, but this changes when fixing a light-cone gauge. Fixing a light-cone gauge breaks the manifest $\mbox{SU(2,2)} \times \mbox{SU}(4)$ bosonic symmetry of the superstring to $\mbox{SU}(2)^4$ by fixing the combination $\mbox{x}^+ \equiv t +\phi$ to be equal to the proper time parametrizing the propagation of the superstring. Naturally, the information on the non-trivial boundary conditions on $t$ and $\phi$ is far from lost; we get the so-called level matching condition from the boundary conditions on $\mbox{x}^- \equiv t -\phi$
\begin{equation}
\label{eq:pwslevelmatching}
p_{ws} = \int_0^{2\pi} d \sigma x^{-\prime} = \alpha_1 - \vartheta_1 + 2 \pi m \, ,
\end{equation}
where $m$ is the winding number of the string.\footnote{This of course nicely matches with the condition $e^{i P} =1$ for the total momentum of the undeformed SYM spin chain.} We would like to emphasize that $p_{ws}$ is the world-sheet momentum of the light-cone gauge fixed superstring with quasi-periodic boundary conditions, not of the superstring on a deformed background. Deformations in the $\mbox{x}^+$ direction are simply world sheet diffeomorphisms.

Next, we should recall that when fixing a light-cone gauge we need to redefine the fermions so that they are not charged under isometries in the $t$ and $\phi$ directions. In the undeformed model this turns the periodic fermionic fields into anti-periodic fields in the odd winding sectors \cite{Arutyunov:2007tc}
\begin{equation}
\eta(\sigma + 2\pi) = (-1)^m \eta(\sigma) \,, \hspace{20pt} \theta(\sigma + 2\pi) = (-1)^m  \theta(\sigma) \, .
\end{equation}
Here $m$ is the winding number of the string and $\eta$ and $\theta$ denote the fermionic fields of the light-cone gauge fixed coset model \cite{Arutyunov:2009ga}. Precisely because of this uncharging the light-cone fermions are insensitive to deformations in the $t$ and $\phi_1$ directions, and only care about the winding number in this regard. As a net result, deformations in the $t$ and $\phi_1$ directions enter our model only through the level matching condition \eqref{eq:pwslevelmatching}.

The remaining boundary conditions corresponding to $\alpha_{2,3}$ and $\vartheta_{2,3}$ can now be put in correspondence with the four remaining Cartan generators of $\mbox{SU}(2)^4$. The easiest way to make this link is to consider the generators of shifts of the angles $\psi_{2,3}$ and $\phi_{2,3}$ in the fundamental of $\mbox{SU}(2,2)$ and $\mbox{SU}(4)$ \cite{Arutyunov:2009ga}
\begin{equation}
\label{eq:CartansC2andC3}
C_2 = \mbox{diag}(-1,1,-1,1) \, , \, \, \, C_3 = \mbox{diag}(-1,1,1,-1)
\end{equation}
and the diagonal embedding of the Cartan generators of the four unbroken $\mbox{SU}(2)$s in $\mbox{SU}(2,2)$ and $\mbox{SU}(4)$, which are each just the usual
\begin{equation}
C = \mbox{diag}(1,-1) \, .
\end{equation}
If we then parametrize the four diagonal elements of $\mbox{SU}(2)$ by $\alpha$ and $\dot{\alpha}$ for anti-de Sitter and $\vartheta$ and $\dot{\vartheta}$ for the sphere we get
\begin{align}
\alpha & = -\frac{\alpha_2+\alpha_3}{2} \, ,  &&\dot{\alpha} = \frac{\alpha_3-\alpha_2}{2} \, , \label{gentwist1}\\
\vartheta & = -\frac{\vartheta_2+\vartheta_3}{2} \, , &&\dot{\vartheta} = \frac{\vartheta_3-\vartheta_2}{2} \,
\label{gentwist2}.
\end{align}
Let us introduce $\varphi \equiv \alpha_1-\vartheta_1$ to parametrize the level matching condition \eqref{eq:pwslevelmatching}. Now we are ready to implement these boundary conditions in the integrable structure of the superstring.

\section{The asymptotic solution}

In chapter \ref{chapter:finitevolumeIQFT} we saw how to generically treat integrable quasi-periodic boundary conditions, and indeed the above boundary conditions are integrable as they correspond to symmetries of the S-matrix (in the sense of eqs. \eqref{eq:genquasiperbcs} and \eqref{eq:BCsymmetryofS}). Soon we will discuss how these boundary conditions precisely affect the mirror TBA equations of the previous chapter, however we will first discuss the asymptotic solution of the model as we will need it to describe excited states, and it will give us twisted asymptotic Bethe equations that were known for many of the above deformations already. To find the asymptotic description of these models we should twist the transfer matrix.

\subsection{The twisted \texorpdfstring{$\ads$}{} transfer matrix}
\label{sec:TTM}

As we discussed in the previous chapter, the S-matrix of the $\ads$ superstring consists of two copies of the centrally extended $\alg{su}(2|2)$ invariant S-matrix and a scalar factor. Up to this scalar factor, the transfer matrix is then the product of two transfer matrices each based on a $\mathfrak{su}(2|2)$ invariant S-matrix. Naturally, the twists we want to implement on the light-cone fields are precisely compatible with this structure.

In general the centrally extended $\mathfrak{su}(2|2)$ invariant S-matrix commutes with elements from $\mbox{SU}(2) \times \mbox{SU}(2)$
\begin{equation}
\label{eq:Scomm}
[\S,G\otimes G] =0\, , \hspace{30pt} G\in \mbox{SU}(2) \times \mbox{SU}(2)\, ,
\end{equation}
so that for any $G\in SU(2)\times SU(2)$ we can define the following twisted transfer matrix
\begin{align}\label{eq:twistedtfgeneral}
T^G(q) = \mathrm{str}_a\,G_a\,\S_{a1}(q,p_1)\,\ldots\,\S_{aK}(q,p_{K^\mathrm{I}})\, ,
\end{align}
We should note that for TsT transformations our twist element $G$ depends on the excitation numbers of the state under consideration so that in general $G$ acts also in the physical space. Within each eigenspace however, the action of $G$ is purely in the auxiliary space. To implement the boundary conditions we just discussed, cf. eqs. \eqref{gentwist1} and \eqref{gentwist2} we should choose
\begin{align}
\label{eq:twist}
G = \mathrm{diag}(e^{i\alpha},e^{-i\alpha},e^{i\vartheta},e^{-i\vartheta})\, \in \mbox{SU}(2) \times \mbox{SU}(2)\,,
\end{align}
for the left copy of the transfer matrix, and the same $G$ with dotted angles for the right one.\footnote{Strictly speaking we defined the boundary conditions oppositely from the convention in chapter \ref{chapter:finitevolumeIQFT} and so should really identify these angles with a minus sign. However, since the physics is invariant under this change of sign and with this minus sign we make direct contact with the equations previously found in the literature, we hope that keeping this slight mismatch does not bother the reader.} These boundary conditions are already diagonal by the way we set up the problem, but we would like to note that considering arbitrary elements in $\mbox{SU}(2) \times \mbox{SU}(2)$ is no generalization because of the $\mbox{SU}(2) \times \mbox{SU}(2)$ symmetry of the S-matrix. To see this, we first write a general $G_1\in \mbox{SU}(2)$ as
\begin{equation}
G_1 = U H U^{-1}\, ,\hspace{10pt} \mbox{where} \hspace{10pt}  U \in \mbox{SU}(2)\,, \hspace{5pt}
H = \begin{pmatrix} e^{i\phi} &0 \\ 0 & e^{-i\phi}\end{pmatrix} \in \mbox{SU}(2).
\end{equation}
Now since $U\in SU(2)$ it is easy to see that
\begin{equation}
T^G(q) = \vec{U}^{-1} \, T^H(q)\, \vec{U}\, , \hspace{10pt} \mbox{where}\hspace{10pt}    \vec{U} = U_1\ldots U_{K^\mathrm{I}}\, ,
\end{equation}
thanks to eqn. \eqref{eq:Scomm}. In other words, any twisted transfer matrix is related to a transfer matrix twisted by an element of the Cartan subgroup via a simple basis transformation on the physical space. Consequently, we see that the eigenvalues of $T^G$ and $T^H$ coincide.

We already encountered a simple twisted transfer matrix in chapter \ref{chapter:finitevolumeIQFT}. There we saw explicitly that the effect of a twist is very simple since it does not affect the fundamental commutation relations but only the eigenvalues of the transfer matrix and the associated Bethe equations. In particular, the different terms which sum up to give the eigenvalue of the transfer matrix are typically eigenstates of the twist element, just with different eigenvalues. For example in eqn. \eqref{eq:XXXTonvacuum} we saw that for the XXX spin chain the eigenvalue of $A$ got twisted by $e^{i\a}$ while that of $D$ got twisted by $e^{-i \a}$. We see that in general we need to carefully keep track of the `charge' of the various terms in the transfer matrix under the twist, but apart from this there is no change in the derivation.

If we look at the derivation of the $\mathfrak{su}(2|2)$ transfer matrix \cite{Martins:2007hb,Arutyunov:2009iq}, we readily realize that the effect of the $\vartheta$ twist on the transfer matrix is very simple while the effect of the $\alpha$ type twist is slightly more involved. The $\vartheta$ twist just directly multiplies two terms in the full eigenvalue, while the $\alpha$ twist modifies the derivation at the auxiliary level of this $\mathfrak{su}(2|2)$ problem, turning an XXX spin chain hiding there into a twisted XXX spin chain. Carefully keeping track of the phase factors the undeformed eigenvalue of eqn. \eqref{eq:FullTransfer} becomes\footnote{A more detailed description of this analysis can be found in \cite{deLeeuw:2012hp}, including also a derivation of the corresponding twists in the generating functional of \cite{Beisert:2006qh}.}
\begin{align}
\label{eq:FullEignvalue}
T_{a,1}(v\,|\,\vec{u}) =&
\left.\prod_{i=1}^{K^{\rm{II}}} {\textstyle{\frac{y_i-x^-}{y_i-x^+} \sqrt{\frac{x^+}{x^-}} }}\right[
{\textstyle{e^{ia\alpha} + e^{-ia\alpha}}} \prod_{i=1}^{K^{\rm{II}}} {\textstyle{\frac{v - \nu_i + \frac{ia }{g} }{v-\nu_i-\frac{ia }{g}}}}
\prod_{i=1}^{K^{\rm{I}}}{\textstyle{\frac{(x^--x^-_i)(1-x^- x^+_i)x^+}{(x^+-x^-_i)(1-x^+ x^+_i)x^-}+ }}\nonumber\\
&{\textstyle{+}}\sum_{k=1}^{a-1}e^{i(a-2k)\alpha} \prod_{i=1}^{K^{\rm{II}}}
{\textstyle{\frac{v-\nu_i + \frac{ia }{g} }{v - \nu_i + \frac{i(a-2k)}{g}}}}
\left\{\prod_{i=1}^{K^{\rm{I}}}{\textstyle{\lambda_+(v,u_i,k)+}}\right.\left.\prod_{i=1}^{K^{\rm{I}}}{\textstyle{\lambda_-(v,u_i,k)}}\right\}\\
& - \sum_{k=0}^{a-1}e^{i(a-2k-1)\alpha}\prod_{i=1}^{K^{\rm{II}}} {\textstyle{\frac{v - \nu_i + \frac{ia }{g} }{v-\nu_i + \frac{i(a-2k)}{g}}}}
\prod_{i=1}^{K^{\rm{I}}}{\textstyle{\frac{x^+-x^+_i}{x^+-x^-_i}\sqrt{\frac{x^-_i}{x^+_i}} \left[1-\frac{\frac{2ik}{g}}{v-u_i+\frac{i(a-1)}{g}}\right]}}\times\nonumber\\
&\times\left. \left\{e^{i\vartheta}\prod_{i=1}^{K^{\rm{III}}}{\textstyle{\frac{w_i-v - \frac{i(a-2k+1)}{g}}{w_i- v - \frac{i(a-2k-1)}{g}} + }}
e^{-i\vartheta}\prod_{i=1}^{K^{\rm{II}}}{\textstyle{\frac{v-\nu_i + \frac{i(a-2k)}{g}}{v-\nu_i+\frac{i(a-2k-2)}{g}}}}
\prod_{i=1}^{K^{\rm{III}}}{\textstyle{\frac{w_i - v - \frac{i(a-2k-3)}{g}}{w_i-v - \frac{i(a-2k-1)}{g}}}}\right\}\right],\nonumber
\end{align}
where the auxiliary roots $y$, and $w$ satisfy the Bethe equations (\ref{eq:FulltwistedBAE2}),(\ref{eq:FulltwistedBAE3}) given below. As in the previous chapter the rapidities $\nu$ are related to $y$ via $\nu = y+1/y$, and the parameters $\lambda_\pm$ are not affected by the twists and still given by eqn. \eqref{eq:lambda-pm}.

\subsection{Bethe-Yang equations}\label{sec:BAE}

To get the Bethe-Yang equations we should recall that the complete S-matrix consists of two copies of the $\mathfrak{su}(2|2)$ invariant S-matrix along with a scalar factor. As usual many terms in (the eigenvalue of) the transfer matrix cancel on solutions of the auxiliary Bethe equations below, and the general Bethe-Yang equation \eqref{eq:BEeigenval} in this case becomes
\begin{align}\label{eq:FulltwistedBAE}
1&= e^{i(\alpha+\dot{\alpha})}\left(\frac{x_k^+}{x_k^-}\right)^J\prod_{l=1,l\neq k}^{K^{\rm{I}}}S_{\sl(2)}(p_{k},p_{l})~
\prod_a \prod_{l=1}^{K^{\rm{II}}_{(a)}}\frac{{x_{k}^{-}-y^{(l)}_{l}}}{x_{k}^{+}-y^{(l)}_{l}}\sqrt{\frac{x^+_k}{x^-_k}}\, .
\end{align}
The auxiliary Bethe equations are given by
\begin{align}
1& = e^{i(\vartheta-\alpha)} \prod_{l=1}^{K^{\mathrm{I}}} \frac{y_{k}-x^{+}_{l}}{y_{k}-x^{-}_{l}}\sqrt{\frac{x^-_l}{x^+_l}} \prod_{l=1}^{K^{\mathrm{III}}_{(l)}} \frac{y_{k}+\frac{1}{y_{k}}-w^{(l)}_{l}+\frac{i}{g}}{y_{k}+\frac{1}{y_{k}} - w^{(l)}_{l}-\frac{i}{g}} \label{eq:FulltwistedBAE2}\\
1&=e^{-2i\vartheta}\prod_{l=1}^{K^{\mathrm{II}}_{(l)}}\frac{w^{(l)}_{k}-y^{(l)}_{l}-\frac{1}{y^{(l)}_{l}}+\frac{i}{g}}{w^{(l)}_{k}-y^{(l)}_{l}-\frac{1}{y^{(l)}_{l}}-\frac{i}{g}}
\prod_{l\neq
k}^{K^{\mathrm{III}}_{(l)}}\frac{w^{(l)}_{k}-w^{(l)}_{l}-\frac{2i}{g}}{w^{(l)}_{k}-w^{(l)}_{l}+\frac{2i}{g}}\label{eq:FulltwistedBAE3},
\end{align}
along with a copy for $y_{(r)}$ and $w_{(r)}$  with dotted angles.

\subsubsection*{The $\mathfrak{su}(2)$ sector}

The above Bethe-Yang equations can be used to find the asymptotic energy of any state upon specifying the corresponding excitation numbers. Since we chose the $\mathfrak{sl}(2)$ vacuum to diagonalize our transfer matrix over, the corresponding equations are simplest when considering states from the $\mathfrak{sl}(2)$ sector; there are no auxiliary excitations. Had we instead diagonalized our transfer matrix over the $\mathfrak{su}(2)$ vacuum, the description of states in this sector would have been simpler. From the point of view of the $\mathfrak{sl}(2)$ sector such states are described by states with $K^{\mathrm{II}}_{(a)}=K^{\mathrm{I}}$ and $K^{\mathrm{III}}_{(a)}=0$, so in this case a description over the $\mathfrak{su}(2)$ vacuum is clearly easier.\footnote{In this sector the number of excitations gives the $J_2$ charge, $J_2 = K^{\mathrm{I}}$ which is also frequently denoted $M$.} The corresponding eigenvalue of the transfer matrix is given by
\begin{align}
T^{\su(2)}_{a,1}(v|\vec{u})=&\,\frac{\sin (a+1)\alpha}{\sin\alpha}\prod_{i=1}^{K^{\rm{I}}}\frac{x^--x^-_i}{x^+-x^-_i}
\sqrt{\frac{x^+}{x^-}}
+\frac{\sin (a-1)\alpha}{\sin\alpha}\prod_{i=1}^{K^{\rm{I}}}\frac{x^--x^+_i}{x^+-x^-_i}
\frac{x_i^--\frac{1}{x^+}}{x_i^+-\frac{1}{x^+}}\sqrt{\frac{x^+}{x^-}}
\nonumber \\
&\,-\frac{\sin a \,\alpha}{\sin\alpha}
\left[e^{i\vartheta}\prod_{i=1}^{K^{\rm{I}}}\frac{x^--x_i^+}{x^+-x^-_i}
\sqrt{\frac{x^+x^-_i}{x^-x^+_i}}
+e^{-i\vartheta}
\prod_{i=1}^{K^{\rm{I}}}\frac{x^--x_i^-}{x^+-x^-_i}\frac{x_i^--\frac{1}{x^+}}
{x_i^+-\frac{1}{x^+}}\sqrt{\frac{x^+x^+_i}{x^-x^-_i}}\right]. \nonumber
\end{align}
and the Bethe-Yang equations are
\begin{equation}
\label{eq:su2twistedBAE}
1 = e^{i (\vartheta+\dot{\vartheta})} \left(\frac{x_k^+}{x_k^-}\right)^{J} \prod_{l=1,l\neq k}^{K^{\rm{I}}}S_{\su(2)}(p_{k},p_{l})\, ,
\end{equation}
where we reexpressed the $\mathfrak{sl}(2)$ S-matrix \eqref{eq:sl2smatrix} in terms of the $\mathfrak{su}(2)$ S-matrix \eqref{eq:su2smatrix}. We can obtain this eigenvalue by explicitly acting with the transfer matrix on the $\mathfrak{su}(2)$ vacuum, or by `dualizing' the $\mathfrak{sl}(2)$ Bethe-Yang equations and transfer matrix.\footnote{The basic idea is to rewrite the auxiliary Bethe equations as a manifest polynomial equation from which we can read off the total number of roots it has. $K^{\mathrm{II}(\mathrm{I})}$ of these are the original $y(w)$ roots, the others are the dual roots $\tilde{y}(\tilde{w})$. The original expression for the polynomial divided by its form as a product of its factors is of course constant, and comparing this (constant) ratio at different arguments allows us to rewrite products over roots into products over dual roots. Substituting $\tilde{y}(\tilde{w})$ as a root of the original expression gives the dualized Bethe equations. For details see e.g. \cite{Arutyunov:2011uz}.} Of course, we can do the same for the generic eigenvalue of the $\mathfrak{sl}(2)$ transfer matrix and the full set of Bethe-Yang equations, but let us not write yet more formulae.

\subsubsection*{Typical spin chain form}

In order to compare the twisted Bethe equations for our models with \cite{Beisert:2005if,Beisert:2005he} we will rewrite them in terms of seven equations corresponding to the underlying $\alg{psu}(2,2|4)$ Dynkin diagram. However, as was shown in \cite{Martins:2007hb,deLeeuw:2007uf} the equations given in \cite{Beisert:2005fw} agree with Bethe equations derived from string theory for $P=0$, while in general they can differ by factors of $e^{iP}$. This is not an issue for physical states in the untwisted case, but here these factors have to be taken carefully into account. Let us follow the procedure outlined in \cite{Martins:2007hb} to bring the equations (\ref{eq:FulltwistedBAE}-\ref{eq:FulltwistedBAE3}) to the form presented in \cite{Beisert:2005fw}. Our equations are in the $\alg{sl}(2)$ grading and hence we will bring them to the form where $\eta = -1$. First, we identify the labels
\begin{align}
\label{eq:KKrelation}
&K^{\rm{I}} = K_4, && K^{\rm{II}}_{(l)} = K_1+K_3, && K^{\rm{II}}_{(r)} = K_5+K_7, && K^{\rm{III}}_{(l)} = K_2,
&& K^{\rm{III}}_{(r)} = K_6,
\end{align}
and we split the products accordingly, {\it i.e.}
\begin{align}
&\prod_{j=1}^{K^{\rm{II}}_{(l)}} \to \prod_{j=1}^{K_1}\ \ \prod_{j=1}^{K_3},
&&\prod_{j=1}^{K^{\rm{II}}_{(r)}} \to \prod_{j=1}^{K_5}\ \ \prod_{j=1}^{K_7}.
\end{align}
Next we relabel the parameters in the following way
\begin{align}
&x^\pm_k = \frac{x_{4,k}}{g}, &&
y_k=\left\{\begin{array}{lc} x_{3,k}/g & \ k=1,\ldots K_3\\ g/x_{1,k} &  \ k=1,\ldots K_1 \end{array}\right.
&&
\dot{y}_k=\left\{\begin{array}{lc} x_{5,k}/g & \ k=1,\ldots K_5\\ g/x_{7,k} &  \ k=1,\ldots K_7 \end{array}\right.
\end{align}
Under this map, the main Bethe equation (\ref{eq:FulltwistedBAE})becomes
\begin{align}
1=e^{-i(p_k L+\alpha+\dot{\alpha})}\prod_{j\neq k}^{K_4}S_{\sl(2)}(p_j,p_k)
\prod_{j=1}^{K_3}\frac{x^+_{4,k}-x_{3,j}}{x^-_{4,k}-x_{3,j}}
\prod_{j=1}^{K_1}\!\frac{1 - \frac{g^2}{x^+_{4,k}x_{1,j}}}{1 - \frac{g^2}{x^-_{4,k}x_{1,j}}}
\prod_{j=1}^{K_5}\frac{x^+_{4,k}-x_{5,j}}{x^-_{4,k}-x_{5,j}}
\prod_{j=1}^{K_7}\!\frac{1 - \frac{g^2}{x^+_{4,k}x_{7,j}}}{1 - \frac{g^2}{x^-_{4,k}x_{7,j}}}\, , \nonumber
\end{align}
where $L=J-\frac{K_1-K_3-K_5+K_7}{2}$. To address the auxiliary Bethe equations we need to introduce the rapidities
\begin{align}
&w^{(l)}_{k}= u_{2,k}/g, && w^{(r)}_{k}= u_{6,k}/g, && u_{i,j} \equiv x_{i,j} + \frac{g^2}{x_{i,j}}.
\end{align}
Then (\ref{eq:FulltwistedBAE3}) is straightforwardly rewritten as (the dotted version is obtained by switching labels $K_{1,2,3}\to K_{7,6,5}$ and parameters similarly)
\begin{align}
1 = e^{2i\vartheta} \prod_{j\neq k}^{K_2}\frac{u_{2,k}-u_{2,j}+2i}{u_{2,k}-u_{2,j}-2i}
\prod_{j}^{K_1}\frac{u_{2,k}-u_{1,j}-i}{u_{2,k}-u_{1,j}+i}
\prod_{j}^{K_3}\frac{u_{2,k}-u_{1,j}-i}{u_{2,k}-u_{1,j}+i},
\end{align}
while (\ref{eq:FulltwistedBAE2}) splits into two sets of equations for roots of type 1 and 3 (we have used the level matching condition $\prod_k\frac{x^+_k}{x^-_k} = e^{i\varphi}$)
\begin{align}
&1= e^{i(\alpha-\vartheta-\varphi/2)}\prod_{j=1}^{K_4}\frac{1-\frac{g^2}{x^-_{4,j}x_{1,k}}}{1-\frac{g^2}{x^+_{4,j}x_{1,k}}}
\prod_{j=1}^{K_2}\frac{u_{1,k}-u_{2,j}-i}{u_{1,k}-u_{2,j}+i}, &&
k = 1,\ldots K_1\\
&1= e^{i(\alpha-\vartheta+\varphi/2)}\prod_{j=1}^{K_4}\frac{x^-_{4,j}-x_{3,k}}{x^+_{4,j}-x_{3,k}}
\prod_{j=1}^{K_2}\frac{u_{3,k}-u_{2,j}-i}{u_{3,k}-u_{2,j}+i}, &&
k = 1,\ldots K_3,
\end{align}
again with similar equations for the dotted indices. We see that written this way our equations are exactly of the form of \cite{Beisert:2005fw} up to the twisting factors. These factors are precisely what was added to the Bethe equations in \cite{Beisert:2005if,Beisert:2005he}; next we will show explicit agreement with their twists for the various deformations.

\section{Explicit Models}
\label{sec:explicitmodels}

We are now ready to discuss specific twisted models and explicitly make a link to existing literature when applicable. We first focus on deformations of the sphere and then consider more general deformations that can involve all of $\ads$.

\subsection{Deformations of \texorpdfstring{$S^5$}{}}

\subsubsection*{Abelian orbifolds}

The twist corresponding to a generic abelian $\mathbb{Z}_S$ orbifold is represented by an element
\begin{align}
\mathrm{diag}(e^{-2\pi i T t_1/S}, e^{2\pi i( t_1-t_2)T/S}, e^{2\pi i( t_2-t_3)T/S}, e^{2\pi i t_3 T/S}),
\end{align}
where $T$ labels the twisted sector. From this we deduce that the angles transform as
\begin{align}
(\delta\phi_1,\delta\phi_2,\delta\phi_3) = \left(-\frac{2\pi t_2 T}{S},-\frac{2\pi(t_1-t_2+t_3)T}{S}, -\frac{2\pi(t_1-t_3)T}{S}\right),
\end{align}
resulting in the twist
\begin{align}\label{eq:TwistOrbifold}
& \alpha=\dot{\alpha}=0, && \varphi = \frac{2\pi t_2 T}{S} \, , \nonumber\\
&\vartheta= \frac{\pi(2t_1-t_2)T}{S}, &&
\dot{\vartheta}= \frac{\pi(2t_3-t_2)T}{S}.
\end{align}
These twist are uniquely fixed by the geometry through the boundary conditions on the fields and in turn fix our Bethe equations (\ref{eq:FulltwistedBAE}-\ref{eq:FulltwistedBAE3}). These should then fully agree with \cite{Beisert:2005he} in the one-loop limit. Actually their result is readily generalized to all loop equations by supplementing \cite{Beisert:2005fw} with the appropriate phase factors.

Following \cite{Beisert:2005he} we should introduce phase factors $e^{2\pi i \frac{Ts_k}{S}}$ in front of the equations for roots of type $k$. In the $\alg{sl}(2)$ grading these twists are of the form \cite{Beccaria:2011qd}
\begin{align}
\nonumber
&s_1 =-t_1, && s_2 = 2t_1-t_2, && s_3 = t_2-t_1, &&s_4=0, && s_5 = t_2-t_3, && s_6 = 2t_3 - t_1, && s_7 = -t_3.
\end{align}
Our Bethe equations already contain phase factors, and comparing the above against our general twisted Bethe equations we find agreement exactly when
\begin{align}
& -\vartheta-\varphi/2 = 2\pi \frac{T s_1}{S}, && 2\vartheta = 2\pi \frac{T s_2}{S},
&& -\vartheta + \varphi/2 =  2\pi \frac{T s_3}{S},
\end{align}
with similar expressions for the dotted versions. Comparing this against (\ref{eq:TwistOrbifold}) we see that they are in perfect agreement.

Since we are considering an orbifold, we should not forget to project onto the orbifold invariant part of the spectrum. This means we should consider states which satisfy \cite{Beisert:2005he}
\begin{align}
\label{eq:orbconstraintsphere}
t_2\, p + t_1\, q_2 +  t_3\, q_1 = 0\mod S.
\end{align}
Naturally this constraint can be phrased in more geometric terms via the relation of the $\alg{su}(4)$ weights $[q_1,p,q_2]$ to the angular momenta of the sphere \cite{Arutyunov:2011uz}
\begin{align}
\label{eq:su4Js}
&J = J_1 = \frac{q_1+2p+q_2}{2},&& J_2= \frac{q_1+q_2}{2},&& J_3 = \frac{q_2-q_1}{2} \,,
\end{align}
so that \eqref{eq:orbconstraintsphere} becomes
\begin{equation}
\sum_{i=1}^3 J_i \, \delta \phi_i|_{T=1} = 0 \mod 2 \pi .
\end{equation}

\subsubsection*{$\gamma$-deformations} As discussed, TsT transformations of the sphere give rise to $\gamma$-deformed theories. Recalling that the twisted boundary conditions are parametrized by three deformation parameters $\gamma_{1,2,3}$ as
\begin{align}
\delta\phi_i=\phi_i(2\pi) - \phi_i(0) = -2\pi\epsilon_{ijk}\gamma_j J_k,
\end{align}
where the $J_i$ are the Noether charges corresponding to the shift isometries in the $\phi_i$ direction on the sphere, these boundary conditions correspond to the twists
\begin{align}
& \alpha=\dot{\alpha}=0, && \varphi = 2\pi (\gamma_2 J_3 - \gamma_3 J_2) ,\nonumber\\
&\vartheta=\pi(\gamma_1 (J_2-J_3) +J_1(\gamma_3-\gamma _2)),&&
\dot{\vartheta}=\pi(J_1(\gamma_2 +\gamma_3) - \gamma_1 (J_2+J_3) ), \label{eq:twistchoicesforgammadef}
\end{align}
For $\gamma_1=\gamma_2=\gamma_3=\beta$ this reduces to the $\beta$-deformed theory for real $\beta$.

We will now compare the Bethe equations following from this twist to those derived in \cite{Beisert:2005if} or equivalently to \cite{Ahn:2010ws}. After a duality transformation to the $\alg{sl}(2)$ grading we find the following set of equations (see also the appendix of \cite{Ahn:2010ws} for the explicit $\alg{sl}(2)$ grading)
\begin{align}
{\scriptsize{
\begin{pmatrix}
-\varphi\\
\alpha-\vartheta-\varphi/2\\
2\vartheta\\
\alpha-\vartheta+\varphi/2\\
-\alpha-\dot{\alpha}\\
\dot{\alpha}-\dot{\vartheta}+\varphi/2\\
2\dot{\vartheta}\\
\dot{\alpha}-\dot{\vartheta}-\varphi/2
\end{pmatrix}=
\begin{pmatrix}
\delta_1 (K_5-2 K_6+K_7)-\delta_3 (K_1-2 K_2+K_3) \\
\delta_3 (K_0+K_2-K_3-K_5+K_6)-\delta_2 (K_5-2 K_6+K_7) \\
(\delta_1+2 \delta_2)(K_5-2 K_6+K_7) + \delta_3(2 K_5-2 K_6-2 K_0-K_1+K_3) \\
\delta_3(K_0+K_1-K_2-K_5+K_6)-(\delta_1 + \delta_2)(K_5-2 K_6+K_7) \\
0 \\
(\delta_2+\delta _3)(K_1-2 K_2+K_3)-\delta_1(K_0+K_2-K_3-K_6+K_7) \\
\delta_1(2 K_0+2 K_2-2 K_3-K_5+K_7)-(2\delta_2 + \delta_3)(K_1-2 K_2+K_3) \\
\delta_2(K_1-2 K_2+K_3) - \delta_1(K_0+K_2-K_3-K_5+K_6)
\end{pmatrix}
}},
\end{align}
where $K_i$ are the excitation labels from the Bethe equations and $K_0\equiv L$. The minus sign in the first term of the left hand side is due to the fact that $P = -(AK)_0$ in the notation of \cite{Beisert:2005if}.

This over-determined system of equations indeed has a unique solution. In order to make contact with the twist discussed above we relate the labels to the conserved charges $J_i$ via \eqref{eq:su4Js}, and use their relation to the excitation numbers $K$ (see also \eqref{eq:KKrelation}) as
\begin{align}
&q_1 = K^{\rm{II}}_{(r)} - 2 K^{\rm{III}}_{(r)}, && q_2 = K^{\rm{II}}_{(l)} - 2 K^{\rm{III}}_{(l)}, && p=J-{\textstyle{\frac{1}{2}}}(K^{\rm{II}}_{(l)} + K^{\rm{II}}_{(r)}) + K^{\rm{III}}_{(l)} + K^{\rm{III}}_{(r)},\nonumber\\
&s_1 = K^{\rm{I}}-K^{\rm{II}}_{(r)}, &&s_2 = K^{\rm{I}}-K^{\rm{II}}_{(l)}.
\end{align}
The parameters $\delta_i$ are related to the parameters $\gamma_i$ as
\begin{align}
&\delta_1 = \pi(\gamma_2 + \gamma_3)\,, &&\delta_2 = \pi(\gamma_1 -\gamma_2)\,, && \delta_3 = \pi(\gamma_2-\gamma_3)\, .
\end{align}
Putting this together correctly reproduces our twist.

\subsection{General deformations}

Let us now consider models involving the most generic set of boundary conditions.

\subsubsection*{General TsT transformations}

The twist for a general TsT transformation can be immediately read off from \eqref{eqn;genTsT} together with \eqref{gentwist1} and \eqref{gentwist2}. It is worthwhile to note that if the TsT transformation involves the time direction that the twist will become energy dependent. In other words the asymptotic Bethe equations pick up phases that depend on $E(p)$. This results in a very involved coupled system of equations.

\subsubsection*{General (abelian) orbifolds}

We can also consider the generic orbifold $\mbox{AdS}_5/\Gamma^a \times \mbox{S}^5 /\Gamma^s$, with $\Gamma^a = \mathbb{Z}_R$ and $\Gamma^s = \mathbb{Z}_S$. Since we do not orbifold the $t$ direction, $\varphi$ is unaffected by $\Gamma^a$ and consequently the twists $\vartheta,\dot{\vartheta}$ and $\varphi$ are again given by \eqref{eq:TwistOrbifold}. We now quickly find the expressions for $\alpha,\dot{\alpha}$ by taking analogous expressions to those for $\vartheta,\dot{\vartheta}$ with the equivalent of $t_2$ put to zero. Labeling the twisted sectors for $\mathbb{Z}_R$ and $\mathbb{Z}_S$ by $\tilde{T}$ and $T$ respectively, this gives
\begin{align}
&\alpha= \frac{2\pi\, r_1 \tl{T}}{R}, && \dot{\alpha}= \frac{2\pi\, r_3 \tl{T}}{R} ,\nonumber\\
&\vartheta= \frac{\pi(2t_1-t_2)T}{S}, &&
\dot{\vartheta}= \frac{\pi(2t_3-t_2)T}{S}, \label{eq:genorbtwist}\\
 & \varphi = \frac{2\pi t_2 T}{S}.\nonumber
\end{align}
Orbifold invariance now requires
\begin{align}
t_2\, p + t_1\, q_2 + t_3\,q_1 + r_1\, s_2 + r_3\ s_1 & = 0 \hspace{-3pt}\mod \mbox{LCM}(S,R)\,,\label{eq:orbconstraint}
\end{align}
where $\mbox{LCM}(S,R)$ denotes the least common multiple of $S$ and $R$ and $s_1$ and $s_2$ are the two spins of the conformal group. As mentioned earlier it is also entirely possible to describe non-abelian orbifolds in this framework. In each twisted sector we still get a twist of the form \eqref{eq:genorbtwist}. The only complication is in determining the physical states we should study, \textit{i.e.} in the analogue of \eqref{eq:orbconstraint}. We refer the interested reader to \cite{Solovyov:2007pw} where some specific examples are discussed in detail.

\section{The twisted mirror TBA}
\label{sec:mirrorTBA}

We can now readily embed our deformations in the framework of the mirror thermodynamic Bethe ansatz of the previous chapter. The effect of our deformations is to modify the boundary conditions on the string fields, and in chapter \ref{chapter:finitevolumeIQFT} we saw that this induces chemical potentials the mirror TBA via a defect operator. Hence the only modification we need to make to the mirror TBA equations for the $\ads$ superstring is to add the eigenvalue of the defect operator on a given particle to its equation.

For the generic deformations introduced above we can immediately read off the eigenvalues of the twist element $g$ on the fundamental particles of the asymptotic Bethe ansatz.\footnote{Note that this twist element $g$ should be taken for the charges of the state we wish to describe. In particular, for $\gamma$ deformations this means that the twist for the ground state only contains the angular momentum $J$, since all other charges vanish. This matches with the twisted $S$-matrix approach where the Drinfeld-Reshetikhin twist of the $\mbox{SU}(2|2)$ invariant $S$-matrix does not contribute to the ground state energy, leaving the state independent boundary condition which depends on $J$ only \cite{Ahn:2011xq}.} The eigenvalue of $g$ acting on a string complex or bound state is of course simply a sum of the eigenvalues on its constituents. For the reader's convenience we have summarized these eigenvalues\footnote{The eigenvalue for $Q$-particles is actually the negative of the chemical potential presented in table \ref{tab:chempot}; with this definition the chemical potentials enter uniformly in our TBA equations where the Y-functions for $Q$-particles are defined inversely with respect to the others.} in table \ref{tab:chempot}; the physical chemical potentials are $\mu_{1|w^\pm}$ and $\mu_{1|vw^\pm}$.

\begin{table}[h!]
\begin{center}
\begin{tabular}{|c|c|}
\hline  $\chi$ & $\mu_\chi$ \\
\hline $M|w^{(l)}$   & $2 i M \vartheta$    \\
\hline $M|vw^{(l)}$   & $2 i M \alpha$    \\
\hline $y^{(l)}$   & $i(\alpha-\vartheta)$ \\
\hline $M|w^{(r)}$   & $2 i M \dot{\vartheta}$  \\
\hline $M|vw^{(r)}$    & $2 i M \dot{\alpha}$  \\
\hline $y^{(r)}$    & $i(\dot{\alpha}-\dot{\vartheta})$ \\
\hline $Q$   & $i Q (\alpha+\dot{\alpha})$   \\
\hline
\end{tabular}
\end{center}
\caption{Chemical potentials for particles of type $\chi$ of the mirror TBA. Note that for TsT deformations $\alpha$ and $\vartheta$ depend on the state under consideration.}
\label{tab:chempot}
\end{table}

\noindent These eigenvalues are defined modulo $2 \pi i$ because they are read off from the eigenvalues $e^{\mu_\chi}$ of the defect operator. The chemical potentials themselves are unambiguously defined because of their direct link to the quasi-periodic boundary conditions for fundamental particles in the string theory.

Before moving on to the TBA equations, we should note that the partition function of the mirror theory with purely imaginary chemical potentials is not well defined. In fact, the twists $\alpha$ and $\vartheta$ should have a small positive imaginary part in order to suppress large $M$ magnonic contributions to the partition function. With this regularization the theory is well defined, and we can consider the canonical ground state TBA equations. The resulting considerations are somewhat subtle, but the upshot is a set of simplified and hybrid TBA equations together with a set of large $u$ asymptotics for the Y-functions that are well defined when the regulator is taken away and can hence be taken as proper tools to study the spectral problem.

\subsection{The TBA equations}

The canonical ground state TBA equations with non-zero chemical potentials are given by
\begin{align}
\log Y^{(a)}_{M|w} =& - \mu^{(a)}_{M|w} + \log\big(1+\frac{1}{Y^{(a)}_{N|w}}\big)\star K_{NM} + \log \frac{1-\frac{1}{Y^{(a)}_-}}{1-\frac{1}{Y^{(a)}_+}} \hstar  K_M \, ,\\
\log Y^{(a)}_{M|vw} =& - \mu^{(a)}_{M|vw} + \log\big( 1+\frac{1}{Y^{(a)}_{N|vw}}\big)\star K_{NM} + \log \frac{1-\frac{1}{Y^{(a)}_-}}{1-\frac{1}{Y^{(a)}_+}}\hstar   K_M \nonumber \\ & \quad - \log (1+Y_Q)\star  K^{QM}_{xv}  \, ,\\
\log {Y^{(a)}_\pm} =& - \mu^{(a)}_\pm  - \log\big(1+Y_Q \big)\star  K^{Qy}_\pm + \log \frac{1+\frac{1}{Y^{(a)}_{M|vw}}}{1+\frac{1}{Y^{(a)}_{M|w}}}\star  K_M\,,\\
\log Y_Q =& \, - \mu_Q - J \, \tilde{\mathcal{E}}_{Q} + \log\big(1+Y_{M} \big) \star K_{\mathfrak{sl}(2)}^{MQ}
\\[1mm]
&\quad + \sum_{(a)=\pm} \log\big(1+ \frac{1}{Y^{(a)}_{M|vw}} \big) \star  K^{MQ}_{vwx} + \log \big(1- \frac{1}{Y^{(a)}_\pm}\big) \hstar   K^{yQ}_\pm   \, , \nonumber
\end{align}
where the $\mu_\chi$ are given in table \ref{tab:chempot} keeping in mind the implicit regulator. The ground state energy is given by
\begin{equation}
\label{eq:gse}
E=-\frac{1}{2\pi}\int\, du\, \frac{d\tilde{p}_Q}{du}\log(1+Y_Q)\, .
\end{equation}

As we will see below, it is useful to transform the canonical TBA equations to an alternate form. In particular we can readily simplify the equations for $w$ and $vw$-strings by application of the kernel $(K+1)^{-1}$.\footnote{The equation for $Q$-particles can of course be simplified, resulting in the untwisted eqs. \eqref{eq:undefsimpTBAQ1} and \eqref{eq:undefsimpTBAQg1}; in hybrid form the cancellation of the chemical potentials is non-trivial.} The chemical potentials are in the null-space of this operator, and hence the simplified equations do not depend explicitly on them, meaning they are just given by their untwisted versions \eqref{eq:undefsimpTBAvw} and \eqref{eq:undefsimpTBAw}. These equations together with the canonical TBA equations allow us to find the asymptotic behavior of the $Y_{M|w}$ and $Y_{M|vw}$ functions as we will show shortly. Then, with this asymptotic behavior and the simplified equations we can eliminate the infinite sums over $w$ and $vw$-strings from the equations for $y$ and $Q$-particles to bring them to their hybrid form. The only modification in the derivation of the hybrid TBA equations is the presence of chemical potentials and we discuss the resulting contribution in \ref{app:hybridTBA}. The upshot is the following set of equations
\begin{align}
\log Y_{M|w}^{{(a)}}  = & \, I_{MN}\log(1+Y_{N|w}^{{(a)}})\star s+
\delta_{M1}\log\frac{1-\frac{1}{Y_{-}^{{(a)}}}}{1-
\frac{1}{Y_{+}^{{(a)}}}}\hstar s \,,
\label{eq:sTBAw}\\
\log Y_{M|vw}^{{(a)}}  = & \,-\log(1+Y_{M+1})\star s+I_{MN}\log(1+Y_{N|vw}^{
{(a)}})\star s+\delta_{M1}\log\frac{1-Y_{-}^{{(a)}}}{1-Y_{+}^{{(a)}}}
\hstar s\,,  \label{eq:sTBAvw}\\
\log\frac{Y_{+}^{{(a)}}}{Y_{-}^{{(a)}}} = &
\log(1+Y_{Q})\star K_{Qy}\,, \label{eq:sTBAyd} \\
\log Y_{-}^{{(a)}}Y_{+}^{{(a)}}  = & \,-\log(1+Y_{Q})\star K_{Q}+
2\log(1+Y_{Q})\star K_{xv}^{Q1}\star s+2\log\frac{1+Y_{1|vw}^{
{(a)}}}{1+Y_{1|w}^{{(a)}}}\star s\,,  \label{eq:sTBAyp}\\
\log Y_Q  = & \,-L\tilde{\mathcal{E}}_Q+\log(1+Y_{Q^\prime})\star
\left(K_{sl(2)}^{Q^\prime Q}+2s\star K_{vx}^{Q^\prime -1,Q}\right)\label{eq:sTBAQ}\\
  & +\sum_{{(a)}=\pm}\biggr[\log\biggl(1+Y_{1|vw}^{{(a)}}\biggr)
\star s\,\hstar K_{yQ}+\log(1+Y_{Q-1|vw}^{{(a)}})\star s \nonumber \\
& \hspace{50pt} -\log
\frac{1-Y_{-}^{{(a)}}}{1-Y_{+}^{{(a)}}}\hstar s\star K_{vwx}^{1Q}
 +\frac{1}{2}\log
\frac{1-\frac{1}{Y_{-}^{{(a)}}}}{1-\frac{1}{Y_{+}^{{(a)}}}}
\hstar K_Q
\nonumber\\
& \hspace{70pt} +\frac{1}{2}\log(1-\frac{1}{Y_{-}^{{(a)}}})
(1-\frac{1}{Y_{+}^{{(a)}}})\hstar K_{yQ}\biggl]\,, \nonumber
\end{align}
From these equations and the large $u$ asymptotics for the Y-functions for $w$ and $vw$-strings we can directly read off the large $u$ asymptotics of all Y-functions.\footnote{They can of course also be derived from the canonical equations.} We see that the chemical potentials have disappeared from this form of the TBA equations, meaning they are unchanged from the original model. Consequently also the $Y$-system is unchanged for these deformations. This also follows from the fact that the chemical potentials satisfy the conditions to obtain an undeformed Y-system, formulated in \cite{Cavaglia:2010nm}. We would like to emphasize again that while the chemical potentials disappear from the simplified TBA and Y-system equations, the Y-functions must satisfy the \emph{canonical} TBA equations, which translates to a set of large $u$ asymptotics on them.

\subsection{The ground state solution}
\label{sec:GS}

The first concrete state we need to consider in a generically deformed theory is the ground state. The Y-functions for this state will give us the concrete large $u$ asymptotics of the Y-functions for any excited state with the same chemical potentials. We start by considering the TBA equations in the asymptotic limit. In this limit the $Y_Q$ functions are exponentially small, and from the canonical TBA equations it follows that $Y_+ = Y_-$ since their chemical potentials are equal. Then then simplified TBA equations for $w$ and $vw$-strings (\ref{eq:undefsimpTBAvw},\ref{eq:undefsimpTBAw}), become simple recursion relations for constant Y-functions
\begin{equation}
Y^{\circ}_{M+1|w} = \frac{Y^{\circ 2}_{M|w}}{1+Y^{\circ}_{M-1|w}} -1 \, ,
\end{equation}
with an identical equation for $vw$-strings, and we note that $Y^{\circ}_{0|(v)w}$ is zero by definition. By this recursion relation all $Y^{\circ}_{M|w}$ are uniquely fixed in terms of the value of $Y^{\circ}_{1|w}$. The constant solution of this equation which also satisfies the canonical TBA equations with zero chemical potentials is given by \cite{Frolov:2009in}
\begin{equation}
Y^{\circ}_{M|w} = M(M+2) \, ,
\end{equation}
which satisfies the recursion relation for the simple reason that $M^2 = (M+1)(M-1)+1$. A clear generalization of this solution is
\begin{equation}
Y^{\circ+}_{M|w} = [M]_{q}[M+2]_{q} \, ,
\end{equation}
where we have introduced $q$-numbers $[n]_q$ as
\begin{equation}
[n]_q = \frac{q^n - q^{-n}}{q-q^{-1}} \, ,
\end{equation}
since $q$-numbers retain the property $[M]_q^2 = [M+1]_q[M-1]_q+1$ for any $q \in \mathbb{C}$. By picking $q$ appropriately, we can obtain any desired constant value for $Y^{\circ}_{1|w}$, and hence this is the general constant solution of the simplified TBA equations. What remains is to determine the value of $q$ such that these Y-functions also satisfy their canonical TBA equations. We do this in appendix \ref{app:hybridTBA}, and by substituting the result in the hybrid equations (\ref{eq:sTBAyd},\ref{eq:sTBAyp},\ref{eq:sTBAQ}) we find the following asymptotic auxiliary Y-functions
\begin{align}
Y^{\circ+}_{M|w} & = [M]_{q_{\vartheta}}[M+2]_{q_{\vartheta}} \,,&   Y^{\circ-}_{M|w} & = [M]_{q_{\dot{\vartheta}}}[M+2]_{q_{\dot{\vartheta}}} \, , \label{eq:YwGSasympt}\\
Y^{\circ+}_{M|vw}&  = [M]_{q_{\alpha}}[M+2]_{q_{\alpha}} \,,& Y^{\circ-}_{M|vw} & = [M]_{q_{\dot{\alpha}}}[M+2]_{q_{\dot{\alpha}}} \, , \label{eq:YvwGSasympt}\\
Y^{\circ+}_{\pm}&  = [2]_{q_{\alpha}}/[2]_{q_{\vartheta}} \,,& Y^{\circ-}_{\pm} & = [2]_{q_{\dot{\alpha}}}/[2]_{q_{\dot{\vartheta}}} \, ,
\label{eq:YpmGSasympt}
\end{align}
while the main Y-functions are
\begin{equation}
\label{eq:YQasympt}
Y^\circ_{Q} =  ([2]_{q_{\alpha}}-[2]_{q_{\vartheta}})([2]_{q_{\dot{\alpha}}}-[2]_{q_{\dot{\vartheta}}})[Q]_{q_{\alpha}}[Q]_{q_{\dot{\alpha}}} e^{-J \tilde{\mathcal{E}}_Q(\tilde{p})}\, ,
\end{equation}
where the $q$ have become phases denoted $q_\theta \equiv e^{i \theta}$. This asymptotic solution is a generalization of the asymptotic solution presented in \cite{deLeeuw:2011rw} and \cite{Ahn:2011xq} for deformations of the sphere. Of course, this asymptotic solution also follows directly from the twisted transfer matrix. For example, the asymptotic $Y_Q$-function is directly given by
\begin{align}
Y^\circ_Q = \chi(\pi(G))\chi(\pi(\dot{G}))e^{-J\tilde{\mathcal{E}}_Q},
\end{align}
where $\chi(\pi(G)) = \mathrm{tr}(\pi(G))$ is the character of the twist element $G$ in the $Q$-particle bound state representation in the transfer matrix (\ref{eq:twistedtfgeneral}); the twisted transfer matrix for zero particles. Naturally this agrees with (\ref{eq:YQasympt}). Just as for the chemical potential, here we should keep in mind that the charges entering the twisted transfer matrix should be those of the ground state.

\subsection{Large \texorpdfstring{$u$}{u} asymptotics and excited states}

In the large $u$ limit, the convolutions on the right hand side of the canonical TBA equations become insensitive to fluctuations of the Y-functions around the origin, and only their constant `background' values play a role. It follows that for any state with a given set of chemical potentials, these constant values should be given by the asymptotic ground state solution. In other words
\begin{equation}
\label{eq:uasymp}
\lim_{u\rightarrow \pm \infty} Y_\chi(u) = Y^\circ_\chi \, ,
\end{equation}
where $Y^\circ_\chi$ denotes the asymptotic ground state values of the Y-function of type $\chi$, as given in equations (\ref{eq:YwGSasympt}-\ref{eq:YQasympt}). Note that while the regulator plays an essential role in the canonical TBA equations, the regulator can be taken away by describing the spectral problem through the simplified TBA equations and the asymptotics (\ref{eq:uasymp}).\footnote{An equivalent frequently encountered characterization of the asymptotics is to give
\begin{equation}
\lim_{M\rightarrow\infty} \frac{Y^{(a)}_{M|(v)w}}{M} = -\mu^{(a)}_{1|(v)w}\, ,
\end{equation}
which for purely imaginary chemical potentials is a slightly awkward statement to make about Y-functions that are real for any finite $M$ upon taking the regulator away. While it is of course correct, it relies essentially on the regulator which is why we prefer to give an unambiguous set of (real) large $u$ asymptotics instead.}

Coming to excited states, as we saw above the twisted transfer matrix can be used to construct the asymptotic Y-functions for the ground state. The same is true for excited states and allows us to apply the contour deformation trick \cite{Arutyunov:2009ax,Arutyunov:2011uz} to obtain excited state TBA equations from the ground state equations. The differences in analytic structure that distinguish the different excited states, i.e. the number and location of poles and zeroes of the Y-functions, can be determined from the asymptotic Y-functions constructed through the twisted transfer matrix. This idea has already been used to find the excited state TBA equations for the $\alg{sl}(2)$ descendant of the Konishi operator upon orbifolding the sphere \cite{deLeeuw:2011rw}. We would like to point out that for orbifolds the asymptotics (\ref{eq:uasymp}) are the same for any state in the same model, but this is not the case for theories obtained by TsT transformations as the chemical potentials depend on the state under consideration.

The simplified ground state TBA equations naturally have real Y-functions as solutions, and indeed the asymptotic solution (\ref{eq:YwGSasympt}-\ref{eq:YQasympt}) is real. Reality of the asymptotic Y-functions for excited states follows by the way the twist parameters enter the twisted transfer matrix \eqref{eq:FullEignvalue}; the terms in the undeformed transfer matrix that are related by conjugation are now multiplied by conjugate phases. The exact Y-functions that solve the excited state TBA equations are therefore also real, since the equations are obtained by the contour deformation trick with a real asymptotic solution.

\section{Wrapping corrections to the ground state energy}
\label{sec:wrappingforgroundstate}

With the asymptotic ground state solution we can perturbatively solve the TBA equations and find the energy of the ground state. From the structure of the $Y^\circ_Q$ functions we see that their expansion starts at order $g^{2J}$, known as single wrapping order for our length $J$ state. Next, expanding the TBA equations it is easy to see that the first correction to the asymptotic form of the $Y_Q$-functions comes in at double wrapping order, here order $g^{4J}$. Hence up to this order we can find the perturbative ground state energy simply by expanding (\ref{eq:gse}) to the desired order in $g$.

The leading order wrapping correction to the ground state energy for general deformations is given by
\begin{align}
\label{eq:GSsinglewrapping}
E(J) =& \frac{(\cos \alpha-\cos \vartheta)(\cos \dot{\alpha}-\cos \dot{\vartheta})}{\sin\alpha\sin\dot{\alpha}}
\frac{\Gamma(J-\frac{1}{2})}{2\sqrt{\pi}\Gamma(J)}\times \\
&\times\left[
\mathrm{Li}_{2J-1} (e^{i(\alpha+\dot{\alpha})}) +
\mathrm{Li}_{2J-1} (e^{-i(\alpha+\dot{\alpha})}) -
\mathrm{Li}_{2J-1} (e^{i(\alpha-\dot{\alpha})}) -
\mathrm{Li}_{2J-1} (e^{-i(\alpha-\dot{\alpha})})\right]\nonumber g^{2J} \, .
\end{align}
Here and below $\mathrm{Li}_{n}$ is the polylogarithm. If we only deform the sphere ($\alpha = \dot{\alpha} =0$) this reduces to
\begin{equation}\label{eq:GSsinglewrappingsphere}
E(J) = - \frac{8 \Gamma(J-\tfrac{1}{2})\zeta(2 J-3)}{
\sqrt{\pi} \Gamma(J)} \sin^2\frac{\vartheta}{2} \sin^2\frac{\dot{\vartheta}}{2} g^{2J}\, .
\end{equation}
Specifying the deformation parameters as in eqs. \eqref{eq:twistchoicesforgammadef} and \eqref{eq:TwistOrbifold} gives the wrapping correction to the ground state in $\gamma$-deformed respectively orbifolded theories. Note that this expression diverges for $J=2$, as we will come back to shortly.

These formulae can be readily extended up to order $g^{4J}$. To do so it is instructive to write the asymptotic Y-function explicitly as function of the mirror momentum $\tilde{p}$
\begin{align}\label{eqn;AsymYQ}
Y^{\circ}_{Q}(\tl{p}) = \chi(G)\chi(\dot{G}) \left[ \frac{2 g}{\sqrt{Q^2+\tl{p}^2}+\sqrt{4g^2 +Q^2+\tl{p}^2}}\right]^{2J}.
\end{align}
To order $g^{4J}$  (\ref{eq:gse}) reduces to
\begin{align}
E = -\frac{1}{2\pi} \sum_{Q=1}^{\infty}\int d\tl{p}\, Y^{\circ}_{Q}(\tl{p}).
\end{align}
Evaluating the integral perturbatively in $g$ we find
\begin{align}
\label{eq:GSfullsinglewrap}
E =- \sum_{m=1}^{\infty}\frac{J (-1)^m (2g)^{2(J + m)}}{4}
\frac{ \Gamma (J+m-\frac{3}{2}) \Gamma (J+m-\frac{1}{2})}{\Gamma (m) \Gamma (2 J+m)} \sum_{Q=1}^{\infty}\frac{\chi(G)\chi(\dot{G})}{Q^{2 (J+ m)-3} }.
\end{align}
The sum over $Q$ yields a term similar to \eqref{eq:GSsinglewrapping}
\begin{align}
\sum_{Q=1}^{\infty}\frac{\chi(G)\chi(\dot{G})}{Q^{x} } =& \frac{(\cos \alpha-\cos \vartheta)(\cos \dot{\alpha}-\cos \dot{\vartheta})}{\sin\alpha\sin\dot{\alpha}} \times\\
&\times\left[
\mathrm{Li}_{x} (e^{i(\alpha+\dot{\alpha})}) +
\mathrm{Li}_{x} (e^{-i(\alpha+\dot{\alpha})}) -
\mathrm{Li}_{x} (e^{i(\alpha-\dot{\alpha})}) -
\mathrm{Li}_{x} (e^{-i(\alpha-\dot{\alpha})})\right]\nonumber.
\end{align}
In the limit of undeformed anti-de Sitter space, (\ref{eq:GSfullsinglewrap}) reduces to the single wrapping correction in equation (5.5) of \cite{Ahn:2011xq}. Going beyond this order requires a significant amount of computation due to the non-trivial corrections to the Y-functions. Nonetheless, the full double wrapping L\"uscher and TBA-type correction to the $\gamma$-deformed ground state energy have been determined and shown to be in full agreement \cite{Ahn:2011xq}. The relevant computations generalize partially to more generic deformations; their results also apply to orbifolds of the sphere. Moreover, in \cite{Ahn:2011xq} the lowest order double wrapping correction to the length three ground state was explicitly found to be
\begin{align}
E_{0}(3)  = & -(2-[2]_{q_{\vartheta}})(2-[2]_{q_{\dot{\vartheta}}})\left(\frac{3}{16}
\zeta_{3}g^{6}-\frac{15}{16}\zeta_{5}g^{8}+\frac{945}{256}
\zeta_{7}g^{10}-\frac{3465}{256}\zeta_{9}g^{12}+\dots\right)\nonumber \\
   & -(2-[2]_{q_{\vartheta}})(2-[2]_{q_{\dot{\vartheta}}})\left([2]_{q_{\vartheta}}+[2]_{q_{\dot{\vartheta}}}-4)
\right)\frac{15}{256}\zeta_{3}\zeta_{5}g^{12} + \dots \nonumber\\
   & +(2-[2]_{q_{\vartheta}})^{2}(2-[2]_{q_{\dot{\vartheta}}})^{2}\left(-
\frac{9}{256}\zeta_{3}^{2}+
\frac{189}{4096}\zeta_{7}\right)g^{12}+\dots \,,
\end{align}
here written in terms of its generic dependence only on $\vartheta$ and $\dot{\vartheta}$. To generalize this result to deformations of the anti-de Sitter space would likely require considerable computational effort, as already the single wrapping correction (\ref{eq:GSsinglewrapping}) is significantly more involved.\footnote{In the final details of their computation, the authors of  \cite{Ahn:2011xq} found a single term through an accurate numerical computation which was then expressed in a transcendental basis using the program EZ-face (documented in \cite{Borwein:1999js}). Given the form of the single wrapping correction for more generic deformations we would likely encounter numbers not even expressible in terms of multiple zeta values, in fact this `number' would be a function of $\alpha$ and $\dot{\alpha}$.} So far we have focussed on completely generic deformations, however in special cases there are certain interesting observations to be made based on the asymptotic ground state solution (\ref{eq:YwGSasympt}-\ref{eq:YQasympt}) and the corresponding energy corrections (\ref{eq:GSsinglewrapping}).

\subsubsection*{Regularizing the ground state energy}

In \cite{Frolov:2009in} the authors discussed the vanishing of the ground state energy of the undeformed theory at any finite size. However they observed a curious divergence of the ground state energy at $J=2$. This divergence was not surprisingly also found subsequently in the case of orbifolds \cite{deLeeuw:2011rw} and $\gamma$-deformations \cite{Ahn:2011xq}.

In order to show vanishing of the ground state energy, we would basically like to show that the asymptotic solution (here for zero twist) is exact, \textit{i.e.} that vanishing $Y_Q$ functions are a solution of the TBA equations. In order to show this the authors of \cite{Frolov:2009in} introduced a regularization in the form of a small chemical potential for the fermionic $y$-particles, denoted $h$, which made the $Y_Q$ functions $O(h^2)$ such that they vanish in the $h \rightarrow 0$ limit. However for $J=2$ this limit does not commute with the infinite sum in the energy formula leading to a divergent energy even though the $Y_Q$ functions vanish. The chemical potential $h$ introduced there directly corresponds to an infinitesimal twist $\vartheta =\dot{\vartheta} =h$, and this divergence is indeed reproduced by our energy formula (\ref{eq:GSsinglewrapping}) which diverges as $\alpha,\dot{\alpha} \rightarrow 0$ for non-zero $\vartheta$ and $\dot{\vartheta}$. However (\ref{eq:GSsinglewrapping}) does not diverge as $\alpha,\dot{\alpha} \rightarrow 0$ for $\vartheta = \dot{\vartheta} = 0$, meaning that a chemical potential in the $\mbox{AdS}$ sector regularizes the ground state solution for the undeformed theory, showing that the ground state energy vanishes without any divergence as it should.

\subsubsection*{The $\gamma$-deformed ground state}

While the construction above provides us with a way to regularize the ground state energy of the undeformed theory, the interpretation or possible resolution of the divergence of the ground state energy\footnote{Here and below, as always, the ground state energy refers to the energy of the state dual to $\mbox{Tr}(Z^J)$, which is the ground state in our twisted boundary condition picture in the light-cone gauge, despite the fact that e.g. $\mbox{Tr}(X^J)$ may have lower energy for particular deformations.} at $J=2$ for certain orbifolds and $\gamma$-deformed theories remains an open question. Of course, as we discussed earlier these deformations break all supersymmetry and may result in broken conformal invariance in the `dual' gauge theory. In particular, it is interesting to note that for $\gamma$-deformed SYM the operator $\mbox{Tr}(Z^J)$ is insensitive to the non-conformality of the theory for $J>2$, with anomalous dimensions precisely matching the energies of eqn. \eqref{eq:GSsinglewrappingsphere} above, while for $J=2$ the result is scheme dependent\cite{Fokken:2013aea}.

\subsubsection*{Residual supersymmetry}

As we saw above, the asymptotic ground state solution is an exact solution of the TBA equations when the corresponding $Y_Q$ functions vanish in some appropriately regularized sense. In addition to the undeformed superstring, this situation arises when we deform parts of anti-de Sitter space and the sphere in an identical fashion. By this we mean a twist where at least two group elements are identical, \textit{i.e.} $\alpha = \pm \vartheta$ or $\dot{\alpha} = \pm \dot{\vartheta}$. Indeed, in this limit the asymptotic $Y^\circ_Q$ functions (\ref{eq:YQasympt}) and correspondingly the ground state energy (\ref{eq:GSsinglewrapping}) vanish. In other words, just as for the undeformed ground state, the corresponding asymptotic solution is an exact solution of the TBA equations. Naturally this is not an accident but owes to the residual supersymmetry left in this situation.

In the undeformed theory the ground state is a half-BPS state, meaning it is annihilated by a set of supercharges
\begin{equation}
Q^{1,2}_\gamma | 0 \rangle  = \bar{Q}^{3,4}_{\dot{\gamma}} | 0 \rangle  = 0 \, , \, \, \, \gamma = 1,2.
\end{equation}
Combined with the fact that it is a highest weight state this allows us to derive a relation between the scaling dimension and other weights of the state from the superconformal algebra in the usual fashion, in particular by considering the anti-commutator $\{ Q, S \}$ \cite{Dolan:2002zh}
\begin{equation}
\label{eq:bpscomm}
\{ Q^i_\gamma, S_j^\delta \} = 4 \delta_i^j (M_\gamma^\delta -\tfrac{i}{2} D) - 4 \delta_\gamma^\delta R^i_j
\end{equation}
where the $M_\gamma^\delta$ are the generators of the conformal group corresponding to the $\mbox{SU}(2)$ subgroup with spin $s_1$ and the $R^i_j$ are the generators of $\mbox{SU}(4)$ and $D$ is the dilatation operator. Indeed, since the left hand side annihilates a highest weight half-BPS state for $i=1,2$ and the action of the $R^i_i$ is expressible via the corresponding weights, we get a relation between the scaling dimension and the $\alg{su}(4)$ weights $[q_1,p,q_2]$. As a result the state cannot have an anomalous dimension; its scaling dimension is protected.

In a deformed theory, the ground state can still be considered half-BPS in the sense that it is annihilated by certain operators, but those operators have lost their meaning and relation to the scaling dimension since the superconformal algebra is broken. This means there are generically no protected operators. However, precisely deformations of the type $\alpha = \pm \vartheta$ or $\dot{\alpha} = \pm \dot{\vartheta}$ preserve part of the superconformal algebra. In particular some of the relations (\ref{eq:bpscomm}) which protect the scaling dimension of the ground state remain. Concretely we preserve the relations (\ref{eq:bpscomm}) for $i=j=1,2$, $\gamma=\delta=2,1,$ if we twist with $\alpha=\vartheta$ and for $i=j=\gamma=\delta=1,2,$ if we twist with $\alpha=-\vartheta$. The relations are similar for the $\bar{Q}$s and $\bar{S}$s with dotted twists. As a result the ground state energy is still a protected quantity in such deformed models.

\subsubsection*{Protected states without supersymmetry}

In addition to the above, there is another class of deformed models with a ground state energy which does not receive corrections. The most interesting representative of this class is $\gamma$-deformed theory with $\gamma_2=\gamma_3=0$. This feature arises because the ground state solution for generic $\gamma$-deformed theories is independent of $\gamma_1$ as immediately follows from our twisted transfer matrix since the ground state has only charge $J=J_1$. This matches with the results of \cite{Ahn:2011xq} since in the twisted $S$-matrix approach the ground state energy is independent of the Drinfeld-Reshetikhin twist, which in this case carries all dependence on $\gamma_1$. Since the $Y_Q$ functions and ground state energy vanish in the limit $\gamma_{2,3} \rightarrow 0$, we have an exact solution of the TBA equations with zero energy. However, except for the cases $\{\gamma_1,\gamma_2,\gamma_3\} = \{\gamma,\pm \gamma,\pm \gamma\}$ (with all possible choices of signs) $\gamma$-deformations break all supersymmetry. Therefore we have no immediate explanation for the apparent `protection' of the ground state energy, and it would be interesting to understand how this arises from the point of view of the deformed theory.\footnote{By mapping it to the undeformed theory which has periodic boundary conditions for the ground state, this result is of course immediate.} Other models of this type would be generic TsT transformed backgrounds, with the parameters that couple to $J$ put to zero.

\section{A twisted S-matrix approach}

Before moving on to the next chapter, we would like to make note of a conceptually different approach to describing $\gamma$-deformed string theory. While we treated strings on $\gamma$-deformed backgrounds by considering quasi-periodic boundary conditions that depend on the state under consideration (also called operatorial boundary conditions), it is possible to take another point of view with equivalent physical results. Rather than taking the (classical) relation between the deformation and string boundary conditions to heart, we could consider the deformed (integrable) field theory directly. The deformation should then show up in the S-matrix, while preserving the Yang-Baxter equation. In \cite{Ahn:2010ws} it was proposed that this S-matrix should be given by a so-called Drinfeld-Reshetikhin twist \cite{DrinfeldTwistRef,drin,twi,Reshetikhin:1990ep} of the undeformed S-matrix.

Essentially, a Drinfeld-Reshetikhin twist takes a solution $S$ of the Yang-Baxter equation and twists it by $F$ which is itself a solution of the Yang-Baxter equation
\begin{equation}
S \rightarrow \tilde{S} \equiv F S F\,,
\end{equation}
ensuring $\tilde{S}$ also satisfies the Yang-Baxter equation. One typical twist that automatically satisfies the Yang-Baxter equation is \cite{Reshetikhin:1990ep}
\begin{equation}
F = e^{ (H_i \otimes H_j - H_j \otimes H_i) \phi_{ij}}\, ,
\end{equation}
where $\phi_{ij}$ is an arbitrary antisymmetric matrix, and the $H_i$ are Cartan elements of the symmetry algebra of our S-matrix.

At this stage a quick counting might give us cause to worry; $\mathfrak{psu}(2|2)_{c.e.}^2$ would only give us a total of six arbitrary parameters, whereas the most general TsT transformation involves fifteen parameters. However, the idea of \cite{Ahn:2010ws} was not just to consider twisting the $\mathfrak{psu}(2|2)_{c.e.}$ invariant factors of the S-matrix individually, but rather consider twisting a hypothetical $\mathfrak{psu}(2,2|4)$ invariant S-matrix by a general twist element corresponding to the Cartan subalgebra of $\mathfrak{psu}(2,2|4)$, and infer its consequences on our S-matrix with reduced symmetry. The result is an `ordinary' Drinfeld-Reshetikhin twist of our S-matrix, together with a boundary condition dependent on the $J$-charge of the state under consideration. One technical complication is that the twist does not factor over the two $\mathfrak{psu}(2|2)_{c.e.}$ factors and hence the resulting S-matrix no longer factorizes. Nonetheless this proposal has been shown to successfully reproduce the $\gamma$-deformed asymptotic Bethe ansatz (eqs. (\ref{eq:FulltwistedBAE}-\ref{eq:FulltwistedBAE3}) with \eqref{eq:twistchoicesforgammadef}), and in fact the full eigenvalues of our twisted transfer matrix (eqn. \eqref{eq:FullEignvalue} with \eqref{eq:twistchoicesforgammadef}).\footnote{In our approach we manifestly introduce state-dependent twists in the Bethe equations, while the twisted S-matrix approach effectively result in these twists. Similar state-dependent twists appear in the context of braided Yang-Baxter algebras \cite{Hlavaty:1994vh,Kundu:1995hr,Fioravanti:2001bx,Fioravanti:2001pt}. As noted in \cite{Ahn:2010ws}, this may indicate a link between such braided algebras and twisted R-matrices.} This twisted S-matrix has since also been nicely matched with the explicit perturbative spin chain S-matrix at weak coupling and world-sheet S-matrix at strong coupling in the near-BMN limit \cite{Ahn:2012hs}, and reproduces the finite size effects observed for the dyonic giant magnon on the Lunin-Maldacena background \cite{Ahn:2012hsa,Ahn:2011dq,Ahn:2011nm}.

At the level of the thermodynamic Bethe ansatz for the ground state both approaches agree from the start, as only the $J$-dependent boundary condition survives from the twisted S-matrix point of view. For excited states there is again a marked difference at the outset however; for $\gamma$-deformations the chemical potentials in table \ref{tab:chempot} depend on the $J_2$ and $J_3$-charge of the state, whereas the manifest chemical potentials in the twisted S-matrix approach only depend on $J$. Now we should not forget that excited states in general have driving terms `due' to the extra charges of the state. In the twisted S-matrix approach these driving terms should in all honesty of course come with the twist included, schematically
\begin{equation}
\log(1+Y) \star \tilde{K} \rightarrow \log(1+Y) \star \tilde{K} + \log \tilde{S}\,,
\end{equation}
and while $\tilde{K} = K$, it should not be hard to convince ourselves that (upon diagonalization) the difference between the twisted and untwisted S-matrix could easily account for the difference in chemical potentials and thereby make the TBA equations agree as well.

\chapter{Twisted excited states}
\label{chapter:excstatesandwrapping}

In this chapter we are going to use the general results of the previous chapter to study excited states and compute finite volume corrections to their energies in twisted theories. In particular, we will discuss the effect of twisting on the excited state TBA equations for (a descendant of) the Konishi state by considering it in an orbifolded version of our string theory. However before doing so we will explicitly compute wrapping corrections for single magnons and the Konishi state in the $\su(2)$ sector of $\beta$-deformed SYM, as these results can be nicely compared to and matched with explicit perturbative computations done in the deformed field theory.

\section{Wrapping corrections in \texorpdfstring{$\beta$}{}-deformed SYM}

\subsection{Single magnons}

A single magnon in the $\su(2)$ sector of  $\beta$-deformed theory has the following Bethe equation (cf. eqs. \eqref{eq:su2twistedBAE} and \eqref{eq:twistchoicesforgammadef})
\begin{equation}
\label{pbetadeformed} 1=e^{i(p+2\pi \beta)J}
~~~~~~\Longrightarrow~~~~~~p=-2\pi \beta+\frac{2\pi n}{J}\, ,~~~~~
n=0,\ldots \Big[\frac{J}{2}\Big]\, .
\end{equation}
Level matching requires $p=-2\pi \beta$ so that a physical magnon has $n=0$. Its asymptotic energy is
\begin{align}
E = J+\sqrt{1+4g^2 \sin^2 \pi\beta}.
\end{align}
On the gauge theory side the corresponding operator is of the form ${\mbox{Tr}}(Z \ldots Z X Z\ldots Z)$ with $J$ $Z$s.

We can find the leading order wrapping correction to the energy from the weak coupling expansion of the exact energy formula \eqref{eq:TBAgeneralstateenergyformula}, giving\footnote{At leading order there is no distinction between $\tilde{p}$ and $v$, but this is important at higher orders; $\frac{d \tilde{p}^Q}{dv} = 1 + 2 g^2 \frac{v^2- Q^2}{(v^2+Q^2)^2}+ \mathcal{O}(g^4)$.}
\begin{align}
\label{eq:LOwrappingE}
E_{LO} = -\frac{1}{2\pi}\sum_{Q=1}^{\infty}\int dv \frac{d\tilde{p}^Q}{dv} Y^{\circ}_{Q}(v).
\end{align}
To compute this we need to expand the twisted $Y_Q$-functions to leading order in $g$. From eqs. \eqref{eq:twistchoicesforgammadef} we see that the two transfer matrices entering in $Y_Q^o$ should be twisted by $\vartheta=\pi \beta$ and $\dot{\vartheta}=(2J-1)\pi \beta$ respectively. Denoting these generically as $T(v|u;\vartheta)$, we obtain the following Y-functions
\begin{align}
Y_Q(v) &=
\frac{g^{2J}}{(Q^2+v^2)^J}\frac{T^{\alg{su}(2)\ell}_{Q,1}(v|u;\beta)T^{\alg{su}(2)\, r}_{Q,1}(v|u;(2J-1)\beta)}{S_0(u,v)}
\, ,
\end{align}
where the coefficient $S_0(u,v)$ is the leading order expansion of the $\mathfrak{sl}(2)$ S-matrix given in appendix \ref{app:excstatesandwrapping}. The leading order twisted transfer matrix in the $\alg{su}(2)$ sector is
{\small
\begin{align}
T_{Q,1}^{\alg{su}(2)}(v|u;\vartheta)=&\frac{g}{\sqrt{Q^2+v^2}} \left[
\frac{(Q-1)\left((Q+1)^2+(v-u)^2\right)}{(i (Q-1)+v-u)(i-u)} +
\frac{Q e^{-i \vartheta }\sqrt{\frac{u+i}{u-i}} (Q^2+(u-v+i)^2)}{(i (1-Q)-v+u)(i-u)} + \right.\nonumber\\
&\left.\qquad\qquad\quad+ \frac{Q e^{i \vartheta } (i
(Q+1)-v+u)}{(i-u)\sqrt{\frac{u+i}{u-i}}} + \frac{(Q+1) (i
(1-Q)+v-u)}{i-u}\right].
\end{align}
}
This transfer matrix starts at order $g$ so that our Y-function and associated corrections to energy and momentum will be of order $g^{2J+2}$. We can now straightforwardly compute the first wrapping correction to the energy by integrating and summing these Y-functions. The result is a sum of $\zeta$-functions\footnote{These results apply for $J>1$, we discuss $J=1$ below. It is also interesting to observe that the leading order wrapping correction vanishes for $J=2$ for any $\beta$. Expanding the asymptotic Y-functions one order higher, a quick calculation shows that this does not persist however. Explicitly this is perhaps most easily seen from the discussion just below regarding the special points $\beta=n/J$ which tells us $E^{\beta=1/2}(2) = \left(2 \zeta(3)-\tfrac{5}{2} \zeta(5)\right)g^8 + \mathcal{O}(g^{10})$.}
\begin{align}
\label{Esu1magnon} E^{\beta}_{\rm LO}(J) =g^{2J+2}
\sum_{n=1}^{\lfloor\frac{J+1}{2}\rfloor} 16 (-1)^n\frac{\Gamma
(J-n+1) \Gamma \left(J-n+\frac{3}{2}\right)}{\sqrt{\pi}\, \Gamma
(J+1) \Gamma (J-2 n+3)}B^{\beta}_n(J) \zeta (2 J-2 n+1) ,
\end{align}
where the coefficients are given by the following expressions
\begin{align}
& B^{\beta}_{1}(J) = -\frac{J}{2}\sin(J\pi\beta)\sin^2(\pi\beta)\sin((J-2)\pi\beta),\\
& B^{\beta}_{n>1}(J) = \sin(J\pi\beta) \sin^{2n}(\pi\beta) \Big[
(n-1) \sin ((J-2n)\pi\beta)-J \cos(\pi\beta)\sin((J-2
n+1)\pi\beta)\Big]\nonumber.
\end{align}
This correction perfectly agrees with the field-theoretic result obtained in \cite{Fiamberti:2008sm,Fiamberti:2008sn}, as well as the results of \cite{Gromov:2010dy} obtained by arbitrarily twisting the asymptotic solution of the Y-system and fixing the twists by comparison with the Bethe equations.

We should pay particular attention when $\beta = \frac{n}{J}$ with $n=0,1,\ldots, J-1$. For these values of the deformation parameter, the naive lowest order in the weak-coupling expansion of the transfer matrix $T_{Q,1}^{\alg{su}(2)}(v|u;(2J-1)\beta)$ vanishes and the expansion really starts two powers of $g$ higher at $g^{2J+4}$. Bearing in mind that the rapidity $u$ is $g$-dependent through eqn.\eqref{eq:uofp} with momentum $p=-2\pi\beta$, we can easily check that the functions $Y_Q$ for these special values of $\beta$ are absolutely the same as for generic $\beta$ provided we shift $J\rightarrow J+1$. Hence we find
\begin{align}
E^{\beta=n/J}_{\rm LO}(J) = g^2 \left.E^{\beta}_{\rm
LO}(J+1)\right|_{\beta=\frac{n}{J}}.
\end{align}

\subsubsection*{Divergences, $\mbox{U}(N)$ versus $\mbox{SU}(N)$, and pre-wrapping}

In section \ref{subsec:tst} of the previous chapter we discussed that the asymptotic Bethe ansatz describes $\mbox{U}(N)$ $\beta$-deformed SYM rather than $\mbox{SU}(N)$ $\beta$-deformed SYM \cite{Frolov:2005iq}. We mentioned that the operator $\mbox{Tr}(XZ)$ ($J=1$) is affected by a so-called pre-wrapping effect which distinguishes between $\mbox{U}(N)$ and $\mbox{SU}(N)$.\footnote{In fact, it is the unique state in the $\mathfrak{su}(2)$ sector for which this distinction matters \cite{Fokken:2013mza}.} $\mbox{Tr}(XZ)$ is in fact protected in the $\mbox{SU}(N)$ theory \cite{Freedman:2005cg}, at least up to two loops \cite{Penati:2005hp}, while it acquires a non-zero anomalous dimension in the $\mbox{U}(N)$ theory. As our string theory is supposed to describe the $\mbox{SU}(N)$ rather than the $\mbox{U}(N)$ theory \cite{Frolov:2005iq} this raises an immediate question because the asymptotic Bethe ansatz for our string predicts a nonzero anomalous dimension
\begin{equation}
E = 1 + 2\sin^2(\pi\beta) g^2  - 2\sin^4(\pi\beta) g^4 + \ldots \,.
\end{equation}
Finite-size effects can and will of course correct this energy, but the conventional wrapping correction discussed above gives a two loop correction for $J=1$, meaning it cannot cancel the one loop contribution of the asymptotic Bethe ansatz. At the same time, taking a look at the above energy correction we see that the leading transcendentality term is $\zeta(2J-1)$. Extending this result to $J=1$ would suggest that the corresponding wrapping correction is divergent, raising questions of its own.

Naively extending our computation to $J=1$ we indeed find a divergent term as in eqn. \eqref{Esu1magnon}, but also a finite piece
\begin{equation}
 E^{\beta}_{\rm LO}(1) = g^4 (4-4 \zeta(1)) \sin^4{\pi \beta} \,.
\end{equation}
This divergence is similar (in form) to that of the leading order wrapping correction for $\mbox{Tr}(Z^2)$ in the $\gamma$-deformed theory. This divergence clearly signals some kind of problem, and its interpretation is an interesting question.

Let us first consider that our string theory really describes the $\mbox{SU}(N)$ theory, meaning that finite-size effects should contain the pre-wrapping effects that cancel the one and two-loop anomalous dimension. It then becomes tempting to consider that there is a technical problem with the naive computation we just considered, and that resolving this would result in the desired cancellations. Of course, we also cannot exclude that for some reason the generalized L\"uscher formula should be adapted in this particular case, but at the moment it would be unclear how. In case our string theory\footnote{As it stands, described by our integrable model with twisted boundary conditions.} is really `dual' to the $\mbox{U}(N)$ theory rather than the $\mbox{SU}(N)$ one, this divergence may be related to the lack of conformality of the former, and could link it to the divergence for $\mbox{Tr}(Z^2)$ in $\gamma$-deformed or non-supersymmetric orbifolded theories mentioned in the previous chapter (see also the discussion in \cite{Fokken:2013aea}). Still, $\beta$-deformed $\mbox{U}(N)$ SYM has the $\mbox{SU}(N)$ theory as its fixed point while the $\gamma$-deformed and non-supersymmetric orbifolded theories have no such fixed point.

From a purely technical point of view we should note that the divergent wrapping corrections we have seen here and in the previous chapter arise for operators of length two, and that it seems unlikely that the asymptotic Y-functions for longer operators would result in divergent energies. Hence, by considering longer operators which are effected by pre-wrapping \cite{Fokken:2013mza} we should be able to gain valuable insight into part of the questions raised here, while side-stepping complicating divergences. Unfortunately these longer operators lie outside the simple $\mathfrak{su}(1|1)$, $\mathfrak{su}(2)$, and $\mathfrak{sl}(2)$ sectors and would require venturing into the $\mathfrak{su}(2|3)$ sector. Provided the corresponding energy corrections are non-divergent, comparing these to anomalous dimensions in the dual gauge theory would presumably give concrete indications regarding the gauge group as well. This point clearly deserves further investigation.

\subsection{\texorpdfstring{$\beta$}{}-deformed Konishi}

Next we would like to reproduce the wrapping energy correction found in \cite{Fiamberti:2008sm} for the Konishi state in the $\alg{su}(2)$ sector of $\beta$-deformed theory. The perturbative computation of the anomalous dimension of $\beta$-deformed $\alg{su}(2)$ Konishi-like states \cite{Fiamberti:2008sm} agrees perfectly with results coming from the asymptotic Bethe equations up to three loops. At four loops there is a discrepancy arising due to wrapping effects, which we should be able to explain via L\"{u}scher formulae. In \cite{Ahn:2010yv} a modified S-matrix was used to obtain an expression for the wrapping correction to the energy, which is in agreement with the explicit perturbative results. Here we would like to approach the problem via our twisted transfer matrix. However, as the result of \cite{Ahn:2010yv} fits in the framework of the subsequently fully developed twisted S-matrix approach discussed at the end of the last chapter, our twisted transfer matrix gives equivalent input and the result is immediate. The energy correction is given by
\begin{equation*}
E^{\beta}_{\mathrm{Konishi}} = g^8 \left[-54(1+\Delta)^3(-5+3\Delta) \zeta(3) -
360 (1+\Delta)^2 \zeta(5) +
\frac{81(1-3\Delta)^2(1+\Delta)^4}{(1+3\Delta)^2}\right],
\end{equation*}
where $\Delta = \tfrac{\sqrt{5+4\cos{4\pi \beta}}}{3}$. Of course explicit calculations based
on the twisted transfer matrix gives the same result. We can compute wrapping corrections to many other states in this way, some showing interesting quantitative features. Explicit results can be found in \cite{Arutyunov:2010gu,deLeeuw:2012hp}.

\section{Orbifolded Konishi}

In this section we will study the $\mathfrak{sl}(2)$ descendant of the Konishi state in a twisted setting. We will consider the simple case of a $Z_S$ orbifold obtained by twisting the two light-cone $\mathfrak{su}(2)$ factors in the sphere equally (or oppositely) by a root of unity; $\vartheta=\pm\dot{\vartheta}$ with $\vartheta=\tfrac{2\pi n}{S}$.\footnote{If we want to study a nontrivial on-shell $\mathfrak{sl}(2)$ state our orbifold can only barely touch level matching due to the required orbifold invariance $J_i \delta \phi_i = 0 \mod 2\pi$ which for states in the $\mathfrak{sl}(2)$ sector gives $J p= 2 \pi \mathbb{Z}$.} In other words our orbifold group is generated by
\begin{equation}
\gamma = e^{-i \frac{4\pi}{S} C_{i}} \in \mbox{SU}(4)\, ,
\end{equation}
where $i=2$ for $\vartheta=\dot{\vartheta}$  and $i=3$ for $\vartheta=-\dot{\vartheta}$  respectively, and $C_2$ and $C_3$ are defined in eqn. \eqref{eq:CartansC2andC3}. The choice of sign does not matter as the resulting theories are physically equivalent, in fact the transfer matrix in the $\mathfrak{sl}(2)$ sector is manifestly symmetric in $\vartheta$ ($\dot{\vartheta}$). This orbifold breaks the R- and super-symmetry completely, meaning that our state is not in the Konishi multiplet anymore. Nonetheless we will refer to it as orbifolded Konishi. Its dual operator is schematically of the form $\mbox{Tr}(\gamma^n Z D^2 Z)$. We can of course also view this state as living in the $\gamma$-deformed theory obtained by taking $\gamma_{3/2}=\frac{1}{S}$ (depending on the choice $\pm$ above).\footnote{Note furthermore that the excited state TBA equations we are finding below based on this asymptotic solution, factor into two independent copies of an $\mathfrak{su}(2|2)$ problem at the asymptotic level. Therefore we can easily take the equations for one of these copies and consider it at a different twist and consider extending the resulting equations beyond the asymptotic limit. This means the results below allow us to find the excited state TBA equations for the $\mathfrak{sl}(2)$  Konishi descendant for \emph{any} orbifold or $\gamma$ deformation of the sphere that does not affect the level matching condition.}

The main ingredient in both the contour deformation trick leading up to excited state TBA equations and L\"uscher's formulae is the asymptotic solution. For states without auxiliary excitations, and an arbitrary twist of the type above (meaning $T^{l}_{Q,1} = T^{r}_{Q,1} = T_{Q,1}$), these are given by
\begin{align}
\label{TS}
T_{Q,1}(v\,|\,\vec{u})=&1+\prod_{i=1}^{K^{\mathrm{I}}} \frac{(x^--x^-_i)(1-x^-
x^+_i)}{(x^+-x^-_i)(1-x^+
x^+_i)}\frac{x^+}{x^-}\\
&\hspace{-1.5cm}-2\cos\vartheta\sum_{k=0}^{Q-1}\prod_{i=1}^{K^{\mathrm{I}}}
\frac{x^+-x^+_i}{x^+-x^-_i}\sqrt{\frac{x^-_i}{x^+_i}}
\left[1-\frac{\frac{2ik}{g}}{v-u_i+\frac{i}{g}(Q-1)}\right]+\sum_{m=\pm}
\sum_{k=1}^{Q-1}\prod_{i=1}^{K^{\mathrm{I}}}\lambda_m(v,u_i,k)\, , \nonumber
\end{align}
where $\lambda_\pm$ is defined in eqn. \eqref{eq:lambda-pm}; note that the transfer matrix is an even function of $\vartheta$.

In addition to the transfer matrix we need the solution of the Bethe-Yang equations for the momenta of the two excitations to find the TBA equations for this excited state from the ground state ones. However, our orbifold does not affect the specific Bethe-Yang equations for our state and hence this solution is just that of the Konishi state in the untwisted theory, already found numerically in \cite{Arutyunov:2009ax}.

In the next section we will discuss how the asymptotic solution for orbifolded Konishi allows us to find a working set of simplified TBA equations. We will then calculate the NLO finite-size correction to the energy and show that L\"uscher's perturbative approach is in perfect agreement with the TBA approach.

\subsection{The TBA equations}
\label{subsec:orbKonTBA}

We will take the approach advocated in chapters \ref{chapter:finitevolumeIQFT} and \ref{chapter:AdS5string} and find the TBA equations for our orbifolded Konishi state by means of the contour deformation trick. For this we need to understand the analytic properties of the asymptotic solution, so we will start there.

\subsubsection{Analytic properties}

\begin{figure}
\begin{center}
 \includegraphics[width=9cm]{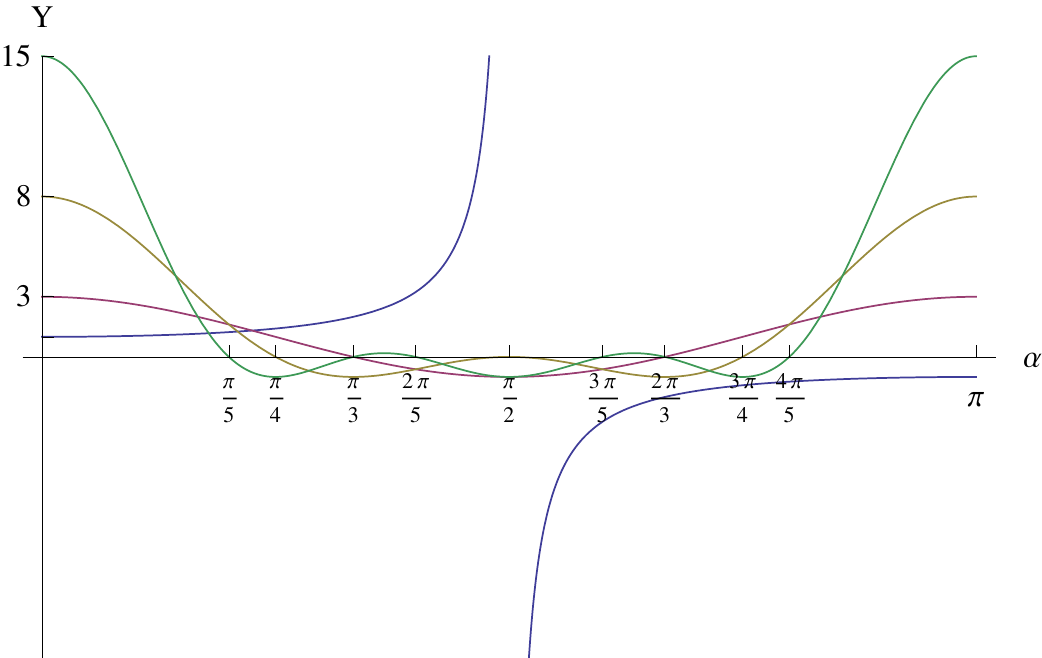}
\end{center}
\caption{The asymptotics of $Y_\pm$ (blue), $Y_{1|w}$ (red), $Y_{2|w}$ (yellow), $Y_{3|w}$(green) as a function of the twist.}
\label{fig:YMwasymptotics}
\end{figure}

Naturally we should expect that the analytic properties of the Y-functions depend rather directly on the twist while their finer details depend on the coupling constant $g$. Firstly, in line with the general discussion in section \ref{sec:GS} the asymptotics of $Y_{M|vw}$ on the mirror line  are $M(M+2)$
\begin{equation}
\lim_{v\rightarrow \infty} Y_{M|vw}(v) = M(M+2),
\end{equation}
while the asymptotics of $Y_{M|w}$ and $Y_\pm$ depend on the twist and indicate interesting behaviour as the asymptote can be both positive and negative. The asymptotic values of both $Y_\pm$ are given by
\begin{equation}
\lim_{v\rightarrow \infty} Y_{\pm}(v) = \sec \alpha,
\end{equation}
while those of $Y_{M|w}$ are given by
\begin{equation}
\label{eq:wasymptotics}
\lim_{v\rightarrow \infty} Y_{M|w}(v) = M + 2\sum_{Q=0}^{M}(M-Q)\cos (2Q+2)\alpha .
\end{equation}
We have plotted the asymptotics of $Y_-$, $Y_{1|w}$, $Y_{2|w}$, and $Y_{3|w}$ in figure \ref{fig:YMwasymptotics}.
It is interesting to note that the asymptotics in the zero twist limit of course smoothly reduce to the known values of $1$ and $M(M+2)$ respectively, but that this also almost happens when the twist parameter is equal to $\pi$,
\begin{align}
\lim_{v\rightarrow \infty} Y^{\alpha = 0,\pi}_{M|w}(v)&  = M(M+2),\\
\lim_{v\rightarrow \infty} Y^{\alpha = 0}_{\pm}(v)& = -\lim_{v\rightarrow \infty} Y^{\alpha = \pi}_{\pm}(v) = 1.
\end{align}
This is of course a very particular value, but still we clearly see that the asymptotic solution is really different from the untwisted case and corresponds to a $\mathbb{Z}_2$ orbifold.

The fact that the asymptotic values of the $Y_{M|w}$-functions can change from positive to negative means that these functions must have roots on the real mirror line for certain values of the twist and coupling, roots which are not present for the untwisted Konishi state \cite{Arutyunov:2009ax}. However, these roots will always become relevant at \emph{finite} values of $g$. As the twist is varied this finite value of $g$ can approach zero, \textit{cf.} figure \ref{fig:YMwasymptotics}. At the same time these roots move away to infinity, so that at weak coupling they do not play a role. Now let us concretely discuss the analytic properties of the Y-functions.

Let us start by recalling the construction of the asymptotic Y-functions in terms of the transfer matrices,
\begin{align}
Y_{M|w} & = \frac{T_{1,M} T_{1,M+2}}{T_{2,M+1}} \, ,  \, \, \, \, \, Y_{-} = -\frac{T_{2,1}}{T_{1,2}} \, ,  \, \, \,\, \,  Y_{+} = - \frac{T_{2,3}T_{2,1}}{T_{1,2}T_{3,2}} \nonumber \, ,\\
\label{eq:YmvwinT}
Y_{M|vw}& = \frac{T_{M,1} T_{M+2,1}}{T_{M+1,2}}  = \frac{T_{M,1} T_{M+2,1}}{T_{M+1,1}^- T_{M+1,1}^+ - T_{M,1} T_{M+2,1}}.
\end{align}
Here $T_{M,Q}^\pm$ denotes $T_{M,Q}$ with its argument shifted by $\pm i/g$. Roots of $Y_{M|w}$ arise from the roots of $T_{1,M}$ and $T_{1,M+2}$, while roots of $Y_{M|vw}$ arise from the roots of $T_{M,1}$ and $T_{M+2,1}$. For $Y_{M|vw}$ it is immediately clear that its roots shifted by $\pm i/g$ give roots of $1+Y_{M-1|vw}$ and $1+Y_{M+1|vw}$,
\begin{equation}
1+Y_{M|vw}(r_{\scriptscriptstyle M}^\pm) = 0, \, \,{\rm where} \, \, \, T_{M,1} (r_{\scriptscriptstyle M-1}) = 0 \, .
\end{equation}
While not obvious from their expression, a similar relation holds for $Y_{M|w}$,
\begin{equation}
1+Y_{M|w}(\rho_{\scriptscriptstyle M}^\pm) = 0, \, \,{\rm where} \, \, \, T_{1,M} (\rho_{\scriptscriptstyle M-1}) = 0.
\end{equation}
The analytic properties for the orbifolded Konishi state are quite involved, in the sense that a large number of roots appear and play a role in the TBA, with many of them displaying critical behavior of a nature that depends on the specific value of the twist. Let us also mention here that this criticality is naturally related to the roots of the $Y_{M|w}$-functions on the real mirror line, indicated above. In the main text we will refrain from discussing these technical complications and focus on the TBA equations at small coupling. A short discussion on the finer details of the analytic properties can be found in appendix \ref{app:criticality}.\footnote{This critical behaviour is also present in an NLIE description, but appears to be simpler since it is limited to a finite set of functions \cite{Suzuki:2012mq}.}

\begin{figure}[h]
\begin{center}
 \includegraphics[width=9cm]{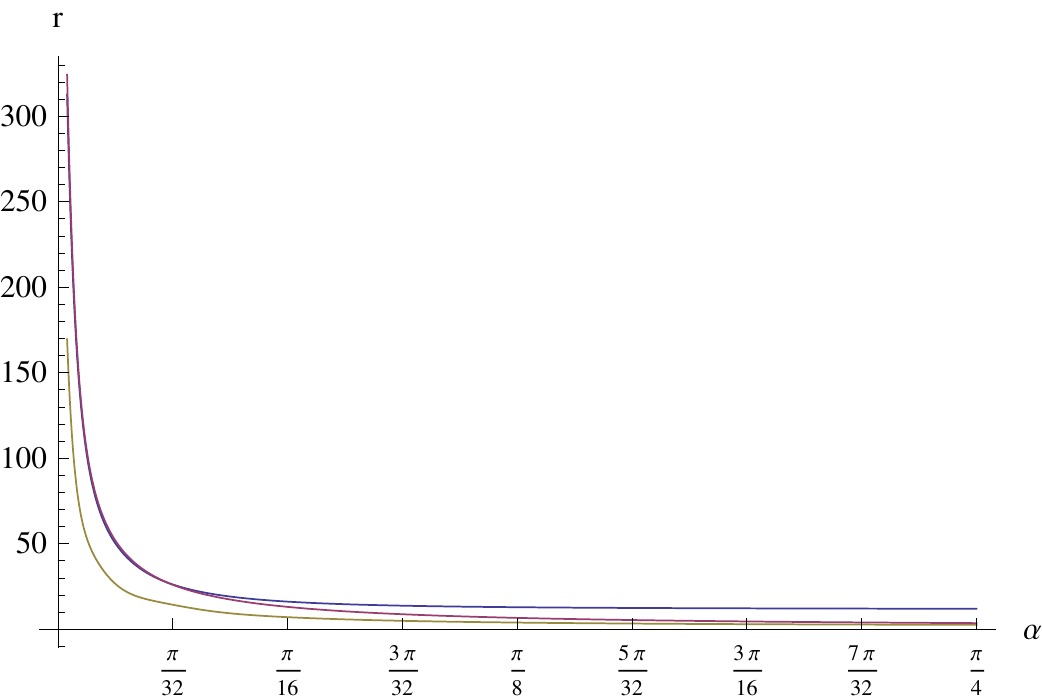}
\end{center}
\caption{The movement of the real root $\mathbf{r}_1$ of $1+Y_{1|vw}(v-i/g)$ with the twist, for $g=\tfrac{1}{10}$ (blue), $g=1$ (red), and $g=10$ (yellow). The root asymptotes to the value of $2$ as the coupling is increased.}
\label{fig:Y1vwroottwistbehaviour}
\end{figure}

At small coupling, the $Y_{M|vw}$-functions have four roots, which we will denote by $\pm \mathbf{r}_{\scriptscriptstyle M\pm1}$.\footnote{As we are discussing the asymptotic solution, these roots of course correspond to roots of the transfer matrices $T_{M,1}$ and $T_{M+2,1}$,  which we denote by $\pm \mathbf{r}_{\scriptscriptstyle M-1}$, and $\pm \mathbf{r}_{\scriptscriptstyle M+1}$ respectively.} These roots were not observed for the true Konishi state, which agrees with our findings since these roots move away towards infinity as the twist is removed, see figure \ref{fig:Y1vwroottwistbehaviour}. As these roots are real, their shifted counterparts, the roots of $1+Y_{M\pm1|vw}$, will always appear in the TBA for orbifolded Konishi.
Just as in the Konishi case, $Y_+$ has poles at $u_i^\pm$, which lead to the appearance of additional driving terms in the TBA equations. While not discussing criticality here, let us briefly note that we reproduce the critical behaviour found in \cite{Arutyunov:2009ax} as the smooth $\vartheta \rightarrow 0 $ limit of our general story, see appendix \ref{app:criticality} for more details.

To summarize, at small coupling the following points are important in the simplified TBA equations,
\begin{equation}
Y_+(u_i^\pm)  = \infty\, , \hspace{10pt}Y_{M|vw}(\mathbf{r}_M^\pm) = Y_{M|vw}(-\mathbf{r}_M^\pm) = -1 \, .
\end{equation}

\subsubsection{The simplified TBA equations}

Following the contour deformation trick, in order for the asymptotic solution to be a solution we take the ground state TBA equations in simplified (and hybrid) form and define the integration contour such that it goes slightly below the line $-i/g$, \textit{i.e.} such that it encloses the poles of $Y_+$ at $u_i^-$ and the roots of $1+Y_{M|vw}$ at $\pm \mathbf{r}_M^-$ between itself and the real line. By taking the integration contour back to the real line we get our explicit TBA equations.
\vspace{10pt}\\
\bigskip
 \noindent
$\bullet$\ $M|w$-strings; $\ M\ge 1\ $, $Y_{0|w}=0$
\vspace{-10pt}
\begin{equation}
\log Y_{M|w} =  \log(1 +  Y_{M-1|w})(1 +Y_{M+1|w})\star s
 + \delta_{M1}\, \log\frac{1-\frac{1}{ Y_-}}{ 1-\frac{1}{ Y_+} }\hstar s + i \pi h_M (\alpha) \,,~~~~~
\end{equation}
These equations are not modified at small coupling, except for the potential appearance of a factor of $i \pi$, corresponding to negative asymptotics for the $Y_{M|w}$ function. $h_M (\alpha)$ is defined accordingly, being one when $Y_{M|w}$ asymptotes to a negative number, and zero otherwise, \textit{cf.} (\ref{eq:wasymptotics}). These equations do get modified by critical behavior, see appendix \ref{app:criticality} for more details. \vspace{10pt}\\
\bigskip
 \noindent
$\bullet$\ $M|vw$-strings; $\ M\ge 1\ $, $Y_{0|vw}=0$
\vspace{-5pt}
\begin{align}
\log Y_{M|vw}(v) = & - \log(1 +  Y_{M+1})\star s +
\log(1 +  Y_{M-1|vw} )(1 +  Y_{M+1|vw})\star
s\\
& - \log{S(\mathbf{r}^{-}_{M\pm1} - v)S(-\mathbf{r}^{-}_{M\pm1} - v)} \nonumber\\
&  + \delta_{M1} ( \log\frac{1-Y_-}{1-Y_+}\hstar s -  \log{S((\pm u_1)^- - v)})\,\nonumber.
\end{align}
Here the driving terms arise from the poles of $Y_+$ and the roots of $1 +  Y_{Q|vw}$; an implicit sum over $\pm$ is implicitly assumed here and below.\vspace{10pt}\\
\bigskip
 \noindent
$\bullet$\   $y$-particles
\begin{align}
\log \frac{Y_+}{ Y_-}(v) = \, &  \log(1 +  Y_{Q})\star K_{Qy} - \log S_{1_*y}(\pm u_1 ,v)  \,,\\
\log {Y_- Y_+}(v) = \, &  2 \log\frac{1 +  Y_{1|vw}}{ 1 +  Y_{1|w} }\star s - \log\left(1+Y_Q \right)\star K_Q + 2 \log(1 +Y_{Q})\star K_{xv}^{Q1} \star s \\
& - 2\log{S(\mathbf{r}_1^- - v)S(-\mathbf{r}_1^- - v)}-  \log \frac{\big(S_{xv}^{1_*1}\big)^2}{ S_2}\star s(\pm u_1,v) \nonumber
\end{align}
In both equations we get contributions from the exact Bethe equation $Y_1(u_{*i}) = -1$, where the star signifies analytic continuation of the rapidity to the string region.\footnote{Similarly, $S_{1_*y}(u_{j},v) \equiv S_{1y}(z_{*j},v)$ is shorthand notation for the S-matrix with the first and second arguments in the string and mirror regions, respectively. The same convention is used for other kernels and S-matrices.} In the second equation the roots of $1 +  Y_{1|vw}$ also contribute.  For the contribution from the exact Bethe equation we have used the following notation
\begin{equation*}
\log  \frac{\big(S_{xv}^{1_*1}\big)^2}{S_2}\star s(u,v) \equiv  \int_{-\infty}^\infty\, dt\, \log  \frac{S_{xv}^{1_*1}(u,t)^2}{ S_2(u-t)}\, s(t-v)
 \,.~~~~~
\end{equation*}
The contribution then follows from the identity
\begin{equation}
\log{S_1(u_j-v)} - 2 \log{S_{xv}^{1_* 1}}\star s(u_j,v) = -\log{\frac{(S_{xv}^{1_* 1})^2}{S_2}}\star s(u_j,v) ,
\end{equation}
\noindent valid for real $u_j$.\vspace{10pt}\\
\bigskip
 \noindent
$\bullet$\ $Q$-particles
\vspace{-5pt}
\begin{align}
\log Y_Q(v) = & - J\, \tH_{Q} + \log \left(1+Y_{Q'} \right) \star (K_{\sl(2)}^{Q'Q} + 2 \, s \star K^{Q'-1,Q}_{vwx} )  - \log S_{\sl(2)}^{1_*Q}(\pm u_1,v)\nonumber\\
&  +  2 \log (1 + Y_{1|vw}) \star s \hstar K_{yQ} + 2 \, \log (1 + Y_{Q-1|vw}) \star s\nonumber\\
&  - 2  \log\frac{1-Y_-}{1-Y_+} \hstar s \star K^{1Q}_{vwx} +  \log \frac{1- \frac{1}{Y_-}}{ 1-\frac{1}{Y_+} } \hstar K_{Q}  +  \log \big(1-\frac{1}{Y_-}\big)\big( 1 - \frac{1}{Y_+} \big) \hstar K_{yQ}  \nonumber\\
& -2 \log{S}\hstar K_{yQ} ((\pm \mathbf{r}_1)^-,v) -2 \log{S}((\pm \mathbf{r}_{Q-1})^-,v)\nonumber\\
& + 2 \log{S}\star_{p.v} K^{1Q}_{vwx} ((\pm u_1)^-,v) - \log{S^{1Q}_{vwx}} (\pm u_1,v) \, .\label{eq:hybrid}
\end{align}
\noindent Note that here we present the hybrid form of the TBA equations for $Q$-particles which is particularly easy to check in the asymptotic limit. Of course $K^{0,Q}_{vwx}=0$ and $Y_{0|vw}=0$ and hence the $\log S ((\pm \mathbf{r}_{0})^-,v)$ terms are not present. The principal value prescriptions are required due to the pole of $S(v)$ at $v=-i/g$.

As discussed in section \ref{sec:psu224symmetry} the length parameter that enters the TBA equations is that of the `longest' state in the super-conformal multiplet of our state under consideration, resulting in $J+2=4$ for the Konishi state in the untwisted theory. Because our orbifold breaks all super-symmetry, different length states split over different multiplets and the relevant length is just $J=2$. This does mean however that the form of the equations changes discontinuously as the twist is removed and super-symmetry is restored. From a different point of view this is perhaps not too surprising since only for strictly zero twist the additional roots of $Y_{M|vw}$ disappear, meaning that the analytic structure is qualitatively different. Of course the full equations and the Y-functions solving them do behave continuously as discussed at the end of appendix \ref{app:criticality}.

On the asymptotic solution, for $Q=1$ eqn. \eqref{eq:hybrid} becomes
\begin{align}
\label{eq:hybridQ=1}
\log{T_{1,1}^2} = & 2 \log (1 + Y_{1|vw}) \star s \hstar K_{y1}\\
&  - 2  \log\frac{1-Y_-}{1-Y_+} \hstar s \star K^{11}_{vwx} +  \log \frac{1- \frac{1}{Y_-} }{ 1-\frac{1}{Y_+} } \hstar K_{1}  +  \log \big(1-\frac{1}{Y_-}\big)\big( 1 - \frac{1}{Y_+} \big) \hstar K_{y1}  \nonumber\\
& -2 \log{S}\hstar K_{y1} ((\pm \mathbf{r}_1)^-,v) + 2 \log{S}\star_{p.v} K^{11}_{vwx} ((\pm u_1)^-,v) - \log{S^{11}_{vwx}} (\pm u_1,v) \, ,\nonumber
\end{align}
which can be readily checked numerically.

\subsubsection{The exact Bethe equation}

If we go beyond the asymptotic regime the rapidity of a particle is determined from the exact Bethe equations $Y_1(u_{*i}) = -1$ rather than the Bethe-Yang equations. The exact Bethe equations are but an analytic continuation away from the TBA equation for a $Q=1$ particle, and by continuing the hybrid equation for $Q=1$ to the string region we immediately get
\begin{align}
\pi i(2n+1) = &  i J\, p_k - \log \left(1+Y_{Q} \right) \star (K_{\sl(2)}^{Q1_*} - 2 \, s \star K^{Q-1,1_*}_{vwx} ) - \, \log S_{\sl(2)}^{1_*1_*}(\pm u_1,u_k)\nonumber\\
&+ 2 \log (1 + Y_{1|vw}) \star ( s \hstar
K_{y1_*} + \ts) - 2  \log\frac{1-Y_-}{ 1-Y_+} \hstar s
\star_{p.v.} K^{11_*}_{vwx} \nonumber \\
& -  \log\frac{1-Y_-}{1-Y_+} \hstar s + \log \frac{1- \frac{1}{Y_-} }{ 1-
\frac{1}{Y_+} } \hstar K_{1} +  \log \big(1-
\frac{1}{Y_-}\big)\big( 1- \frac{1}{Y_+} \big) \hstar
K_{y1_*}\nonumber\\
& -2 \log{S}\hstar K_{y1_*} ((\pm \mathbf{r}_1)^-,u_k) + 2 \log{S}(\pm \mathbf{r}_1-u_k) \label{eq:ExactBethe}\\
& + 2 \, \log {\rm Res}(S)\star K^{11_*}_{vwx}((\pm u_1)^-,u_k) - 2 \sum_j \log{(u_j - u_k - \tfrac{2i}{g})\,\frac{x_j^- -\tfrac{1}{x_k^-}}{x_j^- - x_k^+}} \, .\nonumber
\end{align}
Here we use the notation
\begin{align}
\log {\rm Res}(S)\star K^{11_*}_{vwx} (u^-,v) =&\, \int_{-\infty}^{+\infty}{\rm d}t\,\log\Big[S(u^- -t)(t-u)\Big] K_{vwx}^{11*}(t,v)\,,~~~\nonumber\\
\ts(u)=&\,s(u^-)\nonumber\,,
\end{align}
we recall that the momentum of the magnon is $p = i \tH_{Q}(z_{*})=-i\log\frac{x_s(u+\frac{i}{g})}{ x_s(u-\frac{i}{ g})}$, and note that the second argument in all kernels in (\ref{eq:ExactBethe}) is the Bethe root $u_k$.

Below we will consider the leading order correction to the particle rapidity due to finite-size effects, which means we will be expanding the above equation about the asymptotic solution. In other words we will be considering $\delta \mathcal{R}_k$ with $\mathcal{R}_k$ given by
\begin{align}
\mathcal{R}_k \equiv &  \, 2 \log (1 + Y_{1|vw}) \star ( s \hstar
K_{y1_*} + \ts) - 2  \log\frac{1-Y_-}{1-Y_+} \hstar s
\star_{p.v.} K^{11_*}_{vwx} \nonumber \\
& -  \log\frac{1-Y_-}{
1-Y_+} \hstar s + \log \frac{1- \frac{1}{Y_-}}{ 1-
\frac{1}{Y_+} } \hstar K_{1} + \log \big(1-
\frac{1}{Y_-}\big)\big( 1- \frac{1}{Y_+} \big) \hstar
K_{y1_*}\nonumber\\
& -2 \log{S}\hstar K_{y1_*} ((\pm \mathbf{r}_1)^-,u_k) + 2 \log{S}(\pm \mathbf{r}_1-u_k)\label{eq:Rk} \\
& + 2 \, \log {\rm Res}(S)\star K^{11_*}_{vwx}((\pm u_1)^-,u_k) - 2 \sum_j \log{(u_j - u_k - \tfrac{2i}{g})\,\frac{x_j^- -\tfrac{1}{x_k^-}}{x_j^- - x_k^+}} = 0 \, , \nonumber
\end{align}
\noindent which are the terms in the exact Bethe equation that do not cancel by default when evaluated on the asymptotic solution.

\subsection{Wrapping corrections}

Now that we have the appropriate set of TBA equations we can compute the leading and next-to-leading order (NLO) correction to the energy of the orbifolded Konishi state through both the TBA and generalized L\"uscher approach. At NLO the correction based on generalized L\"uscher formulae does not manifestly agree with the correction coming from the TBA equations, but we will see that both give the same result.

\subsubsection{Leading order}
\label{subsec:LO}

The $Y_Q$-functions are asymptotically given by the generalized L\"uscher's formula \eqref{eq:BJforYQ} which we need to evaluate to lowest order in $g$ to compute the leading order wrapping correction. Let us write this lowest order Y-function by $Y^{\circ}_Q$ as well, leaving the $g$-expansion implicit. Expanding the Y-functions we see that the wrapping correction starts at the two loop level. More precisely, the leading order wrapping correction to the energy ($E_{LO}$) is given by
\begin{align}
E_{LO} = -\frac{1}{2\pi}\sum_{Q=1}^{\infty}\int dv \frac{d\tilde{p}}{dv} Y^{\circ}_{Q}(v).
\end{align}
where the asymptotic Y-function is given by\footnote{Note again that at weak coupling we need to rescale the rapidities by $g$ to keep them finite.}
\begin{align*}
Y^{\circ}_Q(v) = \frac{256}{81}g^4\sin^4\frac{\vartheta}{2}\frac{Q^2}{(Q^2+v^2)^2}\frac{(3(v^2-\frac{1}{3})-Q^2+1)^2}{
f^+_+f^-_+f^+_-f^-_-}, \quad
f^{\pm}_{\mp}=(Q\pm 1)^2+(v \mp \tfrac{1}{\sqrt{3}})^2 \, .
\end{align*}
Integrating and summing these we find the following wrapping correction for orbifolded Konishi
\begin{align}
E_{LO} = -\frac{g^4}{3}\sin^4\frac{\vartheta}{2} \, .
\end{align}

\subsubsection{Next-to-leading order}

Let us begin by computing the NLO correction based on generalized L\"uscher formulae.

\subsubsection*{L\"uscher's approach}

There are two different types of terms that contribute to the NLO wrapping correction. First, we have the expansion of the asymptotic Y-function which we can straightforwardly compute from the explicit expression of the transfer matrix. The other contribution comes from the fact that the Bethe roots receive finite-size corrections, {\it i.e.}
\begin{align}
p \rightarrow p + g^{4}\delta p.
\end{align}
Consequently the asymptotic energy ${\cal E}(p)$ also gets corrected
\begin{align}
\mathcal{E}(p) &= \sqrt{1+4 g^2\sin^2\frac{p + \delta p}{2}} =
\sqrt{1+4 g^2\sin^2\frac{p}{2}} + g^{6}\sin p\, \delta p  +
\mathcal{O}(g^{8}).
\end{align}
We can compute the correction $\delta p$ from the Bethe equations via eqn. \eqref{eq:luscherPcorrection}. For our state this gives
\begin{align}\label{eq:deltaPviaPhi}
\sum_i \frac{\partial BY_k}{\partial p_i}\delta p_i = \Phi_k, \qquad k=1,2,
\end{align}
where at leading order
\begin{align}
BY_k =-\left(\frac{u_k + i}{u_k-i}\right)^2\prod _{j=1}^{2} \frac{u_k-u_j + 2 i}{u_k-u_j-2 i},
\end{align}
and the momentum and (rescaled) rapidity are related via $u = \cot\frac{p}{2}$. $\Phi_k$ is given by eqn. \eqref{eq:LuscherPhidef}, which expanded to lowest order gives
\begin{align}
\Phi_1 = -\Phi_2 =&\, \sum_{Q=1}^{\infty}\int dv\left[\frac{\sqrt{3}-3 v}{Q^2-3 v^2}-\frac{2 (v+\frac{1}{\sqrt{3}})(Q^2+(v+\frac{1}{\sqrt{3}})^2+1)}{((Q-1)^2+(v+\frac{1}{\sqrt{3}})^2)((Q+1)^2+(v+\frac{1}{\sqrt{3}})^2
)}\right]\frac{Y^{\circ}_Q}{2}\nonumber \\
  =&\, \frac{1}{2\pi}\sum_{Q=1}^{\infty} \int{ dv \, \partial_1 Y^{\circ}_{Q}(v)}\, ,
\end{align}
where in the second line the derivative is applied before setting $u_1 = -u_2$ of course. The second form of $\Phi_1$ appears to be quite universal at leading order \cite{Balog:2010vf}. We can now carefully compute the NLO energy correction to the orbifolded Konishi state which results in
\begin{align}
 E_{NLO} = -g^6\left[\frac{1}{12}\sin^4\frac{\vartheta}{2} +  \frac{3}{8}\sin^2\frac{\vartheta}{2}\right].
\end{align}
We can recognize the second term as the leading order wrapping correction in $\beta$-deformed theory \cite{Arutyunov:2010gu,Beccaria:2010kd,deLeeuw:2010ed}.

\subsubsection*{TBA approach}

To compute the NLO contribution in the TBA approach we can follow a similar path. The only difference is that the asymptotic value of the Bethe root $p_i$ receives a correction coming from the main TBA equation (\ref{eq:ExactBethe}) rather than relating it to the derivative of the S-matrix via $\Phi_k$. Let us schematically write (\ref{eq:ExactBethe}) as
\begin{align}
\pi (2n_k+1) = p_k J + \log S(u_k,-u_k) + \mathcal{R}_k.
\end{align}
Varying the above over the asymptotic solution we find
\begin{align}
0 = \delta p_k  J + \delta\mathcal{R}_k.
\end{align}
In order for this to agree with the solution of (\ref{eq:deltaPviaPhi}), $\delta\mathcal{R}_k$ should be related to $\Phi_k$ as
\begin{align}
\Phi_k = -\delta\mathcal{R}_k.
\end{align}
For the Konishi operator this was shown to be the case by relating the linearized problem of computing $\delta{R}_k$ to an analogous problem in the (finite-size) TBA for the XXX-spin chain \cite{Balog:2010xa,Balog:2010vf}, and we can use a similar procedure here.

The crucial observation in the procedure of \cite{Balog:2010xa,Balog:2010vf} is that we can explicitly construct Y-functions for the XXX-spin chain that satisfy an equation which strongly resembles the TBA equation of relevance for us, just with a different source term. Due to the similarities between the TBA equations of the orbifolded model and the undeformed model, we can easily see that $\delta\mathcal{R}_k$ will satisfy the equations discussed in \cite{Balog:2010xa,Balog:2010vf}, albeit with a different source term.

To make contact with an XXX-like model we will consider the following transfer matrix
\begin{align}
t_{M}(v) = (M+1)[3(v-u_1)(v-u_2)-M(M+2)],
\end{align}
which satisfies
\begin{align}
t_{M}(v+i)t_{M}(v-i)=t_{M+1}(v)t_{M-1}(v)+t_{0}(v+(M+1)i)t_{0}(v-(M+1)i).
\end{align}
We can immediately see that $t_M(v)$ has two real roots which exactly coincide with the roots $r$ of $Y_{M|vw}$ if we take $u_1=-u_2=\frac{1}{\sqrt{3}}$. Furthermore we can easily check that $Y_{M|vw}$ expanded to lowest order can be expressed via $t_{M}(v)$. Finally we also find that $Y^{\circ}_{M+1}(v)$ can be written in terms of $t_M(v)$ in the following way
\begin{align}
&Y^{\circ}_{M}(v) = \left.\frac{g^4\sin^4(\vartheta/2)}{(v^2+Q^2)^2}\frac{t_{M-1}(v)^2}{t^+_+ t^+_- t^-_+ t^-_-}\right|_{u_1=-u_2=\frac{1}{\sqrt{3}}}, && t^\pm_\mp = t_{0}(v\pm i(M\mp1)).
\end{align}
This puts us into exactly the same position as \cite{Balog:2010xa,Balog:2010vf}, and directly applying their results we get
\begin{align}
\delta\mathcal{R}_k = \frac{1}{2\pi}\sum_{Q=1}^{\infty} \int{ dv \, \partial_k Y^{\circ}_{Q}(v)}\, ,
\end{align}
in manifest agreement with $\Phi$, proving that both ways of computing the NLO contribution are in perfect agreement.

\section{Summary and outlook}

The upshot of this chapter is that twisted theories can be treated in a fashion entirely analogous to the original theory, just at the expense of possibly increasing the workload by adding one or more parameters to the game.\footnote{It would be interesting to see the detailed implementation of these deformations in the NLIE or quantum spectral curve approaches mentioned at the end of chapter \ref{chapter:AdS5string}, see \cite{Gromov:2013qga} for work in this direction.} Still, as we discussed in this and the previous chapter, there are a number of open problems relating to the physics of our deformed theories.

As discussed at the start of this chapter, it is definitely worthwhile to further investigate particular $\beta$-deformed states that have different anomalous dimensions in the $\mbox{U}(N)$ and $\mbox{SU}(N)$ gauge theory, (thereby) gain further insight into the divergent wrapping correction to $\mbox{Tr}(XZ)$, and get a concrete indication whether our string theory\footnote{Again, as it stands, described by our integrable model with twisted boundary conditions.} really describes the $\mbox{SU}(N)$ theory. Furthermore, coming back to the issues raised in the previous chapter, it would be very interesting to further investigate the (effects of) non-conformality of $\gamma$-deformed SYM from our `dual' perspective. In similar spirit it would be nice to try and obtain exact results that may give us insight into the presence and effects of tachyons in our string theory for non-supersymmetric orbifolds.\footnote{Perhaps the most elegant case to explore here would be the $\mathbb{Z}_2$ orbifold mentioned in section \ref{subsec:orbKonTBA} giving type 0 strings on $\ads$ \cite{Klebanov:1999ch,Nekrasov:1999mn}, in the spirit of \cite{Klebanov:1999um}.}

So far our discussion of the TBA equations has been limited to the ground state and simple excited states with real momenta (in the $\mathfrak{sl}(2)$ sector). Generically we do not expect conceptual changes when leaving the $\mathfrak{sl}(2)$ sector (see e.g. \cite{Sfondrini:2011rr}), however there is one drastically different class of states, those with complex momenta. In the next chapter we will consider such an excited state in the $\mathfrak{su}(2)$ sector of the undeformed theory.

\chapter{Bound states in finite volume}
\label{chapter:boundstates}

The excited states that we have thus far described in the TBA approach are all in the $\mathfrak{sl}(2)$ sector. In this chapter we turn our attention to the $\su(2)$ sector where particles can have complex momenta. As we discussed in section \ref{sec:AdS5stringtheory} (see also section \ref{sec:AdS5mirrortheory}) this means that we can have excited states that approach bound states in the large $J$ limit. We can expect such states to have dramatically different properties from the TBA point of view, which is exactly what we would like to investigate. In this chapter we will see that the analytic properties of such states are quite involved but that nonetheless the contour deformation trick nicely applies, at least for the cases we consider. Because of the technical nature of the results we will begin with a general description of our problem and approach, and summarize the most interesting conceptual features of our results.

\section{Overview}

We will consider the simplest three-particle ($J_2=3$) state in the $\su(2)$ sector that involves complex momenta -
a configuration where the first particle has real (positive) momentum and the other two have complex conjugate momenta, such that the level-matching condition is satisfied.  Indeed, attempting to find solutions to the BY equations in the limit $g\to 0$ we find that such configurations exist; the first one shows up for $J=4$, that is for $L=J+J_2=7$.\footnote{For $L=6$ there is a singular solution composed of a particle with momentum $\pi$ and a two-particle bound state with momentum $-\pi$. Using the general approach developed in this chapter this state can be described by carefully treating it as a limiting case of a regular state in the $\beta$-deformed theory \cite{Arutyunov:2012tx}.}
This solution shows several remarkable related features, namely
\begin{itemize}
\item In the limit $g\to 0$, the complex rapidities $u_2$ and $u_3$ of the second and third particle respectively, lie {\it outside} the analyticity strip, which is the strip bounded by two lines running parallel to the real axis at $\frac{i}{g}$ and $-\frac{i}{g}$.
\item As we increase $g$ $u_2$ and $u_3$ move towards the analyticity strip, more precisely
to the points $-2-\frac{i}{g}$ and $-2+\frac{i}{g}$. If we increase $g$ further the Bethe-Yang equations break down, as the energy of the corresponding configuration becomes complex. This happens before
$u_2$ and $u_3$ reach the boundaries of the analyticity strip.
\item The first three $Y_Q$-functions, $Y_1$, $Y_2$ and $Y_3$, computed on the asymptotic solution, have poles located {\it inside} the analyticity strip. The poles of $Y_2$ are closest to the real line at $u_2^+$ and $u_3^-$.
\end{itemize}
Regarding the first point we should note that we made a wide numerical search for solutions of the one-loop BY equations for three-particle configurations of the type just described, and could not find any solution with $u_2$ and $u_3$ falling inside the analyticity strip. There are however many three-particle solutions with complex rapidities in any of the ($k$th) strips $(k-1)/g < |{\rm Im}(u)| < k/g$, $k=2,3,\ldots$. It is possible to find states with complex Bethe roots within the analyticity strip for four particles. Coming to the second point we expect that while the asymptotic roots $u_2$ and $u_3$ move towards the boundaries of the analyticity strip they cannot cross them because the S-matrix entering the BY equations develops a singularity as $u_3-u_2\to\frac{2i}{g}$. Also, the breakdown of the BY equations should simply be a reflection of their asymptotic nature in comparison to the exact TBA equations. Nevertheless, the exact Bethe equations must coincide with the asymptotic Bethe Ansatz up to the first order of wrapping in the weak coupling expansion, which for an operator of length $L$ from the $\su(2)$ sector means up to order $g^{2L}$. On the third point we would like to emphasize that the fact that the $Y_Q$ can have poles inside the analyticity strip is a new phenomenon compared to the analytic structure of states from the $\sl(2)$ sector which will have important implications when we construct the corresponding TBA equations. We should also mention that as $g$ increases the poles of $Y_2$ move towards the real line; nevertheless  for $g$ sufficiently small $Y_2$ remains small in the vicinity of the real line, {\it i.e.} for these values of $g$ we can trust the asymptotic solution.

The main observation by which we can construct consistent TBA equations is the following. If a $Y_Q$-function has a pole at a point $u_{\infty}$ inside the analyticity strip, then as we see below it must be equal to $-1$ at a point $u_{-1}$ which is located close to the pole. Of course both $u_{\infty}$
and $u_{-1}$ can and will in general depend on $Q$. In the limit $g\to 0$ we can estimate their difference from the asymptotic expression for $Y_1$ as
$$
\delta u=u_{\infty}-u_{-1}\sim g^{2L}\, .
$$
Indeed, the roots will start to differ from each other precisely at $L$-loop order! As we will explain, this guarantees that in the weak coupling expansion the asymptotic Bethe Ansatz agrees with the TBA up to $g^{2L}$.\footnote{Interestingly, a similar analytic structure to the one we encounter here is realized in  the relativistic ${\rm SU}(N)$ principal model for states describing fundamental particles with complex momenta \cite{Kazakov:2010kf,BalogUnpublished}. Also, the fact that these roots and poles lie close to each other may give insight into the `pairing' of singularities observed in \cite{Cavaglia:2011kd} for the ground state Y-functions.}

By understanding the analytic structure of the exact solution in this way, we can find the TBA equations by the contour deformation trick. In what follows we will begin with the canonical TBA equations because there the choice of integration contours can be made most transparent. In particular, in this case the poles of the auxiliary Y-functions play no role, {\it i.e.} we only need to take into account the contributions of their zeroes. Most importantly, we will see that the contours have to enclose all real zeroes of $1+Y_Q$ which are in the string region, and all zeroes and poles related to the complex Bethe roots which are below the real line of the mirror region.\footnote{This means for instance that our contours never enclose the Bethe root $u_3$ which is in the intersection of the string and anti-mirror regions.} We will use the canonical equations to derive the corresponding simplified and hybrid equations.

The driving terms in the resulting TBA equations have quite an intricate structure. As we will see they appear to depend on $u_{2,3}^{(1,2)} $ which are related to singularities of $Y_1$ and $Y_2$, the real root $u_1$, and additional roots $r^M$ related to the auxiliary functions $Y_-$ and $Y_{M|w}$. The exact values of these roots are of course fixed by the corresponding exact Bethe equations. Interestingly, for the state we consider, several apparently different quantization conditions for the Bethe roots arise. For instance for $u_3^{(1)}$ we find
$$
Y_1(u_3^{(1)})=-1~~ \Leftrightarrow ~~Y_1(u_3^{(1)--}) =-1  ~~\Leftrightarrow ~~Y_{1*}(u_3^{(1)})=-1\, .
$$
The first two conditions follow from our assumptions on the analytic structure while the last one involves the analytic continuation of $Y_1$ to the string region ($Y_{1*}$) and is the quantization condition we expect as a finite-size analogue of the BY equation. As we will show, the exact Bethe equations representing these quantization conditions are compatible in a rather non-trivial manner, in particular involving crossing symmetry. This is a strong consistency check of our construction. There are similar quantization conditions involving $Y_2$. For instance, the location of $u_3^{(2)}$ is determined by the
following compatible exact Bethe equations
$$
Y_2(u_3^{(2)-})=-1~~~ \Leftrightarrow ~~~Y_2(u_3^{(2)---})=-1\, .
$$

Next, it turns out that interestingly enough the energy formula is explicitly modified as a consequence of the fact that $1+Y_1$ and $1+Y_2$ have zeroes and poles in the analyticity strip. Combining our choice of integration contours with this we naturally obtain the following energy formula
\begin{align}
E=& \sum_{i=1}^3\mathcal{E}(u_i^{(1)})
-\frac{1}{ 2\pi}\sum_{Q=1}^{\infty}\int_{-\infty}^\infty\, du \frac{d\tilde{p}_Q}{ du}\log(1+Y_Q)
\nonumber\\
&\quad -i\tilde{p}_2(u_2^{(1)+})+i\tilde{p}_2(u_2^{(2)+})
-i\tilde{p}_2(u_3^{(2)-})+i\tilde{p}_2(u_3^{(1)-})\,,
\nonumber
\end{align}
where $u_1^{(1)}\equiv u_1$ and $\mathcal{E}(u)$ is the dispersion relation of a fundamental particle with rapidity variable $u$, while $\tilde{p}_Q$ is the momentum of a mirror $Q$-particle. This energy formula is exact and can be used to compute corrections to the Bethe ansatz energy in the limit $g\to 0$ and $J$ finite, and in the limit $J\to \infty$ and $g$ finite. In the first limit we get the leading wrapping correction given by
\begin{align}
\Delta E^{\rm wrap}=&
-\frac{1}{2\pi}\sum_{Q=1}^{\infty}\int_{-\infty}^\infty\, du\frac{d\tilde{p}_Q}{ du} \, Y_Q
\nonumber
\\
& -i\Big[\, {\rm Res}\Big(\frac{d\tilde{p}_2}{du}(u_2^+)Y_2(u_2^+)\Big)
- {\rm Res}\Big(\frac{d\tilde{p}_2}{du}(u_3^-)Y_2(u_3^-)\Big)\Big]\, .
\nonumber
\end{align}
Note that the last line in the above formula is nothing else but the residue of the integrand for $Y_2$, the function which in comparison to the other $Y_Q$-functions has poles closest to the real line.  The residue terms are of the same order as the integral term.

In the second limit the energy corrections are expected to be exponentially small in $J$ which for simple models or states are given by the generalized L\"uscher formulae discussed in section \ref{sec:luscherandexcitedstatesgeneral}. In particular, in this limit the $Y_Q$-functions are exponentially small and the integral term takes the same form as in the expression for $\Delta E^{\rm wrap}$. This term is usually interpreted as the so-called F-term. However, in our case the situation is more complicated because in the limit $J\to \infty$ the function $Y_2$ develops a double pole on the real line so that we cannot replace $\log(1+Y_2)$ by $Y_2$. Therefore, the large $J$-correction coming from the integral term is not given by the F-term. To our knowledge the $\tilde{p}$-dependent terms in the expression for $E$ are new and as far as we can see cannot be interpreted as L\"uscher's $\mu$-terms. It would be interesting to find the large $J$ expansion of the energy formula.

Finally, to check universality of and generalize our approach we should study other three-particle states. We will not present the details here,\footnote{The interested reader can find the details in appendix A.4 of \cite{Arutyunov:2011mk}.} but investigating an $L=40$ state with complex rapidities $u_2$ and $u_3$ falling inside the third strip reveals the following. Firstly, the analytic structure of the asymptotic and exact Y-functions is very similar to that of the $L=7$ state with the exception that now the first four $Y_Q$-functions have poles inside the analyticity strip; $Y_1$ and $Y_3$ have poles closest to the real line. Moreover, the canonical TBA equations for this state can be found by picking up the same contours as before. This time the driving terms depend on $u^{(2,3)}_{2,3}$ which are related to singularities of $Y_2$ and $Y_3$ rather than $Y_1$ and $Y_2$, and the rapidities $u_{2,3}^{(2,3)}$ are again found from the corresponding exact Bethe equations. The surprising conclusion of these facts is that the `standard' Bethe equations $Y_1(u_{2,3}^{(1)})=-1$ do not play any role in the description of this state, because the TBA equations do not explicitly involve these roots at all!

With two examples at hand a generalization of our construction to a three-particle state with $u_2$ and $u_3$ lying in the $k$th strip seems straightforward. Four functions $Y_{k-2},\ldots, Y_{k+1}$
will have poles in the analyticity strip, with the poles of $Y_{k-2}$ and $Y_k$ lying closest to the real line. The driving terms in the corresponding TBA equations will depend on $u_{2,3}^{(k-1)}$ and $u_{2,3}^{(k)}$ which are fixed by the exact Bethe equations for $Y_{k-1}$ and $Y_k$. The energy formula is then given by
\begin{align}
E=&\
\sum_{i=1}^3\mathcal{E}(u_i^{(1)})
-\frac{1}{ 2\pi}\sum_{Q=1}^{\infty}\int_{-\infty}^\infty\, du \frac{d\tilde{p}_Q}{ du}\log(1+Y_Q)
\nonumber\\
&\quad\quad  -i\tilde{p}_k\Big(u_2^{(k-1)}+{\textstyle{(k-1)}}\tfrac{i}{g}\Big)
+i\tilde{p}_k\Big(u_2^{(k)}+(k-1)\tfrac{i}{g}\Big) \nonumber \\
&\hskip 2.5cm
-i\tilde{p}_k\Big(u_3^{(k)}-(k-1)\tfrac{i}{g}\Big)+i\tilde{p}_k\Big(u_3^{(k-1)}-(k-1)\tfrac{i}{g}\Big)\, .
\nonumber
\end{align}
This completes our discussion of the TBA approach for three-particle states with complex momenta. Briefly coming back to the four particle states mentioned above, it appears that states with rapidities inside the physical strip are clearly distinguished from the type of states we are considering here. Their TBA equations can be constructed essentially as for states with real momenta. Further details as well as the TBA description of a concrete state of this type can be found in \cite{Arutyunov:2011mk}.

In the following sections we derive the above results in some detail. We begin in section \ref{sec:su2states}  with the asymptotic description of three-particle states in the $\su(2)$ sector and discuss the relevant analytic properties of the asymptotic and exact solution for our main state of interest in section \ref{sec:L7}. Section \ref{sec:canL7} is then devoted to the derivation of the energy and momentum formulae and the canonical TBA equations via the contour deformation trick. In section \ref{sec:simpTBA} we cast the canonical equations into simplified and hybrid forms. Then in section \ref{sec:EBandQ} we give the exact Bethe equations and verify various (nontrivial) consistency conditions. There we also discuss the relation of the exact Bethe equations to the asymptotic Bethe Ansatz. We finish with a summary where we indicate some interesting questions and discuss a potential fate of three-particle bound states when $g$ becomes large.

\section{Three-particle states in the \texorpdfstring{$\su(2)$}{} sector}
\label{sec:su2states}

In this chapter we are considering three-particle  $\ads$ superstring excited states with vanishing total momentum which carry two $\su(4)$ charges $J_1=J$ and $J_2=3$. They are dual to operators of length $L=J+3$ from the $\su(2)$ sector of ${\cal N}=4$ SYM. Such states can be composed of either three fundamental particles carrying real momenta or of one particle with a real momentum and two particles with complex momenta which are conjugate to each other at any $L$ for small enough values of the coupling constant $g$. The TBA and exact Bethe equations for states with real momenta are similar to the ones for the $\sl(2)$ states, and in this paper we will discuss only states with complex momenta.

We denote the real momentum of the fundamental particle as $p_1\equiv p$ and assume that it is positive. Then, the complex momenta of two other particles are $p_2=-\frac{p}{2}+iq$ and $p_3=-\frac{p}{2}-iq$, where the parameter $q$ has a positive real part Re$(q)>0$.
It is worth mentioning that for infinite $L$ such a state is a scattering state of a fundamental particle and a two-particle bound state, and that $q$ becomes complex for $g$ exceeding a special value depending on $p$. For these values of $g$ and $p$ the exponentially suppressed corrections to the energy of the string state computed by using the BY equations are complex as well, indicating a breakdown of the BY equations \cite{Arutyunov:2007tc}.

The two independent BY equations in the $\su(2)$-sector \cite{Arutyunov:2004vx} for the state under consideration can be written in the form
\begin{equation}
\begin{aligned}
e^{ip_1 L}\frac{u_1-u_2-\frac{2i}{g}}{u_1-u_2+\frac{2i}{ g}}\,\frac{u_1-u_3-\frac{2i}{ g}}{u_1-u_3+\frac{2i}{ g}}\,\frac{1}{ \sigma(p_1,p_2)^2\sigma(p_1,p_3)^2}=1
\, ,\label{BYE}\\
e^{ip_2 L}\frac{u_2-u_1-\frac{2i}{g}}{u_2-u_1+\frac{2i}{ g}}\,\frac{u_2-u_3-\frac{2i}{ g}}{u_2-u_3+\frac{2i}{ g}}\,\frac{1}{ \sigma(p_2,p_1)^2\sigma(p_2,p_3)^2}=1
\, ,
\end{aligned}
\end{equation}
where we recall that the rapidities $u_k$ are related to the momenta $p_k$ via eqn. \eqref{eq:uofp}. Taking the logarithm  of the BY equations, we get
\begin{equation}
\label{lBYE}
\log( \, {\rm l.h.s.}_1\, )=2\pi i\, n_1\,,\quad \log(\, {\rm l.h.s.}_2\, )=-2\pi i\, n_2\,,
\end{equation}
where $n_1$ and $n_2$ are positive integers because  $p_1$ is positive. Due to the level matching condition they should satisfy the relation $n_2\equiv n=2n_1$. As shown in \cite{Arutyunov:2004vx}, at large values of $g$ the integer $n$ is equal to the string level of the state.

Analyzing solutions of the BY equations, we find that for small values of $g$ there is {\it no} solution with complex roots $u_2$ and $u_3$ lying in the analyticity strip $-1/g < \mbox{Im}(u) < 1/g$. The fact that the complex roots are outside the analyticity strip leads to dramatic changes in the analytic properties of the Y-functions in comparison to the case with real momenta.

Changing the values of $L$ and $n$, we can find solutions with complex rapidities lying in any of the strips $(k-1)/g < |{\rm Im}(u)| < k/g$, $k=2,3,\ldots$. Thus, we can characterize such states not only by $L$ and $n$ but also by the positive integer $k$ which indicates the strips the complex roots $u_2$ and $u_3$ are located in for small values of $g$. Solving the BY equations \eqref{BYE} for increasing values of $g$, we observe that for all solutions the complex roots move towards the boundaries of the analyticity strip, {\it i.e.} the lines $|{\rm Im}(u)|=1/g$. They cannot cross them however because the S-matrix has a pole if Im$(u_3)=-$Im$(u_2)=1/g$. Because of this $u_1$ becomes complex and $u_2$ and $u_3$ are repelled from the lines Im$(u)=\mp 1/g$ when the coupling constant exceeds a critical value. In addition the asymptotic energy of such a state becomes complex clearly demonstrating a breakdown of the BY equations.

In the next sections we discuss one example of the states of this type with $L=7$, $n=2$ and $k=2$ in full detail. We also explicitly investigated a state with $L=40$, $n=2$, $k=3$, the details of which can be found in \cite{Arutyunov:2011mk}. Most of our considerations can be generalized to any $L$, $n$ and $k$.

\section{The \texorpdfstring{$L=7$}{L=7 } state and Y-functions}\label{sec:L7}

We can think of an $\ads$ superstring excited state with complex roots located in the second strip $1/g < |{\rm Im}(u)| < 2/g$ as a finite-size analog of a scattering state of a fundamental particle and a two-particle bound state, because the complex roots of such a state approximately satisfy the bound state condition $u_3-u_2=2i/g$. We will only consider the simplest state of this type with $n=2$ and $L=7$ but our consideration can be applied to any state with $k=2$.\footnote{For $L=7$ we found only one such state with $n=2$ and no state with $n\ge 4$. For large values of $L$, $n$ should be increased to find solutions with $k=2$.}

\begin{figure}[t]
\begin{center}
\includegraphics[width=7cm]{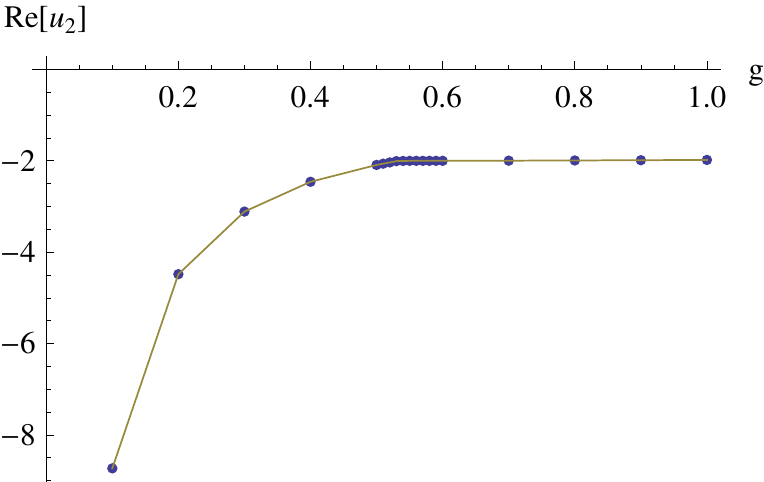}\quad \includegraphics[width=7cm]{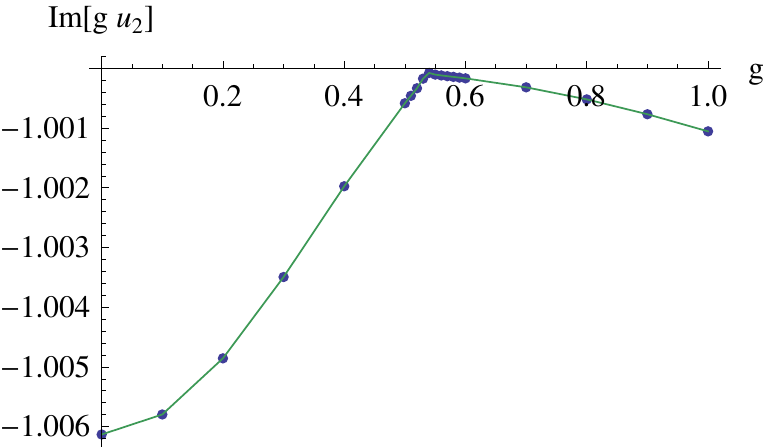}
\caption{The solution to the Bethe Yang equation for $u_2$ at $L=7$. For the imaginary part, the rapidity has been rescaled by a factor of $g$. Note that the rapidity asymptotes to $2-i/g$ before breakdown of the BY equations.}
\label{ReImu2}
\end{center}
\end{figure}

We solved the BY equations \eqref{lBYE} numerically\footnote{The equations can be solved only numerically even at $g=0$.} for $0\le g\le 0.5$ with step size $0.1$, for $0.5< g\le 0.53$ with step size $0.01$, and finally for $g= 0.5301$ and $g= 0.5302$. In table \ref{pqasym} we show the results for $p$ and $q$.
\begin{table}
 {\small
\begin{equation}\nonumber
\hspace{12pt}\begin{array}{|c||c|c||c|c|c|}
\hline
g& p&q&g& p&q\\ \hline
 0. & 2.3129&  0.926075&0.5 & 2.24919 &1.23789 \\
 0.1 & 2.3098&  0.933177&0.51 & 2.24704 & 1.27083\\
 0.2 & 2.30088& 0.955744& 0.52 & 2.2449 & 1.31517\\
 0.3 & 2.28709& 0.99838 & 0.53 & 2.24302 & 1.40691 \\
 0.4 & 2.26953 &  1.0737 &
 0.5301 & 2.24303 & 1.41083  \\  \hline
 ~~~~~~&~~~~~~~~~~~~~~~~~~~&~~~~~~~~~~~~~~~~~~~~&~0.5302 ~& ~2.2431-0.00001 i~&~1.41983-0.001 i ~\\  \hline
\end{array}~~~~~
\end{equation}
}
\caption{Numerical solution of the BY equations for the $L=7$ state.}
\label{pqasym}
\end{table}
We see from the table that $p$ and $q$ become complex at $g=0.5302$, and the BY equations cannot be used anymore. In fact we can probably not even trust the BY equations at $g=0.5301$ because the momentum at this coupling is greater than its value at $g=0.53$, while it has been steadily decreasing up to $g=0.53$. To understand why the BY equations break down it is convenient to analyze the corresponding values of the $u$-plane rapidity variables $u_k$, and their rescaled values $u_k^{rescaled} = g u_k$ which are more convenient for small values of $g$. The results are shown in table \ref{uuasym}.
\begin{table}
{\small
\begin{equation}
\nonumber
\hspace{12pt}\begin{array}{|c||c|c||c|c|}
\hline
g& u_1&u_2&u_1^{rescaled}&u_2^{rescaled}\\\hline
 0. & \frac{0.439807}{ 0}&\frac{-0.865401 - 1.00613 i}{ 0} & 0.439807&-0.865401 - 1.00613 i\\
 0.1 &4.48989&-8.73211 - 10.058 i&0.448989&-0.873211 - 1.0058 i \\
 0.2 &2.37935 &-4.48112 - 5.02428 i & 0.47587& -0.896224 - 1.00486 i \\
 0.3 &1.72919 &-3.11126 - 3.34498i &0.518756 & -0.933377 - 1.00349 i \\
 0.4 & 1.43888&-2.45839 - 2.50493 i &0.575551 & -0.983356 - 1.00197i \\
 0.5 & 1.28853&-2.0896 - 2.00117i  & 0.644265& -1.0448 - 1.00058 i \\
 0.51 &1.27788 & -2.06169 - 1.96168 i&0.651717 & -1.05146 - 1.00046 i \\
 0.52 & 1.26779&-2.03478 - 1.92372i & 0.659252&  -1.05809 - 1.00033 i \\
 0.53 &1.25786 &-2.006 - 1.88712i &0.666668 & -1.06318 - 1.00017i \\
 0.5301 & 1.25772&-2.00538 - 1.88675 i &0.666719 & -1.06305 - 1.00017i \\
 0.5302 &1.26 + 0.00002 i & -2.0041 - 1.88652 i&0.67 + 0.00001i &  -1.06257 - 1.00024 i
 \\\hline
\end{array}~~~~~
\end{equation}
}
\caption{Numerical solution of the BY equation for the $L=7$ state in terms of (rescaled) rapidities. Note that at
$g=0.5302$ the rapidity $u_1$ becomes complex.}
\label{uuasym}
\end{table}
%There is something somewhere in the later part of the paragraph below that generates an error in compiling, but it is strange.
Figure \ref{ReImu2} and table  \ref{uuasym} show us that as $g$ increases $u_2$ approaches $-2-i/g$ which is a branch point of $x(u+i/g)$. It cannot however cross the cut $\mbox{Im}(u)=-1/g$ because the S-matrix has a pole if $\mbox{Im}(u_3)=-\mbox{Im}(u_2)=1/g$. As a result, as soon as $g \gtrsim 0.5301$, $u_1$ becomes complex, and $u_2$ and $u_3$ are repelled from the cuts $\mbox{Im}(u)=\mp 1/g$. Let us also mention that the asymptotic energy of the state at $g = 0.5302$ is complex which makes inapplicability of the BY equations for $g \gtrsim 0.5301$ obvious.

\begin{figure}[t]
\begin{center}
\includegraphics[width=12cm]{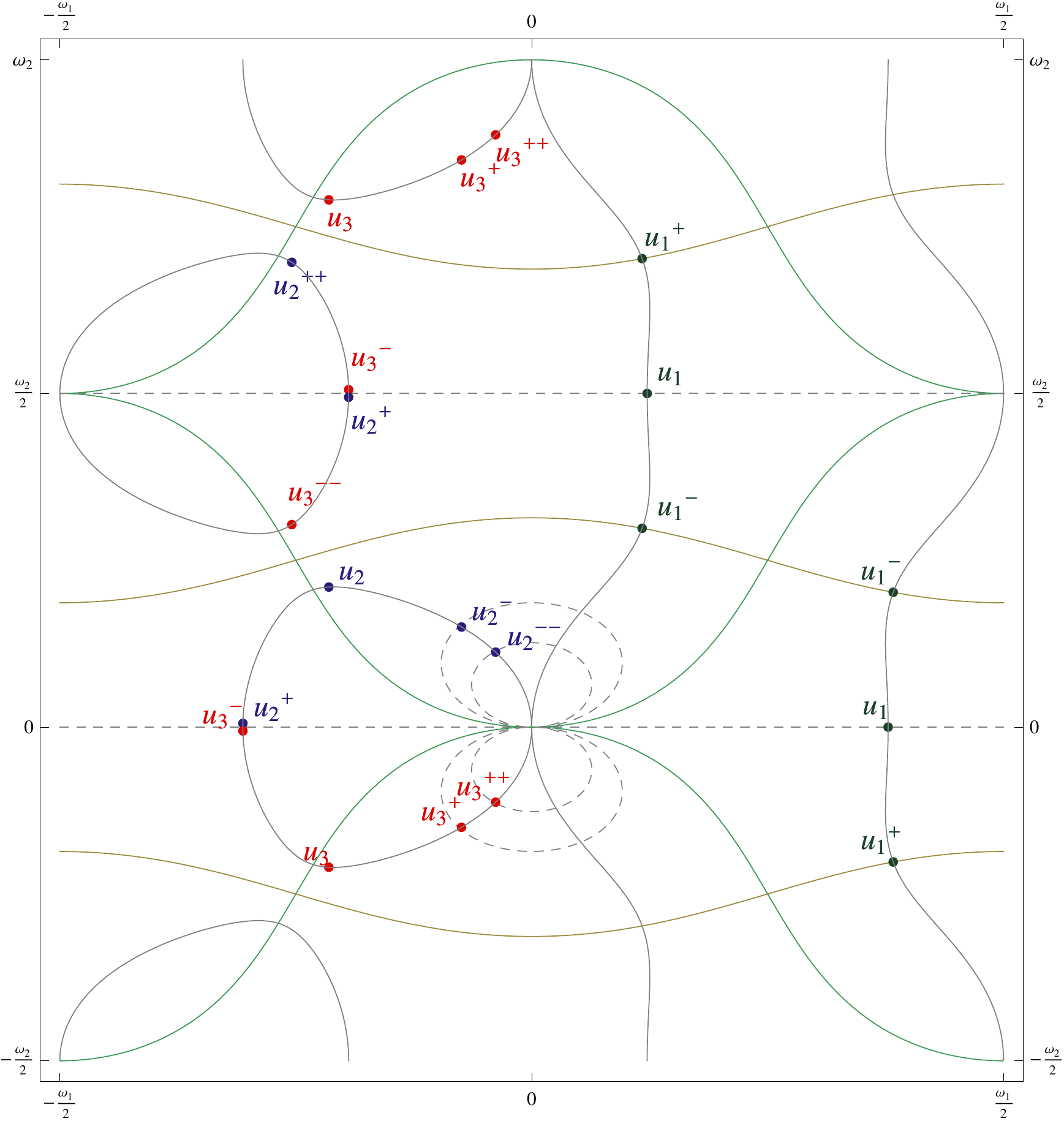}
\caption{The location of the (shifted) rapidities on the $z$-torus ($g=1/2$). The green and yellow lines outline the mirror and string regions respectively, while the gray lines are the contours $\mbox{Re}(u(z)) = \mbox{Re}(u_i)$. The curved dashed gray lines correspond to the lines at $-2 i/g$ and $-3 i/g$ in the mirror $u$-plane. The straight dashed lines are the real mirror and real string line respectively.}
\label{fig:contour}
\end{center}
\end{figure}

To apply the contour deformation trick, it is convenient to know the location of the Bethe roots on the $z$-torus, as we have indicated in figure \ref{fig:contour}. In particular we see that the root $u_2$ is in the intersection of the string and mirror regions. Note that as $g$ increases the root $u_2$ approaches the point of intersection of the boundaries of the mirror and string regions.

\subsection*{Analytic properties of the asymptotic Y-functions}

\begin{figure}[t]
\begin{center}
\includegraphics[width=11cm]{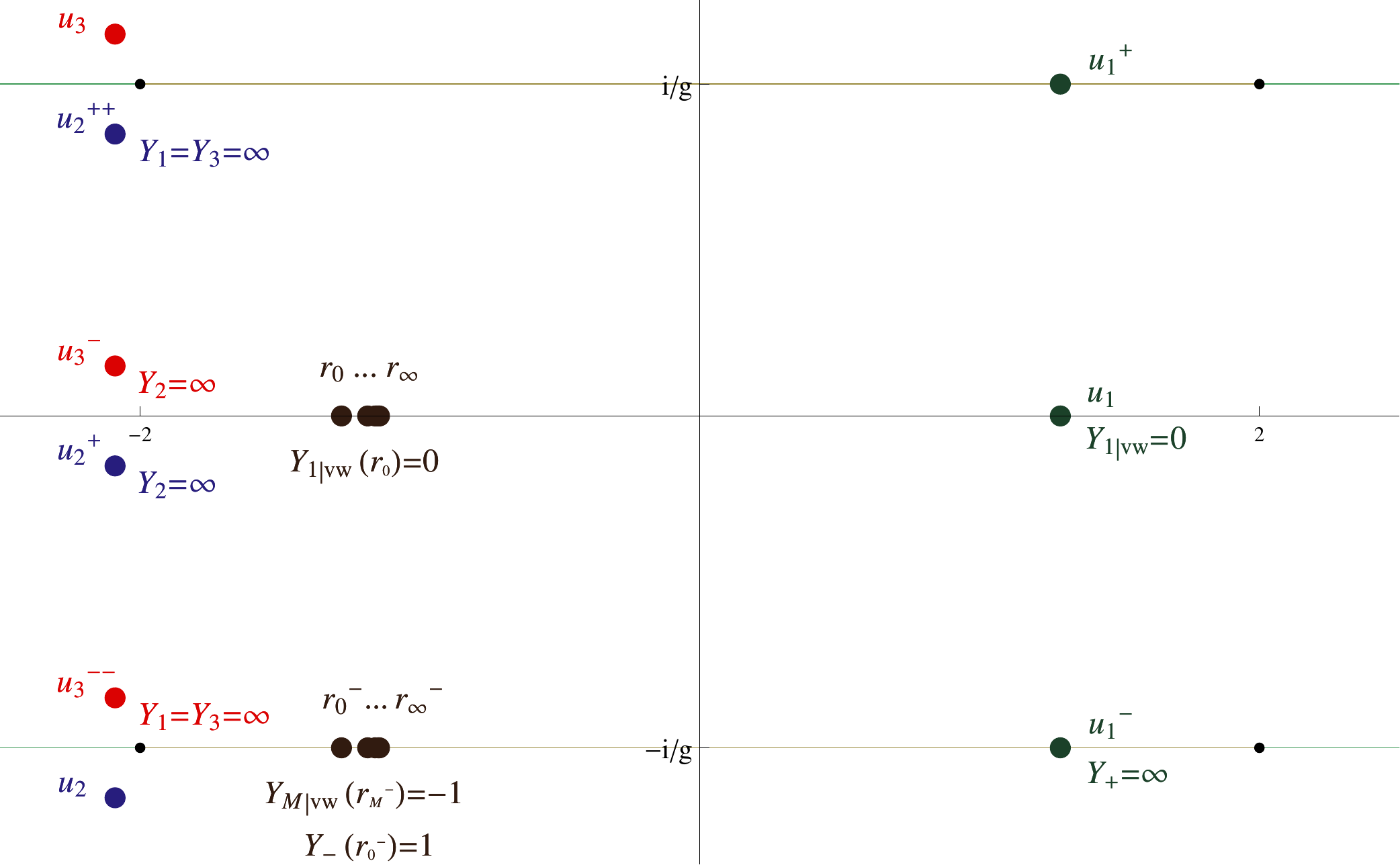}
\caption{Relevant roots and poles of the asymptotic Y-functions on the (mirror) $u$-plane.}
\label{fig:uplanepic}
\end{center}
\end{figure}

We can use the numerical solution for $u_k(g)$ to analyze the analytic properties of asymptotic Y-functions considered as functions of $g$. According to the contour deformation trick, all driving terms in the TBA equations should come from zeroes and poles of $Y$- and $1+Y$-functions. In table \ref{zepo} we only list zeroes and poles relevant for constructing the TBA equations for the state, omitting those which do not appear in the equations. For the reader's convenience we have also schematically indicated their location in the $u$-plane in figure \ref{fig:uplanepic}.
For $Q\ge 4$ the poles of $Y_Q^o$ at $u_3-\frac{i}{g}(Q-1)$ lie below the analyticity strip
and are located on the grey curves associated to the complex rapidities, in the mirror region of figure \ref{fig:contour}. They are close to the points $u_2-\frac{i}{ g}(Q-3)$ but lie on the other side of the line $-\frac{i}{g}(Q-2)$.

\begin{table}
\begin{center}
\begin{tabular}{|c|c|c|}
\hline Y${}^o$-function  & Zeroes   & Poles\\
\hline $Y_{M|w}$   & $r_{M\pm1}$ &  \\
\hline $1+Y_{M|w}$ & $r_M^-\,,\ r_M^+$ & $u_2-(M+1)i/g\,,\ u_3+(M+1)i/g$  \\
\hline $Y_{1|vw}$ &$u_1\,,\ r_0$ & \\
\hline $1+Y_{M|vw}$ && $u_2+(M+1)i/g\,,\ u_3-(M+1)i/g$ \\
\hline $Y_-$      & $u_2^-\,,\ u_3^+$ & $r_1\,,\ u_2^+\,,\ u_3^-$ \\
\hline $Y_+$       & & $r_1\,,\ u_1^-$\\
\hline $1-Y_-$   & $r_0^-\,,\ r_0^+$ & \\
\hline $1-Y_+$   &&\\
\hline $Y_1$   & $r_0$&$u_2^{++}\,,\ u_3^{--}$ \\
\hline $Y_2$   & &$u_2^{+}\,,\ u_3^{-}$ \\
\hline $Y_Q\,, Q\ge 3$   & &$u_2+\frac{i}{ g}(Q-1)\,,\ u_3-\frac{i}{ g}(Q-1)$ \\
\hline
\end{tabular}
\end{center}
\caption{Relevant roots and poles of asymptotic Y-functions within the mirror region.
}
\label{zepo}
\end{table}

For $g=1/2$  the rapidities and the first five roots take the following values
\begin{align*}
&\{ u_1,u_2,u_3\}=\{1.28853, -2.0896 - 2.00117 i, -2.0896 + 2.00117 i\}\,,\\
&\{ r_0,r_1,r_2,r_3,r_4\}=\{-1.28046, -1.18687, -1.16032, -1.14978, -1.14463\}\,.
\end{align*}
We find that all the roots $r_M$ are real and they approach a limiting value at $M\to\infty$.  Next, we note that $u_2^+$, $u_2^{++}$ and $u_3^-$, $u_3^{--}$ are within the analyticity strip $-1/g< {\rm Im}\, u<1/g$. Thus, the first three $Y_Q$-functions have poles in the analyticity strip. This is a drastically different situation compared to all previously studied states. The function $Y_2$ in particular has two complex-conjugate poles located very close to the real line.

\subsection*{Analytic properties of the exact \texorpdfstring{$Y_Q$}{}-functions}
\label{exactYQ}

In the last subsection we pointed out that the asymptotic functions $Y_1$, $Y_2$ and $Y_3$ have poles which lie within the analyticity strip. This leads to a dramatic change in the analyticity structure of the exact $Y_Q$-functions. In particular, we will show that this immediately implies that for small values of $g$ these functions must satisfy the exact Bethe equations $Y_Q(u^{(Q)})=-1$, where $u^{(Q)}$ is located close to a pole of $Y_Q$. The consideration is general and  works either for finite $J$ and small $g$ (which is the case we are interested in) or for finite $g$ and large $J$. To simplify the notations, we drop the index $Q$ and represent the Y-functions in the form
\begin{equation}
Y(u) = \frac{y(u)}{ u-u_\infty}\,,
\end{equation}
where  $y(u)$ is regular and  does not vanish at $u_\infty$ but it may have poles and zeroes elsewhere. Moreover, for any $u$ within the analyticity strip which is not its pole, $y(u)$ is of order $g^{2L-1}$  while $u_\infty$ scales as $1/g$ for small values of $g$.

We want to find $u_{-1}$ close to $u_\infty$ such that $Y(u_{-1})=-1$. We immediately get
\begin{equation}
u_{-1}-u_\infty+y(u_{-1}) = 0
\end{equation}
and expanding $y(u_{-1})$ around $u_\infty$  we obtain
\begin{equation}\label{u1u0}
u_{-1} \approx u_\infty - y(u_\infty) = u_\infty - {\rm Res}\, Y(u_\infty)\,.
\end{equation}
Note that since $y(u_\infty)$ is small, $u_{-1}$ is close to $u_\infty$.

Let us denote by $\tilde{u}_2^{(Q)}\approx u_2$ and $\tilde{u}_3^{(Q)}\approx u_3$ the points which are related to the exact locations of the poles of $Y_Q$ functions in the analyticity strip. The poles can be (and in general are) slightly shifted from their asymptotic positions for small but finite $g$. We assume that all Y-functions are real analytic in the mirror plane, that is $Y(u)^*=Y(u^*)$. Therefore $\tilde{u}_2^{(Q)}$ and $\tilde{u}_3^{(Q)}$ are complex conjugate to each other. Then from table \ref{zepo} we have
\begin{equation}
Y_1(\tu_2^{(1)++})=\infty\,,\quad Y_2(\tu_2^{(2)+})=\infty\,,\quad Y_3(\tu_2^{(3)++})=\infty\,,
\end{equation}
where for definiteness we discuss the pole locations related to $\tilde{u}_2^{(Q)}$ only.

According to the discussion above, there are complex conjugate points $u_2^{(Q)}$ and $u_3^{(Q)}$ which are close to $\tilde{u}_2^{(Q)}$ and $\tilde{u}_3^{(Q)}$ (and to the asymptotic points $u_2\,, u_3$) such that
\begin{equation}
1+Y_1(u_2^{(1)++})=0\,,\quad 1+Y_2(u_2^{(2)+})=0\,,\quad 1+Y_3(u_2^{(3)++})=0\,.
\end{equation}

We now show that the pole locations are determined by the zeroes of the functions $1+Y_Q$. To this end we recall the parametrization of section \ref{subsec:YandTundef} of the Y-functions in terms of T-functions (with an explicit $\Upsilon_Q$)
\begin{equation}
\label{YQexact}
Y_Q={\Upsilon}_Q\,\frac{T_{Q,-1}\,T_{Q,1}}{T_{Q-1,0}T_{Q+1,0}}\,,\quad
{\Upsilon}_Q(v)=e^{-J\widetilde{\cal E}_Q(v)}\,  \prod_{i=1}^{N} S_{\sl(2)}^{Q1_*}(v,u_i)\,, ~~~
\end{equation}
where in the $g\to 0$ limit the T-functions $T_{Q,\pm1}$ reduce to the asymptotic transfer matrices, $T_{0,0}=1$ and $T_{Q,0}$ reduce to one.

Represented this way, the poles of the asymptotic $Y_Q^o$ appear due to poles in ${\Upsilon}_Q$. Now as we mentioned above, the poles of the exact $Y_Q$-functions are shifted from their asymptotic positions for finite $g$. This means that the T-function $T_{Q,0}$ must have a pole at $u_2+\frac{i}{g}Q$ and a zero at $\check{u}_2+\frac{i}{g}Q$ closed to the pole. Thus $T_{Q,0}$
satisfies
\begin{equation}
\label{Tn0exact}
T_{Q,0}(u_2+\frac{i}{ g}Q)=\infty\,,\quad T_{Q,0}(\check{u}_2^{(Q)}+\frac{i}{g}Q)=0\, ,~~~~
\end{equation}
with similar properties for $u_3\,, \check{u}_3^{(Q)}$. In the following we will assume that eqs. \eqref{Tn0exact} hold for any $Q$ and that $\check{u}_2^{(Q)}\neq \check{u}_2^{(Q')}$ for any $Q\neq Q'$.

The zeroes of $T_{Q,0}$ are obviously related to poles of $Y_Q$. In addition they are also related to the zeroes of $1+Y_Q$ as follows from the second representation for $Y_Q$
\begin{equation}\label{YQexact2}
1+Y_Q =  \frac{T_{Q,0}^+\,T_{Q,0}^-}{ T_{Q-1,0}T_{Q+1,0}}\,,
\end{equation}
which follows by the Hirota equation \eqref{eq:Hirota}. Indeed we get
\begin{equation}
\cu_2^{(1)} = \tu_2^{(2)} =u_2^{(1)}\,,\quad \cu_2^{(2)} = \tu_2^{(1)} = \tu_2^{(3)} =u_2^{(2)}\,,\quad \cu_2^{(3)} =u_2^{(3)}\,,
\end{equation}
and in general $\cu_2^{(Q)} =u_2^{(Q)}$. Moreover, the conditions
\begin{equation}\label{Tn0exactb}
T_{Q,0}(u_2^{(Q)}+\frac{i}{ g}Q)=0\,,\quad T_{Q,0}(u_3^{(Q)}-\frac{i}{ g}Q)=0\,~~~~
\end{equation}
imply that in the mirror $u$-plane the function $1+Y_Q$ for $Q\ge 2$ has  zeroes at
$$
 u_2^{(Q)}+\frac{i}{ g}(Q-1)\,, \ u_2^{(Q)}+\frac{i}{ g}(Q+1)\,, \ u_3^{(Q)}-\frac{i}{ g}(Q-1)\,, \ u_3^{(Q)}-\frac{i}{ g}(Q+1)\,,
 $$
 and poles at
$$u_2^{(Q-1)}+\frac{i}{g}(Q-1)\,,\ u_2^{(Q+1)}+\frac{i}{ g}(Q+1)\,,\ u_3^{(Q-1)}-\frac{i}{ g}(Q-1)\,,\ u_3^{(Q+1)}-\frac{i}{ g}(Q+1)\,.$$
Since $1+Y_1$ has just $T_{2,0}$ in its denominator it only has poles at $u_2^{(2)++}$ and $u_3^{(2)--}$ while it has zeroes at
$u_2^{(1)++}$, $u_3^{(1)--}$, $u_2^{(1)}$ and $u_3^{(1)}$. In addition in the string $u$-plane it should have an extra zero at $u_1$ so that $Y_1$ satisfies the exact Bethe equation there. We should also mention that the $Y_Q$-functions have additional poles related to the real Bethe root $u_1$, {\it e.g.} $Y_2$ has a pole at $u_1^-$. These additional poles always lie outside integration contours however and therefore are irrelevant for constructing the TBA equations.

\subsection*{Analytic properties of the auxiliary Y-functions}

The analytic properties of exact auxiliary Y-functions are similar to those of the asymptotic ones.  Basically, all zeroes and poles which depended on $u_2$ and  $u_3$ would now depend on $u_2^{(Q)}$ and $u_3^{(Q)}$.
In fact, all information about $u_2\,,\ u_3$ goes away and all Y-functions can only have singularities related to $u_2^{(Q)}$ and $u_3^{(Q)}$ as can be seen by performing a redefinition of T-functions which removes $\Upsilon_Q$ from $Y_Q$.

Let us recall the general parametrization
\begin{equation}\label{YvT}
Y_{a,s}=\frac{T_{a,s-1}T_{a,s+1}}{ T_{a-1,s}T_{a+1,s}}\,,\quad 1+Y_{a,s} = \frac{T_{a,s}^+T_{a,s}^-}{ T_{a-1,s}T_{a+1,s}}\,,\quad 1+\frac{1}{Y_{a,s}} = \frac{T_{a,s}^+T_{a,s}^-}{ T_{a,s-1}T_{a,s+1}}\,,~~~
\end{equation}
where $Y_{a,s}$ are related to our Y-functions as
\begin{align}
&Y_{1,-1}=-\frac{1}{Y_-^{(l)}}\,, \quad Y_{1,1}=-\frac{1}{Y_-^{(r)}}\,,
\qquad Y_{2,-2}=-Y_+^{(l)}\, ,\quad Y_{2,2}=-Y_+^{(r)}\, ,\\\nonumber
&Y_{Q+1,-1}=\frac{1}{Y_{Q|vw}^{(l)}}\,,\quad
Y_{Q+1,1}=\frac{1}{Y_{Q|vw}^{(r)}}\, ,\qquad
Y_{1,-Q-1}=Y_{Q|w}^{(l)}\,,\quad
Y_{1,Q+1}=Y_{Q|w}^{(r)}\,.~~~~~
\end{align}
For states from the $\su(2)$ sector the auxiliary Y-functions from the left and right wings are equal and we can drop the superscripts ${}^{(a)}$. We want to know how the singularities of Y-functions related to the complex Bethe roots $u_2$ and $u_3$ are shifted due to the presence of $T_{a,0}$-functions in \eqref{YvT}. Thus, we discuss the $s=1$ case which includes $Y_-$ and $Y_{M|vw}$; the singularities of $Y_{M|w}$ are shifted as well but they lie outside the analyticity strip and appear to be irrelevant for the construction of the TBA equations. Concretely, we have

\bigskip
 \noindent
$\bullet$\ $Y_{M|vw}=1/Y_{M+1,1}$, $\ 1+Y_{M|vw} =  \frac{T_{M+1,1}^+T_{M+1,1}^-}{ T_{M+1,0}T_{M+1,2}}$

\medskip

As we know the asymptotic $Y_{M|vw}$ function has poles at $u_2+(M+1)i/g$ and $u_3-(M+1)i/g$.  These poles disappear because $T_{M+1,0}$ has poles there. However new poles at
$u_2^{(M+1)}+(M+1)i/g$ and $u_3^{(M+1)}-(M+1)i/g$ appear because $T_{M+1,0}$  has zeroes there.

\bigskip
 \noindent
$\bullet$\ $Y_{-}=-1/Y_{1,1}$, $\
1-Y_{-} =  \frac{T_{1,1}^+T_{1,1}^-}{T_{1,0}T_{1,2}}$

\medskip

The  poles at $u_2^{+}$ and $u_3^{-}$ are shifted to
$u_2^{(1)+}$ and $u_3^{(1)-}$ because $T_{1,0}$  has zeroes there. Asymptotically $Y_-$ has zeroes at $u_2^-$ and $u_3^+$. The location of these zeroes is shifted too but we do not need them to write the TBA equations for $Y_{M|w}$ because
they lie outside the analyticity strip.

\section{The canonical TBA equations}
\label{sec:canL7}

We will begin our discussion of the TBA equations for the $L=7$, $n=2$ state with their canonical form even though the simplified TBA equations for $Y_{M|w}$, $Y_{M|vw}$ and $Y_{Q}, Q\ge2$ are completely fixed by the zeroes and poles of these functions in the analyticity strip. The main reason for this choice is that in the canonical TBA equations the auxiliary functions $Y_\pm$, $Y_{M|w}$ and $Y_{M|vw}$ appear in the form $1+1/Y$ while the $Y_Q$-functions appear in the form $1+Y_Q$, and therefore the poles of the auxiliary Y-functions and the zeroes of the $Y_Q$-functions do not produce any driving terms, meaning they play no role in the choice of the integration contours. In addition, the kernels appearing in the canonical TBA equations for $Y_\pm$ and $Y_1$ functions have a simpler analytic structure than those in the simplified and hybrid TBA equations which makes the analysis clearer.

\subsection{The integration contour}

There is a choice of integration contours for $Y_Q$-functions which we believe is universal for the type of states under consideration. We suggest that  for any state the integration contours for $Y_Q$ are chosen such that they enclose all the real zeroes of $1+Y_Q$ which are in the string region, and all the zeroes  and poles related to the complex Bethe roots which are below the real line of the mirror region, see figure \ref{fig:contour}. In particular, the contours never go to the anti-mirror region of the $z$-torus. For the $L=7$, $n=2$ state this means that we take into account the poles of $Y_1$ at $u_3^{(2)--}$, of $Y_2$ at $u_2^{(1)+}$ and $u_3^{(3)}-\frac{3i}{g}$, and of $Y_Q, Q\ge3$ at $u_3^{(Q-1)}-\frac{i}{g}(Q-1)$ and $u_3^{(Q+1)}-\frac{i}{ g}(Q+1)$, and then the zeroes of $1+Y_1$ at $u_3^{(1)--}$ and $u_2^{(1)}$, of $1+Y_2$ at $u_2^{(2)+}$ and $u_3^{(2)}-\frac{3i}{g}$, and of $1+Y_Q, Q\ge3$ at $u_3^{(Q)}-\frac{i}{ g}(Q-1)$ and $u_3^{(Q)}-\frac{i}{g}(Q+1)$ in the mirror $u$-plane, and finally the zero of $1+Y_1$ at $u_1$ in the string $u$-plane. The net result of these contributions is discussed in appendix \ref{sec:contcontrib}.  Let us stress that we do {\it not} take into account the complex Bethe root $u_3$ which is in the intersection of the string  and anti-mirror regions. The choice of integration contours is not unique, and we will see that for the $L=7$, $n=2$ state we can make a simpler choice where we only take the contributions of the real zero of $1+Y_1$ in the string $u$-plane, the zeroes $u_2$ and $u_3$  of $1+Y_1$ in the mirror $u$-plane,  and all zeroes and poles of $1+Y_2$ in the analyticity strip of the mirror $u$-plane into account. With this choice the integration contours avoid all other zeroes and poles of $1+Y_Q$, even those which are inside the analyticity strip of the mirror $u$-plane.

The integration contours for all auxiliary Y-functions,  collectively denoted $Y_{\rm aux}$, run along the real line of the mirror region, lie  above the zeroes of Y-functions at real Bethe roots  and below all other real zeroes, and enclose all zeroes of $Y_{\rm aux}$ and $1+ Y_{\rm aux}$ which are inside the analyticity strip of the mirror $u$-plane (including its boundary) but below the real line.

It is worth stressing that the integration contours discussed above are for the canonical TBA equations, and they are different from the contours for the simplified equations. In particular, in the simplified TBA equations
the integration contour for $Y_+$ should enclose the points $u_k^-$ in the mirror $u$-plane for {\it real} Bethe roots $u_k$.

Let us now use the integration contours to derive the energy and momentum formulae, and the canonical TBA equations for the $L=7$, $n=2$ state. We use the kernels and S-matrices  defined in appendix \ref{app:qdefSmatricesandkernels}.

\subsection{The energy and momentum formulae}

According to the contour deformation trick the energy of an excited state is given by the formula
\begin{equation}
E=-\frac{1}{ 2\pi}\int_{C_Q}\, du\, \frac{d\tilde{p}_Q}{ du}\log(1+Y_Q)\,,
\end{equation}
where $C_{Q}$  are the integration contours for  $Y_Q$ functions.

Formula \eqref{YQKQ} can be  used to take the integration contours back to the real line of the mirror $u$-plane.
We can think of $\frac{1}{ 2\pi}\frac{d\tilde{p}_Q}{ du}$ as a kernel with
$i\tilde{p}_Q$ being identified with $\log {\cal S}_Q$ in \eqref{YQKQ}.
It satisfies the discrete Laplace equation, and therefore the energy is
 \begin{align}
E=&\
i\tilde{p}_{1_*}(u_1)+i\tilde{p}_{1}(u_2^{(1)})
-i\tilde{p}_1(u_3^{(1)})-\frac{1}{2\pi}\int_{-\infty}^\infty \, du\, \frac{d\tilde{p}_Q}{ du}\log(1+Y_Q)\nonumber\\
&\quad -i\tilde{p}_2(u_2^{(1)+})+i\tilde{p}_2(u_2^{(2)+})
-i\tilde{p}_2(u_3^{(2)-})+i\tilde{p}_2(u_3^{(1)-})
\,.
\label{Ener}
\end{align}
Taking into account that
\begin{equation}
i\tilde{p}_{1_*}(u_1, v)=\mathcal{E}(u_1)
 \,,\quad i\tilde{p}_1(u_2^{(1)})=\mathcal{E}(u_2^{(1)}) \,,\quad -i\tilde{p}_{1}(u_3^{(1)}) = \mathcal{E}(u_3^{(1)})\,,
\end{equation}
we get
 \begin{align}
E=&\
\sum_{i}\mathcal{E}(u_i^{(1)})
-\frac{1}{ 2\pi}\int_{-\infty}^\infty\, du \frac{d\tilde{p}_Q}{ du}\log(1+Y_Q)
\nonumber\\
&\quad -i\tilde{p}_2(u_2^{(1)+})+i\tilde{p}_2(u_2^{(2)+})
-i\tilde{p}_2(u_3^{(2)-})+i\tilde{p}_2(u_3^{(1)-})\,,
\label{Ener2}
\end{align}
where $u_1^{(1)}\equiv u_1$ and $\mathcal{E}(u)$ is the dispersion relation of a fundamental particle with rapidity variable $u$.

We see that the energy of the state depends only on the singularities of $Y_1$ and $Y_2$. The contributions coming from the other $Y_Q$-functions cancel out, and the rapidity dependent terms
can also be thought of as purely originating from the zeroes of $1+Y_1$ in the string region, and the zeroes and poles of $Y_2$ in the analyticity strip of the mirror $u$-plane. Let us also mention that the terms on the second line can be written as energies of two-particle bound states analytically continued to the mirror region.

Similar consideration can be applied to  the formula for the total momentum (which should vanish for our state) given by \begin{equation}
P=-\frac{1}{ 2\pi}\int_{C_Q}\, du\, \frac{d\tH_Q}{ du}\log(1+Y_Q)\,.
\end{equation}
Since also $\frac{1}{ 2\pi}\frac{d\tH_Q}{ du}$ satisfies  the discrete Laplace equation,
identifying
$i\tH_Q$ with $\log {\cal S}_Q$ in \eqref{YQKQ},  we obtain
\begin{align}
P=&
i\tH_{1_*}(u_1)+i\tH_{1}(u_2^{(1)})
-i\tH_1(u_3^{(1)})-\frac{1}{ 2\pi}\int_{-\infty}^\infty\, du\, \frac{d\tH_Q}{ du}\log(1+Y_Q)
\nonumber\\
&-i\tH_2(u_2^{(1)+})+i\tH_2(u_2^{(2)+})
-i\tH_2(u_3^{(2)-})+i\tH_2(u_3^{(1)-})\,.
\label{momen}
\end{align}
Taking into account that
 \begin{align}\nonumber
i\tH_{1_*}(u_1, v)=p(u_1)\equiv p_1\,,\quad i\tH_1(u_2^{(1)})=p(u_2^{(1)})\equiv p_2 \,,\quad -i\tH_{1}(u_3^{(1)}) = p(u_3^{(1)})\equiv p_3 \,,
\end{align}
we get the following formula for the total momentum
 \begin{align}
P=&
\sum_i\, p_i-\frac{1}{2\pi}\int_{-\infty}^\infty\, du\, \frac{d\tH_Q}{ du}\log(1+Y_Q)
\nonumber\\
&-i\tH_2(u_2^{(1)+})+i\tH_2(u_2^{(2)+})
-i\tH_2(u_3^{(2)-})+i\tH_2(u_3^{(1)-})\,.
\label{momen2}
\end{align}
It was noticed in \cite{Cavaglia:2010nm} that the TBA equations imply a quantization condition for the total momentum. Thus, since the total momentum vanishes as $g\to 0$ and it changes continuously with $g$ the total momentum should vanish for any $g$.

\subsection{The canonical TBA equations}

\subsection*{$w$-strings}

The excited state canonical TBA equations for $w$-strings are given by
\begin{align}\nonumber
\log Y_{M|w} = &\log( 1+\frac{1}{Y_{N|w}})\star_{C_{N|w}} K_{NM} + \log \frac{1-\frac{1}{Y_-}}{1-\frac{1}{Y_+}} \star_{C_{\pm}} K_M \, ,
\end{align}
where $C_{N|w}$, $C_{-}$ and $C_{+}$  are the integration contours for $Y_{N|w}$, $Y_-$ and $Y_+$ functions.
Taking the integration contours back to real line of the mirror $u$-plane, $Y_+$ does not produce any driving term, the zero of $1-Y_-$ at $r_0^-$ produces
$-\log S_{M} (r_0^--v)$, and finally the zeroes of $Y_{N|w}$ at $r_{N-1}$ and $r_{N+1}$, and the zeroes of
$1+Y_{N|w}$ at $r_{N}^-$ give
$$
+\tfrac{1}{2}\sum_{N=1}^{\infty} \log S_{NM} (r_{N-1}-v)S_{NM} (r_{N+1}-v) - \sum_{N=1}^{\infty} \log S_{NM} (r_N^--v)\,,
$$
where $+1/2$ in the first term appears due to the principal value prescription in eqn. (\ref{YwcanTBA}) below.

Taking into account that $S_{NM} (u-v)$ satisfies the discrete Laplace equation
$$
S_{N-1,M} (u-v)S_{N+1,M} (u-v)=S_{NM} (u^--v)S_{NM} (u^+-v) \,,
$$
we can write the canonical TBA equations for $w$-strings
in the form
\begin{align}
\nonumber
\log Y_{M|w} = & \log( 1+\frac{1}{Y_{N|w}})\star_{p.v.} K_{NM} + \log \frac{1-\frac{1}{Y_-}}{1-\frac{1}{Y_+}}\hstar K_M \\
\label{YwcanTBA}
& +\tfrac{1}{2}\sum_{N=1}^{\infty} \log \frac{S_{NM} (r_{N}^+-v)}{S_{NM} (r_{N}^--v)}
+ \tfrac{1}{2}\log S_{1M} (r_0-v)-\log S_{M} (r_0^--v) \, ,
\end{align}
where $\log S_{1M} (r_0-v)$ should be understood as $\log S_{M-1} (r_0-v) +\log S_{M+1} (r_0-v)$.

\subsection*{$vw$-strings}

The excited state  canonical TBA equations for $vw$-strings are given by
\begin{align}\nonumber
\log Y_{M|vw} = & \log( 1+\frac{1}{Y_{N|vw}})\star_{C_{N|vw}} K_{NM} + \log \frac{1-\frac{1}{Y_-}}{1-\frac{1}{Y_+}}\star_{C_{\pm}}  K_M- \log (1+Y_Q)\star_{C_{Q}}  K^{QM}_{xv}  \, ,
\end{align}
where $C_{N|vw}$ and $C_{Q}$  are the integration contours for $Y_{N|vw}$, and $Y_Q$ functions.
Taking the integration contours back to real line of the mirror $u$-plane and using
formula \eqref{YQKQ}, we can bring the canonical TBA equations for $vw$-strings to the form
\begin{align}
\log Y_{M|vw} &=   \log( 1+\frac{1}{Y_{N|vw}})\star_{p.v.} K_{NM} + \log \frac{1-\frac{1}{Y_-}}{1-\frac{1}{Y_+}}\hstar K_M - \log (1+Y_Q)\star K^{QM}_{xv} \nonumber \\
&+ \frac{1}{2} \log \frac{S_{1M} (r_0-v)}{S_{1M} (u_1-v)} - \log S_{M} (r_0^--v)-\log \frac{S^{1M}_{xv} (u_3^{(1)},v)}{S^{1M}_{xv} (u_2^{(1)},v)}\nonumber\\
& \quad \qquad  + \log S^{1_*M}_{xv} (u_1,v) - \log \frac{S^{2M}_{xv} (u_3^{(2)-} ,v)}{S^{2M}_{xv} (u_3^{(1)-},v)}\frac{S^{2M}_{xv} (u_2^{(1)+},v)}{S^{2M}_{xv} (u_2^{(2)+},v)} \, ,
\end{align}
where the first term on the second line appears due to the zeroes of $Y_{1|vw}$ at $u_1$ and $r_0$, the second term arises because of the zero of $1+Y_{-}$  at $r_0^-$.

\subsection*{$y$-particles}

We can use formula \eqref{YQKQ} to write the TBA equation for $Y_+/Y_-$
\begin{align}
\label{ypovym}
\log \frac{Y_+}{ Y_-} = \,   \log(1 +  Y_{Q})\star K_{Qy} -& \sum_i \log S_{1_*y}(u_i^{(1)},v)  +\log \frac{S_{2y}(u_2^{(1)+}, v)}{ S_{2y}(u_2^{(2)+}, v)}\frac{S_{2y}(u_3^{(2)-}, v)}{ S_{2y}(u_3^{(1)-}, v)}\,,
\end{align}
where  we have used that
\begin{equation}\nonumber
S_{1_*y}(u_3^{(1)},v) = 1/S_{1y}(u_3^{(1)},v)\,,\quad S_{1_*y}(u_2^{(1)},v) = S_{1y}(u_2^{(1)},v)
\,.~~~~~
\end{equation}
The driving terms can be also explained by contours which enclose only the zeroes of $1+Y_1$ in the string region, and all zeroes and poles of $Y_2$ in the analyticity strip of the mirror $u$-plane, while avoiding other zeroes and poles of $1+Y_Q$.

To write the canonical TBA equation for $Y_+Y_-$ let us analyze the general equation
\begin{align}
\label{Yfory2}
 &\log {Y_+ Y_-} =  - \log\left(1+Y_Q \right)\star_{C_{Q}}  K_Q
\\\nonumber
&+2\log \big(1+\frac{1}{Y_{M|vw}}
\big)\star_{C_{M|vw}}  K_M - 2\log \big(1+\frac{1}{Y_{M|w}}\big)\star_{C_{M|w}}  K_M\,,
\end{align}
term by term.

\noindent $\bullet$ The term $- \log(1 +  Y_{Q} )\star_{C_{Q}}  K_Q$ produces
\begin{align}
&
\log {S}_{1}(u_1- v)+\log \frac{{S}_1(u_2^{(1)}- v)}{{S}_{1}(u_3^{(1)}- v)}-\log \frac{{S}_2(u_2^{(1)+}- v)}{ {S}_2(u_2^{(2)+}-v)}\frac{{S}_2(u_3^{(2)-}- v)}{ {S}_2(u_3^{(1)-}- v)}
\end{align}

\noindent $\bullet$ The term $2\log (1+\frac{1}{ Y_{M|vw}}) \star_{C_{M|vw}}  K_M$ produces
\begin{align}
&2\log (1+\frac{1}{ Y_{M|vw}}) \star_{p.v.} K_M
-\log {S}_{1}(u_1- v)+\log {S}_{1}(r_0- v)\,,
\end{align}
where we take into account that the contour runs above $u_1$ but below $r_M$.

\noindent $\bullet$ The term $2\log (1+\frac{1}{ Y_{M|w}}) \star_{C_{M|w}}  K_M$ produces
\begin{align}
&2\log (1+\frac{1}{ Y_{M|w}}) \star_{p.v.} K_M +
\log {S}_{M}(r_{M-1}- v){S}_{M}(r_{M+1}- v) -2\log {S}_{M}(r_{M}^- - v)\,,
\end{align}
where we sum over $M$ from 1 to $\infty$. Computing the sum we get
\begin{align}
&2\log (1+\frac{1}{ Y_{M|w}}) \star_{p.v.} K_M +\log {S}_{1}(r_{0}- v)-
\sum_{M=1}^\infty\log \frac{{S}_{M}(r_{M}^-- v)}{ {S}_{M}(r_{M}^+- v) }\,.
\end{align}

Putting this together the canonical TBA equation for $Y_+Y_-$ is
\begin{align}
\label{TBAcanypym} \log {Y_+ Y_-} &=  - \log\left(1+Y_Q
\right)\star K_Q+2\log \frac{1+\frac{1}{ Y_{M|vw}}
}{ 1+\frac{1}{Y_{M|w}}}\star_{p.v.} K_M
 \\\nonumber
&+\log \frac{{S}_1(u_2^{(1)}- v)}{ {S}_{1}(u_3^{(1)}- v)}-\log \frac{{S}_2(u_2^{(1)+}- v)}{ {S}_2(u_2^{(2)+}-v)}\frac{{S}_2(u_3^{(2)-}- v)}{ {S}_2(u_3^{(1)-}- v)}
+
\sum_{M=1}^\infty\log \frac{{S}_{M}(r_{M}^-- v)}{{S}_{M}(r_{M}^+- v) }\,.
\end{align}

\subsection*{$Q$-particles}

The excited state canonical TBA equation for $Y_Q$ can be written in the form
\begin{align}\nonumber
\log Y_Q =&\, -(J+2)\, \tH_{Q} + \log\left(1+Y_{M} \right) \star_{C_M} K_{\sl(2)}^{MQ}
 + 2 \log\left(1+ \frac{1}{Y_{M|vw}} \right) \star_{C_{M|w}}  K^{MQ}_{vwx}
\\[1mm]
&\quad +  \log \frac{1- \frac{1}{Y_-}}{ 1-\frac{1}{Y_+} } \star_{C_\pm}  K_{Q}  +  \log \big(1-\frac{1}{Y_-}\big)\big( 1 - \frac{1}{Y_+} \big) \star_{C_\pm}  K_{yQ} \, .
\end{align}
Taking the integration contours back to the real line of the mirror $u$-plane and using \eqref{YQKQ}, we obtain
\begin{align}\nonumber
\log Y_Q =&\, - (J+2)\, \tH_{Q} + \log\left(1+Y_{Q'} \right) \star K_{\sl(2)}^{Q'Q}
+ 2 \log\left(1+ \frac{1}{Y_{M'|vw}} \right) \star_{p.v.} K^{M'Q}_{vwx}
\\[1mm]\label{YQcanTBA}
&\quad +  \log \frac{1- \frac{1}{Y_-}}{ 1-\frac{1}{Y_+} } \hstar K_{Q}  +  \log \big(1-\frac{1}{Y_-}\big)\big( 1 - \frac{1}{Y_+} \big) \hstar K_{yQ} \\[1mm] \nonumber
&\quad -\log S_{\sl(2)}^{1_*Q}(u_1,v)\frac{S_{\sl(2)}^{1Q}(u_2^{(1)},v)}{
S_{\sl(2)}^{1Q}(u_3^{(1)},v)}+\log \frac{S_{\sl(2)}^{2Q}(u_2^{(1)+},v)}{S_{\sl(2)}^{2Q}(u_2^{(2)+},v)}\frac{S_{\sl(2)}^{2Q}(u_3^{(2)-},v)}{S_{\sl(2)}^{2Q}(u_3^{(1)-},v)}
\\[1mm]\nonumber
&\quad-\log S^{1Q}_{vwx}(u_1,v)+\log S^{1Q}_{vwx}(r_0,v) - \log{S_Q}(r_0^--v) - \log{S_{yQ}}(r_0^-,v)\nonumber\,.
\end{align}
The driving terms dependent on $S_{\sl(2)}^{1Q}$ of the mirror-mirror region can be rewritten in terms of $S_{\sl(2)}^{1_*Q}$ of the string-mirror region by noting that $u_2^{(1)}$ lies in overlap of the string and mirror regions, meaning that $S_{\sl(2)}^{1Q}(u_2^{(1)},v)=S_{\sl(2)}^{1_*Q}(u_2^{(1)},v)$, and that
$u_3^{(1)}$ lies in overlap of anti-string and mirror regions, meaning that we can use crossing relations to replace
$S_{\sl(2)}^{1Q}(u_3^{(1)},v)$ with  $1/S_{\sl(2)}^{1_*Q}(u_3^{(1)},v)$.
\begin{equation}
\label{Cross}
S_{\sl(2)}^{1Q}(u_3^{(1)},v)S_{\sl(2)}^{1_* Q}(u_3^{(1)},v) = \frac{1}{h_Q(u_3^{(1)},v)^2},
\end{equation}
where $h_Q(u,v)$ is defined as  \cite{Arutyunov:2006iu}
\begin{equation}
h_Q(u,v) = \frac{x_s(u-\frac{i}{ g})-x(v+\frac{i}{ g}Q)}{ x_s(u-\frac{i}{ g})- x(v-\frac{i}{ g}Q) }\,  \frac{1 - \frac{1}{x_s(u+\frac{i}{ g})x(v+\frac{i}{ g}Q)}}{  1-\frac{1}{ x_s(u+\frac{i}{ g})x(v-\frac{i}{ g}Q)}}\,.
\end{equation}
We would also like to point out that
 since $u_3^{(1)}$  is in the second strip we can rewrite     $h_Q(u_3^{(1)},v)$ as
\begin{align}
\label{h2identity}
&h_Q^2(u_3^{(1)},v)=\frac{S_{yQ}(u_3^{(1)+},v)}{S_{yQ}(u_3^{(1)-},v)}S_Q(u_3^{(1)+}-v)S_Q(u_3^{(1)-}-v)\,.
\end{align}

\section{The simplified and hybrid TBA equations}\label{sec:simpTBA}

As we did for the ground state before, the canonical TBA equations can be used to derive the simplified and hybrid TBA equations. To do so we apply the operator $(K+1)^{-1}_{NM}$ to both sides of the canonical TBA equations, sum over $N$ and use identities listed in appendix \ref{identY}.

\subsection*{$w$-strings}

We find that the simplified TBA equations for $Y_{M|w}$ are
\begin{align}\label{TBAsimyw}
\log Y_{M|w} =&  \log(1 +  Y_{M-1|w})(1 +
Y_{M+1|w})\star s\\\nonumber
 &+ \delta_{M1}\, \log\frac{1-\frac{1}{ Y_-}}{ 1-\frac{1}{ Y_+} }\hstar s - \log S(r_{M-1}^- - v)S(r_{M+1}^- - v)\, ,~~~~~
\end{align}
where at the simplified level the driving terms appear due to the zero of $1-Y_{-}$ at $r_0^-$ and the zeroes of $1+Y_{M|w}$ at $r_M^-$.

\subsection*{$vw$-strings}

To find the simplified equations for $Y_{M|vw}$ we apply $(K+1)^{-1}$ and subsequently rewrite the result by using the simplified equation for $Y_+/Y_-$ convoluted with $s$. This gives us the following equations for $vw$-strings
\begin{align}
\log Y_{M|vw} = & - \log(1 +  Y_{M+1})\star s +
\log(1 +  Y_{M-1|vw} )(1 +  Y_{M+1|vw})\star
s\nonumber\\\label{TBAsimyvw}
&  + \delta_{M1} \log\frac{1-Y_-}{ 1-Y_+}\hstar s \\
& + \delta_{M1}\Big(
\log\frac{S(u_2^{(2)+} - v)}{S(u_3^{(2)-} - v)} - \log S(u_1^- - v)S(r_{0}^- - v)\Big)\,\nonumber.
\end{align}
The contour deformation trick explains the driving terms at the simplified level for $M=1$ as follows. From $1+Y_2$ we get
$$
\log S(u_2^{(2)+} - v) - \log S(u_2^{(1)+} - v) + \log S(u_3^{(2)---} - v) - \log S(u_3^{(3)---} - v)\,.~~~~
$$
Next, $1-Y_-$ contributes $
+ \log S(u_2^{(1)+} - v)- \log S(r_{0}^- - v)
$,
while $1-Y_+$ contributes
$
- \log S(u_1^- - v)
$.
Finally, $1+Y_{2|vw}$ gives
$
+\log S(u_3^{(3)---} - v)
$.
Summing this up we get the desired driving terms.

The contributions from the poles of $1+Y_{M+1}$ for higher $M$ cancel the contribution from the poles of $1+Y_{M-1|vw}$ and $1+Y_{M+1|vw}$. Note that to explain the remaining driving terms in the simplified equations we would have to take into account the poles of $Y_{M|vw}$ outside the analyticity strip.

\subsection*{$y$-particles}

The simplified TBA equation for the ratio $Y_+/Y_-$ coincides with \eqref{ypovym}, but to derive the equation for $Y_+Y_-$ we need to compute the infinite sums involving the $Y_{M|w}$ and $Y_{M|vw}$-functions which is done in appendix \ref{identY}. Using these formulae, the TBA equation \eqref{ypovym} for $Y_+/Y_-$, and the identities from appendix \ref{identY}, we can bring the TBA equation for $Y_+Y_-$ to the form
\begin{align}\nonumber
 &\log {Y_+ Y_-} =\  2\log \frac{1+Y_{1|vw}
}{ 1+Y_{1|w}}\star s- \log\left(1+Y_Q
\right)\star K_Q
+2\log(1 + Y_{Q}) \star K_{xv}^{Q1}\star s
 \\
 \nonumber
&\qquad\qquad\quad+2\log S(r_{1}^- - v)-2 \log S_{xv}^{1_*1}(u_1,v)\star s+\log {S_2(u_1- v)} \star s
 \\\label{TBAsimypym}
&\qquad\qquad\quad -2 \log \frac{S_{xv}^{11}(u_2^{(1)},v)}{ S_{xv}^{11}(u_3^{(1)},v)}\star s+\log \frac{{S}_1(u_2^{(1)}- v)}{ {S}_{1}(u_3^{(1)}- v)}
  \\\nonumber
 &\qquad\qquad\quad
-\log \frac{{S}_2(u_2^{(1)+}- v)}{ {S}_2(u_2^{(2)+}-v)}\frac{{S}_2(u_3^{(2)-}- v)}{{S}_2(u_3^{(1)-}- v)}
+2 \log \frac{S^{21}_{xv}(u_2^{(1)+}, v)S^{21}_{xv}(u_3^{(2)-}, v)}{ S^{21}_{xv}(u_2^{(2)+}, v)S^{21}_{xv}(u_3^{(1)-}, v)} \star s \,.
\end{align}
We recall that the integration contours run a bit above the real line. We can clearly explain the driving terms in this equation by our choice of integration contours. To be sure that no other driving terms appear the kernel $K_{xv}^{Q1}\star s$ and its S-matrix should be analytically continued to complex points in the mirror and string $u$-planes. This is non-trivial because $K_{xv}^{Q1}$ has poles, and we have not attempted to derive \eqref{TBAsimypym} starting with the simplified equation with deformed contours. We have however checked that $Y_-$ satisfies its Y-system equation which requires a very delicate balance of the driving terms in \eqref{TBAsimypym} and \eqref{ypovym}.

Let us finally present yet another form of the simplified TBA equation
\begin{align}\nonumber
 &\log {Y_+ Y_-} =\  2\log \frac{1+Y_{1|vw}
}{ 1+Y_{1|w}}\star s- \log\left(1+Y_Q
\right)\star K_Q
+2\log(1 + Y_{Q}) \star K_{xv}^{Q1}\star s
 \\
\label{TBAsimypym2}
&\qquad\qquad\quad+ 2\log{S(r_1^- - v)}-  \sum_i \log \frac{\big(S_{xv}^{1_*1}\big)^2}{ S_2}\star s(u_i^{(1)},v) + \log \frac{S(u_2^{(1)}-v)}{ S(u_3^{(1)}-v)}
  \\\nonumber
 &\qquad\qquad\quad
-\log \frac{{S}_2(u_2^{(1)+}- v)}{{S}_2(u_2^{(2)+}-v)}\frac{{S}_2(u_3^{(2)-}- v)}{{S}_2(u_3^{(1)-}- v)}
+2 \log \frac{S^{21}_{xv}(u_2^{(1)+}, v)S^{21}_{xv}(u_3^{(2)-}, v)}{ S^{21}_{xv}(u_2^{(2)+}, v)S^{21}_{xv}(u_3^{(1)-}, v)}\star s  \,,
\end{align}
where we used identities from appendix \ref{identY} to
replace the mirror-mirror S-matrices $S_{xv}^{11}$ with the string-mirror $S_{xv}^{1_*1}$. As before this form indicates that it might be possible to choose the integration contours for $Y_Q$ so that they would only
enclose  the zeroes of $1+Y_1$ in the string region, and all zeroes and poles of $Y_2$ in the analyticity strip of the mirror $u$-plane. Such a choice, however, would require very intricate integration contours for the auxiliary Y-functions which we will not attempt to describe.

\subsection*{$Q$-particles}

Applying $(K+1)^{-1}$ to the canonical equations, the terms which depend on the Y-functions (and involve only the kernels) produce the usual contributions. We can find the contribution of the driving terms through the identities in appendix \ref{identY}. This way we get the following simplified TBA equations for $Q$-particles

\bigskip
\noindent
$\bullet$ $\ Q\ge 4\ $
\begin{equation}
\log Y_{Q}=\log\frac{\left(1 +  \frac{1}{Y_{Q-1|vw}} \right)^2}{ (1 +  \frac{1}{ Y_{Q-1} })(1 +  \frac{1}{ Y_{Q+1} }) }\star s\label{YforQ4}
\,~~~~~~~
\end{equation}
The contributions of the zeroes of $1+Y_{Q}$ cancel each other for $Q\ge 3$, and therefore no driving term appears.

\bigskip
\noindent
$\bullet$ $\ Q=3\ $
\begin{equation}
\log Y_{3}=\log S(u_2^{(2)+}-v)-\log S(u_3^{(2)-}-v)+\log\frac{\left(1 +  \frac{1}{ Y_{2|vw}} \right)^2}{(1 +  \frac{1}{ Y_{2} })(1 + \frac{1}{Y_{4} }) }\star s\label{YforQ3}
\,.~~~~~~~
\end{equation}
The driving terms are explained by the zeroes of $1+Y_2$ at  $u_2^{(2)+}$ and $u_3^{(2)---}$. The contributions of the two zeroes of $1+Y_4$ cancel each
other.

\bigskip
 \noindent
$\bullet$ $\ Q=2\ $
\begin{equation}
\label{YforQ2}
\log Y_{2}=\log S(u_2^{(1)}-v)-\log S(u_3^{(1)}-v)+\log\frac{\left(1 +  \frac{1}{ Y_{1|vw}} \right)^2}{ (1 +  \frac{1}{ Y_{1} })(1 +  \frac{1}{ Y_{3} }) }\star_{p.v} s
\,,~~~~~~~
\end{equation}
The contribution due to the zero of $1+Y_1$ at $u_1$  in the string region is canceled because of the zero of $Y_{1|vw}$ at $u_1$. Next, both $Y_{1|vw}$ and $Y_{1}$ have zeroes at $u=r_0$ and their contributions cancel each other. The contributions of the two zeroes of $1+Y_3$ cancel each other, and we are left with the two driving terms produced by the zeroes of $1+Y_1$ at $u_2^{(1)}$ and $u_3^{(1)--}$.

\bigskip
\noindent
$\bullet$ $\ Q=1\ $
\begin{align}
\log Y_1 &=  \log \left(1-\frac{1}{Y_-}\right)^2\hstar s -  \log \left(1+\frac{1}{Y_2}\right) \star s+ \log \frac{S(u_2^{(2)+} - v)}{S(u_3^{(2)-} - v)} -2 \log S(r_0^--v)
\nonumber\\
&- \check{\Delta} _v{\cstar} s +\log \check{S_1}{\cstar} s (u_1,v) - \log \check{S_1}{\cstar} s (r_0,v)+ 2\log \check{S}{\cstar} s (r_0^-,v)  \label{STbaQ1}
\\
\nonumber
&+\log \check{\Sigma}_{1_*}^2(u_1,v)\frac{\check{\Sigma}_{1}^2 (u_2^{(1)},v)}{ \check{\Sigma}_{1}^2(u_3^{(1)},v)  }{\cstar} s
 -  \log \frac{\check{\Sigma}_2^2  (u_2^{(1)+},v) \check{\Sigma}_2^2 (u_3^{(2)-},v) }{ \check{\Sigma}_2^2  (u_3^{(1)-},v) \check{\Sigma}_2^2 (u_2^{(2)+},v) } {\cstar} s \nonumber
\end{align}
where
\begin{align} \nonumber
\check{\Delta}_v=&L\check{{\cal E}} +\log\left(1-\frac{1}{Y_-}\right)^2\left(1-\frac{1}{Y_+}\right)^2{\hstar}\check{K}+\log\left(1+\frac{1}{Y_{M|vw}}\right)^2 \star_{p.v.}\check{K}_M\\\label{delv}
&+2\log(1+Y_Q)\, \star\, \check{K}^{\Sigma}_Q\, .
\end{align}
Note that the terms $+\log \check{S_1}{\cstar} s (u_1,v) - \log \check{S_1}{\cstar} s (r_0,v)$  on the second line of \eqref{STbaQ1} combine with $\log\left(1+\frac{1}{Y_{M|vw}}\right)^2 \star_{p.v.}\check{K}_M$ and remove the principal value prescription in the integral. We can explain the driving terms in this equation by the zero of $1-Y_-$ at $r_0^-$, the zeroes of $Y_{1|vw}$ at $r_0$ and $u_1$, the zeroes and poles of $1+Y_Q$, and our choice of the integration contours. We could compute the infinite sum involving $Y_{M|vw}$-functions in \eqref{delv} as done in \cite{Arutyunov:2009ux}, which would produce additional driving terms.

Using the identities in appendix  \ref{identY}, we can rewrite the TBA equation for $Y_1$ so that they contain $\check{\Sigma}_{1_*}$-terms similar to the ones for the Konishi state, namely
\begin{align}
\log Y_1 &=  \log \left(1-\frac{1}{Y_-}\right)^2\hstar s -  \log \left(1+\frac{1}{Y_2}\right) \star s+ \log \frac{S(u_2^{(2)+} - v)}{S(u_3^{(2)-} - v)} \frac{S(r_0^+-v)}{ S(r_0^--v)}
\nonumber\\
&- \check{\Delta} _v{\cstar} s +\sum_i \log \check{\Sigma}_{1_*}^2 \check{S_1} (u_i^{(1)},v)\cstar s
 + \sum_{j=2,3}\log\frac{\check{S}(u_j^{(1)+},v)}{\check{S}(u_j^{(1)-},v)}\cstar s
  \label{STbaQ1b}
\\
\nonumber
& +\log \frac{\check{S} (r_0^-,v)}{ \check{S} (r_0^+,v)}{\cstar} s -  \log \frac{\check{\Sigma}_2^2  (u_2^{(1)+},v) \check{\Sigma}_2^2 (u_3^{(2)-},v) }{ \check{\Sigma}_2^2  (u_3^{(1)-},v) \check{\Sigma}_2^2 (u_2^{(2)+},v) } {\cstar} s \,.\nonumber
\end{align}
Using these simplified equations we can prove the real analyticity of the Y-functions.

\subsection*{Hybrid TBA equations for $Y_Q$}

As discussed in earlier chapters, the hybrid form of the TBA equations for $Y_Q$ is derived from the corresponding canonical equations and the simplified equations for $Y_{M|vw}$. To make the presentation transparent, we introduce a function which combines the terms on the right hand side of the hybrid ground state TBA equation
\begin{align}
\label{GQ}
G_Q(v)=& - (J+2)\, \tH_{Q} +\log \left(1+Y_{Q'} \right)
\star (K_{\sl(2)}^{Q'Q}+2 s\star K_{vwx}^{Q'-1,Q})\\
&+  2 \log (1 + Y_{1|vw}) \star s \hstar K_{yQ} +2\log(1+Y_{Q-1|vw})\star s\nonumber \\
&  -  2  \log\frac{1-Y_-}{ 1-Y_+} \hstar s \star K^{1Q}_{vwx} +  \log
\frac{1- \frac{1}{Y_-} }{1-\frac{1}{Y_+} } \hstar K_{Q}  +  \log
\big(1-\frac{1}{Y_-}\big)\big( 1 - \frac{1}{Y_+} \big) \hstar
K_{yQ} \, . \nonumber
\end{align}
With the help of $G_Q$, the hybrid TBA equations for $Y_Q$ reads
\begin{align}
\log Y_Q(v) &= G_Q(v) -\log \frac{S_{\sl(2)}^{1Q}(u_2^{(1)},v)}{S_{\sl(2)}^{1Q}(u_3^{(1)},v)}S_{\sl(2)}^{1_*Q}(u_1,v)\nonumber+\log \frac{S_{\sl(2)}^{2Q}(u_3^{(2)-},v)}{S_{\sl(2)}^{2Q}(u_3^{(1)-},v)}\frac{S_{\sl(2)}^{2Q}(u_2^{(1)+},v)}{S_{\sl(2)}^{2Q}(u_2^{(2)+},v)} \\ \nonumber
& -\log
S^{1Q}_{vwx}(u_1,v) +\log S^{1Q}_{vwx}(r_0,v) -\log S_Q(r_0^--v)S_{yQ}(r_0^-,v)
 \\
&+ 2\log S(u_1^-,v)S(r_0^-,v)\star_{p.v.} K_{vwx}^{1Q}
-2\log \frac{S(u_2^{(2)+},v)}{ S(u_3^{(2)-},v)}\star K_{vwx}^{1Q} \label{HybridQ}\, .
\end{align}
It is worth mentioning that the first two terms on the second line combine nicely with the first term on the third line and give the term $2\log S(u_1^-,v)S(r_0^-,v)\star K_{vwx}^{1Q}$ with the usual integration contour, {\it i.e.} running above $u_1$ but below $r_0$. Finally, we point out that equation (\ref{Cross}) allows us to rewrite  (\ref{HybridQ}) in terms of  the S-matrices $S_{\sl(2)}^{1_*Q}$ which is useful when analyzing the exact Bethe equations and for numerics.

\section{Exact Bethe equations}
\label{sec:EBandQ}
In this section we discuss the exact Bethe equations (quantization conditions) for the roots $u_1$ and $u_i^{(1,2)}$ where $i=2,3$. Let us recall that according to the discussion in section \ref{sec:L7} we can choose the following equations as our quantization conditions
\begin{align}\label{EBEall1}
Y_{1_*}(u_1)=-1\,,\quad Y_1(u_2^{(1)++})=-1\ &\Leftrightarrow&\ Y_1(u_3^{(1)--})=-1\,,\\
\label{EBEall2}
 Y_2(u_2^{(2)+})=-1\ &\Leftrightarrow&\ Y_2(u_3^{(2)-})=-1\,.
\end{align}
This is the simplest set of exact Bethe equations because the complex roots
$u_2^{(Q)++}$ and $u_2^{(Q)+}$ are inside the analyticity strip of the mirror $u$-plane and the analytic continuation of the TBA equations for $Y_Q$ functions to these points is straightforward - all we need to do is to set in \eqref{HybridQ} the variable $v$ to $u_2^{(1)++}$ in the equation for $Y_1$ and to $u_2^{(2)+}$  in the equation for $Y_2$, and then equate the result to $2\pi i n$.
Note that since $u_2^{(2)}\approx u_2^{(1)}$ for small $g$ the mode number appearing in the equation for $u_2^{(2)}$  depends on the one for $u_2^{(1)}$. The exact Bethe equations for $u_3$ are equivalent to those for $u_2$ due to the real analyticity of Y-functions.
%%%%%%%%%%%%%%%%%%%%%%%%%%%%%
%\subsection*{Exact Bethe equation for $u_1$}
%%%%%%%%%%%%%%%%%%%%%%%%%%%%%%%
For the real rapidity $u_1$ the quantization condition is unique and we must
analytically continue the hybrid equation for $Y_1$ to the string region.
Following the derivation in \cite{Arutyunov:2009ax} and using the identities from appendix \ref{app:EBE1}, we get
\begin{align}\nonumber
2\pi i n_1 &= G_{1_*}(u_1) -\log \frac{S_{\sl(2)}^{11_*}(u_2^{(1)},u_1)}{S_{\sl(2)}^{11_*}(u_3^{(1)},u_1)} +\log \frac{S_{\sl(2)}^{21_*}(u_3^{(2)-},u_1)}{S_{\sl(2)}^{21_*}(u_3^{(1)-},u_1)}\frac{S_{\sl(2)}^{21_*}(u_2^{(1)+},u_1)}{S_{\sl(2)}^{21_*}(u_2^{(2)+},u_1)} \\ \nonumber
&-2\log \frac{S(u_2^{(2)+},u_1)}{ S(u_3^{(2)-},u_1)}\star K_{vwx}^{11_*}-\log S_1(r_0^--u_1)S_{y1_*}(r_0^-,u_1)
 \\ \label{EBE1}
&+ 2\log {\rm Res}\, S\star K_{vwx}^{11_*}(r_0^-,u_1)-2\log\left(u_1-r_0-\frac{2i}{g}\right)
\frac{x_s^+(r_0)-x_s^+(u_1)}{x_s^+(r_0)-x_s^-(u_1)}
\\
&
 + 2  \log {\rm Res}\, S\star K^{11_*}_{vwx} (u_1^-,u_1) -2\log\Big(-\frac{2i}{ g}\,
\frac{x_s^-(u_1)-\frac{1}{x_s^-(u_1)}}{x_s^-(u_1)-\frac{1}{x_s^+(u_1)}}\Big)
\nonumber
\, . \end{align}
where $G_{1_*}(u_1)$ is obtained by analytically continuing \eqref{GQ}
\begin{align}
G_{1_*}(u_1) =& i (J+2)\, p_1 +\log \left(1+Y_{Q} \right)
\star (K_{\sl(2)}^{Q1_*}+2 s\star K_{vwx}^{Q-1,1_*})\\
&+ 2 \log (1 + Y_{1|vw}) \star (s \hstar K_{y1_*}+\ts ) \nonumber \\
&  -  2  \log\frac{1-Y_-}{ 1-Y_+} \hstar s \star K^{11_*}_{vwx} +  \log
\frac{1- \frac{1}{Y_-} }{ 1-\frac{1}{Y_+} } \hstar K_{1}  +  \log
\big(1-\frac{1}{Y_-}\big)\big( 1 - \frac{1}{Y_+} \big) \hstar
K_{y1_*} \, , \nonumber
\end{align}
and $\ts(u)=s(u^-)$. Since the root $u_1$ is real while $u_2$ and $u_3$ are complex conjugate to each other the real part of equation \eqref{EBE1} must vanish. We show that this is indeed the case in appendix \ref{app:EBE1}.

We further notice that we can express all $S_{\sl(2)}^{11_*}$ S-matrices via $S_{\sl(2)}^{1_*1_*}$ by using
\begin{equation}
\label{ssid}
\begin{array}{l}
S_{\sl(2)}^{11_*}(u_2^{(1)},u_1)=S_{\sl(2)}^{1_*1_*}(u_2^{(1)},u_1)\, ,\\[1mm]
{S_{\sl(2)}^{11_*}(u_3^{(1)},u_1)}{S_{\sl(2)}^{1_*1_*}(u_3^{(1)},u_1)}=1/h_{1_*}(u_3^{(1)},u_1)^2= h_1(u_1,u_3^{(1)})^2\, ,
\end{array}
\end{equation}
where the last formula  is the analytic continuation of the identity \eqref{Cross}.
The representation of the exact Bethe equations
via  $S_{\sl(2)}^{1_*1_*}$ is useful in proving the vanishing of the real part of equation \eqref{EBE1} and in checking the Bethe-Yang equations in the limit $g\to 0$ as  discussed below.

\subsection*{Equivalence of quantization conditions}
An important fact to emphasize is that the equations \eqref{EBEall1} and  \eqref{EBEall2} are not the only quantization conditions. In addition we should have
\begin{align}\label{EBEall1b}
Y_1(u_2^{(1)})=-1\ &\Leftrightarrow&\ Y_1(u_3^{(1)})=-1\,,\\
\label{EBEall2b}
 Y_2(u_2^{(2)+++})=-1\ &\Leftrightarrow&\ Y_2(u_3^{(2)---})=-1\,,
\end{align}
since these conditions have also been used to derive the TBA equations. These extra quantization conditions obviously have to be equivalent to (\ref{EBEall1}) and (\ref{EBEall2})
respectively, {\it i.e.} we want to verify
\begin{align}
\label{QC1}
Y_1(u_2^{(1)})=-1 &\stackrel{?}{\Longleftrightarrow} & Y_1(u_2^{(1)++})=-1\, ,
\\
\label{QC2}
Y_2(u_2^{(2)+})=-1 &\stackrel{?}{\Longleftrightarrow} & Y_2(u_2^{(2)+++})=-1\, ,
\end{align}
and we will do so by making use of the Y-system.
As can be checked, the Y-functions which solve the TBA equations also solve the corresponding Y-system equations.
In particular $Y_1$ and $Y_2$ satisfy the following equations
\begin{align}
\label{Ys1}
Y_{1}(v^-)Y_{1}(v^{+}) &=\frac{\big(1-\frac{1}{Y_-}\big)^2}{1+\frac{1}{ Y_2}}(v)\,,\\
\label{Ys2}
Y_{2}(v^-)Y_{2}(v^{+}) &=\frac{\big(1+\frac{1}{Y_{1|vw}}\big)^2}{(1+\frac{1}{ Y_1})(1+\frac{1}{ Y_3})}(v)\, ,
\end{align}
which are valid for any $v$ on the mirror $u$-plane (excluding points on its cuts).

Now let us consider the equation for $Y_1$ at $v=u_2^{(1)+}$ and the equation for $Y_2$ at $v=u_2^{(2)++}$. Then  taking into account that
\begin{equation}
\nonumber
Y_2(u_2^{(1)+})=Y_-(u_2^{(1)+})=\infty\, , \quad Y_1(u_2^{(2)++})=Y_3(u_2^{(2)++})=Y_{1|vw}(u_2^{(2)++})=\infty\, ,
\end{equation}
we find
\begin{equation}
Y_{1}(u_2^{(1)})Y_{1}(u_2^{(1)++})=1\,,\quad Y_{2}(u_2^{(2)+})Y_{2}(u_2^{(2)+++})=1\,
\end{equation}
which clearly implies the equivalence of the quantization conditions.

%%%%%%%%%%%%%%%%%%%%%%%%%%%%
\subsubsection*{Mirror and string quantization conditions}
%%%%%%%%%%%%%%%%%%%%%%%%%%%
In addition to the quantization conditions discussed above, we could also expect to have the exact Bethe equation
$Y_{1_*}(u_3^{(1)})=-1$, where $Y_{1_*}$ is the analytic continuation of $Y_1$ to the string region. In other words,
we would then have
\begin{equation}\label{EBEall1c}
Y_1(u_3^{(1)--})=-1\ \Leftrightarrow\ Y_1(u_3^{(1)})=-1\ \Leftrightarrow\ Y_{1_*}(u_3^{(1)})=-1\,.
\end{equation}
The last condition in  \eqref{EBEall1c} is not necessary for our derivation of the TBA equations because the point $u_3^{(1)}$ of the string $u$-plane is not enclosed by the integration contours. Nevertheless, we will show that this condition holds and therefore the exact Bethe equations can be written in precisely the same form as for real momenta
\begin{equation}\label{EBEallr}
Y_{1_*}(u_i^{(1)})=-1\,,\quad i=1,2,3\,,
\end{equation}
where we have also taken into account that $u_2^{(1)}$ lies in the overlap of the mirror and string regions, so that $Y_1(u_2^{(1)})=Y_{1_*}(u_2^{(1)})$.

To show that the quantization condition $Y_1(u_3^{(1)}) = -1$ in the mirror region implies the usual exact Bethe equation $Y_{1_*}(u_3^{(1)}) = -1$ in the string region, we will analytically continue the TBA equation for $Y_1$ to a point $v$ close to $u_3^{(1)}$ in the mirror $u$-plane, and to the same point in the string $u$-plane. The resulting two equations are then added up and used to show that $Y_1(u_3^{(1)}) Y_{1_*}(u_3^{(1)}) =1$. The considerations below require the use of crossing relations for various kernels and S-matrices because the point $u_3^{(1)}$ lies in the overlap of the mirror and anti-string regions.
We find it easier to handle the canonical TBA equation \eqref{YQcanTBA} for $Y_1$ because its kernels and S-matrices have simpler properties under the crossing transformation.

The analytic continuation of the canonical TBA equation \eqref{YQcanTBA} to $v\approx u_3^{(1)}$ of the mirror $u$-plane is straightforward and gives
\begin{align}\nonumber
\log Y_1(v) =&\, - (J+2)\, \tH_{1} + \log\left(1+Y_{Q} \right) \star K_{\sl(2)}^{Q1}
+ 2 \log\left(1+ \frac{1}{Y_{Q|vw}} \right) \star_{p.v} K^{Q1}_{vwx}
\\[1mm]\nonumber
&\quad +  \log \frac{1- \frac{1}{Y_-} }{1-\frac{1}{Y_+} } \hstar K_{1}  +  \log \big(1-\frac{1}{Y_-}\big)\big( 1 - \frac{1}{Y_+} \big) \hstar K_{y1} \\[1mm] \nonumber
&\quad -\log S_{\sl(2)}^{1_*1}(u_1,v)\frac{S_{\sl(2)}^{11}(u_2^{(1)},v)}{
S_{\sl(2)}^{11}(u_3^{(1)},v)}+\log \frac{S_{\sl(2)}^{21}(u_2^{(1)+},v)}{S_{\sl(2)}^{2Q}(u_2^{(2)+},v)}\frac{S_{\sl(2)}^{21}(u_3^{(2)-},v)}{S_{\sl(2)}^{21}(u_3^{(1)-},v)}
\\[1mm]\nonumber
&\quad-\log S^{11}_{vwx}(u_1,v)+\log S^{11}_{vwx}(r_0,v) - \log{S_1}(r_0^--v){S_{y1}}(r_0^-,v)\nonumber\\[1mm]
&\quad-\log(1+Y_2(v^{-}))+2\log\left(1-\frac{1}{Y_-(v^{-})}\right)\label{eq:EBu3mirror}\,,
\end{align}
where the terms on the last line of \eqref{eq:EBu3mirror} appear because of the poles of $K_{\sl(2)}^{21}(t,v)$ and $(K_1 + K_{y1})(t,v)$ at $v = t +i/g$.

The analytic continuation of the canonical TBA equation for $Y_1$ to $v\approx u_3^{(1)}$ in the string region is discussed in detail in appendix \ref{app:EBE1}, and the resulting TBA equation for $v\approx u_3^{(1)}$ is
\begin{align}
&\log Y_{1_*}(v)=\, - (J+2)\, \tH_{1_*} + \log\left(1+Y_{Q} \right) \star K_{\sl(2)}^{Q1_*}+
2 \log\big(1+ \frac{1}{Y_{Q|vw}} \big) \star_{p.v} K^{Q1_*}_{vwx} \nonumber \\
& +  \log \frac{1- \frac{1}{Y_-} }{ 1-\frac{1}{Y_+} } \hstar K_{1}+  \log \big(1-\frac{1}{Y_-}\big)\big( 1 - \frac{1}{Y_+} \big) \hstar K_{y1_*} -\log(1+Y_2(v^-))\nonumber \\
&+2\log\big( 1-\frac{1}{Y_{+_{\hat{*}}}(v^+)}\big) +2\log\left(1+\frac{1}{Y_{1_{\hat{*}}|vw}(v)}\right)
+2\log\left(1+\frac{1}{Y_{2|vw}(v^-)}\right)\nonumber
\\
&-\log \frac{S_{\sl(2)}^{11_*}(u_2^{(1)},v)}{S_{\sl(2)}^{11_*}(u_3^{(1)},v)}S_{\sl(2)}^{1_*1_*}(u_1,v)+\log \frac{S_{\sl(2)}^{21_*}(u_3^{(2)-},v)}{S_{\sl(2)}^{21_*}(u_3^{(1)-},v)}\frac{S_{\sl(2)}^{21_*}(u_2^{(1)+},v)}{S_{\sl(2)}^{21_*}(u_2^{(2)+},v)} \nonumber\\
&-\log S^{11_*}_{vwx}(u_1,v)+\log \frac{S^{11_*}_{vwx}(r_0,v)}{{S_1}(r_0^-v){S_{y1_*}}(r_0^-,v)} \label{eq:EBu3string}\,.
\end{align}
In the above, $Y_{1_{\hat{*}}|vw}$ and $Y_{+_{\hat{*}}}$ are the analytic continuations of $Y_{1|vw}$ and $Y_+$ through their cuts at $i/g$ and $2i/g$ respectively, {\it cf.} appendix \ref{app:EBE1}.

To proceed further we add the right hand sides of equations (\ref{eq:EBu3mirror}) and (\ref{eq:EBu3string}). Then, by using the crossing relations \eqref{crossQ1} for the bound-state dressing factors and other identities from appendix \ref{app:EBE1}, we find for $v\approx u_3^{(1)}$
\begin{align}\nonumber
\log &Y_1(v)Y_{1_*}(v)= -2\log\left(1+Y_{Q} \right)\star
(K_{xv}^{Q1}(v)-K_{Qy} (v^-)) \nonumber \\
&
+2 \log\left(1+ \frac{1}{Y_{Q|vw}} \right) \star_{p.v} K_{Q1}(v)
+  2\log \frac{1- \frac{1}{Y_-} }{ 1-\frac{1}{Y_+} } \hstar K_{1}(v) \nonumber  \\[1mm]\nonumber
&-2\log(1+Y_2(v^-))+2\log\left(1-\frac{1}{Y_-(v^-)}\right)\left(1-\frac{1}{Y_{+_{\hat{*}}}(v^+)}\right) \nonumber
\\
&+2\log\left(1+\frac{1}{Y_{1_{\hat{*}}|vw}(v)}\right)+2\log\left(1+\frac{1}{Y_{2|vw}(v^-)}\right)\nonumber \\
&-\log \frac{S_{\sl(2)}^{11}(u_2^{(1)},v)S_{\sl(2)}^{11_*}(u_2^{(1)},v)}{S_{\sl(2)}^{11}(u_3^{(1)},v)S_{\sl(2)}^{11_*}(u_3^{(1)},v)}S_{\sl(2)}^{1_*1}(u_1,v)S_{\sl(2)}^{1_*1_*}(u_1,v) \nonumber \\
&+\log \frac{S_{\sl(2)}^{21}(u_3^{(2)-},v)S_{\sl(2)}^{21_*}(u_3^{(2)-},v)}{
S_{\sl(2)}^{21}(u_3^{(1)-},v)S_{\sl(2)}^{21_*}(u_3^{(1)-},v)}
+\log \frac{S_{\sl(2)}^{21}(u_2^{(1)+},v)S_{\sl(2)}^{21_*}(u_2^{(1)+},v)}{
S_{\sl(2)}^{21}(u_2^{(2)+},v)S_{\sl(2)}^{21_*}(u_2^{(2)+},v)} \label{YQcanTBA2add}\\
&-\log S_2(u_1-v)+\log \frac{S_{2}(r_0-v)}{S_1(r_0^- -v)^2} \nonumber\,.
\end{align}
To show that the right hand side of this equation in fact vanishes at $v= u_3^{(1)}$, we use the canonical TBA equations for $vw$-strings continued to $v\approx u_3^{(1)}$ through the cut at $i/g$. Noting that $K_{xv}^{Q1_{\hat{*}}}(u,v) = K_{xv}^{Q1}(u,v)-K_{Qy}(u,v^-)$, it reads
\begin{align}
\nonumber
Y_{1_{\hat{*}}|vw}(v)=&\log\left(1+\frac{1}{Y_{Q|vw}}\right)\star K_{Q1}(v)+\log\left(1+\frac{1}{Y_{2|vw}(v^-)}\right)+\\
&+\log\frac{1-\frac{1}{Y_-}}{1-\frac{1}{Y_+}}\hat{\star}K_1-\log(1+Y_Q)\star (K_{xv}^{Q1}(v)-K_{Qy} (v^-)) \label{eq:Y1vwcont}\\
&-\log(1+Y_2(v^-)) + \frac{1}{2}\log\frac{S_2(r_0-v)}{S_2(u_1-v)}-\log S_1(r_0^--v)\nonumber \\
&-\log\frac{S_{xv}^{11}(u_3^{(1)},v)}{S_{xv}^{11}(u_2^{(1)}, v)} + \log\frac{S_{1y}(u_3^{(1)},v^-)}{S_{1y}(u_2^{(1)}, v^-)}  + S_{xv}^{1*1}(u_1,v) -\log S_{1_*y}(u_1^{(1)},v^-)\nonumber \\
&-\log\frac{S^{21}_{xv}(u_2^{(1)+},v)S^{21}_{xv}(u_3^{(2)-},v)}{S^{21}_{xv}(u_2^{(2)+},v)S^{21}_{xv}(u_3^{(1)-},v)} + \log \frac{S_{2y}(u_2^{(1)+}, v^-)}{ S_{2y}(u_2^{(2)+}, v^-)}\frac{S_{2y}(u_3^{(2)-}, v^-)}{S_{2y}(u_3^{(1)-}, v^-)}\nonumber
\end{align}
Using this equation and crossing relations \eqref{crossQ1}, all driving terms and convolution terms cancel and we find a simple result
 \begin{align}\nonumber
Y_1(v)Y_{1_*}(v)= (1+Y_{1_{\hat{*}}|vw}(v))^2\left(1-\frac{1}{Y_-(v^-)}\right)^2\left(1-\frac{1}{Y_{+_{\hat{*}}}(v^+)}\right)^2\, .
\end{align}
It is now straightforward to show that  $Y_1(u_3^{(1)})Y_{1_*}(u_3^{(1)}) =1$.
Firstly, considering the equation for $Y_{1_{\hat{*}}|vw}$ at $u_3^{(1)}$, it is clear that we have
\begin{equation}
Y_{1_{\hat{*}}|vw}(u_3^{(1)})=0 \, ,
\end{equation}
because $S_{1y}(u_3^{(1)},u_3^{(1)-})$ is zero, while the poles of $Y_2$ at $u_3^{(1)-}$ and $S^{21}_{xv}(u_3^{(1)-},v)$ at $u_3^{(1)}$ cancel each other  and all other terms in \eqref{eq:Y1vwcont} are finite. Then, analytically continuing the canonical equations for $y$-particles, we find that after crossing the cut at $2i/g$
\begin{equation}
\log{Y_{+_{\hat{*}}}(u_3^{(1)+})} \sim \log { \left(1+\frac{1}{Y_{1_{\hat{*}}|vw}}\right)} + \mbox{reg.} \, \, \Rightarrow \, \, Y_{+_{\hat{*}}}(u_3^{(1)+}) = \infty \, ,
\end{equation}
so that we obtain the desired result
\begin{align}
Y_1(u_3^{(1)})Y_{1_*}(u_3^{(1)}) = 1\, .
\end{align}

\subsection*{Exact Bethe equations for roots $r_M$}

The TBA equations also depend on additional roots $r_M$. The exact Bethe equations for the roots are just obtained by analytically continuing the equations for $-Y_-$ and $Y_{M|w}$ to $r_0^-$ and $r_M^-$ respectively, and setting the values of these functions to $-1$.

%%%%%%%%%%%%%%%%%%%%%%%%%%%%%%%%%%%
\subsection*{Relation to the asymptotic Bethe Ansatz}
%%%%%%%%%%%%%%%%%%%%%%%%%%%%%%%%%%%
In the asymptotic limits $g\to 0$ with $J$ fixed or $J\to \infty$ with $g$ fixed the exact quantization conditions for the Bethe roots should reduce to the Bethe-Yang equations
{\small
\begin{equation}\label{ebe0}
\pi i (2n_k+1)=ip_kJ - \sum_{j=1}^3\log
S_{\su(2)}^{1_*1_*}(u_j,u_k) \, ,~~~~~~n_k\in {\mathbb Z}\, ,
\end{equation}}
where $S_{\su(2)}^{1_*1_*}$ is the S-matrix in the $\su(2)$-sector related to
$S_{\sl(2)}^{1_*1_*}$ as
\begin{equation}\label{su2S}
S_{\su(2)}^{1_*1_*}(u_j,u_k)=S_{\sl(2)}^{1_*1_*}(u_j,u_k)\, \prod_{j=1}^3\left(\frac{x_k^+ -x_j^-
}{x_k^--x_j^+}\sqrt{\frac{x_j^+ x_k^-}{x_j^- x_k^+}} \right)^2 \,.
\end{equation}
Since in these equations  the S-matrix has both arguments in the string region it is convenient to  express all $S_{\sl(2)}^{11_*}$ S-matrices in the exact Bethe equations via $S_{\sl(2)}^{1_*1_*}$ at the final stage of deriving the Bethe-Yang equations from them.

According to the discussion in section \ref{exactYQ} in the asymptotic limit $u_2^{(2)}\to u_{2}^{(1)}$ and  $u_3^{(2)}\to u_{3}^{(1)}$, and by using \eqref{u1u0} we find
\begin{equation}\label{u2u1}
u_{2}^{(2)}-u_{2}^{(1)}\approx -{\rm Res}\, Y_2(u_2^{(1)+})\, ,\quad
u_{2}^{(1)}-u_{2}^{(2)}\approx -{\rm Res}\, Y_1(u_2^{(2)++})\, ,
\end{equation}
where we have taken into account that $1+Y_1$ has a zero at $u_{2}^{(1)++}$
and a pole at $u_{2}^{(2)++}$ while $1+Y_2$ has a zero at $u_{2}^{(2)+}$
and a pole at $u_{2}^{(1)+}$. Comparing these two expressions we immediately conclude that in the asymptotic limit the residues of $Y_1$ and $Y_2$ must obey the relation
\begin{equation}
{\rm Res}\, Y_2(u_2^{+}) + {\rm Res}\, Y_1(u_2^{++}) = 0\,,
\end{equation}
where we have equated $u_2^{(2)}= u_{2}^{(1)}\equiv u_2$.
This is indeed satisfied, as can be readily verified through the generalized L\"uscher formula \eqref{YQexact} for $Y_Q$ functions.

Restricting ourselves for definiteness to the limit $g\to 0$ with $J$ fixed and rescaling the rapidities $u\to u/g$ so that the rescaled Bethe roots have a finite limit as $g\to 0$,  we  find that the leading term of ${\rm Res}\, Y_2(u_2^{(1)+})$ scales as $g^{2L}$.
Hence, we arrive at the following asymptotic relation for the rescaled rapidities %$u=\frac{\tilde{u}}{g}$
\begin{equation}
u_2^{(2)}-u_2^{(1)}=g^{2L}a+{\cal O}(g^{2L+1})\, ,
\end{equation}
where the constant $a$ can be found either from the TBA equation for $Y_2$ or from the Bajnok-Janik formula \eqref{YQexact}. This formula shows that as expected at weak coupling the corrections to the asymptotic Bethe ansatz start at $L$-loop order.

Taking the limit  $u_2^{(2)}\to u_{2}^{(1)}\equiv u_2$ and  $u_3^{(2)}\to u_{3}^{(1)}\equiv u_3$ and dropping  the subleading terms $\log(1+Y_Q)$
in the exact Bethe equation \eqref{EBE1} for $u_1$ is straightforward, and it is easy to verify numerically that the resulting equation coincides with \eqref{ebe0}.

Considering the asymptotic limit of the exact quantization condition for the complex root $u_2^{(1)}$ is more involved and it is convenient to do this by using the equation $Y_1(u_{2}^{(1)})=-1$ because there the $S_{\sl(2)}^{11}$ S-matrices depend on $u_i^{(1)}$ only.
To write down the exact Bethe equation for $u_2^{(1)}$, we need to analytically continue the
hybrid TBA equation\footnote{Of course we can perform the analytic continuation at the level of the canonical or simplified
equation for $Y_1$ as well. The hybrid form is preferred because it is the simplest one.}  for $Y_1$ to this point. This is done in appendix \ref{app:EBE1} and the resulting exact Bethe equation at $u_2^{(1)}$ is
{\small
\begin{align}
&\log(-1)=\log Y_{1}(u_2^{(1)}) = G_1(u_2^{(1)})
+  2 \log (1 + Y_{1|vw}) \star \tilde{s}  -\log \frac{S_{\sl(2)}^{11}(u_2^{(1)},u_2^{(1)})}{S_{\sl(2)}^{11}(u_3^{(1)},u_2^{(1)})}
S_{\sl(2)}^{1_*1}(u_1,u_2^{(1)}) \nonumber \\
&-2\log \frac{S(u_2^{(2)+}, u_2^{(1)})}{ S(u_3^{(2)-}, u_2^{(1)})}\star
K_{vwx}^{11} -\log S_1(r_0^--u_2^{(1)})S_{y1}(r_0^-,u_2^{(1)})
\nonumber \\
&+\log \frac{
S_{\sl(2)}^{21}(u_3^{(2)-},u_2^{(1)})
}
{     S_{\sl(2)}^{21}(u_3^{(1)-},u_2^{(1)})  }
+\log \frac{{\rm Res}\, S_{\sl(2)}^{21}(u_2^{(1)+},u_2^{(1)})}
{S_{\sl(2)}^{21}(u_2^{(2)+},u_2^{(1)}) \,{\rm Res}\, Y_2(u_2^{(1)+})} \label{ExactBYE1} \\
& +2\log{\rm Res}\, S\star K_{vwx}^{11}(u_1^-,u_2^{(1)})- \log\Big(u_1-u_2^{(1)}-\frac{2i}{g}\Big)^2
\left(\frac{x_s^-(u_1)-\frac{1}{x^-(u_2^{(1)})}}{x_s^-(u_1)-\frac{1}{x^+(u_2^{(1)})}}\right)^2\nonumber\\
& +2\log {\rm Res}\, S\star K_{vwx}^{11}(r_0^-,u_2^{(1)})-\log\Big(r_0-u_2^{(1)}+\frac{2i}{g}\Big)^2
\left(\frac{x_s^+(r_0)-x^+(u_2^{(1)})}{x_s^+(r_0)-x^-(u_2^{(1)})}\right)^2 \, . \nonumber\end{align}
}

Taking the limit  $u_2^{(2)}\to u_{2}^{(1)}$
in this equation is not straightforward because
the S-matrix $S_{\sl(2)}^{21}(u_2^{(2)+},u_2^{(1)}) $ develops a singularity. For $u_2^{(2)}\sim u_{2}^{(1)}$
we have
\begin{equation}
\log S_{\sl(2)}^{21}(u_2^{(2)+},u_2^{(1)}) =\log\frac{{\rm Res}\, S_{\sl(2)}^{21}(u_2^{(1)+},u_2^{(1)})}{u_{2}^{(1)}-u_{2}^{(2)}}+
o(\delta u)\, ,
\end{equation}
where $\delta u=u_2^{(1)}-u_2^{(2)}$. Taking into account \eqref{u2u1}, we get that in the limit  $u_2^{(2)}\to u_{2}^{(1)}$ the
terms on the third line of
equation (\ref{ExactBYE1}) vanish, and therefore equation (\ref{ExactBYE1}) acquires the form
\begin{align}
\nonumber
&\log(-1)= G_1^{\rm asympt}(u_2)+  2 \log (1 + Y_{1|vw}) \star \tilde{s} -\log \frac{S_{\sl(2)}^{11}(u_2,u_2)}{S_{\sl(2)}^{11}(u_3,u_2)}
S_{\sl(2)}^{1_*1}(u_1,u_2)\nonumber \\
&-2\log \frac{S(u_2^+, u_2)}{ S(u_3^-, u_2)}\star
K_{vwx}^{11} -\log S_1(r_0^--u_2)S_{y1}(r_0^-,u_2)
\nonumber  \\
& +2\log {\rm Res}\, S\star K_{vwx}^{11}(u_1^-,u_2)- \log\Big(u_1-u_2-\frac{2i}{g}\Big)^2
\left(\frac{x_s^-(u_1)-\frac{1}{x^-(u_2)}}{x_s^-(u_1)-\frac{1}{x^+(u_2)}}\right)^2\label{ExactBYE1asympt}\\
& +2\log {\rm Res}\, S\star K_{vwx}^{11}(r_0^-,u_2)-\log\Big(r_0-u_2+\frac{2i}{g}\Big)^2
\left(\frac{x_s^+(r_0)-x^+(u_2)}{x_s^+(r_0)-x^-(u_2)}\right)^2\, , \nonumber\end{align}
where $G_1^{\rm asympt}$ is $G_1$ with the subleading terms $\log(1+Y_Q)$ neglected.

It is worth mentioning that our consideration is valid for both the asymptotic limit $g\to 0$ with $J$ fixed, and  $J\to \infty$ with $g$ fixed.
Thus, this formula should coincide with the expression for the
asymptotic Bethe ansatz for {\it any} value of $g$! In other words, if we substitute the asymptotic expressions for the Y-functions in equation (\ref{ExactBYE1asympt}) it should turn into the BY equation \eqref{ebe0} for $u_2$. This is indeed the case as we have verified numerically.

\begin{figure}[t]
\begin{center}
\includegraphics[width=.48\textwidth]{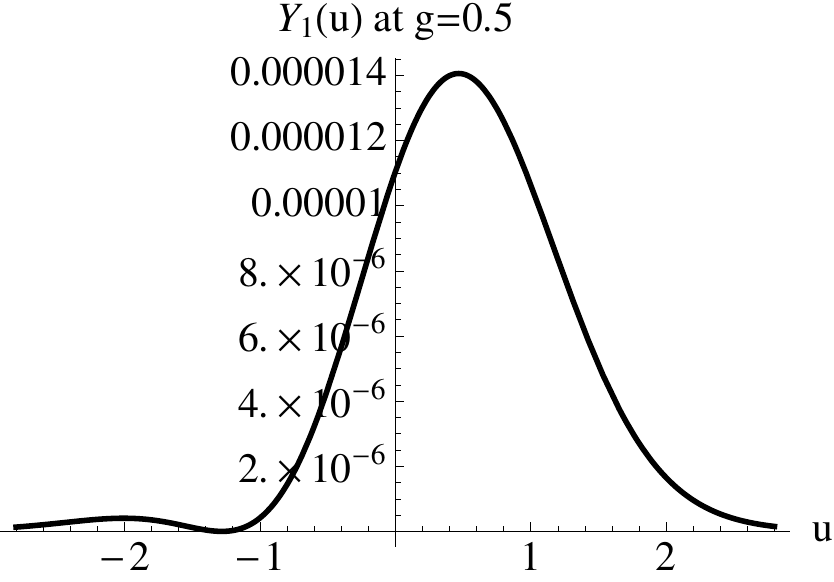}\quad \includegraphics[width=.48\textwidth]{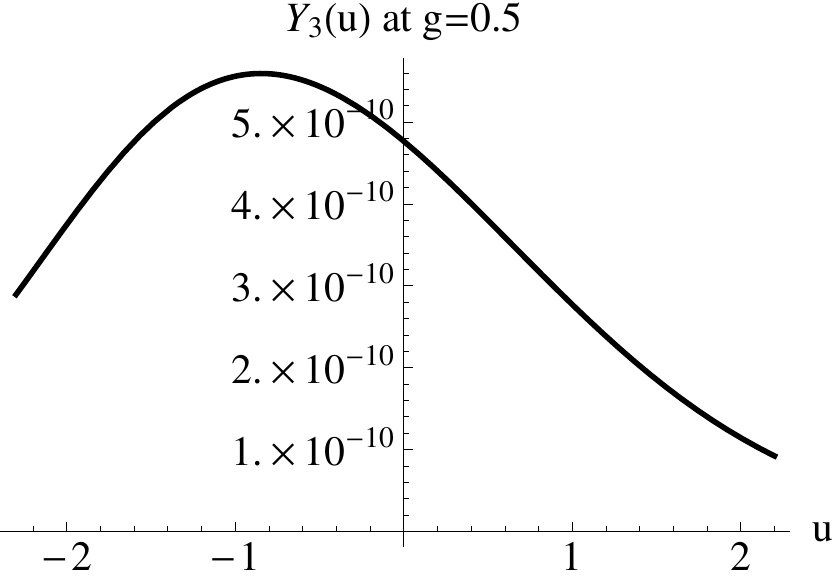}
\end{center}
\caption{The asymptotic $Y_1$- and $Y_3$-functions on the real mirror line at $g=0.5$.  }
\label{figY1Y3}
\end{figure}

\begin{figure}[t]
\begin{center}
\includegraphics[width=.44\textwidth]{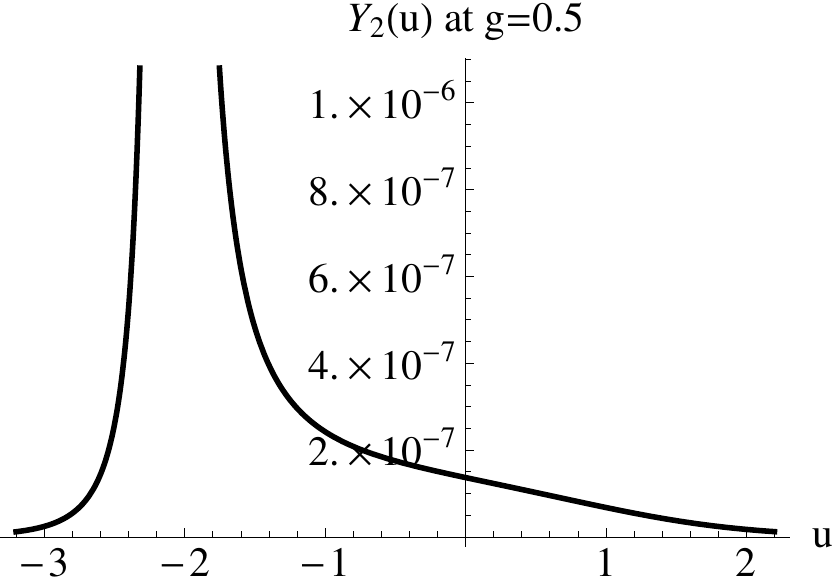}\quad \includegraphics[width=.52\textwidth]{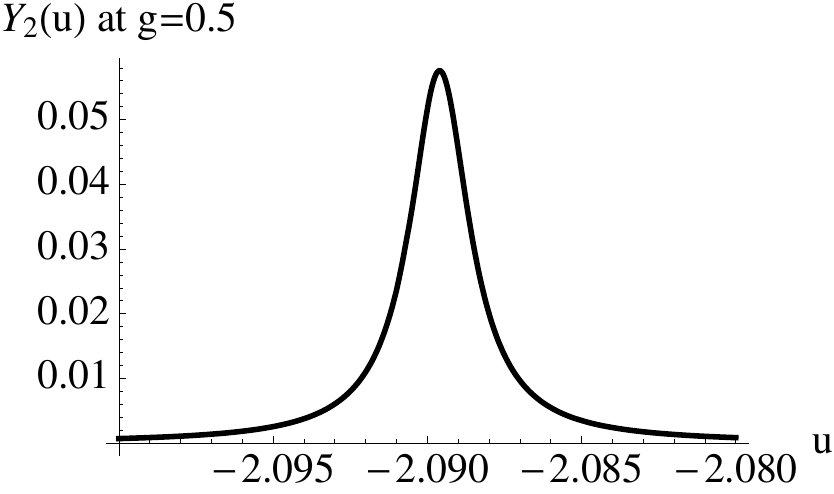}
\end{center}
\caption{The asymptotic $Y_2$-function on the real mirror line at $g=0.5$.  }
\label{figY2}
\end{figure}

\section{Summary and outlook}

In this chapter we developed a description of string excited states with complex momenta. We saw that for suitably small $g$ the asymptotic solution is reliable and the corresponding TBA equations can be constructed by applying the contour deformation trick. However, as soon as $g$ exceeds a certain critical value the description of a state through the BY equations breaks down as its energy becomes complex; in our concrete example this happens for $g\gtrsim 0.53$. Therefore, it is important to understand how the TBA equations may fix this problem and what happens to the state at large values of coupling. The answers to these questions do not appear to be straightforward, requiring detailed analysis of the coupled system of TBA and exact Bethe equations. Based purely on the results of this chapter we can plausibly postulate the following scenario; due to the TBA corrections to the BY equations the motion of the complex Bethe roots towards the boundaries of the analyticity strip slows down so that they actually freeze as $g\to \infty$. Indeed, we find that for $g=0.5$ which is close to the problematic value of $0.53$
the asymptotic $Y_Q$-functions are very small, see figure \ref{figY1Y3} and \ref{figY2}, and approximate the exact
Y-functions with very high precision. At the same time the exact positions of the Bethe roots $u_k$ can change much more noticeably because the roots $u_2$ and $u_3$ are close to the lines $\mbox{Im}(u)=\pm 1/g$ and some of the kernels appearing in the exact Bethe equations develop singularities as Im$(u_{2,3})\to\pm 1/g$, giving large contributions to the r.h.s. of the equations. However it has subsequently been observed in \cite{Arutyunov:2012tx} that at least for the exceptional state considered there (an exact bound state configuration) in the limit of removing the regularizing twist the complex roots move to the branch points $-2 \pm i/g$ and subsequently move towards the exceptional configuration $u_{2/3}=\pm i/g$ while $u_1 \rightarrow 0$. We may expect that this is indicative of general behaviour, whereby the complex roots $u_{2/3}$ should generically reach the branch points at some finite value of $g$, and continue to move to the exceptional configuration \cite{Arutyunov:2012tx}. If true, this would mean that the points $-2$ and $2$, and $0$, $i/g$ and $-i/g$ are universally attractors. While it is tempting to suggest everything should follow this pattern, we should note that states with complex roots in the $k$th strip at weak coupling do not look particularly much like a bound state and are hence far from the exceptional configuration of \cite{Arutyunov:2012tx}. It may well be that the complex roots starting in the $k$th strip at weak coupling remain there, or at least do approach the lines at $\pm i/g$. Confirming the fate of these states is an interesting open question. In this light, it would be very interesting and important to see how states of this type are accounted for in the NLIE or quantum spectral curve approaches mentioned at the end of chapter \ref{chapter:AdS5string}.

\chapter{Quantum deformed TBA}
\label{chapter:quantumTBA}

In the previous chapters we have seen how to derive the TBA equations associated to a $\mathfrak{psu}(2|2)^2$ invariant S-matrix, and how the TBA equations inherit this structure. In this chapter we will describe the corresponding story for a deformation of this S-matrix which will result in deformed TBA equations with a different interesting structure.

The type of deformation we will be considering is a so-called quantum (group) deformation, a particular deformation of the universal enveloping algebra $\,\mathcal{U}(\mathfrak{g})$ of a Lie (super-)algebra $\mathfrak{g}$ modifying the Lie brackets proportional to elements of the Cartan subalgebra. For example, $\mathfrak{sl}_q(2)$\footnote{We will loosely refer to generic quantum deformed algebras $\mathcal{U}_q(\mathfrak{g})$ simply as $\mathfrak{g}_q$.} is the algebra generated by $E$, $F$ and $H$ which satisfy
\begin{equation}
[E,F] = [H]_q \, , \, \,\,  [H,E] = 2E \, , \, \, \, [H,F] = -2F \, ,
\end{equation}
where
\begin{equation}
[n]_q = \frac{q^n - q^{-n}}{q-q^{-1}} \, ,
\end{equation}
are the $q$-numbers we already encountered in chapter \ref{chapter:twistedspectrum}. Note how this algebra reduces to $\mathfrak{sl}(2)$ in the limit $q\rightarrow1$. In principle we can take the deformation parameter $q$ to be any complex number, but in concrete applications it is typically taken to be real or a phase. Instead of insisting that an S-matrix is compatible with some regular Lie algebra symmetry we can insist that it is compatible with some $q$-deformed algebra\footnote{In this case we really require a nontrivial Hopf-algebraic structure; going into details would lead us astray from our main discussion, we refer the reader to \cite{Chari} for an extensive treatment of quantum groups.} and this results in deformations of the original (Lie algebra symmetric) S-matrices that satisfy the Yang-Baxter equation.\footnote{In a pedestrian sense we could view quantum deforming as a solution generating technique for the Yang-Baxter equation, but this does not do the subject justice; quantum groups in fact present a considerable step forward in the classification of solutions to the Yang-Baxter equation.} In other words these deformations preserve integrability in the form of the fundamental commutation relations.\footnote{In fact they naturally lead to structures of the form of the fundamental commutation relations \eqref{eq:FCR}, see e.g. sections two and ten of \cite{Faddeev:1996iy}.} The upshot of considering such a deformation is an S- or R-matrix that lies at the basis of an integrable model. For example, if we consider such deformations for an R-matrix of an integrable spin chain we can obtain other integrable spin chains that in this way can be viewed as `quantum deformations' of simpler ones. The prototypical example of this is the anisotropic XXZ Heisenberg spin chain, a quantum deformation of the isotropic XXX spin chain. As our model will be somewhat technically complicated, and the XXZ spin chain in fact captures the essential change at the level of the TBA equations quite nicely, let us first briefly discuss this.

\section{Point of reference: the XXZ spin chain}

\label{sec:XXZref}

The XXZ spin chain is the anisotropic generalization of the XXX spin chain (eqn. \eqref{eq:XXXham}), with Hamiltonian
\begin{equation}
\label{eq:XXZham}
H = - \frac{J}{4} \sum_{i=1}^{\NF} \left(\sigma^x_i \sigma^x_{i+1}+\sigma^y_i \sigma^y_{i+1}+\Delta\sigma^z_i \sigma^z_{i+1}-1\right) \, ,
\end{equation}
where $2 \Delta = q + 1/q$. The appropriate deformation of the R-matrix is\footnote{We should note that this unitary R-matrix is obtained from the canonical $\su_q(2)$ R-matrix by a simple change of one particle basis. As we will come back to soon, this is a peculiarity that works for $\su_q(2)$ but not in general.}
\begin{equation}
R_{0i}(\W) = \frac{1}{a} \left(\begin{array}{cc:cc} a & 0 & 0 & 0 \\ 0 & b & c & 0 \\ \hdashline 0 & c &b & 0 \\ 0 & 0 & 0 & a \end{array}\right) \, ,
\end{equation}
where
\begin{equation}
a = q^2 \W - \frac{1}{q^2 \W}\, , \hspace{20pt} b = \W-\frac{1}{\W}\, , \hspace{20pt} c = q^2 -\frac{1}{q^2}\, .
\end{equation}
Note that the R-matrix reduces to the permutation matrix at $\W=1$. The Bethe equations that diagonalize the XXZ Hamiltonian and associated transfer matrix are given by
\begin{equation}
\label{eq:XXZBethemult}
\prod_{m=1}^N \frac{q \W_i-1}{\W_i -q} = \prod_{j\neq i}^\NA  \frac{q^2 \W_i-\W_j}{ \W_i-q^2\W_j} \, ,
\end{equation}
while the energy and momentum of the accompanying magnons are
\begin{equation}
\scE = J(\cos(\scp)-\Delta)\, , \hspace{20pt} \scp = -i \log \frac{q \W_i-1}{\W_i -q}\, .
\end{equation}
We will be mostly interested in the case where $q$ is an even principal root of unity, or $q=e^{i \pi /k}$ with $k$ an integer. Note that this means we are considering $-1 < \Delta <1$. We will take $k>2$ since $k=1$, meaning $q=-1$, is equivalent to $q=1$ ($k=\infty$), and $k=2$ corresponds to $\Delta = 0$ which manifestly gives free fermions after appropriately rewriting the remaining terms in the Hamiltonian. Similar to parametrizing $q$ exponentially, it is nice to parametrize $\W$ as $\W \equiv q^{-iw} = e^{\pi w/k}$. In terms of this parametrization the Bethe equations become
\begin{equation}
\label{eq:XXZBetheadd}
\prod_{m=1}^N \frac{\sinh{\frac{\pi}{2k}(v_i -i)}}{\sinh{\frac{\pi}{2k}(v_i +i)}} = \prod_{j\neq i}^\NA  \frac{\sinh{\frac{\pi}{2k}(v_i - v_j -2i)}}{\sinh{\frac{\pi}{2k}(v_i - v_u +2i)}} \, ,
\end{equation}
which also shows they manifestly reduce to the XXX ones in the limit $k \rightarrow \infty$. Of course, this parametrization nicely shows the same for the R-matrix.

Our interest below will lie in the thermodynamics (finite size spectra) of $q$-deformed models, so let us consider the thermodynamics of the XXZ spin chain. As we saw before, the main problem in computing the infinite volume free energy of an integrable model lies in classifying the appropriate solutions to the Bethe equations. For the XXX spin chain we found that these solutions are made out of strings of any number of rapidities spaced evenly and symmetric about the real line. Here we will see that these patterns still take the form of strings, but that their length is not arbitrary. The discussion is simplest if we consider the Bethe equations in the form of eqs. \eqref{eq:XXZBethemult}.

We want to consider the types of solutions eqs. \eqref{eq:XXZBethemult} admit in the limit $N \rightarrow \infty$. To begin with, let us note that since our $q$ lies in the upper half unit circle we have
\begin{equation}
\left | \frac{q \W - 1}{\W - q}  \right | >1 ~~~\mbox{for}~~~\mbox{Im}\, \W>0
\end{equation}
and vice versa. Thus, as before this factor is singular in the limit $N\rightarrow \infty$ unless $\W$ is real. Without loss of generality we can consider a $\W$ root, say $\W_1$, for which the left hand side of the Bethe equations
\begin{equation}
\label{div1}
\prod_{i=1}^{N}\frac{q\W_1-1}{\W_1-q}= \prod_{i=2}^{\NA}\frac{q^2\W_1-\W_i}{\W_1-q^2\W_i}\, ,
\end{equation}
diverges as $N\rightarrow \infty$. This root can only be part of a solution of the Bethe equations if this divergence is compensated by a zero in the second product in the Bethe equation, say
\begin{equation}
\label{w2string}
\W_1 - q^2 \W_2 = 0 \, .
\end{equation}
However we should now also ensure that there are no problems in the equation for $\W_2$
\begin{equation}
\label{div2}
-1=\left(\prod_{i=1}^{N}\frac{q\W_2-1}{\W_2-q}\right)\prod_{i= 1}^{\NA}\frac{\W_2-q^2\W_i}{q^2\W_2-\W_i}\, .
\end{equation}
This can be determined by multiplying the two equations (\ref{div1}) and (\ref{div2}), giving
\begin{equation}
\label{div3}
1=\prod_{i=1}^{N}\frac{q\W_1-1}{\W_1-q}\frac{q\W_2-\W_i}{\W_2-q\W_i}
\prod_{i\neq 1,2}^{\NA}\frac{\W_1-q^2\W_i}{q^2\W_1-\W_i}\frac{\W_2-q^2\W_i}{q^2\W_2-\W_i}\, ,
\end{equation}
so that the divergent term in (\ref{div2}) originating from (\ref{w2string}) cancels out. At this point there are two possibilities. Either the equation for $\W_2$ is finite in the limit $N\to\infty$, which is equivalent to
\begin{equation}
\left|\frac{q \W_1 - \W}{\W_1 - q \W}\frac{q \W_2 - \W}{\W_2 - q \W}\right| = \left|\frac{q^3 \W_2 - \W}{\W_2 - q \W}\right| = 1 \, ,
\end{equation}
giving a string of length two, or this product is greater than one and the process continues to give a longer string. In the latter case we necessarily have to involve another root to find
\begin{equation}
\W_2 - q^2 \W_3 = 0 \, ,
\end{equation}
and analyze its equation in turn. Continuing along these lines we see that a general string configurations of $w$-particles are given by
\begin{equation}
\{\W\}=\{q^{M+1-2j}\W\}\, ,~~~~j=1,\ldots, M\, .
\end{equation}
where $M$ denotes the length of the string. Consistency of the resulting string solution requires that the center of the string is real, as can be seen via fusion. This is not the only constraint however.

As we saw above, in order for our string configuration to keep growing we must keep a divergent product at every stage of the derivation. This means that in order to obtain a string of length $M$ we must have
\begin{equation}
\prod_{k=1}^j \left|\frac{q \W_k - 1}{\W_k - q}\right|= \left |  \frac{q^j(1-q^M \W)}{q^{2j}-q^M \W}\right | > 1 \, , \, \, \mbox{for} \, \, j=1,...,M-1 \, .
\end{equation}
Taking into account that $q=e^{i \frac{\pi}{k}}$, we find that the last condition is equivalent to the following requirement
\begin{equation}
\W \cos{\Big(M \frac{\pi}{k}\Big)}  < \W \cos{\Big((M-2j) \frac{\pi}{k}\Big)}  \, , \, \, \mbox{for} \, \, j=1,...,M-1 \, .\end{equation}
Given these conditions, we can have strings of length one through $k-1$ for a positive center ($\W>0$) and a single type of string of length one with negative center ($\W<0$). We will refer to the positivity or negativity of the string center as positive and negative parity respectively \cite{Takahashi:1972}. In terms of the $w$ parametrization, positive parity strings have real rapidity $w$ and take the form of standard Bethe strings of length one up to $k-1$, while the rapidities of negative parity strings lie on the line $i k$.

Note that the positive parity string of length $k$ lies just on the edge of these conditions and is not allowed; there would have been no divergence to begin with. However, were we to perturb slightly away from $k$ being an integer and let $k$ be $m +\epsilon$, of course it could become an allowed string configuration; correspondingly in the limit where $k$ is exactly an integer this string configuration has fixed momentum $\pi$ and trivial scattering with all other roots, which is another way to see it plays no role in thermodynamics.

The constraints on the length of strings in the XXZ model were first postulated in \cite{Takahashi:1972} and later shown to be equivalent to normalizability of the Bethe wave function of the associated string configuration \cite{hida:1981,Fowler:1981}. We could of course have applied that result here, but we prefer this more intuitive `derivation'.

\begin{figure}[h]
\begin{center}
\includegraphics[width=0.3\textwidth]{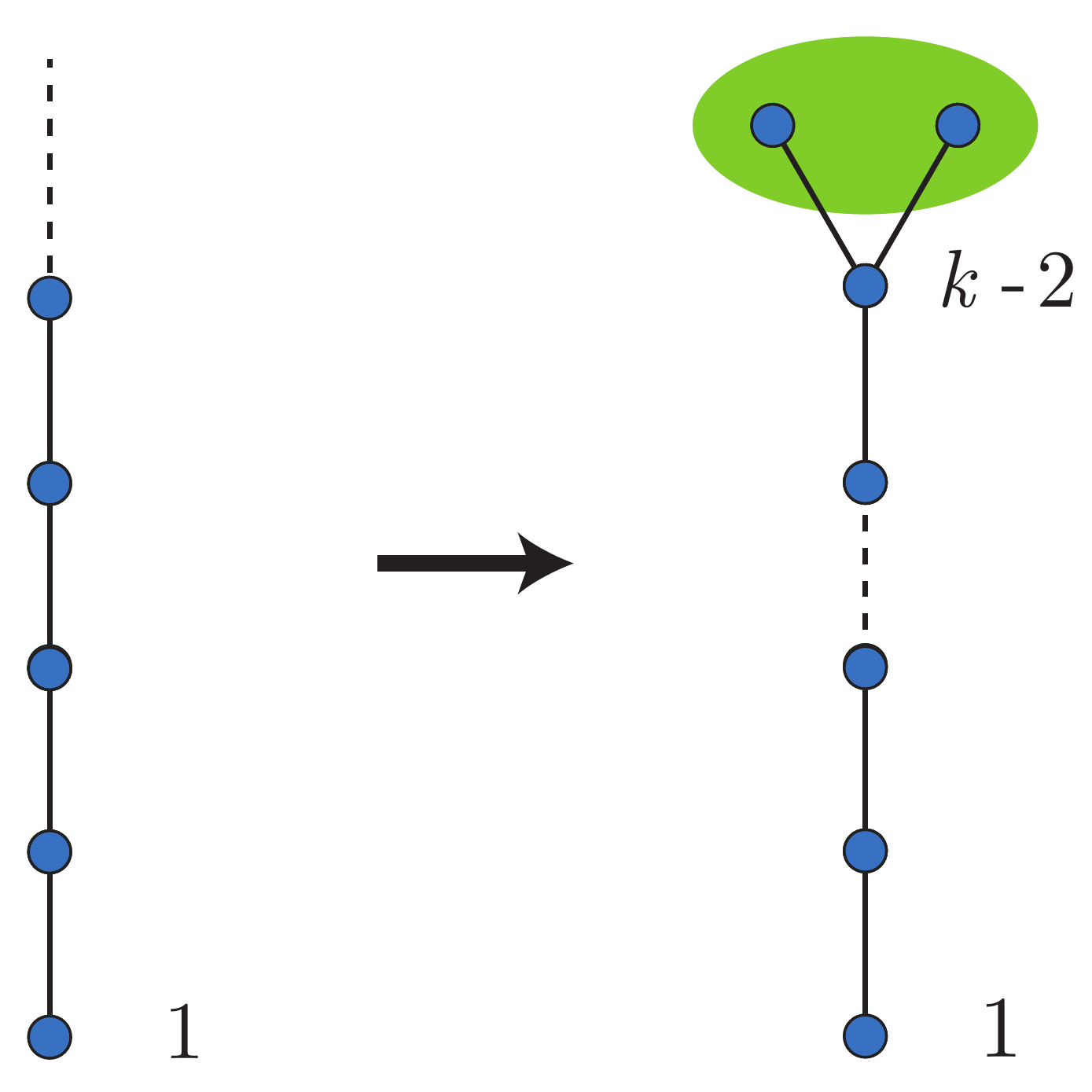}
\caption{The TBA structure of the XXX spin chain and the XXZ spin chain at roots of unity. The XXX spin chain has a standard nearest neighbour coupling between infinitely many Y-functions as we saw in chapter \ref{chapter:finitevolumeIQFT}, while the XXZ model has only a finite number of them. The last two Y-functions $Y_{k-1}$ and $Y_0$ are in general directly related algebraically (represented by the lime-green ellipse) and only couple back to the next to last regular Y-function.}
\label{fig:XXXtoXXZ}
\end{center}
\end{figure}

We can now find the Bethe equations for these string complexes and go through the derivation of section \ref{subsec:TBAgeneral} to find the TBA equations for the XXZ spin chain at roots of unity. Of course, many of the details simply amount to replacing rational functions by appropriate hyperbolic generalizations. However, the key difference is that we have only finitely many different types of particles, and so will only introduce finitely many particle and hole densities, and will end up with only finitely many Y-functions and TBA equations. Very importantly however, if we denote the negative parity string as the \underline{zeroth} type of string, their scattering matrices satisfy
\begin{equation}
S^{0}(v)S^{k-1}(v) = -1\, ,
\end{equation}
and
\begin{equation}
S^{0M}(v)S^{k-1M}(v) = 1\, ,
\end{equation}
where these S-matrices are all obtained by appropriately fusing the elementary S-matrix we here define as
\begin{equation}
S^1 (v) = \frac{\sinh{\frac{\pi}{2k}(v -i)}}{\sinh{\frac{\pi}{2k}(v +i)}}\, .
\end{equation}
Note that $S^0(v) = S^1(v+ik)$. Going through the details of the TBA derivation, we would find that this implies that the associated Y-functions are inverse (up to chemical potential terms we discuss in detail in later sections); $Y_0 Y_{k-1} =1$. At the level of the simplified equations we end up with something that should qualitatively look like the equations for the XXX spin chain as long as $M\ll k$. Also, since $Y_0$ has kernels very similar to $Y_{k-1}$, and $Y_{k-1}$ naturally couples to $Y_{k-2}$, it should not be too surprising that in fact the only modification is to have $Y_{k-1}$ couple doubly to $Y_{k-2}$ (once from $Y_{k-1}$ and once from the original $Y_0$). In short, the simplified TBA equations become
\begin{equation}
\begin{aligned}
\log Y_1 =&\, \log(1+Y_{2}) \star s \, ,\\
\log Y_M =&\,\log(1+Y_{M-1})(1+Y_{M+1}) \star s   \, ,\\
\log Y_{k-2} =&\,\log(1+Y_{k-3})(1+Y_{k-1})^2 \star s  \, ,\\
\log Y_{k-1}=&\,\log(1+Y_{k-2}) \star s \, .
\end{aligned}
\end{equation}
We can summarize the chance in the TBA structure of the XXX spin chain to the XXZ spin chain obtained by $q$-deforming with $q$ a root of unity in the diagrammatic form of figure \ref{fig:XXXtoXXZ}.

Let us now proceed and introduce the more complicated case we are really interested in, the quantum deformation of the $\mathfrak{psu}(2|2)^2$ invariant S-matrix of the superstring, and its associated TBA.

\section{A \texorpdfstring{$\mathfrak{psu}_q(2|2)^2$}{} invariant S-matrix}

\label{sec:psu22qdefintro}

Our construction will be based on a $\mathfrak{psu}_q(2|2)^2$ invariant S-matrix, where $\mathfrak{psu}_q(2|2)$ is the quantum deformation of (the universal enveloping algebra of) centrally extended $\mathfrak{psu}(2|2)$ \cite{Beisert:2008tw,Beisert:2011wq}. The corresponding $\mathfrak{psu}_q(2|2)^2$ invariant S-matrix satisfies the Yang-Baxter equation, and gives a quantum deformation of the $\ads$ world-sheet S-matrix. In this chapter we will focus on the case where the deformation parameter $q$ is a phase, in particular a root of unity.

Apart from an intrinsically interesting deformation of the superstring S-matrix, when we take $q=e^{i\frac{\pi}{k}}$ the full $q$-deformed S-matrix \cite{Hoare:2011wr,Hoare:2012fc} interpolates between the S-matrix of the original string and mirror theory at $q=1$, and (modulo some subtleties discussed below) that of the Pohlmeyer reduced version of the $\ads$ superstring \cite{Hoare:2011fj,Hoare:2011nd,Hoare:2013ysa} in the limit of infinite coupling \cite{Beisert:2010kk}.\footnote{The Pohlmeyer reduced $\ads$ superstring \cite{Grigoriev:2007bu,Mikhailov:2007xr} is a fermionic extension of a gauged Wess-Zumino-Witten model with an integrable potential which is classically equivalent to the $\ads$ superstring, with coupling constant (level) $k$.} We can also anticipate that the deformed model has very interesting intrinsic properties particularly when we take the deformation parameter to be a root of unity. Indeed, as we just saw the XXZ spin chain already shows a very interesting structure in its TBA equations when $q$ is a root of unity. With this in mind, in the present paper we will derive the TBA equations for the quantum deformed theory at $q=e^{i\frac{\pi}{k}}$. As discussed above, the XXZ model with $k=2$ is equivalent to free fermions, and also in our model the point $k=2$ is degenerate, so we will consider $k>2$.\footnote{There is no principal objection to treating the case where $q$ is a generic phase along the lines of the XXZ model \cite{Takahashi:1972,Takahashi:book}, but already there the technicalities become somewhat involved which would lead to a truly undesirable level of technical details here.} Because the deformation affects the theory at the fundamental level of the dispersion relation already, we have to carefully go through the procedure outlined in chapter \ref{chapter:finitevolumeIQFT} to derive the TBA equations for the ground state of the deformed model at roots of unity.

Before proceeding we should note a subtlety regarding the unitarity of the $q$-deformed S-matrix we will be working with. The $\mathfrak{psu}_q(2|2)^2$ invariant S-matrix of course satisfies all usual requirements of integrability, but its physical unitarity was not fully investigated until recently. We found that as it stands this S-matrix and the $\mathfrak{psu}_q(2|2)$ invariant $R$-matrix out of which it is built are not physically unitary but rather physically pseudo-unitary,\footnote{The notion of pseudo-unitarity is defined in appendix \ref{app:qdefmatrixSmatrix}, see eqn. \eqref{eq:pseudounitarity}. For the notion of pseudo-Hermitian quantum mechanics, see e.g.  \cite{Mostafazadeh:2002hb}.} in particular in a local fashion.\footnote{In other words, the Hermitian automorphism involved in the pseudo-unitarity factors over the one particle states, \emph{cf.} eqs. \eqref{eq:pseudounitarity} and \eqref{eq:pseudounitaritylocal}.} After the work we are describing in this chapter was fully completed \cite{Arutyunov:2012zt,Arutyunov:2012ai} it was found that the appropriate basis of scattering states in the Pohlmeyer reduced superstring lies not in the vertex but rather in the so-called IRF (interaction-round-the-face) picture \cite{Hoare:2013ysa}. In this basis the scattering states are kinks rather than solitons, and carefully performing this change of basis for the $\mathfrak{psu}_q(2|2)$ invariant S-matrix gives a physically unitary S-matrix in the IRF picture \cite{Hoare:2013ysa}. We should note however that there are some minor subtleties in considering the undeformed limit of this S-matrix to obtain the light-cone superstring S-matrix. In any case, the work we are describing in this chapter is based on the $\mathfrak{psu}_q(2|2)$ invariant S-matrix in the standard vertex picture, and is unequivocally a direct deformation of the light-cone S-matrix. Whether and to what extent the TBA equations we will be deriving in this chapter would be affected by this change of picture is not entirely clear at this moment. As the structure of the TBA equations nicely matches with restrictions arising in the representation theory of $\mathfrak{psu}_q(2|2)$, it seems quite reasonable to assume that this will not be affected. However, it may very well be that the vertex-to-IRF transformation comes accompanied by a set of driving terms or chemical potentials.

Coming back to the pseudo-unitarity of the vertex S-matrix we are basing our present discussion on, we should note that in general pseudo-unitary (pseudo-Hermitian) models come in two classes; the Hamiltonian has either a (non-trivially) self-conjugate, or a real spectrum. In the latter case the model is quasi-unitary (quasi-Hermitian) \cite{Mostafazadeh:2001nr}. Our uniformized two body S-matrix generically satisfies generalized pseudo-unitary for complex arguments, and has a unitary spectrum for both the string and mirror theory. This quasi-unitary structure does not appear to be compatible with locality however, and as such does not necessarily extend to the many-body scattering theory. In fact, the many-body S-matrix in the vertex picture makes a clear distinction between the string and mirror theory; the string theory many body S-matrix has non-unitary eigenvalues and is therefore only pseudo-unitary, while the mirror theory many body S-matrix appears to remain quasi-unitary.\footnote{In other words, the `lack of unitarity' in the vertex picture for the mirror theory is quite mild.} This difference means that we can expect the mirror theory to have a stronger sense of reality than the string theory, and  is in fact what we will observe soon; these properties translate to the string hypothesis and TBA equations.

\begin{figure}[h]
\begin{center}
\includegraphics[width=0.75\textwidth]{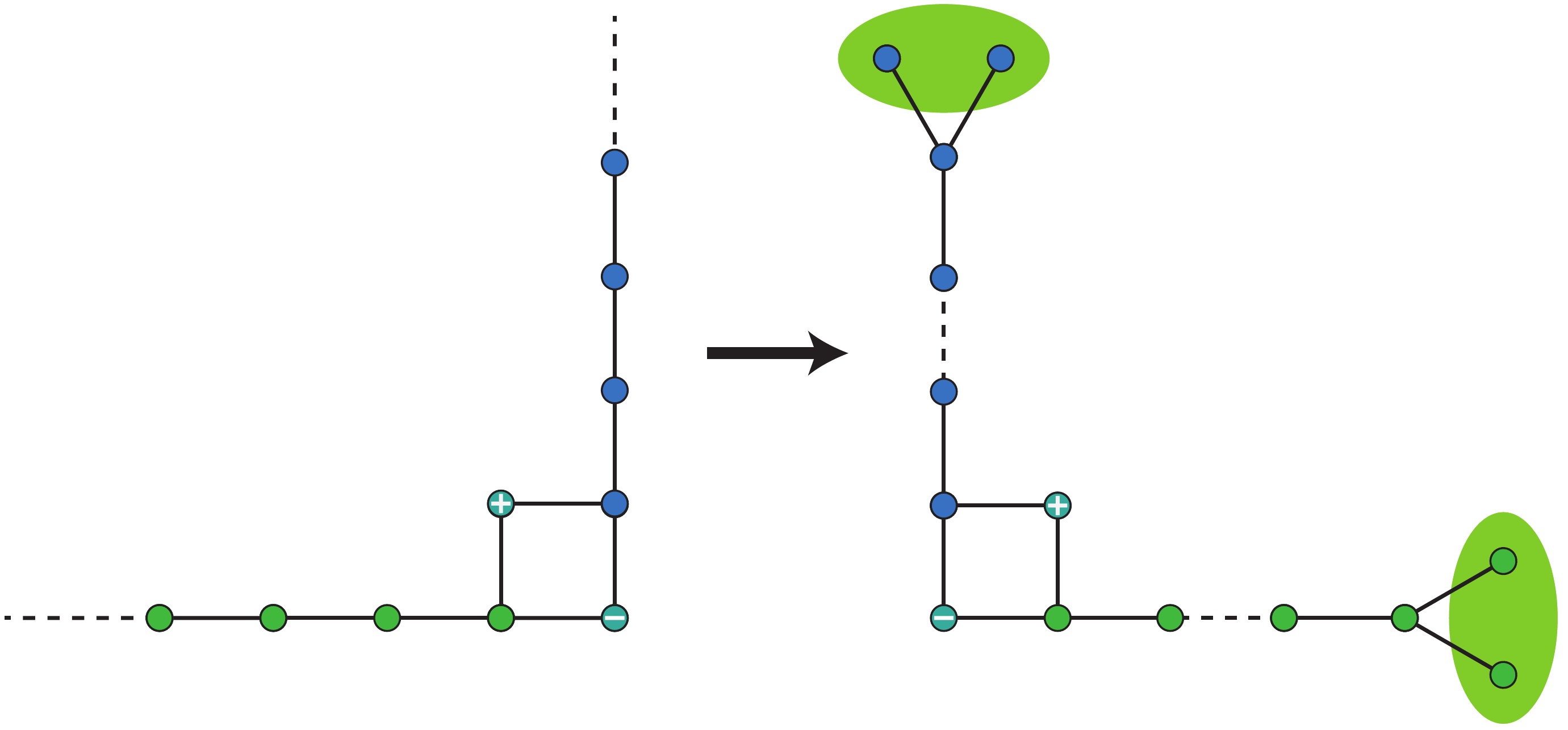}
\caption{The TBA structure of the Hubbard model (left) and the deformed Hubbard model at roots of unity (right). The green, teal (with $\pm$) and blue dots indicate what we call $Y_{M|w}$, $Y_{\pm}$, and $Y_{M|vw}$ functions respectively; their coupling is nearest neighbour apart from the indicated coupling near the corner. The deformation of this structure is just an XXZ type deformation on each of the $\su_q(2)$ wings. (The reflection of the diagram is for purely aesthetic purposes.)}
\label{fig:HubbtoDefHubb}
\end{center}
\end{figure}

In general the question of a $q$-deformed TBA at roots of unity is notoriously difficult to answer. It might be surprising to learn that the TBA for a general $\su_q(N)$ spin chain at roots of unity is in fact not known. One of the difficulties is related to the fact that the associated local Hamiltonian is non-Hermitian\footnote{By a quick investigation they (naturally) appear to be pseudo-Hermitian, but not quasi-Hermitian, perfectly in line with our discussion. Perhaps the vertex-to-IRF transformation also has interesting applications here.} for $N>2$, highlighting the special status of the XXZ model. For $\su_q(3)$ work has been done on the TBA for complex Toda theories \cite{Saleur:2000bq}, involving interesting but rather unusual non-unitary string complexes, but the general story apparently remains unknown. Nevertheless, as we will shortly see, likely due to the quasi-unitarity of the mirror theory we appear to be able to correctly account for the thermodynamics of the quantum deformed Hubbard model\footnote{The $\mathfrak{psu}_q(2|2)$ invariant R-matrix of \cite{Beisert:2008tw} gives models which are closely related to the Alcaraz-Bariev model \cite{Alcaraz:1999}, however the detailed relation is not yet fully understood in all cases. The Alcaraz-Bariev model can be viewed as a quantum deformed version of the Hubbard model and is sometimes referred to as such. Here we loosely refer to our $\mathfrak{psu}_q(2|2)$ invariant auxiliary system as the quantum deformed Hubbard model.} with a `real' string hypothesis. This string hypothesis results in an elegant structure highly analogous to the XXZ case, which in essence can be understood as follows. The undeformed Hubbard TBA structure is that of two $\mathfrak{su}(2)$ (XXX-like) wings coupled via two extra Y-functions, making it only natural that at roots of unity the deformation results in the XXZ-like modification of each of these wings independently illustrated in figure \ref{fig:HubbtoDefHubb}.\footnote{The TBA diagram for the generically deformed Hubbard model would have infinite extent like the undeformed case.} Thus it appears that the \emph{mirror} $\mathfrak{psu}_q(2|2)$ theory is the next simplest case beyond the XXZ model.

\begin{figure}[h]
\begin{center}
\includegraphics[width=0.75\textwidth]{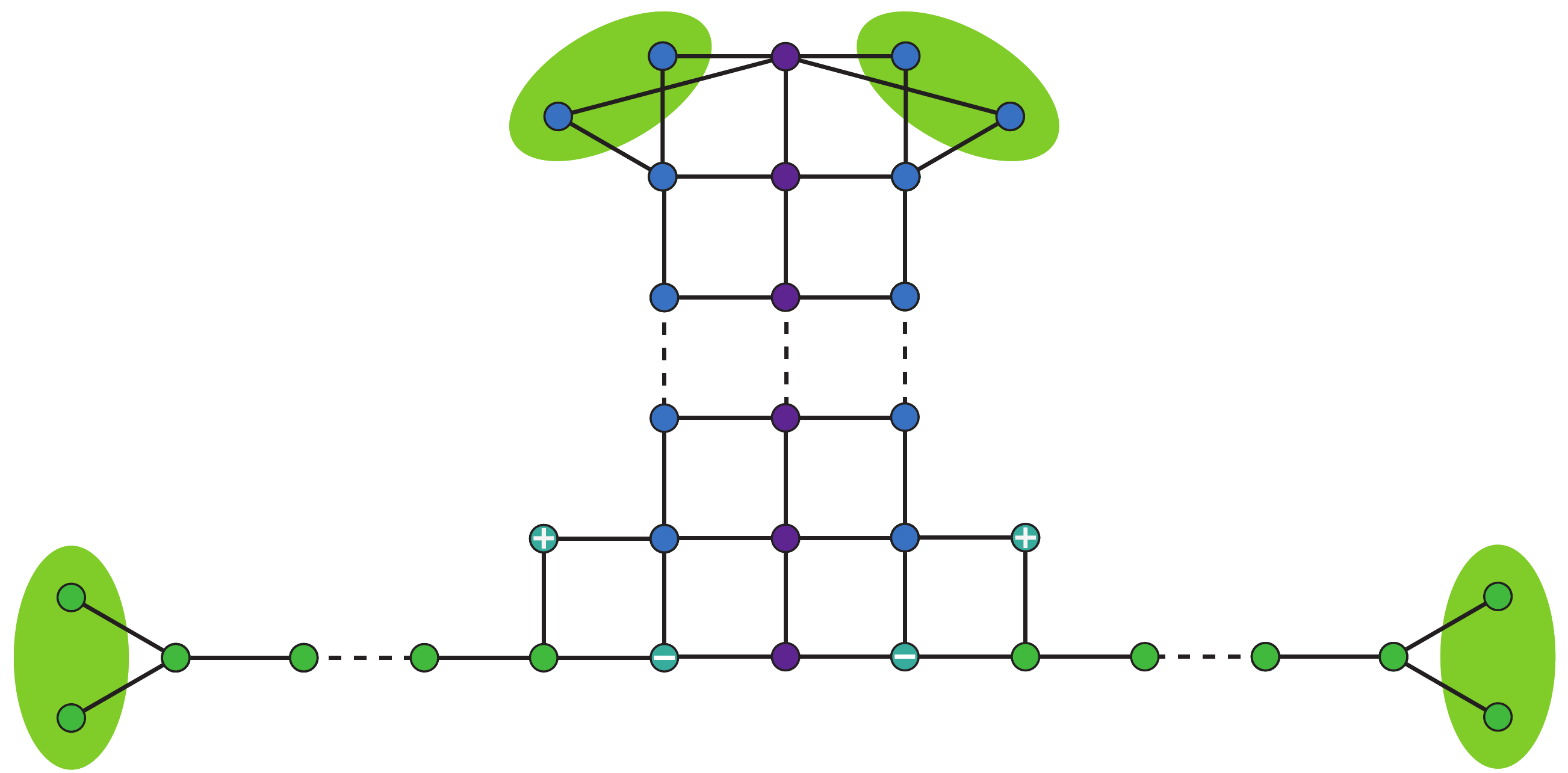}
\caption{The structure of the full deformed Y-system. The coupling between the two deformed Hubbard systems is via $k$ momentum carrying $Y_Q$-functions, where the $k$th one precisely couples to the end structure of the deformed Hubbard model, notably in a local fashion.}
\label{fig:DefYsys}
\end{center}
\end{figure}

Coming to the full $\ads$ model then, in the undeformed case the structure of the TBA equations is the one illustrated in figure \ref{fig:Thookundef} of chapter \ref{chapter:AdS5string}. As we will discuss below, the bound state structure of the full quantum deformed model precisely fits with the deformed Hubbard models and gives the nice coupling illustrated in figure \ref{fig:DefYsys}.

We will see that the deformed model shows various interesting features in addition to the fact that it can be described through a finite number of TBA equations. To start with, we already indicated that the S-matrix upon which our TBA equations are based is not unitary but pseudo-unitary. Nonetheless the TBA equations are real, which is no doubt due to the not only pseudo- but apparent quasi-unitarity of the mirror theory, briefly discussed in appendix \ref{app:qdefmatrixSmatrix}. Interestingly, it also turns out that the Y-system associated to this model is not strictly universal anymore; it picks up a dependence on the excitation numbers of the state under consideration. This appears to be the first instance of a model with such a dependence. As this effect appears to depend crucially on the combination of the deformation with the fact that we have a nested system, perhaps this is not too surprising. The Y-system also depends on chemical potentials associated to twisted boundary conditions, but this type of dependence is already seen in the XXZ spin chain \cite{Takahashi:1972}. Furthermore, the fact that it is possible to obtain a simplified TBA and Y-system equation for length $k$ mirror bound states is due to quite nontrivial cancellations, in particular involving the crossing relation as discussed in section \ref{sec:Ysysgeneral} and appendix \ref{app:sTBAYk}.

Below we will start by discussing the parametrization of the problem, followed by the Bethe-Yang equations. Then in section \ref{sec:boundstates} we discuss the bound states of the mirror model, followed by the complete string hypothesis and associated canonical TBA equations in section \ref{sec:string hypothesis}. We then give a very generic discussion of excited states in this deformed TBA, and show how this leads to a mildly non-universal Y-system. In section \ref{sec:Asymptoticsolution} we discuss the asymptotic solution of our Y-system and how the fusion relations of the asymptotic transfer matrices provide nontrivial checks of our equations. We would like to emphasize that our string hypothesis in principle contains the string hypothesis of the quantum deformed Hubbard model at roots of unity, which has not been investigated before.

\section{Kinematics of the \texorpdfstring{$q$}{}-deformed (mirror) model}
\label{sec:kinematics}

In this section we will discuss the deformation of the $z$-torus of chapter \ref{chapter:AdS5string} that parametrizes the fundamental representation of the centrally extended $\mathfrak{psu}_q(2|2)$ algebra and uniformizes dispersion relations of the $q$-deformed model and its mirror. Of course we will end up with deformed $x$-functions on the $u$-plane that we will use in practice.

\subsection*{The rapidity torus}

\begin{figure}[t]
\begin{center}
\includegraphics[width=0.8\textwidth]{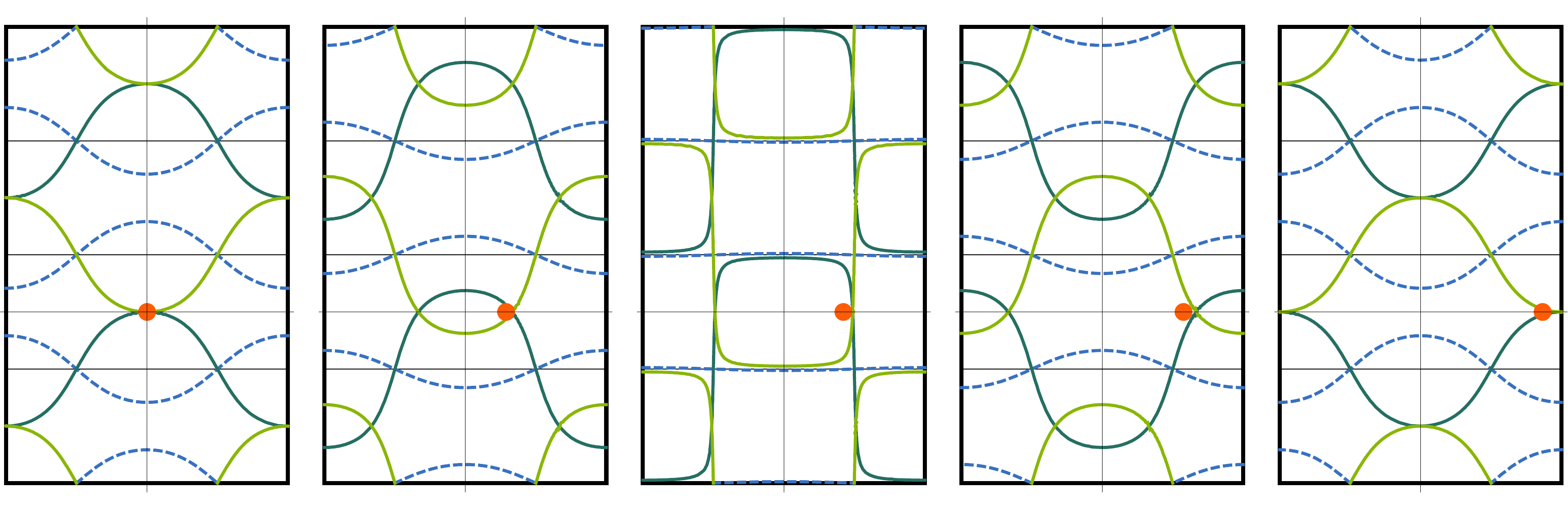}
\caption{
The division of the torus by the curves $|x^{\pm}|=1$ (blue, dashed) and $\mbox{Im}(x^{\pm})=0$ (two greens, solid) for five different values of $q$ on the unit circle. The orange dot on each plot indicates the position of the corresponding $z_0$; the middle plot corresponds to $z_0$ just below the value for which $q=i$. The shape of the `string' region ($|x^{\pm}|=1$) smoothly changes from the ``fish" to the ``hour-glass" shape as $z_0$ runs from $0$ to $\omega_1(\kappa)/2$, {\it i.e.} $q$ from $q=1$ to $q=-1$, but it does interchange the location of its crests and troughs at $q=i$. For the `mirror' region ($\mbox{Im}(x^{\pm})=0$) the change around and the situation at $q=i$ is much more dramatic as the figure shows.}
\label{fig:stringtorusqvariable}
\end{center}
\end{figure}

As in the undeformed case we can introduce $x^\pm$ variables to parametrize the fundamental representation of our now $q$-deformed algebra. These deformed $x^{\pm}$ variables satisfy the following constraint \cite{Beisert:2008tw,Beisert:2011wq}
\begin{equation}
\label{fc}
\frac{1}{q}\left(x^++\frac{1}{x^+}\right)-q\left(x^-+\frac{1}{x^-}\right)=\left(q-\frac{1}{q}\right)\left(\xi+\frac{1}{\xi}\right)\, ,
\end{equation}
where the parameter $\xi$ is related the coupling constant $g$ as
\begin{equation}
\xi=-\frac{i}{2}\frac{g(q-q^{-1})}{\sqrt{1-\frac{g^2}{4}(q-q^{-1})^2}}\, .
\end{equation}
As before, this constraint (which will become the dispersion relation) can be uniformized on an elliptic curve. This elliptic curve has real period $2\omega_1(\kappa)=4{\rm K}(m)$ and imaginary period $2\omega_2(\kappa)=4i{\rm K}(1-m)-4{\rm K}(m)$, where ${\rm K}(m)$ is the elliptic integral of the first kind considered as a function of the elliptic modulus $m=\kappa^2$. It is convenient for us to parametrize the deformation parameter $q$ by a variable $z_0$, so that
\begin{equation}
q=e^{i {\rm am}(2 z_0)}=\frac{\cso+i\dno}{\cso-i\dno}\, .
\end{equation}
Here and below we will use a concise notation for Jacobi elliptic functions; no subscript denotes the free variable in the equation, while the zero subscript indicates evaluation at $z_0$. For instance, ${\rm cs}(z_0)\equiv {\rm cs}_0$.
The modulus $\kappa$ is related to the coupling constant $g$ as
\begin{equation}
g=-\frac{i\kappa}{2 \dno}\sqrt{1-\kappa^2 {\rm sn}^4_0}\, .
\end{equation}
In the limit $q\to 1$, {\it i.e.} $z_0\to 0$, we recover the familiar relation $\kappa^2=-4g^2$. With these conventions, the variables $x^{\pm}$ are the following meromorphic functions on the $z$-torus
\begin{equation}
x^{+}(z)=\kappa\, {\rm sn}_0^2 \frac{\cs +\cso}{\dn -\dno}
\frac{\dn -i\cso}{\cs -i\dno} \, , ~~~~
x^{-}(z)=\frac{1}{\kappa\, {\rm sn}_0^2}\frac{\dn+\dno}{\cs-\cso}
\frac{\cs-i\dno}{\dn-i\cso} \, ,
\end{equation}
while $\xi$ is given by
\begin{equation}
\xi=-i\kappa\frac{\sno\cno}{\dno}\, .
\end{equation}
Since we want to take $q$ to be a root of unity we restrict ourselves to real $z_0$. In fact, $q(z_0)$ covers the unit circle once for
\begin{equation}
-\omega_1(\kappa)/2 \leq z_0 \leq \omega_1(\kappa)/2\, .
\end{equation}
Next, we take the modulus $\kappa$ to be purely imaginary with $\mbox{Im}\, \kappa>0$, so that the coupling constant $g$ is positive. Accordingly, $\xi$ lies between zero and one. Note that for real $z_0$ conjugation of the $x^\pm$ functions on the torus takes the form
\begin{equation}
x^{\pm}(z)^*=x^{\mp}(z^*)\,.
\end{equation}
As already indicated in the introduction we will take $q=e^{i \frac{\pi}{k}}$ with $k$ an integer greater than two. The special nature of $k=2$ can also be seen in figure \ref{fig:stringtorusqvariable}, where we see that the `string' and `mirror' type regions `flip their character' precisely at $q = i$.
The central charges of the fundamental representation are\footnote{The central elements $U$ and $V$ are related to the central elements $\mathbb{H}$ and $\mathbb{P}$ of the undeformed algebra as $V = q^{\mathbb{H}/2}$ and $U=e^{i \mathbb{P}/2}$.}
\begin{align}
U^2=& \frac{1}{q} \frac{x^+ + \xi}{x^- + \xi}=\frac{\cs+i\dno}{\cs-i\dno}=e^{i({\rm am} (z+z_0)+{\rm am}(z-z_0))}\, , \\
V^2=& q \frac{x^+}{x^-} \frac{x^- + \xi}{x^+ + \xi}=\frac{\cso+i\dn}{\cso-i\dn}=e^{i({\rm am} (z+z_0)-{\rm am}(z-z_0))}\,
\end{align}
and the shortening condition reads
\begin{equation}
\Big(\frac{V-V^{-1}}{q-q^{-1}}\Big)^2-\frac{g^2}{4}(1-U^2V^2)(V^{-2}-U^{-2})=1\, .
\end{equation}
The typical division of the torus by the lines $|x^\pm|=1$ and $\mbox{Im}(x^\pm)=0$ is illustrated in figure \ref{fig:torusvsplane} below, where we introduce the rapidity variable $u$.

\subsection*{The dispersion relations}

For the $\ads$ superstring the dispersion relation is naturally fixed by the shortening condition since the central elements are directly related to the energy, momentum and coupling constant of the string. As we are considering a deformed S-matrix without knowing of a concrete underlying interpolating theory we do not have such a direct relation here. However, as a deformation of the $q=1$ case it is natural to make the following identification
\begin{equation}
\label{eq:UandVtoEandP}
V=q^{\frac{H}{2}}\, , ~~~~~~U=e^{i\frac{p}{2}}\, ,
\end{equation}
where $H$ and $p$ are the energy and momentum of the model. For real $z$, the energy is a non-negative periodic function, while the momentum takes values in the interval $(-\pi,\pi)$ as $z$ runs along $(-\tfrac{\omega_1}{2},\tfrac{\omega_1}{2})$ as in the undeformed case. In the limit $q\to 1$ we smoothly obtain the undeformed string theory result  $p=2\, {\rm am}\, z$ and $H=\dn\, z$. In terms of $H$ and $p$ the shortening condition turns into the following dispersion relation
\begin{equation}
\label{Sdispersion}
e^2 [H/2]^2_{q}- g^2\sin^2\frac{p}{2}=\left[1/2\right]_q^2\, ,
\end{equation}
where we have introduced a manifestly positive ``coupling constant" $e$
\begin{equation}
e^2\equiv1-\frac{g^2}{4}(q-q^{-1})^2 = 1+ g^2 \sin^2 \frac{\pi}{k}\, .
\end{equation}
To obtain the dispersion relation for the mirror theory we perform a double Wick rotation, i.e.
\begin{equation}
\label{eq:mirrortf}
H\to i\tilde{p}\, , ~~~~p\to i\tilde{H}\, ,
\end{equation}
where $\tilde{p}$ and $\tilde{H}$ are the momentum and energy of the mirror model. In this way we find
\begin{equation}
\tilde{H}=2\, {\rm arcsinh}\frac{[1/2]_q}{g}\sqrt{1-e^2[i\tilde{p}]^2_{q^{1/2}}}\, .
\end{equation}
In particular, for  $q=e^{\frac{i\pi}{k}}$ the formula takes the form
\begin{equation}
\tilde{H}=2\, {\rm arcsinh}\left(\frac{1}{g}\frac{\sin \frac{\pi}{2k}}{\sin\frac{\pi}{k}}\sqrt{1+e^2\, \frac{\sinh^2\frac{\pi}{2k}\tilde{p}}{\sin^2\frac{\pi}{2k}}}\right)\, .
\end{equation}
Very importantly, the mirror momentum is real on the line at $\tfrac{\omega_2}{2}$ in the green region of the torus (see figure \ref{fig:torusvsplane} below), but \emph{not} on this line in the yellow region. This is perhaps surprising, but most certainly not a problem since this interval on the torus already covers the whole real line of mirror momenta. On this same interval the energy is positive and bounded from below. Let us note that as in chapter \ref{chapter:AdS5string} the mirror transformation on the rapidity torus is given by a shift of the $z$ variable by a quarter of the imaginary period.

To discuss the relativistic limit of our theory we should rescale the energy and momentum as $\tilde{H} \rightarrow  \tfrac{\tilde{H}}{g}$ and $\tilde{p} \rightarrow \tfrac{k}{\pi} \tfrac{\tilde{p}}{g}$ and take the limit $g\to \infty$. This gives the relativistic dispersion
\begin{equation}
\tilde{H}^2 - \tilde{p}^2 = \frac{1}{\cos^2{\tfrac{\pi}{2k}}}\,
\end{equation}
which agrees with \cite{Hoare:2012fc} up to a rescaling of $g$ by a factor of two. Interestingly, if we instead take the limit $g\to \infty$ without rescaling $\tilde{H}$ or $\tilde{p}$ we get a linear (phononic) dispersion
\begin{equation}
{\tilde H}=\frac{\pi}{k}|\tilde{p}|\, .
\end{equation}

\subsection*{The \texorpdfstring{$u$}{}-plane}

Let us finish the discussion of parametrization in one go and introduce the analogue of the $u$-plane and the corresponding deformations of $x(u)$ and $x_s(u)$. Of course these are the obvious deformations which as we will shortly see match nicely with the bound state structure of the deformed model.

\begin{figure}[h]
\begin{center}
\includegraphics[width=0.8\textwidth]{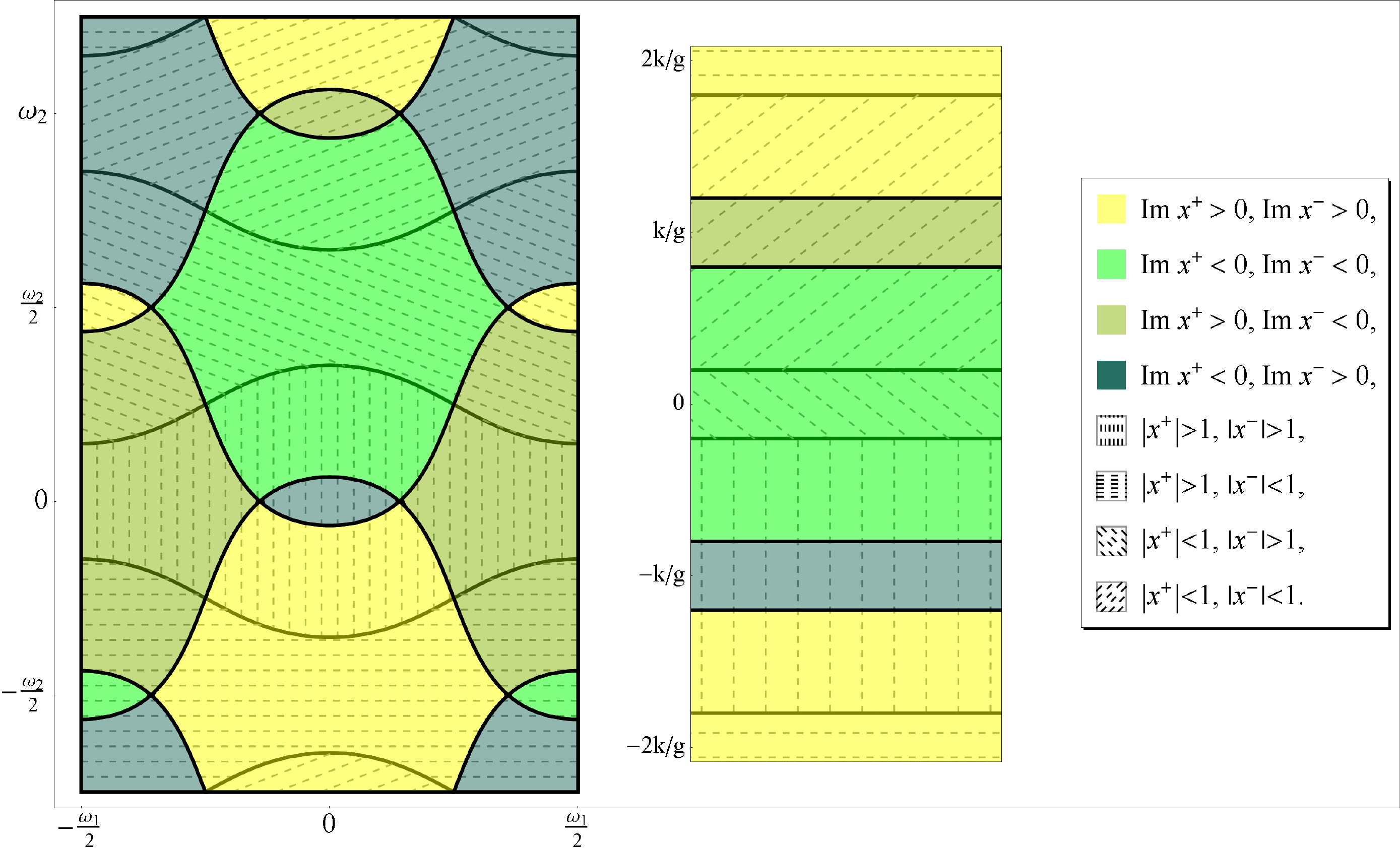}
\caption{The torus and the mirror $u$-plane. The left figure shows the torus with the distinguishing properties of the $x^\pm$ functions defined on it. The right figure depicts the mirror $u$-plane, with the same indications regarding $x^\pm$, illustrating the map between the mirror $u$-plane and the torus.}
\label{fig:torusvsplane}
\end{center}
\end{figure}

First, let us introduce a multiplicative evaluation parameter $\UU\equiv \UU(x)$ defined as
\begin{equation}
\label{ev}
\UU =-\frac{x+\frac{1}{x}+\xi+\frac{1}{\xi}}{\xi-\frac{1}{\xi}}\, ,
\end{equation}
so that we can determine $\UU(x^+)=q^2\,  \UU(x^-)$ and $x^{\pm}$ in terms of a single variable $x$ as
\begin{equation}
\UU(x^+)=q\,\UU(x)\, , ~~~~~\UU(x^-)=q^{-1}\UU(x)\, .
\end{equation}
On the torus we can express it as
\begin{equation}
\UU=\frac{\cso\dno+\cs\,\dn }{\cso\dno-\cs\,\dn}\, .
\end{equation}
Inverting eqn. (\ref{ev}) one solution is given by
\begin{align}
\label{eq:multxmirror}
x(\UU) = \frac{\xi-\tfrac{1}{\xi}}{2}\Big[\frac{1+\xi^2}{1-\xi^2}-\UU+i\sqrt{\Big(\UU-\frac{1-\xi}{1+\xi}\Big)\Big(\frac{1+\xi}{1-\xi}-\UU\Big)}\Big]\, .
\end{align}
This function has two branch points $0<\frac{1-\xi}{1+\xi}<1$ and $1<\frac{1+\xi}{1-\xi}$ since $\xi$ is positive and less than one. We chose the solution $x(\UU)$ such that the cuts lie on $(-\infty,\frac{1-\xi}{1+\xi})\cup (\frac{1+\xi}{1-\xi},+\infty)$.  As $\UU$ runs along the interval $(\frac{1-\xi}{1+\xi},\frac{1+\xi}{1-\xi})$ $x(\UU)$ spans the unit half-circle in the lower half plane. Correspondingly, $1/x(\UU)$ spans the upper half-circle. We can express the variables $x^{\pm}$ via $x(\UU)$ simply as\footnote{The parameters $x^{\pm}$ are two complex variables obeying one (complex) relation (\ref{fc}). Expressing $x^{\pm}$ in terms of $\UU$, we explicitly resolve this relation.}
\begin{equation}
x^{\pm}=x(q^{\pm 1} \UU)\, .
\end{equation}
This $x$ function maps the $\UU$-plane onto the green regions of the torus indicated in figure \ref{fig:torusvsplane}. As the branch cuts of $x(\UU)$ are straight lines, the branch cuts of $x^\pm (\UU)$ necessarily intersect and in this fashion cut of part of the $\UU$-plane. It is precisely the disconnected green region on the torus which corresponds to this disconnected region of the $\UU$-plane.

While the multiplicative $\UU$ parameter has its uses, typically it will be nicer for us to work in the additive setting of the $u$-plane parametrization $\UU=q^{-igu}=e^{\frac{\pi g u}{k}}$.\footnote{We will see shortly that this is exactly the analogue of the conventional hyperbolic parametrization of the XXZ model with $|\Delta|<1$.}  Since the shift $u\to u+\frac{2ik}{g}$ leads to the same value of $\UU$, the $u$-plane is an infinitely-sheeted cover of the $\UU$-plane. For this parametrization the map $x(\UU)$ turns into\footnote{We would like to note that this expression can be written more elegantly by rescaling $u\to k\tilde{u}/\pi g$ and introducing $g \sin \pi/k = \sinh\rho$, giving $x(\tilde{u}) = \frac{e^{\tilde{u}}-\cosh 2 \rho-i e^{\tilde{u}/2} \sqrt{2 \cosh 2 \rho -2 \cosh\tilde{u}}}{2 \cosh \rho \sinh \rho}$. In terms of this notation the branch points (see below) also greatly simplify to $\tilde{u}_b = \pm 2 \rho$. This notation slightly obscures the undeformed limit $k\to\infty$ however.}
\begin{equation}
\label{eq:qdefxmirror}
x(u)=\frac{e^{\frac{\pi g u}{2k}}\Big(\sinh\frac{\pi g u}{2k}-i\,  \sqrt{g^2\sin^2\frac{\pi}{k}-\sinh^2\frac{g\pi u}{2k} }\Big)-g^2\sin^2\frac{\pi}{k} }{g\sin\frac{\pi}{k}\sqrt{1+g^2\sin^2\frac{\pi}{k}}}\, ,
\end{equation}
where we have replaced $e^{\frac{\pi g u}{k}}$ which originally appears under the square root by $e^{\frac{\pi g u}{2k}}$ in front of it, removing a square root ambiguity. Because of this the function is no longer periodic with period $2ik/g$, but rather
\begin{equation}
\label{shift}
x\left(u+\tfrac{2ik}{g}\right)=\frac{1}{x(u)}\,
\end{equation}
meaning $x(u)$ is periodic on the $u$-plane with period $\frac{4 i k}{g}$. In other words, by resolving a square root ambiguity we have extended our $x$ function beyond the original mirror theory $\UU$-plane. Extended in this fashion the mirror $x$ function covers a full vertical band of the torus, illustrated in figure \ref{fig:torusvsplane}.
In fact, although we have yet to discuss it, eqn. (\ref{shift}) is nothing but the crossing transformation. Moreover, the scattering matrix of the $q$-deformed model (see appendix \ref{app:qdefmatrixSmatrix}) can be put on two copies of the $u$-plane where it is compatible with crossing symmetry and its matrix part is periodic with period $4ik/g$ in either argument.

\begin{figure}[h]
\begin{center}
\includegraphics[width=0.4\textwidth]{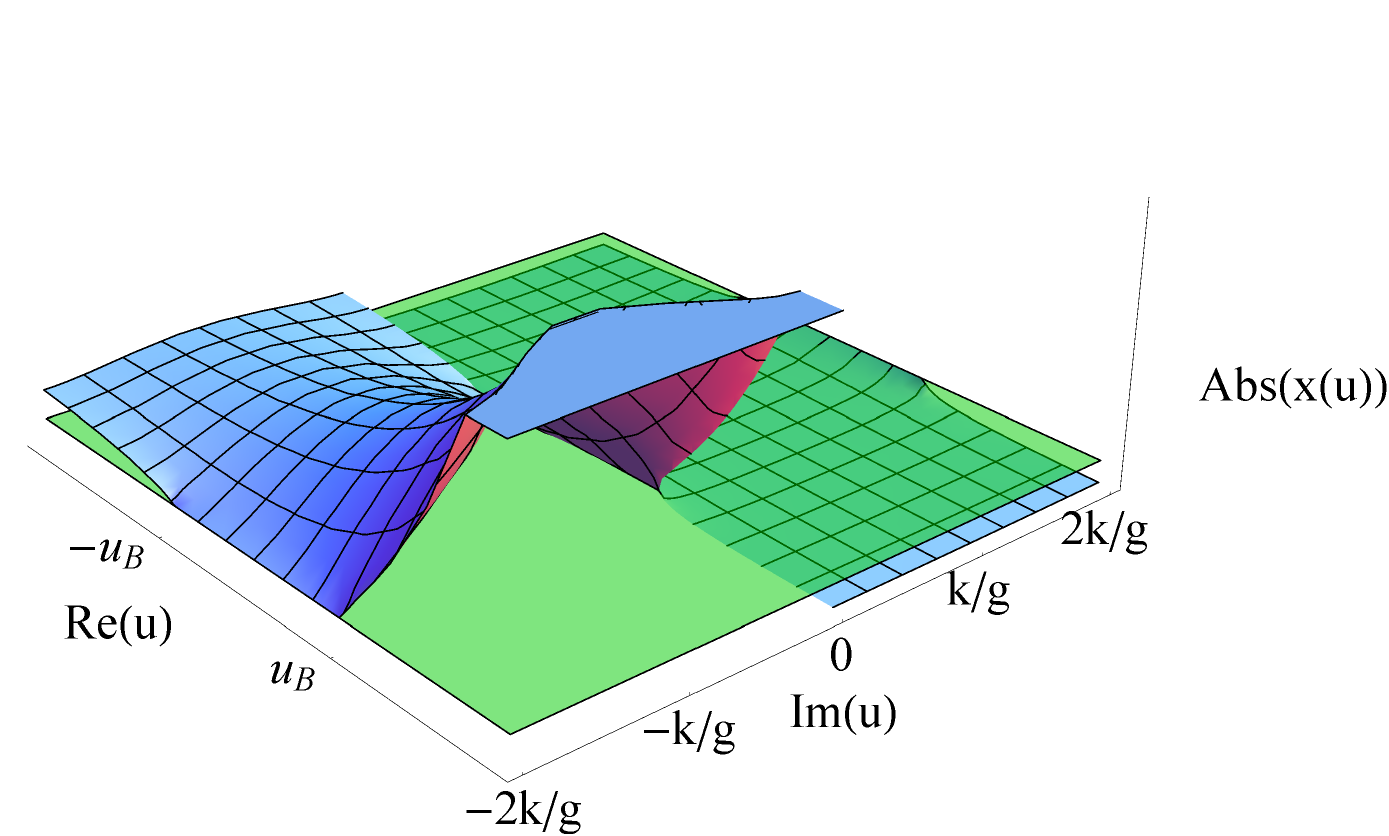}\hspace{15pt}
\includegraphics[width=0.4\textwidth]{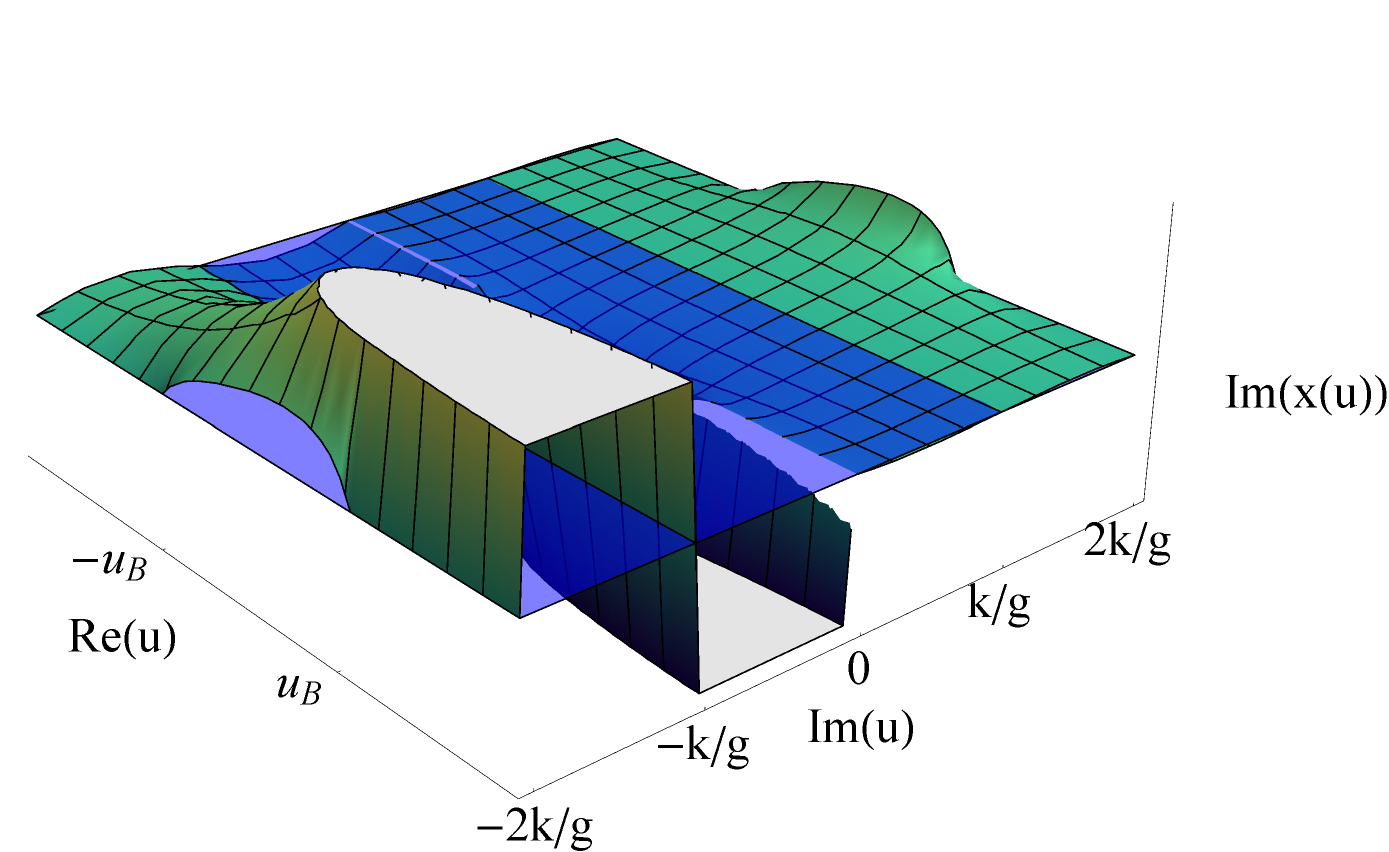}
\caption{The absolute value and imaginary part of $x(u)$ on the $u$-plane. The green surface distinguishes $|x|>1$ from $|x|<1$ in the left plot, while the blue surface distinguishes the sign of the imaginary part in the right plot.}
\label{fig:xmirror}
\end{center}
\end{figure}

The $x$ function has branch cuts on the $u$-plane running outward along the real line from $\pm u_b$ where
\begin{equation}
u_b = \frac{k}{\pi g} \log\frac{1+\xi}{1-\xi}=\frac{2k}{\pi g}{\rm arcsinh}\Big(g\sin\frac{\pi}{k}\Big)\, ,
\end{equation}
as well as outward from any of the points $\pm u_b + 2 \mathbb{Z} ik/g$ as illustrated in figure \ref{fig:xmirror}.
In the limit $k\to\infty$, the variable $x(u)$ tends to the standard mirror function of the undeformed theory (eqn.  \eqref{eq:xmirror}) and naturally has the conjugation property of the standard mirror function $x(u)^* = \frac{1}{x(u^*)}$. Note that the interval on the torus corresponding to real mirror momenta corresponds to the whole real line of the (mirror) $u$-plane.

Let us finally also introduce $x_s$, the $x$-function with complementary branch cuts to $x$. In terms of the $u$ variable $x_s$ is given by
\begin{equation}
\label{eq:qdefxstring}
x_s(u)=\frac{ \left(e^{\frac{\pi g u}{2 k}} \sinh\frac{\pi  g u}{2 k}-g^2 \sin^2\frac{\pi}{k}\right) \left(1+\sqrt{1-\frac{e^{\frac{\pi g u}{k}} g^2 \sin^2\frac{\pi }{k}  \left(1+g^2\sin^2\frac{\pi}{k}\right)}{\left(e^{\frac{\pi  g u}{2k}}\sinh \frac{\pi g u}{2 k}-g^2 \sin^2\frac{\pi }{k}\right)^2}}\right)}{\sqrt{\left(1+g^2 \sin^2\frac{\pi }{k}\right)g^2\sin^2\frac{\pi }{k}}}\, .
\end{equation}
Of course this function reduces to the one in eqn. \eqref{eq:xstring} in the undeformed limit.

\section{Bethe-Yang equations for the mirror model}

\label{sec:BetheYang}

As mentioned in the introduction, the S-matrix of our model has $\mathfrak{psu}_q(2|2)\oplus \mathfrak{psu}_q(2|2)$ symmetry and up to a scalar factor factorizes into two smaller S-matrices each invariant under $\mathfrak{psu}_q(2|2)$. As in the undeformed case we can try constrain this scalar factor by crossing and unitarity. Of course this does not fix the scalar factor uniquely, but a very natural generalization\footnote{Taking a different scalar factor corresponding to a different solution to the crossing equation would change the S-matrix. However, insisting on the proper $q \rightarrow 1$ limit we expect the modification not to change the bound state picture we describe below. Therefore the only modification would be a simple prefactor in the $\mathfrak{sl}(2)$ S-matrix which should not affect our results in any further way.} of the undeformed scalar factor that satisfies these requirements has been given in \cite{Hoare:2011wr}. We will take this scalar factor for our full S-matrix and proceed from there.

The mirror Bethe-Yang equations associated to this $q$-deformed S-matrix are given by
\begin{align}
\label{eq:qdefBYmain}
1& = e^{i\tilde{p}_l R} \prod_{i\neq l}^{K^{\mathrm{I}}} S_{\sls(2)}(x_l,x_i)\prod_{\alpha=1}^2\prod_{i=1}^{K^{\mathrm{II}}_{(\alpha)}} \sqrt{q}\frac{y_i^{(\alpha)} - x_l^-}{y_i^{(\alpha)} - x_l^+}\sqrt{\frac{x_l^+}{x_l^-}},\\
\label{eq:qdefBYy}
1&= \prod_{i=1}^{K^{\mathrm{I}}}\sqrt{q}\frac{y_m^{(\alpha)} - x^-_i}{y_m^{(\alpha)} - x^+_i}\sqrt{\frac{x^+_i}{x^-_i}}
\prod_{i=1}^{K^{\mathrm{III}}_{(\alpha)}}\frac{\sinh{\frac{\pi g}{2k}\big(v_m^{(\alpha)} - w_i^{(\alpha)}-\frac{i}{g}\big)}}{\sinh{\frac{\pi g}{2k}\big(v_m^{(\alpha)} - w_i^{(\alpha)}+\frac{i}{g}\big)}}\, ,\\
\label{eq:qdefBYw}
-1&= \prod_{i=1}^{K^{\mathrm{II}}_{(\alpha)}} \frac{\sinh{\frac{\pi g}{2k}\big(w_n^{(\alpha)} - v_i^{(\alpha)} + \frac{i}{g}\big)}}{\sinh{\frac{\pi g}{2k}\big(w_n^{(\alpha)} - v_i^{(\alpha)} - \frac{i}{g}\big)}}\prod_{j=1}^{K^{\mathrm{III}}_{(\alpha)}}\frac{\sinh{\frac{\pi g}{2k}\big(w_n^{(\alpha)} - w_j^{(\alpha)} - \frac{2i}{g}\big)}}{\sinh{\frac{\pi g}{2k}\big(w_n^{(\alpha)} - w_j^{(\alpha)} + \frac{2i}{g}\big)}}\, ,
\end{align}
where $\alpha=1,2$ and $\tilde{p}$ is the mirror momentum defined in (\ref{eq:UandVtoEandP},\ref{eq:mirrortf}), and
\begin{equation}
e^{\frac{\pi g v}{k}}= \V = -\frac{y+\tfrac{1}{y} + \xi +\tfrac{1}{\xi}}{\xi - \tfrac{1}{\xi}}\,.
\end{equation}
We will come back to the scalar factor $S_{\sls(2)}$ when discussing bound states in the next section; the full deformed matrix S-matrix and its properties are summarized in appendix \ref{app:qdefmatrixSmatrix} (see also appendix \ref{app:qdefdressingphase} for details on the dressing phase).

The auxiliary problem corresponds to two copies of the quantum deformed Hubbard model and the associated transfer matrix is given below (eqn. \eqref{eq:qdeffulltransfer}). The auxiliary Bethe equations follow from analyticity of this S-matrix. Let us note that the second set of auxiliary equations is identical to those of the inhomogeneous Heisenberg XXZ spin chain; the limit $v_i \rightarrow 0$ gives the homogeneous XXZ spin chain. Of course by construction these equations reduce to the undeformed eqs. (\ref{eq:mirrorBYmain}-\ref{eq:mirrorBYw}) in the limit $k \to \infty$.

\section{Bound states of the mirror theory}
\label{sec:boundstates}

In order to discuss the thermodynamics of the mirror theory in the infinite volume limit, we need to determine the spectrum of excitations that make up the thermodynamic ensemble in infinite volume. In this section we discuss the spectrum of physical excitations of the infinite volume mirror model, while in the next section we join these with the spectrum of auxiliary excitations whose analysis we have already partially done at the start of this chapter, resulting in our string hypothesis.

\subsection*{Two-particle bound states}

To discuss the physical bound states of the mirror theory we need to analyze consistency of the Bethe equations in the limit $R\to \infty$. In the absence of auxiliary roots, the main Bethe equation (\ref{eq:qdefBYmain}) takes the form
\begin{align}
\label{eq:mainagain}
1& = e^{i\tilde{p}_lR} \prod_{i\neq l}^{K^{\mathrm{I}}} S_{\sls(2)}(x_l,x_i)\, ,
\end{align}
where the S-matrix corresponding to the $q$-deformed analogue of the $\sls(2)$ sector of string theory is given by
\begin{equation}
\label{sl2}
S_{\sls(2)}(x_1,x_2)=\sigma^{-2}\frac{x_1^+-x_2^-}{x_1^--x_2^+}\frac{1-\frac{1}{x_1^-x_2^+}}{1-\frac{1}{x_1^+x_2^-}}=
\frac{\sinh\frac{\pi g}{2k} (u_1-u_2+\frac{2i}{g})}{\sinh\frac{\pi g}{2k}(u_1-u_2-\frac{2i}{g})}\left(\frac{1-\frac{1}{x_1^-x_2^+}}{1-\frac{1}{x_1^+x_2^-}}\sigma^{-1}\right)^2\, ,~~
\end{equation}
and $\sigma$ is the dressing factor. An explicit formula for $\tilde{p}(u)$ is given below in eqn. \eqref{eq:ptildeofu}. For complex values of momenta the S-matrix exhibits a pole at $x_1^-=x_2^+$. At the level of the Bethe equations this pole is accompanied by a divergence of the momentum factor in the limit $R\to\infty$. Indeed, if the first particle has a momentum with positive imaginary part, the factor $e^{i\tilde{p}_1R}$ in \eqref{eq:mainagain} goes to zero, but the total equation remains finite when accompanied by a pole at $x_1^-=x_2^+$. If the second particle has a momentum with negative imaginary part such that the resulting total momentum $\tilde{p}_1+\tilde{p}_2$ is real, we obtain a two-particle bound state.

\begin{figure}[h]
\begin{center}
\includegraphics[width=0.6\textwidth]{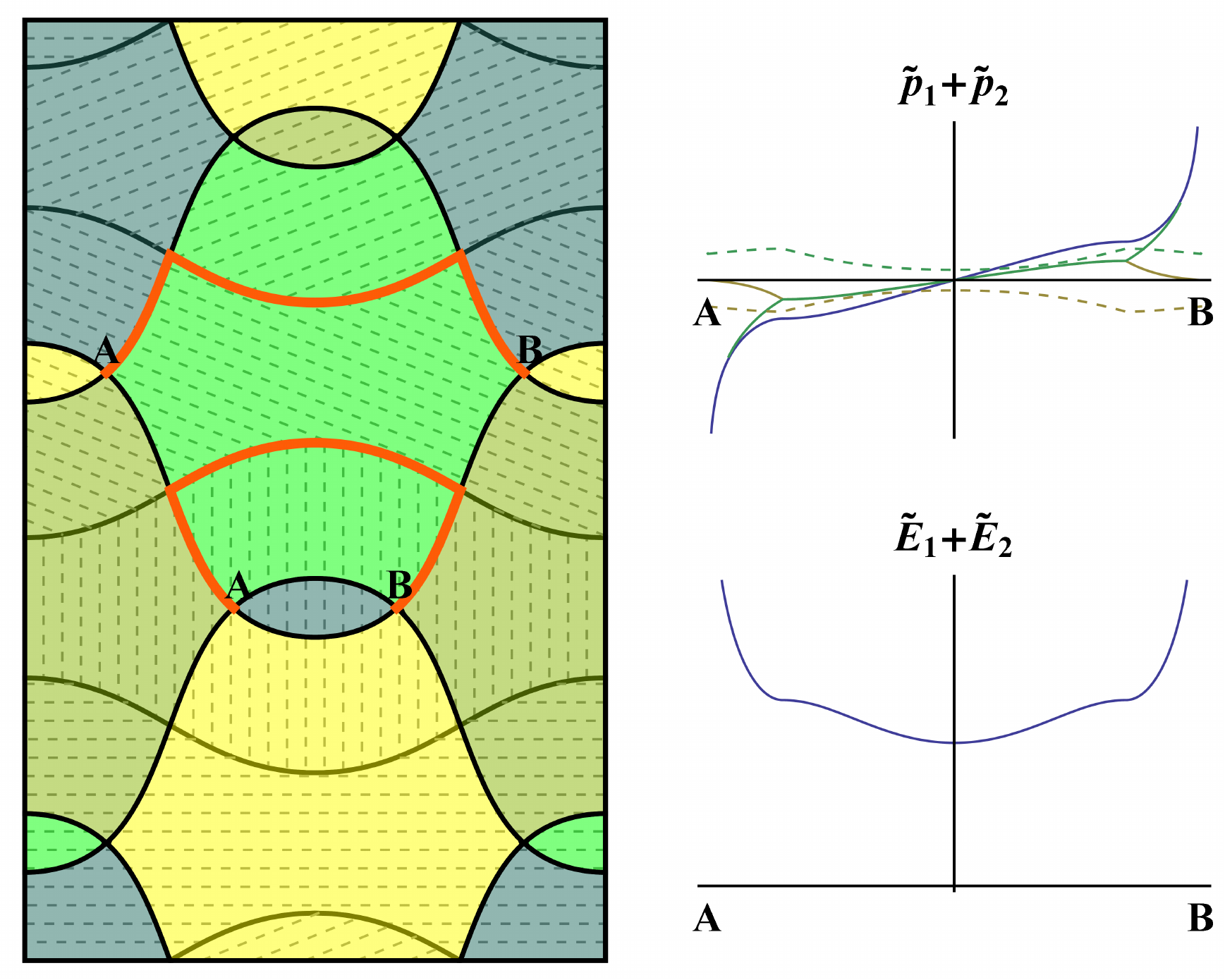}
\caption{The two-particle bound state on the torus with its momentum and energy. The constituents of the bound state lie along the orange curves, where $\tilde{p}=\tilde{p}_1+\tilde{p}_2$ increases from $-\infty$ to $\infty$ as we move from point $A$ to $B$. The kinks corresponds to critical values of the momentum, beyond which the two constituent momenta are no longer complex conjugate. This is shown in the upper right graph where the green and yellow lines show the real (solid) and imaginary (dashed) parts of the constituent momenta.}
\label{fig:2partboundstate}
\end{center}
\end{figure}

The two-particle bound state condition can be solved in terms of the torus $z$-variable and the solution corresponding to the simplest two-particle bound state is shown in figure \ref{fig:2partboundstate}. As in the undeformed model, there is a critical value of the particle momenta; below the critical value  the rapidities of the constituent particles are conjugate to each other (with respect to the real line of the mirror theory) while beyond the critical value the conjugation property is lost. Because of this, in addition to the solution indicated in figure  \ref{fig:2partboundstate} there is another solution corresponding to
reflecting the picture in the mirror line.  This behavior is completely analogous to the undeformed case \cite{Arutyunov:2007tc},
and has been observed in the present context in \cite{Hoare:2012fc}.

The critical value of the momenta for the two-particle bound state is given by $\tilde{p}_{cr}=\tilde{p}_2(u_b)$. Explicit computation gives
\begin{equation}
\tilde{p}_{cr}=\frac{k}{\pi}\log\frac{\cos^2\frac{\pi}{k}+g^2\sin^2\frac{\pi}{k} +\rho_1\sin^2\frac{\pi}{k}+\rho_2\sin\frac{\pi}{k} }{1+g^2\sin^2\frac{\pi}{k}}\, ,
\end{equation}
where
\begin{align}
\rho_1=&\sqrt{1+4g^2+4g^4\sin^2\frac{\pi}{k}}\, ,  \\
\rho_2=&\sqrt{-2\cos^2\frac{\pi}{k} +2g^2\sin^2\frac{\pi}{k} +4g^4\sin^4\frac{\pi}{k} +2(\cos^2\frac{\pi}{k}+g^2\sin^2\frac{\pi}{k})\rho_1 }  \, .
\end{align}
In the limit $k\to \infty$ we get $\tilde{p}_{cr}=\sqrt{-2+2\sqrt{1+4g^2}}$ which coincides with the result obtained in \cite{Arutyunov:2007tc}.

Note that the two-particle mirror bound state trajectory, together with part of the two-particle string bound state trajectory which lies along the boundary of the small blue region in figure \ref{fig:2partboundstate} isolates a region of the torus which can be naturally called the physical mirror region of the theory.

\subsection*{Multi-particle bound states}

In case the momentum of the two-particle bound state is not real we will necessarily have to involve a third particle to render the first two equations finite; the bound state grows. Without loss of generality we can take $\tilde{p}_1+\tilde{p}_2$ to have positive imaginary part\footnote{Otherwise we could have equivalently analyzed the poles of $S^{-1}$ starting with the Bethe-Yang equation for the second particle.}, giving a zero that is cancelled by the pole $x_2^-=x_{3}^+$.  At this point either the total momentum is real again, giving a three-particle bound state, or we add a fourth particle and continue the process. In this way, we can obtain a $Q$-particle bound state defined by the conditions
\begin{equation}
x_1^-=x_2^+\, , ~~~x_2^-=x_3^+\, , \,\, \ldots \,\, , x_{Q-1}^-=x_Q^+\, .
\end{equation}
The question now becomes which bound states are physical.

As in the undeformed case, we can insist that the bound state condition has a unique solution in the physical region of our theory. The physical mirror region referred to just above, the big green region of figures \ref{fig:torusvsplane} and \ref{fig:2partboundstate}, has this property, and contains bound states length of up to and including $k$. Were we to go beyond length $k$ we would first enter the lower blue and upper olive-green regions of the torus, as will become clear shortly when we consider bound states on the $u$-plane (\emph{cf.} eqn. \eqref{uj} just below). These regions contain in particular fundamental particles and on their boundaries two particle bound states of the string and anti-string theory, making it undesirable to consider proceeding into them. Were we to continue even further, we would enter what we can by now refer to as the anti-mirror region and we would manifestly lose the uniqueness of the solution to the bound state equation. Moreover, this would correspond to constructing bound states containing both particles and anti-particles. For these reasons we define the physical region of the mirror theory as the large green region on the torus of figures \ref{fig:torusvsplane} and \ref{fig:2partboundstate}. On the $u$-plane this corresponds to the strip $|\mbox{Im}(u)|\leq k-1$, as is clear from figure \ref{fig:torusvsplane}.

\begin{figure}[h]
\begin{center}
\includegraphics[width=0.4\textwidth]{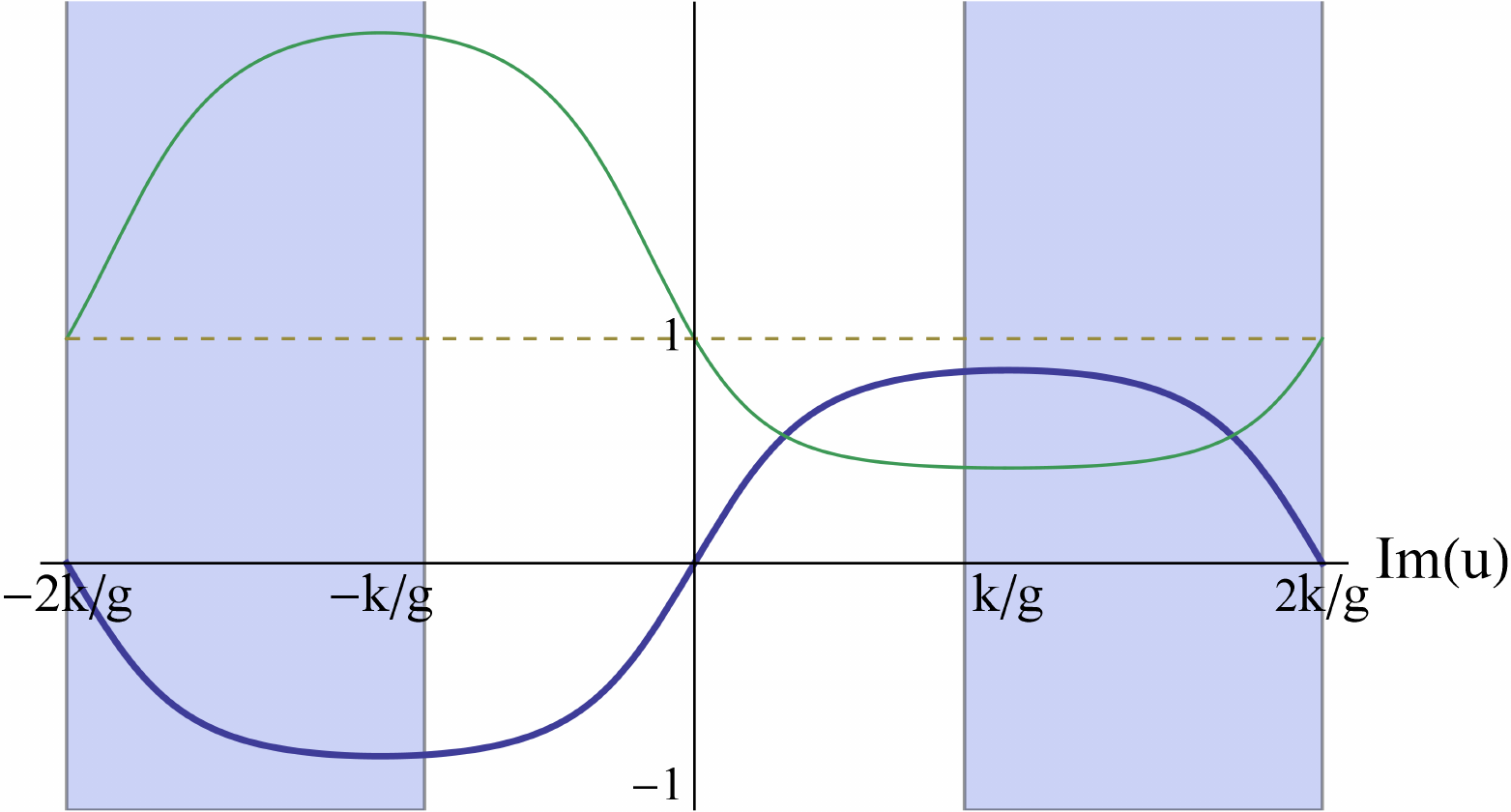}\hspace{15pt}\includegraphics[width=0.4\textwidth]{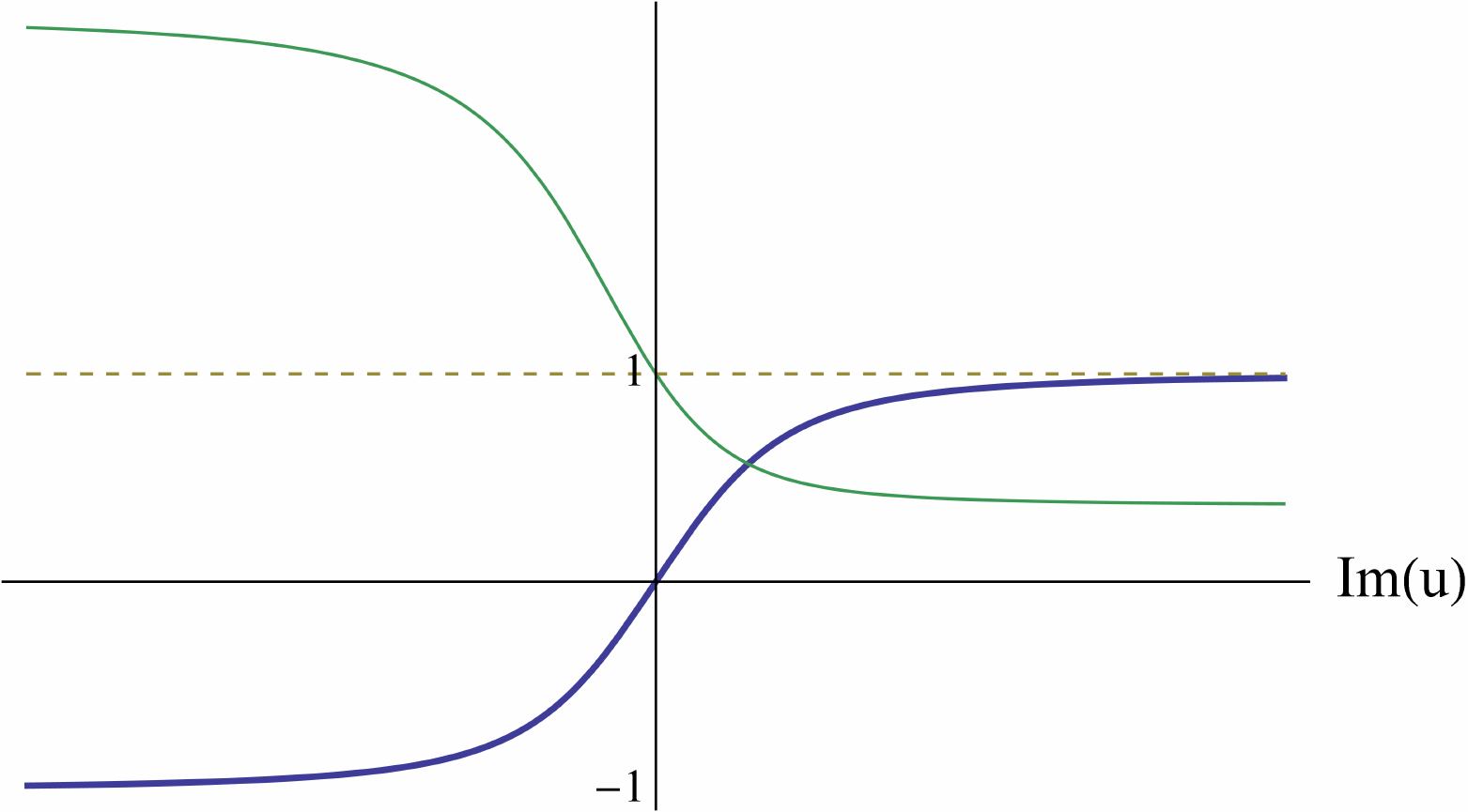}
\caption{The left plot shows the imaginary part of $\tilde{p}(u)$ in blue for some representative value of $\mbox{Re}(u)$, the thin green line shows $|e^{i \tilde{p}}|$. The shaded area of the plot lies outside the physical mirror region. The right plot shows the same functions for the undeformed model. The maximal value of ${\rm Im}\, \tilde{p}$ in reached at $\mbox{Im}(u)=k/g$ in the deformed model and at $\mbox{Im}(u)=\infty$ in the undeformed one.}
\label{fig:immom}
\end{center}
\end{figure}

For a more concrete discussion let us put our bound states on the $u$-plane. There the set of roots making up a $Q$-particle bound state takes the standard form of the Bethe string
\begin{equation}
\label{uj}
u_j=u+\frac{i}{g}(Q+1-2j)\, , ~~~j=1,\ldots, Q\, .
\end{equation}
The pole structure of these bound states is compatible with the Bethe-Yang equations since the mirror momentum $\tilde{p}(u)$ has positive imaginary part in particular for $0<\mbox{Im}(u)<k-1$ as illustrated in figure \ref{fig:immom}. This means that the imaginary part of the momentum of the first half of the particles is positive for all bound states we consider.\footnote{At this point the attentive reader might be slightly concerned that by these considerations higher length bound states also appear to be allowed, at least as far as the imaginary part of the mirror momentum is concerned, while we just argued they should not be. The \emph{only} reason we are even pondering this question at this point is that due to the deformation the $u$-plane can naturally cover both the mirror and anti-mirror region, making it possible to smoothly consider the mirror momentum along the full imaginary direction of the torus in one go. In the undeformed limit, we could construct a very similar picture, but (un)fortunately it can only be obtained by gluing two entire planes together. Colloquially speaking, fusing through the line $ik/g$ is like fusing through infinity in the undeformed case.}

Hence in the limit $R\to\infty$ we have $Q$-particle bound states of length up to and including $k$ defined by the following equations\footnote{Note that due to the periodicity introduced by the deformation, this type of string configuration for $Q = k$ necessarily implies $x_1^+ = 1/x_k^-$. This appears to contradict the string hypothesis since the improved dressing phase $\Sigma(x_1,x_k)\equiv \tfrac{1-1/{x_1^+x_k^-}}{1-1/{x_1^-x_k^+}}\sigma(x_1,x_2)$ has an apparent zero there. However, as the name suggests the improved dressing phase is perfectly finite at these points; the zero is precisely cancelled by a pole in $\sigma(x_1,x_2)$.}
\begin{equation}
x_1^-=x_2^+\, , ~~~x_2^-=x_3^+\, , ~~ \ldots\, , x_{Q-1}^-=x_Q^+\, .
\end{equation}
These $Q$-particle bound states transform in short representations of the symmetry algebra with central charges
\begin{equation}
U_Q^2 =  \frac{1}{q^Q} \frac{x^+ + \xi}{x^- + \xi}\, , ~~~~~~~~
V_Q^2 =  q^Q \frac{x^+}{x^-} \frac{x^- + \xi}{x^+ + \xi}\, .
\end{equation}
Here the variables $x^\pm = x(u\pm i Q/g)$ satisfy the relation
\begin{equation}
\label{fcb}
\frac{1}{q^Q}\left(x^++\frac{1}{x^+}\right)-q^Q\left(x^-+\frac{1}{x^-}\right)=\left(q^Q-\frac{1}{q^Q}\right)\left(\xi+\frac{1}{\xi}\right)\, .
\end{equation}
The associated mirror momentum and energy are given by
\begin{equation}
\tilde{p}_Q =  -\frac{k}{\pi} \log V_Q^2\, ,\vspace{20pt}
\tilde{\mathcal{E}}_Q =  - \log U_Q^2\, .
\end{equation}
Both the energy and momentum are real quantities for real values of $u$, as we can easily see from
\begin{equation}\label{cr}
\Big[x^+(u)\Big]^*=\frac{1}{x^-(u^*)}\, .
\end{equation}
On the $u$-plane they are explicitly given by
\begin{align}
\label{eq:ptildeofu}
\tilde{p}_Q=&\,-\frac{k}{\pi}\log\frac{\cosh\frac{\pi g}{2k}(u+\tfrac{iQ}{g})-i\sqrt{g^2\sin^2\frac{\pi}{k} -\sinh^2\frac{\pi g}{2k}(u+\tfrac{iQ}{g})}}
{\cosh\frac{\pi g}{2k}(u-\tfrac{iQ}{g})-i\sqrt{g^2\sin^2\frac{\pi}{k} -\sinh^2\frac{\pi g}{2k}(u-\tfrac{iQ}{g})}} \, ,\\
\label{eq:Etildeofu}
\tilde{\cal E}_Q=&\,-\log\frac{\sinh\frac{\pi g}{2k}(u+\tfrac{iQ}{g})-i\sqrt{g^2\sin^2\frac{\pi}{k} -\sinh^2\frac{\pi g}{2k}(u+\tfrac{iQ}{g})}}
{\sinh\frac{\pi g}{2k}(u-\tfrac{iQ}{g})-i\sqrt{g^2\sin^2\frac{\pi}{k} -\sinh^2\frac{\pi g}{2k}(u-\tfrac{iQ}{g})}} \, .
\end{align}

\section{The string hypothesis and TBA equations}
\label{sec:string hypothesis}

In this section we describe the spectrum of the infinite volume mirror model and derive the TBA equations describing thermodynamics of the mirror theory.

\subsection{Physical excitations of the mirror model}

In the previous section we analyzed the bound states of the mirror theory. We also encountered the string hypothesis for the XXZ model in section \ref{sec:XXZref} already. To consider what happens in our model we should consider the full set of coupled Bethe-Yang equations (\ref{eq:qdefBYmain}-\ref{eq:qdefBYw}). Looking at these equations we realize that just as in the undeformed case there cannot be bound states of $Q$ and $y$-particles, leaving us with just the auxiliary problem. As we did for the XXZ spin chain, let consider the auxiliary equations in the multiplicative parametrization
\begin{align}
\label{eq:auxy}
1&= \prod_{i=1}^{K^{\mathrm{I}}}\sqrt{q}\frac{y_m - x^-_i}{y_m - x^+_i}\sqrt{\frac{x^+_i}{x^-_i}}
\prod_{j=1}^{K^{\mathrm{III}}}\frac{\V_m - q \W_j}{q \V_m - \W_j}\, ,\\
\label{eq:auxw}
-1&= \prod_{i=1}^{K^{\mathrm{II}}} \frac{q \W_n - \V_i}{\W_n - q \V_i} \prod_{j=1}^{K^{\mathrm{III}}}\frac{\W_n - q^2 \W_j}{ q^2 \W_n - \W_j }\, ,
\end{align}
where $\V = q^{-igv}$ and $\W = q^{-igw}$. Analyzing the equations for $y$-particles we realize that they are non-singular in the limit $K^{\mathrm{I}} \rightarrow \infty$ if $|y|=1$, which means $v \in (-u_b,u_b)$ and in particular $\V>0$. Looking back at the analysis we did for the XXZ spin chain, we realize that it directly applies in the presence of $y$-particles which have $\V>0$ so that our $\W(w)$-strings should be those of the XXZ model. However, for bound states involving $y$-particles the reasoning becomes a bit involved because there are two $y$s associated to every $\V$; we present the corresponding analysis in detail in appendix \ref{subsec:StringHypo-vw}. The result is a very natural mix of the string hypothesis for the Hubbard model and that of the XXZ spin chain; we can have bound states of $y$-particles and $w$-particles in the form of $vw$-strings for which the $w$-particles are constrained exactly as in the XXZ spin chain.

Putting this together with the bound states of the previous section, we propose that the spectrum of excitations of the $q$-deformed mirror model in the thermodynamic limit is given by
\begin{enumerate}
\item{Bound states of fundamental particles; $Q$-particles,}
\item{Complexes of $y$ and $w$-particles; $M|vw$-strings,}
\item{Complexes of $w$-particles; $M|w$-strings,}
\item{Single $y$-roots; $y$-particles.}
\end{enumerate}
The significant difference with the string hypothesis for the undeformed string mirror model is that the length and type of such complexes is constrained.

\paragraph{\underline{$Q$-particles}} are bound states of fundamental particles, which in the present context can have a maximal length of $Q=k$. The rapidities representing possible $Q$-particle bound state are given by
\begin{equation}
\{ u \} = \{u +\frac{i}{g}(Q+1-2j) \, | \, j=1,\ldots,Q \} \, , \, \, \, Q =1,\ldots,k \, .
\end{equation}

\paragraph{\underline{$vw$-strings}} are most naturally classified in as two types; string of length less than $k$ and strings of length one with a particular type of complex rapidity. As in the XXZ spin chain, we will refer to these as strings with positive and negative parity respectively. The roots constituting possible positive parity $M|vw$ string are given by
\begin{eqnarray}
\{w\}&=&\{ v + \frac{i}{g}(M+1-2j)\, | \, j=1,\ldots,M\}\, ,
\nonumber \\
\{v\}&=&\{v + \frac{i}{g}(M-2j)\, | \, j=1,\ldots,M-1\}\cup\{v + \frac{i}{g} M , v - \frac{i}{g} M \}\, ,
\end{eqnarray}
for $M=1, \ldots, k-1$ and $v$ real, whereas the negative parity length one string has its center on the line $i k/g$.\footnote{This is not a strange type of solution, merely a consequence of parametrization. In multiplicative $\UU$-type variables this simply corresponds to negative roots which are perfectly allowed in the Bethe equations.} To each of the $v$ rapidities in the first set of the second line we associate two $y$-roots, one with $|y|>1$ and one with $|y|<1$, while the single $y$-roots associated to $v+ i M/g$ and $v- i M/g$ have $|y|<1$ and $|y|>1$ respectively. In terms of the mirror $x$ function we have $y_{\pm M} = x(v \pm i M/g)$ for positive parity strings, while $y_{1} = x(v+i(k+1)/g)$ and $y_{-1} = \frac{1}{x(v+i(k-1)/g)}$ with $v$ real for negative parity strings.

\paragraph{\underline{$w$-strings}} come in the same types as $vw$-strings, which is no coincidence as we explain in the appendix. The set of $w$-particles making up a $M|w$ string is given by
\begin{equation}
\{w\}=\{w + i (M+1-2j)/g\}\, ,~~~~j=1,\ldots, M\, ,
\end{equation}
where $M=1, \ldots, k-1$ and $w$ real or $M=1$ with $w$ on the line $i k/g$ for positive respectively negative parity strings.

\paragraph{\underline{$y$-particles}} are characterized by the property that $|y|=1$.  This means their associated rapidities $v$ run over the interval $(-u_b,u_b)$. In terms of the mirror $x$ function we have $y^- = x(v)$ while $y^+ = \frac{1}{x(v)}$, where the $\pm$ denotes the sign of the imaginary part of $y$.

\paragraph{\underline{Negative parity strings}} turn out to disappear rather naturally from the problem; their Y-functions are almost inverse to those of strings of length $k-1$. This fact already manifests itself at the level of the string hypothesis; by fusing a length one negative parity $(v)w$-string with a length $k-1$ $(v)w$ string we get a configuration with momentum $\pi$. Thinking about this for a second we realize this means the product of the Y-functions carries no dynamics. This immediately follows when we realize that the negative parity strings scatter inversely to positive parity length $k-1$ strings
\begin{align}
\label{eq:Kproperties11}
\frac{d}{du} \log S_\chi^{0N}(u,v) & = - \frac{d}{du} \log S_\chi^{k-1,N}(u,v) \, ,\\
\label{eq:Kproperties12}
\frac{d}{du} \log S_\chi^{M0}(u,v) & = - \frac{d}{du} \log S_\chi^{M,k-1}(u,v) \, ,\\
\label{eq:Kproperties13}
\frac{d}{du} \log S^{00}(u,v) & = \frac{d}{du} \log S^{11}(u,v) \, ,
\end{align}
where the last equation holds simply by the definition of type $0$ $w$ and $vw$-strings, and $\chi$ is a generic label denoting any relevant S-matrix.

Given this string hypothesis and the Bethe-Yang equations (\ref{eq:qdefBYmain}-\ref{eq:qdefBYw}) we can follow the general derivation in chapter \ref{chapter:finitevolumeIQFT} to find the TBA equations. The only subtlety we would encounter is that because of relations (\ref{eq:Kproperties11}-\ref{eq:Kproperties13}) the counting functions for negative parity strings need to be defined oppositely from those of positive parity strings to have positive densities. Going through the derivation while carefully keeping track of the consequences of  (\ref{eq:Kproperties11}-\ref{eq:Kproperties13}) we get a set of coupled integral equations for a set of Y-functions, one for each particle type in the spectrum we just discussed; $k$ $Y_Q$ functions for $Q$-particles, $Y_\pm$ functions for $y$-particles with positive, respectively negative imaginary part, and $k$ $Y_{M|(v)w}$ functions for $(v)w$-strings. Of course we have a set of $Y_\pm$ and $Y_{M|(v)w}$ functions for both values of $\alpha$. The index on the $Y_{M|(v)w}$ functions runs from \underline{zero} to $k-1$, where we denote negative parity strings separately as the \underline{zeroth} type of strings.

Note that given relations (\ref{eq:Kproperties11}-\ref{eq:Kproperties13}) we will only have integration kernels with indices greater than zero unless specifically indicated. Hence in what follows the indices $M$,$N$ and $L$ run from one to $k-1$, while $Q$,$P$ and $R$ run from one to $k$; repeated indices indicate a standard sum. We also define our convolutions in accordance with the slightly moved branch points on the $u$ plane, namely
\begin{align}
f\star h(u,v)=\,&\int_{-\infty}^{\infty}\, dt\, f(u,t)h(t,v) \, , \\
f\, \hat{\star}\,  h(u,v)=\,&\int_{-u_b}^{u_b}\, dt\, f(u,t)h(t,v)\, , \\
f\, \check{\star}\, h(u,v)=\,&\int_{-\infty}^{-u_b}\, dt\, f(u,t)h(t,v)+\int_{u_b}^{\infty}\, dt\, f(u,t)h(t,v)\, .
\end{align}

\subsection{The canonical TBA equations and free energy}

\paragraph{The TBA equations for $Q$-particles} are

\begin{align}
\label{eq:cTBAQ0}
\log Y_Q = \, &  - \mu_{Q} -L\, \tilde{\mathcal{E}}_{Q} + \log \left(1+Y_P \right)\star K_{\mathfrak{sl}(2)}^{PQ} + \sum_a \log \left(1+\tfrac{1}{Y_{M|vw}^{(a)}} \right)\star  K^{MQ}_{vwx} \\ \nonumber
&  + \sum_a \log\left(1+\tfrac{1}{Y_{0|vw}^{(a)}} \right) \star  K^{k-1,Q}_{vwx} + \sum_a \log\left(1-\tfrac{1}{Y_{\beta}^{(a)}} \right)\, \hat{\star}\, K^{yQ}_\beta\, ,
\end{align}
where we have an implicit sum over $\beta=\pm$.

\paragraph{The TBA equations for $y$-particles} are
\begin{equation}
\label{eq:cTBAypm}
\log{Y_{\pm}^{(a)}} = - \mu^{(a)}_{\pm} -\log\left(1+Y_{Q}\right)\star K^{Qy}_{\pm} + \log\frac{1+\tfrac{1}{Y_{M|vw}^{(a)}}}{1+\tfrac{1}{Y_{M|w}^{(a)}}}\star K_{M}+ \log\frac{1+\tfrac{1}{Y_{0|vw}^{(a)}}}{1+\tfrac{1}{Y_{0|w}^{(a)}}}\star K_{k-1} \, .
\end{equation}

\paragraph{The TBA equations for $w$-strings} are
\begin{align}
\label{eq:cTBAwM}
\log{Y_{M|w}^{(a)}}=&  - \mu^{(a)}_{M|w}+ \,\log\left(1+\tfrac{1}{Y_{N|w}^{(a)}}\right)\star K_{NM}+\log\left(1+\tfrac{1}{Y_{0|w}^{(a)}}\right)\star K_{k-1,M}\\
& \, +\log\frac{1-\frac{1}{Y_-^{(a)}}}{1-\frac{1}{Y_+^{(a)}}}\,\hat{\star}\,K_M\, ,\nonumber\\
\label{eq:cTBAw0}
\log{Y_{0|w}^{(a)}}=& - \mu^{(a)}_{0|w} - \mu^{(a)}_{k-1|w} - \log{Y_{k-1|w}^{(a)}}\, .
\end{align}

\paragraph{The TBA equations for $vw$-strings} are
\begin{align}
\label{eq:cTBAvwM}
\log{Y_{M|vw}^{(a)}}
=& - \mu^{(a)}_{M|vw}+\log\left(1+\tfrac{1}{Y_{N|vw}^{(a)}}\right)\star K_{NM} + \log\left(1+\tfrac{1}{Y_{0|vw}^{(a)}}\right)\star K_{k-1,M} \\
& \quad \quad +\log\frac{1-\frac{1}{Y_-^{(a)}}}{1-\frac{1}{Y_+^{(a)}}}\,\hat{\star}\,K_M - \log\left(1+Y_Q\right)\star K^{QM}_{xv}\,, \nonumber\\
\label{eq:cTBAvw0}
\log{Y_{0|vw}^{(a)}}=& - \mu^{(a)}_{0|w} - \mu^{(a)}_{k-1|vw} - \log{Y_{k-1|vw}^{(a)}}\, .
\end{align}

\paragraph{The free energy} of the mirror model allows us to find the energy of the deformed `string' theory as
\begin{equation}
\label{eq:Energy}
E(L) =-\int {\rm d}u\, \sum_{Q=1}^{k}\frac{1}{2\pi}\frac{d\tilde{p}^Q}{du}\log\left(1+Y_Q \right)\,.
\end{equation}
Note that we have only a finite number of TBA equations and a finite number of Y-functions contributing to the energy.

\section{The simplified TBA equations and Y-system}
\label{sec:Ysysgeneral}

Typically the Y-system associated to a set of TBA equations is universal in the sense that it is the same for all excited states and independent of chemical potentials associated to symmetries of the theory. However, already at this stage it is far from clear that the chemical potential dependence in the functional relation between $Y_{0|(v)w}$ and $Y_{k-1|(v)w}$,
\begin{equation}
Y_{0|w}^{(a)} Y_{k-1|w}^{(a)} =e^{-\mu^{(a)}_{0|w}-\mu^{(a)}_{k-1|w}}\, ,
\end{equation}
can be removed from the Y-system, and as we will shortly see this is indeed not possible. Furthermore, our Y-system will not quite by the same for all excited states but will pick up a mild dependence on the excitation numbers. Both of these properties are intimately related to having a finite number of TBA equations.\footnote{While atypical, chemical potentials (magnetic fields) already appear in the Y-system of the XXZ spin chain at roots of unity \cite{Takahashi:1972}.} To see how excited states can influence the Y-system, let us investigate the effect of considering an excited state at the level of the canonical TBA equations, on the simplified TBA equations and associated Y-system. We will discuss this explicitly for the simple case of $w$-strings.

\subsection{Excited states in the deformed mirror TBA}

The canonical ground state TBA equations for $w$-strings are given by eqs. \eqref{eq:cTBAwM} and \eqref{eq:cTBAw0}, but let us temporarily introduce kernels with indices $0$ and reinstate them appropriately to get
\begin{align}
\log{Y_{M|w}^{(a)}}=&  - \mu^{(a)}_{M|w}+ \,\log\left(1+\tfrac{1}{Y_{N|w}^{(a)}}\right)\star K_{NM}-\log\left(1+\tfrac{1}{Y_{0|w}^{(a)}}\right)\star K_{0M}\\
& \, +\log\frac{1-\frac{1}{Y_-^{(a)}}}{1-\frac{1}{Y_+^{(a)}}}\,\hat{\star}\,K_M\, ,\nonumber\\
\log{Y_{0|w}^{(a)}}=&  - \mu^{(a)}_{0|w} + \,\log\left(1+\tfrac{1}{Y_{N|w}^{(a)}}\right)\star K_{0M}-\log\left(1+\tfrac{1}{Y_{0|w}^{(a)}}\right)\star K_{00}\\
& \, +\log\frac{1-\frac{1}{Y_-^{(a)}}}{1-\frac{1}{Y_+^{(a)}}}\,\hat{\star}\,K_0\,.\nonumber
\end{align}
Now we recall that the $S$-matrices the above kernels are derived from have the properties
\begin{equation}
\label{eq:basicSmatrels}
S^{k-1} S^{0} = -1 \, , \, \, \, \, S^{Mk-1}S^{M0} = 1.
\end{equation}
This immediately implies \eqref{eq:cTBAw0} for the ground state. For excited states however, we should be more careful.

As we discussed in chapter \ref{chapter:finitevolumeIQFT} the TBA equations for an excited state are typically the same as those for the ground state up to a set of driving terms. These driving terms depend on the state under consideration and are different for each particle type's TBA equation. Let us for simplicity assume that we have a state with $K_w$ roots of $1-Y_-$ that enter the TBA equations and no further roots. Considering the above form of the TBA equations, this would result in driving term contributions of the form
\begin{equation}
\sum_{i=1}^{K_w} \log S^M(r_i^{(a)} - u - \tfrac{i}{g})\, ,
\end{equation}
in the equation for $\log Y^{(a)}_{M|w}(u)$ ($M$ can be zero here). Because of the first relation in eqn. \eqref{eq:basicSmatrels}, this results in a relative sign between the driving terms for $\log Y_{k-1|w}$ and $\log Y_{0|w}$, exactly in line with their Y-functions being inverse, as for the ground state. However, due to the minus sign in the first of eqs. \eqref{eq:basicSmatrels} we also get an extra factor of $K_w i \pi$.\footnote{Of course the reader might argue that an S-matrix-like driving term $\log S(u,v)$ based on a kernel $K$ is defined up functions of $v$ only. This ambiguity should then be fixed by comparison to an asymptotic solution for the Y-functions. We do not have a direct asymptotic form for $Y_{0|w}$, making this impossible. However, let us firstly note that the chosen driving terms are the completely natural choice. Moreover, as we will discuss below the resulting equations are compatible with the available asymptotic form for the other Y-functions, while other potential forms of the driving terms are not. Hence there is no true ambiguity.} This means that while
\begin{equation}
Y_{0|w}^{(a)} Y_{k-1|w}^{(a)} =e^{-\mu^{(a)}_{0|w}-\mu^{(a)}_{k-1|w}}\, \, \, \, \, \, \mbox{for even}\, \, K_w\, ,
\end{equation}
we have
\begin{equation}
Y_{0|w}^{(a)} Y_{k-1|w}^{(a)} = -e^{-\mu^{(a)}_{0|w}-\mu^{(a)}_{k-1|w}} \, \, \, \, \, \, \mbox{for odd}\, \, K_w\, .
\end{equation}
Now let us come back to our original assumption. If we were to consider additional roots of $Y_\pm$ or $1-Y_+$, clearly $K_w$ above should be replaced by the total number of all these roots, taking relative signs between driving terms into account. Furthermore, due to the plus sign in the second relation in eqn. \eqref{eq:basicSmatrels} additional roots of $1+Y_{M|w}$ do not further affect this relation.

Next, these driving terms also play a direct and important role in the simplified TBA equations and Y-system for $Y_{k-1|w}$. Conventionally, we can rewrite the TBA equations in simplified form by noting identities between various kernels. As convolutions with convoluted kernels turn into convolutions with the standard kernel $s$, the driving terms in the simplified TBA equations turn into $\log S$ terms, where $S$ is the standard $S$-matrix associated to $s$
\begin{equation}
s = \frac{1}{2 \pi i} \frac{d}{du} \log S \, .
\end{equation}
Not surprisingly, when acting with $s^{-1}$ to obtain the Y-system these driving terms are annihilated. As discussed in section \ref{subsec:TBAgeneral}, up to potential subtleties with branch cuts these identities between nontrivial S-matrices and the subsequent annihilation of the remainder by $s^{-1}$ can be summarized by the statement that these S-matrices satisfy the discrete Laplace equation
\begin{equation}
\frac{S^M (u,v+i/g)S^M (u,v-i/g)}{S^{M+1}(u,v)S^{M-1} (u,v)} =1\, ,
\end{equation}
which holds for any of the relevant S-matrices in what follows. Since we have a boundary however, the identities we use there are equivalent to getting a final contribution of
\begin{equation}
\label{eq:bndrydrivingterm}
\tfrac{1}{2} \log \frac{S^{k-1}(u,v+i/g)S^{k-1} (u,v-i/g)}{S^{k-2} (u,v)} = \tfrac{1}{2} \log S^{k} (u,v)
\end{equation}
in the simplified TBA equation for $Y_{k-1|w}$. Of course $S^{k}(u,v)$ is constant such that its associated kernel vanishes. However, with the auxiliary problem being $2k$ periodic, $S^k$ is not quite one, but rather minus one. This means that the simplified TBA equation for $Y_{k-1|w}$ gets an extra factor of $K_w i \pi/2$. Repeating the above analysis for the canonical TBA equations for $vw$-strings it is easy to see that the relation between $Y_{0|vw}$ and $Y_{k-1|vw}$ should be sensitive to both the roots associated to $Y_\pm$ and those to associated to $Y_Q$, as $S^{Q0}_{xv}S^{Qk-1}_{xv}=-1$ as well. Let us denote the number additional roots due to $Y_Q$ by $K_{vw}$.

At this point we might wonder whether the number of these roots effectively follows a pattern. To understand and motivate this, we should realize that physically speaking the difference between the driving terms for the $0|(v)w$ and $k-1|(v)w$ string comes about because of their different charges; they feel the presence of the excited state differently. This hints that the number of roots could be directly related to the charge of the state, in particular to their excitation numbers.

To continue along this track, let us start by considering generic states in the extensively studied $\mathfrak{sl}(2)$ sector of the undeformed theory. Based on various studies of the TBA equations for excited states in this sector \cite{Gromov:2009bc,Arutyunov:2009ax,Balog:2010vf,AFunpublished} it seems reasonable to assume that here the number of relevant roots is restricted to those of $1+Y_Q$ corresponding to the fundamental excitations in the string theory. This would mean that $K_{vw} = K^{\rm I}$ and $K_w = 0$ in this sector. Next, if we consider states with nonzero $K^{\rm II}_{(a)}$ with real momenta such as those studied in \cite{Sfondrini:2011rr} the natural generalization of this statement is that $K_{vw} = K^{\rm I}$ and $K_w = K^{\rm II}_{(a)}$.\footnote{Our excitation numbers refer to the $\mathfrak{sl}(2)$ grading.} At the level of the number of roots this simple picture is not quite true anymore for for example an orbifolded magnon \cite{Arutyunov:2010gu} or states with complex momenta in the $\mathfrak{su}(2)$ sector \cite{Arutyunov:2011mk}. However, we should note that the relation between the Y-functions only depends on whether $K_w$ and $K_w + K_{vw}$ are even or odd, and that in this sense any currently investigated state matches the described pattern. Assuming this picture not to change qualitatively in our deformed model, we then have the natural conjecture that in general
\begin{equation}
\label{eq:excYreln}
\begin{aligned}
Y_{0|w}^{(a)} Y_{k-1|w}^{(a)} & = (-1)^{K^{\rm II}_{(a)}}e^{-\mu^{(a)}_{0|w}-\mu^{(a)}_{k-1|w}}\, , \\
Y_{0|vw}^{(a)} Y_{k-1|vw}^{(a)} & = (-1)^{K^{\rm I}+K^{\rm II}_{(a)}}e^{-\mu^{(a)}_{0|w}-\mu^{(a)}_{k-1|w}}\, .
\end{aligned}
\end{equation}

We might wonder whether or not the excitation number $K^{\rm III}_{(a)}$ should make an appearance. To understand let us first look at the simple case of $w$-strings again. In essence there the above story tells us that the `length of the spin chain' associated to a particular particle type (nested level) determines the appearance of the minus sign. However, then we would only expect a minus sign for $vw$-strings depending on whether $K^{\rm I}$ is even or odd, apparently contradicting the above. The reason for the appearance of $K^{\rm II}_{(a)}$ in this picture is that in deriving the 'canonical' TBA equations for $vw$-strings the canonical TBA equations for $w$-strings were used to exchange the presence of infinitely many $Y_{M|w}$ functions for the presence of $Y_\pm$ \cite{Arutyunov:2009zu}. If this had not been done, the direct dependence would be on $K^{\rm I}$ only, where the dependence on $K^{\rm II}_{(a)}$ would come in when eliminating the $Y_{M|w}$ functions from the simplified TBA or Y-system. From this point of view the presence of $w$-particles (nonzero $K^{\rm III}_{(a)}$) cannot result in any potential minus signs because there is no deeper nested level.

While the precise form of the general relations \eqref{eq:excYreln} is a conjecture, it is completely natural to expect simple relations of this type by the physical picture that they are caused by the difference in charges of the respective string solutions. Of course, this conjecture is also supported by the asymptotic Y-system we will discuss in section \ref{sec:Asymptoticsolution}. Finally, state dependence is clearly present by the above discussion, regardless of the precise form.

Before continuing, let us specific our chemical potentials to those associated with boundary conditions in the physical theory, as in table \ref{tab:chempot} (where $\mu_{0|(v)w}=\mu_{1|(v)w}$ as both strings consist of one magnon). This means that the sum of chemical potentials in eqs. \eqref{eq:excYreln} can be written as
\begin{equation}
\mu^{(a)}_{0|(v)w}+\mu^{(a)}_{k-1|(v)w} = k\mu^{(a)}_{1|(v)w}\, .
\end{equation}
Clearly, also the dependence on the boundary conditions arises due to a mismatch between the charge and scattering properties of the negative parity string.

\subsection{The simplified TBA equations}

The above discussion clearly shows us that the simplified TBA equations and Y-system pick up an explicit dependence on the chemical potential and (the excitation numbers of) the state under consideration, in addition to conventional driving terms that generically still enter the simplified TBA equations for excited states. Part of this dependence comes about through the boundary associated with having finitely many TBA equations, while the other part comes in explicitly by eliminating $Y_{0|(v)w}$ from the TBA equations. We could choose not to eliminate $Y_{0|(v)w}$ from the TBA equations, but since this would not remove the dependence on excitation numbers and chemical potential completely in any case and we would have to keep extra careful track of the origin of driving terms, it seems best to eliminate $Y_{0|(v)w}$. This way the TBA equations can be treated as in the undeformed case, with an asymptotic solution for the remaining Y-functions and the contour deformation trick applying directly and straightforwardly. The only modification is that we additionally have the manifest dependence on the excitation numbers and chemical potential. Let us note again that this dependence does derive properly from the contour deformation trick and the full set of canonical TBA equations as discussed above; it is simply easier to manifest it before continuing.

With this in mind we can write down the simplified TBA equations for excited states with nonzero chemical potentials, up to standard $\log S$ terms following from the analytic properties of the \emph{independent} Y-functions. Let us begin with the auxiliary equations.

\subsubsection*{$w$ and $vw$-strings}

Using \eqref{eq:simpKM}, \eqref{eq:simpKMN} and \eqref{eq:simpKQMxv} and the canonical equations for $\log{\frac{Y_+}{Y_-}}$ we obtain the following simplified TBA equations for $w$ and $vw$-strings
\begin{align}
\log{Y_{M|vw}} = \, & \log{(1+Y_{M+1|vw})(1+Y_{M-1|vw})} \star s -
\log\left(1+Y_{M+1}\right)\star s + \delta_{M,1}
\log{\frac{1-Y_-}{1-Y_+}}  \hat{\star}  s \nonumber\\
\log{Y_{k-2|vw}}  = \, & \log{(1+Y_{k-3|vw})(1+Y_{k-1|vw})(1+e^{\chi_{vw}}Y_{k-1|vw})} \star s  - \log\left(1+Y_{k-1}\right)\star s\, ,\nonumber \\
\log{Y_{k-1|vw}}  = \, & \log{(1+Y_{k-2|vw})} \star s - \log\left(1+Y_{k}\right)\star s - \tfrac{\chi_{vw}}{2}\, ,\nonumber \\
\log{Y_{M|w}}  = \, &\log{(1+Y_{M+1|w})(1+Y_{M-1|w})} \star s + \delta_{M,1}
\log{\frac{1-Y^{-1}_-}{1-Y^{-1}_+}}  \hat{\star}  s\, ,\\
\log{Y_{k-2|w}}  = \, &\log{(1+Y_{k-3|w})(1+Y_{k-1|w})(1+e^{\chi_{w}}Y_{k-1|w})} \star s\, , \nonumber\\
\log{Y_{k-1|w}}  = \, &\log{(1+Y_{k-2|w})} \star s - \tfrac{\chi_{w}}{2}\, , \nonumber
\end{align}
where $Y_{0|(v)w}=0$ and
\begin{align*}
\chi^{(a)}_{vw} & = k \hspace{1pt} \mu^{(a)}_{1|vw} + i \pi (K^{\rm I}+K^{\rm II}_{(a)})\, ,\\
\chi^{(a)}_{w} & = k \hspace{1pt} \mu^{(a)}_{1|w} + i \pi K^{\rm II}_{(a)}\,.
\end{align*}

\subsubsection*{$y$-particles}

As in the undeformed case the $y$-particles do not have simplified TBA equations. Their hybrid form is
\begin{align}
\log\frac{Y_{+}^{(a)}}{Y_{-}^{(a)}} = &
\log(1+Y_{Q})\star K_{Qy}\,,\\
\log Y_{-}^{(a)}Y_{+}^{(a)}  = & \,-\log(1+Y_{Q})\star K_{Q}+
2\log(1+Y_{Q})\star K_{xv}^{Q1}\star s+2\log\frac{1+Y_{1|vw}^{
{(a)}}}{1+Y_{1|w}^{(a)}}\star s\,,
\end{align}

\subsubsection*{$Q$-particles}

Applying $(K+1)^{-1}$ defined in \eqref{eq:Kp1inv} and using the identities in appendix \ref{app:qdefkernels}, we find the simplified TBA equation for $Y_1$
\begin{equation}
\log Y_1=\sum_a \log\left(1-\tfrac{1}{Y_-^{(a)}}\right) \hstar s - \log \left(1+\tfrac{1}{Y_2}\right) \star s -\check{\Delta}\check{\star} s\, ,
\end{equation}
where
\begin{align}
\check{\Delta}=&L\check{\cal E}+\sum_a \log \left(1-\tfrac{1}{Y_-^{(a)}}\right)\left(1-\tfrac{1}{Y_+^{(a)}}\right)\star \check{K}+2\log(1+Y_Q)\star
\check{K}^{\Sigma}_Q\\
\nonumber
&\quad\quad\quad +\sum_a \log\left(1-\tfrac{1}{Y_{M|vw}^{(a)}}\right)\star \check{K}_M+ \log\left(1-Y_{k-1|vw}^{(a)}\right)\star \check{K}_{k-1}\, .
\end{align}
The kernels appearing in the expression above are defined in appendix \ref{app:qdefkernels}. Analogously, for $Q=2,\ldots,k-1,$ we find that the simplified TBA equations are given by
\begin{align}
\log Y_Q = \, & \log\frac{{Y_{Q+1}Y_{Q-1}}}{(1+Y_{Q-1})(1+Y_{Q+1})}\star s + \sum_a \log \left(1+\tfrac{1}{Y_{Q-1|vw}^{(a)}}\right) \star s\, .\label{eq:sTBAQg1}
\end{align}
For $Q=k$ we cannot simply apply the usual operator
\begin{equation}
(K+1)^{-1}_{Pk}=\delta_{Pk} - s (\delta_{Pk-1}+\delta_{Pk+1})\, ,
\end{equation}
since there is no $k+1$st equation. Instead, we apply the operator $\delta_{Q,k} - 2 \delta_{Q,k-1} s$. As we discuss in appendix \ref{app:sTBAYk}, using special identities and the TBA equations for $vw$-strings we obtain
\begin{equation}
\label{eq:sTBAQk}
\log Y_k =\, 2\log{Y_{k-1}}\star s -\log(1+Y_{k-1})\star s + \sum_a \log \left(1+\tfrac{1}{Y^{(a)}_{k-1|vw}}\right)\left(1+\tfrac{
e^{-\chi^{(a)}_{vw}}}{Y^{(a)}_{k-1|vw}}\right) \star s \, .
\end{equation}
Naively we might have expected to get terms like $\left(1+\tfrac{1}{Y_{k-1|vw}}\right)\left(1+e^{\chi_{vw}}Y_{k-1|vw}\right)$ in the above equation. Because it is exactly analogous to what happens in the equations for $vw$-strings it is very natural that we get a doubled contribution instead, though it arises in the far less obvious fashion shown in appendix \ref{app:sTBAYk}.

\subsection{The Y-system}

By applying $s^{-1}$ to the above equations we immediately read off the Y-system\footnote{Of course also in our deformed model $Y_+$ has no Y-system equation.}

\paragraph{$w$-strings}
\begin{align}
Y_{1|w}^+ Y_{1|w}^- & =(1+Y_{2|w})\left(\frac{1-Y_-^{-1}}{1-Y_+^{-1}}\right)^{\theta(u_B-|u|)}  \\
Y_{M|w}^+ Y_{M|w}^- & =(1+Y_{M-1|w})(1+Y_{M+1|w}) \, , \\
Y_{k-2|w}^+ Y_{k-2|w}^- & = (1+Y_{k-3|w})(1+Y_{k-1|w})(1 + e^{\chi_w}Y_{k-1|w}) \, ,\\
Y_{k-1|w}^+ Y_{k-1|w}^- & = e^{-\chi_w}(1+Y_{k-2|w}) \, .
\end{align}

\paragraph{$vw$-strings}
\begin{align}
Y_{1|vw}^+ Y_{1|vw}^- & =\frac{1+Y_{2|vw}}{1+Y_2}\left(\frac{1-Y_-}{1-Y_+}\right)^{\theta(u_B-|u|)}  \\
Y_{M|vw}^+ Y_{M|vw}^- & =\frac{(1+Y_{M-1|vw})(1+Y_{M+1|vw})}{1+Y_{M+1}}\, ,\\
Y_{k-2|vw}^+ Y_{k-2|vw}^- & = \frac{(1+Y_{k-3|vw})(1 + Y_{k-1|vw})
(1 +e^{\chi_{vw}} Y_{k-1|vw})}{1+Y_{k-1}}\, ,\\
Y_{k-1|vw}^+ Y_{k-1|vw}^- & = e^{-\chi_{vw}}\frac{1+Y_{k-2|vw}}{1+Y_k}\, .
\end{align}

\paragraph{$y$-particles}
\begin{equation}
Y_{-}^{+}Y_{-}^{-} = \frac{1+Y_{1|vw}}{1+Y_{1|w}}\frac{1}{1+Y_1}\, .
\end{equation}

\paragraph{$Q$-particles\\}
\begin{align}
\frac{Y_1^+ Y_1^-}{Y_2} & = \frac{\displaystyle{\prod_a}
\left(1-\frac{1}{Y^{(a)}_{-}}\right)}{1+Y_2}\, , \, \, \, \, \, \mbox{for} \, \,
|u|<u_b\, .\\
\frac{Y_Q^+ Y_Q^-}{Y_{Q+1}Y_{Q-1}} & = \frac{\displaystyle{\prod_a} \Bigg(
1+\frac{1}{Y_{Q-1|vw}^{(a)}}\Bigg)}{(1+Y_{Q-1})(1+Y_{Q+1})} \, , \\
\frac{Y_k^+ Y_k^-}{Y_{k-1}^2} & = \frac{\displaystyle{\prod_a}
\Bigg(1 +\frac{1}{Y_{k-1|vw}^{(a)}}\Bigg)
\Bigg(1 +\frac{e^{-\chi_{vw}}}{Y_{k-1|vw}^{(a)}}\Bigg)}{1+Y_{k-1}}\,.
\end{align}

The above equations have an explicit dependence on $\chi$ that cannot be removed by simply redefining the Y-functions. However, let us note that by the redefinition
\begin{equation}
Y_{k-1|(v)w} \rightarrow e^{-\tfrac{\chi_{(v)w}}{2}} Y_{k-1|(v)w}\,
\end{equation}
the above equations do become manifestly symmetric in $\chi$, and in particular in the twist. Since we would expect the spectrum to be symmetric in the twist, this is only natural.

As in the undeformed case the Y-system admits analytic continuation onto the complex $u$-plane with cuts where the TBA equations fix the corresponding jump discontinuities of the Y-functions. Both the TBA equations and the Y-system presented here manifestly tend to their undeformed cousins in the limit $k\to \infty$.

\subsubsection*{Periodicity}

When an integrable model has finitely many Y-functions, these functions may turn out to be periodic on the $u$-plane as a consequence of the Y-system \cite{Zamolodchikov:1991et}. While not all Y-functions in our model appear to be periodic, we would still like to briefly address this interesting phenomenon.

To illustrate how a Y-system can imply periodicity of its Y-functions let us consider a very simple Y-system with only one Y-function, the Y-system of the scaling Lee-Yang model \cite{Zamolodchikov:1989cf}
\begin{equation}
Y(u+i)Y(u-i) = 1+Y(u)\, .
\end{equation}
Extending this relation into the complex plane we readily see
\begin{equation}
\begin{aligned}
Y(u+2i) &= \frac{1+Y(u+i)}{Y(u)}\,,\\
Y(u+3i) &= \frac{1+Y(u)+Y(u+i)}{Y(u)Y(u+i)}\,,\\
Y(u+4i) &= \frac{1+Y(u)}{Y(u+i)}\,,\\
Y(u+5i) &= Y(u)\,.\\
\end{aligned}
\end{equation}
In other words, the Y-functions have period $5i$! In a similar fashion it is not hard to convince ourselves that the Y-functions for the XXZ spin chain at $q=e^{i \pi /k}$ have periodicity $(k+2)i$. In general, such periodicity is known to be present in Y-systems associated to Dynkin diagrams of type A, D and E \cite{Zamolodchikov:1991et}. Our model is not of this type, and it appears that the fermionic Y-functions do not allow for periodicity of all Y-functions.

To begin with let us note that if we formally consider the asymptotic limit $Y_Q=0$, at the auxilialy level we get a simple Y-system where $Y_+$ equals $Y_-$ and so the Y-system equations for $w$ and $vw$-strings just become those of the XXZ spin chain with periodicity $(k+2)i/g$. It is then not hard to determine that the periodicity of $Y_-$ is $4(k+2)i/g$.\footnote{Generalizing this particular problem slightly, we would like to determine the periodicity of a function $Y$ for which the combination $Y(u+ia) Y(u) \equiv f(u)$ ($a\in \mathbb{N}$) has periodicity $i n$. Writing out readily gives $Y(u+2mia)/Y(u) = \prod_{j=1}^{m} f(u+(2(m-j)+1)i a)/f(u+2(m-j)i a)$, which upon using $f(u+in)=f(u)$ to cancel the first $\lceil m/2 \rceil$ denominators against the last $\lceil m/2 \rceil$ numerators and then cancelling the remainder gives $Y(u+2nia)/Y(u)=1$.} Since $Y_Q$ is not zero however, $Y_+$ is not equal to $Y_-$ and only the $Y_{M|w}(u)$ functions remain manifestly periodic for $|u|>u_b$. Beyond this point the question of periodicity gets complicated by the fact that $Y_+$ has no Y-system equation, making periodicity look unlikely. To try to get around this we could consider the T-system instead of the Y-system. However, we should realize that while T-function periodicity implies Y-function periodicity, the converse is not true due to the gauge dependence of the T-functions. Already in the XXZ case we would only find periodicity in the T-system in a very special class of gauges, and we have not attempted to exhaustively check periodicity of the T-system in generic gauges for our more complicated model. A quick look at some nice gauge choices that proved to be useful in the undeformed model (see e.g. \cite{Gromov:2011cx}) appears to give Y-systems without periodicity; the full Y-system appears to have no periodicity with the exception of the $Y_{M|w}(u)$ functions for $|u|>u_b$.

\subsection{The asymptotic ground state solution}

As we will show in this section, the ground state TBA equations of the deformed model have basically the same solution as the undeformed model in the small $Y_Q$ limit; only the boundary Y-functions are different. In particular this means that in the limit of zero twist the ground state energy and associated $Y_Q$ functions are zero.

In the asymptotic limit, the constant bulk Y-system for the auxiliary problem is given by
\begin{align}
Y_+^2 = Y_-^2 & = \frac{1+Y_{1|vw}}{1+Y_{1|w}}\, ,\\
Y_{M|w}^2 & = (1+Y_{M-1|w})(1+Y_{M+1|w})\, ,\\
Y_{M|vw}^2 & = (1+Y_{M-1|vw})(1+Y_{M+1|vw})\, ,
\end{align}
where potential $M=0$ terms on the right hand side are zero. The general solution to these recurrence relations for the bulk $Y_{M|w}$ and $Y_{M|vw}$ is given by
\begin{align}
\label{eq:gssolnbasic}
Y_{M|w}& = [M]_{e^{i b}} [M+2]_{e^{i b}}\, , \\
Y_{M|vw} & = [M]_{e^{i a}} [M+2]_{e^{i a}} \, ,
\end{align}
where $a$ and $b$ are undetermined complex numbers for the moment. Of course, we have yet to take into account the nontrivial boundary of the Y-system. These boundary equations read
\begin{align}
\left([k-2]_{e^{i b}} [k]_{e^{i b}}\right)^2& = (1+[k-3]_{e^{i b}}[k-1]_{e^{i b}})(1+Y_{k-1|w})(1+e^{k \mu_{1|w}}Y_{k-1|w})\, , \\
Y_{k-1|w}^2& = e^{-k \mu_{1|w}}(1+[k-2]_{e^{i b}}[k]_{e^{i b}}) \, ,
\end{align}
with a similar equation for $Y_{k-1|vw}$. These equations do not have a unique solution for arbitrary $\mu_{1|(v)w}$ and $b (a)$. However, insisting that these equations satisfy the canonical TBA equations we immediately find that $a$ and $b$ are precisely $\alpha$ and $\vartheta$ respectively as in table \ref{tab:chempot}, and we get
\begin{align}
\label{eq:gssolnbndry}
Y_{k-1|w} & = e^{-i k \vartheta} [k-1]_{e^{i\vartheta}}\, ,\\
Y_{k-1|vw} & = e^{-i k \alpha} [k-1]_{e^{i\alpha}}\, .
\end{align}
With this solution we know $Y_\pm$ by the Y-system equation above and can then find the $Y_Q$ functions by their canonical TBA equations. The resulting $Y_Q$ functions are formally identical to the $Y_Q$ functions of the undeformed twisted ground state and are given by\footnote{For more details see \cite{Arutyunov:2012zt,Arutyunov:2012ai}.}
\begin{equation}
\label{eq:gssolnYQ}
Y_{Q} =  ([2]_{e^{i\alpha}}-[2]_{e^{i\vartheta}})([2]_{e^{i\dot{\alpha}}}-[2]_{e^{i\dot{\vartheta}}})[Q]_{e^{i\alpha_+}}[Q]_{e^{i\alpha_-}} e^{-J \tilde{\mathcal{E}}_Q(\tilde{p})}\, ,
\end{equation}
As before, this asymptotic solution captures the large $u$ asymptotics of the Y-functions. While the simplified TBA equations do manifestly depend on the chemical potential, we just saw that this was not enough to fix the solution, and so the simplified TBA equations or Y-system need to be supplemented with these large $u$ asymptotics.\footnote{The alternative characterization in terms of the large $M$ limit of the $Y_{M|(v)w}$ functions we saw before does not apply here of course.} As the entire solution except the boundary $w$ and $vw$ Y-functions formally coincides with the undeformed twisted ground state solution we discussed in section \ref{sec:GS}, this solution of the deformed TBA equations smoothly turns into the twisted ground state solution of the undeformed model in the simplest way possible.

Computing the ground state energy with this twisted solution at leading order in $g$ gives complicated and not particularly insightful expressions. We would however like to note that it converges to the undeformed result as $\tfrac{1}{k^2}$ at large $k$. Explicitly
\begin{equation}
 E(k,J) = E(J) - \frac{J\pi^2}{3k^2} \left(E(J) - \frac{J-2}{4(J-1)} E(J-1)\right) + \mathcal{O}\left(\frac{1}{k^3}\right)\, ,
\end{equation}
where $E(J)$ is the ground state energy of the undeformed twisted model given in eqn. \eqref{eq:GSsinglewrapping}.

\section{The asymptotic solution for excited states}
\label{sec:Asymptoticsolution}

As in the undeformed model, we would like to have the solution for the Y-system in the asymptotic limit $Y_Q\rightarrow Y^\circ_Q\sim 0$. In this section all Y-functions will be understood in the asymptotic limit and to avoid clutter we will identify all Y-functions with their asymptotic limit and drop the superscript $\circ$. As we saw in the previous chapters, this asymptotic solution should be given in terms of transfer matrices, in this case built out of the $q$-deformed S-matrix and a general twist.\footnote{Note that the quantum group symmetry of the $q$-deformed S-matrix contains the Cartan subgroup of the original symmetry group as a `regular' symmetry so that we can still readily twist by it.} The derivation of the eigenvalues of these transfer matrices is of course a close variation on the undeformed derivation. Before stating the result, let us introduce the following shorthand notation
\begin{align}
&\SH^A_B =  \prod_{i=1}^{K^A}\sinh\frac{\pi g}{2k}(u-u_i ^{(A)} + B\frac{i}{g}),
&& R^\pm_A = \prod_{n=1}^{K^{\mathrm{I}}}  \frac{x(u+\frac{Ai}{g}) - x^{\mp}_{n}
}{q^\frac{A\mp1}{4}\sqrt{x^{\mp}_{n} }} ,
&& B^\pm_A = \prod_{n=1}^{K^{\mathrm{I}}}  \frac{\frac{1}{x(u+\frac{Ai}{g})} -
x^{\mp}_{n}}{q^\frac{A\mp1}{4}\sqrt{x^{\mp}_{n} }} \,,\nonumber
\end{align}
where $u^{(\mathrm{I})}_i =u_i$, $u^{(\mathrm{II})}_i =v_i$ and $u^{(\mathrm{III})}_i = w_i$. In terms of these the general eigenvalue is then given by \cite{Arutyunov:2012ai}
\begin{align}
\label{eq:qdeffulltransfer}
T_{a,1}=& e^{i\alpha
a}\frac{\SH^\mathrm{II}_a}{R^+_aB^-_a} \prod_{i=1}^{K^{\mathrm{II}}}
q^{\frac{a}{2}}\frac{y_i - x^-}{y_i
-x^+}\sqrt{\frac{x^+}{x^-}}\left\{\sum_{m=0}^{a}
\frac{R^{+}_{a-2m}B^{-}_{a-2m}}{e^{2i\alpha m}\,\SH^\mathrm{II}_{a-2m}}
+
\sum_{m=1}^{a-1}
\frac{R^{-}_{a-2m}B^{+}_{a-2m}}{e^{2i\alpha m}\,\SH^\mathrm{II}_{a-2m}}
\right.\nonumber\\
& - \sum_{m=0}^{a-1}\!
\frac{R^{-}_{a-2m}B^{-}_{a-2m}}{e^{i\alpha(2m+1)}\SH^\mathrm{III}_{a-2m-1}}
\left.
\left[\!
e^{i\vartheta}
\frac{\SH^\mathrm{III}_{a-2m+1}}{\SH^\mathrm{II}_{a-2m}}
+e^{-i\vartheta}
\frac{\SH^\mathrm{III}_{a-2m-3}}{\SH^\mathrm{II}_{a-2m-2}}
\right]
\right\},
\end{align}
where the auxiliary parameters $y_m$ and $w_n$ satisfy the auxiliary Bethe
equations (\ref{eq:qdefBYw}) and (\ref{eq:qdefBYw})

As in the undeformed case the $Y_Q$ functions should have the asymptotic form of eqn. \eqref{eq:BJforYQ}
\begin{equation}
Y_Q^o = \Upsilon_Q \, T^{(l)}_{Q,1}(v|\{u_j\})\, T^{(r)}_{Q,1}(v|\{u_j\}) \, ,
\end{equation}
where
\begin{align}
\Upsilon_Q \equiv e^{-J\tH_Q}\prod_{i=1}^{\KK{I}}S^{Q 1_*}_{\alg{sl}(2)}(v,u_i)\,.
\end{align}
And correspondingly the auxiliary Y-functions are generically again given by eqs. \eqref{eq:YtoTasympt}
\begin{align}
& Y^{(a)o}_{M|vw} = \frac{T^{(a)}_{M+2,1}T^{(a)}_{M,1}}{T^{(a)}_{M+1,2}}\, , && Y^{(a)o}_{M|w} = \frac{T^{(a)}_{1,M+2}T^{(a)}_{1,M}}{T^{(a)}_{2,M+1}}\, ,\\
& Y^{(a)o}_{+} = -\frac{T^{(a)}_{2,1}T^{(a)}_{2,3}}{T^{(a)}_{1,2}T^{(a)}_{1,3}}\,, &&
 Y^{(a)o}_{-} = - \frac{T^{(a)}_{2,1}}{T^{(a)}_{1,2}} \, .
\end{align}
However, this standard construction clearly does not apply for $Y_{k-1|vw}$ and $Y_{k-1|w}$ as their contribution to the equations for $Y_{k-2|vw}$ and $Y_{k-2|w}$ respectively are not of standard type. Still, we can determine an explicit expression for $Y_{k-1|(v)w}$ from the asymptotic Y-system equation for $Y_{k-2|(v)w}$ insisting on the correct large $u$ asymptotics.\footnote{As it turns out it is also possible to define an object similar in form to the transfer matrix above and use it to write a concrete expression for $Y_{k-1|(v)w}$, see \cite{Arutyunov:2012ai} for details.} The nontrivial fact is then that this explicit expression still needs to solve its own Y-system equation. The asymptotic solution constructed in this way precisely does so for $q^k=-1$ and presents a nontrivial test of our entire procedure, including in particular the string hypothesis.

There is a second nontrivial consistency check hiding in the equation for $Y_k$. From
the asymptotic equation for $Y_{k-1}$ we can immediately see that $Y_k$ is in
fact given by the bulk expression at $Q=k$, while it satisfies an equation of a
different type itself. The fact that its equation is satisfied
nonetheless, relies crucially on the factor $\Upsilon_Q$. Indeed, since
$\Upsilon_Q$ satisfies the discrete Laplace equation
\begin{align}
\Upsilon_Q^+\Upsilon_Q^- = \Upsilon_{Q+1}\Upsilon_{Q-1},
\end{align}
we see that the dependence of $\Upsilon_Q$ does not drop out from the right hand side of its asymptotic Y-system equation
\begin{equation}
\frac{Y_k^+ Y_k^-}{Y_{k-1}^2} = \prod_{r=\pm}
\left(1 +\frac{1}{Y_{k-1|vw}^{(r)}}\right)
\left(1 +\frac{e^{-\chi^{(r)}_{vw}}}{Y_{k-1|vw}^{(r)}}\right)\, .
\end{equation}
Using the arguments of appendix \ref{app:sTBAYk} we can however derive the following
identity
\begin{align}
\frac{\Upsilon_k^+\Upsilon_k^-}{\Upsilon_{k-1}^2} = (S^{10}_{xy})^2 =
\left[\prod^{\KK{I}}_{i=1}q\frac{x^{[k+1]}-x^-_i}{x^{[k+1]}-x^+_i}
\frac{1-\frac{1}{x^{[k-1]}x^-_i}}{1-\frac{1}{x^{[k-1]}x^+_i}}\right]^2\, ,
\end{align}
and this nontrivial factor indeed makes the asymptotic equation for $Y_k$ hold. Interestingly enough, the derivation of the above identity involves crossing, and so it appears we have found an instance where the fusion relations of the transfer matrices `know' about the crossing property of the S-matrix.\footnote{This of course has a (very indirect) link to the ideas worked out in \cite{Janik:2008hs}.}

\section{The relativistic limit}
\label{sec:Relativisticlimit}

Keeping $k$ fixed and taking $g\rightarrow \infty$ in an appropriate fashion, we obtain a set of TBA equations based on the vertex form of the S-matrix of the Pohlmeyer reduced superstring. The appropriate limit to take is to rescale our rapidities $u \rightarrow \frac{\tilde{u}}{g}$ and take $g \rightarrow \infty$ keeping $\tilde{u}$ fixed. In what follows by conventional abuse of notation we drop the tilde. In this limit the full S-matrix and consequently all S-matrices entering the Bethe-Yang equations become of difference form as appropriate for a relativistic theory.

At the level of the simplified TBA equations this relativistic limit is implemented in a very simple fashion. In the interpolating theory we encounter three types of convolutions, namely
\begin{align}
f\star s(u,v)=\,&\int_{-\infty}^{\infty}\, dt\, f(u,t)s(t-v) \, , \\
f\, \hat{\star}\,  s(u,v)=\,&\int_{-u_b}^{u_b}\, dt\, f(u,t)s(t-v)\, , \\
f\, \check{\star}\,  s(u,v)=\,&\int_{-\infty}^{-u_b}\, dt\, f(u,t)s(t-v)+\int_{u_b}^{\infty}\, dt\, f(u,t)s(t-v)\, .
\end{align}
where
\begin{equation}
s(u)=\frac{g}{4 \cosh\frac{g \pi u}{2}}\, ,
\end{equation}
and
\begin{equation}
u_b = \frac{k}{\pi g} \log\frac{1+\xi}{1-\xi}=\frac{2k}{\pi g}{\rm arcsinh}\Big(g\sin\frac{\pi}{k}\Big)\, .
\end{equation}
Rescaling the rapidities and taking the infinite coupling limit the points $\pm \, g \, u_b$ go to positive and negative infinity respectively, and consequently the limit can be summarized as
\begin{align}
f\star s & \rightarrow f\star s \, , \\
f\, \hat{\star}\, s & \rightarrow f\star s\, , \\
f\, \check{\star}\,  s& \rightarrow 0\, ,
\end{align}
where on the right hand side we of course have the properly rescaled kernel
\begin{equation}
s(u) = \frac{1}{4 \cosh\frac{\pi u}{2}}\, .
\end{equation}

In this limit the discontinuities of the Y-functions following from the TBA equations have disappeared completely. The dressing phase in particular only enters in defining the analytic structure of the Y-functions but does not enter in the TBA equations explicitly anymore. This is because before taking the relativistic limit the dressing phase enters the simplified TBA equations with a $\check{\star}$ convolution as we have carefully proven for the deformed dressing phase in appendix \ref{app:qdefdressingphase}, and this goes to zero in the relativistic limit. The only `coupling constant' left in the game is the level $k$, and as in the interpolating theory its effect comes in through the boundaries on the Y-system itself rather than the discontinuity relations the Y-system is supplemented with. It should not be surprising that the kernels in the simplified TBA equations do not depend explicitly on the coupling constant; this is a common feature of relativistic models. The information on the value of the coupling constant comes in through the kernels of the canonical TBA equations which do depend on it, which in turn define the asymptotics of the Y-functions.

There is one special case however, which is $Y_+$. Before taking the limit, $Y_+$ can be obtained as the analytic continuation of $Y_-$ through its cut along the real line. Now in the relativistic limit this cut disappears and $Y_+$ is a completely independent function that truly does not have a Y-system. It can be expressed in terms of $Y_-$ and $Y_Q$ as
\begin{equation}
Y_+(u) = Y_-(u) e^{\log\left(1+Y_Q \right) \star K_{Qy} (u)} \, ,
\end{equation}
where in the relativistic limit
\begin{equation}
K_{Qy} (u) = \frac{1}{2\pi i} \frac{d}{du} \log \frac{\sinh\frac{\pi}{4k}(u-i Q)}{\sinh\frac{\pi}{4k}(u+i Q)}\frac{\cosh\frac{\pi}{4k}(u+i Q)}{\cosh\frac{\pi}{4k}(u-i Q)}\, .
\end{equation}
In general the relation between $Y_+$ and $Y_-$ will have state dependent driving terms because of the convolution involving $Y_Q$. Applying $s^{-1}$ to the equation for $Y_+$ does not appear to give more insight. Apart from this subtlety, the Y-system becomes a finite set of algebraic relations between meromorphic functions on the $u$-plane. Of course, also in this limit the Y-system depends on the state and twist in the manner described above. Explicit expressions for the transfer matrices in this limit can be found in \cite{Arutyunov:2012ai}.

\section{Summary and outlook}

In this chapter we constructed TBA equations based on a $q$-deformation of the superstring S-matrix. At the level of excited states we found that the Y-system depends on (the excitation numbers of) the state under consideration, albeit in a mild way. This feature depends crucially on the root of unity deformation and the fact that we have a nested system, and to our knowledge has not been observed before. In the asymptotic limit our equations can be verified explicitly and hold due to special fusion relations between the $\mathfrak{psu}_q(2|2)$ transfer matrices. In principle we can use this asymptotic solution to find excited state TBA equations. It would be interesting to see to what extent the deformation qualitatively affects the analytic properties of the Y-functions for an excited state on top of the new effects we already see at the level of the Y-system. Aside from this it would also be interesting to work out the precise consequences of the vertex-to-IRF transformation for the considerations of this chapter.

We should mention that there has been very recent and promising progress towards finding a Lagrangian description of these $q$-deformed theories with $q$ a root of unity \cite{Hollowood:2014fha,Hollowood:2014rla}, starting from the non-Abelian T-dual of the original sigma model and introducing a deformation via a gauged Wess-Zumino-Witten model, so that the Poisson structure of the model interpolates between the one of the original sigma model and the Pohlmeyer reduced theory. It would be very exciting to see this story concretely realized at the quantum level, including fermions.

Interestingly, the interpolating Poisson structure was already used shortly before to define another class of integrable deformations of the string sigma model \cite{Delduc:2013qra,Delduc:2014kha},\footnote{For earlier and related work on deformed sigma models see e.g. \cite{Cherednik:1981df,Klimcik:2008eq,Klimcik:2002zj,Delduc:2013fga,Kawaguchi:2011ub,Kawaguchi:2012ve,Kawaguchi:2012gp,Kameyama:2013qka,Sfetsos:2013wia}.}  that surprisingly turns out to be described by real $q$ instead. Indeed in a light-cone gauge this deformed sigma model appears to be described precisely by our $q$-deformed model with $q$ real \cite{Arutyunov:2013ega}, and in fact the full model has $\mbox{PSU}_q(2,2|4)$ symmetry \cite{Delduc:2014kha}. Based on this structure, the thermodynamic Bethe ansatz has been worked out for these models as well \cite{Arutynov:2014ota}.\footnote{As we might expect, these equations differ less dramatically from their undeformed counterparts than the phase-deformed ones we have been discussing. This is in line with the situation in the XXZ spin chain \cite{Takahashi:book,Gaudin:1971gt} where $q$ real corresponds to $\Delta>1$.} Interestingly, it turns out these models exhibit a feature called `mirror-duality', whereby the deformed model at one value of the (sigma-model) deformation parameter is actually equal to the mirror version of the deformed model at a `dual' value of the deformation parameter (upon suitably identification of charges and units) \cite{Arutynov:2014ota,Arutyunov:2014cra}. In particular we can get \emph{the undeformed mirror model} that we extensively discussed in the previous chapters, as a natural `maximal' deformation limit of this deformed model, and complete its geometry to a solution of type IIB supergravity \cite{Arutyunov:2014cra}. We should mention that the maximal deformation limit was actually expected to yield a sigma model on $\mathrm{dS}_5 \times \mathrm{H}_5$ \cite{Delduc:2013qra}, which can indeed be realized by considering a suitable different real form of $\mbox{PSU}(4|4)$ \cite{Delduc:2014kha}. The mirror limit is a direct limit on the geometry however, requiring a specific rescaling of fields, and is related to $\mathrm{dS}_5 \times \mathrm{H}_5$ by a combinations of a time-like and a space-like T-duality instead.\footnote{Another maximal deformation limit was considered in \cite{Hoare:2014pna}, which results in a (non-real) geometry that is related to $\mathrm{dS}_5 \times \mathrm{H}_5$ by two space-like T-dualities.} Finally, in light of the discussion above we might wonder whether there is meaning in the analytic continuation from real $q$ to $q$ being a phase. Indeed this appears to be the case \cite{Hoare:2014pna} (up to questions regarding unitarity, cf. the discussion of section \ref{sec:psu22qdefintro}), in particular in the lower dimensional cases of $\mathrm{AdS}_3 \times \mathrm{S}^3$ and $\mathrm{AdS}_2 \times \mathrm{S}^2$ where there is no B-field that would become imaginary \cite{Hoare:2014pna}. It would be interesting to investigate these points further, especially in relation to the deformed models of \cite{Hollowood:2014rla}.

The construction of \cite{Delduc:2013qra} can also be adapted to so-called non-standard $q$-deformations \cite{Kawaguchi:2014qwa}, actually corresponding to Yangian (`undeformed') symmetry. In this way it appears possible to reproduce geometries such as the Lunin-Maldacena background and its generalizations \cite{Matsumoto:2014nra} that we were considering in chapter \ref{chapter:twistedspectrum}, and the method at least extends to other geometries obtained by TsT transformations such as the background dual to non-commutative SYM \cite{Matsumoto:2014gwa} found in \cite{Maldacena:1999mh}. Understanding a possible (partial) classification of deformed geometries preserving integrability based on the algebraic construction of \cite{Delduc:2013qra,Kawaguchi:2014qwa,Delduc:2014kha} is an interesting open question.

Finally, from the point of view of condensed matter physics both the real and phase-deformation are interesting deformations of the Hubbard and related models.\footnote{The Hubbard model is one of four parity invariant Hermitian lattice models \cite{Frolov:2011wg} that can be obtained from Shastry's R-matrix.} The thermodynamics of these models can be described with minor modifications of the equations in this chapter \cite{DeformedHubbardInProgress}.

\chapter*{Acknowledgements}

The author would like to thank G. Arutyunov, Z. Bajnok, J. Balog, J. Fokken, S. Frolov, W. Galleas, A. Hegedus, B. Hoare, A. Kl\"umper, M. de Leeuw, C. Sieg, A. Sfondrini, R. Suzuki, C. Toldo and M. Wilhelm for insightful discussions, and G. Arutyunov, Z. Bajnok, S. Frolov, B. Hoare, C. Sieg, C. Toldo and M. Wilhelm for useful comments on the manuscript. The author is supported by Einstein Foundation Berlin in the framework of the research project "Gravitation and High Energy Physics". This work is part of the ERC Advanced grant research programme No. 246974,
\emph{Supersymmetry: a window to non-perturbative physics"} and was performed in the framework of the Netherlands Organization for Scientific Research (NWO) VICI grant 680-47-602. The author also wishes to acknowledge further support from the People Programme (Marie Curie Actions) of the European Union's Seventh Framework Programme FP7/2007-2013/ under REA Grant Agreement No 317089.

\appendix

\chapter{Appendices}

\addtocounter{section}{2}

\section{Chapters \ref{chapter:finitevolumeIQFT}, \ref{chapter:AdS5string} and \ref{chapter:quantumTBA}}

\label{app:qdefandgeneral}

We would like to ask readers coming to this appendix from chapters \ref{chapter:finitevolumeIQFT} and \ref{chapter:AdS5string} to note that in this appendix we generically discuss the $q$-deformed or hyperbolic analogue of the S-matrices and kernels used in chapters \ref{chapter:finitevolumeIQFT} and \ref{chapter:AdS5string}. Concretely this deformation amounts to replacing rational functions of the rapidity by an appropriate ratio of hyperbolic functions. The results appropriate for chapters \ref{chapter:finitevolumeIQFT} and \ref{chapter:AdS5string} are found in the limit $q\rightarrow1$. As the limit $q\rightarrow1$ (or equivalently $k\rightarrow \infty$) is really quite trivially taken in almost every expression, we hope that this provides an adequate means of conciseness. We have indicated the appropriate $q\rightarrow 1$ limit where not completely obvious. Note that $x$ and $x_s$ should simply be interpreted as those of eqs. \eqref{eq:xmirror} and \eqref{eq:xstring} in the undeformed case and as those in eqs. \eqref{eq:qdefxmirror} and \eqref{eq:qdefxstring} when $q=e^{i \pi/k}$, and that $u_b \rightarrow 2$ in the limit $q\rightarrow 1$.

\subsection{The S-matrix}
\label{app:qdefmatrixSmatrix}

In this appendix we summarize the properties of the ($q$-deformed) S-matrix. We use $E_{ij}$ to denote the $4\times 4$ $(i,j)$ matrix unity, {\it i.e.} a matrix with a one in the $(i,j)$th entry and zeroes everywhere else. Next, we introduce the following definition
\begin{equation}
E_{kilj}=(-1)^{\epsilon(l)\epsilon(k)}E_{ki}\otimes E_{lj}\, ,
\end{equation}
where $\epsilon(i)$ denotes the parity of the index, equal to $0$ for $i=1,2$ (bosons) and to $1$ for $i=3,4$ (fermions). The matrices $E_{kilj}$
can be used to write down invariance with respect to the action of two copies of $\su_{(q)}(2)$. If we introduce
\begin{align}
\Lambda_1=&E_{1111}+\frac{q}{2}E_{1122}+\frac{1}{2}(2-q^2)E_{1221}+\frac{1}{2}E_{2112}+\frac{q}{2}E_{2211}+E_{2222}\, ,\nonumber\\
\Lambda_2=&\frac{1}{2}E_{1122}-\frac{q}{2}E_{1221}-\frac{1}{2q}E_{2112}+\frac{1}{2}E_{2211}\, , \nonumber \\
\Lambda_3=&E_{3333}+\frac{q}{2}E_{3344}+\frac{1}{2}(2-q^2)E_{3443}+\frac{1}{2}E_{4334}+\frac{q}{2}E_{4433}+E_{4444} \, , \nonumber\\
\Lambda_4=&\frac{1}{2}E_{3344}-\frac{q}{2}E_{3443}-\frac{1}{2q}E_{4334}+\frac{1}{2}E_{4433}\, , \nonumber\\
\Lambda_5=&E_{1133}+E_{1144}+E_{2233}+E_{2244}\, ,\\
\Lambda_6=&E_{3311}+E_{3322}+E_{4411}+E_{4422}\, , \nonumber\\
\Lambda_7=&E_{1324}-qE_{1423}-\frac{1}{q}E_{2314}+E_{2413}\, , \nonumber\\
\Lambda_8=&E_{3142}-qE_{3214}-\frac{1}{q}E_{4132}+E_{4231}\, , \nonumber\\
\Lambda_9=&E_{1331}+E_{1441}+E_{2332}+E_{2442}\, , \nonumber\\
\Lambda_{10}=&E_{3113}+E_{3223}+E_{4114}+E_{4224}\, , \nonumber
\end{align}
the S-matrix of our ($q$-deformed) `string' is given by
\begin{equation}
S_{12}(p_1,p_2)=\sum_{k=1}^{10}a_k(p_1,p_2)\Lambda_k\, ,
\end{equation}
where the coefficients are
\begin{align}
a_1=&1\, ,  \nonumber \\
a_2=&-q+\frac{2}{q}\frac{x^-_1(1-x^-_2x^+_1)(x^+_1-x^+_2)}{x^+_1(1-x^-_1x^-_2)(x^-_1-x^+_2)}\nonumber \\
a_3=&\frac{U_2V_2}{U_1V_1}\frac{x^+_1-x^-_2}{x^-_1-x^+_2}\nonumber \\
a_4=&-q\frac{U_2V_2}{U_1V_1}\frac{x^+_1-x^-_2}{x^-_1-x^+_2}+\frac{2}{q}\frac{U_2V_2}{U_1V_1}\frac{x^-_2(x^+_1-x^+_2)(1-x^-_1x^+_2)}{x^+_2(x^-_1-x^+_2)(1-x^-_1x^-_2)}\nonumber \\
a_5=&\frac{x^+_1-x^+_2}{\sqrt{q}\, U_1V_1(x^-_1-x^+_2)}
\\
\nonumber
a_6=&\frac{\sqrt{q}\, U_2V_2(x^-_1-x^-_2)}{x^-_1-x^+_2} \\
a_7=&\frac{ig}{2}\frac{(x^+_1-x^-_1)(x^+_1-x^+_2)(x^+_2-x^-_2)}{\sqrt{q}\, U_1V_1(x^-_1-x^+_2)\gamma_1\gamma_2}
\nonumber \\
a_8=&\frac{2i}{g}\frac{U_2V_2\,  x^-_1x^-_2(x^+_1-x^+_2)\gamma_1\gamma_2}{q^{\frac{3}{2}} x^+_1x^+_2(x^-_1-x^+_2)(x^-_1x^-_2-1)}\nonumber \\
a_9=&\frac{(x^-_1-x^+_1)\gamma_2}{(x^-_1-x^+_2)\gamma_1} \nonumber \\
\nonumber
a_{10}=&\frac{U_2V_2 (x^-_2-x^+_2)\gamma_1}{U_1V_1(x^-_1-x^+_2)\gamma_2}\, .
\end{align}
Here the central charges are given by
\begin{equation}
U_i^2=\frac{1}{q}\frac{x^+_i+\xi}{x^-_i+\xi}\, , ~~~~V^2_i=q\frac{x^+_i}{x^-_i}\frac{x^-_i+\xi}{x^+_i+\xi} \, ,
\end{equation}
and the parameters $\gamma_i$ are
\begin{equation}
\gamma_i=q^{\frac{1}{4}}\sqrt{\frac{ig}{2}(x^-_i-x^+_i)U_iV_i}\, .
\end{equation}
The dependence of the S-matrix on the variables $\gamma_i$, $i=1,2$, is gauge-like. Indeed, introducing the diagonal matrix
$\Gamma_i={\rm diag}(1,1,\gamma_i,\gamma_i)$, we find
\begin{equation}
\Big[ \Gamma_1\otimes \Gamma_2\Big]\,  S_{12}^{\gamma_i=1}(z_1,z_2)\, \Big[\Gamma_1^{-1}\otimes \Gamma_2^{-1}\Big]=S_{12}(z_1,z_2)\, ,
\end{equation}
where $S_{12}^{\gamma_i=1}$ is the S-matrix where $\gamma_1$ and $\gamma_2$ are set to one.

\medskip

Let us summarize the most important properties of the S-matrix. The S-matrix satisfies
\begin{itemize}
\item The Yang-Baxter equation;

\item The unitarity condition
\begin{equation}
S_{21}(z_2,z_1)S_{12}(z_1,z_2)=1;
\end{equation}

\item The transposition property
\begin{equation}
S^t(z_1,z_2)=\mI^g \Omega \, S(z_1,z_2) \, \Omega^{-1} \mI^g\, ,
\end{equation}
where $\mI^g=(-1)^{\epsilon_i\epsilon_j}E_i^i \otimes E_j^j$ is the graded identity and $\Omega$ is given by
\begin{equation}
\Omega=\exp \frac{i\pi}{2}\Big(E_1\otimes F_1+F_1\otimes E_1+E_3\otimes F_3+F_3\otimes E_3\Big)\, .
\end{equation}
Here $E_i,F_i$ are positive and negative roots of two $\su(2)$s. The graded identity commutes with $\Omega$. In the limit
$q\to 1$ one finds
$$
\lim_{q\to 1}\Big[\Omega\, S(z_1,z_2) \, \Omega^{-1}\Big]= \lim_{q\to 1}S(z_1,z_2) \, .
$$
Note also the formula $x^+(z,q)=-x^{-}(-z,1/q)$.

\item Generalized physical pseudo-unitarity

\begin{equation}
\label{eq:pseudounitarity}
S(z_1,z_2)^\dagger = B S^{-1}(z_1^*,z_2^*) B^{-1}
\end{equation}
where\footnote{Our matrix $B$ is Hermitian as is required for pseudo-unitarity \cite{Mostafazadeh:2002hb}.}
\begin{equation}
\label{eq:pseudounitaritylocal}
B = A\otimes A \, , \, \, \mbox{with} \, \, A = \mbox{diag}(\sigma_1,\sigma_1)\,
\end{equation}
and $\sigma_1 = \left(\begin{array}{cc} 0 & 1 \\ 1& 0\end{array}\right)$ is the first Pauli matrix. In the limit $q\rightarrow1$ the S-matrix commutes with $B$ and is (physically) unitary.

In addition, for $q= e^{i \pi/k}$ the S-matrix has a unitary spectrum on the real line of the string and mirror theory, but is generically not unitarizable by a local basis transformation. In fact the many body S-matrix on the real string line has non-unitary eigenvalues. Nonetheless, on the mirror line the many body S-matrix appears to have unitary eigenvalues always.\footnote{We have checked this numerically for the first few $n$-body S-matrices.} This is equivalent to being quasi-unitary, where quasi-unitarity means that the matrix $B$ above is positive definite; $B=O O^\dagger$ \cite{Mostafazadeh:2001nr}. However, there does not appear to be a matrix $B$ of this form which is also factorizable over the one-particle basis, and as such we have not rigorously proved quasi-unitarity on the mirror line. Of course, as noted in chapter \ref{chapter:quantumTBA} after a change from the vertex to an IRF picture the S-matrix is unitary \cite{Hoare:2013ysa}.

\item For coincident arguments the S-matrix reduces to the (graded) permutation.

\item It is compatible with crossing symmetry. If we introduce the following charge conjugation matrix
{\small
$$
C=\left(\begin{array}{cccc}  0 & -i q^{1/2} & 0 & 0\\
i q^{-1/2} & 0 & 0 & 0 \\
0 & 0 & 0 & q^{1/2} \\
0 & 0 & -q^{-1/2} & 0
\end{array}\right)\, ,
$$
}
the crossing relation reads\footnote{Under the crossing transformation $x^{\pm}\to 1/x^{\pm}$ so that the central charges transform as
$U^2\to 1/U^2$ and $V^2\to 1/V^2$.}
\begin{equation}
S_{12}(z_1,z_2)C_2 S_{12}^{t_2}(z_1,z_2-w_2)C_2^{-1}=\frac{1}{q}
\frac{(x_1^+-x_2^-)\Big(1-\frac{1}{x_1^+x_2^+}\Big)}{(x_1^--x_2^-)\Big(1-\frac{1}{x_1^-x_2^+}\Big)}\, .
\end{equation}
\end{itemize}

\vskip 0.3cm

The complete S-matrix comprising two copies of $S$ can be written in the form\footnote{In \cite{Hoare:2011wr} the factor $\frac{x_1^+ x_2^-}{x_1^-x_2^+}$ in the S-matrix was replaced by $\frac{U_1^2}{U_2^2}=\frac{x_1^++\xi}{x_1^-+\xi}\frac{x_2^-+\xi}{x_2^++\xi}$. This is a minor change which leads to the corresponding modification of the crossing equation for $\sigma$.}
\begin{equation}
\hspace{-1.5cm}
{\mathbf S}=S_{\su(2)} S\hat{\otimes} S\, , ~~~S_{\su(2)}=\frac{1}{\sigma(z_1,z_2)^2}\frac{x_1^+}{x_1^-}\frac{x_2^-}{x_2^+}\cdot
\frac{x_1^--x_2^+}{x_1^+-x_2^-}\frac{1-\frac{1}{x_1^-x_2^+}}{1-\frac{1}{x_1^+x_2^-}}\, ,
\end{equation}
where $\hat{\otimes} $ stands for the graded tensor product and $\sigma$ is the dressing phase \cite{Arutyunov:2004vx,Beisert:2006ib,Beisert:2006ez}. Substituting this representation for ${\mathbf S}$  into the crossing equation \cite{Janik:2006dc}, we deduce that the dressing phase must obey the following equation
\begin{equation}
\label{cross1}
\sigma(z_1,z_2)\sigma(z_1,z_2-\omega_2)=q^{-1}\frac{x_1^+}{x_1^-}\frac{x_1^--x_2^+}{x_1^--x_2^-}\frac{1-\frac{1}{x_1^+x_2^+}}{1-\frac{1}{x_1^+x_2^-}}\, .
\end{equation}
Using the unitarity relation the previous formula also implies that
\begin{equation}
\label{cross2}
\sigma(z_1+\omega_2,z_2)\sigma(z_1,z_2)=q^{-1}\frac{x_2^-}{x_2^+}\cdot \frac{x_1^--x_2^+}{x_1^--x_2^-}\frac{1-\frac{1}{x_1^+x_2^+}}{1-\frac{1}{x_1^+x_2^-}}\, .
\end{equation}
The formulae (\ref{cross1}) and (\ref{cross2})
are $q$-deformed analogue of eqs. (2.8), (2.9) and (2.10) from \cite{Arutyunov:2009kf}. A natural solution of this equation has been obtained in \cite{Hoare:2011wr}, where an explicit expression for $\sigma$ can be found, generalizing the DHM representation of the undeformed dressing phase \cite{Dorey:2007xn}.

The dressing factor $\sigma^{PQ}$ which describes scattering of $P$- and $Q$-particle bound states can be obtained from the dressing factor
for fundamental particles by means of fusion. The corresponding crossing equations have almost the same form as in the undeformed case ({\it cf.} eqn.(2.14) in \cite{Arutyunov:2009kf})
\begin{align}
\sigma^{PQ}(z_1,z_2)\sigma^{PQ}(z_1,z_2-\omega_2)=&q^{-PQ}\left(\frac{x_1^+}{x_1^-}\right)^Q h^{PQ}(z_1,z_2)\, ,\\
\sigma^{PQ}(z_1,z_2)\sigma^{PQ}(z_1+\omega_2,z_2)=&q^{-PQ}\left(\frac{x_2^-}{x_2^+}\right)^P h^{PQ}(z_1,z_2)\, .
\end{align}
Here $x_1$ and $x_2$ solve the bound state conditions
\begin{align}
\frac{1}{q^P}\left(x^+_1+\frac{1}{x^+_1}\right)-q^P\left(x^-_1+\frac{1}{x^-_1}\right)=&\left(q^P-\frac{1}{q^P}\right)\left(\xi+\frac{1}{\xi}\right)\, \, ,\\
\frac{1}{q^Q}\left(x^+_2+\frac{1}{x^+_2}\right)-q^Q\left(x^-_2+\frac{1}{x^-_2}\right)=&\left(q^Q-\frac{1}{q^Q}\right)\left(\xi+\frac{1}{\xi}\right)\, ,
\end{align}
and we have introduced the crossing function
\begin{equation}
\label{crossfunc}
h^{PQ}=\frac{x_1^--x_2^+}{x_1^--x_2^-}\frac{1-\frac{1}{x_1^+x_2^+}}{1-\frac{1}{x_1^+x_2^-}} \prod_{j=1}^{P-1} S_{Q-P+2j}\, ,
\end{equation}
where $S_Q$ is defined in \eqref{eq:SM}. We discuss the solution of these crossing equations in detail in \ref{app:qdefdressingphase} below.

\subsection{S-matrices and kernels}

\label{app:qdefSmatricesandkernels}

We would like to ask readers coming here from chapters \ref{chapter:finitevolumeIQFT} and \ref{chapter:AdS5string} to consider the paragraph at the start of section \ref{app:qdefandgeneral}, and in line with this consider the expressions below in the limit $q\rightarrow 1$ ($k\rightarrow \infty$) and disregard any mention of so-called `negative parity' strings.

\subsection*{Fusion of S-matrices}

In what follows, unless otherwise indicated the indices $M$, $N$ and $L$ run from one to $k-1$ while the indices $Q$,$P$ and $R$ run from one to $k$.

\subsection*{$\boldsymbol S_M$ and $\boldsymbol S_{MN}$}

The basic S-matrix $S_1$
\begin{equation}
S_1(u-v) \equiv \frac{\sinh{\frac{\pi g}{2k}(u - v - i/g)}}{\sinh{\frac{\pi g}{2k}(u - v + i/g)}} \, ,
\end{equation}
fuses as
\begin{equation}
\label{eq:SM}
S_M (u-v) = \frac{\sinh{\frac{\pi g}{2k}(u - v - M i/g)}}{\sinh{\frac{\pi g}{2k}(u - v + M i/g)}}\,.
\end{equation}
In the undeformed limit $k\rightarrow \infty$ relevant for chapters \ref{chapter:finitevolumeIQFT} and \ref{chapter:AdS5string} this of course reduces to the expression
\begin{equation}
S_M (u-v) = \frac{u - v - M i/g}{u - v + M i/g}\,.
\end{equation}
and we note that in terms of the S-matrices in chapter \ref{chapter:finitevolumeIQFT} we have $S_M = (S^{fM})^{-1} = (S^{Mf})^{-1}$.

Note that the fundamental scattering matrix for negative parity particles is almost inverse to the one for a $k-1$ positive parity string, namely
\begin{equation}
\label{eq:negparSM}
S_0 (u-v) S_{k-1} (u-v) = -1\,.
\end{equation}

Next, by fusion over the previously untouched argument we get
\begin{eqnarray}
S^{MN}(u-v)=S_{M+N}(u-v)S_{|M-N|}(u-v)\prod_{m=1}^{\min{(M,N)}-1}S_{|M-N|+2m}^2(u-v)\, ,
\end{eqnarray}
which in this form is manifestly symmetric under interchange of $M$ and $N$. Negative parity particles scatter with positive parity strings with
\begin{equation}
S_{0M} (u-v) \equiv S_{1M} (u+ik/g-v) = S^{-1}_{M+1} (u-v) S^{-1}_{M-1} (u-v)  \,,
\end{equation}
while between themselves they scatter as positive parity particles meaning
\begin{equation}
S_{00} (u-v) \equiv S_{11} (u-v) \,.
\end{equation}
Note that here we have
\begin{equation}
\label{eq:negparSMN}
S_{0M} (u-v) S_{k-1,M} (u-v) = 1\,.
\end{equation}
Finally, note that these S-matrices are trivial when one of the indices is equal to $k$
\begin{equation}
S_k = -1 \, , \, \, \, S_{Mk} = S_{kM} =1\,.
\end{equation}
This shows that at $k$ these solutions to the discrete Laplace equation have a natural boundary.

\subsection*{$\boldsymbol S^{yQ}$, $\boldsymbol S^{Qy}$ and $\boldsymbol S^{QM}_{xv}$}

Fusing the scattering matrix of $y^\pm$ particles with fundamental particles over a $Q$-particle bound state directly gives
\begin{align}
S_-^{yQ}(u,v)&= q^{Q/2} \, \frac{x(u) -x^-(v)}{x(u)-x^+(v)}\sqrt{\frac{x^+(v)}{x^-(v)}} \, , \\
S_+^{yQ}(u,v) &= q^{Q/2} \, \frac{\frac{1}{x(u)} -x^-(v)}{\frac{1}{x(u)}-x^+(v)}\sqrt{\frac{x^+(v)}{x^-(v)}} \, ,
\end{align}
where $x^\pm$ are the parameters for a $Q$-particle bound state; $x^\pm(v) = x(v \pm i Q/g)$, and the subscript $\pm$ in $S^{Qy}_\pm$ denotes the sign of the imaginary part of the $y$-particle under consideration. Analogously we define the S-matrices for scattering of bound states with $y$-particles as
\begin{align}
S_-^{Qy}(u,v)&= q^{Q/2} \, \frac{x^-(u)-x(v)}{x^+(u)-x(v)}\sqrt{\frac{x^+(u)}{x^-(u)}} \, , \\
S_+^{Qy}(u,v) &= q^{Q/2} \, \frac{x^-(u)-\frac{1}{x(v)}}{x^+(u)-\frac{1}{x(v)}}\sqrt{\frac{x^+(u)}{x^-(u)}} \, .
\end{align}
The scattering matrix of a $Q$-particle bound state with an $M|vw$-string of positive parity is given by
\begin{equation}
\label{eq:SQMxv}
S^{QM}_{xv}(u,v) \equiv q^{Q}\frac{x^-(u)-x^+(v) }{x^+(u)-x^+(v) }\frac{x^-(u)-x^-(v) }{x^+(u)-x^-(v)}\frac{x^+(u)}{x^-(u)}
\prod_{i=1}^{M-1} S_{Q+M-2i}(u-v)
\end{equation}
where $x^\pm(v) = x(v \pm i M/g)$, while for length one $vw$-strings with negative parity we have simply
\begin{equation}
S^{Q0}_{xv}(u,v) \equiv q^{Q}\frac{x^-(u)-x(v+i(k+1)/g) }{x^+(u)-x(v+i(k+1)/g) }\frac{x^-(u)-\frac{1}{x(v+i(k-1)/g)} }{x^+(u)-\frac{1}{x(v+i(k-1)/g)}}\frac{x^+(u)}{x^-(u)} \, .
\end{equation}
Again there is a special relation between these S-matrices, namely
\begin{equation}
\label{eq:negparSQMxv}
S^{Q0}_{xv}(u,v) S^{Qk-1}_{xv}(u,v) = (-1)^Q \, .
\end{equation}

\subsection*{$\boldsymbol S_{\mathfrak{sl}(2)}^{QP}$}

Introducing the same split as in the undeformed case we write
\begin{equation}
S_{\mathfrak{sl}(2)}(x_1,x_2) = S_2^{-1} \Sigma(x_1,x_2)^{-2}\, ,~~
\end{equation}
where we have introduced the improved dressing factor $\Sigma$
\begin{equation}
\Sigma(x_1,x_2)\equiv \frac{1-\frac{1}{x_1^+x_2^-}}{1-\frac{1}{x_1^-x_2^+}}\sigma(x_1,x_2)\, .
\end{equation}
This S-matrix fuses as
\begin{equation}
S^{QM}_{\mathfrak{sl}(2)}= S_{QM}^{-1} \Sigma_{QM}^{-2}\, ,~~
\end{equation}
where $\Sigma_{QM}$ is discussed in section \ref{app:qdefdressingphase} below.

\subsection*{Kernels and their properties}

\label{app:qdefkernels}

\subsection*{$\boldsymbol K_M$ and $\boldsymbol K_{MN}$}

As in the main text, we have
\begin{equation}
\label{eq:KM}
K_M(u) \equiv \frac{1}{2\pi i} \frac{d}{du} \log{S_M(u)} = \frac{g}{2k} \frac{\sin{\frac{M\pi}{k}}}{\cosh{\frac{\pi g u}{k}}-\cos{\frac{M\pi}{k}}}\, ,
\end{equation}
and
\begin{equation}
K_{MN}(u) \equiv \frac{1}{2\pi i} \frac{d}{du} \log{S_{MN} (u)} =\mathbf{K}_{M+N}+\mathbf{K}_{|M-N|} + 2 \sum_{j=1}^{\min{(M,N)}-1}\mathbf{K}_{|M-N|+2j}\, .
\end{equation}
Note that these kernels are positive and in the limit $k \rightarrow \infty$ are also the kernels used in chapter \ref{chapter:finitevolumeIQFT}; $K^{Mf}=K^{fM}=K_M$ and $K^{MN}=K_{MN}$.

The kernels involving scattering of the negative parity particles can be defined in terms of the above via \eqref{eq:negparSM} and \eqref{eq:negparSMN}, but their definition is explicitly given where used in the main text. Of course had it been possible for $M$ to exceed $2k$ we would have to take into account the $2k$ periodicity of the S-matrix and take $M \mod 2k$ in the above formulae. In order to simplify the TBA equations we would like to understand the integral identities satisfied by these kernels.

For real rapidities the kernel $K_M$ is real and positive and has the following Fourier transform
\begin{equation}
\hat{K}_M(\omega) \equiv \int_{-\infty}^{\infty} du e^{i g \omega u} K_M(u)= \frac{\sinh{(k-M)\omega }}{\sinh{k\omega}}\, ,
\end{equation}
where we defined the Fourier transform with an unconventional factor of $g$. The Fourier transform of the kernel $K_{MN}$ is then
\begin{equation}
\hat{K}_{MN}(\omega)= \frac{\coth{\omega}}{\sinh{k\omega}} \left(\cosh{(|M-N|-k)\omega}-\cosh{(M+N-k)\omega}\right) - \delta_{MN} \, .
\end{equation}
These kernels satisfy the following properties
\begin{align}
& \hat{K}_N (\delta_{N,M} - I_{NM} \hat{s}) = \hat s \delta_{M,1} \, , \label{eq:simpKM} \, ,\\
& \hat{K}_{ML}(\delta_{L,N} - I_{LN} \hat{s}) = \hat{s}I_{M,N} \, .\label{eq:simpKMN}\, ,
\end{align}
where $\hat{s}(\omega)=\frac{1}{2\cosh{\omega}}$ is the Fourier transform of $s$
\begin{equation}
s(u)=\frac{g}{2\pi}\int_{-\infty}^{\infty} d\omega \frac{e^{-i g \omega u}}{2\cosh{\omega}}=\frac{g}{4 \cosh\frac{g \pi u}{2}}\, ,
\end{equation}
and $I_{MN} = \delta_{M,N-1} + \delta_{M,N+1}$ is the incidence matrix, to be appropriately interpreted on the boundary as $I_{N,k-1} = \delta_{N,k-2}$.\footnote{Note that we have $K_{k}=K_{Mk}=0$.} Since we will need it again, let us define
\begin{equation}
\label{eq:Kp1inv}
(K+1)^{-1}  \equiv 1 - I \star s\, .
\end{equation}

\subsection*{$\boldsymbol K^{yQ}$, $\boldsymbol K^{Qy}$, $\boldsymbol  K^{QM}_{xv}$ and $\boldsymbol K^{MQ}_{vwx}$}

In line with the conventions of \cite{Arutyunov:2009ux} in the main text we used the following positive kernels
\begin{align}
K^{QM}_{xv}(u,v) & \equiv \frac{1}{2\pi i} \frac{d}{du} \log S^{QM}_{xv} (u,v)\, ,\\
K^{MQ}_{vwx}(u,v) & \equiv - \frac{1}{2\pi i} \frac{d}{du} \log S^{QM}_{xv}(v,u)\, ,\\
K^{Qy}_{\beta}(u,v) & \equiv \frac{1}{2\pi i} \frac{d}{du} \log S^{Qy}_{\beta} (u,v) \, ,\\
K^{yQ}_{\beta}(u,v) & \equiv \beta \frac{1}{2\pi i} \frac{d}{du} \log S^{Qy}_{\beta} (v,u) \, .
\end{align}
The scattering kernels for negative parity strings can be defined in terms of the above kernels via \eqref{eq:negparSQMxv}. We will mainly work with linear combinations of the last two kernels, namely
\begin{align}
K^{Qy}_{-}(u,v) - K^{Qy}_{+}(u,v) & \equiv K_{Qy}(u,v) = K(u+iQ/g,v) - K(u-iQ/g,v)\, , \\
K^{Qy}_{-}(u,v) + K^{Qy}_{+}(u,v) & = K_Q(u,v)\, , \\
K^{yQ}_{-}(u,v) - K^{yQ}_{+}(u,v) & = K_Q(u,v)\, , \label{eq:KyQKQdef}\\
K^{yQ}_{-}(u,v) + K^{yQ}_{+}(u,v) & \equiv K_{yQ}(u,v) = K(u,v+iQ/g) - K(u,v-iQ/g)\, ,\label{eq:KyQKyQdef}
\end{align}
where
\begin{equation}
K(u,v) = \frac{1}{2\pi i} \frac{d}{du} \log \frac{x(u)-\frac{1}{x(v)}}{x(u)-x(v)} \, ,
\end{equation}
and $K_Q$ is defined in \eqref{eq:KM}. These kernels satisfy
\begin{align}
K^{QN}_{xv}(\delta_{N,M}-I_{NM}\star s) & = \delta_{Q-1,M} s + \delta_{M,1} K_{Qy}\, \hat{\star}\,  s  \, , \label{eq:simpKQMxv} \\
K^{MP}_{vwx}(\delta_{P,Q}-I_{PQ}\star s) & = \delta_{M+1,Q} s + \delta_{Q,1}\check{K}_M\,  \check{\star} \, s \, ,\label{eq:simpKMQvwx} \\
K_{yP}(\delta_{P,Q}-I_{PQ}\star s)& = \delta_{Q,1}(2\check{K}\, \check{\star}\,  s + s)\, ,\label{eq:simpKyQ}  \\
K_{P}(\delta_{P,Q}-I_{PQ}\star s)& = \delta_{Q,1} s\, ,\label{eq:simpKQ}
\end{align}
where we note again that for $M$ (not $Q$-particle) type indices $I_{N,k-1} = \delta_{N,k-2}$ while for $Q$ type indices this incidence matrix is fine for $Q=k-1$, but at $Q=k$ we need slightly different identities. Namely
\begin{align}
K^{MP}_{vwx}(\delta_{P,k}-2\delta_{P,k-1}\star s) & = \delta_{M,k-1} s - K_{M,k-1} \star s \, ,\label{eq:simpKMQvwxbndry} \\
K_{yP}(\delta_{P,k}-2\delta_{P,k-1}\star s)& = 0\, .\label{eq:simpKyQbndry}
\end{align}
The kernels entering in the above identities are
\begin{equation}
\check{K}(u,v) = \theta(|u|-u_b) \frac{1}{2\pi i} \frac{d}{du} \log \frac{x(u)-\frac{1}{x_s(v)}}{x(u)-x_s(v)} \, ,
\end{equation}
and
\begin{equation}
\check{K}_M(u,v) \equiv \check{K}(u+i M/g,v)+ \check{K}(u-i M/g,v)\, .
\end{equation}
Let us also define
\begin{equation}
\label{eq:checkE}
\check{\cal E}=\log\frac{x_s+\xi}{\frac{1}{x_s}+\xi}\, .
\end{equation}

\subsection*{$\boldsymbol K_{\mathfrak{sl}(2)}^{QP}$}

The main kernel $K_{\mathfrak{sl}(2)}$ has the following structure
\begin{equation}
K_{\mathfrak{sl}(2)}^{QP}(u,v) = - K_{QP}(u-v) - 2 K^\Sigma_{QP}(u,v)\,.
\end{equation}
As proven below in section \ref{app:qdefdressingphase}, it satisfies
\begin{equation}
K^\Sigma_{QP}\star (K + 1)^{-1}_{PR}=\delta_{1,R}\check{K}_{Q}^\Sigma \check{\star} s \,,
\end{equation}
where the kernel $\check{K}_{Q'}^\Sigma(u,v)$ vanishes for $|v|<u_b$, and $(K+1)^{-1}$ is defined in (\ref{eq:Kp1inv}).

\subsection{The dressing phase}

\label{app:qdefdressingphase}

\subsubsection*{The dressing phase for fundamental particles of the 'string' theory}

The crossing equation that follows from the $q$-deformed $R$-matrix has a solution which is a natural deformation of the $\ads$ dressing phase \cite{Arutyunov:2004vx,Beisert:2006ez,Dorey:2007xn}. This solution, which we will denote $\tilde{\sigma}$, was found in \cite{Hoare:2011wr}. The dressing phase in our conventions, $\sigma$, is related to $\tilde{\sigma}$ as
\begin{equation}
\label{eq:dressingvsdressing}
\sigma^2(x_1,x_2)  = \tilde{\sigma}^2 (x_1,x_2) \frac{x_1^+}{x_1^-} \frac{x_2^-}{x_2^+} \frac{x_1^- +\xi}{x_1^+ +\xi}\frac{x_2^+ +\xi}{x_2^- +\xi} \equiv  \tilde{\sigma}^2 (x_1,x_2) \frac{P(x_1)}{P(x_2)}\, .
\end{equation}
Both $\sigma$ and $\tilde{\sigma}$ solve the $\ads$ crossing equation \cite{Janik:2006dc} in the limit $q\rightarrow1$. The dressing phase $\tilde{\sigma}$ is conventionally written in the form
\begin{equation}
\tilde{\sigma}(z_1,z_2) \equiv e^{i \tilde{\theta}(z_1,z_2)} = \exp { i\left(\chi(x_1^+,x_2^+) - \chi(x_1^-,x_2^+) - \chi(x_1^+,x_2^-) + \chi(x_1^-,x_2^-)\right) } \, ,
\end{equation}
where when both particles are in the string region, the $\chi$-functions are given by
\begin{equation}
\chi(x_1,x_2) = i \oint_{|z|=1} \frac{dz}{2 \pi i} \frac{1}{z-x_1}\oint_{|w|=1} \frac{dw}{2 \pi i} \frac{1}{w-x_2} \log  \frac{\Gamma_{q^2} (1+\frac{ig}{2}(u(z)-u(w)))}{\Gamma_{q^2} (1-\frac{ig}{2}(u(z)-u(w)))}\,.
\end{equation}
Here $\Gamma_q$ is the $q$-analogue of the $\Gamma$ function, which satisfies
\begin{equation}
\label{eq:Gammaqdef}
\Gamma_{q^2}(1+x) = \frac{1-q^{2x}}{1-q^2} \Gamma_{q^2}(x)\, .
\end{equation}
Under analytic continuation to other regions of the rapidity torus the expression for the $\chi$-function changes. The above double integral is commonly denoted $\Phi$
\begin{equation}
\label{eq:phidef}
\Phi(x_1,x_2) \equiv i \oint_{|z|=1} \frac{dz}{2 \pi i} \frac{1}{z-x_1}\oint_{|w|=1}  \frac{1}{w-x_2}\frac{dw}{2 \pi i} \log  \frac{\Gamma_{q^2} (1+\frac{ig}{2}(u(z)-u(w)))}{\Gamma_{q^2} (1-\frac{ig}{2}(u(z)-u(w)))}\, ,
\end{equation}
and is equal to the $\chi$-function in the string region. The $\Phi$-function has a discontinuity on the edge of the string region ($|x^\pm|=1$), and to properly define the $\chi$-function beyond it we will need further terms, which leads us to introduce the $\Psi$-function
\begin{equation}
\label{eq:psidef}
\Psi(x_1,x_2) \equiv i \oint_{|z|=1} \frac{dz}{2 \pi i} \frac{1}{z-x_2} \log  \frac{\Gamma_{q^2} (1+\frac{ig}{2}(u_1-u(z)))}{\Gamma_{q^2} (1-\frac{ig}{2}(u_1-u(z)))} \, .
\end{equation}
With these definitions we are ready to give the expression for the dressing phase in the currently relevant regions of the torus, defined as
\begin{equation}
\mathcal{R}_{0}: |x^\pm|>1\, , \, \, \, \mathcal{R}_{1}: |x^+|<1,|x^-|>1\, , \, \, \,  \mathcal{R}_{2}: |x^+|<1,|x^-|<1\, .
\end{equation}
$\mathcal{R}_{a,b}$ denotes regions on the product of two rapidity tori in the obvious fashion. We will be most interested in the (bound state) dressing phase on the real line of the mirror theory, which lies in region $\mathcal{R}_{1,1}$. For completeness let us first briefly repeat the explicit proof that $\sigma$ satisfies the crossing equation, given in \cite{Hoare:2011wr}. In order to do so, we need to analytically continue the dressing phase to region $\mathcal{R}_{2,0}$.

\subsubsection*{Proof of crossing}

\paragraph{The dressing phase in $\mathcal{R}_{1,0}$\\}
Continuing $\Phi$ through $|x_1^+|=1$ to $|x_1^+|<1$ gives

\begin{align}
{\cal R}_{1,0}:\quad \chi(x_1^+, x_2^\pm) =& \,\Phi(x_1^+, x_2^\pm) - \Psi(x_1^+, x_2^\pm)\,,\\
 \chi(x_1^-, x_2^\pm) =& \,\Phi(x_1^-, x_2^\pm)\,.
\end{align}

\paragraph{The dressing phase in $\mathcal{R}_{2,0}$\\}
With $|x_1^\pm|<1$ we have

\begin{align}
{\cal R}_{2,0}:\quad \chi(x_1^+, x_2^\pm) =& \,\Phi(x_1^+, x_2^\pm) - \Psi(x_1^+, x_2^\pm) +\frac{1}{i}\log \frac{\frac{1}{x_1^-}-x_2^\pm}{x_1^--x_2^\pm}\,,\\
\chi(x_1^-, x_2^\pm) =& \,\Phi(x_1^-, x_2^\pm) -
\Psi(x_1^-, x_2^\pm)\,.
\end{align}

\paragraph{Identities I\\}

In order to prove the crossing relation between $\mathcal{R}_{2,0}$ and $\mathcal{R}_{0,0}$ we will need some identities. First of all we have\footnote{These identities follow by changing variables from $z$ to $z^{-1}$ in the second integral on the left hand side.}
\begin{align}
\Phi(x_1,x_2) + \Phi(1/x_1,x_2) =&\, \Phi(0,x_2)\,,\label{eq:phibasic}\\
\Psi(x_1,x_2) + \Psi(x_1,1/x_2) =&\, \Psi(x_1,0)\,. \label{eq:psibasic}
\end{align}
Secondly we need
\begin{align}
\nonumber \Psi(1/x_1^-,x_2^+)-& \Psi(1/x_1^+, x_2^+) +\Psi(1/x_1^+, x_2^-)
-\Psi(1/x_1^-, x_2^-) =\\
& \frac{1}{i}\log\frac{1-\frac{1}{x_1^-x_2^+}}{1-\frac{1}{x_1^-x_2^-}}\frac{1-\frac{1}{x_1^+x_2^+}}{1-\frac{1}{x_1^+x_2^-}} + i \log \frac{1+\frac{\xi}{x_2^+}}{1+\frac{\xi}{x_2^-}}\, .
\end{align}
This identity follows from the identity
\begin{equation}
\label{eq:basicpsiid}
\Psi(1/x_1^-,x_2)-\Psi(1/x_1^+, x_2) = -i \log \frac{x_2 - \frac{1}{x_1^+}}{x_2}\frac{x_2 - \frac{1}{x_1^-}}{x_2+\xi} \, ,
\end{equation}
whose derivation plays an important role in the fusion of the mirror dressing phase below as well, so let us discuss it in some detail. To prove this identity, we begin by combining the two integrals
\begin{align}
\Psi(1/x_1^-,x_2)-& \Psi(1/x_1^+, x_2) =\\
& = i \oint_{|z|=1} \frac{dz}{2 \pi i} \frac{1}{z-x_2} \log  \frac{\Gamma_{q^2} (1+\frac{ig}{2}(u_1^- -u(z)))}{\Gamma_{q^2} (1-\frac{ig}{2}(u_1^--u(z)))}\frac{\Gamma_{q^2} (1-\frac{ig}{2}(u_1^+ -u(z)))}{\Gamma_{q^2} (1+\frac{ig}{2}(u_1^+-u(z)))}\nonumber
\end{align}
Next we use the defining property of the $\Gamma_q$ function given in eqn. \eqref{eq:Gammaqdef} to find
\begin{align}
\label{eq:basicpsiidstep2}
\Psi(1/x_1^-,x_2)-& \Psi(1/x_1^+, x_2) = \, i \oint_{|z|=1} \frac{dz}{2 \pi i} \frac{1}{z-x_2} \log \frac{1- q^{i g(u_1^- - u(z))}}{1-q^2}\frac{1- q^{i g(u(z) - u_1^+)}}{1-q^2}\, \nonumber\\
= & \, i \oint_{|z|=1} \frac{dz}{2 \pi i} \frac{1}{z-x_2} \log \left(1- \tfrac{z + \frac{1}{z} + \xi +\frac{1}{\xi}}{x_1^- + \frac{1}{x_1^-} + \xi +\frac{1}{\xi}}\right)\left(1- \tfrac{x_1^+ + \frac{1}{x_1^+} + \xi +\frac{1}{\xi}}{z + \frac{1}{z} + \xi +\frac{1}{\xi}}\right) \, , \\
= & \, i \oint_{|z|=1} \frac{dz}{2 \pi i} \frac{1}{z-x_2}
\log
\left(-\tfrac{x_1^-\xi}{(x_1^-+\xi)(x_1^-\xi+1)}
\tfrac{(z-x_1^-)\left(z-\tfrac{1}{x_1^-}\right)(z-x_1^+)\left(z-\tfrac{1}{x_1^+}\right)}{z (z+\xi)(z+\frac{1}{\xi})}
\right) \, ,\nonumber
\end{align}
where we note that the $\log(1-q^2)$ terms do not contribute to the integral as the integration contour can be shrunk to nothing. After integrating by parts\footnote{Note again that $|x_2|>1$.} we find
\begin{align}
\Psi(1/x_1^-,x_2)& \,-  \Psi(1/x_1^+, x_2) = \nonumber \\
& \, -i \oint_{|z|=1} \frac{dz}{2 \pi i} \log(z-x_2) \left( - \tfrac{1}{z} + \tfrac{1}{z-x_1^+} + \tfrac{1}{z-x_1^-} + \tfrac{1}{z-\frac{1}{x_1^+}} + \tfrac{1}{z-\frac{1}{x_1^-}} - \tfrac{1}{z+\xi} - \tfrac{1}{z+\frac{1}{\xi}}\right) \nonumber \\
& \, \hspace{67pt} = -i \log \frac{x_2 - \frac{1}{x_1^+}}{x_2}\frac{x_2 - \frac{1}{x_1^-}}{x_2+\xi} \, ,\label{eq:basicpsiidstep3}
\end{align}
by summing the poles within the unit circle. Here we used the fact that $|x_1^\pm|>1$. Had it been different, the respective $x$ would have to be replaced by $1/x$ in the above expression. Note that $\Psi$ is invariant under inversion of its first argument since $u(x) = u(1/x)$, so that the above also directly applies to $\Psi(x_1^-,x_2)-  \Psi(x_1^+, x_2)$.

In the construction of the mirror bound state dressing phase we will need a further identity,\footnote{In these derivations we will not be very careful about factors of $i \pi$ - the typical ambiguity in defining $\tfrac{1}{2} \log(-1)^2$ - as they do not affect $\log \sigma^2$.} namely
\begin{align}
\Psi(1/x_1^-,0) \,-  \Psi(1/x_1^+,0) =& \, i\log\frac{-\xi^2}{(1-q^2)^2} +i \log x_1^- x_1^+ \frac{x_1^-}{x_1^- + \xi}\frac{1}{\xi x_1^- + 1} \label{eq:basicpsiidx2zero}\\
 = & \, i\log\frac{-\xi^2}{(1-q^2)^2} +i \log x_1^- x_1^+  - i \log q^{-i g (u_1 -i/g)} (1-\xi^2)\, .\nonumber
\end{align}
which follows by carefully integrating \eqref{eq:basicpsiidstep2}, noting that now the constant log terms in the integral \emph{cannot} be dropped. In the limit $q\rightarrow1$, get the undeformed result
\begin{equation}
\Psi(1/x_1^-,0) \,-  \Psi(1/x_1^+,0) = i\log\frac{g^2}{4} +i \log x_1^- x_1^+
\end{equation}
since for $q\rightarrow1$, $\xi \rightarrow 0$ while $\tfrac{-\xi^2}{(1-q^2)^2}\rightarrow \tfrac{g^2}{4}$.

\paragraph{Crossing \\}

Putting the above together we can directly prove crossing between $\mathcal{R}_{0,0}$ and $\mathcal{R}_{2,0}$. This firstly makes use of the identity
\begin{equation}
\Phi(x_1,x_2) + \Phi(1/x_1,x_2) = \Phi(0,x_2)\,,
\end{equation}
valid for $|x_1| \neq 1$. Using this we find that
\begin{equation}
\Delta\tilde{\theta} \equiv \tilde{\theta}(z_1,z_2)+\tilde{\theta}(z_1+\omega_2,z_2)\, ,
\end{equation}
is given by
\begin{align}
\Delta\tilde{\theta} =& \,\Psi(1/x_1^-,x_2^+)-\Psi(1/x_1^+, x_2^+) +\Psi(1/x_1^+, x_2^-)
-\Psi(1/x_1^-, x_2^-)\nonumber  \\
&~~~~~~~~~~~~~~~~~+ \frac{1}{i}\log\frac{x_1^--x_2^+}{\frac{1}{x_1^-}-x_2^+}\frac{\frac{1}{x_1^-}-x_2^-}{x_1^--x_2^-}\,.
\end{align}
Then we use the identity
\begin{align}
\nonumber \Psi(1/x_1^-,x_2^+)-& \Psi(1/x_1^+, x_2^+) +\Psi(1/x_1^+, x_2^-)
-\Psi(1/x_1^-, x_2^-) =\\
& \frac{1}{i}\log\frac{1-\frac{1}{x_1^-x_2^+}}{1-\frac{1}{x_1^-x_2^-}}\frac{1-\frac{1}{x_1^+x_2^+}}{1-\frac{1}{x_1^+x_2^-}} + i \log \frac{1+\frac{\xi}{x_2^+}}{1+\frac{\xi}{x_2^-}}\, .
\end{align}
to finally find the crossing equation
\begin{align}
\nonumber \Delta\tilde{\theta} =& \, \frac{1}{i}\log\frac{1-\frac{1}{x_1^-x_2^+}}{1-\frac{1}{x_1^-x_2^-}}\frac{1-\frac{1}{x_1^+x_2^+}}{1-\frac{1}{x_1^+x_2^-}} +\frac{1}{i}\log\frac{x_1^--x_2^+}{\frac{1}{x_1^-}-x_2^+}\frac{\frac{1}{x_1^-}-x_2^-}{x_1^--x_2^-}+ i \log \frac{1+\frac{\xi}{x_2^+}}{1+\frac{\xi}{x_2^-}}\nonumber \\
= & \,
\frac{1}{i} \log \Big[\frac{x_2^- +\xi}{x_2^+ +\xi} h(x_1,x_2)\Big]\,.
\end{align}
In other words
\begin{equation}
\tilde{\sigma}(x_1,x_2) \tilde{\sigma}(1/x_1,x_2) = \frac{x_2^- +\xi}{x_2^+ +\xi} h(x_1,x_2)\, .
\end{equation}
If we now rewrite this in terms of $\sigma$ we get\footnote{Note that the dispersion implies $\frac{x_1^- +\xi}{x_1^+ +\xi}\frac{\frac{1}{x_1^-} +\xi}{\frac{1}{x_1^+} +\xi} = \frac{1}{q^2}$.}
\begin{equation}
\sigma(x_1,x_2)\sigma(1/x_1,x_2) = \frac{1}{q} \frac{x_2^-}{x_2^+} h(x_1,x_2)\, ,
\end{equation}
which is of course precisely the crossing equation $\sigma$ is supposed to solve.

\subsubsection*{The improved bound state dressing phase of the mirror theory}

We would like to construct the dressing phase for bound states of the mirror theory. In order to do so we take the approach taken in \cite{Arutyunov:2009kf} in the undeformed case, and take the constituents of the mirror bound state such that only the first particle lies in region $\mathcal{R}_1$, i.e. $|x_1^-|>1$ but $|x_1^+|<1$, while all other particles lie in region $\mathcal{R}_0$ with $|x^\pm|>1$ \cite{Bajnok:2008bm}. In order to sum up all contributions to the bound state dressing phase of the mirror theory, in addition to the above we need the dressing phase in $\mathcal{R}_{1,1}$.

\paragraph{The dressing phase in $\mathcal{R}_{1,1}$\\}

Continuing the $\chi$-functions to the region with $|x_1^+|<1$, $|x_1^-|>1$, $|x_2^+|<1$ $|x_2^-|>1$ means we have to add additional terms to the expressions in region $\mathcal{R}_{1,0}$, corresponding to crossing $|x_2^+|=1$. This means the two $\Phi$-functions give extra $\Psi$ contributions, but also the already present $\Psi$-function gives an additional contribution. The full $\chi$-functions are then given by
\begin{align}
\nonumber {\cal R}_{1,1}:\quad  \chi(x_1^+, x_2^+) =&\,
\Phi(x_1^+, x_2^+)+ \Psi(x_2^+, x_1^+) - \Psi(x_1^+,
x_2^+)\\\nonumber &~~~~~~~~~~~~+i \log\frac{\Gamma_{q^2} (1+\frac{ig}{2}(x_1^++\frac{1}{x_1^+}-x_2^+-\frac{1}{x_2^+}))}
{\Gamma_{q^2} (1-\frac{ig}{2}(x_1^++\frac{1}{x_1^+}-x_2^+-\frac{1}{x_2^+}))}\,,\\\nonumber
\chi(x_1^+, x_2^-) =&\, \Phi(x_1^+, x_2^-) - \Psi(x_1^+, x_2^-)
\,,\\
\chi(x_1^-, x_2^+) =&\, \Phi(x_1^-, x_2^+)+ \Psi(x_2^+, x_1^-)\,,\nonumber\\
\chi(x_1^-, x_2^-) =&\, \Phi(x_1^-, x_2^-)\,,~~~~~~~~
\end{align}

\paragraph{The bound state dressing phase I\\}

As already indicated, the constituents of a mirror bound state can be taken to lie in the regions $\mathcal{R}_{1,1}$, $\mathcal{R}_{1,0}$, $\mathcal{R}_{0,1}$, and $\mathcal{R}_{0,0}$. By antisymmetry of the dressing factor the above sections provide the appropriate dressing factors for the constituents, which can then be fused. At this point we would like to introduce the improved dressing phase $\Sigma$
\begin{align}
\Sigma(x_1,x_2) \equiv & \, \sigma(x_1,x_2) \frac{1-\frac{1}{x_1^+x_2^-}}{1-\frac{1}{x_1^-x_2^+}} \\
=& \, e^{i\left(\chi(x_1^+,x_2^+) - \chi(x_1^-,x_2^+) - \chi(x_1^+,x_2^-) + \chi(x_1^-,x_2^-)\right)}  \sqrt{P(x_1)/P(x_2)} \frac{1-\frac{1}{x_1^+x_2^-}}{1-\frac{1}{x_1^-x_2^+}} \, , \nonumber
\end{align}
where $P$ is defined in \eqref{eq:dressingvsdressing}. The chosen representative constituents for a $Q$-particle bound are parametrized as
\begin{align}
x_j^-(u) = &\, x_s(u+(Q-2j)\frac{i}{g}) \, , \, \, \, j= 1,\ldots,Q\, , \\
x_1^+(u) = &\, \frac{1}{x_s(u+ i Q/g)}\, , \, \, \, x_j^+(u) = x_s(u+(Q-2j+2)\frac{i}{g})\, .
\end{align}
We then denote the final mirror theory bound state parameters as
\begin{equation}
y_1^\pm = x_{1/Q}^\pm(u) \, , \, \, \, y_2^\pm = x_{1/Q^\prime}^\pm(v) \, .
\end{equation}

Summing up all contributions for the $Q$-particle bound state - $Q^\prime$-particle bound state improved mirror dressing phase using the appropriate form of the $\chi$-functions in each region, we directly find
\begin{align}
-i\log\Sigma^{QQ'}(y_1,y_2) =& \,
\Phi(y_1^+,y_2^+)-\Phi(y_1^+,y_2^-)-\Phi(y_1^-,y_2^+)+\Phi(y_1^-,y_2^-)
\nonumber\\ &\,-
\Psi(y_1^+,y_2^+)+\Psi(y_1^+,y_2^-)+\Psi(y_{2}^+,y_1^+)-\Psi(y_{2}^+,y_1^-)
\nonumber\\ &\,-i\log \frac{\Gamma_{q^2}\left(1-\frac{ig}{2}\left(y_1^++\frac{1}{y_1^+}-y_2^+-\frac{1}{y_2^+}\right)\right)}
{\Gamma_{q^2}\left(1+\frac{ig}{2}\left(y_1^++\frac{1}{y_1^+}-y_2^+-\frac{1}{y_2^+}\right)\right)} P(y_1)^{Q^\prime/2}P(y_2)^{-Q/2}
\nonumber\\
&\,-i\log\frac{1- \frac{1}{y_1^+y_2^-}}{1-\frac{1}{y_1^-y_2^+}}\prod_{j=1}^{Q-1} \frac{ 1- \frac{1}{x_j^-y_2^-}}{1-\frac{1}{x_j^-y_2^+}} \prod_{k=1}^{Q'-1}\frac{ 1- \frac{1}{y_1^+z_k^-}}{ 1-\frac{1}{y_1^-z_k^-}}\,,\label{eq:improveddressingphasedirect}
\end{align}
where we emphasize that
\begin{equation}
|y_1^+| <1\, , \, \, \, |y_1^-| >1\, , \, \, \, |y_2^+| <1\, , \, \, \, |y_2^-| >1\,.
\end{equation}
This result has an apparent dependence on the bound state constituents. To manifestly remove this dependence we need a few more identities.

\paragraph{Identities II\\}

Firstly we have
\begin{align}
\Psi(y_{1}^+, y_2^-) -\Psi(y_{1}^-, y_2^-) =& \, i\log\frac{y_2^- - y_1^+}{y_2^-} \frac{y_2^- -\frac{1}{y_1^-}}{y_2^- + \xi} \prod_{j=1}^{Q-1}\frac{y_2^--\frac{1}{x_j^-}}{y_2^-}\frac{y_2^--\frac{1}{x_j^-}}{y_2^- + \xi} \label{eq:psiboundprop}\\
= & \,i\log \left(\frac{y_2^-}{y_2^- + \xi}\right)^Q \left(1-\frac{y_1^+}{y_2^\pm} \right)\left(1-\frac{1}{y_1^- y_2^-}\right)\prod_{j=1}^{Q-1}\left(1-\frac{1}{x_j^-y_2^-}\right)^2 \, , \nonumber
\end{align}
which follows by applying the defining property of $\Gamma_q$ $Q$ times in the derivation (\ref{eq:basicpsiid},\ref{eq:basicpsiidstep2},\ref{eq:basicpsiidstep3}) and using the fact that $|y_1^+|<1$ while $|y_{1,2}^-|>1$. With both $y_1$ and $y_2$ referring to particles in region $\mathcal{R}_1$ this also immediately implies
\begin{equation*}
\Psi(y_2^+,y_1^-) -\Psi(y_2^-,y_1^-) =i\log \left(\frac{y_1^-}{y_1^- + \xi}\right)^{Q^\prime} \left(1-\frac{y_2^+}{y_1^-} \right)\left(1-\frac{1}{y_1^- y_2^-}\right)\prod_{j=1}^{Q^\prime-1}\left(1-\frac{1}{x_j^-y_1^-}\right)^2 \, .
\end{equation*}
Then we also have the bound-state analogue of \eqref{eq:basicpsiidx2zero} for a particle in region $\mathcal{R}_1$
\begin{align*}
\Psi(y_1^-,0) \,-  \Psi(y_1^+,0) =& \, i Q \log\frac{-\xi^2}{(1-q^2)^2} +i \log \frac{y_1^-}{y_1^+}
 \prod_{j=1}^{Q-1} (x_j^-)^2 - i Q \log q^{-i g(u-\tfrac{i}{g})} (1-\xi^2)\, .
\end{align*}
By applying the basic property \eqref{eq:psibasic} we can reexpress terms of the type $\Psi(y_1^+,y_2^+)$ and apply the above type of identities to find
\begin{align}
\Psi(y_1^+,y_2^+) - \Psi(y_1^-,y_2^+) = & \, \Psi(y_1^-,\tfrac{1}{y_2^+}) - \Psi(y_1^+,\tfrac{1}{y_2^+}) + \Psi(y_1^+,0) - \Psi(y_1^-,0)\nonumber \\
= & \, i Q \log (\xi y_2^+ +1) (1-q^2)^2(1-\xi^{-2}) q^{-i g(u-\tfrac{i}{g})} + \nonumber \\
 & \, - i\log (y_2^+ -y_1^-)\left(y_2^+-\frac{1}{y_1^+}\right)\prod_{j=1}^{Q-1}(x_j^--y_2^+)^2 \nonumber\\
= &\, i Q \log (\xi y_2^+ +1) (1-q^2)^2(1-\xi^{-2}) q^{-i g(u-\tfrac{i}{g})}+ \nonumber \\
 & \, - i\log (y_2^+ -y_1^-)\left(y_2^+-\frac{1}{y_1^+}\right)\prod_{j=1}^{Q-1}\left(1-\frac{1}{x_j^-y_2^+}\right)^{-2} \\
 &\, - i \log \prod_{j=1}^{Q-1}(\xi^{-1} - \xi)^2 \left(q^{-i g(u+(Q-2j)\tfrac{i}{g})}-q^{-ig(v+Q^\prime\tfrac{i}{g})}\right)^2\, .\nonumber
\end{align}
Similarly we find\footnote{Here the parameters $x_j^-$ of course refer to bound state number $Q^\prime$ and center $v$.}
\begin{align}
\Psi(y_2^+,y_1^+) - \Psi(y_2^-,y_1^+)
= &\, i Q^\prime \log (\xi y_1^+ +1) (1-q^2)^2(1-\xi^{-2}) q^{-i g(v-\tfrac{i}{g})} + \nonumber \\
 & \, - i\log (y_1^+ -y_2^-)\left(y_1^+-\frac{1}{y_2^+}\right)\prod_{j=1}^{Q^\prime-1}\left(1-\frac{1}{x_j^-y_1^+}\right)^{-2} \\
 &\, - i \log \prod_{j=1}^{Q^\prime -1}(\xi^{-1} - \xi)^2 \left(q^{-ig(v+(Q^\prime-2j)\tfrac{i}{g})}-q^{-i g(u+Q\tfrac{i}{g})}\right)^2\nonumber
\, .
\end{align}
Putting the above identities together with \eqref{eq:psiboundprop} we have
\begin{align}
-\Psi(y_1^+,y_2^+)& \, +\Psi(y_1^+,y_2^-)+\Psi(y_{2}^+,y_1^+)-\Psi(y_{2}^+,y_1^-)
-i\log \prod_{j=1}^{Q-1} \frac{1- \frac{1}{x_j^-y_2^-}}{1-\frac{1}{x_j^-y_2^+}} \prod_{k=1}^{Q'-1}\frac{ 1- \frac{1}{y_1^+z_k^-}}{ 1-\frac{1}{y_1^-z_k^-}} \nonumber \\
 = & \,\frac{1}{2}\left(-\Psi(y_1^+,y_2^+)\, +\Psi(y_1^+,y_2^-)+\Psi(y_{2}^+,y_1^+)-\Psi(y_{2}^+,y_1^-)\right) \nonumber \\
 & \, + \frac{1}{2}\left(-\Psi(y_1^-,y_2^+)\, +\Psi(y_1^-,y_2^-)+\Psi(y_{2}^-,y_1^+)-\Psi(y_{2}^-,y_1^-)\right)\nonumber \\
 & \, + i \log \frac{\prod_{j=1}^{Q-1} \left(q^{-i g(u+(Q-2j)\tfrac{i}{g})}-q^{-ig(v+Q^\prime\tfrac{i}{g})}\right)/(1-q^2)}{\prod_{j=1}^{Q^\prime-1} \left(q^{-ig(v+(Q^\prime-2j)\tfrac{i}{g})}-q^{-i g(u+Q\tfrac{i}{g})}\right)/(1-q^2)} -\frac{i}{2} \log \frac{y_1^+}{y_1^-} \frac{y_2^-}{y_2^+}\nonumber \\
 & \, + \frac{i Q^\prime}{2} \log q^{-ig(v-\tfrac{i}{g})} (\xi y_1^+ + 1) \frac{y_1^- + \xi}{y_1^-} - \frac{i Q}{2} \log q^{-ig(u-\tfrac{i}{g})} (\xi y_2^+ + 1) \frac{y_2^- + \xi}{y_2^-} \nonumber \\
 & \, + i \frac{Q-Q^\prime}{2} \log(\xi^2-1)\, .
\end{align}
Then we can use the defining property of the $\Gamma_q$ function to find
\begin{align}
& \frac{\prod_{j=1}^{Q^\prime-1} \left(q^{-ig(v+(Q^\prime-2j)\tfrac{i}{g})}-q^{-i g(u+Q\tfrac{i}{g})}\right)/(1-q^2)}{\prod_{j=1}^{Q-1}   \left(q^{-i g(u+(Q-2j)\tfrac{i}{g})}-q^{-ig(v+Q^\prime\tfrac{i}{g})}\right)/(1-q^2)}\frac{\Gamma_{q^2}\left(1- \frac{ig}{2}(u(y_1^+) - u(y_2^+))\right)}{\Gamma_{q^2}\left(1+ \frac{ig}{2}(u(y_1^+) - u(y_2^+))\right)}\nonumber\\
& \hspace{50pt} = \frac{\prod_{j=1}^{Q^\prime-1} q^{-ig(v+(Q^\prime-2j)\tfrac{i}{g})}}{\prod_{j=1}^{Q-1}   q^{-i g(u+(Q-2j)\tfrac{i}{g})}} \frac{\Gamma_{q^2}\left(Q^\prime - \frac{ig}{2}(u(y_1^+) - u(y_2^+))\right)}{\Gamma_{q^2}\left(Q+ \frac{ig}{2}(u(y_1^+) - u(y_2^+))\right)} \nonumber\\
& \hspace{50pt} = q^{-ig (Q^\prime-1) v} q^{ig (Q-1) u} \frac{\Gamma_{q^2}\left(Q^\prime - \frac{ig}{2}(u(y_1^+) - u(y_2^+))\right)}{\Gamma_{q^2}\left(Q+ \frac{ig}{2}(u(y_1^+) - u(y_2^+))\right)} \, .
\end{align}

\paragraph{The bound state dressing phase II\\}

By using the above identities, the improved dressing phase given in eqn. \eqref{eq:improveddressingphasedirect} can be written as
\begin{align}
-i\log\Sigma^{QQ'}(y_1,y_2) =& \, \Phi(y_1^+,y_2^+)-\Phi(y_1^+,y_2^-)-\Phi(y_1^-,y_2^+)+\Phi(y_1^-,y_2^-) \label{eq:improveddressingphase} \\ &\, -\frac{1}{2}\left(\Psi(y_1^+,y_2^+)+\Psi(y_1^-,y_2^+)-\Psi(y_1^+,y_2^-)-\Psi(y_1^-,y_2^-)\right) \nonumber \\
&\, +\frac{1}{2}\left(\Psi(y_{2}^+,y_1^+)+\Psi(y_{2}^-,y_1^+)-\Psi(y_{2}^+,y_1^-)
-\Psi(y_{2}^-,y_1^-) \right) \nonumber \\
&\, - i \log \frac{i^Q \, \Gamma_{q^2}\left(Q^\prime - \frac{ig}{2}(u(y_1^+) - u(y_2^+))\right)}{i^{Q^\prime} \Gamma_{q^2}\left(Q+ \frac{ig}{2}(u(y_1^+) - u(y_2^+))\right)}\frac{1- \frac{1}{y_1^+y_2^-}}{1-\frac{1}{y_1^-y_2^+}} \sqrt{\frac{y_1^+}{y_1^-} \frac{y_2^-}{y_2^+}} \nonumber \\
& \, + \frac{i}{2} \log q^{Q-Q^\prime } q^{-i g (Q+Q^\prime -2)(u-v)}\nonumber \, .
\end{align}
This version of the improved mirror bound state dressing phase manifests its proper fusion, as any apparent dependence on the bound state constituents has been removed.

To see that the improved dressing phase is unitary it is important to realize that the $\log \Gamma_{q^2}$ combinations entering in $\Phi$, $\Psi$ and explicitly in the formula above, are \emph{not} real. Rather their imaginary parts cancel precisely against the manifestly imaginary terms in the above prescription. One way to show this is to realize that the above expression is independent of the constituent particles, and so should agree with the improved bound state dressing phase obtained by fusing over a manifestly mirror theory conjugation symmetric bound state configuration. The latter expression can be easily proven to be unitary by using the crossing equation and the generalized unitarity of the dressing phase about the string line. We have also verified unitarity of the above expression explicitly numerically. The further relevant analytic properties of the dressing phase are covered in the section below. Note that this expression for the dressing phase manifestly reduces to the improved mirror dressing phase of the undeformed theory.

\subsubsection*{Action of the discrete Laplace operator}

The mirror-mirror bound state dressing phase is a holomorphic function in the intersection of the region $R_{1,1}$ and the mirror region, which in particular contains the mirror line. This follows by the discontinuity relations for the $\Psi$-functions in their first argument, which we discuss below. Holomorphicity and fusion mean that the dressing phase is immediately annihilated by the discrete Laplace operator $\Delta_{Q^\prime,P} = \delta_{Q^\prime,P}s^{-1} -( \delta_{Q^\prime +1,P}+\delta_{Q^\prime-1,P})$ for $P\geq 1$, since we are far from any branch cut ambiguities of the arguments of the dressing phase. For $P=1$ this is no longer the case; acting with the operator
\begin{equation}
\label{eq:Kplus1invQ1}
\Delta_{Q^\prime,1}(v) = \delta_{Q^\prime,1}s^{-1}(v) - \delta_{Q^\prime,2}\, ,
\end{equation}
only annihilates the mirror-mirror bound state dressing phase for $v \in (-u_b,u_b)$.\footnote{The significance of the interval $(-u_b,u_b)$ lies in the fact that $x_m(u)$ covers the unit circle precisely as $u$ runs over this interval.} This is the reason the $Y$-system for $Y_1$ only exists on this interval of the real mirror line. Because we do not encounter any branch cuts of the arguments of the dressing phase for $v \in (-u_b,u_b)$, annihilation within this interval still follows immediately by holomorphicity.

To show that the dressing phase is holomorphic in the intersection of $\mathcal{R}_{1,1}$ and the mirror region, in addition to the formulae already derived we will need the discontinuity of the $\Psi$-function $\Psi(x_1,x_2)$ in its first argument. The relevant discontinuity is at $u_1 = u + 2 i /g$ with $u\in (-u_b,u_b)$. To find it we proceed as in the undeformed case \cite{Arutyunov:2009kf} and integrate by parts to write the $\Psi$-function as
\begin{equation*}
\Psi(x_1,x_2) = -\frac{g}{2} \oint_{|z|=1} \frac{dz}{2 \pi i} \log(z-x_2)\frac{d u}{dz}\left( \psi_{q^2}(1+\tfrac{i g }{2}(u_1-u(z)))+\psi_{q^2}(1-\tfrac{i g }{2}(u_1-u(z)))\right)\, ,
\end{equation*}
where $\psi_q$ is the $q$-digamma function, the logarithmic derivative of the $\Gamma_q$ function. From the defining relation of the $\Gamma_q$ function it is easy to see that the $q$-digamma function still has simple poles at the negative integers with residue negative one. This makes the discontinuities immediately clear as two poles\footnote{Recall that $u(z) = u(1/z)$.} hit the integration contour when $u_1 = u + 2 i /g$. The corresponding discontinuity is then simply
\begin{equation}
\label{eq:Psidiscfirstarg}
\lim_{\epsilon \rightarrow 0} \Psi(e^\epsilon x_1,x_2) - \Psi(e^{-\epsilon} x_1,x_2) = - i \log \frac{x_m(u) - x_2}{\frac{1}{x_m(u)} -x_2} \, ,
\end{equation}
following from the two pole contributions just inside and just outside the unit circle\footnote{Effectively giving $x_m$ and $1/x_m$ respectively.} respectively. Analogous formulae apply for discontinuities at $u_1 = u + 2 n i/g$, exactly as in the undeformed case \cite{Arutyunov:2009kf}. For example for the discontinuity through $u_1 = u - 2 i/g$ we get
\begin{equation}
\lim_{\epsilon \rightarrow 0} \Psi(e^\epsilon x_1,x_2) - \Psi(e^{-\epsilon} x_1,x_2) =  i \log \frac{x_m(u) - x_2}{\frac{1}{x_m(u)} -x_2} \, ,
\end{equation}
where the different sign arises because the poles cross the integration contour in the opposite direction. This shows that the cuts of combinations like $\Psi(y_2^+,y_1^+)+\Psi(y_2^-,y_1^+)$ precisely cancel in the intersection of $R_{1,1}$ and the mirror region, making the dressing phase a holomorphic function there as mentioned above.

With this identity we can also directly compute the action of $\Delta_{Q^\prime,1}(v)$ on $\Sigma^{QQ^\prime}(u,v)$ almost identically to the undeformed case \cite{Arutyunov:2009ux}, and show explicitly that it indeed vanishes when $v \in (-u_b,u_b)$. Acting on the $\Phi$-functions, noting that the $\Psi$-function is precisely its discontinuity when its second argument crosses the unit circle (cf. eqn. \eqref{eq:phidef}), we get
\begin{equation}
\left(\Phi(y_1^+,y_2^+) -\Phi(y_1^+,y_2^-)-\Phi(y_1^-,y_2^+)+\Phi(y_1^-,y_2^-)\right)\Delta_{Q^\prime,1}(v) = \Psi(x_m(v),y_1^-)-\Psi(x_m(v),y_1^+)\, .\nonumber
\end{equation}
For the first line of the $\Psi$-functions in \eqref{eq:improveddressingphase} we get
\begin{align}
-\frac{1}{2}\left(\Psi(y_1^+,y_2^+) +\Psi(y_1^-,y_2^+)-\Psi(y_1^+,y_2^-)-\Psi(y_1^-,y_2^-)\right)\Delta_{Q^\prime,1}(v)& =  \\
= -\frac{i}{2}\log q^{-ig(Q-2)(u-v)} S_Q + i \log i^Q & \frac{\Gamma_q^2\left(\tfrac{Q}{2}-\tfrac{i g}{2}(u-v)\right)}{\Gamma_q^2\left(\tfrac{Q}{2}+\tfrac{i g}{2}(u-v)\right)}\nonumber
\end{align}
which simply follows from the integral representation for $\Psi$ of eqn. \eqref{eq:psidef} when its second argument crosses the unit circle. $\Delta$ effectively acts on the first argument of the second line of $\Psi$-functions, giving
\begin{align}
\frac{1}{2}\left(\Psi(y_2^+,y_1^+) +\Psi(y_2^-,y_1^+)-\Psi(y_2^+,y_1^-)-\Psi(y_2^-,y_1^-)\right)\Delta_{Q^\prime,1}(v) & = \nonumber\\
= \Psi(x_m(v),y_1^+)-\Psi(x_m(v),y_1^-) - \frac{i}{2} \frac{y_1^+ - x_m(v)}{y_1^+ - \frac{1}{x_m(v)}}&\frac{y_1^- - \frac{1}{x_m(v)}}{y_1^- - x_m(v)}\, .
\end{align}
where we used the discontinuity of the $\Psi$-function we just computed, given by eqn. \eqref{eq:Psidiscfirstarg}. This can be partly rewritten as
\begin{equation}
-\frac{i}{2} \log \frac{y_1^+ - x_m(v)}{y_1^+ - \frac{1}{x_m(v)}}\frac{y_1^- - \frac{1}{x_m(v)}}{y_1^- - x_m(v)} = i
\log \frac{y_1^+ - \frac{1}{x_m(v)}}{y_1^- - \frac{1}{x_m(v)}}\sqrt{\frac{y_1^-}{y_1^+}} +\frac{i}{2} \log q^{-Q}S_Q(u,v)\, ,
\end{equation}
which follows from the general identity
\begin{equation}
\frac{1 - \tfrac{1}{x^-_i x^-_j}}{1- \frac{1}{x^+_i x^+_j}}\frac{x^-_i - x^-_j}{x^+_i - x^+_j} = q^{-(Q+M)} S_{Q-M} \, ,
\end{equation}
where particles $i$ and $j$ have bound state numbers $Q$ and $M$ respectively. Finally the last terms give
\begin{align}
-i \log \frac{i^Q \, \Gamma_{q^2}\left(Q^\prime - \frac{ig}{2}(u(y_1^+) - u(y_2^+))\right)}{i^{Q^\prime} \Gamma_{q^2}\left(Q+ \frac{ig}{2}(u(y_1^+) - u(y_2^+))\right)}\frac{1- \frac{1}{y_1^+y_2^-}}{1-\frac{1}{y_1^-y_2^+}} \sqrt{\frac{y_1^+}{y_1^-} \frac{y_2^-}{y_2^+}} \Delta_{Q^\prime,1}(v) &=\nonumber\\
= -i \log i^Q \frac{\Gamma_q^2\left(\tfrac{Q}{2}-\tfrac{i g}{2}(u-v)\right)}{\Gamma_q^2\left(\tfrac{Q}{2}+\tfrac{i g}{2}(u-v)\right)} - i \log \frac{y_1^+ - \frac{1}{x_m(v)}}{y_1^- - \frac{1}{x_m(v)}} & \sqrt{\frac{y_1^-}{y_1^+}}
\end{align}
and
\begin{equation}
\frac{i}{2} \log q^{Q-Q^\prime } q^{-i g (Q+Q^\prime -2)(u-v)}\Delta_{Q^\prime,1}(v) = \frac{i}{2} \log q^Q  q^{-i g (Q-2)(u-v)}
\end{equation}
respectively. Adding everything up we get zero as promised. In summary, this means
\begin{equation}
\Sigma^{Q Q^\prime} \Delta_{Q^\prime,1}(u,v) = 0 \, , \, \, \, \, \, \, \, \mbox{for} \, \, v \in (-u_b,u_b)\,.
\end{equation}
When $v$ is outside this interval the result of applying $\Delta$ is nonzero. In the undeformed case the resulting kernel can be cast in a simple form \cite{Arutyunov:2009ux}, we will not pursue a simple version of the resulting deformed kernel here.

\section{Chapter \ref{chapter:twistedspectrum}}

\subsection{The ground state solution and hybrid equations}
\label{app:hybridTBA}

In this appendix we show how the ground state solution is fixed by the canonical TBA equations and how the chemical potentials naturally disappear from the hybrid form of the TBA equations for $Y_+ Y_-$ and $Y_Q$. We will discuss the canonical equations for $w$-strings; for $vw$-strings the story is identical. When coming here from chapter \ref{chapter:finitevolumeIQFT} we simply need to realize that $Y_0$ ($1/Y_f$) is small at large $u$ (the energy grows with $u$), and in this limit the canonical TBA equations for the $Y_Q^{ch. 2}$ become of the same form as those of $Y_{M|(v)w}$.

\subsubsection*{Canonical equations}

For constant solutions, the ground state canonical TBA equations for $w$-strings in exponential form are given by
\begin{equation}
Y^\circ_{M|w} = e^{-\mu_{M|w}} \left(\prod_{N=1}^{M-1}(1+ \frac{1}{Y^\circ_{N|w}})^{2N} \right) (1+ \frac{1}{Y^\circ_{M|w}})^{2M-1} \left(\prod_{N=M+1}^{\infty}(1+ \frac{1}{Y^\circ_{N|w}})^{2M}\right) \, .
\end{equation}
Next we write $q=e^z$ and write the asymptotic solution (\ref{eq:YwGSasympt}) as
\begin{equation}
Y^\circ_{M|w} = \frac{\sinh{Mz}\sinh{(M+2)z}}{\sinh^2{z}} \, .
\end{equation}
Now we can readily evaluate the following partial product
\begin{align}
\prod_{N=1}^{M-1}(1+ \frac{1}{Y^\circ_{N|w}})^{2N} & = \frac{\sinh^{2M}{Mz}}{\sinh^2{z}\sinh^{2M-2}{(M+1)z}} \, ,\\
\prod_{N=1}^{k}(1+ \frac{1}{Y^\circ_{N|w}}) & = \frac{\sinh{2 z}}{\sinh{z}} \frac{\sinh{(k+1)z}}{\sinh{(k+2)z}} \, .
\end{align}
Taking the limit $k\rightarrow \infty$ on the second product for $\mbox{Re}(z)<0$ gives
\begin{equation}
\prod_{N=1}^{\infty}(1+ \frac{1}{Y^\circ_{N|w}})^{2M} = 4^M \cosh^{2M}{z} \, e^{2Mz}\, ,
\end{equation}
and now putting everything together we get
\begin{align}
Y^\circ_{M|w} \, = & \, \, e^{-\mu_{M|w}}  \frac{\sinh^{2M}{Mz}}{\sinh^2{z}\sinh^{2M-2}{(M+1)z}} \left(\frac{\sinh^{2}{(M+1)z}}{\sinh{Mz}\sinh{(M+2)z}}\right)^{2M-1} \nonumber \\
 & \,\, \quad \quad \frac{\sinh^{2M}{z}}{\sinh^{2M}{2 z}} \frac{\sinh^{2M}{(M+2)z}}{\sinh^{2M}{(M+1)z}} 4^M \cosh^{2M}{z} e^{2Mz} \nonumber \\
 = & \,\, 4^M \cosh^{2M}{z} \, \sinh^{2M-2}{z} \, \sinh^{-2M}{2z} \, \sinh{Mz}\, \sinh{(M+2)z} \, \, e^{2Mz-\mu_{M|w}} \nonumber \\
 = & \,\,Y^\circ_{M|w} \, e^{2Mz-\mu_{M|w}} \, ,
\end{align}
showing that since $\mu_{M|w} = 2 i M (\alpha+i \epsilon)$ we should take $z = i (\alpha+i \epsilon)$. For $vw$-strings an identical computation gives $z = i (\beta + i \epsilon)$, and so we obtain the asymptotic solution (\ref{eq:YwGSasympt}-\ref{eq:YQasympt}).

\subsubsection*{Hybrid equations}

In order to derive the hybrid equations, we follow \cite{Frolov:2011wg} and extend the considerations of \cite{Arutyunov:2009ax} to the case of non-zero chemical potentials, and indicate the extra terms appearing in the derivation. To derive the hybrid equations we need to compute infinite sums of the form
\begin{equation}
\log \big(1 + \frac{1}{Y_{M|(v)w}}\big) \star \mathcal{K}_M \, ,
\end{equation}
where for our applications $\mathcal{K}_M$ is some kernel which satisfies the schematic identity
\begin{equation}
\label{eq:kernelid}
(K+1)^{-1}_{MN} \star \mathcal{K}_N \equiv \mathcal{K}_M - I_{MN} s \star \mathcal{K}_N = \delta\mathcal{K_M} \, .
\end{equation}
We start by rewriting the simplified equations for $w$-strings
\begin{equation}
\log Y_{M|w} = \, I_{MN}\log(1+Y_{N|w})\star s+
\delta_{M1}\log\frac{1-\frac{1}{Y_{-}}}{1-
\frac{1}{Y_{+}}}\hat{\star}s \,,
\end{equation}
in the form
\begin{align}
\log Y_{M|w} - I_{MN}\log(Y_{N|w}) \star s  =& \, I_{MN}\log\big(1+\frac{1}{Y_{N|w}}\big)\star s+
\delta_{M1}\log\frac{1-\frac{1}{Y_{-}}}{1-\frac{1}{Y_{+}}}\hat{\star}s \,, \\
=& \log Y_{M|w}^{r} - I_{MN}\log(Y^{r}_{N|w}) \star s \, ,
\end{align}
where we have introduced regularized Y-functions $Y_{M|w}^{r} = Y_{M|w} e^{\mu_{M|w}}$ which asymptote to one at large $M$ \textit{cf.} (\ref{eq:YwGSasympt}) given our regulator, without changing the equality. The reason for introducing these regularized Y-functions is that we would like to convolute this equation with $\mathcal{K}_M$ and sum over $M$. For the deformed Y-functions the left hand side of the resulting equation would be the sum of two divergent terms whose sum is regular; the regularized Y-functions simply make this manifest. Using the identity (\ref{eq:kernelid}) we obtain
\begin{equation}
\log Y_{Q|w}^{r} \star \delta \mathcal{K}_Q = \log\big(1+ \frac{1}{Y_{M|w}}\big) \star \mathcal{K}_M - \log \big(1+ \frac{1}{Y_{Q|w}}\big) \star \delta \mathcal{K}_Q +
\log\frac{1-\frac{1}{Y_{-}}}{1-\frac{1}{Y_{+}}}\hstar s \star \mathcal{K}_1 \, .
\end{equation}
Combining the terms appropriately we get
\begin{equation}
\log\big(1+ \frac{1}{Y_{M|w}}\big) \star \mathcal{K}_M  = \log(1+ Y_{Q|w}) \star \delta \mathcal{K}_Q -
\log\frac{1-\frac{1}{Y_{-}}}{1-\frac{1}{Y_{+}}}\hstar s \star \mathcal{K}_1  +  \mu_{Q|w} \star \delta \mathcal{K}_Q\, .
\end{equation}
So we see that the only modification to the identity for the infinite sum over $Y_{M|w}$ functions is the addition of the term $\mu_{Q|w} \star \delta \mathcal{K}_Q$ to the $\log(1+ Y_{Q|w}) \star \delta \mathcal{K}_Q$ contribution. Similarly for $vw$-strings we have to add the term $\mu_{Q|vw} \star \delta \mathcal{K}_Q$ to the $\log(1+ Y_{1|vw}) \star \delta \mathcal{K}_Q$ term.

All that now remains is to identify the kernels $\delta \mathcal{K}$ relevant in the derivation of the hybrid equation for $Y_+ Y_-$ and the one for $Y_Q$, and add the corresponding terms along with the chemical potentials originally in the equation to the undeformed hybrid equations. In the case of $Y_+ Y_-$, $\delta \mathcal{K}_Q = \delta_{Q,1} s$ \textit{cf.} the simplified equation (\ref{eq:sTBAyp}). This means we have to add $2(\mu_{1|vw}-\mu_{1|w})\star s  = \mu_{1|vw}-\mu_{1|w}$ to the chemical potential $-2\mu_y$ of the (doubled) $y$-particles. Now noting that $2\mu_y = \mu_{1|vw}-\mu_{1|w}$ as follows from table \ref{tab:chempot}, we see that the total contribution of the chemical potentials to the hybrid equation for $Y_+ Y_-$ is zero. Next, for $Y_Q$ we have $\delta \mathcal{K}_N = \delta_{N,Q-1} s + \delta_{N,1} s \hstar K_{yQ}$, meaning we add $\mu_{Q-1|vw} \star s + \mu_{1|vw}\star s \hstar K_{yQ} = \tfrac{1}{2} (\mu_{Q-1|vw} + \mu_{1|vw}) = \tfrac{1}{2} \mu_{Q|vw} $ for both the left and right sectors to the chemical potential $-\mu_Q$. This gives a total of $-\mu_Q + \tfrac{1}{2} (\mu^+_{Q|vw} + \mu^-_{Q|vw})$ which is indeed zero for the chemical potentials in table \ref{tab:chempot}.

\section{Chapter \ref{chapter:excstatesandwrapping}}

\label{app:excstatesandwrapping}

When computing wrapping corrections we need the weak coupling expansion of the asymptotic $Y_Q$ functions, which in turn involves the weak coupling expression of the scalar factor of our S-matrix in the string-mirror region, $S_{\sl(2)}^{1_*Q}$. This factor can be found in \cite{Arutyunov:2009kf} (see also \cite{Bajnok:2009vm}) and it has the following weak coupling expansion
\begin{equation*}
S_{\sl(2)}^{1_*Q}(u,v)=S_0(u,v)+g^2 S_{2}(u,v)+\ldots \, ,
\end{equation*}
where {\small
\begin{align}
S_0(u,v) = -\frac{\big[(v-u)^2+(Q+1)^2\big]\big[Q-1 + i (v-u)
)\big]}{(u-i)^2 \big[Q-1-i( v-u)\big]}.
\end{align}
} and {\small
\begin{align}
S_2(u,v)& =-S_0(v,u)
\frac{2\big[2Q(u-i)+(u+i)(v^2+Q^2+2v(u-i))\big]}{(v^2+Q^2)
(1+u^2)}+\\
\nonumber &
\frac{S_0(v,u)}{1+u^2}\Big[4\gamma+\psi\left(1+\frac{Q+iv}{2}\right)+\psi\left(1-\frac{Q+iv}{2}\right)+\psi\left(1+\frac{Q-iv}{2}\right)+
\psi\left(1-\frac{Q-iv}{2}\right)\Big]\, .
\end{align}}

\subsection{Critical behaviour of the asymptotic solution}
\label{app:criticality}

There is a considerable number of additional roots that plays a role in the TBA for orbifolded Konishi as the coupling is increased, and their behaviour changes as the twist is varied. We will give an overview of this behaviour, but refrain from e.g. presenting the sets of driving terms these roots would result in, as their derivation is immediate once the roots are known. Let us start with the roots for $Y_{M|vw}$-functions.

\subsection*{$Y_{M|vw}$-functions}

\begin{figure}
\begin{center}
 \includegraphics[width=3.5in]{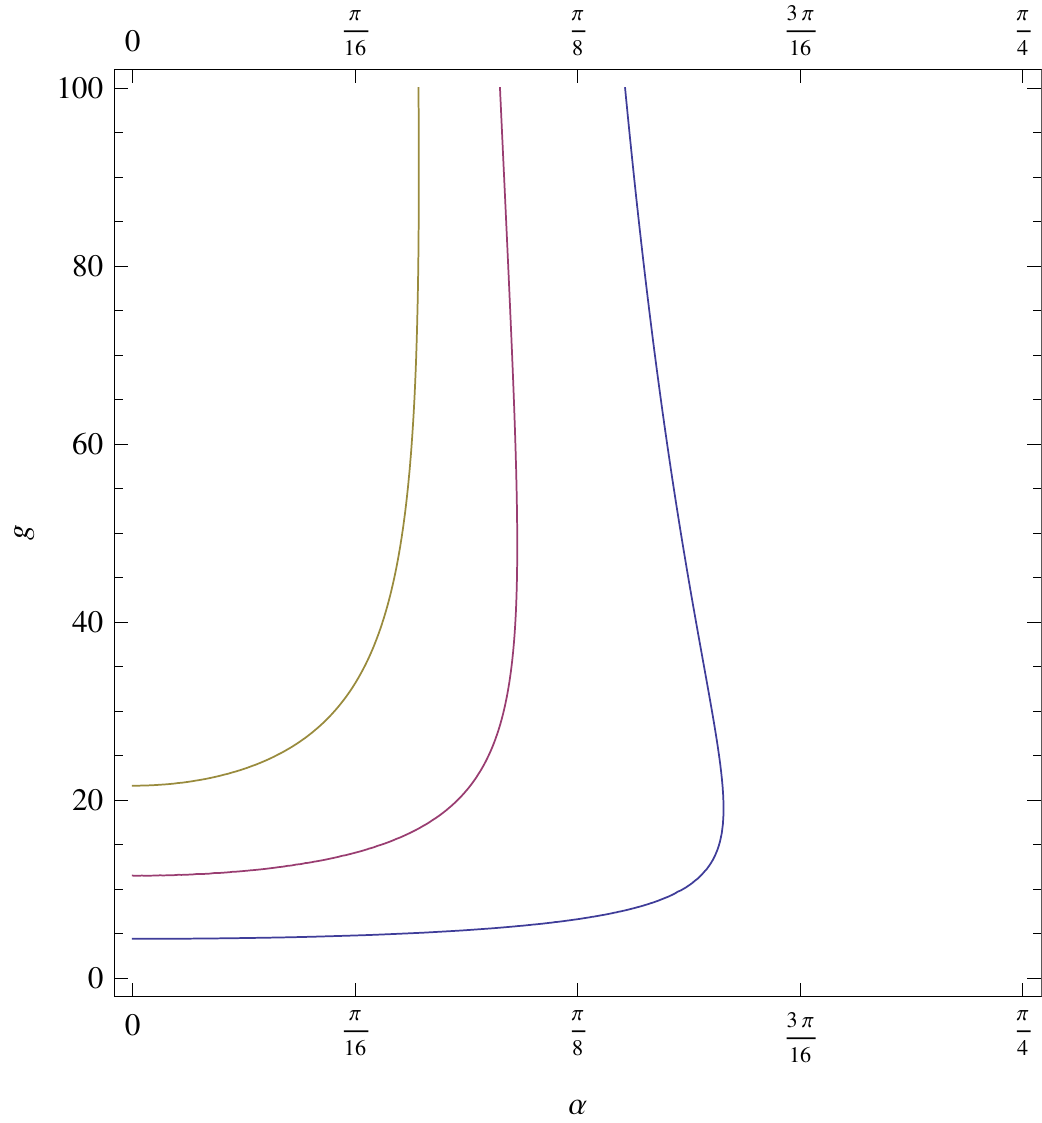}
\end{center}
\caption{The asymptotic critical values corresponding to $Y_{1|vw}$ (blue), $Y_{2|vw}$ (red), and $Y_{3|vw}$ (yellow).}
\label{fig:YMvwcrit}
\end{figure}

The critical behaviour of roots of $1+Y_{M|vw}$ is largely analogous to that observed for the asymptotic solution of the undeformed model \cite{Arutyunov:2009ax}, only there are twice the number of roots. In the undeformed case, $1+Y_{M|vw}$ has two roots that cross the integration contour at a certain value of the coupling, resulting in driving terms above the so called critical value, $g_{crit}$. These roots are imaginary and move towards the lines $\pm i/g$, where they each split in two roots lying symmetrically on the lines $i/g$ and $-i/g$. For the asymptotic solution, the first few critical values were found to be $4.429$, $11.512$ and $21.632$ corresponding to $Y_{1|vw}$, $Y_{2|vw}$ and $Y_{3|vw}$ respectively. For our orbifolded model, we find that $1+Y_{M|vw}$ in general has four extra roots, in addition to the noncritical ones mentioned in the main text. Two of these correspond directly to the roots present in the undeformed case, and we can plot their critical values as a function of twist. The result, immediately requiring some further discussion, is presented in figure \ref{fig:YMvwcrit}.
Note that these plots directly reproduce the critical values observed for the undeformed model. The main interesting feature is that the plot of the critical value as presented, is not really a function of the twist; it is multi-valued. However, looking at a specific twist, this simply corresponds to having two critical values. As the reader might have guessed by now, this second critical value simply corresponds to the critical value for the second pair of roots present in our orbifolded model. Beyond the point where the two critical values merge, the roots no longer play a role in the TBA; they do not cross the integration contour.

The interesting question that remains is whether the upper part of these curves actually touches the $g$-axis at a finite (but clearly large) value of the coupling, meaning that the second pair of roots should also play a role in the undeformed model, though only at very large coupling. This is in fact not the case, as we can show nicely that these curves should close in on the $g$-axis at infinite coupling; let us do so explicitly for the  curve corresponding to $Y_{1|vw}$.

The critical value curve for $Y_{1|vw}$ corresponds to the curve $T_{21}(\pm i/g)=0$, which follows by construction of the Y-functions, \textit{cf.} (\ref{eq:YmvwinT}). Considering that in the  large $g$ limit, the solution of the BY equation is given by \cite{Arutyunov:2009ax}
\begin{equation}
p = \sqrt{\tfrac{2\pi}{g}} - \tfrac{1}{g},
\end{equation}
the expansion of $T_{21}(\pm i/g)$ around $g = \infty$ gives
\begin{equation}
T_{21}(\pm i/g) = 4(1-\cos{\alpha}) + \sqrt{\mathcal{O}(\tfrac{1}{g})},
\end{equation}
with the first correction proportional to $c + \cos(\alpha)$\footnote{$T_{Q1}(\pm i/g)$ gives $2Q(1-\cos{\alpha})$, with corrections proportional to something of the form $\tilde{c} + \cos(\alpha)$.}, where $c$ is some constant. This shows that at zero twist, the second zero of $T_{21}(\pm i/g)$ lies at infinite coupling.

\subsection*{$Y_{M|w}$-functions}

\begin{figure}
\begin{center}
 \includegraphics[width=3.5in]{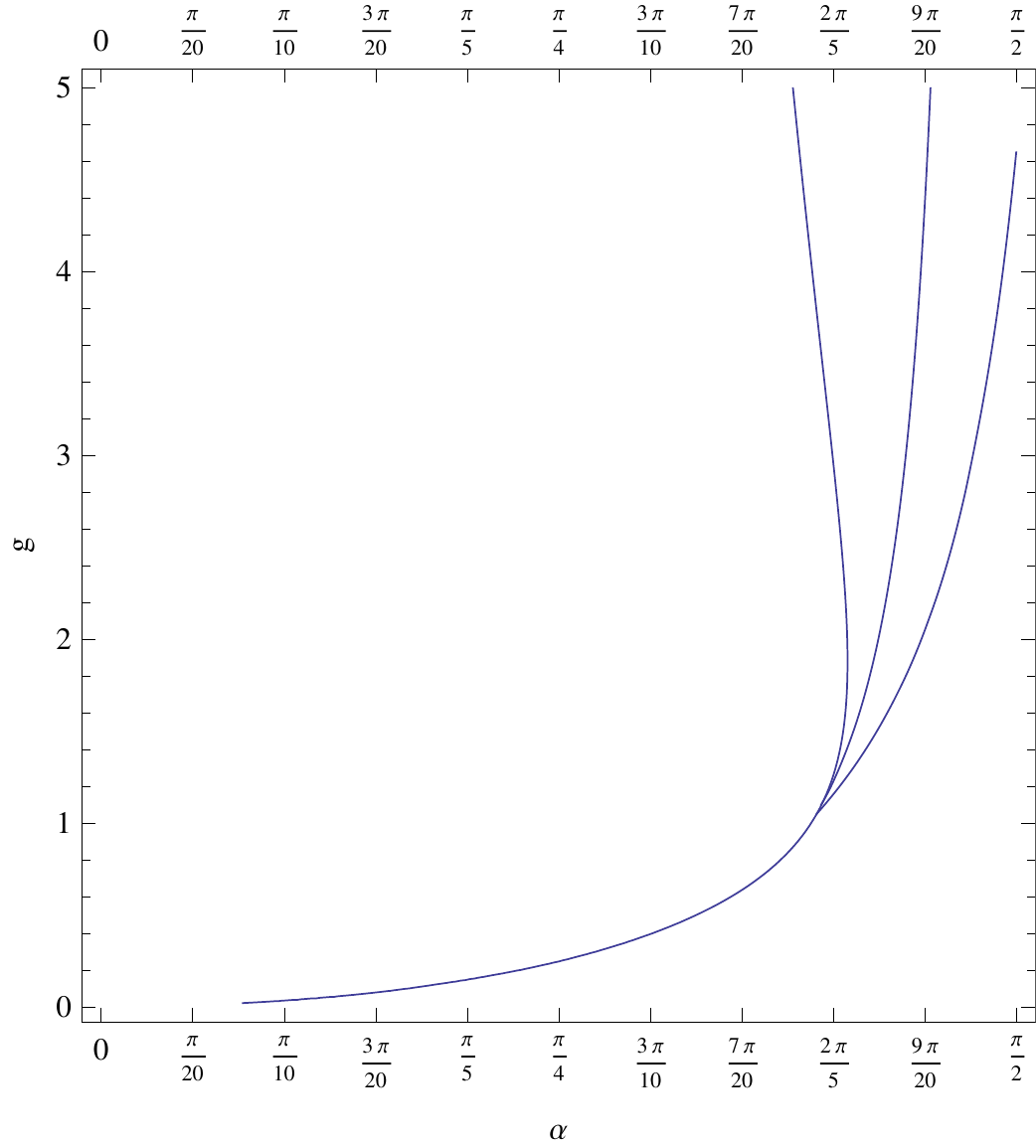}
\end{center}
\caption{The asymptotic critical values corresponding for $Y_-$. Note the possible existence of four critical values.}
\label{fig:YMwcrit1}
\end{figure}

While the analytic structure of the $Y_{M|vw}$-functions was still similar to the undeformed model, the case of $Y_{M|w}$-functions, where we include $Y_-$ as $-Y_{0|w}$, is qualitatively different and more involved. In general, these functions have four roots, and these four roots can show three different types of critical behaviour. The type of critical behaviour depends on the index $M$ of the Y-function, and the specific value of the twist, and can be quite intricate. Let us illustrate the general discussion with a plot of the critical values for $Y_-$ in figure \ref{fig:YMwcrit1}.
The way to interpret the above plot is to consider a specific twist, and look at the number of critical values. This can be one, two, three, or four, depending on the twist chosen. The rightmost curve corresponds to $T_{11}(-i/g)=0$, analogous to the above discussion for the $Y_{M|vw}$-functions. The following three types of critical behaviour can then occur:

\bigskip

\noindent \textbf{Type I:}

\smallskip

\noindent Provided we cross the critical value corresponding to the curve $T_{11}(-i/g)=0$ first (\textit{i.e.} $\alpha$ being smaller than the intersection point of the curves), there are simply two roots that enter the equations, and another two that enter after crossing the second critical value, just as for $Y_{M|vw}$.

\bigskip

\noindent \textbf{Type II:}

\smallskip

\noindent Beyond this point however, four roots will generically enter the equations after crossing the lowest critical value. A second set of roots will approach from outside the integration contour, and merge with the first set of roots exactly at the second critical value, pinching the integration contour. Immediately after, these merged roots will split in two again, where now half the contribution from all \emph{eight} roots needs to be taken into account, in accordance with the pinching.

Provided the twist is such that we can still cross the curve $T_{11}(-i/g)=0$, the first crossing corresponds to the second set of roots moving away again after merging at $\pm i/g$, leaving us with half the contribution from four roots, only to be completed again to half the contribution from all eight roots after crossing the curve a second time.

\bigskip

\noindent \textbf{Type III:}

\smallskip

\noindent As type II, but beyond the point where we can cross the curve $T_{11}(-i/g)=0$.

\bigskip

\begin{figure}[h]
\begin{center}
\includegraphics[width=4in]{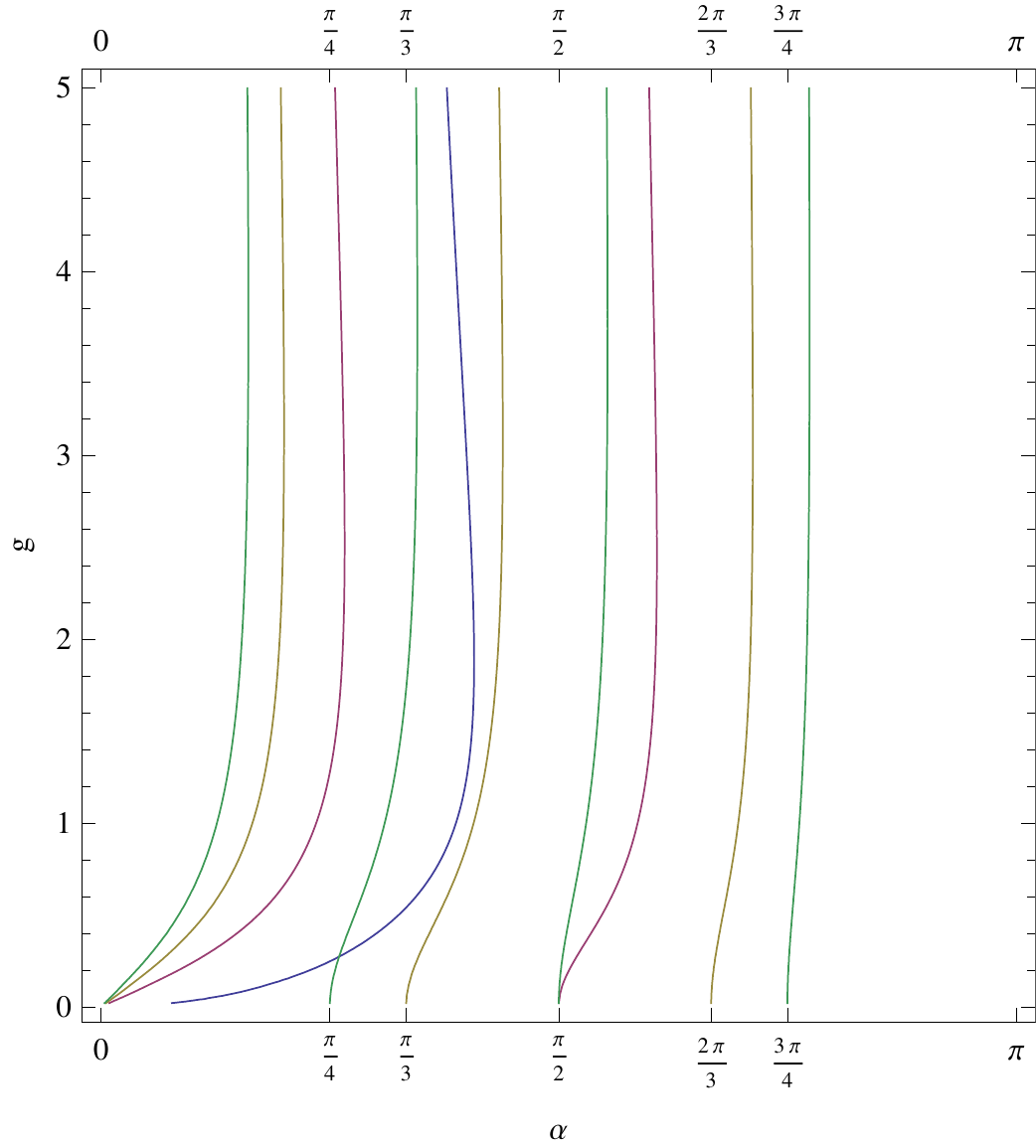}
\end{center}
\caption{The curves $T_{1j}(-i/g)=0$ for $j=1,2,3,4$, corresponding to critical values for $Y_-$(blue), $Y_{1|w}$(red), $Y_{2|w}$(yellow), and $Y_{3|w}$(green) respectively.}
\label{fig:YMwcrit2}
\end{figure}

Similar behaviour occurs for the $Y_{M|w}$-functions, with the change that there are multiple separate curves corresponding to the critical values; in general the critical value for a $Y_{M|w}$ function starts from zero $M+1$ times, at $\alpha = 0\mod\pi/(M+1)$. This has been illustrated in figure \ref{fig:YMwcrit2}. The complete picture of the asymptotic criticality is then obtained by adding additional curves to figure \ref{fig:YMwcrit2}, so that a picture qualitatively the same as figure \ref{fig:YMwcrit1} arises from each point where the critical value is zero. We refrain from presenting these curves here, as they provide no insight, and are time-consuming, though conceptually simple, to find.

Finally we would like to come back to the continuity of the hybrid equations, considered as a function of twist. While the relationship between the length parameter in the TBA and $J$ is discontinuous in the zero twist limit, the full equation, equation \ref{eq:hybridQ=1}, is of course continuous. The extra $2 \tH$ appears as the contribution from the extra roots of $1 - Y_-$ discussed above. In the zero twist limit, these roots move towards $\pm \infty -i/g$, while their critical value goes to zero simultaneously, \textit{cf.} figure \ref{fig:YMwcrit1}. Analyzing the resulting driving terms, the only term that survives in the limit of roots with large real parts is the term $\tfrac{1}{2}\log{\tfrac{S_{yQ}(r^-,v)S_{yQ}(-r^-,v)}{S_{yQ}(r^+,v)S_{yQ}(-r^+,v)}}$, where $r$ is the real part of any of the four roots. Taking the limit $r\rightarrow \infty$ in the expression for this kernel gives exactly $ -2 i \log{\tfrac{x^+}{x^-}} = 2 \tH$.

\section{Chapter \ref{chapter:boundstates}}

This appendix contains the technical details omitted from the main text of chapter \ref{chapter:boundstates}.

\subsection{Contribution of \texorpdfstring{$\log(1+Y_Q)\star_{C_Q} {\cal K}_Q$}{}}
\label{sec:contcontrib}

In this section of the appendix we consider the contribution of the terms of the form $\log(1+Y_Q)\star_{C_Q} {\cal K}_Q$ where ${\cal K}_Q$ is an arbitrary kernel and ${\cal S}_Q$ is the corresponding S-matrix. Taking the contour described in the main text back to the real line, we obtain the following contributions from the different $Y_Q$ functions\\

 \noindent
$\bullet$\ $Q=1$
\begin{align}
\log {\cal S}_1(u_3^{(2)--},v)-\log {\cal S}_1(u_3^{(1)--}, v)-\log {\cal S}_1(u_2^{(1)}, v)-\log {\cal S}_{1_*}(u_1, v)
\end{align}

 \noindent
$\bullet$\ $Q=2$
\begin{align}
&+\log {\cal S}_2(u_2^{(1)+}, v)+\log {\cal S}_2(u_3^{(3)---}, v)\nonumber\\
&-\log {\cal S}_2(u_2^{(2)+}, v)-\log {\cal S}_2(u_3^{(2)---}, v)
\end{align}

 \noindent
$\bullet$\ $Q\ge 3$
\begin{align}\label{YQKQ3}
&+\log {\cal S}_Q(u_3^{(Q-1)}-\frac{i}{ g}(Q-1), v)+\log {\cal S}_Q(u_3^{(Q+1)}-\frac{i}{ g}(Q+1), v)\nonumber\\
&-\log {\cal S}_Q(u_3^{(Q)}-\frac{i}{ g}(Q-1), v)-\log {\cal S}_Q(u_3^{(Q)}-\frac{i}{ g}(Q+1), v)
\end{align}

\noindent Now let us assume that ${\cal S}_Q$ satisfies the discrete Laplace equation
$$
{\cal S}_{Q-1} (u,v){\cal S}_{Q+1} (u,v)={\cal S}_{Q} (u^-,v){\cal S}_{Q} (u^+,v) \,.
$$
Then we take a sum over $Q\ge3$ of the terms in \eqref{YQKQ3} and get
\begin{align}
&\sum_{Q=3}^\infty \log \frac{{\cal S}_Q(u_3^{(Q-1)}-\frac{i}{ g}(Q-1), v){\cal S}_Q(u_3^{(Q+1)}-\frac{i}{g}(Q+1), v)}{ {\cal S}_Q(u_3^{(Q)}-\frac{i}{ g}(Q-1), v){\cal S}_Q(u_3^{(Q)}-\frac{i}{ g}(Q+1), v)}
=\log \frac{{\cal S}_3(u_3^{(2)--}, v)}{ {\cal S}_2(u_3^{(3)---}, v) }\,.
\end{align}
Adding the contributions from $Q=1,2$,  we finally get the driving terms originating from
$\log(1+Y_Q)\star_{C_Q} {\cal K}_Q$
 \begin{align}\label{YQKQ}
&-\log {\cal S}_{1_*}(u_1, v)-\log\frac{ {\cal S}_{1}(u_2^{(1)}, v)}{ {\cal S}_1(u_3^{(1)}, v)}
+\log \frac{{\cal S}_2(u_2^{(1)+}, v)}{ {\cal S}_2(u_2^{(2)+}, v)}\frac{{\cal S}_2(u_3^{(2)-}, v)}{ {\cal S}_2(u_3^{(1)-}, v)}\,.
\end{align}
In the asymptotic limits $g\to 0$ with $J$ fixed or $J\to \infty$ with $g$ fixed the  last term in \eqref{YQKQ} goes to 0.
Using the discrete Laplace equation, equation (\ref{YQKQ}) can be also written in the form
\begin{equation}
-\log {\cal S}_{1_*}(u_1, v)+\log \frac{{\cal S}_2(u_2^{(1)++}, v)}{{\cal S}_2(u_3^{(1)--}, v)}\frac{{\cal S}_2(u_3^{(2)-}, v)}{ {\cal S}_2(u_2^{(2)+}, v)}\,.
\end{equation}
It is worth mentioning that all the driving terms in \eqref{YQKQ} depend only on the singularities of $Y_1$ and $Y_2$ functions, and in fact they can also be explained by the integration contours which pick up the contribution of the real zero of $1+Y_1$ in the string $u$-plane, the zeroes $u_2^{(1)}$ and $u_3^{(1)}$  of $1+Y_1$ in the mirror $u$-plane, and all zeroes and poles of $1+Y_2$ in the analyticity strip of the mirror $u$-plane, but avoid all the other zeroes and poles of $1+Y_Q$ even those which are inside the analyticity strip of the mirror $u$-plane. The generalization of these identities to states with rapidities in the $k$th strip can be found in appendix  A.1 of \cite{Arutyunov:2011mk}.

\subsection{Identities to simplify the TBA equations}\label{identY}

In this section we collect the identities necessary to derive the simplified TBA equations from the canonical ones. Before listing the specific identities, let us discuss a frequently encountered situation; integrating $\log f$ of a complex function $f$ over an integration contour which runs either a bit above or below the real line. For any function $f(t)$ which has real zeroes at $u_i^o$ and real poles at $u_j^\infty$ we define
$\log f$ as
\begin{equation}\label{logf}
\log f(t) \equiv \log \Big(f(t)\, \frac{\prod_j (t-u_j^\infty)}{ \prod_i (t-u_i^o)}\Big) +\sum_i \log(t-u_i^o) - \sum_j \log(t-u_j^\infty)\,.
\end{equation}
Since $\tilde f(t)\equiv f(t)\, \frac{\prod_j (t-u_j^\infty)}{\prod_i (t-u_i^o)} $ has no real zeroes or real poles, the cuts of $\log \tilde f$ can and must be chosen so that they do not intersect the real line. With such a choice of the cuts of $\log \tilde f$ the imaginary part of $\log f^{p.v}$ is continuous on the real line where the function $f^{p.v.}$ is defined as $f^{p.v.}(t)\equiv  f(t)\, {\prod_j {\rm sign}(t-u_j^\infty)\,\prod_i {\rm sign}(t-u_i^o)}$. If $f(t)$ is real for real $t$ and $f(\infty)>0$ then  $f^{p.v.}(t)= |f(t)|$. The function $f^{p.v.}$ is used to define the principal value prescription by the formula
\begin{equation}\label{logf2}
\log f\star_{p.v.}K \equiv \log f^{p.v.}\star K\,,
\end{equation}
where  on the right hand side the Cauchy principal value of the integral is computed over the real line. This definition is a generalization of the one used in \cite{Arutyunov:2009ax} to complex functions $f$. The formulae \eqref{logf} and \eqref{logf2} are also used if some of the zeroes or poles coincide, {\it e.g.} if $f$ has a real double pole at $u^\infty$ then $\log f$ is understood as
\begin{equation}
\log f(t)  \equiv \log \big(f(t)(t-u^\infty)^2\big) - 2\log (t-u^\infty )\, ,
\end{equation}
and a similar expression if $f$ has a real double zero.

In all formulae below we define two actions of the operator $(K+1)^{-1}_{NM}$ on any set of functions $\log f_N$. The first one  is defined as
\begin{equation}
\log f_N\, (K+1)^{-1}_{NM} \equiv \log f_M - \log f_{M-1}\star s -\log f_{M+1}\star s\,,
\end{equation}
where the integration contour for the $\star$-convolution runs a bit above the real line to deal with zeroes and poles of $f_{M-1}$ and $f_{M+1}$ on the real line.

The second action explicitly takes into account the real zeroes and  poles  by using the principal value prescription defined above
\begin{equation}
\log f_N\star_{p.v.}  (K+1)^{-1}_{NM} \equiv \log f_M^{p.v.} - \log f_{M-1}\star_{p.v.} s -\log f_{M+1}\star_{p.v.} s\,.
\end{equation}

To simplify the notation, in chapter \ref{chapter:boundstates} and this appendix we often use the conventions
\begin{equation}
f(u-v)\star {\rm K} \equiv \int \, dt \, f(u-t){\rm K}(t,v)\,,\quad g(u,v)\star {\rm K} \equiv \int \, dt \, g(u,t){\rm K}(t,v)\,,
\end{equation}
where $f$, $g$, and ${\rm K}$ are arbitrary kernels or functions. Notice that according to our other conventions
\begin{equation}
f(u-v)\star {\rm K} \equiv f\star {\rm K}(u,v) \,,\quad g(u,v)\star {\rm K} \equiv g\star {\rm K}(u,v)\,.
\end{equation}

\subsection*{Identities  for $Y_{M|w}$}
Firstly we have
\begin{equation}
\sum_{N=1}^{\infty}\log S_{1N} (r_0-v) (K+1)^{-1}_{NM}=\delta_{2,M}\log S(r_0-v)\, ,\end{equation}
and
\begin{equation*}
\tfrac{1}{2}\sum_{P,N=1}^{\infty} \log \frac{S_{PN} (r_{N}^+-v)}{S_{PN} (r_{N}^--v)}\star_{p.v.} (K+1)^{-1}_{NM}=
\begin{cases} -\log |S(r_{M-1}^--v)S(r_{M+1}^--v)| & \mbox{for } M\neq 1 \, ,\\
-\log |S(r_{2}^--v)|& \mbox{for } M= 1 \, ,
\end{cases}
\end{equation*}
where the p.v. prescription has been used to deal with zeroes and poles of
$S_{N-1,N}$ and $S_{N+1,N}$ at $v=r_{N}$. Moreover, we need the sum
\begin{equation}
\sum_{N=1}^{\infty}\log S_{N} (r_0^--v)\star_{p.v.} (K+1)^{-1}_{NM}=\left\{\begin{array}{cl} 0 &~~ {\rm for}~~M\geq 3 \,,\\
\tfrac{1}{2}\log S(r_0-v)&~~ {\rm for}~~M= 2 \,,\\
\log |S(r_{0}^--v)| &~~ {\rm for}~~M= 1 \, .\\
\end{array}\right.\,~~~~~
\end{equation}

\subsection*{Identities for $Y_{M|vw}$}

In addition to the identities for $Y_w$ we also need
\begin{align*}
\log S^{1_*Q}_{xv}\star_{p.v} (K+1)^{-1}_{QM} (u_1,v) = &\delta_{M,1}\big( \log S_{1_*y}\star s (u_1,v) - \log S(u_1^- - v)\big)+ \frac{1}{ 2}\delta_{M,2}\log S(u_1 - v)\,,\\
\log \frac{S^{1Q}_{xv} (u_2,v)}{S^{1Q}_{xv} (u_3,v)} \star (K+1)^{-1}_{QM} =&  \delta_{M,1} \Big(\log S_{1_*y}\hstar s(u_2,v)+\log S_{1_*y}\hstar s(u_3,v) +\log \frac{S(u_2^+ -v)}{ S(u_3^- -v)}\Big)\, ,
\end{align*}
where we have rewritten $S_{1y}$ in terms of $S_{1_*y}$, and
\begin{align}
\log \frac{S^{2Q}_{xv} (u_2^+,v)}{S^{2Q}_{xv} (u_3^-,v)}(K+1)^{-1}_{QM} = & \delta_{M,1} \Big( \log \frac{S_{2y}(u_2^+,v)}{S_{2y}(u_3^-,v)}\hstar s +\log \frac{S(u_2^+ -v)}{ S(u_3^- -v)}\Big) \, .
\end{align}

\subsection*{Identities  for $Y_+Y_-$ }
To simplify the canonical TBA equation for $Y_+Y_-$ we need to compute the infinite sums involving $Y_{w}$ and $Y_{vw}$ functions.
Using the method from section 8.4 of \cite{Arutyunov:2009ax}  which we modify slightly due to the presence of zeroes of Y-functions and driving terms, we get the following two formulae
\begin{align}
\label{identityYMvw}
\sum_{M=1}^\infty& \log (1 + \frac{1}{Y_{M|vw}}) \star_{p.v.} {K}_M =
\log (1 + Y_{1|vw}) \star s
\\[1mm]
&\qquad\qquad\qquad -  \log\frac{1-Y_-}{ 1-Y_+} \hstar s\star {K}_1+ \sum_{M=1}^\infty \log(1 + Y_{M+1}) \star s \star {K}_M
\notag
\\[1mm]
&\qquad\qquad\qquad-\Big(
\log\frac{S(u_2^{(2)+} - v)}{S(u_3^{(2)-} - v)} - \log S(u_1^- - v)S(r_{0}^- - v)\Big)\star_{p.v.} {K}_1\,,
\notag
\end{align}
and
\begin{align}\label{identityYMw}
\sum_{M=1}^\infty \log (1 + \frac{1}{Y_{M|w}}) \star_{p.v.} {K}_M &=
\log (1 + Y_{1|w}) \star s
-  \log\frac{1-\frac{1}{Y_-}}{1-\frac{1}{ Y_+}} \hstar s\star {K}_1
\\[1mm]
&+\sum_{M=1}^\infty  \log S(r_{M-1}^- - v)S(r_{M+1}^- - v) \star_{p.v.} {K}_M\,.
\notag
\end{align}
The sum on the second line of equation \eqref{identityYMw} can be transformed to the form
\begin{align}\notag
\sum_{M=1}^\infty  &\log S(r_{M-1}^- - v)S(r_{M+1}^- - v) \star_{p.v.} {K}_M =
 \log S(r_{0}^- - v) \star_{p.v.} {K}_1\\[1mm]
&\label{sumYMw}
\qquad\qquad\qquad\qquad\qquad -\log S(r_{1}^- - v) +\frac{1}{2}
\sum_{M=1}^\infty\log \frac{{S}_{M}(r_{M}^-- v)}{ {S}_{M}(r_{M}^+- v) }
\,.
\end{align}
Then we also use
\begin{align}
& \log S(u_1^- - v)\star_{p.v.} {K}_1= \log |S_1(u_1^- - v)|\star s\,,
 \\[1mm]%\nonumber
& \log S_{1_*y}(u_1,v)
\hstar {K}_1 = \log {S^{1_*1}_{xv}(u_1, v)} - \log {S_1(u_1^{+}- v)}
\,, \\[1mm]
&
\log |S_1(u_1^- - v)|\star s
+\log {S_1(u_1^{+}- v)} \star s = \frac{1}{2}\log {S_2(u_1- v)} \star s
\,, \\[1mm]
%\nonumber
& \log S_{xv}^{1_*1}(u_1,v)\star s-\frac{1}{2}\log {S_2(u_1- v)} \star s = \frac{1}{2}\log \frac{S_{xv}^{1_*1}(u_1,v)^2}{ S_2(u_1- v)} \star s
  \,,\\[1mm]\nonumber
&\nonumber
\log\frac{S(u_2^{(2)+} - v)}{S(u_3^{(2)-} - v)}\star {K}_1 = \log\frac{S_1(u_2^{(2)+} - v)}{S_1(u_3^{(2)-} - v)}\star s \,,
\end{align}
\begin{align}
%\nonumber
 & \log \frac{S_{1y}(u_2^{(1)},v)}{  S_{1y}(u_3^{(1)},v)}
\hstar {K}_1= \log \frac{S^{11}_{xv}(u_2^{(1)}, v)}{ S^{11}_{xv}(u_3^{(1)}, v)} - \log \frac{S_1(u_2^{(1)+}- v)}{ S_1(u_3^{(1)-}- v)}
\,,\\[1mm]
&\log \frac{S_{2y}(u_2^{(1)+}, v)}{ S_{2y}(u_3^{(1)-}, v)}\hstar {K}_1 = \log \frac{S^{21}_{xv}(u_2^{(1)+}, v)}{S^{21}_{xv}(u_3^{(1)-}, v)} - \log \frac{S_1(u_2^{(1)+}- v)}{S_1(u_3^{(1)-}- v)}
\,, \\[1mm]
&%\nonumber
 \log {S^{11}_{xv}(u_3^{(1)}, v)}
 =- \log {S^{1_*1}_{xv}(u_3^{(1)}, v)} + \log {S_2(u_3^{(1)}- v)}\,,
\\[1mm]
& \nonumber \log \frac{{S}_1(u_2^{(1)}- v)}{ {S}_{1}(u_3^{(1)}- v)}-2 \log \frac{S_{xv}^{11}(u_2^{(1)},v)}{ S_{xv}^{11}(u_3^{(1)},v)}\star s
 =-\log \frac{S_{xv}^{1_*1}(u_2^{(1)},v)^2}{ S_2(u_2^{(1)}- v)} \frac{S_{xv}^{1_*1}(u_3^{(1)},v)^2}{ S_2(u_3^{(1)}- v)} \star s  \\
 \nonumber
 & \hspace{6cm} + \log \frac{S(u_2^{(1)}- v)}{S(u_3^{(1)}- v)}  \nonumber  \,.
\end{align}

%%%%%%%%%%%%%%%%%%%%%%%%%%
\subsection*{Identities  for $Y_Q$}
Let us start by recalling that the $S_{\sl(2)}^{1_*Q}$ S-matrix has the following structure
\begin{align}
\log S^{1_*Q}_{sl(2)}(u,v)=&-2\log \Sigma_{1_*Q}(u,v) -\log S_{1Q}(u-v)\\\nonumber
=&-2\log \Sigma_{1_*Q}(u,v) -\log S_{Q-1}(u-v)-\log S_{Q+1}(u-v)\, .
\end{align}
Thus identities involving $S_{\sl(2)}^{1_*M}$ follow from the corresponding identities for $\Sigma_{1_*Q}$ and $S_{Q}$. Firstly for $S_Q$  with the both arguments in the analyticity strip we have the following identity
\begin{equation*}
\log S_M\star (K+1)^{-1}_{MQ}
=\log S_Q -\log S_{Q-1}\star s-\log S_{Q+1}\star s=\delta_{Q1}\log S(u-v)\,,\quad u\in \mathbb{R}\,.~~~~
\end{equation*}
Analytically continuing this identity in the variable $u$ outside the analyticity strip to the locations of $u_2$ and $u_3$ we get
\begin{align*}
\log S_M\star (K+1)^{-1}_{MQ} (u_2,v)
=&\delta_{Q1}\log S(u_2-v)+\delta_{Q2}\log S(u_2^+-v)\,,~~~~\\
\log S_M\star (K+1)^{-1}_{MQ} (u_3,v)
=&\delta_{Q1}\log S(u_3-v)+\delta_{Q2}\log S(u_3^--v)\,.
\end{align*}
The next identity is for the dressing factor $\Sigma_{1_*Q}$ with the first argument on the real line of the string $u$-plane, {\it e.g.} equal to $u_1$
\begin{equation}%\nonumber
\log \Sigma_{1_*M}\star (K+1)^{-1}_{MQ} (u_1,v)
=\delta_{Q1}\log \check{\Sigma}_{1_*}\cstar s(u_1,v)\,.~~~~
\end{equation}
It was derived in \cite{Arutyunov:2009ax} where the precise definition of $\check{\Sigma}_{1_*}$ can be found, see equation (8.24) there. It is worth mentioning that the expression (8.24) in  \cite{Arutyunov:2009ax} for  $\check{\Sigma}_{1_*}(u,v)$ is valid for
any real $v$ and any complex $u$ on the string $u$-plane because the cuts there are inside the vertical strip $-2\le {\rm Re}\,  u\le 2$.

It is convenient to use the canonical TBA equations in the form \eqref{YQcanTBA}, meaning we also need identities for $\Sigma_{1Q}$ with the both arguments in the mirror $u$-plane. Since $\Sigma_{1Q}$ is a holomorphic function in the mirror region these identities have the same form for any value of the first argument, {\it i.e.} also for $u_2$ and $u_3$
\begin{equation}%\nonumber
\log \Sigma_{1M}\star (K+1)^{-1}_{MQ} (u_i,v)
=\delta_{Q1}\log \check{\Sigma}_{1}\cstar s(u_i,v)\,.~~~~
\end{equation}
We also need the standard identities
\begin{equation}
\log \Sigma_{2M}(K+1)^{-1}_{MQ} = \delta_{Q,1} \log \check{\Sigma}_2 {\cstar} s \, ,
\end{equation}
and
\begin{equation}
\log S_{2M}(K+1)^{-1}_{MQ} = (\delta_{Q,1}+\delta_{Q,3}) \log S \, ,
\end{equation}
which can be applied directly since the relevant arguments lie within the analyticity strip.
Finally we need the following identities for the auxiliary S-matrices
\begin{equation}\nonumber
\log S^{1M}_{vwx}\star (K+1)^{-1}_{MQ} (u,v)
=\delta_{Q2}\log S(u-v)-\delta_{Q1}\log \check S_1\cstar s(u,v)\,,\quad u,v\in \mathbb{R}\,,~~~~~~
\end{equation}
\begin{align}
\log S_{yM}\star_{p.v.} (K+1)^{-1}_{MQ}(r_0^-,v) =&  \delta_{Q,1} (\log S(r_0^--v)-2\log \check{S}\cstar s(r_0^-,v))  \nonumber\\
& +\frac{1}{2} \delta_{Q,2} \log S(r_0 -v) \,.
\end{align}
We also need
\begin{equation}
 \check{\Sigma}_{1}(u_2,v) =  \check{\Sigma}_{1_*}(u_2,v)\,,\quad
1/\check{\Sigma}_{1}(u_3,v) =  \check{\Sigma}_{1_*}\check{S}_{1}(u_3,v)\,,
\end{equation}
to replace $ \check{\Sigma}_{1}$ by $ \check{\Sigma}_{1_*}$ in the simplified TBA equation for $Y_1$. The last identity uses
\begin{equation}
\check{S}_{1}(u_3,v)=\check{S}(u_3^+,v)/\check{S}(u_3^-,v)\,,
\end{equation}
Other useful identities are
\begin{equation}
\check{S}_{1}(u_1,v)=\check{S}(u_1^+,v)\check{S}(u_1^-,v)\,,\quad \check{S}_{1}(u_2,v)=\check{S}(u_2^-,v)/\check{S}(u_2^+,v)\,.
\end{equation}

\subsection{Identities for the exact Bethe equations}
\label{app:EBE1}

\subsection*{Identities for the exact Bethe equation for $u_1$}

The derivation of the exact Bethe equation for the real root $u_1$ requires the following  identities
\begin{align}
& 2\log S\star_{p.v.} K_{vwx}^{11}(u_1^-,v)-\log S^{11}_{vwx}(u_1,v)= \nonumber \\
&~~~~~~~~~~~2\log{\rm Res}\, S\star K_{vwx}^{11}(u_1^-,v)- 2\log\left(u_1-v-\frac{2i}{g}\right) \label{Resu1} \frac{x_s^-(u_1)-\frac{1}{x^-(v)}}{x_s^-(u_1)-\frac{1}{x^+(v)}}\, , \\
&2\log S\star_{v.p.}  K_{vwx}^{11}(r_0,v)+\log S^{11}_{vwx}(r_0,v) =\nonumber\\
&~~~~~~~~~~~2\log{\rm Res}\, S\star K_{vwx}^{11}(r_0^-,v)-2\log\left(v-r_0-\frac{2i}{g}\right) \frac{x_s^+(r_0)-x^+(v)}{x_s^+(r_0)-x^-(v)}\, ,
\label{Resro}
\end{align}
where we use the notation
\begin{equation}
\log\mbox{Res}\, S\star K_{vwx}^{11_*}(u^-,v)=\int_{-\infty}^{\infty}{\rm d}t\, \log \Big[S(u^--t)(t-u)\Big]K^{11_*}_{vwx}(t+i0,v)\, .~~~~~
\end{equation}

%There is something somewhere in the first two formulas below that generates an error in compiling, but it is strange.

Next, to show that the real part of equation \eqref{EBE1} vanishes we use the following identities valid for real $t$ and $v$
\begin{align}
&\mbox{Re}\Big(2 s\star K_{vwx}^{Q,1_*}(t,v) \Big)= K_Q(t-v)\\
&\mbox{Re}\Big(K_{\mathfrak{sl}(2)}^{Q1_*}(t,v)+2 s\star K_{vwx}^{Q-1,1_*}(t,v)\Big) = -K_{yQ} \hstar K_1(t,v)
\end{align}
which allow us to prove that
\begin{align}
\mbox{Re}\Big(G_{1_*}(u_1) \Big)= & -\log \left(1+Y_{Q} \right)\star K_{yQ} \hstar K_1(u_1)+ \log \frac{Y_+}{Y_-} \hstar K_1(u_1)  \\ \nonumber
= &- \sum_i \log S_{1_*y}(u_i^{(1)},u_1)\hstar K_1  + \log \frac{S_{2y}(u_2^{(1)+},u_1)}{S_{2y}(u_3^{(1)-},u_1)} \frac{S_{2y}(u_3^{(2)-}, u_1)}{S_{2y}(u_2^{(2)+},u_1)} \hstar K_1\,.
\end{align}
To handle the driving terms in \eqref{EBE1} we further use \eqref{ssid} to write
\begin{equation}
\log\frac{S_{\sl(2)}^{11_*}(u_2^{(1)},u_1)}{S_{\sl(2)}^{11_*}(u_3^{(1)},u_1)} = \log\frac{S_{\sl(2)}^{1_*1_*}(u_2^{(1)},u_1)S_{\sl(2)}^{1_*1_*}(u_3^{(1)},u_1)}{ h_1(u_1,u_2)h_1(u_1,u_3)}+ \log \frac{h_1(u_1,u_2)}{ h_1(u_1,u_3)}\,.~~~~~~
\end{equation}
It can be shown that the first term on the r.h.s. is imaginary while the second one is real.  Then we need the identities
\begin{equation*}
{\rm Re}\frac{S_{\sl(2)}^{21_*}(u_2^{+},u_1)}{ S_{\sl(2)}^{21_*}(u_3^{-},u_1)}= \frac{S_{xv}^{21}(u_3^{-},u_1)}{ S_{xv}^{21}(u_2^{+},u_1)} \, ,\quad
{\rm Re}\Big(2\log \frac{S(u_2^{+},u_1)}{S(u_3^{-},u_1)}\star K_{vwx}^{11_*}\Big) = \log \frac{S_1(u_2^{+}-u_1)}{ S_1(u_3^{-}-u_1)}\,.~~~~~~
\end{equation*}
By using these identities it is then straightforward to check numerically that the real part of \eqref{EBE1} vanishes.

\subsection*{Exact Bethe equation for $u_3$ in the string region}

\begin{figure}[t]
\begin{center}
\includegraphics[width=.45\textwidth]{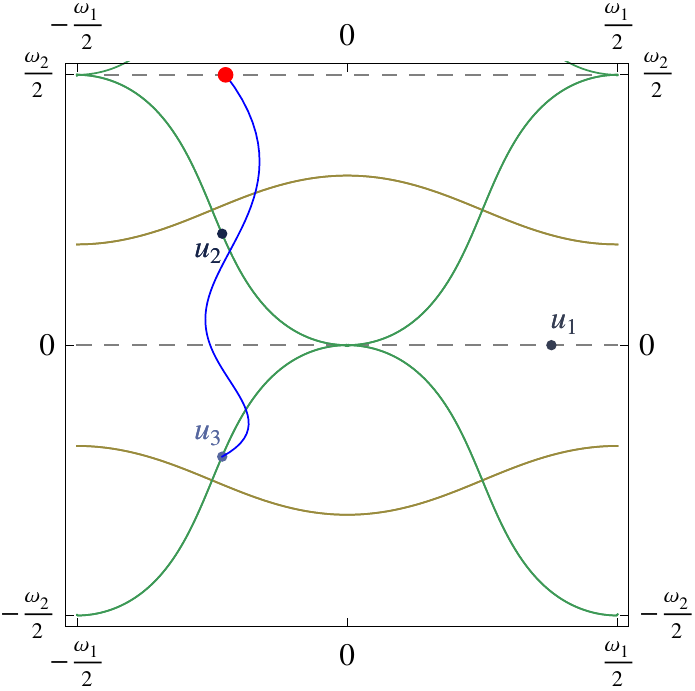}\qquad\includegraphics[width=.45\textwidth]{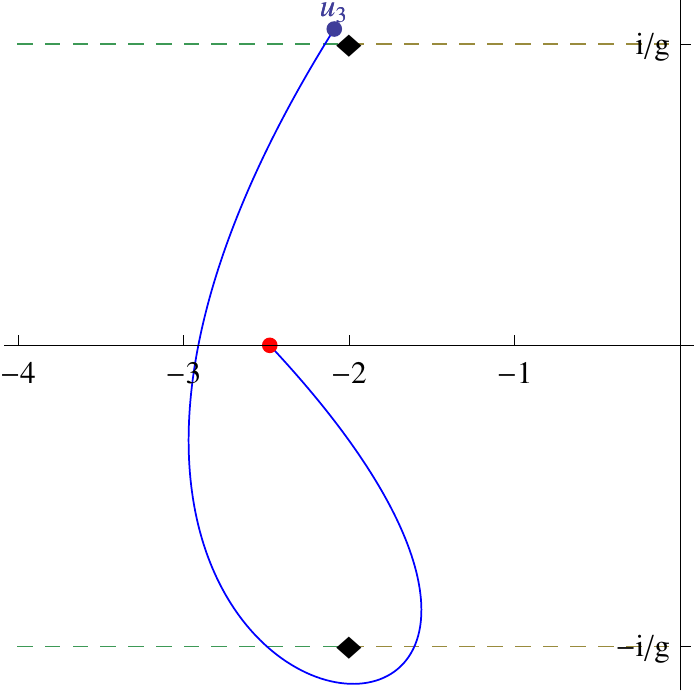}
\caption{The left picture depicts a continuation path connecting
a point on the real line of the mirror theory to the point $u_3^{(1)}$
in the string region. The right picture shows the same (homotopic) path on the $u$-plane.}
\label{fig:contpicture}
\end{center}
\end{figure}

Here we  discuss the continuation of the canonical TBA equation for $Y_1$ to $v\approx u_3^{(1)}$ in the string region. Since Im$\,u_3>1/g$ once we are in the string region we need to cross the real line and the line  ${\rm Im}\,  v=1/g$ from below and \emph{outside} of $(-2+i/g,2+i/g)$, as illustrated in figure \ref{fig:contpicture}. Let us consider the continuation of the relevant terms individually.
\begin{itemize}
\item
$\log\left(1+Y_{Q} \right) \star K_{\sl(2)}^{Q1}(v)$. Continuation of this term gives
$$\log\left(1+Y_{Q} \right) \star K_{\sl(2)}^{Q1}(v)\to \log\left(1+Y_{Q} \right) \star K_{\sl(2)}^{Q1*}(v)
-\log(1+Y_2(v^-))\, .$$
Note that the line ${\rm Im}\,  v=-1/g$ is crossed twice during the continuation, giving vanishing net contribution, while the line  ${\rm Im}\,  v=1/g$ is crossed once.

\item $2 \log\left(1+ \frac{1}{Y_{Q|vw}} \right) \star_{p.v} K^{Q1}_{vwx} (v)$. Nothing happen to the kernels
upon crossing the line ${\rm Im}\,  v=-1/g$. However entering the string region we have to change
\begin{equation*}
2 \log\left(1+ \frac{1}{Y_{Q|vw}} \right) \star_{p.v} K^{Q1}_{vwx} (v)\to
2 \log\left(1+ \frac{1}{Y_{Q|vw}} \right) \star_{p.v} K^{Q1_*}_{vwx} (v)\,.
\end{equation*}
Continuing further we cross the real line where $K^{11*}_{vwx}$ exhibits a pole, and  crossing the line  ${\rm Im}\,  v=1/g$ we encounter a pole of  $K^{21*}_{vwx}$
\begin{equation*}
K^{11^*}_{vwx}(u,v)=-\frac{1}{2\pi i}\frac{1}{u-v} + \mbox{reg.} \, ,\quad
K^{21*}_{vwx}(u,v)=-\frac{1}{2\pi i}\frac{1}{u-v+i/g} + \mbox{reg.}
\end{equation*}
Resolving these singularities gives
\begin{align}
\nonumber
&2 \log\left(1+ \frac{1}{Y_{Q|vw}} \right) \star_{p.v} K^{Q1}_{vwx} (v)\to
2 \log\left(1+ \frac{1}{Y_{Q|vw}} \right) \star_{p.v} K^{Q1_*}_{vwx} (v)\\
&\hspace{1cm }+2\log\left(1+ \frac{1}{Y_{1_{\hat{*}}|vw} (v)}\right) +2\log\left(1+ \frac{1}{Y_{2|vw} (v^-)}\right)\,,
\end{align}
where $Y_{1_{\hat{*}}|vw}$ is the analytic continuation of $Y_{1|vw}$ across its cut at ${\rm Im}\,  v=1/g$.

\item $\log\big(1- \frac{1}{Y_-} \big) \hstar (K_{1}  +  K_{y1}).$ Taking into account the pole of $(K_1 + K_{y1})(t,v)$ at $v = t -i/g$ we get
\begin{equation*}
\hspace{-10pt} \log\big(1- \frac{1}{Y_-} \big) \hstar (K_{1}  +  K_{y1})\to \log\big(1- \frac{1}{Y_-} \big) \hstar (K_{1}  +  K_{y1_*})
+ 2\log\big(1-\frac{1}{Y_{+_{\hat{*}}}(v^+)} \big)
\end{equation*}
Here $Y_+(v^+)$ appears because we continue $Y_-$ across its cut on the real line; $Y_{+_{\hat{*}}}$ denotes  $Y_+$ analytically continued  across  its cut at $2i/g$.

\item $\log\big(1- \frac{1}{Y_+} \big) \hstar (-K_{1}  +  K_{y1}).$  This term does not produce any extra term.
\end{itemize}
The resulting analytic continuation of the TBA equation to the string region for $v\approx u_3^{(1)}$  is given by \eqref{eq:EBu3string}.

To proceed further, we recall that the  crossing relations for the bound-state dressing factors imply the following identity for the $S_{\sl(2)}^{Q1}$ S-matrix \cite{Arutyunov:2008zt}
\begin{equation}\label{crossQ1}
S_{\sl(2)}^{Q1}(v_1,v_2)S_{\sl(2)}^{Q1_*}(v_1,v_2)=\left(\frac{x_1^+-x_2^+}{x_1^--x_2^+}\frac{1-\frac{1}{x_1^+x_2^-}}{1-\frac{1}{x_1^-x_2^-}}\right)^2=\left(\frac{S_{Qy}(v_1,v_2^-)}{S^{Q1}_{xv}(v_1,v_2)}\right)^2\, .
\end{equation}
This identity in its turn leads to the following crossing relations
\begin{align}
K_{\sl(2)}^{Q1}(t,v) & +K_{\sl(2)}^{Q1_*}(t,v) =2 K_{Qy}(t,v^-)- 2 K^{Q1}_{xv}(t,v) \, .
\end{align}
Then we can easily check that for $v\approx u_3^{(1)}$
\begin{align*}
K_{y1}(t,v) & + K_{y1_*} (t,v)=0 \, ,\quad
K^{Q1}_{vwx}(t,v)   + K^{Q1_*}_{vwx}(t,v)  = K_{Q1}(t-v) \, .
\end{align*}
Thus, adding the right hand sides of equations (\ref{eq:EBu3mirror}) and (\ref{eq:EBu3string}) and using the above crossing relations, we find equation \eqref{YQcanTBA2add} for $v\approx u_3^{(1)}$.

%%%%%%%%%%%%%%%%%%%%%%%%%%%%%
\subsection*{Asymptotic limit of the exact Bethe equation for $u_2$}
%%%%%%%%%%%%%%%%%%%%%%%%%%%%%%%
Here we analytically continue the
hybrid TBA equation for $Y_1$ to the point $u_2^{(1)}$.
Recall that
$u_2^{(1)}$ is in the intersection of the string and mirror regions and it lies below the line $-\frac{i}{g}$ in the mirror theory. We have
$$\log\left(1+Y_{Q} \right) \star
K_{\sl(2)}^{Q1} = -\log\left(1+Y_{Q} \right) \star
K_{Q1} -2\log\left(1+Y_{Q} \right) \star
K^{\Sigma}_{Q1} \,.$$
Since the dressing kernel is holomorphic in the region containing the continuation path \cite{Arutyunov:2009kf}, it is sufficient to consider
$-\log\left(1+Y_{Q} \right) \star K_{Q1}$. Since $K_{n1}=K_{n+1}+K_{n-1}$
and at $u - v-\frac{i}{ g}Q\sim 0$
\begin{equation}
K_Q(u-v) = \frac{1}{ 2\pi i}\frac{1}{ u - v-\frac{i}{ g}Q} +\ldots\,,
\end{equation}
we conclude that only the term with $K_{21}$ containing $K_1$ plays  a role for analytic continuation to $u_2^{(1)}$.
Continuing beyond the line
Im$(v)=-1/g$ from above produces the term $\log\left(1+Y_{2} \right)(v+i/g) $.
Taking this into account we get for Im$(v)<-1/g$
\begin{equation}
-\log\left(1+Y_{2} \right) \star K_{21}(v) \to -\log\left(1+Y_{2}(v+\frac{i}{ g})  \right) -\log\left(1+Y_{2} \right) \star K_{21}(v)\,.~~~~~~
\end{equation}
Thus  the continuation to $v=u_2^{(1)}$ produces an extra term $-\log(1+Y_{2}(u_2^{(1)+}))=-\log \infty $ which is
actually divergent! However, the equation (\ref{HybridQ}) contains the driving term
$
+\log \frac{S_{\sl(2)}^{21}(u_2^{(1)+},v)}{S_{\sl(2)}^{21}(u_2^{(2)+},v)}\, .
$
Since
\begin{equation} S_{\sl(2)}^{QQ'}(u,v)=S^{QQ'}(u-v)^{-1}\Sigma_{QQ'}(u,u')^{-2}\, , \end{equation}
upon continuation of this term to
$u_2^{(1)}$ we get another divergent contribution arising due to the S-matrix $S^{21}$
$$
\log {S_{\sl(2)}^{21}(u_2^{(1)+},u_2^{(1)}) }\to -\log S^{21}(u_2^{(1)+}-u_2^{(1)})=-\log S^{21}(i/g)=-\log 0=\log\infty\, ,$$
which precisely cancels  the infinity coming from $-\log\left(1+Y_{2} \right) \star K_{21}$. Therefore, it makes
sense to combine these divergent terms into a regular expression
\begin{equation}
\lim_{v\to u_2^{(1)}}\log \frac{S_{\sl(2)}^{21}(u_2^{(1)+},v)}
{1+Y_2(v^+)} =\log \frac{{\rm Res}\, S_{\sl(2)}^{21}(u_2^{(1)+},u_2^{(1)})}
{{\rm Res}\, Y_2(u_2^{(1)+})} \, .\end{equation}
Continuation of all the other terms in equation (\ref{HybridQ}) goes without any difficulty and as a result we find the exact Bethe equation \eqref{ExactBYE1} .

\section{Chapter \ref{chapter:quantumTBA}}

\subsection{\texorpdfstring{$q$}{}-deformed \texorpdfstring{$vw$}{}-strings}

\label{subsec:StringHypo-vw}

Let us take a look at the equations for $y$ roots in the limit $K^{\mathrm{I}} \rightarrow \infty$. To analyze this limit, we need to analyze the product over $Q$-particle excitations in eqn. \eqref{eq:qdefBYy}. This product consists of terms of the form
\begin{equation}
S^{Qy}=q^{Q/2}\,\frac{x^--y}{x^+-y}\sqrt{\frac{x^+}{x^-}}\,,
\end{equation}
where $x^{\pm}$ describes a $Q$-particle bound state with real rapidity $u$. Noting that the ratio $x^+/x^-$ is real, the modulus of $S^{Qy}$ is given by
\begin{equation}
\left|S^{Qy}\right|^2=\frac{y^*-\frac{1}{x^+}}{y^*-\frac{1}{x^-}}\frac{y-x^-}{y-x^+}\frac{x^+}{x^-}=
\frac{|y|^2\frac{x^+}{x^-}+1-y^*x^+-\frac{y}{x^-}}{|y^2|+\frac{x^+}{x^-}-y^*x^+-\frac{y}{x^-}}\, .
\end{equation}
Hence this modulus is smaller than one provided
\begin{equation}
\left(1-\frac{x^+}{x^-}\right)(1-|y|^2)<0\, .
\end{equation}
Now for any bound state number $Q=1,\ldots, 2k-1$ and real $u$ we have $x^+/x^-<1$. So $|S^{Qy}|<1$ for $|y|>1$ and vice versa. This implies that in the limit $K^{\mathrm{I}} \rightarrow \infty$ the product over the momentum carrying particles in \eqref{eq:auxy} diverges when $|y_k|<1$ and goes to zero when $|y_k|>1$. Auxiliary roots that have $|y_k|=1$ are identified as $y$-particles; here we assume that  $|y_k|\neq1$.

Without loss of generality we take $|y_1|<1$. Then in the limit $K^{\mathrm{I}} \rightarrow \infty$ eqn. \eqref{eq:qdefBYy} becomes
\begin{equation}
\label{eq:auxyzero}
1  =\frac{1}{0} \times \prod_{i=1}^{K^{\mathrm{III}}}\frac{\sinh{\frac{\pi g}{2k}(v_k-w_i-\frac{i}{g})}}{\sinh{\frac{\pi g}{2k}( v_k-w_i + \frac{i}{g})}} \,,
\end{equation}
which requires one of the numerators in the product on the right hand side of the equation to vanish. This can be achieved by taking for instance
\begin{equation}
\label{eq:1stVWcondition}
v_1 - w_1 -\frac{i}{g} = 0 \, .
\end{equation}
We should also satisfy the equation for $w_1$ however, which now gives
\begin{equation}
1  =0 \times \prod_{i=2}^{K^{\mathrm{II}}} \frac{\sinh{\frac{\pi g}{2k}(w_1 - v_i+\frac{i}{g})}}{\sinh{\frac{\pi g}{2k}(w_1 - v_i - \frac{i}{g})}} \prod_{j=1}^{K^{\mathrm{III}}}\frac{\sinh{\frac{\pi g}{2k}(w_1 - w_j-\frac{2i}{g})}}{\sinh{\frac{\pi g}{2k}(w_1 - w_j+\frac{2i}{g})}}\, ,
\end{equation}
so that we should take\footnote{Taking a denominator in the product over $w$ roots to vanish would give $v_1 + \frac{i}{g}= w_2$ which contradicts the original considerations in \eqref{eq:auxyzero}, since the zero in the equation for $w_1$ by assumptions scales as the one in \eqref{eq:auxyzero}.}
\begin{equation}
\label{eq:1vwstring}
w_1 -v_2 - \frac{i}{g} = 0\, .
\end{equation}
Now $y_2$ has become involved in the game, and there are two options. If $|y_2|>1$ the equation for $y_2$ is `satisfied' and we get a $1|vw$-string.

In case $|y_2|<1$ we get further conditions which we explicitly work out one further level before giving the general pattern. Assuming $|y_2| <1$, analogous to \eqref{eq:auxyzero} and \eqref{eq:1stVWcondition} we take
\begin{equation}
\label{eq:2ndWinVW}
v_2 - w_2 - \frac{i}{g}=0\, .
\end{equation}
However we now have the equation for $w_2$ as well, which tells us
\begin{equation}
1  = 0 \times \prod_{i\neq 2}^{K^{\mathrm{II}}} \frac{\sinh{\frac{\pi g}{2k}(w_2 - v_i+\frac{i}{g})}}{\sinh{\frac{\pi g}{2k}(w_2 - v_i - \frac{i}{g})}} \prod_{j=1}^{K^{\mathrm{III}}}\frac{\sinh{\frac{\pi g}{2k}(w_2 - w_j-\frac{2i}{g})}}{\sinh{\frac{\pi g}{2k}(w_2 - w_j+\frac{2i}{g})}}\, ,
\end{equation}
so that $w_2 - v_3 -\frac{i}{g} = 0$. However we should not forget that from \eqref{eq:1vwstring} and \eqref{eq:2ndWinVW} we now also get
\begin{equation}
w_2 =  w_1 -\frac{2i}{g}\, ,
\end{equation}
meaning we should additionally have a numerator vanish\footnote{Note that we have $w_2 = w_1 - \frac{2i}{g}$ only `after' imposing \eqref{eq:2ndWinVW}; it cannot fix the original vanishing denominator.} by imposing
\begin{equation}
w_2 - v_4 + \frac{i}{g} = 0\, .
\end{equation}
At this point we have the configuration
\begin{equation}
\label{eq:2vwstring}
v_1 - \frac{2i}{g} =  v_2 = w_1 -\frac{i}{g} =  w_2 + \frac{i}{g} = v_3 + \frac{2i}{g}= v_4 \, ,
\end{equation}
The configuration \eqref{eq:2vwstring} forms a $2|vw$-string if $|y_{3,4}|<1$. If $|y_3|>1$ or $|y_4|>1$ we get more conditions, and in general we can end up with an $M|vw$-string given by the following configuration of roots
\begin{eqnarray}
\{ w \}&=&\left\{ v+(M-1)\frac{i}{g},\ldots ,  v -(M-1)\frac{i}{g}\right\}\, ,
\nonumber \\
\label{vwsol}
\{v\}&=&\left\{v +M\frac{i}{g},\,  \underbrace{v+(M-2)\frac{i}{g},\ldots , v-(M-2)\frac{i}{g}}_{M-1},\, v-M\frac{i}{g}\right\}\, .
\end{eqnarray}
Note that there are two $y$-roots $y_j,1/y_j$ associated to each of the underbraced rapidities
$$
v_j=v+(M-2j),~~~~ j=1,\ldots\, , M-1
$$
in a $M|vw$-string. We denote the roots associated with the first and the last rapidities in eqn. (\ref{vwsol}) by $y_M$ and $y_{-M}$ respectively. According to our construction, $|y_M|<1$ and $|y_{-M}|>1$. In total there are $2M$ $y$-roots.

As usual the reality conditions for the string centers follow by fusing the S-matrices and insisting that the S-matrix for the total $M|vw$-string is unitary. Fusing the S-matrix $S^{Qy}$ over the constituents $y$-roots of an $M|vw$-string gives
\begin{align}
\prod_{i\in M|vw} S^{Qy}(u,v) & = q^{QM}\frac{x^--y_M}{x^+-y_M}\frac{x^--y_{-M}}{x^+-y_{-M}}\frac{x^+}{x^-}
\prod_{i=1}^{M-1} \frac{x^--y_i}{x^+-y_i}\frac{x^--\frac{1}{y_i}}{x^+-\frac{1}{y_i}}\frac{x^+}{x^-}  \nonumber \\
& = q^{QM}\frac{x^--y_M}{x^+-y_M}\frac{x^--y_{-M}}{x^+-y_{-M}}\frac{x^+}{x^-}
\prod_{i=1}^{M-1} \frac{x^-+\frac{1}{x^-}-y_i-\frac{1}{y_i}}{x^++\frac{1}{x^+}-y_i-\frac{1}{y_i}}\nonumber \\
& = q^{QM}\frac{x^--y_M}{x^+-y_M}\frac{x^--y_{-M}}{x^+-y_{-M}}\frac{x^+}{x^-}
\prod_{i=1}^{M-1} \frac{q^{-Q}e^{\frac{\pi g u}{k} }-q^{M-2i}e^{\frac{\pi g v}{k} }}{q^Q e^{\frac{\pi g u}{k} }-q^{M-2i}e^{\frac{\pi g v}{k} }}  \nonumber \\
& = q^{QM}\frac{x^--y_M}{x^+-y_M}\frac{x^--y_{-M}}{x^+-y_{-M}}\frac{x^+}{x^-}
\prod_{i=1}^{M-1} q^{M-Q-2i} S_{Q+M-2i}(u-v) \nonumber \\
& = q^{Q}\frac{x^--y_M}{x^+-y_M}\frac{x^--y_{-M}}{x^+-y_{-M}}\frac{x^+}{x^-}
\prod_{i=1}^{M-1} S_{Q+M-2i}(u-v)\, .
\end{align}
Requiring this S-matrix to be unitary requires that $y_M^*=\frac{1}{y_{-M}}$ and $u$ to be real modulo $i k/g$.

At this point we do not have a restriction on the string length yet. This is because at every stage there are two $y$-particles associated to each rapidity, which means we can always choose the one with the desired properties. However, going beyond length $k$ it is not hard to see that we must necessarily have coincident roots which means the solution is not physical (in particular the associated wave-function would vanish). For $w$-particles however, we have no length $k$ string either. Again here the peculiarities of $y$-roots do not allow us to dismiss the length $k$ string; we can obtain a bona-fide divergence and repeat the arguments above. However, similarly to $w$-strings the would-be $k|vw$-string has fixed `momentum' (in the spin chain sense) and scatters trivially with everything else. This means these strings do not contribute to the physics and we can restrict our strings to have length up to $k-1$.

Coming back to the location of the center, it is easy to see that by translating a given root configuration by $2ik/g$ the set of all but the outer two $y$-roots is immediately invariant. Now since we have the constraints $|y_M|<1$ and $y_M^*=\frac{1}{y_{-M}}$, also the outermost $y$ roots cannot change either, and hence the center of the string can be taken to lie either on the real line or on the line $ik/g$.

At this point it appears we can have negative parity $vw$-strings of length up to $k-1$, while for $w$-strings this was not the case. Also this strange fact arises due to the curious property of the $y$-particles mentioned earlier; we can naively choose the correct $y$ roots to make our string hypothesis work. However, the root configurations that lead to negative parity strings of length greater than one are not consistent at the at the level of $w$ roots, where they violate the conditions we derived when discussing the XXZ spin chain\footnote{Of course the conditions on the $w$-particles could even be used to skip a considerable part of the above analysis done directly for the $y$-particles.}. This in particular implies that the Bethe wave-function associated to a negative parity $vw$-string of length greater than one in the deformed Hubbard model subsystem would not be normalizable. Hence we are led to exclude negative parity string of length greater than one, and end up with a string content exactly analogous to the second auxiliary level of $w$ roots.

\subsection{The simplified TBA equation for \texorpdfstring{$Y_k$}{}}

\label{app:sTBAYk}

To derive the simplified TBA equation for $Y_k$ we act with $\delta_{Q,k} - 2 \delta_{Q,k-1} s$ on the canonical TBA equations for $Q$-particles and make use of the special identities \eqref{eq:simpKMQvwxbndry} and \eqref{eq:simpKyQbndry} and definitions \eqref{eq:KyQKQdef} and \eqref{eq:KyQKyQdef} to get
\begin{align}
\label{eq:sTBAQktemp}
\log Y_k = \, & -\mu_k + \mu_{k-1} + 2\log{Y_{k-1}}\star s + \sum_a \log \left(1+\tfrac{1}{Y_{k-1|vw}^{(a)}}\right)\left(1+\tfrac{1}{Y_{0|vw}^{(a)}}\right) \star s \nonumber \\
& - \sum_a \log\left(1 + \tfrac{1}{Y^{(a)}_{M|vw}}\right) \star K_{M,k-1}\star s - \sum_a\log\left(1 + \tfrac{1}{Y^{(a)}_{0|vw}}\right) \star K_{k-1,k-1}\star s\\
& -\sum_a\log \frac{1-\tfrac{1}{Y_-^{(a)}}}{1-\frac{1}{Y_+^{(a)}}}\,
 \hat{\star}\,  K_{k-1} \star s  +\log \left(1+Y_P \right)\star (K_{\mathfrak{sl}(2)}^{Pk}- 2 K_{\mathfrak{sl}(2)}^{Pk-1}\star s)  \nonumber\, .
\end{align}
Note that the energy $\tilde{\cal E}_Q$ is annihilated by the operator $\delta_{Q,k} - 2 \delta_{Q,k-1} s\,$ because $\mathcal{E}_{k+1} = \mathcal{E}_{k-1}$ and it satisfies the discrete Laplace equation by construction. Examining eqn.(\ref{eq:sTBAQktemp}) we immediately recognize a large part of the canonical TBA equation for $Y^{(a)}_{k-1|vw}$, albeit integrated with $s$ and without the $Y_Q$ term. Rewriting then gives
\begin{align}
\label{eq:sTBAQktemp2}
\log Y_k =\, & 2\log{Y_{k-1}}\star s + \sum_a\log\left(1+\tfrac{1}{Y_{k-1|vw}^{(a)}}\right)\left(1+\tfrac{1}{Y_{0|vw}^{(a)}}\right)\frac{e^{-\chi^{(a)}_{vw}}}{Y^{(a)}_{k-1|vw}}\star s \\
& \quad + \log \left(1+Y_P \right)\star (K_{\mathfrak{sl}(2)}^{Pk}- 2 K_{\mathfrak{sl}(2)}^{Pk-1}\star s - 2 K^{Pk-1}_{xv} \star s)  \nonumber\, ,
\end{align}
where we have exchanged the chemical potentials (including those coming from the equation for $Y^{(a)}_{k-1|vw}$) for the $\chi^{(a)}_{vw}$s which also include possible factors of $i\pi$ coming from the contour deformation trick, in line with our choice to incorporate such effects manifestly in the simplified equations. The kernel $K_{\mathfrak{sl}(2)}$ satisfies the discrete Laplace equation, meaning that if we introduce the would-be fused $k+1$st kernel  we should have
\begin{equation*}
K_{\mathfrak{sl}(2)}^{Pk}- 2 K_{\mathfrak{sl}(2)}^{Pk-1}\star s = K_{\mathfrak{sl}(2)}^{PQ}(K+1)^{-1}_{Q,k} - K_{\mathfrak{sl}(2)}^{Pk-1}\star s + K_{\mathfrak{sl}(2)}^{Pk+1}\star s = (K_{\mathfrak{sl}(2)}^{Pk+1}-K_{\mathfrak{sl}(2)}^{Pk-1}-\delta_{Pk-1})\star s\,.
\end{equation*}
In this way we reduce the equation for $Y_k$ to the following form
\begin{align}
\label{eq:sTBAQktemp2int}
\log Y_k =\, & 2\log{Y_{k-1}}\star s + \sum_a \log\left(1+\tfrac{1}{Y_{k-1|vw}^{(a)}}\right)\left(1+\tfrac{e^{-\chi^{(a)}_{vw}}}{Y_{k-1|vw}^{(a)}}\right)\star s -\log(1+Y_{k-1})\star s\\
&\quad \quad + \log \left(1+Y_P \right)\star (K_{\mathfrak{sl}(2)}^{Pk+1}-K_{\mathfrak{sl}(2)}^{Pk-1} - 2 K^{Pk-1}_{xv} ) \star s \nonumber\, ,
\end{align}
where we have also eliminated $Y_{0|vw}$ in favour of $Y_{k-1|vw}$ via eqn. \eqref{eq:excYreln}. Finally, the combination of the kernels in eqn.(\ref{eq:sTBAQktemp2int}) can be evaluated as follows. Let us consider the corresponding combination of S-matrices
\begin{equation}
\nonumber
\frac{S_{\sls(2)}^{Pk+1}(u,v)}{S_{\sls(2)}^{Pk-1}(u,v)}\Big(S_{xv}^{Pk-1}(u,v)\Big)^{-2}=S_{\mathfrak{sl}(2)}^{P1}\Big(u,v+\tfrac{ ik}{g}\Big)S_{\mathfrak{sl}(2)}^{P1}\Big(u,v-\tfrac{ik}{g}\Big)\left(S^{Pk-1}_{xv}(u,v)\right)^{-2} .\end{equation}
Here we have used the fact that $S^{PQ}_{\sls(2)}$ is obtained by fusing the S-matrices $S_{\sls(2)}^{P1}$. Now
we immediately realize that the first two terms are related by crossing. Using fusion in the first argument
it is not difficult to show that
\begin{align}
&S_{\mathfrak{sl}(2)}^{P1}\Big(u,v+\tfrac{ ik}{g}\Big)S_{\mathfrak{sl}(2)}^{P1}\Big(u,v-\tfrac{ik}{g}\Big)=\\
\nonumber
&\hspace{1cm}=\left(\frac{x_1^+}{x_1^-}\right)^2\Big(\sigma^{P1}\Big(u,v+\tfrac{ ik}{g}\Big)\sigma^{P1}\Big(u,v-\tfrac{ ik}{g}\Big)\Big)^{-2}=
q^{2P}h^{P1}\Big(u,v+\tfrac{ ik}{g}\Big)^{-2}\, ,
\end{align}
where $\sigma^{P1}$ is the dressing factor and we have used the crossing equation describing scattering of a $P$-particle bound state
with the fundamental particle. Here
\begin{equation*}
h^{PQ}(u,v) = \frac{x(u-i P/g) - x(v+i Q/g)}{x(u-i P/g) - x(v-i Q/g)}\frac{1- \frac{1}{x(u+i P/g)x(v+i Q/g)}}{1- \frac{1}{x(u+i P/g)x(v-i Q/g)}}
\prod_{j=1}^{P-1} S_{Q-P+2j}(u-v)\,,
\end{equation*}
as defined in appendix \ref{app:qdefmatrixSmatrix}. Thus, we are left to compute the quantity
\begin{equation}
\label{eq:crossingtoSQMxv}
q^{-P}h^{P1}(u,v+ik/g) S^{Pk-1}_{xv}(u,v) \, .
\end{equation}
To proceed we will denote $x(u\pm iP/g)$ and $x(v\pm i(k-1)/g)$ as $x^\pm_1$ and $x^\pm_2$. Note that $x(v\pm i/g + ik/g) = x(v\mp i(k-1)/g)^{\mp1} = (x_2^\mp)^{\mp1}$. Working out the product (see also eqn. \eqref{eq:SQMxv}), immediately removing the shift on $v$  in $h$ as just indicated, we get
\begin{align}
& q^{-P}h^{P1}(u,v+ik/g) S^{Pk-1}_{xv}(u,v)  = \nonumber \\
& \hspace{40pt} =
\frac{x^-_1 - \tfrac{1}{x^-_2}}{x^-_1 - x^+_2}\frac{1- \frac{x^-_2}{x^+_1}}{1- \frac{1}{x^+_1 x^+_2}} \frac{x^-_1-x^+_2 }{x^+_1-x^+_2 }\frac{x^-_1-x^-_2 }{x^+_1-x^-_2}\frac{x^+_1}{x^-_1} \nonumber \\
& \hspace{60pt} \times \prod_{j=1}^{P-1} S_{1-P+2j}(u,v+i k/g) \prod_{m=1}^{k-2} S_{P+1-k+2m}(u,v) \nonumber\\
& \hspace{40pt} =  \frac{1 - \tfrac{1}{x^-_1x^-_2}}{1- \frac{1}{x^+_1 x^+_2}}\frac{x^-_1-x^-_2 }{x^+_1-x^+_2} \prod_{j=1}^{P-1} S_{1-P+2j}(u,v+i k/g) \prod_{m=1}^{k-2} S_{P+1-k+2m}(u,v) \nonumber \\
& \hspace{40pt} =  -q^{-(P-1)} S_{P+1-k} (u,v)\prod_{j=1}^{P-1} S_{1-P+2j}(u,v+i k/g) \prod_{m=1}^{k-2} S_{P+1-k+2m}(u,v) \nonumber \\
& \hspace{40pt} =  (-1)^{P}q^{-(P-1)}\, .
\end{align}
In the third equality we used the general identity
\begin{equation}
\frac{1 - \tfrac{1}{x^-_i x^-_j}}{1- \frac{1}{x^+_i x^+_j}}\frac{x^-_i - x^-_j}{x^+_i - x^+_j} = q^{-(P+Q)}S_{P-Q} \, ,
\end{equation}
where $P$ and $Q$ refer to the bound state numbers of $i$ and $j$ respectively. In the fourth we used
\begin{align}
& \prod_{j=1}^{M-1} S_{1-P+2j} = S_{P-1}\, , \\
& S_P(u\pm i k/g) = -S_{P-k}(u)\, \end{align}
and
\begin{equation}
\prod_{j=1}^k S_{P+2j} = (-1)^{P+k+1}\, .
\end{equation}
Thus, we have proven that the combination of the S-matrices we are interested in is constant and therefore the corresponding combination of the kernels vanishes.
With this result eqn. \eqref{eq:sTBAQktemp2} simplifies further, now properly deserving the name, giving
\begin{equation}
\log Y_k = \, 2\log{Y_{k-1}}\star s -\log(1+Y_{k-1})\star s + \sum_a \log\left(1+\tfrac{1}{Y_{k-1|vw}^{(a)}}\right)\left(1+\tfrac{e^{-\chi^{(a)}_{vw}}}{Y_{k-1|vw}^{(a)}}\right)\star s \, .
\end{equation}
Note that the contribution of $Y_{k-1|vw}$ has doubled up here, exactly as it did in the equations for $vw$-strings, though admittedly in a less obvious fashion.

\cleardoublepage \addcontentsline{toc}{chapter}{Bibliography}

\chapter*{Bibliography}

\makeatletter
\renewenvironment{thebibliography}[1]
     {%\section*{\bibname}%
      \@mkboth{\MakeUppercase\bibname}{\MakeUppercase\bibname}%
      \list{\@biblabel{\@arabic\c@enumiv}}%
           {\settowidth\labelwidth{\@biblabel{#1}}%
            \leftmargin\labelwidth
            \advance\leftmargin\labelsep
            \@openbib@code
            \usecounter{enumiv}%
            \let\p@enumiv\@empty
            \renewcommand\theenumiv{\@arabic\c@enumiv}}%
      \sloppy
      \clubpenalty4000
      \@clubpenalty \clubpenalty
      \widowpenalty4000%
      \sfcode`\.\@m}
     {\def\@noitemerr
       {\@latex@warning{Empty `thebibliography' environment}}%
      \endlist}
\makeatother

\bibliographystyle{JHEP}

\bibliography{phdrev}

\newpage
\thispagestyle{empty}

\end{document}